\documentclass[12pt]{oxarticle}
\pdfoutput=1

\usepackage[usenames, dvipsnames]{xcolor}
\usepackage{graphicx, float, array, xspace, amscd, amsthm, amssymb, latexsym, longtable, bbm}
\usepackage{fancyhdr, ifthen}
\usepackage[letterpaper, body={6.5in, 9.25in}, top=1.0in, left=1.5in]{geometry}
\usepackage[bbgreekl]{mathbbol}
\usepackage[utf8]{inputenc}
\usepackage[sort&compress, comma, square, numbers]{natbib}
\usepackage[titletoc]{appendix}
\usepackage{cjhebrew}

\DeclareSymbolFontAlphabet{\mathbb}{AMSb}
\DeclareSymbolFontAlphabet{\mathbbl}{bbold}

\newcommand{\Gr}{\color{gray}}

\let\SS=\S 

\renewcommand{\a}{\alpha}
\renewcommand{\b}{\beta}
\newcommand{\g}{\gamma}\newcommand{\G}{\Gamma}
\renewcommand{\d}{\delta}\newcommand{\D}{\Delta}
\newcommand{\e}{\epsilon}\newcommand{\ve}{\varepsilon}
\newcommand{\z}{\zeta}

\renewcommand{\th}{\theta}\newcommand{\Th}{\Theta}\newcommand{\vth}{\vartheta}

\renewcommand{\k}{\kappa}
\renewcommand{\l}{\lambda}\renewcommand{\L}{\Lambda}
\newcommand{\m}{\mu}
\newcommand{\n}{\nu}
\newcommand{\x}{\xi}

\newcommand{\p}{\pi}\renewcommand{\P}{\Pi}\newcommand{\vp}{\varpi}
\renewcommand{\r}{\rho}
\newcommand{\s}{\sigma}\renewcommand{\S}{\Sigma}

\newcommand{\ph}{\phi}\newcommand{\vph}{\varphi}
\newcommand{\ch}{\chi}
\newcommand{\ps}{\psi}
\renewcommand{\o}{\omega}\renewcommand{\O}{\Omega}

\DeclareFontFamily{OT1}{pzc}{}
\DeclareFontShape{OT1}{pzc}{m}{it}{<-> s * [1.200] pzcmi7t}{}
\DeclareMathAlphabet{\mathpzc}{OT1}{pzc}{m}{it}

\newcommand{\cC}{\mathcal{C}}

\newcommand{\cF}{\mathcal{F}}

\newcommand{\cH}{\mathcal{H}}\newcommand{\ccH}{\mathpzc H}

\newcommand{\cL}{\mathcal{L}}

\newcommand{\cO}{\mathcal{O}}
\newcommand{\cP}{\mathcal{P}}

\newcommand{\cR}{\mathcal{R}}
\newcommand{\cS}{\mathcal{S}}

\newcommand{\cU}{\mathcal{U}}\newcommand{\ccU}{\mathpzc U}

\DeclareFontFamily{U}{bbold}{}
\DeclareFontShape{U}{bbold}{m}{n}
 {  <-5.5> s*[1.05] bbold5
    <5.5-6.5> s*[1.05] bbold6
    <6.5-7.5> s*[1.05] bbold7
    <7.5-8.5> s*[1.05] bbold8
    <8.5-9.5> s*[1.05] bbold9
    <9.5-11.5> s*[1.05] bbold10
    <11.5-16> s*[1.05] bbold12
    <16-> s*[1.05] bbold17
 }{}

\newcommand{\IC}{\mathbbl{C}}

\newcommand{\IF}{\mathbbl{F}}

\newcommand{\IP}{\mathbbl{P}}
\newcommand{\IQ}{\mathbbl{Q}}
\newcommand{\IR}{\mathbbl{R}}
\newcommand{\IS}{\mathbbl{S}}

\newcommand{\IZ}{\mathbbl{Z}}

\newcommand{\Ialpha}{\mathbbl{\bbalpha}}

\newcommand{\IDelta}{\mathbbl{\Delta}}

\newcommand{\one}{\mathbbm{1}}

\font\eightrm=cmr8 at 8pt
\font\csc=cmcsc10

\newcommand{\beq}{\begin{equation}}
\newcommand{\eeq}{\end{equation}}
\newcommand{\beqnn}{\begin{equation*}}
\newcommand{\eeqnn}{\end{equation*}}
\newcommand{\bea}{\begin{eqnarray}}
\newcommand{\eea}{\end{eqnarray}}
\newcommand{\bean}{\begin{eqnarray*}}
\newcommand{\eean}{\end{eqnarray*}}

\newcommand{\fref}[1]{Figure~\ref{#1}}
\newcommand{\tref}[1]{Table~\ref{#1}}
\newcommand{\sref}[1]{\SS\ref{#1}}

\newcommand{\pd}[2]{\frac{\partial #1}{\partial #2}}

\newcommand{\norm}[1]{\left\| #1\right\|}
\newcommand{\ee}{\text{e}}
\newcommand{\ii}{\text{i}}
\newcommand{\dd}{\text{d}}

\newcommand{\place}[3]{\vbox to0pt{\kern-\parskip\kern-7pt
                             \kern-#2truein\hbox{\kern#1truein #3}
                             \vss}\nointerlineskip}
\newcommand{\smallfrac}[2]{\frac{\scriptstyle #1}{\scriptstyle #2}}

\DeclareFontFamily{U}{wncy}{}
\DeclareFontShape{U}{wncy}{m}{n}{<->wncyr10}{}
\DeclareSymbolFont{mcy}{U}{wncy}{m}{n}
\DeclareMathSymbol{\sha}{\mathord}{mcy}{"58}

\newcommand{\capt}[3]{\parbox{#1}{\renewcommand{\baselinestretch}{1.0}
                                                           \caption{\label{#2}\small\it #3}}}

\newcommand{\cy}{Calabi-Yau\xspace}

\newcommand{\cys}{Calabi-Yau manifolds\xspace}
\newcommand{\K}{K\"ahler\xspace}
\newcommand{\cicy}[2]{\begin{matrix} #1\end{matrix}\!\left[\begin{matrix}#2 \end{matrix}\right]}

\newcommand{\+}{\phantom{-}}

\newcommand{\tr}{\,\text{Tr}\,}

\newcommand{\Teich}{\text{Teich}}

\renewcommand{\=}{\;=\;}
\newcommand{\hodgenos}{(h^{1,1}, h^{2,1})}

\newcommand{\Rodland}{R{\symbol{'034}}dland\xspace}

\newcommand{\tildeccU}{\hbox{$\hskip5pt\widetilde{\hskip-5pt\ccU}$}}
\newcommand{\hatccU}{\hbox{$\hskip5pt\widehat{\hskip-5pt\ccU}$}}

\newcommand{\Fr}{\hbox{Fr}}
\newcommand{\fr}{\hbox{fr}}
\newcommand{\notdiv}{\not\kern2pt|\kern2pt}
\newcommand{\Gp}{\G_{\!\! p}}

\makeatletter
\g@addto@macro\bfseries{\boldmath}

\def\blindfootnote{\xdef\@thefnmark{}\@footnotetext}
\makeatother

\renewcommand{\baselinestretch}{1.1}
\numberwithin{equation}{section}
\newif\iftables
\begin{document}
\proofmodefalse
\tablestrue
\pagestyle{empty}      

\begin{center}
\null\vskip0.1in
{\Huge Local Zeta Functions\\[0.2in]
From \cy Differential Equations\\[0.4in]}
{\csc Philip Candelas$^{*\,1}$, Xenia de la Ossa$^{*\,2}$\\
and\\
Duco van Straten$^{\dagger \,3}$\\[0.4in]}
\parbox[c]{0.4\textwidth}{\centering
\it $^*$Mathematical Institute\hphantom{$^*$}\\
University of Oxford\\
Andrew Wiles Building\\
Radcliffe Observatory Quarter\\
Oxford, OX2 6GG, UK}
\hspace{1cm}
\parbox[c]{0.4\textwidth}{
\centering 
\it $^\dagger$Fachbereich 08\hphantom{$^\dagger$}\\ 
AG Algebraische Geometrie\\
Johannes Gutenberg-Universit\"at\\
D-55099 Mainz\\ 
Germany}
\blindfootnote{$^1\,$candelas@maths.ox.ac.uk \hfill 
$^2\,$delaossa@maths.ox.ac.uk \hfill
$^3\,$straten@mathematik.uni-mainz.de\kern20pt}
\vfill
{\bf Abstract}
\end{center}
\vskip-7pt
\begin{minipage}{\textwidth}
\baselineskip=14.5pt
\noindent It was observed by Dwork that the $\z$-function of a manifold is closely related to 
and sometimes can be calculated completely in terms of its periods. We report here on a practical 
and computationally rapid implementation of this procedure for families of \cys  with one 
complex structure parameter~$\vph$. Although partly conjectural, it turns out to be possible 
to compute the matrix of the Frobenius map on the third cohomology group of $X_{\vph}$ directly 
from the Picard-Fuchs differential operator of the family. To illustrate our method, we compute 
tables of the quartic numerators of the $\z$-functions for six manifolds of increasing complexity 
as the parameter $\vph$ varies in $\IF_p$. For four of these manifolds, we do this for the $500$ 
primes $p=5,7,\ldots,3583$, while for two manifolds we extend the calculation to 1000 primes. 
The tables for $5\,{\leqslant}\; p \;{\leqslant} \;97$ are part of this article while the remaining tables 
are attached in electronic form. Interest attaches to the cases for which the numerators factorise. Some of these factorisations can be associated with parameter values for which the underlying 
manifold becomes singular. For the cases we consider here, the singularities are all of conifold 
or hyperconifold type. In these cases the numerator degenerates to a cubic and this factorises into 
the product of a linear and a quadratic factor. As has been noted elsewhere, the quadratic term 
contains a coefficient that is the $p$'th coefficient of a  modular form.  Some of our examples 
have singularities when the parameter satisfies a polynomial equation that does not factorise 
over~$\IQ$. When this happens, the corresponding forms are modular forms with neben type or 
Hilbert modular forms. The numerator can also factorise into two quadrics. This happens when 
the Hodge structure of the manifold splits, sometimes this happens for algebraic values of the 
parameter and we identify, in this way, attractor points of rank two of the parameter space.
\vspace*{10pt}
\end{minipage}
%
%
\newpage
{\baselineskip=14pt
\tableofcontents}

\newpage
\setcounter{page}{1}
\pagestyle{fancy}
\setlength{\headheight}{14.5pt}
\renewcommand{\headrulewidth}{0pt}
\fancyfoot{}
\lhead{\ifthenelse{\isodd{\value{page}}}{\thepage}{\ifproofmode\eightrm draft: \today, \hourandminute \else {}\fi}}
\rhead{\ifthenelse{\isodd{\value{page}}}{\ifproofmode\eightrm draft: \today, \hourandminute \else {}\fi}{\thepage}}

\section{Introduction}\label{sec:intro}
\vskip-10pt
\subsection{Preamble}
\vskip-10pt
Consider a \cy threefold $X$, defined over $\IQ$; by this is meant that $X$ is defined by 
polynomials with rational coefficients. By multiplying out denominators, we may take the 
polynomials to have coefficients in $\IZ$. It now makes sense to reduce these coefficients 
modulo a prime $p$ and to consider the manifold over the field $\IF_{p}$. The fundamental 
quantities of interest are the numbers, $N_r$, of points of $X$, with coordinatess in the
fields $\IF_{\! p^r},~r{\=}1,2,\ldots\;{}$ and the zeta-function, which is a generating 
function for these
\beq
N_r \= \#X\!\left(\IF_{\! p^r}\right)~;~~~~\z_X(T) \= 
\exp\left( \sum_{r=1}^\infty \frac{N_r\, T^r}{r}\right)~.
\notag\eeq
In the case that the Picard group is generated by divisors defined over the ground field $\IF_p$, 
the general form of $\z_X$ is dictated by the Weil conjectures
\beq
\z_X \= \frac{R(T)}{(1-T)(1-pT)^{h^{11}}(1-p^2T)^{h^{11}}(1-p^3T)}~. 
\label{eq:zetafunction}\eeq

We will be concerned with one-parameter families of such manifolds, $X_\vph$, and use $\vph$ 
to denote the parameter. Since the manifold depends on $\vph$, so do the $N_r$. However, this 
dependence is generally suppressed in the notation.
 
The function $\z_X$ has remarkable properties in virtue of the Weil conjectures. Among these is the fact that it is a rational function, as a function of $T$, with the degrees of the numerator and denominator governed by the Betti numbers of $X$. Since it is a rational funnction, $\z_X$ may be found from a knowledge of the first few numbers $N_r$, and these may be found, in principle, by counting points, as per the definition. In practice, however, it is frequently difficult to count the points directly, even for low values of $r$. A case when this is so is the frequent case when the manifold is defined as a quotient. It is therefore both a remarkable and useful fact that the $N_r$, and so $\z_X$, can be computed in terms of the periods of $X$. The periods vary with $\vph$ and satisfy certain differential equations, the Picard-Fuchs equations. Thus we come to a fact that is well known although surprising on first acquaintance: that~$\z_X$ may be found by solving differential equations. This observation was exploited to great effect by Dwork in his researches relating to the $\z$-function and the Weil conjectures. 

We will here be concerned with the computation of the numerator $R(T)$. When $X$ is smooth, $R$ 
has the form
\beq
R\= 1 + a\,T + b\,p T^2 + a\, p^3 T^3 + p^6 T^4~,
\notag\eeq
and is determined by two integers $a$ and $b$ that depend on $\vph$.  
This quantity may be calculated as a determinant
\beq
R(\vph,T)\= \det\big( \one - T U(\vph) \big)~~~\text{where}~~~ U(\vph)\= E^{-1}(\vph^p) U(0) E(\vph)\label{eq:DworkForm}\eeq
and $E$ and $U$ are $4\times 4$ matrices, with $E$ a wronskian of periods of $X$ and $U$ the matrix that represents the inverse of the Frobenius map. We will use expansions around a point $0$ of
maximal unipotent monodromy and a priori it is not clear what matrix $U(0)$ to use, as the 
corresponding Calabi-Yau space is very degenerate. We conjecture a precise form for $U(0)$ that
worked in all cases considered.

When $X$ is singular, the degree of $R$ is reduced. For a conifold or hyperconifold singularity, 
for example, $R$ reduces to a cubic and this factorises into a linear factor and a quadric, and 
these have interesting  forms that are dictated by a modular group. We will also be interested in the circumstance that $R$ factorises over $\IZ$ into two quadrics, when such a factorisation is determined by algebraic values of the parameter, independently of the prime $p$.

This paper is largely concerned with the practical evaluation of the matrix $U(\vph)$ and so of $R(\vph,T)$. A first remark is that, while the matrix $E$ has logarithms, these cancel in the process of forming $U$, which can therefore be understood as a matrix of power series in $\vph$ with rational coefficients. We wish to evaluate $U$ for, say, $\vph$ an integer in the range $1{\,\leqslant\,}\vph{\,\leqslant\,}p{-}1$. As series of real numbers, these do not converge. But the series \emph{do} converge, if they are regarded as $p$-adic series. However, although the series converge, in this sense, they do so only slowly. Naive procedures for summing the series typically require summing to $p^6$ terms in order to identify the integers $a$ and $b$ that appear in $R$, and this becomes impractical already for moderate values of $p$. It is a key observation of A.\,Lauder~\cite{lauder2004} that the convergence of the series for $U$ is greatly improved by regarding them as a limit of a sequence of rational functions of the parameter $\vph$. 

In practice this means the following: we fix good prime $p$ and a p-adic accuracy, say $\cO(p^4)$, 
since this order is sufficient to identify the integers $a$ and $b$ for $p\geqslant5$. To indicate 
the process, start by taking $p{\;<\;}100$, say, and expand $U(\vph)$ as a matrix of series to 1000 terms. The series in~$\vph$ have coefficients that are p-adic integers, mod $p^4$, hence also integers mod $p^4$. 

The Picard-Fuchs operator takes the form
\beq
\cL\= S_4\,\vth^4 + S_3\,\vth^3 + S_2\, \vth^2 + S_1\, \vth + S_0~;~~~\vth\=\vph\frac{\dd}{\dd\vph}~.
\label{eq:PFoperator}\eeq
with coefficient functions $S_j$ that are polynomials in $\vph$. The coefficient $S_4$ we call the 
discriminant of the differential equation and plays a special role.  If we multiply the matrix of
power series $U(\vph)$ by successive powers of $S_4(\vph)$, we get a succession of such series.  
However when we multiply by $S_4(\vph)^p$ we find that the series, that consisted previously of 
1000 terms, shorten dramatically to polynomials of degree a small multiple of $p$. In other words 
$U(\vph)$ has the form, mod $p^4$, of a matrix of rational functions of the parameter~$\vph$.
\beq
U(\vph)\= \frac{\ccU(\vph)}{S_4(\vph)^p} + \cO(p^4)~.
\label{eq:WanLauderForm}\eeq 
Given this form, one may immediately evaluate $U(\vph)$ for values of the parameter 
other than the singular values of the differential equation, which are the zeros of the
polynomial $S_4(\vph)$. It is an important observation, however, that the behaviour of the 
series for $U(\vph)$ is even better behaved, in many cases, owing to cancellations between 
the numerator and denominator in the above expression.

A first remark is that there seems to be an important distinction depending on whether or
not the differential equation has \emph{apparent singularities}, which are singularities of
the differential equation that are not singularities of the manifold. Their occurrence is 
quite frequent and, when they occur,  they turn out to play a significant role in the calculation 
of the zeta-function. 
We denote by $\D$ the discriminant of the manifold, so for the case that there is no apparent 
singularity, this can be taken to be equal to the discriminant of the equation,
\beq
S_4(\vph) \= \D(\vph)~.
\notag\eeq
For the cases that there is a generic apparent singularity at $\vph{\=}\vph_0$, we have instead 
the relation
\beq
S_4(\vph) \= (\vph - \vph_0)^2\D(\vph)~.
\notag\eeq
If there are no apparent singularities, then the zeros of $S_4$ correspond to the genuine 
singularities of the manifold, the mirror of the quintic threefold being the prime example.
 For the cases of conifold, or hyperconifold, singularities, which are the most common form 
of singularities, there is complete cancellation of the denominator in \eqref{eq:WanLauderForm},
 at least for the examples considered here. In these cases, we are left with  
\beq
U(\vph)\= \tildeccU(\vph) + \cO(p^4)~,
\label{eq:GenuineSings}\eeq
      where $\tildeccU$ is a matrix of polynomials of degree of order of magnitude $p$. We will be more precise about the degree of $\tildeccU$ in the following. Now we may use this form to evaluate $U(\vph)$ and do this also for values of $\vph$ for which $X$ have hyper conifold singularities. 

If, on the other hand, the differential equation has an apparent singularity $\vph_0$ (it can have more than one apparent singularity, but the examples we study here have at most one) then, surprisingly, there is a cancellation of the true singularities and we are left with
\beq
U(\vph)\= \frac{\hatccU(\vph)}{(\vph-\vph_0)^{2p}}+ \cO(p^4)~,
\label{eq:ApparentSings}\eeq
where $\hatccU$ denotes another matrix of polynomials of degree of order of magnitude $p$. We may use this expression to evaluate $U(\vph)$ for all values of $\vph$, apart from the apparently singular value~$\vph_0$. In particular, this expression allows us to evaluate the factor $R(\vph, T)$ for the values of $\vph$ for which the manifold $X$ is genuinely singular. It has been observed that for conifold or hyperconifold points, which are the singularities that we encounter here, the factor $R$ degenerates to the form
\beq
R \= (1 - p\ch T) (1 - \b_p T + p^3 T^2)~,
\label{eq:Rfacs}\eeq
where $\ch{\=}\pm1$ is a character. Moreover, for the case that the discriminant $\D(\vph)$ factors over~$\IQ$, the quantity $\b_p$ is the p'th coefficient in the $q$-expansion of a weight 4 modular form for a principle congruence subgroup $\G_0(N)\subset\text{SL}(2,\IZ)$, for an integer $N$, whose prime factors are a subset of the bad primes. 

Two of our examples, the mirror of a complete intersection in the Grassmannian $G(2,5)$ and the Pfaffian Calabi-Yau  in $\IP^6$ considered by \Rodland, have discriminants that do not factor over $\IQ$. For the case of the mirror of a hypersurface in $G(2,5)$, the roots of the discriminant are $-\smallfrac{11}{32}{\;\pm\;}\smallfrac{5}{32}\sqrt{5}$ and so only exist in $\IF_p$ if 5 is a square mod $p$. By quadratic reciprocity, this is when  $p{\=}{\pm1}$ mod 5. For these manifolds we observe that, when the discriminant factors in $\IF_p$, the quantity $R$ again factors as in \eqref{eq:Rfacs}. The coefficients $\a_p$, when this happens, have been identified, in the thesis of A.\,Thorne~\cite{ThorneThesis}, with coefficients of a Hilbert modular form. 
      
Another interesting factorisation over the integers arises, for suitable values of $\vph$, in all 
our examples:
\beq
R \= (1 - p\a_p T + p^3 T^2) (1 - \b_p T + p^3 T^2)~.
\label{eq:Rquadfacs}\eeq  
      Note the `extra' factor of  $p$ associated with $\a_p$ in this expression, so the first factor can be considered to be
\beq
1 - \a_p (pT) + p\,(pT)^2
\notag\eeq
      which has the form of the numerator of the $\z$-function of an elliptic curve. 
      
The frequency of such factorisations varies significantly between manifolds, as is evident from Figures~\ref{fig:AESZ34} and \ref{fig:Quintic} which compares these frequencies for the manifold AESZ34 with the mirror quintic. Many of these factorisations occur for unexplained, or perhaps random reasons. However sometimes, and this is the case for AESZ34, they correspond to a \emph{persistent factorization}: one that occurs for $\vph$ the root of a polynomial $G(\vph)$ with integer coefficients. When this happens, it makes sense to reduce $G(\vph)$ mod $p$ and the polynomial will have roots in $\IF_p$ for infinitely many, and in fact a nonzero proportion of all primes. Such values of $\vph$ correspond to a splitting of the Hodge structure of the manifold and the roots of $G(\vph)$ are attractor points of rank two, in the sense of type II supergravity. This phenomenon arises with respect to the manifold AESZ34 and we have examined this in detail in a parallel publication~\cite{Candelas:2019llw}.
\begin{figure}[!t]
\begin{center}
\includegraphics[width=\textwidth]{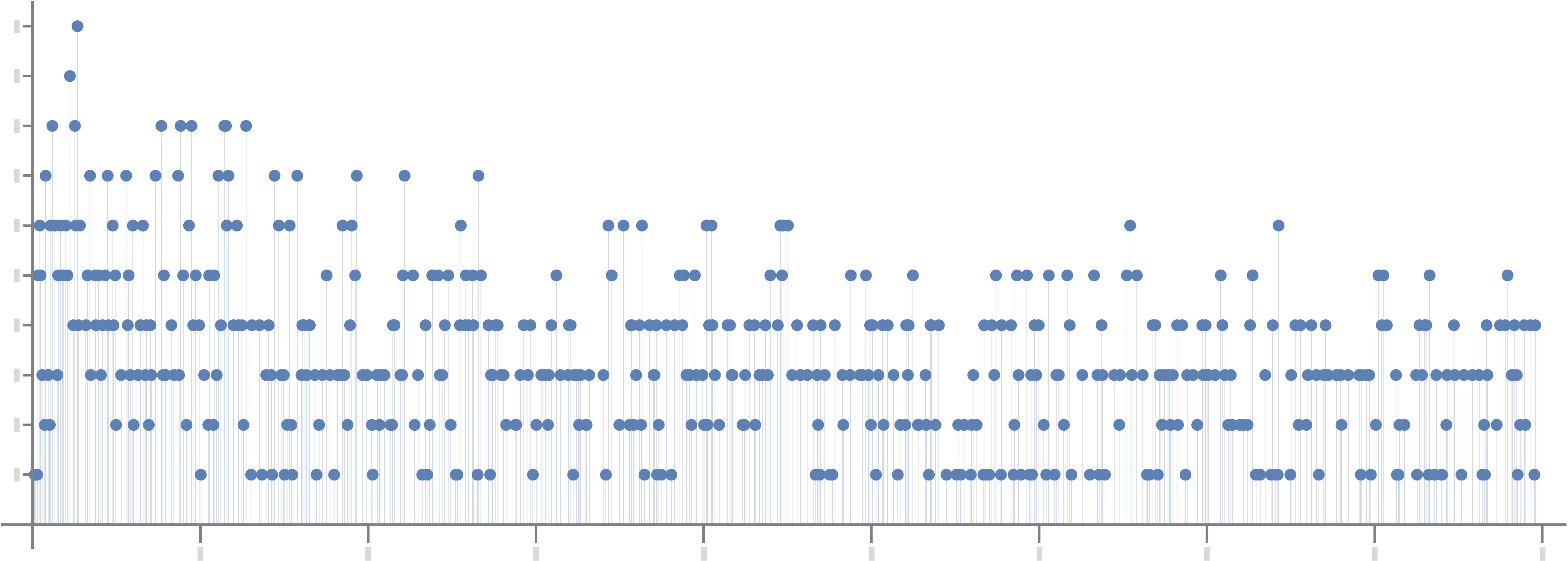}
\vskip0pt 
\place{-0.05}{0.54}{\scriptsize 1}
\place{-0.05}{0.75}{\scriptsize 2}
\place{-0.05}{0.96}{\scriptsize 3}
\place{-0.05}{1.16}{\scriptsize 4}
\place{-0.05}{1.37}{\scriptsize 5}
\place{-0.05}{1.57}{\scriptsize 6}
\place{-0.05}{1.78}{\scriptsize 7}
\place{-0.05}{1.99}{\scriptsize 8}
\place{-0.05}{2.20}{\scriptsize 9}
\place{-0.07}{2.41}{\scriptsize 10}
\place{0.73}{0.15}{\scriptsize 400}
\place{1.43}{0.15}{\scriptsize 800}
\place{2.10}{0.15}{\scriptsize 1200}
\place{2.80}{0.15}{\scriptsize 1600}
\place{3.49}{0.15}{\scriptsize 2000}
\place{4.19}{0.15}{\scriptsize 2400}
\place{4.89}{0.15}{\scriptsize 2800}
\place{5.59}{0.15}{\scriptsize 3200}
\place{6.29}{0.15}{\scriptsize 3600}
\vskip0pt
\includegraphics[width=\textwidth]{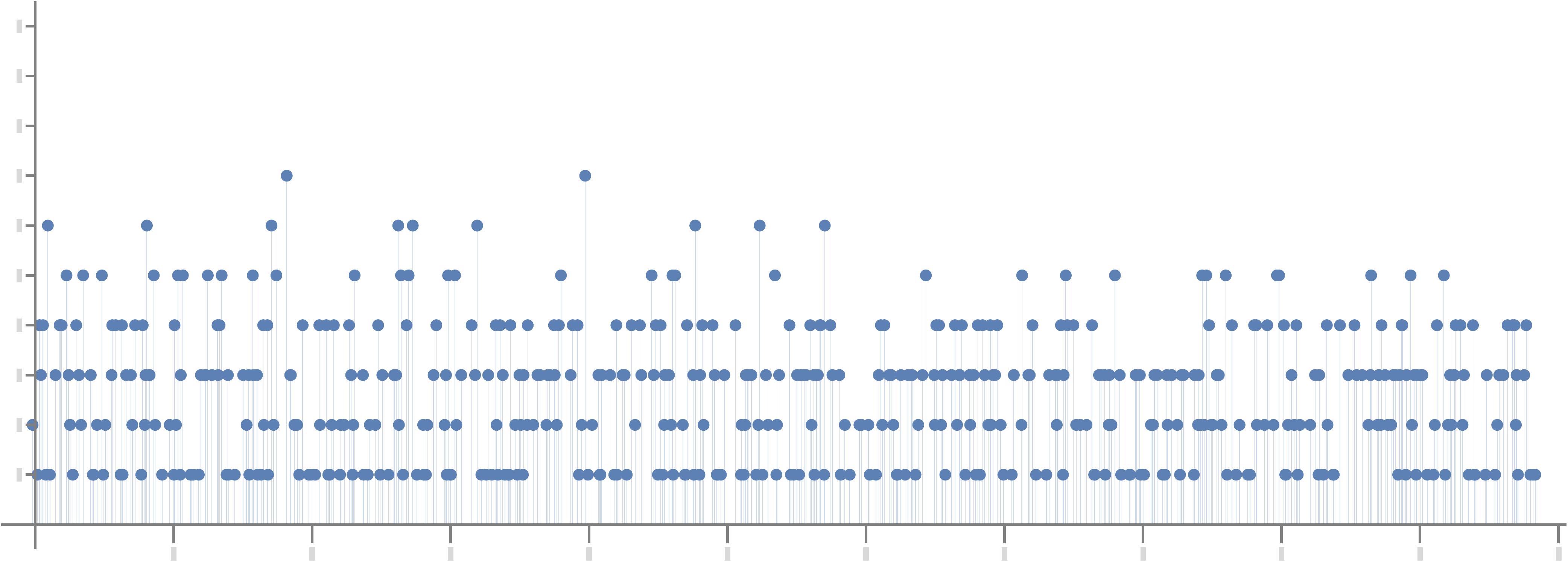}
\vskip0pt 
\place{-0.05}{0.54}{\scriptsize 1}
\place{-0.05}{0.75}{\scriptsize 2}
\place{-0.05}{0.96}{\scriptsize 3}
\place{-0.05}{1.16}{\scriptsize 4}
\place{-0.05}{1.37}{\scriptsize 5}
\place{-0.05}{1.57}{\scriptsize 6}
\place{-0.05}{1.78}{\scriptsize 7}
\place{-0.05}{1.99}{\scriptsize 8}
\place{-0.05}{2.20}{\scriptsize 9}
\place{-0.07}{2.41}{\scriptsize 10}
\place{0.02}{0.15}{\scriptsize 3600}
\place{0.60}{0.15}{\scriptsize 4000}
\place{1.18}{0.15}{\scriptsize 4400}
\place{1.73}{0.15}{\scriptsize 4800}
\place{2.32}{0.15}{\scriptsize 5200}
\place{2.90}{0.15}{\scriptsize 5600}
\place{3.48}{0.15}{\scriptsize 6000}
\place{4.05}{0.15}{\scriptsize 6400}
\place{4.62}{0.15}{\scriptsize 6800}
\place{5.19}{0.15}{\scriptsize 7200}
\place{5.77}{0.15}{\scriptsize 7600}
\place{6.35}{0.15}{\scriptsize 8000}
\vskip-10pt
\capt{6in}{fig:AESZ34}{These plots show the number of factorisations into two quadrics, for AESZ34, as $\vph$ varies over each $\IF_p$, for the 1000 primes $5\leqslant p\leqslant3583$ and $3593\leqslant p\leqslant7933$.}
\end{center}
\end{figure}
\begin{figure}[!t]
\begin{center}
\includegraphics[width=\textwidth]{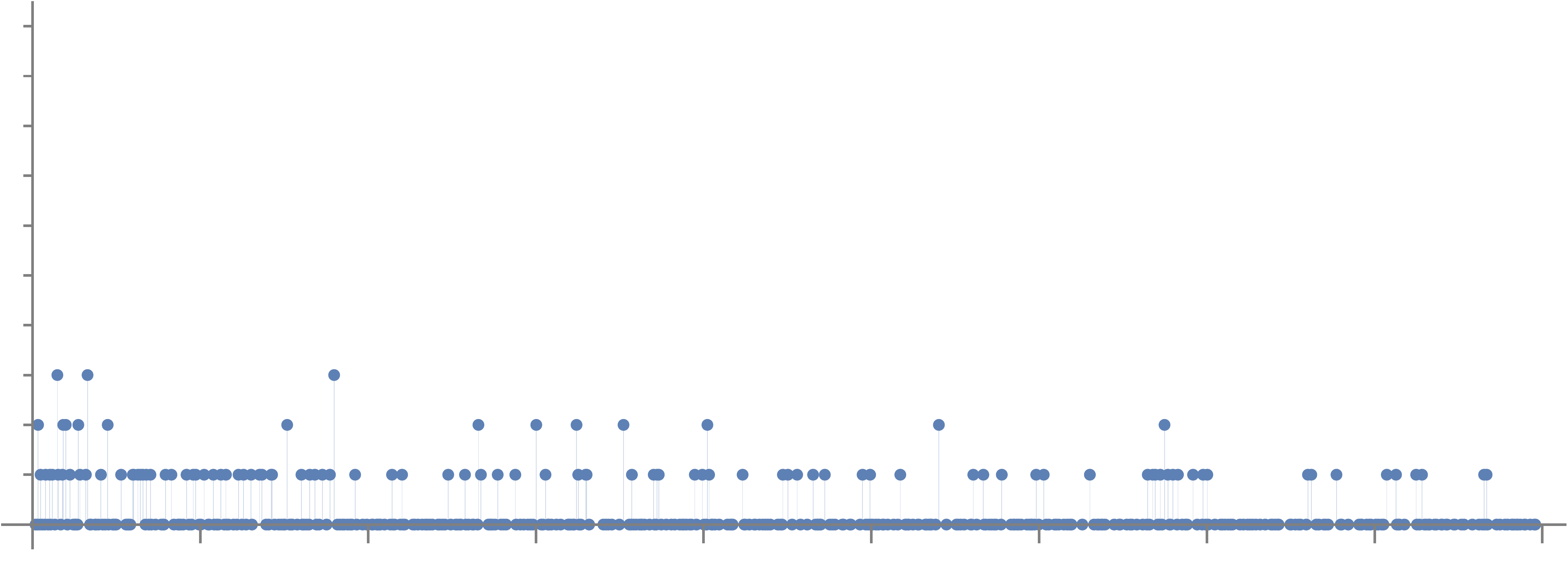}
\vskip0pt 
\place{-0.05}{0.54}{\scriptsize 1}
\place{-0.05}{0.75}{\scriptsize 2}
\place{-0.05}{0.96}{\scriptsize 3}
\place{-0.05}{1.16}{\scriptsize 4}
\place{-0.05}{1.37}{\scriptsize 5}
\place{-0.05}{1.57}{\scriptsize 6}
\place{-0.05}{1.78}{\scriptsize 7}
\place{-0.05}{1.99}{\scriptsize 8}
\place{-0.05}{2.20}{\scriptsize 9}
\place{-0.07}{2.41}{\scriptsize 10}
\place{0.73}{0.15}{\scriptsize 400}
\place{1.43}{0.15}{\scriptsize 800}
\place{2.10}{0.15}{\scriptsize 1200}
\place{2.80}{0.15}{\scriptsize 1600}
\place{3.49}{0.15}{\scriptsize 2000}
\place{4.19}{0.15}{\scriptsize 2400}
\place{4.89}{0.15}{\scriptsize 2800}
\place{5.59}{0.15}{\scriptsize 3200}
\place{6.29}{0.15}{\scriptsize 3600}
\vskip0pt
\includegraphics[width=\textwidth]{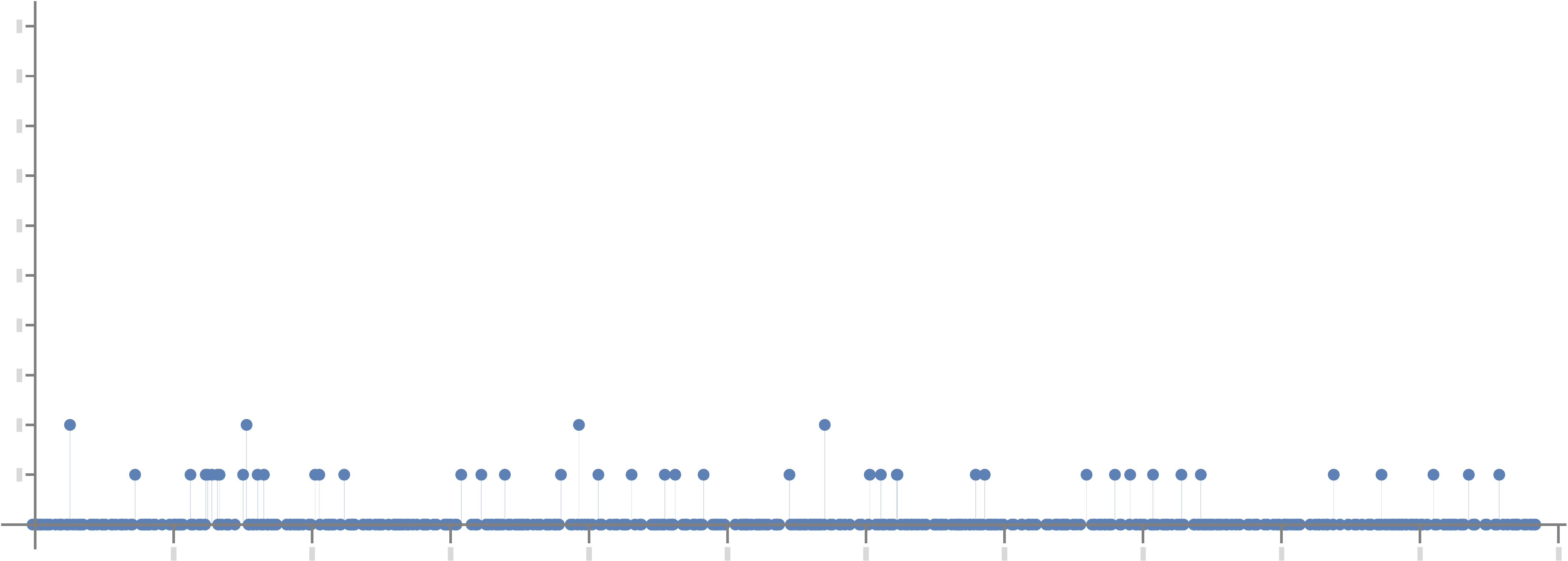}
\vskip0pt 
\place{-0.05}{0.54}{\scriptsize 1}
\place{-0.05}{0.75}{\scriptsize 2}
\place{-0.05}{0.96}{\scriptsize 3}
\place{-0.05}{1.16}{\scriptsize 4}
\place{-0.05}{1.37}{\scriptsize 5}
\place{-0.05}{1.57}{\scriptsize 6}
\place{-0.05}{1.78}{\scriptsize 7}
\place{-0.05}{1.99}{\scriptsize 8}
\place{-0.05}{2.20}{\scriptsize 9}
\place{-0.07}{2.41}{\scriptsize 10}
\place{0.02}{0.15}{\scriptsize 3600}
\place{0.60}{0.15}{\scriptsize 4000}
\place{1.18}{0.15}{\scriptsize 4400}
\place{1.73}{0.15}{\scriptsize 4800}
\place{2.32}{0.15}{\scriptsize 5200}
\place{2.90}{0.15}{\scriptsize 5600}
\place{3.48}{0.15}{\scriptsize 6000}
\place{4.05}{0.15}{\scriptsize 6400}
\place{4.62}{0.15}{\scriptsize 6800}
\place{5.19}{0.15}{\scriptsize 7200}
\place{5.77}{0.15}{\scriptsize 7600}
\place{6.35}{0.15}{\scriptsize 8000}
\vskip-10pt
\capt{6in}{fig:Quintic}{These plots shows the number of factorisations into two quadrics, for the mirror quintic, as $\vph$ varies over each $\IF_p$, for the 999 primes $7\leqslant p\leqslant3583$ and $3593\leqslant p \leqslant 7793$.}
\end{center}
\end{figure}
\subsection{Utility of the tables}
\vskip-10pt
There is interesting information in the tables of Appendix~\ref{sec:tables}, that are the principal output of our calculations. Each of the examples that follow have conifold, or hyperconifold, singularities for certain values of the parameters, that are defined by algebraic equations. For these values of the parameters, the polynomial $R(\vph, T)$ becomes a cubic and factors as in Eq~\eqref{eq:Rfacs}. The coefficients $\b_p$ can be read off from the tables and these identify a weight-four modular form. This is, as we have indicated, a well-known story, but our tables provide many examples of this modular behaviour and the method can generate many more.

Plots such as those of \fref{fig:AESZ34} and \fref{fig:Quintic} are a crude summary of aspects of the full tables, yet even here there is additional information about the possible existence of rank two attractor points. If we assume that, for a given manifold, there are finitely many rank two attractor points and that these correspond to values of the parameter that are the roots of a polynomial $G(\vph)$, with rational, and so integer, coefficients. The polynomial $G(\vph)$ may be reducible over $\IQ$, if so, let us suppose that it has $k$ irreducible factors.
      The following result, which is a known, but perhaps not widely known, consequence of the Cebotar\"{e}v density theorem is relevant. A proof is reviewed in~Appendix~\ref{sec:Chebotarev}. 
      
{\bf Theorem:} \emph{Let $f(\vph)$ be a polynomial with integer coefficients, that is irreducible over $\IQ$, and let $\n_p$ be the number of roots of $f(\vph)$ over $\IF_{\! p}$. Further let $S$ be a large set of primes. Then, subject to some mild assumptions about the set\footnote{For example, $S$ could be the set of all $p{\,\leqslant\,}p_\text{max}$.} $S$,
\beq
\lim_{|S|\to\infty} \frac{1}{|S|} \sum_{p\in S} \n_p \= 1~.
\notag\eeq}

Applying this theorem to each of the irreducible factors of $G$, we see that the expected number of roots of $G$ over $\IF_{\! p}$ is $k$. If now $n_p$ is the number of factorisations over $\IF_p$ then $n_p\geqslant\n_p$ and~so
\beq
\lim_{|S|\to\infty} \frac{1}{|S|} \sum_{p\in S} n_p\; \geqslant \; k~.
\notag\eeq
If the inequality remains valid for $S$ large, but not infinite, then, since the average is easily computed from the data of plots such as those of Figures\,\ref{fig:AESZ34} and \ref{fig:Quintic}, we obtain bounds on $k$. 

For the mirror quintic we see from the second plot, even by eye, that the average of $n_p$ is less than one. If we compute the average for primes in bins of 125 primes we find averages that are given in \fref{fig:BarChartQ}. The case is strong that $k{\=}0$, and that there are no rank two attractor points for the mirror~quintic.

For the manifold AESZ34 the situation, with regard to the existence of rank two attractor points, is more interesting. We plot the running averages of the $n_p$, for the 1000 primes $5\leqslant p\leqslant 7933$ in \fref{fig:BarChartAESZ34}. The data for 500 primes corresponds to the first four bars, which might have been consistent with $k{\=}3$, but by including the extra primes it is compelling that in fact $k\leqslant 2$.
      In \cite{Candelas:2019llw} it is shown that $G$ has at least two irreducible factors, so there is a compelling statistical case that $k{\=}2$ for this manifold.
\begin{figure}[!t] 
\begin{center}
\includegraphics[width=3.5in]{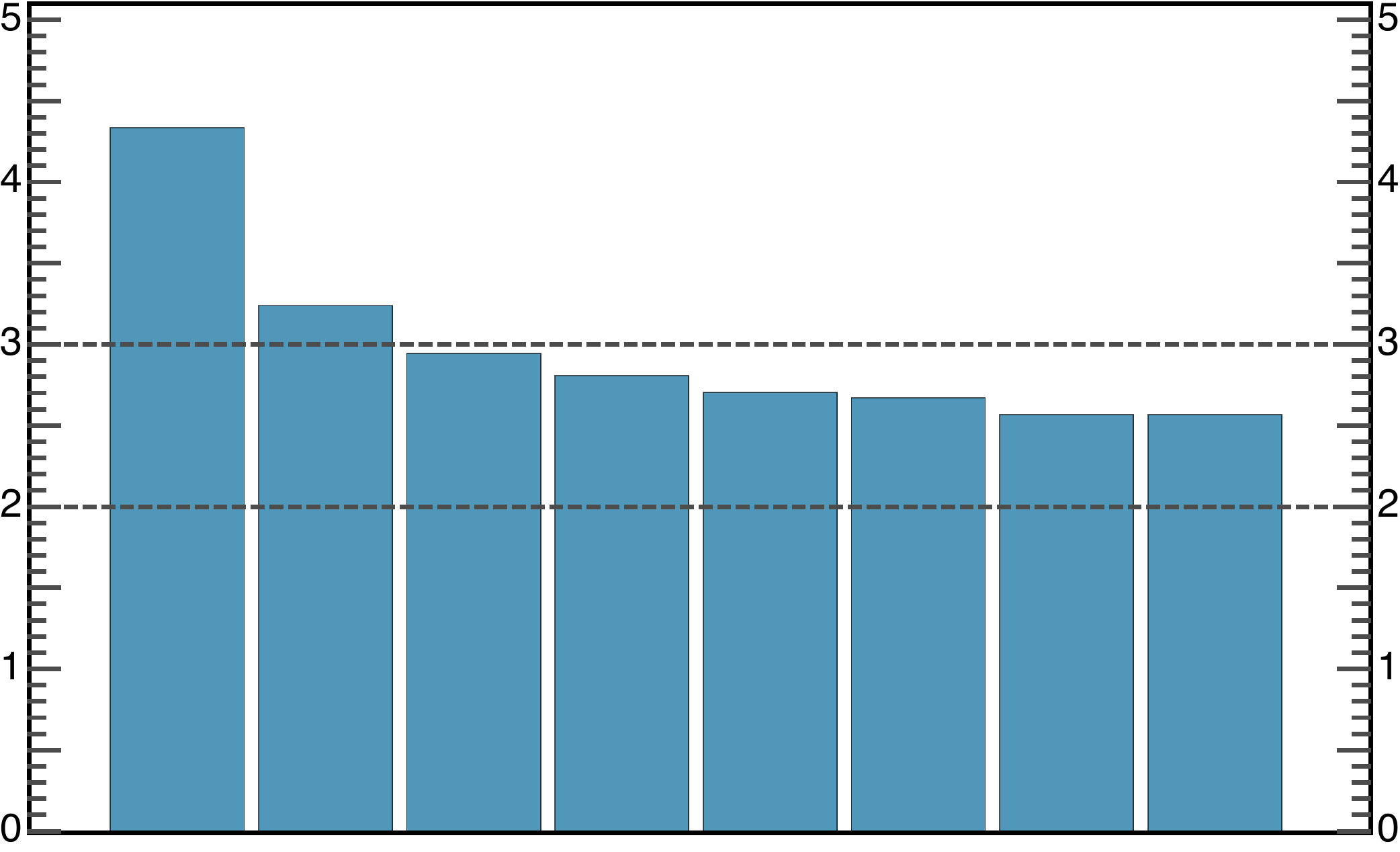}
\capt{5.3in}{fig:BarChartAESZ34}{Running averages for the data of AESZ34 from \fref{fig:AESZ34}. The averages are taken for bins of 125 primes for the 1000 primes $5\leqslant p\leqslant 7933$.}
\end{center}
\end{figure}
\begin{figure}[!t] 
\begin{center}
\includegraphics[width=3.65in]{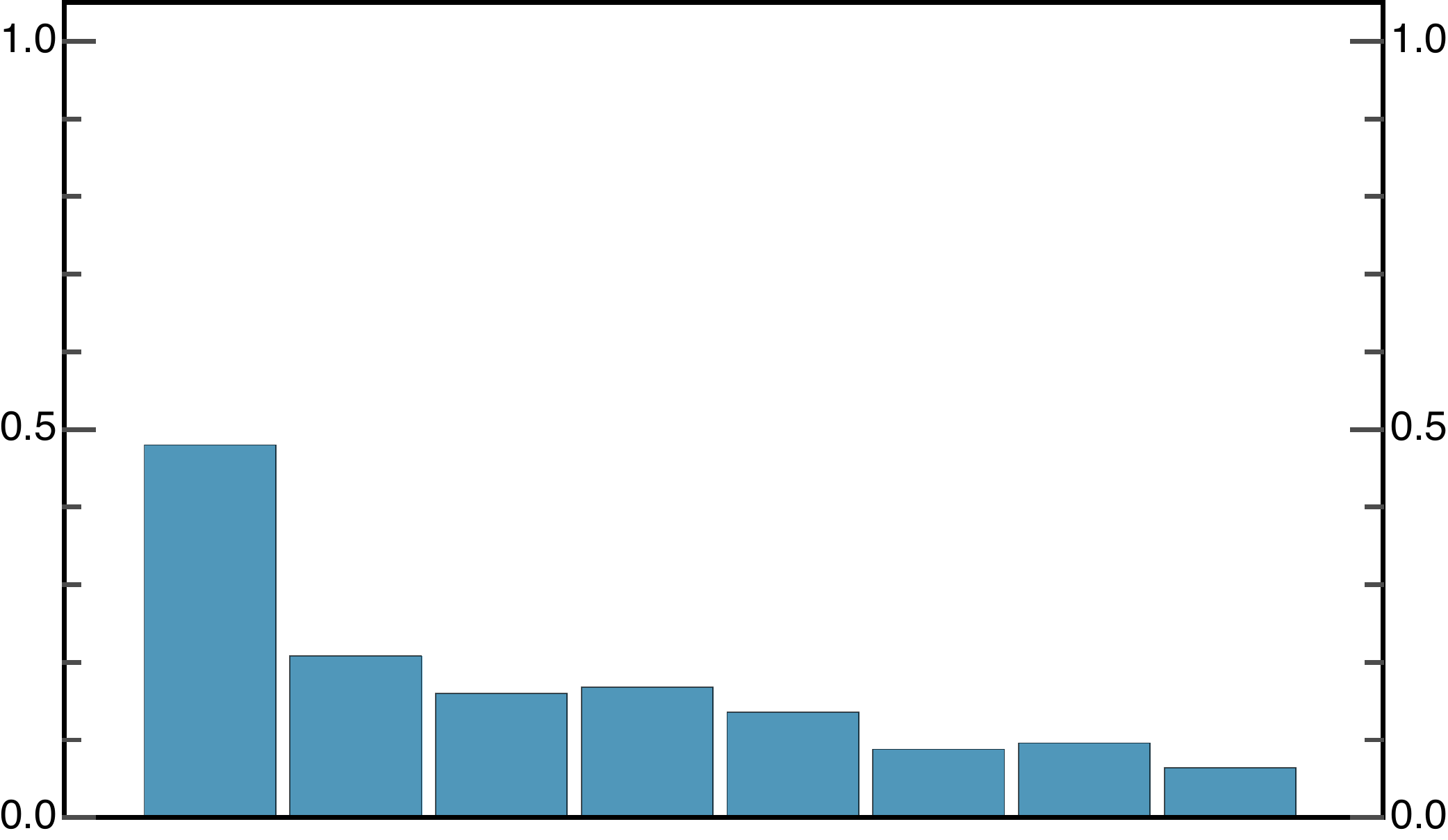}
\capt{5.3in}{fig:BarChartQ}{Running averages for the mirror quintic data of \fref{fig:Quintic}. The averages are taken for bins of 125 primes for the 999 primes $7\leqslant p\leqslant 7933$. Note that the vertical scale on this bar chart is very different from that of \fref{fig:BarChartAESZ34}. }
\end{center}
\end{figure}
\subsection{Layout of the article}
\vskip-10pt
The principal concern of this article is not directly with the utility of the $\z$-function so much as with how to compute it as a practical matter. We compute this function from the periods and these are determined by the Picard-Fuchs operator. We therefore start in \sref{sec:basic} with a discussion of the method of Frobenius and a basis of the periods. From the periods, we can calculate the matrix $E_i{}^j$, that appears in \eqref{eq:DworkForm}. We discuss here also certain number-theoretic aspects of the Picard-Fuchs equation.
     The equation has a discriminant that vanishes for parameter values for which the equation is singular. It can happen that two singular values $\vph_i$ and $\vph_j$, that are generically distinct, coincide mod $p$ for certain primes $p$ and this confluence of singularities affects the form of the $\z$-function. The study of this phenomenon devolves upon the study of the discriminant of the discriminant of the Picard-Fuchs equation, as a polynomial, and we call this the \emph{hyperdiscriminant}. This has, generically, simple zeros, however, owing to the confluence of roots for certain bad primes, the hyperdiscriminant can develop repeated roots when considered mod $p$.
     
The observation of Dwork that the Frobenius polynomial $R$ can be calculated in terms of a matrix $U(\vph)$ that corresponds to the inverse of the Frobenius map as in \eqref{eq:DworkForm} is is central to our calculation. The computation of the Frobenius matrix $F(\vph){\=}U^{-1}(\vph)$ proceeds in two stages. First, the matrix is calculated for a particular value $\vph_0$ of $\vph$ and then $F(\vph)$ is extended to a region containing $\vph_0$ by means of the relation 
\beq
F(\vph) \= E^{-1}(\vph) F(\vph_0) E(\vph^p)~,
\notag\eeq
      where $E(\vph)$ is a matrix constructed from the periods of $X$. We examine this process for the case that 
      $\vph_0$ is a value of the parameter for which $X$ is smooth in \sref{sec:Frobenius1} and for the case that 
      $\vph_0{\=}0$ is the point of maximal unipotent monodromy, for which $X$ is highly singular, but is the case we actually use in calculation, in \sref{sec:Frobenius2}. The Frobenius matrix has an analogue for the infinite prime and this is the matrix of complex conjugation $C(\vph)$.  We compute this matrix and see that it is indeed closely analogous to $F(\vph)$.
      
In \sref{sec:Frobenius2} we calculate the Frobenius matrix taking the point of maximal unipotent monodromy as the point about which we expand. Although clearly inspired by the procedure for the case that $\vph_0$ is a point for which $X$ is smooth, expanding about $\vph{\=}0$ is nevertheless a significant adaptation.
      We discuss in this section also the form of the central matrix $U(0)$. A notable fact is that while the periods contain logarithms, as can be seen in \eqref{eq:periods}, nevertheless these logarithms cancel from the matrix $U(\vph)$, so this is a matrix of power series. We find also that the $U(0)$ that we conjecture, is given by a $p$-adic gamma class. So the conjectured form, which is supported by the calculations that produce our zeta-function tables, amounts to a $p$-adic gamma-class conjecture. We compute also the charge conjugation matrix and the closely related matrix corresponding to the natural hermitian form on $X$ and observe that these are closely analogous to $U(0)$.
      
In \sref{sec:ComputingU} we consider the series expansion of $U(\vph)$. These series converge, but do so only slowly. Summing the series is rendered practical by the observation of Lauder~\cite{lauder2004} that, after fixing a $p$-adic accuracy, the series can be resummed to rational functions, thereby permitting their rapid evaluation.  We review also, in this section, the `unit root method'. It was observed by Dwork that one of the roots of the Frobenius polynomial is a $p$-adic unit and this root can, moreover, be computed as a ratio of partial sums of the fundamental period. A knowledge of one root  can be used to fix the Frobenius polynomial, when this is irreducible, and to fix one of the quadratic factors, when the Frobenius polynomial factorises. We discuss the utility and the limitations of this process. Finally, in this section, we discuss the bounds on the $a,\,b$ coefficients required by the Weil conjectures. These coefficients have a natural parametrisation that we also discuss.

In \sref{sec:manifolds} we give a telegraphic account of the manifolds that are our examples. These are all taken from the AESZ list and are chosen to exhibit increasing complication, by which we mean that the respective Picard-Fuchs equations become more complicated. These operators all have the form \eqref{eq:PFoperator} but the coefficient polynomials $S_j$ become of increasingly higher degree.  The first manifold is the mirror quintic, which is AESZ1 and has coefficient polynomials that are linear in the parameter $\vph$, while our last two examples have coefficient polynomials that are of degree 8 and 7, respectively. As the degree of the $S_j$ increases, two phenomena begin to arise. The first is that the discriminants $S_4$ have increasing numbers of roots, so the corresponding manifolds have increasing numbers of singularities, though, it turns out that not all the singularities of the differential operator correspond to singularities of the manifold. The singularities of the operator, that are not singularities of the manifold, are termed apparent singularities. These apparent singularities play an important, though still somewhat mysterious, role with respect to the computation of the rational functions of the parameter that lead to the identification of the Frobenius polynomial.

Our first example is, as we have observed, the mirror quintic. This must be the most studied of all \cys. We use this manifold to set out the procedure that we will apply to all our examples. We turn next to a manifold defined by the complete intersection of three polynomials of degrees $(1,2,2)$ in the Grassmanian $G(2,5)$. This space has a Picard Fuchs equation with coefficient polynomials of degree 2. There is no apparent singularity, but the discriminant does not factor over $\IQ$. The roots of the discriminant exist in $\IF_{\! p}$ only when 5 is a square mod $p$. A consequence of this is that while the conifold singularities of the manifold exhibit modular behaviour, the modular forms are not classical modular forms for $\G_0(N)$, for some $N$, but rather modular forms with nebentype and Hilbert modular forms. 

The third example is the Hulek-Verrill manifold. The coefficient polynomials for the corresponding Picard-Fuchs operator are of degree three. A striking feature of this example is the remarkable frequency with which the Frobenius polynomial factorises into two quadrics as in \eqref{eq:Rquadfacs}. This can be appreciated from Figures \ref{fig:AESZ34} and \ref{fig:Quintic}, the latter being much closer to the general case. 

The fourth example is the \Rodland manifold. This is a manifold whose definition does not involve toric geometry, and which has many remarkable properties. Among these is the fact that the moduli space of this manifold contains two large complex structure points and so can be said to correspond to two different mirror manifolds. The coefficient polynomials of the Picard-Fuchs equation are of degree five and the discriminant has the form
\beq
S_4\= (\vph - 3)^2 \left(\vph^3-289 \vph^2-57 \vph +1\right)~.
\notag\eeq
This exhibits two complications: the factor $(\vph {-} 3)^2$ corresponds to an apparent singularity, the moduli space is smooth at $\vph{\=}3$ and the monodromy about this point is trivial; and the second factor of the discriminant does not factor over $\IQ$. The roots of this cubic factor are conifold points and, at these points the Frobenius polynomial factors in the form \eqref{eq:Rfacs} with a quadratic factor that we believe to be modular, in the sense of corresponding to a Hilbert modular form, though we have not yet identified the modular form.

Our last two examples are somewhat similar. The first of these is a manifold which arises as the manifold of a three generation heterotic model of string theory. For this case the coefficient polynomials have degree eight,
      and $S_4(\vph){\=}(\vph{-}3/2)^2 \D(\vph)$, with $\vph{\=}3/2$ corresponding to an apparent singularity. The parameter space for this manifold has six conifold~points.
      
The last example concerns the remarkable manifolds, discovered by V.~Braun~\cite{Braun:2011hd}, that have Hodge numbers $h^{1,1}{\=}h^{2,1}{\=}1$. In many ways this example is similar to the previous one.
      The coefficient polynomials are, in this case, of degree seven and there is also an apparent singularity. The parameter space corresponding to this manifold also has six conifold points. 
      
In \sref{sec:slopes} we return to \eqref{eq:GenuineSings} and \eqref{eq:ApparentSings} and enquire as to the degree of the polynomials in the matrices $\tildeccU$ and $\hatccU$, for our examples. We find that these degrees are determined in a simple manner by the indices corresponding to $\infty$ that appear in the Riemann symbols for the differential equations.

In \sref{sec:HigherOrder} we indicate briefly how the process of expressing $U(\vph)$ as a rational function is modified if a higher p-adic accuracy is specified.

Two appendices deal with ancillary matters: in Appendix~\ref{sec:Chebotarev} we indicate a proof of a corollary to the Frobenius-Cebotar\"{e}v theorem and in Appendix~\ref{sec:AppendixZeta} we give a telegraphic review of the theory of the $p$-adic $\G$ and $\z$-functions.
\newpage
\section{The Picard-Fuchs Equation}\label{sec:basic}
\vskip-10pt
\subsection{The large p-adic structure expansion}\label{sec:PFeqAndLargeStructure}
\vskip-10pt
The Picard-Fuchs operators listed in the AESZ database all take the general form \eqref{eq:PFoperator} with coefficient functions $S_k$ that are polynomials in $\vph$ with integer coefficients. Our examples differ in the precise form of the coefficient polynomials $S_k$ and, in particular, in the degrees of these polynomials. All the operators have a point of large p-adic structure by which we mean a point of maximal unipotent monodromy. The parameter $\vph$ is chosen such that this point corresponds to $\vph{\=}0$. 

We take a Frobenius solution of the form
\beq
\vp(\vph,\e)\= \sum_{n=0}^\infty A_n(\e)\vph^{n+\e}~;~~~A_0(\e)\= 1~.
\label{eq:FrobPer}\eeq
It is a consequence of $\vph{\=}0$ being a point of maximum unipotent monodromy that the indicial equation is $\e^4{\=}0$. 
Periods $\vp_k(\vph)$ are obtained by expanding the Frobenius period in powers of $\e$
\beq
\vp(\vph,\e)\= \sum_{k=0}^3 \frac{\e^k}{k!}\,\vp_k(\vph)~,
\notag\eeq
with the periods $\vp_k$, $k=0,\ldots,3$, themselves given by expressions of the form
\beq\begin{split}
\vp_0(\vph)&\= f_0(\vph) \\
\vp_1(\vph)&\= f_0(\vph)\log\vph + f_1(\vph) \\
\vp_2(\vph)&\= f_0(\vph)\log^2\!\vph + 2f_1(\vph)\log\vph + f_2(\vph) \\
\vp_3(\vph)&\= f_0(\vph)\log^3\!\vph+ 3f_1(\vph)\log^2\!\vph+ 3f_2(\vph)\log\vph + f_3(\vph)~, \\
\end{split}\label{eq:periods}\eeq
where the $f_j(\vph)$ are regular series in $\vph$.
A consequence of taking the coefficient $A_0(\e) {\,=\,}1$ is that $f_0(0){\=}1$ and $f_j(0){\=}0$ for $j{\=}1,2,3$, and this fixes this basis uniquely. We will term this basis of periods the \emph{arithmetic Frobenius basis}. There are other bases of periods that are useful and commonly used, so we pause to list these and indicate the relations between them. These are mostly elementary, but it is worth being explicit about the relations between different sets of conventions. 

First we give two bases that are very simple variations of the arithmetic Frobenius basis.

\begin{itemize}
\item
\emph{The complex Frobenius basis:}
\beq
\widehat{\vp}_j \= \frac{\vp_j}{(2\p\ii)^j}
\notag\eeq
This is used in relating the arithmetic basis with the integral basis.

\item
A widely used basis that, for want of a better name, we can term \emph{the sophisticated basis:}
\beq
\vp^\sharp_j \= \frac{1}{j!}\vp_j~.
\notag\eeq

\item
\emph{The integral basis:} a particular choice within $H^3(X,\IZ)$ that is adapted to mirror symmetry and the large complex structure limit, see, for example \cite{Candelas:2019llw} and references cited therein. 
      Briefly put: a symplectic basis $\{\a_a,\b^b\}$, $a,b{\=}1,2$, is chosen for $H^3(X,\IZ)$ and the corresponding periods are defined by writing the holomorphic three form in terms of this basis
\beq
\O\= z^a\a_a - \cF_b(z) \b^b
\notag\eeq
and it can be shown that there is a function $\cF(z)$, called the prepotential, such that $\cF_b{\=}\partial \cF/\partial z^b$.

In a matrix notation, in which the four integral periods are gathered into a vector
\beq
\P\= \begin{pmatrix} \cF_a \\[3pt] z^b \end{pmatrix}
\notag\eeq 
and the four $\vp_j$ are denoted by a vector $\vp$, we have
\beq
\P\= \widehat{\r}\,\widehat{\vp}~,
\label{eq:rhohat}\eeq
with
\beq
\widehat{\rho}\= 
\left( \begin{array}{ccrr@{\hskip13pt}}
- \frac{1}{3} Y_{000} & -\frac{1}{2} Y_{001} & 0 &\+{\hskip13pt}\frac{1}{6} Y_{111}{\hskip-13pt}\\[3pt]
 -\frac{1}{2} Y_{001} & - \hphantom{\frac{1}{2}}Y_{011} & -\frac{1}{2} Y_{111}{\hskip-13pt} & 0\\[3pt]
 1 & 0 & 0 & 0 \\[3pt]
 0 & 1 & 0 & 0
\end{array} \right) ~.
\notag\eeq
The quantities $Y_{abc}$ are constants that appear in the prepotential, which has an expansion
\beq
\cF\= -\frac{1}{3!}\frac{Y_{abc}z^a z^b z^c}{z^0} + \ldots
\notag\eeq
where the terms indicated by the ellipsis are of order $\cO(\ee^{2\p\ii z^1/z^0})$.

By choice of symplectic basis, the quantities $Y_{abc}$ are related to invariants of the mirror manifold~$\widetilde{X}$. It is believed that here is a choice of basis such that
\beq\begin{split}
Y_{111}&\= \int_{\widetilde{X}}\! e^3 \\[4pt]
Y_{011}&\;\in\;\left\{0,\smallfrac12\right\}\\[4pt]
Y_{001}&\;= -\frac{1}{12}\int_{\widetilde{X}}\! c_2\, e\\[4pt]
Y_{000}&\;= -3\frac{\z(3)}{(2\p\ii)^3}\,\chi\big(\widetilde{X}\big)
\end{split}\notag\eeq
where $e$ is the generator of $H^2\big(\widetilde{X},\IZ\big)$. It is perhaps intuitive that the coefficients $Y_{011}$ should be given by the integral of $c_1 e^2$ and so vanish. However, this is not quite true. Rather $Y_{011}$ can, by choice of basis, be made to take either the value 0 or $\frac12$. For the case of one parameter, the rule is simple and depends on whether $Y_{111}$ is even or odd. If $Y_{111}$ is even, then $Y_{011}$ can be taken to vanish, and if $Y_{111}$ is odd, it can be taken to~be~$\frac12$. 

\item
Note that the matrix $\widehat\r$ has rational elements, apart from the element $Y_{000}$. We can remove this element by a further change of basis. This brings us to \emph{the modified complex Frobenius basis}
      or, in the slightly shorter form, the {\it modified complex basis}. This basis differs from $\widehat{\vp}_j$ only when $j{\=}3$
\beq
\widetilde{\widehat\vp}_j \= 
\begin{cases} 
\widehat{\vp}_j~, &\text{for}~j=0,1,2~,\\[5pt]
\widehat{\vp}_3 - 2\frac{\raisebox{2pt}{$Y_{000}$}}{\raisebox{-3pt}{$Y_{111}$}}\, \widehat{\vp}_0 ~,&\text{for}~ j=3~.
\end{cases}
\notag\eeq
This basis is related to the integral basis $\P$ by a matrix $\tilde{\hat\r}$ 
\beq
\P \= \widetilde{\widehat\r}\,\widetilde{\widehat\vp}
\notag\eeq
with
\beq
\widetilde{\widehat\rho}\= \left( \begin{array}{ccrr@{\hskip13pt}}
         0                       & -\frac{1}{2} Y_{001} & 0 &\+{\hskip13pt}\frac{1}{6} Y_{111}{\hskip-13pt}\\[3pt]
 -\frac{1}{2} Y_{001} & - \hphantom{\frac{1}{2}}Y_{011} & -\frac{1}{2} Y_{111}{\hskip-13pt} & 0\\[3pt]
 1 & 0 & 0 & 0 \\[3pt]
 0 & 1 & 0 & 0
\end{array} \right)~. 
\notag\eeq

\item
Finally, we will require also a {\it modified arithmetic Frobenius basis}, or for short a {\it modified arithmetic basis}, which we shall denote by $\widetilde{\vp}_j$ with
\beq
\widetilde{\vp}_j \=
\begin{cases}
\vp_j~, &\text{for}~j=0,1,2~,\\[5pt]
\vp_3 + 6 \z(3) 
\frac{\raisebox{3pt}{$\displaystyle \chi(\widetilde X)$}}{\raisebox{-3pt}{$\displaystyle y$}}~,
&\text{for}~ j=3~.
\end{cases}
\notag\eeq
where here and in the following we write $y$ for the topological Yukawa coupling $Y_{111}$.
\end{itemize}
\subsection{Discriminants and hyperdiscriminants}\label{sec:discriminants}
\vskip-10pt
Let us denote the roots of $\D$ by $\vph_i$, $i{\=}1,\ldots,k$, say. For the cases considered here, these are simple zeros. It is important to observe that there are bad primes for the manifold and for  the Picard-Fuchs operator, owing to confluence of the roots $\vph_i$. This arises because, for some $p$, two roots $\vph_i$ and $\vph_j$, that are generically distinct, will coincide mod $p$. Also, if a root $\vph_i$ is rational, and $p$ divides the numerator of $\vph_i$, then $\vph_i$ coincides, mod $p$, with the point of maximal unipotent monodromy, $\vph{\=}0$. Furthermore, if $p$ divides the denominator of a root, the root moves to $\infty$ where it may undergo confluence with other roots, or with a preexisting singularity. It seems best to take the parameter space of $\vph$ mod $p$ to be the projective space $\IP\IF_{\! p}$, so $\infty$ is an allowed value.

Thus we will be interested in the discriminant of $\D$ as a polynomial. For a polynomial
\beq
h_n\vph^n + h_{n-1}\vph^{n-1}+\ldots + h_0~,
\notag\eeq
      with roots $\vph_i$ this is understood to be
\beq
h_n^{2n-2}\prod_{i<j}(\vph_i - \vph_j)^2~,
\label{eq:IDelta}\eeq
      which vanishes whenever two roots coincide. The overall constant is conventional and, with this choice, the discriminant of a quadratic equation is the familiar $h_1^2{\,-\,}4h_0h_2$. Since we are now dealing with the discriminant of the discriminant, we shall refer to this quantity as the \emph{hyperdiscriminant of the manifold}. We will abuse notation by redefining $\D$, and related quantities in relation to thinking of $\vph{\=}(u {\,:\,} v)$  in terms of projective coordinates. We now understand $\D$ as the homogeneous polynomial
\beq
\D\= h_n u^n + h_{n-1} u^{n-1}v+\ldots + h_0 v^n~,
\notag\eeq   
      of course, setting $u{\=}\vph$ and $v{\=}1$ returns us to the previous definition. The new definition however has the advantage that the value of the hyperdiscriminant is invariant under the action of the M\"{o}bius group on the coordinates $(u {\,:\,} v)$. We will also understand by \emph{hyperdiscriminant}  the discriminant of the new definition of $\D$ and we will denote this quantity~by~$\IDelta$.
     
For the case that there is an apparent singularity, it is of interest to know when $\vph_0$ coincides, mod $p$, with one of the genuine singularities. To this end we consider the homogenised discriminant of $(\vph-\vph_0)\D$, note that the first factor is linear, so this is not the same as $S_4{\=}(\vph-\vph_0)^2\D$. We denote this quantity by $\IS_4$
\beq
\IS_4\= \text{discr}\Big( (\vph-\vph_0)\D \Big)
\notag\eeq 
and refer to it as the \emph{hyperdiscriminant of the Picard-Fuchs equation}. This hyperdiscriminant corresponds to a product analogous to \eqref{eq:IDelta}, but where the index $i$ also runs over the value~0.
      
The quantities $\IDelta$ and $\IS_4$ do not reveal the whole story of the bad primes of the differential equation since it can also happen that a prime can be bad for all parameter values, we will refer to such primes as \emph{notorious primes}. The mirror quintic furnishes an example of this phenomenon. For this case the discriminant is linear
\beq
\D\= v - 5^5 u
\notag\eeq
      and $\IDelta{\=}1$, independent of $\vph{\=}(u{\,:\,}v)$. Nevertheless we want to take $5$ to be a bad prime for the following reason. Take the defining equation of the mirror quintic to be $F{\=}0$, with
\beq
F(x)\= \sum_{i=1}^5 x_i^5 - \tilde{\psi}\, x_1 x_2 x_3 x_4 x_5~.
\notag\eeq 
Then the five derivatives
\beq
\pd{F}{x_i} \= 5x_i^4 - \tilde{\psi}\prod_{j\neq i} x_j
\notag\eeq 
all vanish, mod 5, at points where any two coordinates vanish.  So the manifold is singular mod 5, irrespective of the value of the parameter. 

It is notable and indicative of the fact that the Picard-Fuchs equations have an important number-theoretic significance that, in the examples we study here, no new bad primes are introduced by passing from $\IDelta$ to $\IS_4$.
\newpage
\section{The Frobenius and Complex Conjugation Matrices I}\label{sec:Frobenius1}
\vskip-10pt
In this section we set up the basic formalism related to Dwork's deformation
method. We examine the formulation of the Frobenius map and the complex conjugation matrix for parameter values for which the variety $X_{\vph}$ is smooth. The Frobenius map is an automorphism of every manifold defined over $\IQ_{p}$.
  For the infinite prime, so for manifolds defined over $\IR$, there is an analogous map which is complex conjugation.  For this reason it is appropriate to discuss these maps together.   The discussion proceeds by picking a point $\vph=\vph_{0}$  for which $X_{\vph_{0}}$ is smooth and then studying the maps for $\vph$ in a neighbourhood of $\vph_{0}$. 
 
  In the following section we extend the technique to  parameter values of $\vph$ in the neighbourhood of the 
  point of maximal unipotent monodromy (MUM) $\vph = 0$, for which the variety is very singular.  It is the technique developed in that section the one used to compute the data for the $\z$-function.

\subsection{Complex conjugation and the Frobenius map}
\vskip-10pt
The field $\IQ$ of rational numbers carries, for each $r\in\IQ$ and prime number 
$p$, a $p$-adic absolute value $\norm{r}_p$ as well as the usual absolute 
value $|r|{\=}\norm{r}_\infty$, which is sometimes said to belong to the \emph{infinite
prime}. The completions with respect to these absolute values give the
field $\IQ_p$ of $p$-adic numbers and the field $\IR$ of real numbers. 
For each $r \in \IQ$ the {\em product formula}
\[ 
\norm{r}_{\infty}\hskip-5pt\prod_{p\;\textup{prime}}\hskip-5pt \norm{r}_p \= 1
\]
holds, which underlines that all primes, including the infinite prime,
have an equal footing.
The absolute values $\norm{*}_p$ is {\em non-archimedean}
\[ 
\norm{x+y}_p \leqslant \max \big( \norm{x}_p,\,\norm{y}_p \big)~.
\]
As a result, the set of $x \in \IQ_p$ for which $\norm{x}_p \leqslant 1$ is closed under 
multiplication and addition and forms the ring $\IZ_p$ of $p$-adic
integers and computing mod $p$ in $\IZ_p$ brings us to the finite field $\IF_p$.
The analogous set $|x| \leqslant 1$, in $\IR$, is closed under multiplication, but not under addition, so there is no complete analog of the ring $\IZ_p$ for the infinite prime.

Consider now an algebraic variety $X$. For sake of
concreteness, let us suppose it is defined by a polynomial 
$g(x){\,:=\,}g(x_1,x_2,\ldots,x_n)$ with coefficients in $\IZ$. We then can look for 
the solution set $X(\cR)$ to the equation
\beq
g(x_1,x_2,\ldots,x_n)\=0~, \label{eq:giszero}
\eeq
where all the coordinates $x_i$ belong to a ring $\cR$. The equation \eqref{eq:giszero} may not admit any real solutions, but there is always a non-empty manifold of complex solutions $X(\IC)$. As the coefficients of $g$ are in $\IZ$, so in particular real, one has
\[
\overline{g(x)} \= g (\overline{x})~,
\]
so that {\em complex conjugation} defines a map
\[ 
\fr_{\infty}:~X(\IC) \to X(\IC)~;\quad x \mapsto \overline{x}
\]
and the fixed point set is nothing but $X(\IR)$. We can try to
do something similar for each prime~$p$. We can look for solutions
to $\eqref{eq:giszero}$ in $\IF_p$, the field with $p$ elements, or in extension fields
$\IF_{\! p^k}$, or even its algebraic closure ${\IF}_{\! p}^{\text{alg}}$. By Fermat's little theorem, we have 
\[ 
g(x)^p \= g(x^p) \mod p~,
\]
so that taking the $p$'th power of the coordinates defines the {\em Frobenius map}
\[ 
\fr_p: X\big({\IF}_{\! p}^{\text{alg}}\big) \to  X\big({\IF}_{\! p}^{\text{alg}}\big)~;\quad x \mapsto x^p ~.
\]
The fixed points of this map $\fr_p$ form precisely the set $X(\IF_{\! p})$, and 
similarly, the fixed point set of $(\fr_p)^k$ is $X(\IF_{\! p^k})$.

The maps $\fr_{\infty}$ and $\fr_p$ act on the cohomology groups of the corresponding 
varieties. For the complex manifold $X(\IC)$ one can consider singular 
cohomology or the deRham cohomology with real coefficients. As the precise
choice is of no importance here, we will just denote them by $H^k\big(X(\IC)\big)$
and get linear maps induced by complex conjugation:
\[
\Fr_{\infty}:~H^k\big(X(\IC)\big) \to H^k\big(X(\IC)\big)~.
\]
To do do something similar for $X({\IF}_{\! p}^{\text{alg}})$, one has to invent 
a good cohomology theory for varieties over a finite field. Dwork \cite{Dwork1964}, Washnitzer-Monsky (see \cite{VanDerPut1986}), Grothendieck, Berthelot (see \cite{Kedlaya2006}) and others constructed $p$-adic cohomology spaces, which are finite dimensional vector spaces over $\IQ_p$. 
Details will not be important here, but all constructions involve {\em lifting} the variety over $\IF_p$ to $\IZ_p$ and considering the result over the field $\IQ_p$ and taking the deRham complex of this lift, the cohomology of which is essentially independent of the choices of the lifts that are made.  We will denote the cohomology groups simply by $H^k\big(X({\IF}_{\! p}^{\text{alg}})\big)$. 
The action of Frobenius can be lifted and induces Frobenius maps
acting on the $p$-adic cohomology
\[
\Fr_p :~H^k\!\left(X\big(\IF_{\! p}^\text{alg}\big)\right) \to H^k\!\left(X\big(\IF_{\! p}^\text{alg}\big)\right)~.
\]

If $\psi:X \to X$ is a self-map of a topological space $X$, then the 
{\em Lefschetz trace formula} expresses the alternating sum of cohomological
traces 
\[
\sum_{k=0}^{2d} (-1)^k \tr\big(\psi | H^k(X)\big)
\]
in terms of the fixed point set of $\psi$. Weil famously imagined this to be applicable 
to the Frobenius map acting on appropriate cohomology groups, which were
constructed subsequently, leading to the relation
\[ 
\sum_{k=0}^{2d} (-1)^k \tr\big(\Fr_{\! p^n}| H^k(X)\big) \= \#\big(X(\IF_{\! p^n})\big)~.   
\]
It was shown by Dwork that this also holds for his p-adic cohomology theory.
By standard linear algebra, this leads to the representation of the zeta-function in cohomological terms
\[
\exp\left(\sum_{r=1}^{\infty} \#X(\IF_{\! p^r})\,\frac{T^r}{r}\right) \= \frac{R_1(T)R_{3}(T)\ldots R_{2d-1}(T)}{R_0(T)R_{2}(T)\ldots R_{2d}(T)} 
\]
where
\[ 
R_k(T):=\; \det\left(1 - T\,\Fr_p^{-1}|H^k(X)\right)~.
\]

We refer to \cite{Monsky1970} for a nice introduction, \cite{Candelas:2007mb} for
a physics oriented account, \cite{Kedlaya2008} for a computationally orientied
introduction and \cite{LeStum2007RigidCohomology,Kedlaya2010} for more details.  In the situation we consider in this paper $d=3$, and
and the Betti numbers $b^{1}$ and $b^{5}$ vanish.  Therefore, the factors $R_{1}$  and $R_{5}$ are trivial and we are left with $R_{3}$ and we henceforth omit the index. If the variety is smooth the form of the $\z$-function is  given by equation \eqref{eq:zetafunction}
whenever the Picard group is generated by divisors which are defined over $\IF_{\! p}$. In the contrary case that the Picard group is not generated by divisors defined over $\IF_{\! p}$, but rather over an extension $\IF_{\! p^k}$, then the factor $(1-pT)^{h^{11}}(1-p^2T)^{h^{11}}$ in the denominator of the $\z$-function assumes the more general form
\beq
(1-pT)^{h^{11}-m_p} (1+pT)^{m_p}\, (1-p^2T)^{h^{11}-m_p} (1+p^2T)^{m_p} ~,
\notag\eeq
where $m_p$ is an integer, $0\leqslant m_p\leqslant h^{11}$ that can depend on $p$.
\subsection{The deformation method}
\vskip-10pt
We will also consider families of varieties $X_\vph$, parametrised by
$\vph$, defined by a polynomial equation 
\beq 
g(x,\vph)\= 0~, 
\notag\eeq
where are assume, as before, that the coefficients of $g$ are in $\IZ$.
Note that
\[ 
\overline{g(x,\vph)} \= g(\overline{x},\overline{\vph})~,
\]
and this gives rise to a map
\[ 
\fr_\infty(\vph):~X_\vph(\IC) \to X_{\overline{\vph}}(\IC)~,
\]
which is a self-map only for real $\vph$. 

Analogously, since
\[ 
g(x,\vph)^p \= g(x^p,\vph^p) \mod p~,
\]
we obtain maps
\[  
\fr_p(\vph):~X_\vph\big({\IF}_{\! p}^{\text{alg}}\big) \to  X_{\vph^p}\big({\IF}_{\! p}^{\text{alg}}\big)~,
\]
which is a self-map only for $\vph$ in the ground field $\IF_{\! p}$.
We assume the manifolds $X_{\vph}$ are smooth for $\vph \in S$, where $S$ is
$\IP^1$, minus a finite set of singular values. Then the cohomology groups 
$H^n\big(X_\vph(\IC)\big)$ form a vector bundle $\ccH$ on $S$, such that
\[ 
\ccH_\vph^{\, n} \= H^n\big(X_\vph(\IC)\big)~.
\]
It carries a canonical {\em Gauss-Manin} connection: for each vector 
field $\vth$ on the base we obtain a covariant derivative in the direction 
$\vth$, giving a map
\[ 
\nabla_{\! \vth}:~\ccH \to \ccH~. 
\]
In the following, we will always take the logarithmic derivative
\[ 
\vth \= \vph \frac{\partial}{\partial \vph}
\]
as vector field and write simply $\nabla$ for $\nabla_{\vth}$.

If we fix a frame $e^k$ for $\ccH$, the information of the connection is given
by a connection matrix $B{\=}(B_j{}^k)$
\[
\nabla e^k \= e^j B_j{}^k~.
\]
The reason for the `transposed' convention associated with the connection matrix $B$ is that, if we think of the periods $\vp_j$ as forming
a column vector, then a relation such as the following, that we shall come across in the next subsection, 
\beq
\O \= \vp^\sharp_j\,f^j~,
\notag\eeq
with $\O$ the holomorphic three-form, the $\vp^\sharp_j$ the `sophisticated' periods from \sref{sec:PFeqAndLargeStructure} and the $f^j$ a certain cohomology basis, shows that the $f^j$ should be thought of as forming a row vector. 
The entries of the matrix $B$ are rational functions of $\vph$, with rational
coefficients, that have poles in the complement of $S$. The matrix $B$ 
can be computed effectively, \cite{Dwork1964,Griffiths1969Periods,Movasati2007,lauder2004,Lairez2016},
in fact we shall do this presently.

A similar construction can be done in $p$-adic cohomology for a prime of good reduction.
This then leads to the `same' matrix $B$, only now the 
coefficients are interpreted as belonging to $\IQ_p$. In this situation, it was 
realised by Dwork \cite{Dwork1962ICM} that the Frobenius transformation 
$\Fr(\vph):\cH_\vph \to \cH_{\vph^p}$ is {\em compatible with the Gauss-Manin 
connection}. The map $\Fr{\=}\Fr(\vph)$ has the fundamental commutation relation
\beq 
\nabla\, \Fr \= p\,\Fr\,\nabla ~.
\label{eq:FundamentalEquation}\eeq
In this way we end up with a mathematical structure that is commonly referred 
to as a {\em crystal}~\cite{NKatz1973}; it consists of a free module $\ccH$ over 
a certain $p$-adic ring  $\cR \subset \IZ_p[\hskip-1.5pt[\vph]\hskip-1.5pt]$, together with two 
operations $\nabla:\ccH \to \ccH$ and $\Fr:\ccH\to \ccH$ that satisfy \eqref{eq:FundamentalEquation}.
For the coefficient ring $\cR$ one has two similar operations $\cR{\,\to\,}\cR$
\beq
\vth:\, c(\vph) \to \vph\frac{\partial c(\vph)}{\partial \vph} \quad \text{and} \quad
\fr:\, c(\vph) \to c(\vph^p) 
\notag\eeq
that satisfy the similar relation
\[ 
\vth\, \fr \= p\, \fr\, \vth~.
\]
The operators $\nabla$ and $\Fr$ are compatible with these: $\nabla$ satifies 
the appropriate Leibniz rule, whereas $\Fr$ is Frobenius-linear. For example
\beq
\Fr\big( c(\vph) e^k \big) \= c(\vph^p)\, \Fr(e^k)~.
\notag\eeq

If we write out the action of $\Fr$ on the frame $e^k$ as
\[ 
\Fr(\vph)(e^k) \= e^j F_j{}^k(\vph)
\]
and consider, say, the quantity $\nabla\Fr\, (e^{k})$ and use the identity \eqref{eq:FundamentalEquation}
we find that
\beq
\vth(F(\vph))\= p F(\vph) B(\vph^p) - B(\vph)F(\vph)~, 
\label{eq:FrobDiffEq}\eeq
which can be interpreted as a differential equation for $F(\vph)$. 
The upshot is that if we know the Frobenius matrix $F(\vph)$ at some point  $\vph_0 \in S$, then we can, in principle, use the 
above differential equation to determine $F$ at other points. 
This is the core of the  {\em deformation method}, which was worked out 
by Dwork in special cases. In \cite{Dwork1964}, Dwork took initial 
points~$\vph_0$ corresponding to Fermat-varieties, for which the Frobenius 
matrix can be written down in  closed form, as was first done by Weil~\cite{Weil1949NumbersOfSolutions}.

It is easy to solve the differential equation \eqref{eq:FrobDiffEq} 
in terms of  the {\em fundamental solution} $E(\vph)$ for the matrix 
differential equation defined by $B$. The matrix $E$ is the unique matrix series solution to the equation
\[ 
\vth E(\vph)\= E(\vph)\,B(\vph)
\]
with the property that
\[ 
E(\vph_0)\= \one~.
\]
Then the series
\beq
F(\vph) \= E(\vph)^{-1} F(\vph_0) E (\vph^p)
\label{eq:Frelation}\eeq
solves \eqref{eq:FrobDiffEq}. Indeed, 
\[
\begin{split}
\vth\big(F(\vph)\big)
&\= -E(\vph)^{-1} E(\vph) B(\vph) E(\vph)^{-1} F(\vph_0) E(\vph^p) + p E(\vph)^{-1} F(\vph_0) E(\vph^p) B(\vph^p) \\[5pt]
&\= -B(\vph)F(\vph)+pF(\vph)B(\vph^p)~,
\end{split}
\]
which is again \eqref{eq:FrobDiffEq}.

The crystals arising from the cohomology of smooth $d$-dimensional
projective varieties have also additional structure. There is the 
Poincar\'e intersection pairing $(\ast,\ast): H^d \times H^d \to H^{2d}$, symmetric
for $d$ even and alternating for $d$ odd. The connection
$\nabla$ and Frobenius map $\Fr$ are also compatible with respect to this
pairing. In fact one has 
\begin{subequations}\label{eq:Crystal} 
\begin{align}
\vth (\x, \eta) &\=(\nabla \x,\eta) + (\x,\nabla \eta) \\[5pt]
(\Fr\,\x, \Fr\,\eta) &\= p^d \,\Fr \bigl(( \x,\eta )\bigr)~.
\end{align}
Furthermore, one can introduce a {\em Hodge-filtration}
\beq
\text{Fil}^d \subset \text{Fil}^{d-1} \subset \ldots \subset \text{Fil}^0 = \ccH\,,
\quad\text{with}~~
\nabla(\text{Fil}^i) \subset \text{Fil}^{i-1}~\text{and}~~ \Fr(\text{Fil}^i) \subset\, p^i \ccH~.
\eeq
\end{subequations}
Augmenting the crystal structure by requiring also the compatibility (\ref{eq:Crystal}\,a) and (\ref{eq:Crystal}\,b) yields a structure that may be called a {\em polarised F-crystal}, and
if the filtration is also included, then we have a {\em polarised divisible Hodge 
F-crystal} also known as a {\it Fontaine-Lafaille crystal}, we will call it a {\em CY-crystal} 
for short. 

Let us now compute the matrix of complex conjugation, which we will denote by $C(\vph)$, for a basis $e^{k}$ which is assumed to be holomorphic.  This matrix is defined by
\beq
\overline{e^k} \= e^j C_j{}^k(\vph)~.
\notag\eeq
Note that the matrix $C$ cannot be holomorphic. 
A first relation follows from the identity $\nabla\,\overline{e^k}{\=}0 $. On writing out the derivative, we have
\beq
\nabla\,\overline{e^k} \= \nabla \left(e^j C_j{}^k \right) \= e^j\left( B_j{}^i C_i{}^k + \vth C_j{}^k\right)~.
\notag\eeq
So
\beq
\vth C \;= -BC~.
\notag\eeq
The antiholomorphic derivative of $C$ is determined by the identity 
$\overline{\strut\nabla e^k}{\=}\overline{\strut\nabla}\,\overline{\strut e^k}$, which yields the relation
\beq
\overline{\vth}C \= C \overline{B}~.
\notag\eeq
These equations are solved by writing
\beq
C(\vph) \= E^{-1}(\vph) C(\vph_0) E(\overline{\vph})
\label{eq:Crelation}\eeq
and the condition $C(\vph)\overline{C(\vph)}{\=}\one$ is satisfied for all $\vph$, if it is satisfied for $\vph{\=}\vph_0$.
The complete analogy between \eqref{eq:Crelation} and \eqref{eq:Frelation} is apparent.
\newpage
\section{The Frobenius and Complex Conjugation Matrices II}\label{sec:Frobenius2}
\vskip-10pt
\subsection{Structure near a MUM-point}
\vskip-10pt

It is of great interest to adapt the discussion of the previous section to deform about singular points. 
We will now study these structures for a family of Calabi-Yau threefolds near a MUM-point, that is
a point of {\em maximal unipotent monodromy}, also known, depending on context, as a large complex
structure limit, or a large $p$-adic structure limit. We will always assume that $h^{2,1}{\=}1$, so that the cohomology group 
$H^3$ is four-dimensional.   
  
Our aim is to obtain the power-series expansions for the inverse matrix of the Frobenius action on $H^3(X_{\vph})$ and the 
charge conjugation matrix about this point, something that was already explored by Dwork~\cite{Dwork1969} and Lauder \cite{Lauder2011Degenerations} for the Frobenius matrix. 

We follow Dwork and Lauder in conjecturing that this matrix can be written in an analogous form to that of the previous section
\beq 
U(\vph) \= E^{-1}(\vph^p) U(0) E(\vph)~.
\label{eq:UnearMUMpoint}\eeq

The MUM-point $\vph{\=}0$ corresponds, however, to a variety $X_0$ that is highly singular. Owing to the singularity the matrix $E(\vph)$ contains logarithms and is itself singular at $\vph{\=}0$. So while $U(\vph)$ can be written in the form above for some matrix $U(0)$ as a consequence of the differential equation \eqref{eq:FrobDiffEq} it is not clear, a priori, that $U(\vph)$ has a finite limit as $\vph\to 0$ and that $U(\vph)$ tends to the matrix that we have denoted by $U(0)$. However, if we take $U(0)$ to be consistent with the CY-crystal structure then,  as we will see, the logarithms that are present in $E(\vph)$ cancel and $U(\vph)$ has an expansion about $\vph{\=}0$ as a matrix of power series and so tends to a well defined matrix $U(0)$ as $\vph{\,\to\,}0$. We are led to further conjecture the precise form of the matrix $U(0)$ in terms of the $p$-adic $\G$-function. In fact it is compelling to conjecture that the Frobenius matrix is governed by a $p$-adic version of the $\G$-conjectures. We set this out in \sref{conjecturesU(0)}.

Consider the sophisticated basis of periods $\vp_j^\sharp$ defined in \sref{sec:PFeqAndLargeStructure}  and its Wronskian matrix $E$
\beq
\vp_j^\sharp \= \frac{1}{j!}\,\vp_j~,\qquad E_j{}^k \= \vth^k \vp_j^\sharp~.
\notag\eeq
The PF equation in first order form is
\beq
\vth E \= E B \quad \text{with} \quad 
B\= \begin{pmatrix}
0&0&0&\hskip-3pt-S_0/S_4\\
1&0&0&\hskip-3pt-S_1/S_4\\
0&1&0&\hskip-3pt-S_2/S_4\\
0&0&1&\hskip-3pt-S_3/S_4
\end{pmatrix}~.
\label{eq:PFinFirstOrderForm}\eeq
We will show presently that these matrices $E(\vph)$ and $B(\vph)$ thus defined, are the appropriate analogues to the matrices $E(\vph)$ and $B(\vph)$ of the previous section. As has been noted,  owing to the presence of logarithms, $E(\vph)$ is singular at $\vph{\=}0$. 

Let $\O$ be the holomorphic three-form and define a basis in cohomology $f^j$ by 
\beq
\O\= \vp_j^\sharp\,f^j~.
\notag\eeq
Note that the $f^j$ are related to the integral basis by a constant matrix, which we will calculate later explicitly,  so  $\nabla f^j{\=}0$.
Let us define also a matrix $F(0)$ by
\beq
\Fr\,f^k \= f^j F_j{}^k(0)~.
\notag\eeq

Now take a moving cohomology basis
\beq
e^k \= \nabla^k \O \= f^j\, E_j{}^k~.
\notag\eeq
It would be better to take a basis that transforms covariantly under gauge transformations and so adapt crystalline cohomology to special geometry~\cite{Strominger:1990pd,candelas1991moduli}, but we leave aside this interesting topic, to which we hope to return elsewhere.

To continue, we define also a matrix $F(\vph)$ by
\beq
\Fr\,e^k \= e^j F_j{}^k(\vph)~.
\notag\eeq

By the previous relation, we have
\beq
\Fr \, e^k 
\= f^i F_i{}^j(0)\, E_j{}^k(\vph^p)\\
\= e^\ell  E^{-1}(\vph)_\ell{}^i \, F_i{}^j(0)\, E_j{}^k(\vph^p)~.
\notag\eeq
Thus
\beq
F(\vph)\= E^{-1}(\vph)\, F(0)\, E(\vph^p) \quad \text{and} \quad
U(\vph)\= F^{-1}(\vph)\=  E^{-1}(\vph^p)\, U(0)\, E(\vph)~.
\notag\eeq
It remains to show that $F(\vph)$, as just defined, tends to $F(0)$ as $\vph{\,\to\,}0$. This is true, but not obvious a priori owing to the fact that the $E$-matrices contain logarithms of $\vph$ and are singular as $\vph{\,\to\,}0$. We will show presently that the logarithms cancel in the product and that $F(\vph)$ is indeed continuous at $\vph{\=}0$.

The calculation of the charge conjugation matrix is somewhat more delicate than that for the Frobenius matrix, so is deferred until after we discuss the form of the wronskian matrix.
\subsection{The form of the matrix $E$}
\vskip-10pt
Consider the matrix $E^0$, corresponding to the leading terms of $E$ as $\vph{\,\to\,}0$
\beq
E^0{}_j{}^k\= \frac{1}{j!}\vth^k\log^j\!\vph\=
\begin{pmatrix}
1 & 0 & 0 & ~~~0 \\[3pt]
\hphantom{\frac12}\log\vph & 1 & 0 & ~~~0 \\[3pt]
\frac{1}{2}\log^2\!\vph &\hphantom{\frac12}\log\vph & 1 & ~~~0 \\[3pt]
\frac{1}{6}\log^3\!\vph & \frac{1}{2}\log^2\!\vph & \log\vph & ~~~~1~\\
\end{pmatrix}~.
\notag\eeq
and in terms of this define
\beq
\widetilde{E}\= \big(E^0\big)^{-1}E~.
\notag\eeq
The prefactor $\big(E^0\big)^{-1}$ removes the logarithms from $E$ and the result is formally the same as performing the differentiations to compute $E$ and then setting $\log\vph{\=}0$
\beq
\widetilde{E}_j{}^k \= \frac{1}{j!}\vth^{k}\,\vp_j\Big|_{\log\vph=0}\=
\sum_{s=0}^{\min(j,k)}\frac{k!}{(j-s)!\, (k-s)!\, s!}\,\vth^{k-s}f_{j-s}~.
\notag\eeq
Note that, when $\vph{\=}0$, the only terms to survive in the sum above are those for which $s{\=}k{\=}j$ and that these surviving terms are unity. So we see that
\beq
\widetilde{E}(0) \= \one~.
\notag\eeq

Return now to the Picard-Fuchs equation \eqref{eq:PFinFirstOrderForm}.  Note that it is a general property that the terms $S_j$  vanish at $\vph{\=}0$ for $0\leqslant j\leqslant 3$,  while $S_{4}(0)$ is non-zero. Also, let us denote the matrix $B(0)$ by $\e$.
\beq
\e \= B(0) \=  
\begin{pmatrix}0&\+ 0&\+ 0&\+ 0\\
                       1&\+ 0&\+ 0&\+ 0\\
                       0&\+ 1&\+ 0&\+ 0\\
                       0&\+ 0&\+ 1&\+ 0
\end{pmatrix}~. 
\notag\eeq
Notice that $\e^4{\=}0$ and also that $B(0)$ generates the monodromy of $E$ around $\vph{\=}0$. Furthermore, $E$ and the vector $\vp^\sharp$ have the same monodromy. Thus
\beq
\vp^\sharp \to \ee^{2\p\ii\e}\,\vp^\sharp
\notag\eeq
and we see that we can identify the matrix $\e$ with the $\e$ that appears in the Frobenius period.

We can now show that, in virtue of the CY-crystal relations, we may replace $E$ by $\widetilde{E}$, for the purpose of computing $U(\vph)$, in \eqref{eq:DworkForm}. Note first that setting $\vph{\=}0$ in \eqref{eq:FrobDiffEq} yields an identity that we conjecture holds for $U(0)$
\beq
p\,\e\, U(0) \= U(0)\,\e~.
\label{eq:CommutationOfUWithEpsilon}\eeq
We observe also that
\beq
E^0(\vph)\= \vph^\e~.
\notag\eeq
Thus we have
\beq
U(0)\, E^0(\vph) \= \vph^{p\e}\, U(0)\= E^0(\vph^p)\, U(0)~.
\notag\eeq
Hence
\beq
\big(E^0(\vph^p)\big)^{-1}\, U(0)\, E^0(\vph) \= U(0)
\notag\eeq
and it follows that
\beq
E^{-1}(\vph^p)\, U(0)\, E(\vph)\= \widetilde{E}^{-1}(\vph^p)\, U(0)\, \widetilde{E}(\vph)~.
\notag\eeq
It is now immediate, since $\widetilde{E}(\vph)$ is smooth at $\vph{\=}0$ and $\widetilde{E}(0){\=}\one$,
that $F(\vph){\,\to\,}F(0)$ and $U(\vph){\,\to\,}U(0)$ as $\vph{\,\to\,}0$.

\subsection{The form of $U(0)$}
\vskip-10pt
Let us now seek to write down the explicit form of $U(0)$. The commutation relation \eqref{eq:CommutationOfUWithEpsilon} restricts $U(0)$ to the form
\beq
U(0) \=
\left(\begin{array}{cccc}
u &   0 &     0 & 0 \\
a & p u &     0 & 0 \\
b & p a & p^2 u & 0 \\
c & p b & p^2 a & p^3 u
\end{array}\right)~.
\notag\eeq 

There are further restrictions to be made on this matrix. If we impose
the condition (\ref{eq:Crystal}\hspace{2pt}c) we find that the $i$'th row of $U(0)$ 
should be divisible by $p^i$, $0\leqslant i \leqslant 3$ so that we can write
\[
a \= p u \alpha,\quad b \= p^2 u \beta, \quad c \= p^3 u \gamma~.
\]
Now we have
\beq
U(0)\=
u \left(\begin{array}{cccc}
1 & 0 & 0 & 0 \\
p\,\a & p & 0 & 0 \\
p^2 \b & p^2 \a & p^2 & 0 \\
p^3 \g & p^3 \b & p^3 \a  & p^3 \end{array}\right)~.
\notag\eeq
In order to constrain the form of $U(0)$ yet further, we should pause to examine the antisymmetric and hermitian forms that exist for the case of immediate interest for which the complex dimension is three. 

There is a natural antisymmetric form $(\ast,\ast) : H^3{\times}H^3 {\,\to\,}H^6$
\beq
(\x, \eta) \= \int_X \x\wedge\eta
\notag\eeq
There is also an associated hermitean form, that is sesquilinear on the first argument
\beq
\langle \x, \eta \rangle \= \ii (\overline{\x}, \eta)~.
\notag\eeq
It is useful to have explicit matrices corresponding to these quadratic forms. For the cohomology group $H^3(X)$ we have already introduced the integral symplectic basis $\{\a_a, \b^b\}$, which we can take together and denote by $\{\Ialpha^m\}$. Thus we can write
\beq
\O \= \vp^\sharp_k f^k \= \P_m \Ialpha^m~.
\label{eq:fbasis}\eeq
If we take a matrix
\beq
\l \= \text{diag}\left(1, \,\frac{1!}{2\p\ii}, \,\frac{2!}{(2\p\ii)^2}, \,\frac{3!}{(2\p\ii)^3} \right)~,
\notag\eeq
then
\beq
\P \= \r^\sharp \vp^\sharp~; \quad \r^\sharp \= \widehat{\r}\, \l~,
\notag\eeq
where $\widehat{\r}$ is the matrix defined in \eqref{eq:rhohat}. The $f^k$ are related to the $\Ialpha^m$ by the matrix $\r^\sharp$ which is the ``matrix of constants'' alluded to earlier
\beq
f^k \= \Ialpha^m \r^\sharp{}_m{}^k~.
\label{eq:fInTermsOfIalpha}\eeq
We have
\beq
(\Ialpha^m, \Ialpha^n) \= \S^{mn}~;\quad 
\S\=\left(\!\!\begin{array}{rrrr}
0 & 0 &\+ 1 & 0 \\
0 & 0 & 0 &\+ 1 \\
-1& 0 & 0 & 0 \\
0 &-1 & 0 & 0 
\end{array}\right)~.
\notag\eeq 

Let us denote the matrix corresponding to the antisymmetric product in the $\{f^j\}$-basis by~$\s$. Then
\beq
\s^{jk} \= (f^j, f^k)~; \quad 
\s \;= - \frac{y}{(2\p\ii)^3}\left(\!\!\begin{array}{rrrr}
0 & 0 & 0 & \+ 1 \\
0 & 0 &-1 & 0 \\
0 &\+ 1 & 0 & 0 \\
-1 & 0 & 0 & 0 
\end{array}\right)~.
\notag\eeq 

Now from (\ref{eq:Crystal}\hspace{2pt}c) we see that we should have
\beq
U(0)\,\s\,U(0)^T \= p^3\,\s
\notag\eeq
and this has the consequence
\beq
u^2 \= 1\quad \text{and} \quad \b \= \frac12 \a^2~.
\notag\eeq
We have that $u{\=}\pm1$ and we come to the form
\beq
U(0) \= u\, \L\left(\one + \a\e +\b\e^2 +\g\e^3\right)~,
\label{eq:Uzero}\eeq
where $\b{\=}\a^2/2$ and
\beq
\L \= 
\text{diag}(1, p\,, p^2, p^3)~.
\notag\eeq

\subsection{Conjectures relating to the form of $U(0)$}\label{conjecturesU(0)}
\vskip-7pt
Based on many calculations, we make the following
\vskip10pt
{\bf Conjecture:} {\it For \cy threefolds in a one parameter family,}
\vspace*{-5pt}
\begin{itemize}
\item
\it the matrix $U(0)$ is as in \eqref{eq:Uzero}, with $u{\=}1$, $\a{\=}\b{\=}0$ and
\beq
\g\= \frac{\chi(X)}{y}\, \z_p(3)~,
\notag\eeq
where $\chi$ is the Euler number of the manifold, $y$ denotes the large complex structure value of the Yukawa coupling, so the triple intersection value of the generator of $H^2$, for the \emph{mirror manifold}, and $\z_p(3)$ denotes the p-adic analog of $\z(3)$, which we calculate in terms of the p-adic $\G$-function by means of the formula
\beq
\z_p(3)\;= -\frac12 \Big( \G_{\!\!p}'''(0) -  \G_{\!\!p}'(0)^3\Big)~.
\label{eq:ZetapOf3}\eeq

\item 
\it Moreover, for good primes, $U(\vph){\=}E^{-1}(\vph^p)\, U(0)\, E(\vph)$ is a matrix of power series in $\vph$ with coefficients that are $p$-adically integral.
\end{itemize}

Some special cases of these conjectures have been proved. The statement that $u{\=}1$ is proved, for the 14 hypergeometric, one-parameter families of \cy threefolds, in the dissertation of 
      K.\,Samol~\cite{DissertationSamol} and the expression \eqref{eq:ZetapOf3} has been derived, for the case of the mirror quintic, by Shapiro~\cite{IShapiro2009Quintic} and by Thorne~\cite{ThorneThesis}.

In \eqref{eq:ZetapOf3} $\G_p$ denotes the $\G$-function that is named for Morita~\cite{Morita1975}. Though, on the matter of attribution, Dwork, writing under the pseudonym of Boyarski~\cite{Boyarski1980}, pointed out that this function was known to him much earlier~\cite{Dwork1964}.

There is a degree of choice in the definition of the $p$-adic zeta function and so in the definition of $\z_p(3)$. We give a derivation of the formula above for $\z_p(3)$, and discuss these matters further, in 
      Appendix\hspace{2pt}\ref{sec:AppendixZeta}. 
      
One can consider our conjecture as a manifestation of a $p$-adic version of the  
$\Gamma$-conjectures (see \cite{Hosono:1994ax,LibgoberChernClasses,Halverson:2013qca,HosonoCentralCharges,%
DissertationHofmann,GalkinGolyshevIritani2016}), which are still to be formulated in full generality. Let us nevertheless pursue these matters a little further.
      
Having taken $\a{\=}0$ the form for $U(0)$ is
\beq
U(0) \= \L\left( \one + \frac{\chi(X)}{y(\widetilde{X})} \z_p(3) \,\e^3 \right)~.
\notag\eeq

Consider the following identity
\beq
\ee^{-\l\, \Gp'(0)\,\e}\; \Gp(\l \e) \= \one - \frac13\, \l^3 \z_p(3)\,\e^3
\notag\eeq
This follows easily by expanding the left hand side in powers of $\e$ and using the identity 
$\Gp''(0){\=}\Gp'(0)^2$ as well as \eqref{eq:ZetapOf3}. It is also an immediate consequence of \eqref{eq:LogGpExpansion} from \hbox{Appendix\hspace{2pt}\ref{sec:AppendixZeta}}.

In order to perform a characteristic class calculation we write $\l_j$, $j{\=}1,2,3$ for the eigenvalues of the
quantity $\Th/\o$, where $\Th$ denotes the curvature matrix and $\o$ denotes the \K form of $\widetilde{X}$. On taking the product of three instances of the expression above we find
\beq
\prod_{j=1}^3  \ee^{-\l_j\, \Gp'(0)\,\e}\; \Gp(\l_j \e) \= 
\one - \frac13 \left(\sum_{j=1}^{3}\l_j^3\right)\z_p(3)\e^3~,
\notag\eeq
Writing the power sum on the right in terms of the symmetric polynomials we have
\beq
\sum_{j=1}^{3}\l_j^3 \= \s_1^3 - 3\s_1\s_2 + 3\s_3~.
\notag\eeq
We have $\s_1{\=}c_1/\o{\=}0$ and 
\beq
\s_3\= \frac{c_3(\widetilde{X})}{y(\widetilde{X})} \;= -\frac{\ch{(X)}}{y(\widetilde{X})}
\notag\eeq
The upshot is that
\beq
U(0) \= \L\, \prod_{j=1}^3  \ee^{-\l_j\, \Gp'(0)\,\e}\; \Gp(\l_j \e)~.
\notag\eeq
Note that 
\beq
\prod_{j=1}^3  \ee^{-\l_j\, \Gp'(0)\,\e} \= \ee^{-\s_1\, \Gp'(0)\,\e}
\notag\eeq
and that $\s_1$ is set to zero. Thus the factors $\ee^{-\l_j\, \Gp'(0)\,\e}$, while useful for simplifying intermediate expressions, can in fact be omitted. Hence we may write
\beq
U(0) \= \L\, \widehat{\G}_{\! p} \qquad \text{with} \qquad \widehat{\G}_{\! p}\= 
\prod_{j=1}^3 \Gp(\l_j \e)~.
\notag\eeq

Our interest is primarily with threefolds, but since we have the expansion \eqref{eq:LogGpExpansion} to hand, we may write out the expansion of $\widehat{\G}_{\! p}$ in a form that allows us to pick out the terms appropriate to higher dimensions.
\beq\begin{split}
\prod_j \Gp(\l_j \e) \= 
&\one 
- \z_p(3)\,\s_3 \, \e^3 
- \z_p(5) \left(\s_5 - \s_2 \s_3\right) \, \e^5 
+\smallfrac{1}{2} \z_p(3)^2 \s _3^2 \, \e^6   \\[-3pt]
& \hskip-1.75cm - \z_p(7)\Big(\s_3 \left(\s_2^2 - \s_4\right) - \s_2 \s_5 + \s_7\Big) \e^7
+ \z_p(3) \z_p(5)\, \s_3 \left(\s_5 - \s_2\s_3\right) \e^8 \\[5pt]
& \hskip-1.75cm - \Big[\, \smallfrac{1}{6} \z_p(3)^3 \s_3^3  + 
\z_p(9)\Big(\smallfrac13 \s_3^2 - \s_3\s_2^3 + \s_5\s_2^2 + (2\s_3\s_4 - \s_7)\s_2  - \s_4\s_5 - \s_3\s_6 + \s_9\Big)\Big] \,\e^9 \\[5pt]
&\hskip-1.75cm + \ldots ~.
\end{split}\notag\eeq
\subsection{The matrix of complex conjugation}
\vskip-10pt
As we have indicated, complex conjugation can be considered as an analogue of the Frobenius maps, corresponding to the infinite prime. A complication is that while the Frobenius matrix $F(\vph)$ is regular at the MUM-point, the same is not true of the complex conjugation matrix $C(\vph)$. Nevertheless we may proceed as follows.

We define a matrix $\cC^0$ relative to the cohomology basis $\{f^k\}$ of \eqref{eq:fbasis}
\beq
\overline{f^k} \= f^j\, \cC^0{}_j{}^k~.
\notag\eeq
As before, we also have the relations
\beq
e^k\= f^j E_j{}^k(\vph) \quad \text{and} \quad \overline{e^k} \= e^j\, C_j{}^k(\vph)~.
\notag\eeq
 We also have
\beq\begin{split}
\overline{e^k} 
&\= \overline{f^j} E_j{}^k(\overline{\vph}) \\[3pt]
&\= f^i\, \cC^0{}_i{}^j E_j{}^k(\overline{\vph}) \\[3pt]
&\= e^\ell E^{-1}{}_\ell{}^i(\vph)\, \cC^0{}_i{}^j E_j{}^k(\overline{\vph})~.
\end{split}\notag\eeq
So
\beq
C(\vph) \= E^{-1}(\vph)\, \cC^0 E(\overline{\vph})~.
\notag\eeq
But it is not asserted that $C(\vph){\,\to\;}\cC^0$ as $\vph{\,\to\,}0$.

To compute $\cC^0$, note that by means of \eqref{eq:fInTermsOfIalpha} we can relate the $\{f^k\}$-basis to the real $\{\Ialpha^m\}$-basis. So we may write
\beq
\overline{f^k}
\= \Ialpha^m\, \overline{\r^\sharp{}_m{}^k} 
\= f^j \bigl((\r^\sharp)^{-1}\bigr){}_j{}^m\, \overline{\r^\sharp{}_m{}^k}~.
\notag\eeq
So
\beq
\cC^0 \=  (\r^\sharp)^{-1}\, \overline{\r^\sharp} \= 
\left(\begin{array}{rrrr}
1 & 0 & 0 & 0 \\
0 &-1& 0 & 0 \\
0 & 0 &\+ 1 & 0 \\
\!\!- \frac{2\chi}{y}\z(3)\hspace*{-12pt} & 0 & 0 &-1
\end{array}\right)~.
\notag\eeq
Note the consonance with the observed form for the matrix $U(0)$. The diagonal entries show that for the infinite prime one has to replace $p$ by $-1$. The extra factor of $2$, associated with the lower left corner of the matrix, however, remains currently unexplained.

As for the behaviour of $C(\vph)$ as $\vph$ tends to zero. Note that
\beq
\cC^0\,\e \;= -\e\; \cC^0~.
\notag\eeq
Thus 
\beq
C(\vph)\;\sim\; \vph^{-\e}\,\cC^0\, \left(\overline{\vph}\right)^\e \= \cC^0\, |\vph|^{2\e} \= \cC^0 +
\left(\begin{array}{rrrr}
0 & 0 &\+ 0 &\+ 0 \\[2pt]
-L &0& 0 & 0 \\[2pt]
\frac12 L^2\hspace*{-4pt} & L &\ 0 & 0 \\[2pt]
\hspace*{-5pt}-\frac16 L^3 \hspace*{-5pt}& \hspace{4pt}\frac12 L^2\hspace*{-4pt} & -L & 0
\end{array}\right)~,
\notag\eeq
where, in the last expression, $L{\=}\log|\vph|^2$. We see that, in some sense, $\cC^0$ is the regular part of $C(\vph)$ at $\vph{\=}0$.

Finally, in this subsection, let us write down the matrix $h$ corresponding to the hermitean form.
We have
\beq
h^{jk} \= \langle f^j , f^k \rangle \= \ii (f^\ell\, \cC^0{}_\ell{}^j, f^k) 
\;= -\ii \s^{k\ell}\,\cC^0{}_\ell{}^j~.
\notag\eeq
Hence
\beq
h \= -\ii\big(\s\, \cC^0\big)^T \=
\frac{y}{(2\pi)^3}
\left(\begin{array}{rrrr}
\frac{2\chi}{y}\z(3)\hspace*{-12pt} &\+ 0 & \+ 0 & \+1 \\
0 & 0& 1 & 0 \\
0 & 1 & 0& 0 \\
1 & 0 & 0  & 0\end{array}\right)~.
\notag\eeq
We could also have come to this same expression for $h$ by noting that
\beq
h \= \ii (\r^\sharp)^\dag\, \S\, \r^\sharp~.
\notag\eeq
\newpage
\section{Computing $U(\vph)$}\label{sec:ComputingU}
\vskip-10pt
We wish to calculate the determinant $R(\vph, T)$ as a function of the parameter $\vph$.  Which amounts to computing the coefficients $a$ and $b$. We will discuss two methods of evaluating these coefficients, which differ in important ways with regard to their practical application. 

\subsection{Evaluation of $U(\vph)$ at Teichm\"uller points}
\vskip-10pt
It is our aim to compute
\[ 
R(\vph, T) \= \det\big(\one - T\, U(\vph) \big)
\]
for values $\phi \in \IF_p$. We have seen that the matrix can be expressed in the form
\[ 
U(\phi) \= E^{-1}(\phi^p) U(0) E(\phi) \= \widetilde{E}^{-1}(\phi^p) U(0) \widetilde{E}(\phi)~.
\]
If we want to evaluate this series at some point $\phi \in \IF_p$, we 
first form the Teichm\"uller lift $\phi:=\Teich(\vph) \in \IZ_p$ and the issue
of the convergence of the series $U(\phi)$ at this point becomes relevant.
Note that by definition $\phi{\=}\phi^p$, so $\norm{\phi}_p {\=}1$ if $\phi {\,\neq\,} 0$.
The periods and therefore the matrix $\widetilde{E}$ converge in the disk 
$\norm{\phi}_p{<\;}1$, so it is not permissible to try to substitute the 
Teichm\"{u}ller value directly into $\widetilde{E}$. Indeed, if it were permissible 
to do this, then the matrix $U(0)$ would be merely 
conjugated by $E(\ph)$ and $R(\ph,T)$ would be independent of~$\ph$.
Instead, we have to expand $U(\vph)$ as a matrix of p-adic series. 
These series converge on the larger disk $\norm{\vph}_p{\leqslant\;}1+\d$ for some $\d{\;>\;}0$: 
the series for the $U$-matrix are said to be {\em overconvergent}. This important property,
discovered by Dwork and proved in great generality by Berthelot, implies that
it \emph{is} permissible to substitute the Teichm\"{u}ller value into these 
series and that the corresponding limit is correct.






\subsection{$U(\vph)$ as a sequence of rational functions}
\vskip-10pt
The matrix $U(\vph)$ is, as we have observed, a power series in $\vph$ and so also are the coefficients $a$ and $b$. The problem that arises, however, is that, while the series for $U(\vph)$ converge, they do so only slowly. The aim, when working with these series is therefore to rearrange the terms so as to achieve faster convergence.

Let us fix a $p$-adic accuracy. We will see in \sref{sec:FormOfR}, that a knowledge of $a$ and $b$ mod $p^3$ is sufficient to identify these coefficients exactly, for $p\geqslant 5$. However, we lose a power of $p$ when passing from $U$ to $b$, owing to the second relation in \eqref{eq:abcoeffs}, so, to start, we work mod~$p^4$ with~$U$. 

We find that, mod $p^4$, the matrix $U$ is a matrix of rational functions of $\vph$ of the form stated in \eqref{eq:WanLauderForm}. This much seems to be true in all cases and already permits us to calculate $R$ for the values of $\vph$ for which $X$ is smooth. We have stated already in \sref{sec:ComputingU}, that in many cases the above relation performs better than advertised owing to cancellations between the numerator and denominator. For the examples that we examine here, we are able to calculate the Frobenius polynomial $R$ for all $\vph{\,\neq\,}0,\infty$, apart from apparent singularities. The difficulty that we find with apparent singularities is puzzling. Let us pause to explain the nature of the difficulty.

The form of the expression \eqref{eq:ApparentSings} suggests that $U(\vph)$ has a singularity at an apparent singularity $\vph_0$. However this cannot be so, since the manifold is smooth at $\vph_0$. Indeed one can check that the numerator $\hatccU(\vph)$ is $\cO(p^4)$ at $\vph{\=}\vph_0$. This being so, it is natural to try to evaluate $U(\vph_0)$ by taking a limit as $\vph{\,\to\,}\vph_0$. In order to discuss this, let us first note that, instead of \eqref{eq:ApparentSings}, we can~write
\beq
U(\vph)\= \frac{\ccU^\sharp(\vph)}{\big(\vph_{\phantom{0}}^p - \vph_0^p \big)^2} + \cO(p^4)~,
\notag\eeq
This form does not extend to higher p-adic accuracy, for which we have to proceed as in \sref{sec:HigherOrder}
but it does work mod $p^4$ for the apparent singularities that we study here. The numerator $\ccU^\sharp(\vph)$ is again $\cO(p^4)$ when $\vph{\=}\vph_0$. The problem, however, is now clear: if we take two derivatives of both the numerator and denominator, to compute a limit as $\vph{\,\to\,}\vph_0$, then a factor of $p^2$ appears in the denominator and the accuracy with which we can calculate the coefficients is correspondingly reduced. The upshot is that we can only calculate $a$ mod $p$ and $b$ mod 1. This is disappointing, we have no information about $b$ and there are easier ways of calculating $a$ mod p. One such is the formula
\beq
a\; = - \sum_{n=0}^{p - 1} a_n \vph^n \mod p~,
\notag\eeq
which is the reduction mod $p$ of the unit root, which we discuss in the following subsection.

\subsection{Calculation of the unit root}
\vskip-10pt
It has already been noted that, as a consequence of the Weil conjectures, the eigenvalues $\l_i$ of the matrix $U$ have norm $p^\frac32$, as complex numbers. Since $U$ is computed as a matrix of p-adic numbers, the meaning of this statement is perhaps not immediately obvious. The polynomial $R(\vph,T)$ however is the characteristic polynomial of $U$ and this is a polynomial in $T$ with coefficients that are rational integers and the eigenvalues $\l_i$ are the (inverse) roots of this polynomial, so are complex numbers. We may also solve for the inverse roots of $R$ as p-adic numbers and  we assume that the four roots are 
      $\cO(1),\,\cO(p),\,\cO(p^2),\,\cO(p^3)$ as p-adic numbers.
      
We may label the eigenvalues so that $\l_1$ is the eigenvalue that is $\cO(1)$. Dwork observes that this eigenvalue may be computed as the limit, as $\vph$ tends to a Teichm\"{u}ller value, of 
\beq
\l_1\= \frac{\vp_0(\vph)}{\vp_0(\vph^p)}~,
\notag\eeq
where $\vp_0$ is the fundamental period defined in \eqref{eq:periods}. Again, the series expansion of the period $\vp_0$ has radius of convergence $\norm{\vph}_p{\,<\,}1$ but the ratio of periods has a radius of convergence $1{+}\d$ for some $\d{\;>\;}0$.

The unit eigenvalue may also be computed, more conveniently, as the limit  
\beq
\l_1\= \lim_{s\to\infty}\frac{{}^{(s)}\vp_0(\vph)}{{}^{(s-1)}\vp_0(\vph)}~~~\text{where}~~~
{}^{(r)}\vp_0(\vph)\= \sum_{n=0}^{p^r - 1} a_n \vph^n~.
\notag\eeq
In fact we observe, for the examples that we study here, that
\beq
\frac{{}^{(s)}\vp_0(\vph)}{{}^{(s-1)}\vp_0(\vph)}\= \l_1 + \cO(p^s)~.
\notag\eeq
If $R$ is irreducible over the integers then a calculation of $\l_1$ for $s{\=}6$ is likely to give $R$ unambiguously.  Given $\l_1$, it is straightforward to construct a table of possible values $R(a,b,\l_1)$ for values of $a$ and $b$ corresponding to the allowed region of \fref{fig:KirasPlot} and there are approximately $\smallfrac{32}{3}p^\frac72$ such values. We then find $(a,b)$ as the pair of integers for which $R(a,b,\l_1)$ vanishes to highest p-adic order.

This procedure works well for small values of $p$, say for $p{\;\leqslant\;}19$, but, since the number of $a_n$ coefficients that need to be calculated is $p^6$, becomes rapidly impractical for larger values of $p$. We can however use this method to calculate $R$ for $p{\=}2,3,5$, for which our previous method does not apply. We can also use it as a check, for values of $p$ for which both methods are practical, and to fill in values for apparent singularities, for which our primary method~fails. 

On the other hand, the unit eigenvalue method has limitations for those cases that $R$ factorises. For example, at conifold points we observe a factorisation of the form
\beq
R\= (1 - p\chi\, T) (1 - \b\, T + p^3 T^2)~,
\label{eq:ConifoldFactorisation}\eeq
where $\chi{\=}\pm1$ is a character. In these cases we are led to a root of the quadratic factor, and so find a value for $\b$, but we have no information about the character $\chi$. Similarly when $R$ factors in to a product of two quadrics
\beq
R\= (1- p\a\,T+p^3T^2)(1-\b\,T + p^3T^2)~,
\notag\eeq
and we are particularly interested in factorisations of this form in relation to attractor points, then the method provides us with a root of the second factor, but no information about the coefficient $\a$ of the first factor.
\subsection{The form of $R(\vph,T)$ and bounds on the coefficients $a, b$}\label{sec:FormOfR}
\vskip-10pt
The considerations of this subsection are, in part, adapted from the thesis of Kira Samol~\cite{DissertationSamol}.

We have already observed that $R$ has the form
\beq
R\= \det(\one - T U)\= 1 + a\, T + b\, pT + a\, p^3 T^3 + p^6 T^4~,
\notag\eeq
where the third expression holds when the manifold is smooth. The coefficients $a$ and $b$ are most easily calculated as
\beq
a\;= -\text{Tr}(U)~;~~~b\= \frac{1}{2p}\Big( (\text{Tr}\,U)^2 - \text{Tr}\big(U^2\big)\Big)~.
\label{eq:abcoeffs}\eeq

Denoting the eigenvalues of $U$ by $\l_i,~i=1,..,4$ we have
\beq
a\;= -\sum_i \l_i ~~~\text{and}~~~b p\= \sum_{i<j} \l_i \l_j ~,
\notag\eeq
It follows from the Weil conjectures that, as complex numbers, $|\l_i|{\=}p^{3/2}$. These eigenvalues are the inverse roots of $R$, so come in complex conjugate pairs. If we write $p^{3/2}\ee^{\pm\ii \th_1}$ and $p^{3/2}\ee^{\pm\ii \th_2}$ for the four eigenvalues,  and also set $a{\=}p^{3/2}\tilde{a}$ and $b{\=}p^2\tilde{b}$, we have
\beq
\tilde{a} \= -2(\cos\th_1 + \cos\th_2)~;~~~\tilde{b} \= 2 + 4\cos\th_1 \cos\th_2~.
\notag\eeq
From the first relation we have immediately that 
\beq
|\tilde{a}|\;\leqslant 4~~~\text{and so}~~~|a|\;\leqslant 4p^{3/2}~.
\label{eq:alim}\eeq
On eliminating $\cos\th_2$, the second relation becomes
\beq
\tilde{b} \= 2 - 4\cos\th_1\left(\cos\th_1 + \frac{\tilde{a}}{2}\right)~.
\notag\eeq
An upper bound is obtained by maximising the right hand side, while lower bounds are obtained from the limiting values 
$\cos\th_1{\=}\pm1$. In this way, we find
\beq
-2 + 2|\tilde{a}| \;\leqslant\; \tilde{b} \;\leqslant\; 2 + \frac{\tilde{a}^2}{4}~.
\notag\eeq
This region, which we denote by $\widetilde\cS$, is sketched, on the right, in \fref{fig:KirasPlot}. 

Let $\tilde{b}_\text{mid}$ be the midpoint of the maximum and minimum values of $\tilde{b}$, for given $\tilde{a}$:
\beq
\tilde{b}_\text{mid} \= 2\left( \left( 1 + \frac{|\tilde{a}|}{4}\right)^2 -1\right)~.
\notag\eeq
Then
\beq
\left| \tilde{b} - \tilde{b}_\text{mid} \right| \;\leqslant\; 2\left(1 - \frac{|\tilde{a}|}{4} \right)^2 \;\leqslant\; 2
\notag\eeq
and, given $a$, we have
\beq
| b - b_\text{mid} |\; \leqslant 2p^2 +1~,~~~\text{where}~~~
b_\text{mid} \= \left[ 2p^2\left( \left( 1 + \frac{a}{4p^{3/2}}\right)^2  - 1\right) \right]
\notag\eeq
and, in this last relation $[...]$ denotes the nearest integer function. In virtue of the relations above, we are able to identify $a$ and $b$ from a knowledge of their values mod~$p^3$, for $p{\;\geqslant\;}5$.

For values of $\vph$ for which the manifold is singular, we expect one of the roots of the quartic to go to zero and for the resulting cubic to factor in the form
\beq
R~=~(1 -  \chi\, pT)(1 - \b\, T + p^3 T^2)
\label{eq:ConifoldForm}\eeq
where $\chi=\pm1$ is a character. Given this form, the quantities $\chi$ and $\a$ may be reconstructed from a knowledge of the linear and quadratic terms, as a polynomial in $T$. The roots of the quadratic factor have complex norm $p^{3/2}$, and $|\b|{\;\leqslant\;}2p^{3/2}$ so a calculation mod~$p^3$ serves in these cases also.

There is a natural way to place coordinates on the region $\widetilde\cS$. Consider the factorisation
\beq
1+ a\,T + b\,p T^2 + a\,p^3 T^3 +p^6T^4 \= (1-\a\,pT+p^3T^2)(1-\b\,T+p^3T^2)~.
\notag\eeq
We are most interested in the cases that the coefficients $a,b,\a,\b$ are integers, but let us temporarily take them to be merely real. Over $\IR$, factorisation, as above, is always possible and we have the relations
\begin{align}
a &\;= -(p\a + \b)~,& b &\= 2p^2 + \a\b~,\label{eq:maptoS}\\
\intertext{which we rewrite as}
\tilde{a}&\;= -(\tilde{\a} + \tilde{\b})~,& \tilde{b}&\= 2 +\tilde{\a}\tilde{\b}~,\label{eq:maptoStilde} 
\end{align}
with
\beq
\hskip1.3cm\tilde{\a}\=\frac{\a}{p^{1/2}} \;= -2\cos\th_1~,\hskip2.1cm
\tilde{\b}\=\frac{\b}{p^{3/2}} \;= -2\cos\th_2~.
\notag\eeq
and we see that $\tilde{\a}$ and $\tilde{\b}$ are simply related to the parameters $\th_1$ and $\th_2$ introduced previously.

Owing to the symmetry of \eqref{eq:maptoStilde} under interchange of $\tilde{\a}$ and $\tilde{\b}$, the map to $\widetilde\cS$ is generically 2--1 with $(\tilde{\a},\tilde{\b})$ and $(\tilde{\b},\tilde{\a})$ mapping to the same point. We can divide the square into two triangles by the diagonal $\tilde{\a}{\=}\tilde{\b}$; and choose the lower triangle, say, to be a fundamental region for parametrising the points $(\tilde{a},\tilde{b})$. We denote this fundamental region by $\widetilde{\cS}_0$, this region is sketched on the left in \fref{fig:KirasPlot}. 

\begin{figure}[!p]
\begin{minipage}[c]{\textwidth}
\framebox[\textwidth]{\begin{minipage}[c]{\textwidth}
\vspace{20pt}
\hspace*{-5pt}\includegraphics[width=6in]{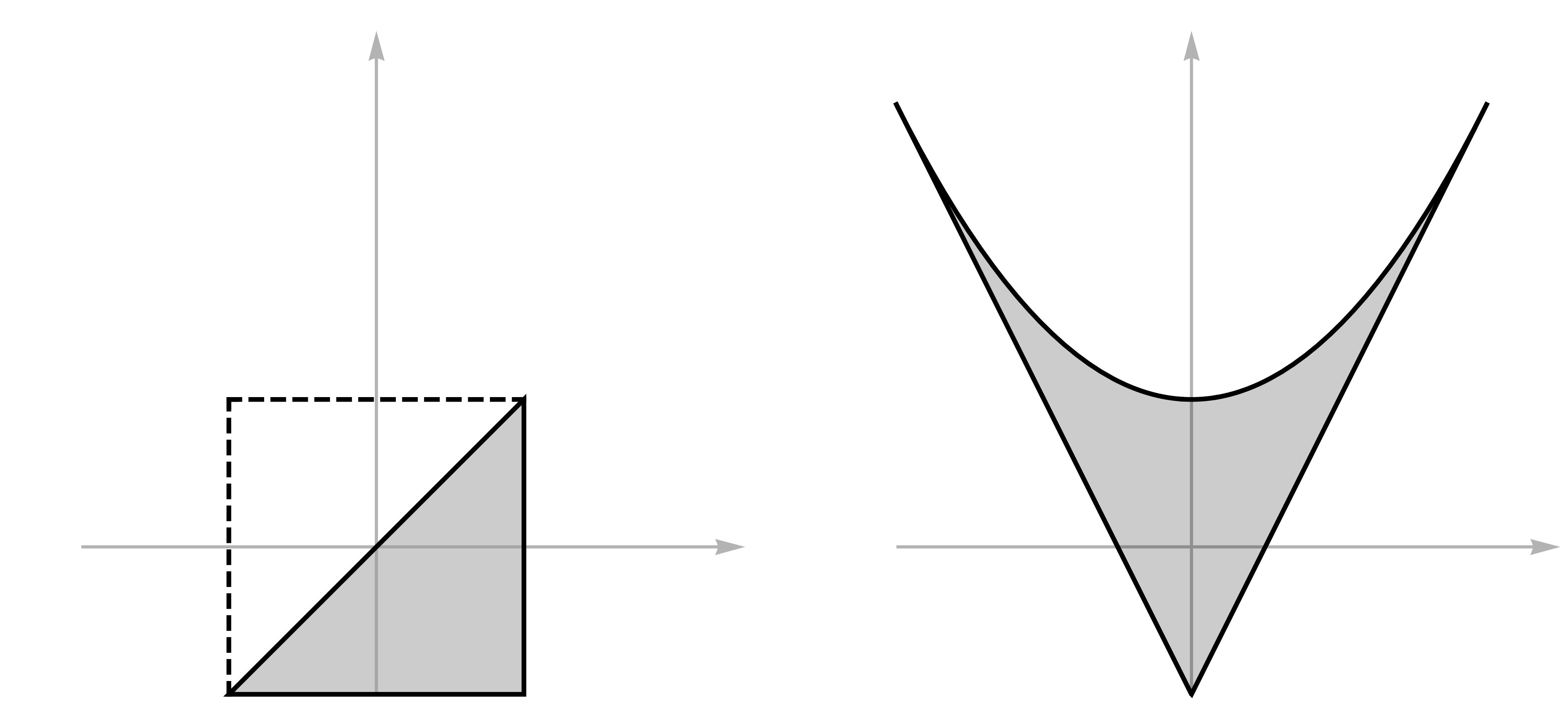}
\vspace{20pt}
\end{minipage}}
\vskip12pt 
\place{4.25}{0.45}{$(0,-2)$}
\place{4.3}{1.95}{$(0,2)$}
\place{3.05}{3.05}{$(-4,\,6)$}
\place{5.45}{3.05}{$(4,\,6)$}
\place{5.94}{1.23}{\large$a/p^{3/2}$}
\place{4.37}{3.32}{\large$b/p^2$}
\place{4.25}{1.45}{\large$\widetilde\cS$}
\place{0.45}{0.4}{$(-2,-2)$}
\place{1.7}{0.4}{$(2,-2)$}
\place{1.75}{1.9}{$(2,\,2)$}
\place{2.8}{1.23}{\large$\a/p^{1/2}$}
\place{1.22}{3.32}{\large$\b/p^{3/2}$}
\place{1.6}{0.9}{\large$\widetilde{\cS}_0$}
\centering
\capt{6.0in}{fig:KirasPlot}{The allowed region $\widetilde\cS$ for the coefficients $(a,b)$ is shown on the right and the fundamental region $\widetilde{\cS}_0$ for the parameters $(\a,\b)$, defined by relations \eqref{eq:maptoStilde}, is shown on the left.}
\null\vskip1.2cm
\framebox[\textwidth]{\begin{minipage}[c]{\textwidth}
\vspace{20pt}
\includegraphics[width=6in]{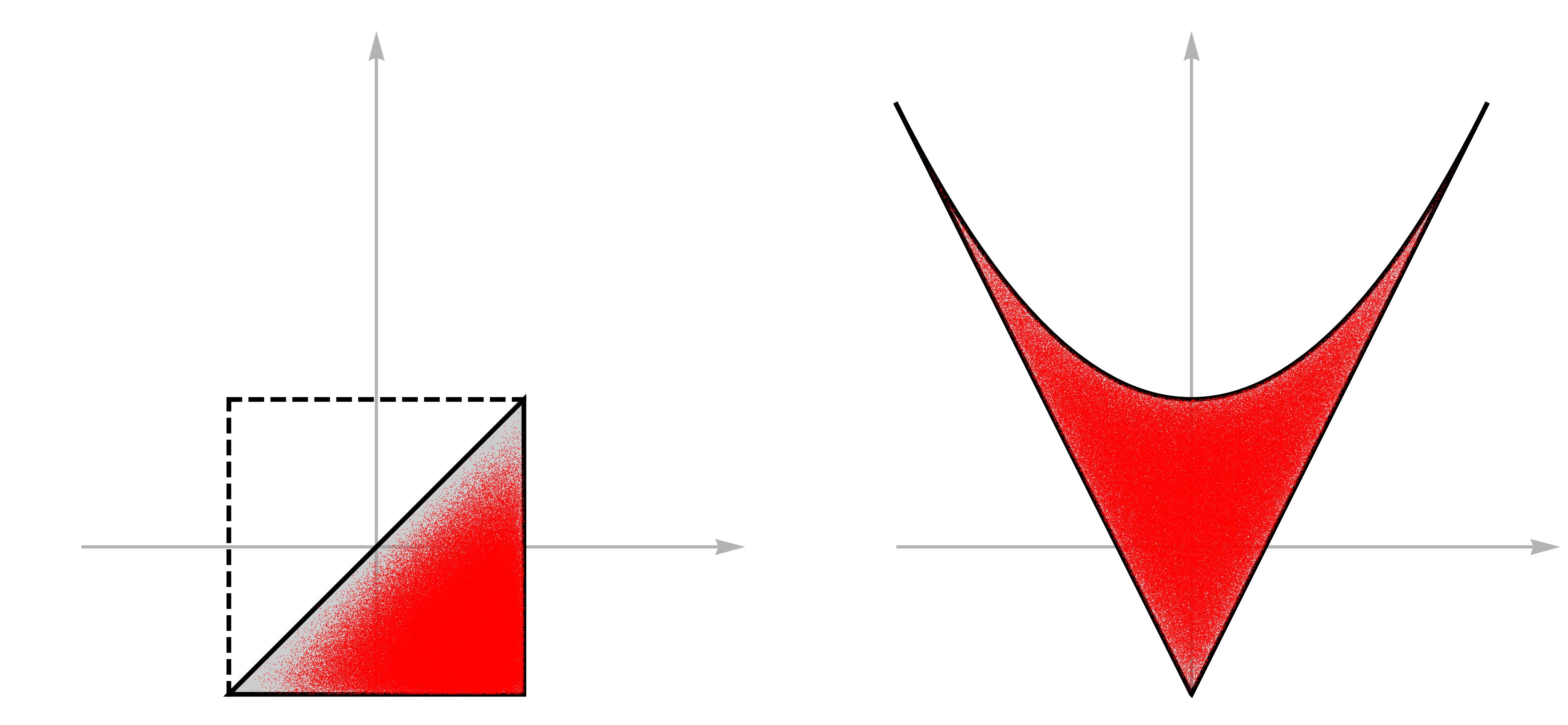}
\vspace{20pt}
\end{minipage}}
\vskip12pt
\centering
\capt{6.0in}{fig:CumulativePlots}{On the right is shown a plot of the pairs $(a,b)$ for the 75 primes $p_{427}\leqslant p\leqslant p_{502}$ and all values of $\vph$ for which the manifold AESZ34 is smooth. On the left, these points are mapped back onto the fundamental region. Note how the density decreases near the diagonal boundary.}
\end{minipage}
\end{figure}
\begin{figure}[!th]
\centering
\includegraphics[width=3.5in]{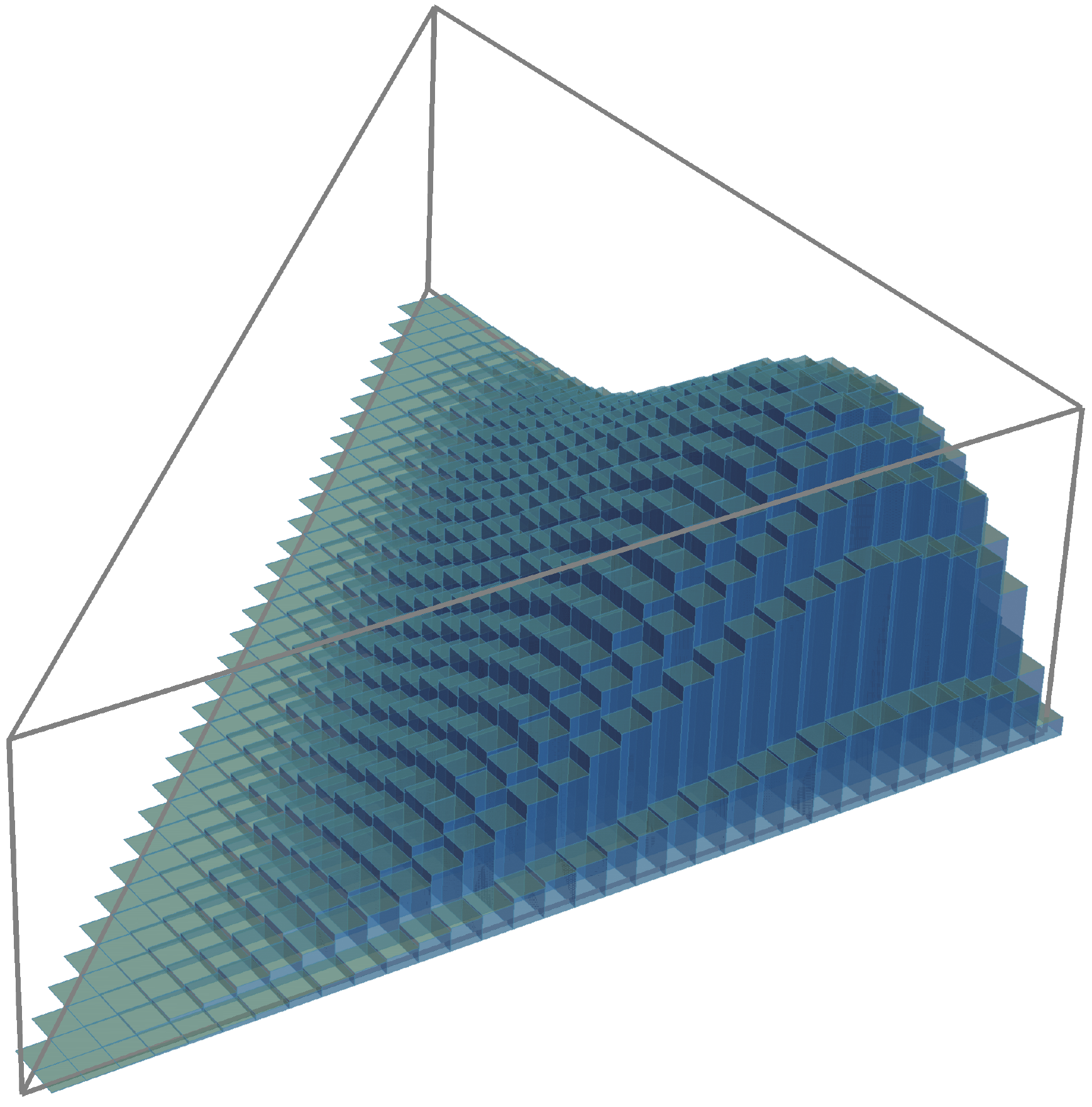}
\capt{5.8in}{fig:histogram}{A histogram showing the density of pairs $(a,b)$, mapped back to the region $\widetilde{\cS}_0$. The data shown is for the manifold AESZ34 and all values of $\vph$ for which the manifold is smooth, for the 1000 primes $p_3\leqslant p\leqslant p_{1002}$.}
\vspace{10pt}
\end{figure}
In \fref{fig:CumulativePlots} we indicate the distibution of the pairs $(a,b)$ as they map to $\widetilde\cS$ and to $\widetilde{\cS}_0$. A~histogram corresponding to the 1000 primes $p_3\leqslant p\leqslant p_{1002}$, and all values of $\vph$ for which the manifold is smooth, is presented for the manifold AESZ34 in \fref{fig:histogram}. The histogram is remarkably close to the probability distribution function $f\,\dd\th_1 \dd\th_2$ with $f$ given by
\beq
f \= \frac{16}{\p^2} \sin^2\th_1 \sin^2\th_2\, (\cos\th_1 - \cos\th_2)^2~.
\notag\eeq
This probability distribution function corresponds to the eigenvalue distribution of USp(4) matrices that are distributed randomly with respect to the Haar measure. We have conjectured~\cite{Candelas:2019llw} that this the correct distribution function for AESZ34 and we conjecture that this also the correct distribution function for all one-parameter examples. 

\subsection{Slopes}\label{sec:slopes}
\vskip-10pt
The polynomials $\widetilde{\cU}$ and $\widehat{\cU}$ from \eqref{eq:GenuineSings} and \eqref{eq:ApparentSings} show interesting regularities with respect to degree and which admit of some refinement. Considering first our examples that have no apparent singularities, let us write
\beq
U(\vph)\= \widetilde{\cU}_j(\vph) + \cO(p^j)~;~~~j\= 1,2,3,4~.
\notag\eeq
We term $\deg\,\widetilde{\cU}_4$ the \emph{slope} of the corresponding manifold $X$ and, more generally, refer to the vector $\deg\,\widetilde{\cU}_j$, $j{\= 1},\ldots,4$ as the \emph{slope vector}. We give these slope vectors in \tref{tab:slopes}.

Let us take the three examples without apparent singularities first. For these we make the observation that the coefficients of $p$ in the slope vectors coincide, in each case, with the indices of the corresponding differential operators at infinity. For the three examples with apparent singularities the corresponding statement is that the coefficients of $p$ differ from the corresponding index at infinity by 2. The difference of 2 accounts for the denominator that appears in \eqref{eq:ApparentSings}. 

While the slope vectors that appear for the examples without apparent singularities are exact, for the range of $p$ for which we have data. We have to make caveats for the examples with apparent singularities. For the \Rodland manifold the stated expressions for the slope vectors are exact (always for the range of primes for which we have data). For the three generation manifold with Hodge numbers $(4,1)$, the first component of the slope vector is $3p{\,-\,}1$, apart from for $p{\=}11$, for which it is $3p{\,-\,}2$. The second component is $4p{\,-\,}1$, apart from $p{\=}139$, for which it is $4p{\,-\,}2$, while the remaining components are exact. For the manifold with Hodge numbers $(1,1)$, the first three components of the slope vector are exact, while the fourth component is $5p{\,-\,}1$, apart from for $p{\=}11$, for which it is $5p{\,-\,}2$. These caveats do not seem to change the relation between the slope vectors and the indices at infinity.\vskip10pt
\begin{table}[!bth]
\renewcommand{\arraystretch}{2.0}
\begin{center}
\begin{tabular}{|>{}l<{}||>{$}c<{$}|>{$}c<{$}|>{$}c<{$}|>{$}c<{$}||>{$}c<{$}|}
\hline
Manifold & j=1 & j=2 & j=3 & j=4 & \text{Indices at}~\infty\\[7pt]
\hline\hline
Quintic & \left[\frac{p}{5}-\frac12\right] 
           & \left[\frac{2p}{5}-\frac12\right] 
           & \left[\frac{3p}{5}-\frac12\right] 
           & \left[\frac{4p}{5}-\frac12\right] 
           & \frac15,\,\frac25,\,\frac35,\,\frac45\\[7pt]
\hline
AESZ25 & \left[\frac{p}{2} - 1 \vphantom{\frac12}\right] 
             & \left[\,\frac{p}{2} - 1 \vphantom{\frac12}\hskip1pt\right] 
             & \left[\frac{3p}{2} - 1 \right] 
             & \left[\frac{3p}{2} - 1 \right] 
             & \frac12,\,\frac12,\,\frac32,\,\frac32\\[7pt]
\hline
AESZ34 & p - \frac32 + \frac12 \! \left(\frac{p}{15}\right)
            & p-1 
            & 2p - \frac32 -\frac12 \! \left(\frac{p}{15}\right) 
            & 2p-2 
            & 1,\,1,\,2,\,2\\[7pt]
\hline\hline
\Rodland & 3p-1 
              & 3p-1 
              & 3p-1 
              & 3p-1
              & 1,\,1,\,1,\,1 \\[7pt]
\hline 
(4,1) Mfld. & 3p-1
             & 4p-1
             & 5p-1             & 6p-1
             & 1,\,2,\,3,\,4 \\[7pt]
\hline
(1,1) Mfld. & 3p-1
               & 4p-2
               & 4p-1
               & 5p-1
               & 1,\,2,\,2,\,3 \\[7pt]
\hline
\end{tabular}
\capt{5.5in}{tab:slopes}{The slopes for our six examples. The first three have no apparent singularities, while the remaining three do have them. Square brackets denote the round function, with the convention that, for $n{\,\in\,}\IZ$, $[n+\frac12]$ rounds to the nearest even integer. In the third row $(\frac{p}{15})$ denotes the Jacobi symbol. The last column gives the indices at infinity, these are taken from the corresponding Riemann symbols.}
\end{center}
\end{table}
\subsection{Evaluating $U$ to higher $p$-adic order}\label{sec:HigherOrder}
\vskip-10pt
We have seen that for the purpose of evaluating the numerator of the $\z$-function, for $p{\,>\,}5$, it is sufficient to work mod $p^4$. We ask now how the calculation of $U(\vph)$ changes if we work mod $p^n$, for $n>4$. We have looked only at the case of the quintic, in this regard, and then only for $p{\=}7$ and $p{\=}3$. We want to examine exponents $n$ that are in the vicinity of powers $p^k$ for a few values of $k$, and this is only currently practicable for small $p$. Let us start with $p{\=}7$, for which we examine powers $n{\,\leqslant\,}360$, that is $n{\,\leqslant\,}p^3{\,+\,}17$. The relation that we find, in this case, is
\beq
U(\vph) \= \frac{\widetilde{\cU}_n(\vph)}{\D(\vph)^{p(n - 4 + \e_n)}} + \cO\left( p^n \right)~,
\notag\eeq
where
\beq
\e_n\= \a_{n_0} + \b_{n_0}\d_{n_1, 6} + \g_{n_0}\d_{n_1, 6}\,\d_{n_2, 6}
\notag\eeq
where the $n_j$ are the first three p-adic digits of $n{\,-\,}5$ and $\a$, $\b$ and $\g$ are the following related vectors
\beq\begin{split}
\a &\= (0,0,1,1,1,1,0) \\[2pt]
\b &\= (0,2,1,1,1,0,0) \\[2pt]
\g &\= (3,1,1,1,0,0,0)~.
\end{split}\notag\eeq
As the degree of the denominator increases, the degree of the numerator $\widetilde{\cU}_n(\vph)$ also increases and, for the range of $n$ that we examine, the degree of the numerator differs from the degree of the denominator by a constant independent of $n$.

We can go to powers $n$ in the vicinity of higher powers $p^k$ for $p{\,=\,}3$. For this prime the form that $U$ takes is
\beq
U(\vph) \= \frac{\widetilde{\cU}_n(\vph)}{\D(\vph)^{p(n - 3 + \e_n)}} + \cO\left( p^n \right)~.
\notag\eeq
Note the power of $\D$ in the denominator. One consequence is that there are already inverse powers of $\D$ if we work mod $p^4$. 

The quantities $\e_n$ naturally form a hypercubical array, as is evident from the following $3^5$ values for the range $3\leqslant n\leqslant 247$, which should be read by proceeding in the natural order within each matrix.
\newenvironment{smallarray}[1]{\null\vcenter\bgroup\footnotesize\renewcommand{\arraystretch}{0.8}%
  \arraycolsep=.17em\hbox\bgroup$\array{@{}#1@{}}}{\endarray$\egroup\egroup\null}
\beq
\left\{\,\left[
\begin{smallarray}{ccc}
 \left(
\begin{smallarray}{ccc}
 0 & 0 & 0 \\
 0 & 1 & 1 \\
 1 & 1 & 1 \\
\end{smallarray}
\right) & \left(
\begin{smallarray}{ccc}
 0 & 0 & 0 \\
 0 & 1 & 1 \\
 1 & 1 & 1 \\
\end{smallarray}
\right) & \left(
\begin{smallarray}{ccc}
 0 & 0 & 1 \\
 2 & 2 & 2 \\
 2 & 2 & 1 \\
\end{smallarray}
\right) \\[14pt]
 \left(
\begin{smallarray}{ccc}
 0 & 0 & 0 \\
 0 & 1 & 1 \\
 1 & 1 & 1 \\
\end{smallarray}
\right) & \left(
\begin{smallarray}{ccc}
 0 & 0 & 0 \\
 0 & 1 & 1 \\
 1 & 1 & 1 \\
\end{smallarray}
\right) & \left(
\begin{smallarray}{ccc}
 0 & 0 & 1 \\
 2 & 2 & 2 \\
 2 & 2 & 1 \\
\end{smallarray}
\right) \\[14pt]
 \left(
\begin{smallarray}{ccc}
 0 & 0 & 0 \\
 0 & 1 & 1 \\
 1 & 1 & 1 \\
\end{smallarray}
\right) & \left(
\begin{smallarray}{ccc}
 0 & 0 & 0 \\
 0 & 1 & 1 \\
 1 & 1 & 1 \\
\end{smallarray}
\right) & \left(
\begin{smallarray}{ccc}
 0 & 2 & 3 \\
 3 & 3 & 3 \\
 3 & 2 & 1 \\
\end{smallarray}
\right)
\end{smallarray}
\right],\;\left[
\begin{smallarray}{ccc}
 \left(
\begin{smallarray}{ccc}
 0 & 0 & 0 \\
 0 & 1 & 1 \\
 1 & 1 & 1 \\
\end{smallarray}
\right) & \left(
\begin{smallarray}{ccc}
 0 & 0 & 0 \\
 0 & 1 & 1 \\
 1 & 1 & 1 \\
\end{smallarray}
\right) & \left(
\begin{smallarray}{ccc}
 0 & 0 & 1 \\
 2 & 2 & 2 \\
 2 & 2 & 1 \\
\end{smallarray}
\right) \\[14pt]
 \left(
\begin{smallarray}{ccc}
 0 & 0 & 0 \\
 0 & 1 & 1 \\
 1 & 1 & 1 \\
\end{smallarray}
\right) & \left(
\begin{smallarray}{ccc}
 0 & 0 & 0 \\
 0 & 1 & 1 \\
 1 & 1 & 1 \\
\end{smallarray}
\right) & \left(
\begin{smallarray}{ccc}
 0 & 0 & 1 \\
 2 & 2 & 2 \\
 2 & 2 & 1 \\
\end{smallarray}
\right) \\[14pt]
 \left(
\begin{smallarray}{ccc}
 0 & 0 & 0 \\
 0 & 1 & 1 \\
 1 & 1 & 1 \\
\end{smallarray}
\right) & \left(
\begin{smallarray}{ccc}
 0 & 0 & 0 \\
 0 & 1 & 1 \\
 1 & 1 & 1 \\
\end{smallarray}
\right) & \left(
\begin{smallarray}{ccc}
 0 & 2 & 3 \\
 3 & 3 & 3 \\
 3 & 2 & 1 \\
\end{smallarray}
\right)
\end{smallarray}
\right],\;\left[
\begin{smallarray}{ccc}
 \left(
\begin{smallarray}{ccc}
 0 & 0 & 0 \\
 0 & 1 & 1 \\
 1 & 1 & 1 \\
\end{smallarray}
\right) & \left(
\begin{smallarray}{ccc}
 0 & 0 & 0 \\
 0 & 1 & 1 \\
 1 & 1 & 1 \\
\end{smallarray}
\right) & \left(
\begin{smallarray}{ccc}
 0 & 0 & 1 \\
 2 & 2 & 2 \\
 2 & 2 & 1 \\
\end{smallarray}
\right) \\[14pt]
 \left(
\begin{smallarray}{ccc}
 0 & 0 & 0 \\
 0 & 1 & 1 \\
 1 & 1 & 1 \\
\end{smallarray}
\right) & \left(
\begin{smallarray}{ccc}
 0 & 0 & 0 \\
 0 & 1 & 1 \\
 1 & 1 & 1 \\
\end{smallarray}
\right) & \left(
\begin{smallarray}{ccc}
 0 & 0 & 1 \\
 2 & 2 & 2 \\
 2 & 2 & 1 \\
\end{smallarray}
\right) \\[14pt]
 \left(
\begin{smallarray}{ccc}
 0 & 0 & 0 \\
 0 & 1 & 1 \\
 1 & 1 & 1 \\
\end{smallarray}
\right) & \left(
\begin{smallarray}{ccc}
 0 & 0 & 0 \\
 0 & 1 & 1 \\
 1 & 1 & 1 \\
\end{smallarray}
\right) & \left(
\begin{smallarray}{ccc}
 3 & 4 & 4 \\
 4 & 4 & 4 \\
 3 & 2 & 1 \\
\end{smallarray}
\right) 
\end{smallarray}
\right]\,\right\}
\notag\eeq

The lines of a small matrix fit together to form a square, and three squares form a cube. The matrices bounded by square brackets are four-cubes, and three of these form a five-cube. Notice that the first two areas are the same while the third is different, and the first two cubes are the same while the third is different. This structure repeats: the first two four-cubes are the same, while the third is different, and the differences reside in the final cube of each four-cube. We do not however have a good formula to describe this.
\newpage
\section{The Manifolds}\label{sec:manifolds}
\vskip-10pt
We describe here in telegraphic form the manifolds that are our examples and give references to more complete descriptions.
\subsection{The mirror of the quintic threefold, AESZ\hskip2pt 1}\label{sec:quintic}
\vskip-10pt
The quintic threefold and its mirror must be the most studied of all \cys and the following analysis can be found in many references, however we take the opportunity to use this case to describe a procedure that can be applied to all our examples. This proceeds as follows:
\begin{itemize}
\item Given a presentation of the manifold by projective or toric polynomial, calculate a sufficient number of the coefficients, say 50, of the fundamental period, so as to be able to write
\beq
\vp_0(\vph) \= \sum_{n=0}^N a_n\,\vph^n + \cO(\vph^{N+1})
\notag\eeq
with known coefficients, for some sufficiently large $N$.
\item Choose a degree $k_\text{max}$ for the coefficient polynomials $S_j$ and set 
\beq
S_j\= \sum_{k=0}^{k_\text{max}} s_{jk}\,\vph^k
\notag\eeq
then use the equation $\cL \vp_0{\=}0$ to solve for the $s_{jk}$. If $k_\text{max}$ is chosen too small, then the only solution is $s_{jk}{\=}0$. So we increase $k_\text{max}$ until a non-trivial solution appears. In order to have sufficient conditions to solve for the $s_{jk}$ we have to take the ``sufficient number'' $N$ of known coefficients $a_n$ to satisfy $N{\;>\;}5k_\text{max}$.
\item Once the Picard-Fuchs operator $\cL$ has been found, it is a simple matter to seek a Frobenius solution
\beq
\vp(\vph,\e)\= \sum_{n=0}^\infty A_n(\e) \vph^{n+\e}~~~\text{with}~~~A_0(\e)\=1~,
\notag\eeq
and find a recurrence relation for the coefficients $A_n(\ve)$. On setting
\beq
A_n(\e)\= a_n + b_n\e + \frac{1}{2!} c_n\e^2 + \frac{1}{3!} d_n\e^3 + \cO(\e^4)
\notag\eeq
and expanding the recurrence relation we find recurrence relations for the coefficients $a_n,\,b_n,\,c_n,\,d_n$ of the series $f_0,\,f_1,\,f_2,\,f_3$. Given these functions, we construct the matrix $E$ and the inverse Frobenius matrix $U$.
\end{itemize}

The mirror quintic is realised by taking a quotient of the hypersurface in $\IP^4$
\beq
\sum_{i=1}^5 X_i^5 - 5\psi\prod_{i=1}^5 X_i \= 0
\notag\eeq
by the automorphism $x_i {\;\to\;} \a^n_i X_i$, with $\a$ a nontrivial fifth root of unity and $\sum_i n_i{\=}0 \mod 5$. 

By a series of standard manoeuvres the fundamental period can be represented by the integral
\beq
\vp_0 \= \frac{1}{(2\p\ii)^5}
\int_\g \frac{\dd^5X}{\prod_{i=1}^5 X_i}\left(1 - \frac{\sum_{i=1}^5 X_i^5}{5\psi\prod_{i=1}^5 X_i}\right)^{-1}~,
\label{eq:QuinticFundPer}\eeq
where now the variables $X_i$ are viewed as taking values in $\IC^5$, the contour $\g$ is a product of five small circles enclosing the loci $X_i{\=}0$, and $\psi$ is taken sufficiently large. The integral is evaluated by expanding the bracket in inverse powers of $\psi$ and evaluating term by term by residues. The only terms that contribute are the terms independent of $X$ in the powers
\beq
\left(\frac{\sum_{i=1}^5 X_i^5}{5\psi\prod_{i=1}^5 X_i}\right)^m~.
\notag\eeq
Such a term is zero unless $m{\=}5n$ and in that case is $(5n)!/(n!)^5$. So, for the mirror quintic,
\beq
\vp_0(\vph) \= \sum_{n=0}^\infty \frac{(5n)!}{(n!)^5}\, \vph^n~~~\text{where we have taken}~~~
\vph\=\frac{1}{(5\psi)^5}~.
\notag\eeq
In this case, we have been able to evaluate the coefficients of the fundamental period at one blow. In more difficult cases we will often not be able to find a closed form expression for these coefficients. However, it will always be the case that there is a Laurent polynomial $F(X)$ in variables $X_i$ such that 
\beq
a_n\= \left[F(X)^n\right]_0~,
\notag\eeq
where $[\ldots]_0$ denotes the operation of taking the part independent of the coordinates $X_i$, and we will need to evaluate the $a_n$, for sufficiently many $n$, in order to get started.

Returning to the mirror quintic, we see from \eqref{eq:QuinticFundPer} that the fundamental period satisfies the differential equation $\cL\vp_0{\=}0$, with
\beq
\cL\= \vth^4 - 
5^5\vph\left(\vth + \frac15\right)\left(\vth + \frac25\right)\left(\vth + \frac35\right)\left(\vth + \frac45\right)~.
\notag\eeq
This operator can, of course, be written in the form \eqref{eq:PFoperator}. The coefficient polynomials $S_j$ are then linear in $\vph$.
\beq
\cL\= (1 - 3125\,\vph)\,\vth^4 - 6250\,\vph\,\vth^3 - 4375\,\vph\,\vth^2  - 1250\,\vph\,\vth - 120\,\vph~.
\notag\eeq

This operator corresponds to the Riemann symbol
\beq
\cP\left\{
\begin{tabular}{ccc}
~0~~&~$5^{-5}$\hskip-5pt{}&~$\infty$~~ \\[1pt]
\noalign{\hrule height1pt width3cm}
\vrule height0pt width0pt\\[-12pt]
0&0&$\smallfrac15$\\[2pt]
0&1&$\smallfrac25$\\[2pt]
0&1&$\smallfrac35$\\[2pt]
0&2&$\smallfrac45$
\end{tabular}
\hskip5pt\vph~\right\}~.
\notag\eeq

If we now seek a Frobenius solution of the form \eqref{eq:FrobPer} we find a recurrence relation for the coefficients $A_n(\e)$ of the form
\beq
(n + \e)^4 A_n(\e) - 
5^5\left(n + \e - \smallfrac15\right) \left(n + \e - \smallfrac25\right) \left(n + \e - \smallfrac35\right)
\left(n + \e - \smallfrac45\right) A_{n-1}(\e) \= 0~.
\notag\eeq
In this case, we have a two term recurrence relation which we could solve directly in terms of $\G$-functions. Instead we will follow the procedure and expand this last relation in powers of $\e$. This yields the following relations
\beq\begin{split}
n^4 a_n &\= \+ 5(5n-1)(5n-2)(5n-3)(5n-4)a_{n-1}\\[10pt]
n^4 b_n &\= -4n^3 a_n + 1250(2n-1)(5n^2 - 5n +1) a_{n-1}\\
&\hskip20pt +5(5n-1)(5n-2)(5n-3)(5n-4) b_{n-1}\\[8pt]
n^4 c_n &\= -12n^2 a_n+ 1250(30n^2 - 30n + 7) a_{n-1}\\
&\hskip20pt -8 n^3 b_n + 2500(2n-1)(5n^2 - 5n +1)b_{n-1}\\
&\hskip20pt + 5(5n-1)(5n-2)(5n-3)(5n-4)c_{n-1}\\[8pt]
n^4d_n &\= -24n a_n + 37500(2n - 1)a_{n-1}\\
&\hskip20pt  - 36n^2 b_n + 3750(30n^2 - 30 n + 7)b_{n-1}\\
&\hskip20pt  - 12n^3c_n + 3750(2n-1)(5n^2 - 5n +1)c_{n-1}\\
&\hskip20pt + 5(5n-1)(5n-2)(5n-3)(5n-4)d_{n-1}~,
\end{split}\notag\eeq
subject to the initial conditions implied by $A_0(\e){\=}1$, so $a_0{\=}1$ and $b_0{\=}c_0{\=}d_0{\=}0$.

For uniformity of presentation we give the Hodge numbers
\beq
h^{pq}\=\begin{array}{ccccccc}
& & &1& & &\\
& &0&&0& & \\
&0&&101&&0& \\ 
1&&1&&1&&1~. \\
&0&&101&&0& \\ 
& &0&&0& & \\
& & &1& & &\\
\end{array}
\notag\eeq
For the quintic, so the manifold with $h^{11}{\=1}$, we have the numerical invariants
\beq
\chi\;= -200~,~~c_2 H\= 50~,~~H^3\= 5~.
\notag\eeq
The figure showing the numbers of factorisations into two quadrics appears as \fref{fig:Quintic} of~\sref{sec:intro}.

The discriminant for this manifold is simply $\D{\=} v - 5^5 u$ and the hyperdiscriminant is 
\beq
\IDelta\= 1~.
\notag\eeq
The only bad prime is the notorious prime $p{\=}5$, for which, as discussed in \sref{sec:discriminants}, the manifold is singular mod 5, for all values of $\vph$.
\newpage
\subsection{The mirror of a Calabi-Yau hypersurface in G(2,5), AESZ\hskip2pt 25}\label{sec:G25mfld}
\vskip-10pt
In \cite{Batyrev:1998kx} Batyrev, Ciocan-Fontanine, Kim and one of the present authors studied Calabi-Yau complete 
intersections in Grassmanians and their mirror manifolds.  One of the cases is the manifold $X$ with $h^{11}{\=}1,h^{21}{\=}61$,
realised as the intersection of hypersurfaces of degree $(1,2,2)$ in the six dimensional Grassmanian 
$G(2,5)$, in its Pl\"ucker embedding. Its characteristic numbers are:
\[
H^3\= 20,\;\;\;c_2H\= 68,\;\;\;\chi\;=-120~.
\]
The mirror manifold $\widetilde{X}$ can be described in terms of a toric Laurent polynomial 
\beq
F_0 \= 1-\vph f g~,
\notag\eeq 
\vspace*{-10pt}
with
\beq\begin{split}
f(Y)&\= \frac{(1+Y_1)^2 (1+Y_2)^2}{Y_1 Y_2}~,\\[8pt]
g(Y)&\= \left(1+\frac{1}{Y_3}\right) \left(1+\frac{1}{Y_4}\right)(1+ Y_3 + Y_4)~.
\end{split}\notag\eeq
The Newton polyhedron of $F_0$ is reflexive and has twenty vertices and nine facets. Five of the facets are hexahedra and four are prisms over a pentagonal base. The way in which these facets meet is sketched in \fref{fig:GraphicsRow}. The hexahedra are first stacked to form a column, then the top and bottom faces of the column are identified to form a ring. The five pentagonal prisms are also stacked and the top and bottom faces identified to form a second ring. The two rings are linked, as in the sub-figure on the right. The figure is deceptive insofar as it suggests that the Newton Polyhedron has a symmetry generated by rotating each of the two rings. There are, however, no matrices that act on the points of the Newton polyhedron that correspond to these operations.

The facets of the dual polyhedron are sketched in \fref{fig:cluster}. There are twenty facets and nine vertices. The twenty facets are all tetrahedra. One way to think of how these facets are arranged is to first join the tetrahedra in fives so as to form diamonds, as in the subfigure on the left, and then to join the four diamonds into a cluster, as on the right of the figure. The apparently exposed faces are then all identified in pairs. 

The variety corresponding to $F_0$ is singular with 6 nodes. A subdivision of the quadrilateral two-faces corresponds to a conifold resolution of these nodes and yields a manifold $X^*$ with the following Hodge~numbers. Thus $\chi(X^*){\=}2(h^{11}-h^{21}){\=}120$. 

\vspace*{-5pt}
\beq
h^{pq}(X^*)\=\begin{array}{ccccccc}
& & &1& & &\\
& &0&&0& & \\
&0&&61&&0& \\ 
1&&1&&1&&1~. \\
&0&&61&&0& \\ 
& &0&&0& & \\
& & &1& & &\\
\end{array}
\notag\eeq

\setlength{\fboxsep}{0.5cm} 
\begin{figure}[H]
\begin{center}
\framebox[\textwidth][l]{
\begin{minipage}{0.95\textwidth}
\vspace*{-0.3cm}\hskip-1cm
\includegraphics[width=6.3in]{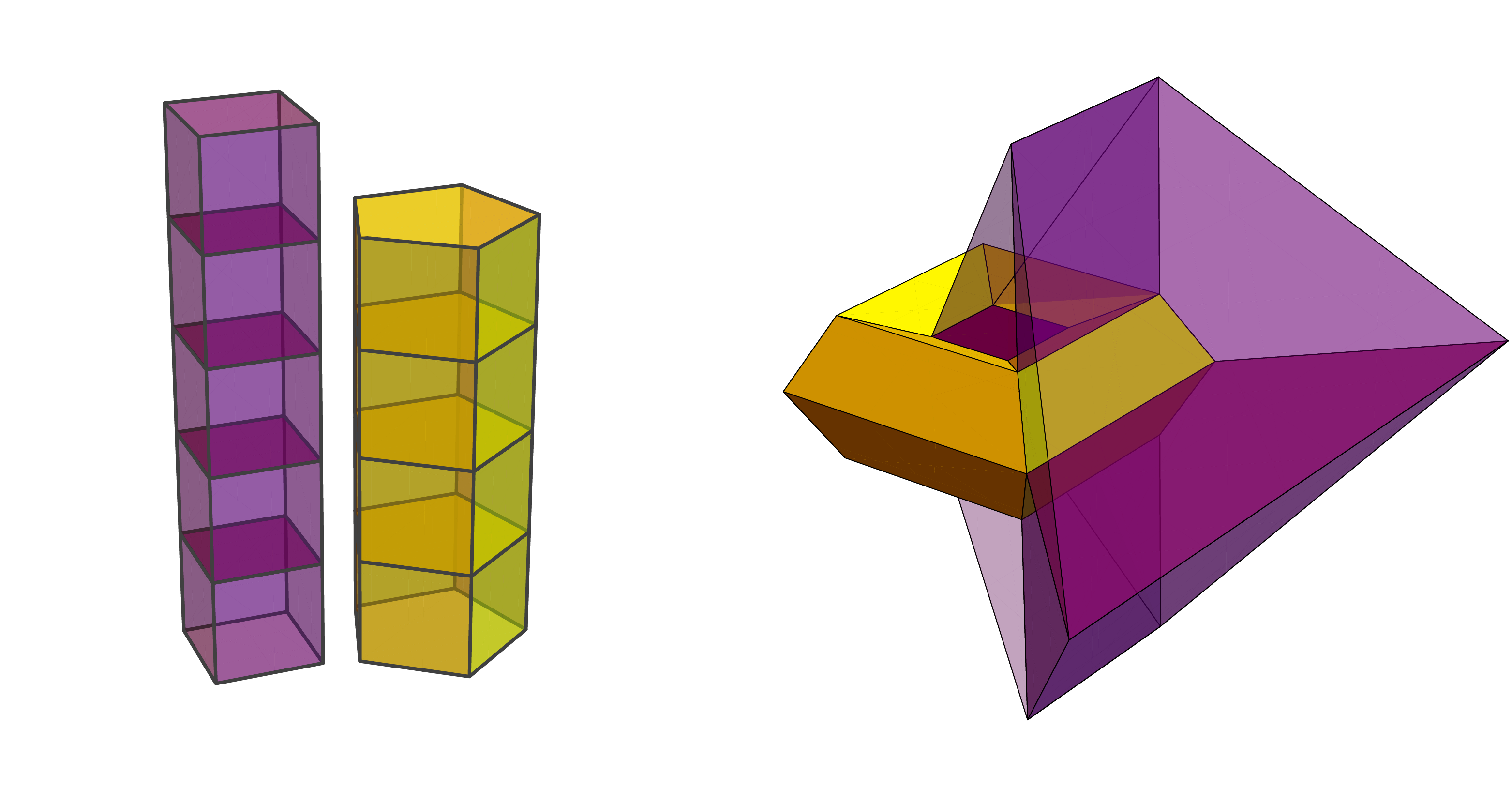}
\vspace*{-0.3cm}
\end{minipage}}
\capt{6in}{fig:GraphicsRow}{A sketch indicating how the facets of the toric polyhedron for $X$ fit together. There are five facets that are hexahedra and four that are prisms of pentagonal section.}
\end{center}
\end{figure}
\begin{figure}[H]
\begin{center}
\framebox[\textwidth][l]{
\hskip0.3cm
\raisebox{1.75cm}{\includegraphics[width=1.7in]{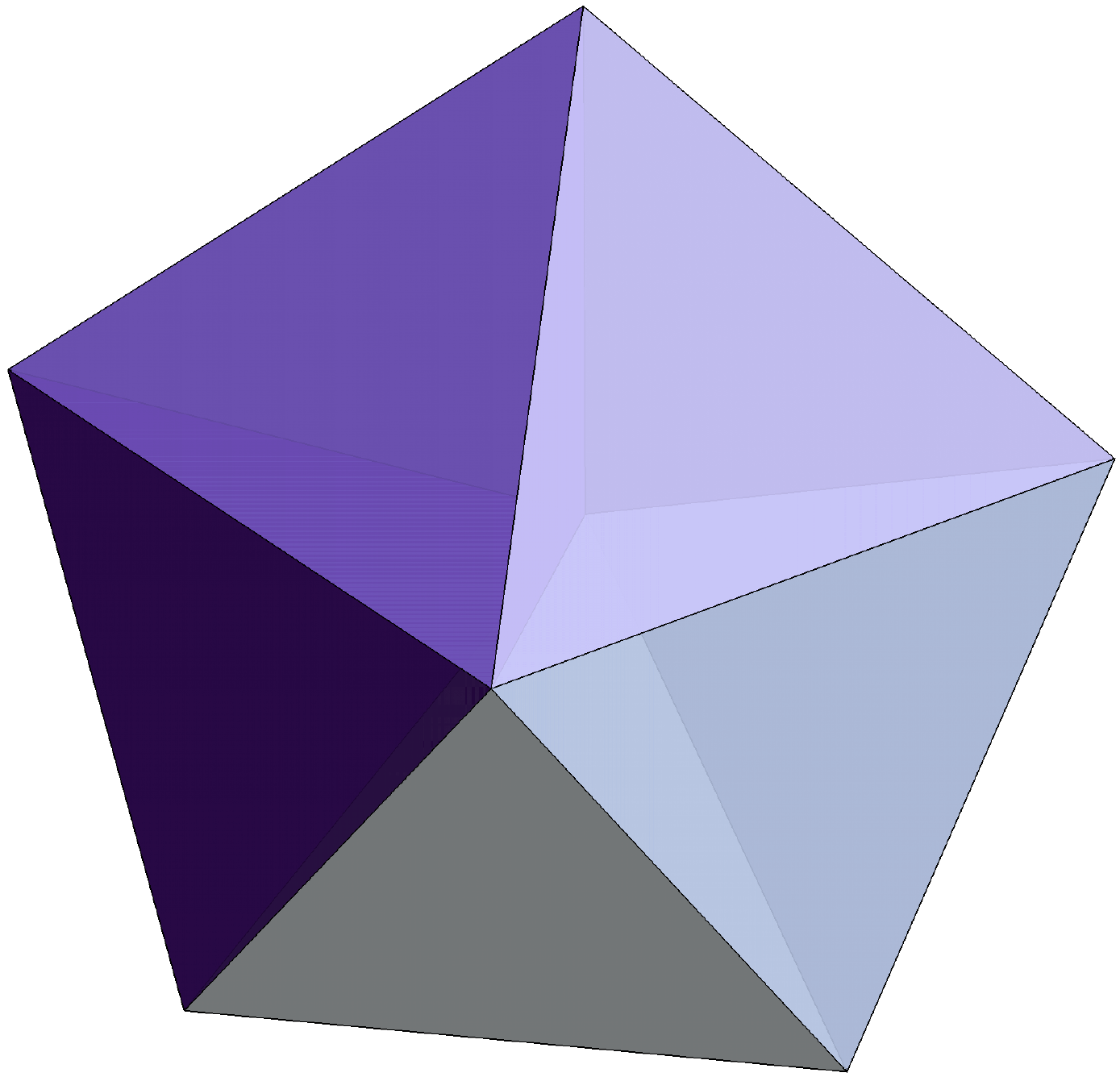}}\hskip2.0cm
\includegraphics[width=3.2in]{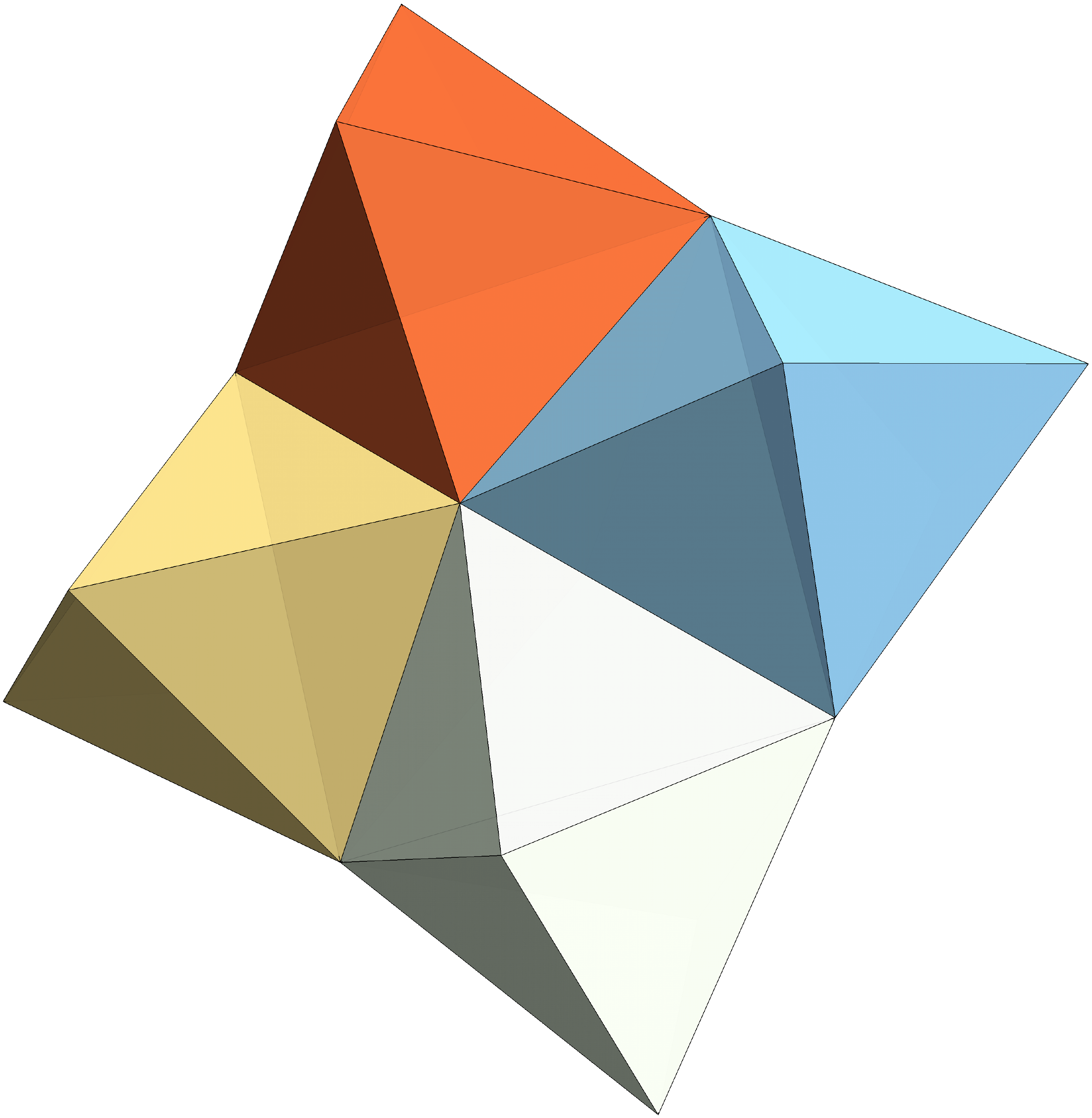}}
\capt{6in}{fig:cluster}{A sketch indicating how the facets of the dual polyhedron for $X$ fit together. There are now twenty facets that are all tetrahedra. These fit together in fives to form diamonds, as in the figure on the left, and four diamonds fit together as indicated in the figure on the right. Apparently exposed faces are identified, so that, in reality, there is no boundary.}
\end{center}
\end{figure}

The coefficients for the fundamental period, for this case, admit an interesting closed form
\beq
a_n\= \binom{2n}{n}^2\sum_{k=0}^n \binom{n}{k}^2 \binom{n+k}{n} ~.
\notag\eeq
which equals the famous Ap\'ery numbers (for $\zeta(2)$), multiplied by the square of the central binomial coefficients.
They satisfy the following three term recurrence relation
\beq
n^4 a_n \= 4 (2n-1)^2 (11n^2 - 11n +3)\, a_{n-1} + 16 (2n - 1)^2 (2n - 3)^2\, a_{n-2}~.
\notag\eeq 
The corresponding Picard-Fuchs operator
\beq
\cL\= \vth^4 - 4 \vph\left(2\vth + 1\right)^2 \left(11\vth^2 + 11\vth+3\right)-16\vph^2 \left(2\vth + 1\right)^2\left(2\vth + 3\right)^2~.
\notag\eeq
was listed as number $25$ in \cite{almkvist2005tables}. 
Its Riemann symbol is
\beq
\cP\left\{
\begin{array}{ccccc}
~0~&\hphantom{_{+}}\vph_{+}&\hphantom{_{-}}\vph_{-}&~\infty~\\[3pt]
\noalign{\hrule height 1pt width4.0cm}\\[-13pt]
0&0&0&\frac12\\[3pt]
0&1&1&\frac12\\[3pt]
0&1&1&\frac32\\[3pt]
0&2&2&\frac32\\[3pt]
\end{array}
\hskip5pt\vph~\right\}
\notag\eeq
The coefficient functions of the Picard-Fuchs operator are quadrics and 
$\vph_\pm{\=}\frac{1}{32}(-11{\pm}5\sqrt{5})$ are the roots of $S_4$. These exist in $\IF_p$ only if 5 is a square mod p, so when $p\;=\pm1$ mod 5.

The coefficient functions of the Picard-Fuchs operator are
\beq\begin{split}
S_4 &\= 256\,  \vph^2  + 176\,  \vph - 1~. \\[3pt]
S_3 &\= 32\vph (32\vph + 11)\\[3pt]
S_2 &\= \hphantom{1}4\vph (352\vph + 67) \\[3pt]
S_1 &\= \hphantom{1}4\vph (192\vph + 23)\\[3pt]
S_0 &\= 12\vph (12\vph + 1)~.
\end{split}\notag\eeq

The discriminant of the manifold is given by the homogenisation of $S_4$, above, after setting $\vph{\=}u/v$. The hyperdiscriminant is
\beq
\IDelta\= 2^8 5^3 v^2
\notag\eeq
so the bad primes are 2 and 5.
If we reduce $\D$ mod 2 we find
\beq
\D\= v^2 \mod 2
\notag\eeq
which has a repeated zero when $\vph{\=}\infty$. 
If we reduce mod 5 we have
\beq
\D\=(u-2v)^2 \mod 5~,
\notag\eeq
which has a repeated root when $\vph{\=}2$.\enlargethispage{\baselineskip}
\vskip30pt
\begin{figure}[H]
\begin{center}
\includegraphics[width=\textwidth]{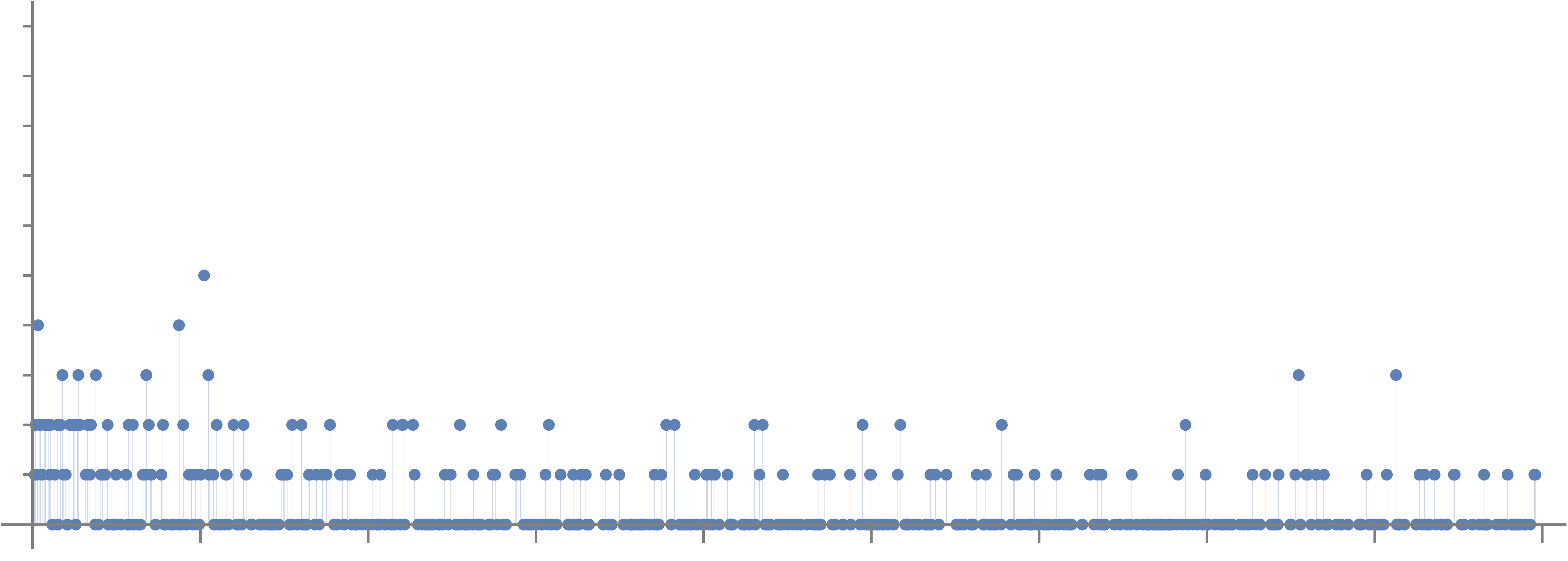}
\vskip0pt 
\place{-0.05}{0.54}{\scriptsize 1}
\place{-0.05}{0.75}{\scriptsize 2}
\place{-0.05}{0.96}{\scriptsize 3}
\place{-0.05}{1.16}{\scriptsize 4}
\place{-0.05}{1.37}{\scriptsize 5}
\place{-0.05}{1.57}{\scriptsize 6}
\place{-0.05}{1.78}{\scriptsize 7}
\place{-0.05}{1.99}{\scriptsize 8}
\place{-0.05}{2.20}{\scriptsize 9}
\place{-0.07}{2.41}{\scriptsize 10}
\place{0.73}{0.15}{\scriptsize 400}
\place{1.43}{0.15}{\scriptsize 800}
\place{2.10}{0.15}{\scriptsize 1200}
\place{2.80}{0.15}{\scriptsize 1600}
\place{3.49}{0.15}{\scriptsize 2000}
\place{4.19}{0.15}{\scriptsize 2400}
\place{4.89}{0.15}{\scriptsize 2800}
\place{5.59}{0.15}{\scriptsize 3200}
\place{6.29}{0.15}{\scriptsize 3600}
\capt{5.2in}{fig:G}{The plot shows the number of factorisations into two quadrics as $\vph$ varies over each $\IF_p$, $7\leqslant p\leq3583$, for the mirror of the hypersurface in $G(2,5)$.}

\end{center}
\end{figure}
\vskip5pt
\begin{figure}[H] 
\begin{center}
\includegraphics[width=3.6in]{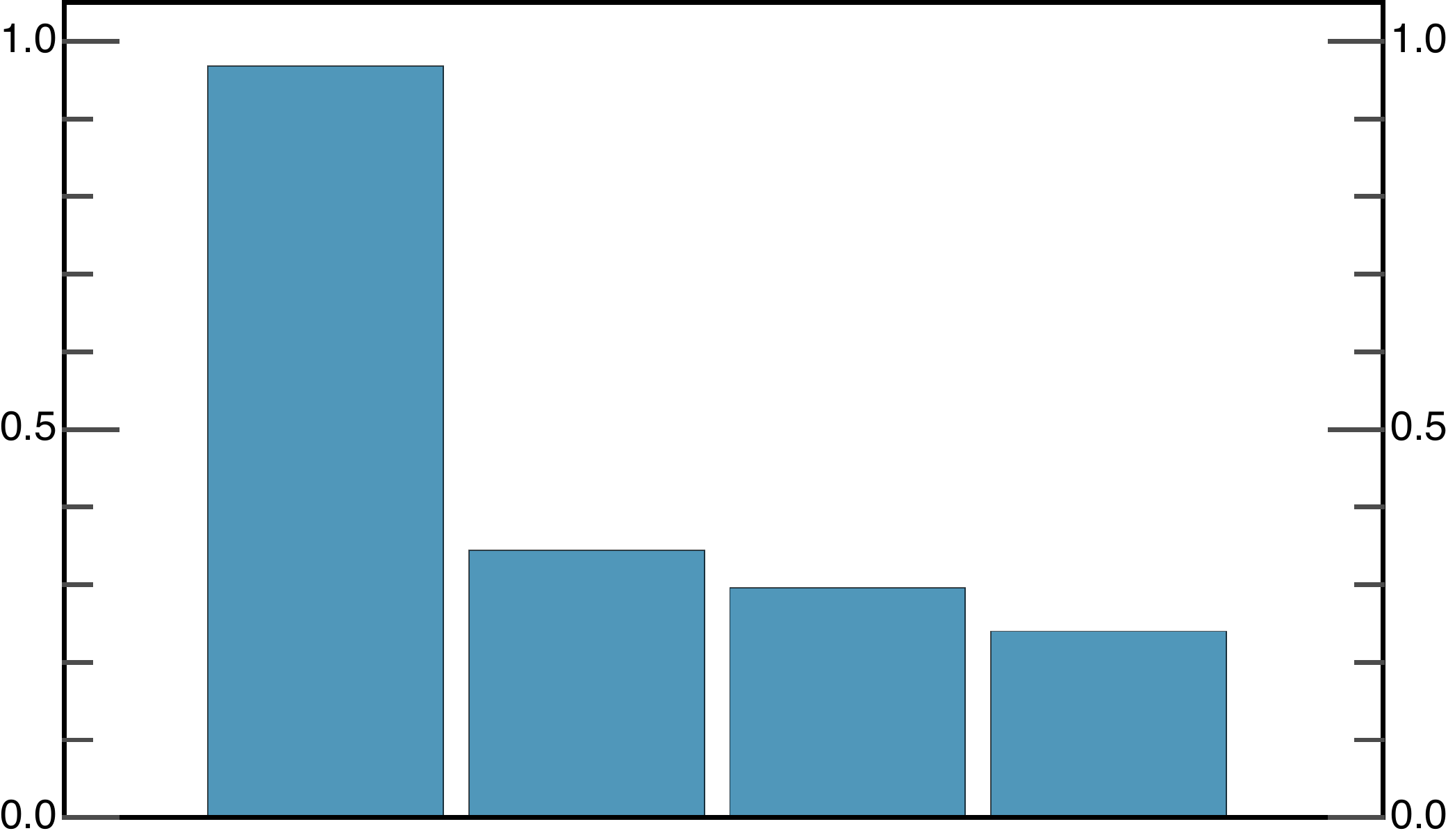}
\capt{5.2in}{fig:BarChartG}{Running averages for data taken from \fref{fig:G}. The averages are taken for bins of 125 primes.}
\end{center}
\end{figure}

The manifold has conifold singularities when the discriminant $S_4$ vanishes. These roots are 
\beq
\vph\;= -\frac{11}{32} \pm \frac{5\sqrt{5}}{32}
\notag\eeq
and exist in $\IF_{\! p}$ only when 5 is a square mod $p$, so, by quadratic reciprocity, when $p$ is a square mod 5, which is when $p{\=}1~\text{or}~4 \mod 5$. For these cases, the Frobenius polynomial factorises in the form \eqref{eq:ConifoldFactorisation} and we record the data corresponding to this factorization in \tref{tab:conifoldsG25}. The Table records the values of the character $\chi$ and the coefficients $\b_\pm$, corresponding to the roots $\vph_\pm$ of the discriminant. While $\b_{+}$ corresponds to $\vph_{+}$ and $\b_{-}$ to $\vph_{-}$, the labels $\pm$, however, have no intrinsic meaning and are assigned simply on the grounds that $\vph_{-}\,{<}\,\vph_{+}$, considered as integers. It was established by 
      A.\hspace{3pt}Thorne~\cite{ThorneThesis} that the $\b_\pm$ coefficients correspond to a [4,4] Hilbert modular form of weight [256,16,16]. 
      
      We are grateful to J.~Voight for pointing out to us the following simplification. The coefficients $\b_\pm$ agree up to sign, and for $p{\=}1~\text{or}~9\mod 20$ they have the same sign, so $\b_{+}{\=}\b_{-}$. Moreover, for $p{\=}1~\text{or}~4 \mod 5$, the $\b_\pm$ agree up to sign with the corresponding coefficients of the weight 4 and level 20 classical modular form with LMFDB denomination {\bf 20.4.c.a}. The coefficients in this expansion involve the quantity $\g{\=}2\sqrt{-19}$.
\beq\begin{split}
&f_{\bf 20.4.c.a}\=\\[3pt]
&q - \g q^{3} + (7 + \g) q^{5} + \g q^{7} - 49 q^{9} + 20 q^{11} + 6\g q^{13} + 
(76 - 7\g) q^{15} - 8\g q^{17} - 84 q^{19} + 76 q^{21} - \\[3pt]
&7\g q^{23} - (27 - 14\g) q^{25} + 22\g q^{27} + {\bf 6\,} q^{29} - 224 q^{31} - 20\g q^{33} -
(76 - 7\g) q^{35} - 14\g q^{37} + \\[3pt]
 & 456 q^{39} + {\bf 266\,} q^{41} + 35\g q^{43} - 49(7 + \g) q^{45} - 43\g q^{47} + 267 q^{49} - 608 q^{51} +42\g q^{53} + \\[3pt] 
 & 20(7 + \g) q^{55} + 84\g q^{57} - 28 q^{59} + {\bf 182\,} q^{61} - 49\g q^{63} -
   6 (76 - 7\g) q^{65} - 49\g q^{67} - 532 q^{69} + \\[3pt]
 &  408 q^{71} - 124\g q^{73} + (1064 + 27\g) q^{75} + 20\g q^{77} + 
 48 q^{79} + 349 q^{81} + 23\g q^{83} + 8(76 - 7\g) q^{85} - \\[3pt]
 & 6\g q^{87} - {\bf 1526\,} q^{89} - 456 q^{91} + 224\g q^{93} - 84(7 + \g) q^{95} - 64\g q^{97} - 
 980 q^{99} + {\bf 1246\,} q^{101} + \\[3pt] 
 & 97\g q^{103} + 76 (7+\g) q^{105} + 147 \g  q^{107} + {\bf 902\,} q^{109} +   \cO(q^{111}) ~.\\[-10pt]
\end{split}\notag\eeq
Notice that the coefficients of $q^p$, for $p{\=}29,\,41,\,61,\,89,\,101,\,109$ do indeed agree, up to sign, with the corresponding coefficients of \tref{tab:conifoldsG25}. Since, for $p{\=}1~\text{or}~9\mod 20$, we have $\b_{+}{\=}\b_{-}$, there is a well defined sign, which we denote by $\chi_{f,g}$, that relates the sign of the coefficient of $q^p$ in the modular form $f_{\bf 20.4.c.a}$ to that of the Hilbert modular form $g$ of the Table. This quantity is recorded in the third column of the Table. 

If $p{\;=}\pm1 \mod 5$, then, as has already been observed, 5 is a square mod $p$. Consider, in relation to this, the quantity
\beq
\tilde\chi \;= -5^\frac{p-1}{4} \mod p
\notag\eeq
mod $p$, we have $5{\=}z^2$, say. Apart from a sign, the right hand side of the above expression is $z^\frac{p-1}{2}$, which for $p{\,>\,}2$ is a square root of unity so $\pm1$ mod $p$. The values of $\tilde\chi$ are listed in the fourth column of the Table. From the Table, we see that $\tilde\chi$ repeats modulo 20. The set $\IZ/20\IZ$ has generators $\langle 11,\, 17\rangle$ so $\tilde\chi$ can be extended to all $p$, as a Dirichlet character, by assigning the values 
\beq
\tilde\chi(11)\;= -1~, \quad \tilde\chi(17)\= 1~,
\notag\eeq
which are consistent with the Table. We can similarly extend $\chi$, up to complex conjugation by assigning the values
\beq
\chi(11)\= 1~, \quad \chi(17) \= \ii~.
\notag\eeq

Now for $p{\=}1$ or $9$ mod 20. The coefficients $\b_\pm$ have the same sign and we can compare this to the sign of the corresponding coefficient in the modular form $f_{\bf 20.4.c.a}$. Let us define a quantity $\chi_{f,g}$ by taking
     $-\chi_{f,g}$ to be this ratio. The notation recalls that we are comparing the signs of the coefficients of the modular form $f$, with those of a Hilbert modular form $g$ of the Table. The quantity $\chi_{f,g}$ is recorded in the third column of the Table. Note that, where we know $\chi_{f,g}$, we have the relation
\beq
\chi_{f,g}\= \chi \tilde\chi~.
\notag\eeq 
\enlargethispage{\baselineskip}
\begin{table}
\def\skip{\hskip10pt}
\renewcommand{\arraystretch}{1.07}
\begin{center}
\begin{tabular}[H]{|c|c|c|c|c|>{\bfseries}c<{}|c|>{\bfseries}c<{}|}
\hline
\vrule height17pt depth12pt width0pt \skip$p$\skip{} & \skip$\chi$\skip{} & \hskip5pt$\chi_{f,g}$\hskip5pt{} 
& \skip$\tilde\chi$\skip{} & \skip\mbox{$\vph_{-}$}\skip{} 
& \skip$\b_{-}$\skip{} & \skip\mbox{$\vph_{+}$}\skip{} & \skip$\b_{+}$\skip{} \\
\hline\hline
 11 & -1\+ && -1\+ & 2 & 20 & 9 & -20\+ \\\hline
 19 &  1     && -1\+ & 3 & -84\+ & 7 & 84 \\\hline
 29 & -1\+ & -1\+ & 1 & 5 & 6 & 7 & 6 \\\hline
 31 & -1\+ && -1\+ & 19 & 224 & 21 & -224\+ \\\hline
 41 & 1 & 1 & 1 & 13 & -266\+ & 35 & -266\+ \\\hline
 59 & 1 && -1\+ & 15 & 28 & 47 & -28\+ \\\hline
 61 & 1 & 1 & 1 & 28 & -182\+ & 59 & -182\+ \\\hline
 71 & -1\+ && -1\+ & 60 & -408\+ & 68 & 408 \\\hline
 79 & 1 && -1\+ & 25 & -48\+ & 78 & 48 \\\hline
 89 & -1\+ & -1\+ & 1 & 36 & -1526\+ & 69 & -1526\+ \\\hline
 101 & 1 & -1\+ & -1\+ & 13 & 1246 & 81 & 1246 \\\hline
 109 & -1\+ & 1 &-1\+& 10 & -902\+ & 37 & -902\+ \\\hline
 131 & -1\+ &&-1\+& 36 & -2940\+ & 37 & 2940 \\\hline
 139 & 1 &&-1\+& 58 & -364\+ & 89 & 364 \\\hline
 149 & -1\+ & 1 &-1\+& 3 & -254\+ & 136 & -254\+ \\\hline
 151 & -1\+ &&-1\+& 46 & -2360\+ & 76 & 2360 \\\hline
 179 & 1 &&-1\+& 133 & -1972\+ & 146 & 1972 \\\hline
 181 & 1 & -1\+ &-1\+& 52 & -1330\+ & 117 & -1330\+ \\\hline
 191 & -1\+ &&-1\+& 87 & 1728 & 163 & -1728\+ \\\hline
 199 & 1 &&-1\+& 167 & -1512\+ & 193 & 1512 \\\hline
 211 & -1\+ &&-1\+& 23 & 1644 & 95 & -1644\+ \\\hline
 229 & -1\+ & -1\+ & 1 & 103 & 3934 & 111 & 3934 \\\hline
 239 & 1 &&-1\+& 124 & -3856\+ & 189 & 3856 \\\hline
 241 & 1 & 1 & 1 & 76 & -994\+ & 89 & -994\+ \\\hline
 251 & -1\+ &&-1\+& 5 & 6300 & 10 & -6300\+ \\\hline
 269 & -1\+ & 1 & -1\+ & 22 & 5082 & 95 & 5082 \\\hline
 271 & -1\+ &&-1\+& 130 & 2800 & 225 & -2800\+ \\\hline
 281 & 1 & 1 & 1 & 146 & 1254 & 187 & 1254 \\\hline
 311 & -1\+ &&-1\+& 98 & 5208 & 154 & -5208\+ \\\hline
 331 & -1\+ &&-1\+& 46 & -7828\+ & 305 & 7828 \\\hline
\end{tabular}
\vskip5pt
\capt{6.5in}{tab:conifoldsG25}{The discriminant $S_4$ factors in $\IF_p$ when $p{\,=\,}\pm1 \mod 5$. These roots are denoted by $\vph_{-}$ and $\vph_{+}$. The corresponding $\b$--coefficients are denoted by $\b_{-}$ and $\b_{+}$. These are the $p$'th coefficients of a [4, 4] Hilbert modular form of level [256, 16, 16]. The table gives also the values for the character $\chi$, and also, for $p{\;=\;}1~\text{or} ~9 \!\mod 20$, the relative sign $-\chi_{f,g}$ of the $\b$ coefficients in this table and in the modular form $f_{\bf 20.4.c.a}$.}
\end{center}
\end{table}
\newpage
\subsection{The Hulek--Verrill manifold, AESZ34}
\vskip-10pt
Hulek and Verrill in \cite{hulek_verrill_2005} consider the family of \cys that are birational to the variety defined on $T\= \IP^4\setminus\{X_1X_2X_3X_4X_5\=0\}$ by the equation
\beq
(X_1+X_2+X_3+X_4+X_5)
\left(\frac{\m_1}{X_1}+\frac{\m_2}{X_2}+\frac{\m_3}{X_3}+\frac{\m_4}{X_4}+\frac{\m_5}{X_5}\right)
\= \m_6~.
\notag\eeq
A multiplication of the coefficients $\m_j,\,j{\=}1,\ldots,6$ by a common scale has no effect, so superficially this equation defines a five parameter family of manifolds. The equation defines a reflexive polyhedron, in the sense of Batyrev. This polyhedron has, as noted by Hulek and Verrill, 10 facets that are tetrahedra and 20 that are prisms with triangular section. The way in which these fit together and a similar description of the dual polyhedron can be found in~\cite{hulek_verrill_2005} and in Appendix A of \cite{Candelas:2019llw}. Analysis of the polyhedron reveals that the superficial count of complex structure parameters is in fact correct and that the Hodge numbers for a generic member of the family are given by
\beq
h^{pq}\=\begin{array}{ccccccc}
& & &1& & &\\
& &0&&0& & \\
&0&&45&&0& \\ 
1&&5&&5&&1~. \\
&0&&45&&0& \\ 
& &0&&0& & \\
& & &1& & &\\
\end{array}
\notag\eeq
Thus $\chi{\=}2(h^{11}-h^{21}){\=}80$. 

If we consider now a 1-parameter subfamily where $\m_j{\=}1,\;j{\=}1,\ldots,5$ and $\m_6{\=}1/\vph$ then the manifold admits symmetries with $\IZ/5\IZ$ and $\IZ/2\IZ$ generators
\beq
X_i \;\to\; X_{i+1}~~~\text{and}~~~X_i \;\to\;1/X_i~,
\notag\eeq
where the indices are understood mod 5. It is easy to see that these symmetries are fixed point free if 
$\vph{\;\neq\;}1,\,\frac19,\,\frac{1}{25}$.  
Taking the quotient by either the $\IZ/5\IZ$ or $\IZ/10\IZ$ symmetry yields a family of smooth manifolds with one complex structure parameter and Hodge numbers
\beq
h^{pq}\=\begin{array}{ccccccc}
& & &1& & &\\
& &0&&0& & \\
&0&&\hskip-8pt 4\k{+}1\hskip-8pt{}&&0& \\ 
1&&1&&1&&1~,\\
&0&&\hskip-8pt 4\k{+}1\hskip-8pt{}&&0& \\ 
& &0&&0& & \\
& & &1& & &\\
\end{array}
\notag\eeq
where $\k{\=}1,\,2$ according as the quotient is taken by a group of order 10 or 5.

The coefficients of the fundamental period have a closed form expression
\[
a_n \;= \sum_{i+j+k+l+m=n} \left(\frac{n!}{i! j!k!l! m!}\right)^2~,
\]
which satisfy the recurrence relation
\beq\begin{split}
n^4 a_n \= &\left(35 n^4-70 n^3+63 n^2-28 n+5\right) a_{n-1} \\
                 & - (n-1)^2 \left(259 n^2 - 518 n + 285\right)  a_{n-2}\\
                 & +\,225  (n-1)^2 (n-2)^2 a_{n-3}~.\\
\end{split}\notag\eeq
corresponding to the Picard-Fuchs operator
\beq
\cL = \vth^4 - \vph \big(35 \vth^4{+}70 \vth^3{+}63 \vth^2{+}28 \vth{+}5\big) +
\vph^2(\vth+1)^2 \big(259 \vth^2 {+} 518 \vth{ +} 285\big) - 225 \vph^3  (\vth+1)^2 (\vth+2)^2~.
\notag\eeq

It was listed as operator 34 in \cite{almkvist2005tables}.
The Riemann symbol is
\beq
\cP\left\{
\begin{array}{ccccc}
0&\frac{1}{25}&\frac{1}{9}&~1~&\hskip-4pt\infty\hskip-4pt{}\\[3pt]
\noalign{\hrule height1pt width3.3cm}
\\[-14pt]
0&0&0&0&1\\[1pt]
0&1&1&1&1\\[1pt]
0&1&1&1&2\\[1pt]
0&2&2&2&2\\[3pt]
\end{array}
\hskip5pt\vph~\right\}
\notag\eeq
and the singular points at $\vph{\,=\,}\smallfrac{1}{25}, \smallfrac19, 1$ correspond to hyperconifold singularities. 
The coefficient functions of the Picard-Fuchs operator are the cubics
\beq\begin{split}
S_4 &\= (\vph - 1)(9\vph - 1)(25\vph - 1) \\[3pt]
S_3 &\= 2 \vph ( 675 \vph^2 - 518 \vph + 35 ) \\[3pt]
S_2 &\= \hphantom{2}\vph ( 2925 \vph^2 - 1580 \vph + 63 ) \\[3pt]
S_1 &\= 4 \vph ( 675 \vph^2 - 272 \vph + 7 ) \\[3pt]
S_0 &\= 5 \vph (180 \vph^2 - 57 \vph + 1 )~. 
\end{split}\notag\eeq

The figure showing the numbers of factorisations into two quadrics has been given in \sref{sec:intro}. 

We observe for this manifold, factorisations of the form \eqref{eq:ConifoldForm}, corresponding to conifold and hyperconifold singularities, and note that the $\b_p$ coefficients derive from modular forms of weight 4. These facts were known to Hulek and Verrill~\cite{hulek2005modularity} and to Meyer~\cite{DissertationMeyer}.

The discriminant of the manifold is given by the homogenisation of $S_4$, above. The hyperdiscriminant is
\beq
\IDelta\= 2^{20} 3^2 v^6~,
\notag\eeq
and we see that 2 and 3 are the bad primes. If we reduce $\D$ modulo these primes, we find
\beq
\D\=(u-v)^3 \mod 2~,
\notag\eeq
so $\D$ has a cubic root mod 2, and
\beq
\D\= 2v(u-v)^2 \mod 3~,
\notag\eeq
and we see that $\D$ has a double root mod 3.
\begin{table}
\begin{center}
\renewcommand{\arraystretch}{1.05}
\renewcommand{\+}{\hspace*{14pt}}
\begin{tabular}[H]{| >{$~} c <{~$} || >{$~} c <{~$} | >{$~} c <{~$}| >{$~} c <{~$}|}
\hline
\vrule height14pt width0pt\text{Singularity} & \frac19   & \frac{1}{25} & 1       \\[3pt]\hline
\vrule height14pt width0pt\text{Meyer label}        & 6/1         &     30/1         &   6/1  \\[3pt]\hline
\vrule height14pt width0pt\text{LMFDB}       &\bf 6.4.a.a &\bf 30.4.a.a &\bf 6.4.a.a \\[3pt]\hline\hline
p=5\hphantom{0}  &\bf 6  &\bf *  &\bf 6\\ \hline
p=7\hphantom{0}  &\bf -16\+  &\bf 32  &\bf -16\+\\ \hline
p=11  &\bf 12  &\bf -60\+  &\bf 12\\ \hline
p=13  &\bf 38  &\bf -34\+  &\bf 38\\ \hline
p=17  &\bf -126\+  &\bf 42  &\bf -126\+\\ \hline
p=19  &\bf 20  &\bf -76\+  &\bf 20\\ \hline
p=23  &\bf 168  &\bf 0  &\bf 168\\ \hline
p=29  &\bf 30  &\bf 6  &\bf 30\\ \hline
p=31  &\bf -88\+  &\bf -232\+  &\bf -88\+\\ \hline
p=37  &\bf 254  &\bf 134  &\bf 254\\ \hline
p=41  &\bf 42  &\bf 234  &\bf 42\\ \hline
p=43  &\bf -52\+  &\bf -412\+  &\bf -52\+\\ \hline
p=47  &\bf -96\+  &\bf -360\+  &\bf -96\+\\ \hline
p=53  &\bf 198  &\bf 222  &\bf 198\\ \hline
p=59  &\bf -660\+  &\bf 660  &\bf -660\+\\ \hline
p=61  &\bf -538\+  &\bf -490\+  &\bf -538\+\\ \hline
p=67  &\bf 884  &\bf 812  &\bf 884\\ \hline
p=71  &\bf 792  &\bf 120  &\bf 792\\ \hline
p=73  &\bf 218  &\bf 746  &\bf 218\\ \hline
p=79  &\bf -520\+  &\bf 152  &\bf -520\+\\ \hline
p=83  &\bf -492\+  &\bf -804\+  &\bf -492\+\\ \hline
p=89  &\bf 810  &\bf -678\+  &\bf 810\\ \hline
p=97  &\bf 1154  &\bf 194  &\bf 1154\\ \hline
p=101  &\bf -618\+  &\bf 798  &\bf -618\+\\ \hline
p=103  &\bf 128  &\bf 1088  &\bf 128\\ \hline
p=107  &\bf -1476\+  &\bf 1716  &\bf -1476\+\\ \hline
p=109  &\bf 1190  &\bf -970\+  &\bf 1190\\ \hline
p=113  &\bf -462\+  &\bf 426  &\bf -462\+\\ \hline
p=127  &\bf -2536\+  &\bf 200  &\bf -2536\+\\ \hline
p=131  &\bf 2292  &\bf 60  &\bf 2292\\ \hline
p=137  &\bf -726\+  &\bf 642  &\bf -726\+\\ \hline
\end{tabular}
\vskip15pt
\capt{6in}{tab:AESZ34}{\it For the manifold corresponding to AESZ34, the coefficient $\b_p$ for the characteristic polynomial of Frobenius for the cases that the manifold has (hyper-) conifold singularities.}
\end{center}
\end{table}
%
%
\begin{table}
\begin{center}
\begin{tabular}[H]{lr}
\begin{minipage}{2.2in}
\begin{center}
\renewcommand{\arraystretch}{0.98}
\begin{tabular}[H]{|c|>{\bfseries}c|>{\bfseries}c|}
\hline
\multicolumn{3}{|c|}{\vrule height16pt depth10pt width0pt$\vph\;= -\frac17$}\\
\hline
\vrule height13pt depth8pt width0pt$~~p~~$ & $~~\a~~$ & $\b$ \\\hline\hline
 5 & 0 & -14\+ \\\hline 
 7 &  &   \\\hline 
 11 & 0 & -28\+ \\\hline  
 13 & -4\+ & 18 \\\hline  
 17 & 6 & 74 \\\hline  
 19 & 2 & 80 \\\hline  
 23 & 0 & -112\+ \\\hline  
 29 & -6\+ & 190 \\\hline  
 31 & -4\+ & 72 \\\hline  
 37 & 2 & -346\+ \\\hline  
 41 & 6 & 162 \\\hline  
 43 & 8 & -412\+ \\\hline  
 47 & -12\+ & 24 \\\hline  
 53 & 6 & 318 \\\hline  
 59 & -6\+ & -200\+ \\\hline  
 61 & 8 & -198\+ \\\hline  
 67 & -4\+ & -716\+ \\\hline  
 71 & 0 & 392 \\\hline  
 73 & 2 & 538 \\\hline  
 79 & 8 & 240 \\\hline  
 83 & -6\+ & -1072\+ \\\hline  
 89 & -6\+ & 810 \\\hline  
 97 & -10\+ & 1354 \\\hline  
 101 & 0 & -1358\+ \\\hline  
 103 & -4\+ & -832\+ \\\hline  
 107 & 12 & 444 \\\hline  
 109 & 2 & 1870 \\\hline  
 113 & 6 & 1378 \\\hline  
 127 & -16\+ & 1944 \\\hline  
 131 & 18 & -848\+ \\\hline  
 137 & 18 & -2966\+ \\\hline 
 \end{tabular}
\end{center}
\end{minipage}
&
\begin{minipage}{2.2in}
\begin{center}
\renewcommand{\arraystretch}{0.98}
\begin{tabular}[H]{|c|>{\bfseries}c|>{\bfseries}c|}
\hline
\multicolumn{3}{|c|}{\vrule height16pt depth10pt width0pt$\vph\;=\; 33\pm 8\sqrt{17}$}\\
\hline
\vrule height13pt depth8pt width0pt$~~p~~$ & $~~\a~~$ & $\b$ \\\hline\hline
 13 & 2 & -42\+ \\\hline
 17 & -6\+ & 34 \\\hline  
 19 & -4\+ & 60 \\\hline  
 43 & -4\+ & 508 \\\hline  
 47 & 0 & -136\+ \\\hline  
 53 & 6 & 318 \\\hline  
 59 & 12 & 300 \\\hline  
 67 & -4\+ & -676\+ \\\hline  
 83 & -12\+ & -1132\+ \\\hline  
 89 & 6 & -350\+ \\\hline  
 101 & -6\+ & -1218\+ \\\hline  
 103 & 8 & 8 \\\hline  
 127 & -16\+ & -1216\+ \\\hline  
 137 & -18\+ & 1954 \\\hline  
 149 & 6 & -1010\+ \\\hline  
 151 & 8 & -968\+ \\\hline  
 157 & 14 & 1654 \\\hline  
 179 & 12 & -980\+ \\\hline  
 191 & 0 & 952 \\\hline  
 223 & -16\+ & -712\+ \\\hline  
 229 & -22\+ & 5230 \\\hline  
 239 & 0 & 2040 \\\hline  
 251 & -12\+ & -5868\+ \\\hline  
 257 & 6 & -4646\+ \\\hline  
 263 & 24 & -6472\+ \\\hline  
 271 & -16\+ & 8312 \\\hline  
 281 & 18 & -518\+ \\\hline  
 293 & 6 & -6402\+ \\\hline  
 307 & 20 & -3516\+ \\\hline  
 331 & -4\+ & 2892 \\\hline  
 349 & -34\+ & 5270 \\\hline 
\end{tabular}
\end{center}
\end{minipage}
\end{tabular}
\vskip1pt
\capt{6.2in}{tab:AttractorCoeffs}{The $(\a,\,\b)$-coefficients for the attractor points $\vph{\;=}-\frac17$ and $\vph{\;=\;}33{\;\pm\;}8\sqrt{17}$. For $\vph{\;=}-\frac17$ the $\a_p$ are the p'th coefficients of a weight 2 modular form, with LMFDB designation {\bf 14.2.a.a}, for $\G_0(14)$. The coefficients $\b_p$ are the p'th coefficients of a weight four modular form, with designation {\bf 14.4.a.a}, also for $\G_0(14)$. For $\vph{\;=\;}33\pm 8\sqrt{17}$, with the exception of $p{\;=\;}17$, the correspondence is for primes such that  $\vph^2 {\,-\,} 66\vph{\,+\,}1$ factorises in $\IF_p$. For these primes, the $\a_p$ are the p'th coefficients of a weight two modular form, with designation {\bf 34.2.b.a} and the $\b_p$ are the p'th coefficients the weight 4 modular form {\bf 34.4.4.a}, both for the group~$\G_1(34)$.}
\end{center}
\end{table}
\newpage
\subsection{The \Rodland manifold, AESZ 27}\label{sec:Rmfld}
\vskip-10pt
The \Rodland manifold,which is defined by the vanishing of the $7$ cubic Pfaffians of a generic anti-symmetric $7 \times 7$ matrix of linear forms in $7$ variables, is a Calabi-Yau threefold $X$ with $h^{11}{\=}1$ and $h^{21}{\=}440$ and characteristic numbers
\[
H^3 \= 378,\;\;\;c_2H\;=-252,\;\;\;\chi\;=-882~.
\]
\Rodland in ref.~\cite{Rodland:1998pm}, which also describes the manifold,  constructed a mirror manifold $\widetilde{X}$ by an orbifold construction and determined the associated Picard-Fuchs equation. This turned out the be the same operator as obtained in \cite{Batyrev:1998kx} for the mirror of Calabi-Yau obtained intersection of $7$ linear hyperplanes in the $10$-dimensional Grassmanian $G(2,7)$. So the very different Pfaffian and Grassmanian Calabi-Yau threefolds turn out to have the same mirror manifold, which led to to the suspicion that these manifolds have equivalent categories of coherent sheaves, a fact proved in \cite{BorisovCaldararu}.

The coefficient functions of the differential equation take the following form
\beq\begin{split}
&S_4\= (\vph - 3)^2 \left(\vph^3-289 \vph^2-57 \vph +1\right) \\[2pt]
&S_3\= 4\vph\, (\vph -3) \left(\vph^3-149 \vph^2+867 \vph +85\right) \\[2pt]
&S_2\= 2 \vph  \left(3 \vph^4-239 \vph^3+2353 \vph^2-7597 \vph -408\right) \\[2pt]
&S_1\= 2 \vph  \left(2 \vph^4-87 \vph^3+675 \vph^2-4773 \vph -153\right) \\[2pt]
&S_0\= \hphantom{2} \vph  \left(\vph^4-26 \vph^3+12 \vph^2-2166\vph -45\right) \\[2pt]
\end{split}\notag\eeq

The Riemann symbol is
\beq
\cP\left\{
\begin{array}{cccccc}
~0~&~3~&~\raisebox{1pt}{$\vph_1$}~&~\raisebox{1pt}{$\vph_2$}~&~\raisebox{1pt}{$\vph_3$}~&~\infty~ \\[1pt]
\noalign{\hrule height1pt width 5.8cm}\\[-14pt]
0&0&0&0&0&1\\[2pt]
0&1&1&1&1&1\\[2pt]
0&3&1&1&1&1\\[2pt]
0&4&2&2&2&1\\[2pt]
\end{array}
\hskip5pt\vph~\right\}~.
\notag\eeq
where the $\vph_j$, $j=1,2,3$, are the roots of $\vph^3-289 \vph^2-57 \vph +1$. Note that the Picard-Fuchs operator has two points of maximal unipotent monodromy: the degeneration at $0$ corresponds to the mirror of the Pfaffian, the point $\infty$ corresponds to the mirror of the Grassmanian Calabi-Yau manifold.

The coefficients of the fundamental period satisfy the following recurrence relation
\beq\begin{split}
9n^4\, a_n\=&\hskip15pt 3\left(173 n^4-352n^3+290 n^2-114 n+18\right)\, a_{n-1}\\
&+2 \left(1129 n^4-4000 n^3+4501 n^2-1359 n-267\right)\, a_{n-2}\\
&-2 \left(843 n^4-7488 n^3+24223 n^2-33531n+16485\right)\, a_{n-3}\\
&+\left(295 n^4-4112 n^3+21502 n^2-49986 n+43586\right)\, a_{n-4}\\
&-(n-4)^4\, a_{n-5}~.
\end{split}\notag\eeq

\vskip30pt
\begin{figure}[H]
\begin{center}
\includegraphics[width=\textwidth]{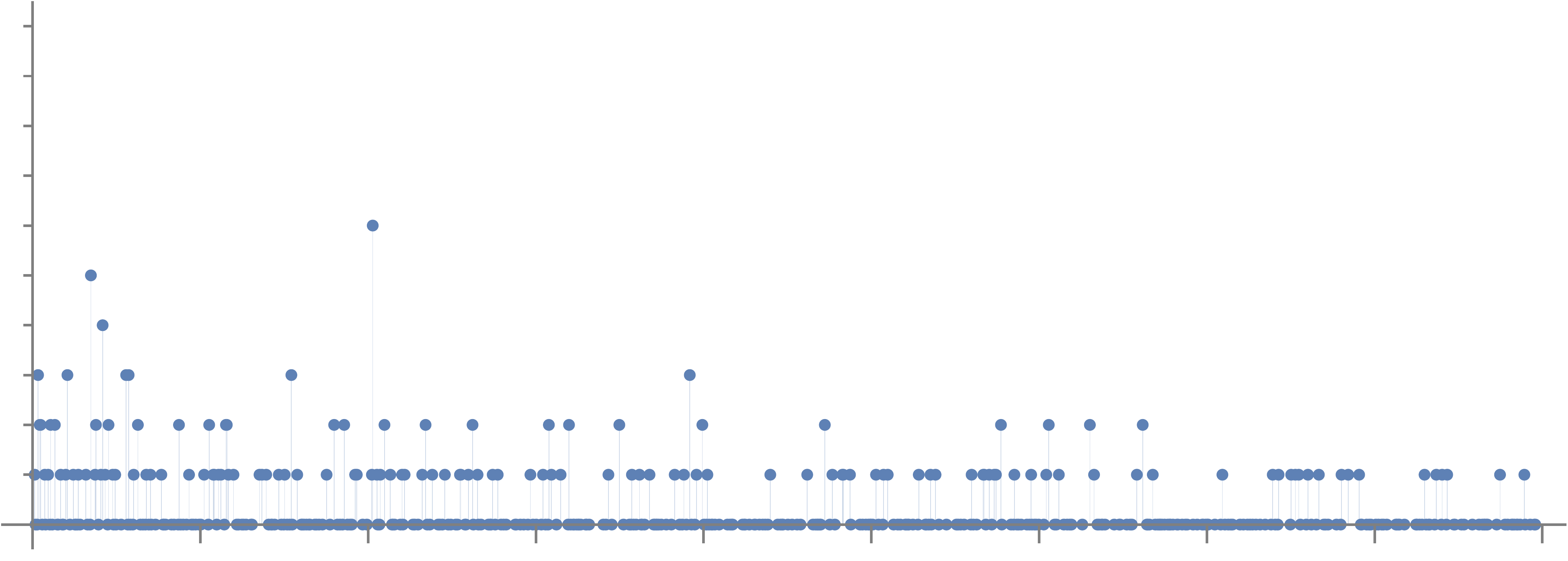}
\vskip0pt 
\place{-0.05}{0.54}{\scriptsize 1}
\place{-0.05}{0.75}{\scriptsize 2}
\place{-0.05}{0.96}{\scriptsize 3}
\place{-0.05}{1.16}{\scriptsize 4}
\place{-0.05}{1.37}{\scriptsize 5}
\place{-0.05}{1.57}{\scriptsize 6}
\place{-0.05}{1.78}{\scriptsize 7}
\place{-0.05}{1.99}{\scriptsize 8}
\place{-0.05}{2.20}{\scriptsize 9}
\place{-0.07}{2.41}{\scriptsize 10}
\place{0.73}{0.15}{\scriptsize 400}
\place{1.43}{0.15}{\scriptsize 800}
\place{2.10}{0.15}{\scriptsize 1200}
\place{2.80}{0.15}{\scriptsize 1600}
\place{3.49}{0.15}{\scriptsize 2000}
\place{4.19}{0.15}{\scriptsize 2400}
\place{4.89}{0.15}{\scriptsize 2800}
\place{5.59}{0.15}{\scriptsize 3200}
\place{6.29}{0.15}{\scriptsize 3600}
\vskip-5pt
\capt{4.8in}{fig:Rodland}{The figure shows the number of factorisations of the numerator of the $\z$-function into two factors, for the \Rodland manifold.}
\end{center}
\end{figure}
\begin{figure}[H] 
\begin{center}
\includegraphics[width=3.6in]{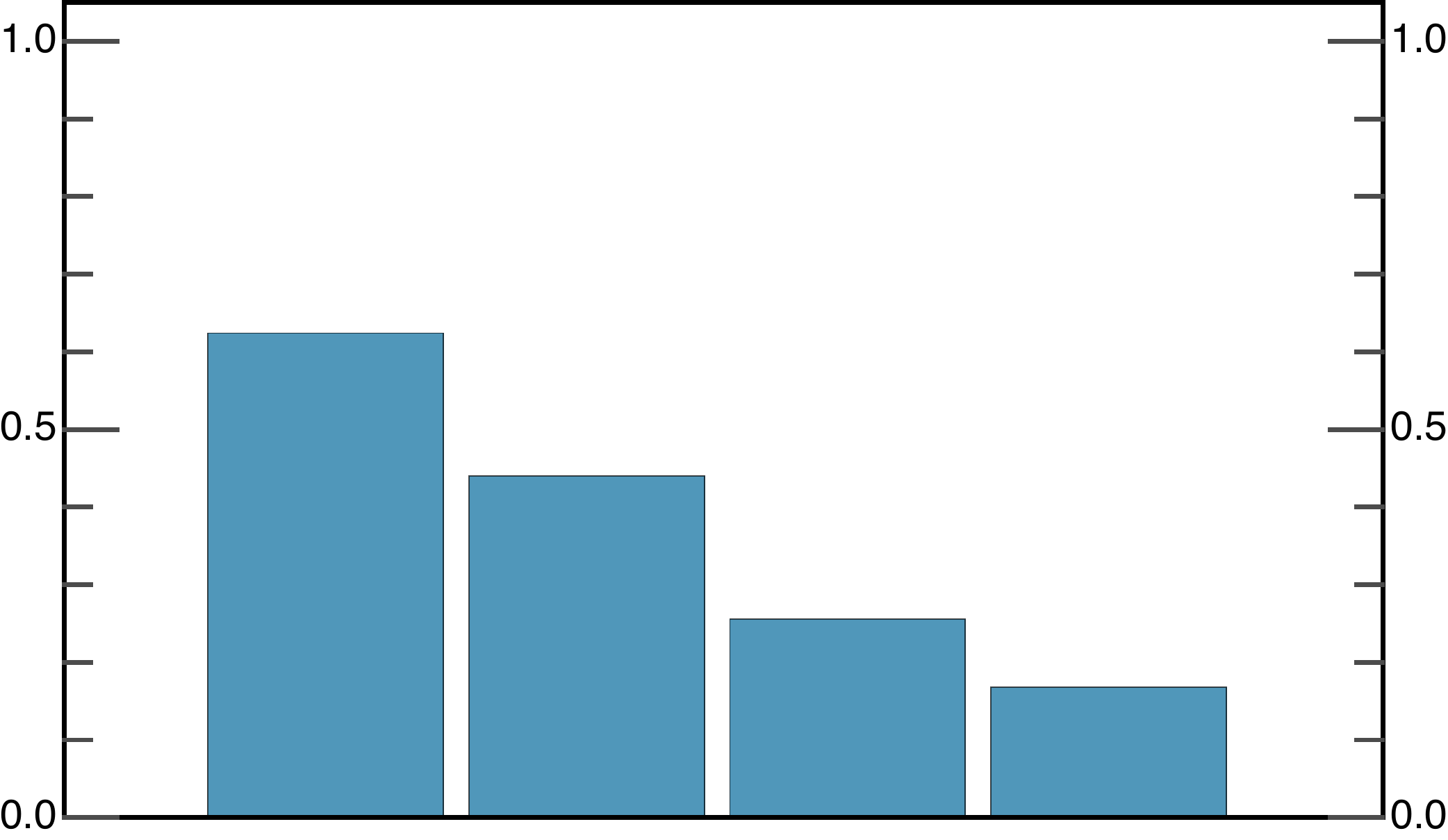}
\vskip-5pt
\capt{4.8in}{fig:BarChartR}{Running averages for the data of the three generation manifold taken from 
     \fref{fig:Rodland}. The averages are taken for bins of 125 primes for the 500 primes $5\leqslant p\leqslant 3583$.}
\end{center} 
\end{figure}
\begin{table}[H]
\def\skip{\hskip10pt}
\renewcommand{\arraystretch}{1.1}
\begin{center}
\begin{tabular}[H]{|>{$}c<{$}|>{$}c<{$}|>{$}c<{$}|}
\hline
\vrule height15pt depth10pt width0pt \skip p \skip{} & 3  & \skip R \skip{} \\
\hline\hline
5   & 3 & 1 - 2 T - 20 pT - 2p^3 T^3 + p^6 T^4 \\\hline
7   & 3 & \text{no unit root}                              \\\hline
11 & 3 &  1 + 34 T + 238 pT + 34 p^3 T^3 + p^6 T^4\\\hline
13 & 3 &  1 + 14 T + 142 pT + 14 p^3 T^3 + p^6 T^4\\\hline
17 & 3 &  1 + 58 T + 396 pT + 58 p^3 T^3 + p^6 T^4\\\hline
\end{tabular}
\capt{4.8in}{tab:UnitRootsR}{The Frobenius polynomials for the primes $5,\,7,\,11,\,13$ at the apparent singularity $\vph{\;=\;}3$. These are calculated by the unit root method.}
\end{center}
\end{table}

With this manifold we come to an example that has an apparent singularity. The hyperdiscriminant of the manifold is
\beq
\IDelta\= 2^6 7^8 v^6~,
\notag\eeq
while the hyperdiscriminant of the differential equation is
\beq
\IS_4\= 2^{12} 7^{14} v^{12}~.
\notag\eeq
Reducing $S_4$ modulo the bad primes, we find
\beq\begin{split}
S_4&\= (u-\phantom{3}v)^5 \mod 2~, \\[3pt]
&\= (u-3v)^5 \mod 7~.
\end{split}\notag\eeq

Finally, we wish to make some remarks regarding the splitting field of the discriminant
\beq
\D(\vph) \= \vph^3 - 289\vph^2 - 57\vph +1
\notag\eeq
since these are relevant to understanding the factorisations of the Frobenius polynomial at the three conifold points.

First note that setting
\beq
\vph \= \frac{-2\x + 1}{4\x+5}
\notag\eeq
brings the discriminant to the form
\beq
\D \;= -2744 \frac{g(\x)}{(4\x+5)^3} \quad \text{where} \quad g(\x)\= \x^3 - \x^2 - 2\x +1~.
\notag\eeq
So the three roots $\vph_j$, $j{\=}1,2,3$, of $\D$ are given by $\vph_j{\=}\vph(\x_j)$, in terms of the three roots $\x_j$ of $g$. It is easy to check that the $\x_j$ are
\beq
2\cos\frac{5\p}{7},\; 2\cos\frac{3\p}{7},\; 2\cos\frac{\p}{7}~.
\notag\eeq
Let $\z$ denote \emph{any} nontrivial seventh root of unity, then it is a quick check that the quantities
\beq
-(\z+\z^{-1}),\; -(\z^2+\z^{-2}),\; -(\z^3+\z^{-3})
\notag\eeq
are the three roots above, up to cyclic order. We learn that the splitting field of $g$ is $\IQ(\z+\z^{-1})$ and that the Galois group of $g$ is $\IZ/3\IZ$. While, for a generic cubic, the Galois group is $S_3$, the permutation group on three objects.

Note also that the bad prime 7 factors nontrivially in this field, since if  $\n_j {\=} 2 + \x_j$ we have
\beq
7 \= \n_1 \n_2 \n_3~.
\notag\eeq
Each $\n_j$ is an integer of the field and, by the above relation, has field-norm 7. Since the norm of an integer of the field is a rational integer, the $\n_j$ cannot themselves factor nontrivially and so are primes of the field. In fact the $\n_j$ are equal, up to units, so the factorisation of 7 is best expressed in terms of ideals. The ideals $(\n_j)$ are all equal so let ${\mathfrak P}$ denote this common ideal, and let $\cO$ denote the ring of integers of the field, then
\beq
7\,\cO \= {\mathfrak P}^3~.
\notag\eeq
On the other hand, the prime 2 does not factor nontrivially.

J.\hspace{3pt}Voight has recognised the $\b$-coefficients of \tref{tab:Rodland} as corresponding to a Hilbert modular form for the field $\IQ(\z+\z^{-1})$, with weight $[4,4,4]$ and level $2{\mathfrak P}$.
\begin{table}
\begin{center}
\renewcommand{\arraystretch}{1.05}
\begin{tabular}[H]{|c||c||c|>{\bfseries}c||c|>{\bfseries}c||c|>{\bfseries}c|}
\hline
\vrule height14pt depth10pt width0pt $p$ & $~\chi~$ & $\vph_1$ & $\b_1$ & $\vph_2$ & $\b_2$ & $\vph_3$ & $\b_3$ \\[3pt]
\hline\hline
 13 & 1 & 2 & -42\+ & 5 & 42 & 9 & -14\+ \\ \hline
 29 & 1 & 12 & -306\+ & 17 & -110\+ & 28 & 282 \\ \hline
 41 & 1 & 23 & 70 & 27 & -70\+ & 34 & -434\+ \\ \hline
 43 & 1 & 1 & 128 & 36 & -264\+ & 37 & -68\+ \\ \hline
 71 & 1 & 5 & 604 & 25 & -180\+ & 46 & -376\+ \\ \hline
 83 & 1 & 10 & -1148\+ & 14 & 1372 & 16 & 84 \\ \hline
 97 & 1 & 2 & 854 & 19 & -546\+ & 74 & -798\+ \\ \hline
 113 & 1 & 42 & -54 & 65 & -1818\+ & 69 & -642\+ \\ \hline
 127 & 1 & 19 & 1556 & 68 & -404\+ & 75 & 2536 \\ \hline
 139 & 1 & 41 & -812\+ & 119 & -756\+ & 129 & -2100\+ \\ \hline
 167 & 1 & 60 & 112 & 110 & -3696\+ & 119 & -1288\+ \\ \hline
 181 & 1 & 45 & -1498\+ & 102 & -3850\+ & 142 & 2030 \\ \hline
 197 & 1 & 68 & -1762\+ & 93 & 3138 & 128 & 2 \\ \hline
 211 & 1 & 111 & 2396 & 188 & 1024 & 201 & -3680\+ \\ \hline
 223 & 1 & 9 & 3864 & 130 & -2912\+ & 150 & 392 \\ \hline
 239 & 1 & 107 & 4832 & 188 & 2872 & 233 & -7320\+ \\ \hline
 251 & 1 & 46 & 5236 & 57 & -5460\+ & 186 & 1148 \\ \hline 
 281 & 1 & 130 & -3078\+ & 219 & 2214 & 221 & -138\+ \\ \hline
 293 & 1 & 25 & -8050\+ & 59 & 1638 & 205 & -1722\+ \\ \hline
 307 & 1 & 112 & -8036\+ & 180 & 5796 & 304 & 10164 \\ \hline
 337 & 1 & 174 & 1794 & 220 & 11006 & 232 & -6242\+ \\ \hline
 349 & 1 & 169 & 10346 & 173 & -9310\+ & 296 & -5306\+ \\ \hline
 379 & 1 & 182 & -2980\+ & 194 & -12388\+ & 292 & -1020\+ \\ \hline
 419 & 1 & 170 & 13524 & 262 & 6804 & 276 & -4676\+ \\ \hline
 421 & 1 & 73 & 14002 & 293 & 4398 & 344 & 9298 \\ \hline
 433 & 1 & 23 & 2562 & 126 & 10094 & 140 & -11914\+ \\ \hline
 449 & 1 & 24 & -1398\+ & 349 & 366 & 365 & -7866\+ \\ \hline
 461 & 1 & 126 & 8106 & 214 & 12250 & 410 & -2002\+ \\ \hline
 463 & 1 & 128 & -6844\+ & 229 & 2172 & 395 & 9032 \\ \hline
 491 & 1 & 159 & 9116 & 186 & 5980 & 435 & 13428 \\ \hline
 503 & 1 & 109 & 18144 & 310 & -2744\+ & 373 & -8064\+ \\ \hline
 547 & 1 & 28 & 25356 & 308 & -11492\+ & 500 & 7324 \\ \hline
 587 & 1 & 27 & -11004\+ & 116 & -3444\+ & 146 & 140 \\ \hline
\end{tabular}
\vskip11pt
\capt{5.2in}{tab:Rodland}{For the \Rodland manifold, the character, which is in all cases trivial, together with the coefficients $\b_p$, for the three roots of the discriminant for the cases that the roots exist in $\IF_p$, which is when $p{\;=}\pm1 \mod 7$.}
\end{center}
\end{table}
\newpage
\subsection{Three-generation manifolds with $\hodgenos{\=}(4,1)$}
\vskip-10pt 
These manifolds may be realised as the mirror manifolds of quotients of hypersurfaces in $\text{dP}_6{\times}\text{dP}_6$, where $\text{dP}_6$ is the del Pezzo surface of degree 6, by a freely acting group $G$ of order 12. There are, in reality, two variants of this `manifold' since there are two choices for $G$, which is either $\IZ/12\IZ$ or the nonabelian group $\text{Dyc}_3$. Before taking the mirror, this space can be realised as the CICY~\cite{Candelas:1987kf, Hubsch:1992nu}
\beqnn
\cicy{\IP^2\\ \IP^2\\ \IP^2\\ \IP^2\\}{1& 1& 1& 0& 0\\
                                                          0& 0& 1& 1& 1\\
                                                          1& 1& 1& 0& 0\\
                                                          0& 0& 1& 1& 1\\}^{(1,4)}\hskip-22pt\raisebox{-30pt}{/G}
\eeqnn
or as a reflexive polyhedron, corresponding to the fact that the covering space is a hypersurface in the toric variety $\text{dP}_6{\times}\text{dP}_6$. A detailed description of the manifolds can be found in
      Refs.~\cite{Braun:2009qy, Candelas:2008wb}. For the basic invariants of $X_\vph$ we have
\beq
\hodgenos \= (4,1)~; \quad\ch\= 6~;\quad c_2 H\= 12~; \quad H^3\= 18~.
\notag\eeq

From~\cite{Braun:2009qy} we have that the fundamental period is given by the following integral representation
\beqnn
\vp_0(\vph)\= \frac{1}{(2\p\ii)^4}\int\frac{\dd^4 t}{t_1 t_2 t_3 t_4\, \big(1 - \vph F(t)\big)}
\eeqnn
where 
\begin{equation}
  F(t)\= W(t_1,t_2) + W(t_3,t_4) ~~~\text{with}~~~W(t_1,t_2)~=~t_1 + \frac{1}{t_1} + t_2 +
  \frac{1}{t_2} + \frac{t_1}{t_2} + \frac{t_2}{t_1}~,
  \label{eq:Wdef}\end{equation}
and the contour, for the integral, consists of a product of four loops enclosing the poles $t_j{\,=\,}0$.
From the integral we see that the coefficients of the fundamental period are given, as anticipated in \sref{sec:quintic}, by
\beq
a_n\= \big[F(t)^n]_0~.
\notag\eeq

The question arises as to how to calculate say the first 50 coefficients $a_n$. Naive computation of $\left[F(t)^n\right]_0$ becomes very slow as $n$ increases. In order to compute these coefficients more efficiently, we set $W_k{\=}\left[W(t_1,t_2)^k\right]_0$ and note that
\beq
\big[F(t)^n\big]_0\= \sum_{r=0}^n \binom{n}{r}\, W_{n-r}W_r \= \begin{cases}
\displaystyle 2\sum_{r=0}^{\frac{1}{2}(n-1)}\!\binom{n}{r}\, W_{n-r}W_r~;& \text{if $n$ is odd}\\[18pt] 
\displaystyle 2~\sum_{r=0}^{\frac{n}{2}-1}\;\binom{n}{r}\, W_{n-r}W_r + 
\binom{n}{\frac{n}{2}}\,W_{\frac{n}{2}}^2~; & \text{if $n$ is even}~.\end{cases}
\notag\eeq
We can calculate the quantities $W_k$ explicitly by collecting powers of $t_1$
\beq
W(t_1,t_2)\= t_1\left( 1 + \frac{1}{t_2}\right) + \left( t_2 + \frac{1}{t_2}\right) + \frac{1}{t_1}(1+t_2)
\notag\eeq
and picking out the terms in the expansion of $S(t_1,t_2)^k$ that are independent of $t_1$ and then of $t_2$.
In this way we find
\beq
W_k\= \sum_{\raisebox{-3pt}{$\scriptstyle r=0$}}^{\raisebox{3pt}{$\scriptstyle \lfloor \frac{k}{2}\rfloor$}}
\hskip5pt\sum_{s=\max\left(0,\lceil \frac{k-3r}{2}\rceil\right)}^{\min\left(k-2r,\lfloor\frac{k-r}{2}\rfloor\right)}
\frac{k!\, (2r)!}{(r!)^2 \, s!\, (3r + 2s - k)!\, (k - r -2s)!\, (k - 2r -s)!}~.
\notag\eeq 
Availing ourselves of these expressions, there is no difficulty in computing the first few hundred $a_n$.

If we now seek the differential equation that the fundamental period satisfies, we find the following coefficient functions $S_j$:
\beq\begin{split}
S_4&\= 69120\,\left(\vph -\smallfrac32\right)^2 \left(\vph +\smallfrac14\right)
\left(\vph +\smallfrac15\right) \left(\vph +\smallfrac16\right) 
\left(\vph -\smallfrac{1}{12}\right) \left(\vph -\smallfrac14\right) \left(\vph -\smallfrac13\right) \\[15pt]  
S_3&\= 8\vph \smash{\left(\vph - \smallfrac32\right)} \left(86400 \vph^6-158976 \vph^5-1512 \vph^4+15964 \vph^3+160 \vph^2-345 \vph -5\right)\\[20pt]
S_2&\= \vph  \big(2419200 \vph^7-8581248 \vph^6+7771104 \vph^5+274360 \vph^4-552220
   \vph^3-5250 \vph^2\\[-3pt]
&\hskip11.5cm+6917 \vph +39\big) \\[5pt] 
S_1&\= \vph  \big(3456000 \vph^7-12745728 \vph^6+12372480 \vph^5+166288 \vph^4-679952
   \vph^3-1584 \vph^2\\[-3pt]
&\hskip11.5cm+5532 \vph +9\big) \\[5pt]
S_0&\= 48 \vph^2 \left(34560 \vph^6-130464 \vph^5+132120 \vph^4+284 \vph^3-6182\vph^2
+9 \vph +36\right) \\
\end{split}\notag\eeq

The differential operator defined by these coefficient functions appears, after transformation to a new coordinate
\beq
\widetilde{\vph} \= \frac{3\vph}{3-2\vph}
\notag\eeq
as operator 6.24, under the revised naming convention, in the AESZ list. In terms of the variable $\widetilde{\vph}$ the differential operator is slightly simplified and the new coefficient functions have degree 6, rather than 8 as above. The designation 6.24 indicates that this is the 24th operator in the list whose coefficient functions have degree 6.

From the ratio $S_3/S_4$ we may compute the Yukawa coupling
\beq\begin{split}
y_{\vph\vph\vph}&\= 
\frac{1}{\vph^3}\,\exp\left( -\frac12 \int\! \frac{d\vph}{\vph}\, \frac{S_3(\vph)}{S_4(\vph)}\right)\\[10pt]
&\= \frac{\vph -\smallfrac32}{1440\vph^3 \left(\vph +\smallfrac14\right)
\left(\vph +\smallfrac15\right) \left(\vph +\smallfrac16\right) 
\left(\vph -\smallfrac{1}{12}\right) \left(\vph -\smallfrac14\right) \left(\vph -\smallfrac13\right)}~.
\end{split}\eeq
Notice that the coupling has a zero at $\vph{\=}3/2$.

The Riemann symbol for the Picard-Fuchs operator is
$$
\cP\left\{\begin{array}{rccrrrcccc}
0 &~\infty~ &~\frac32~ &-\frac14 &-\frac15 &-\frac16 &~~\frac{1}{12}~ &~\frac14~ &~\frac13~&\\[2pt]
\noalign{\hrule height1pt width8.0cm}\\[-14pt]
0 & 1           & 0             & 0         & 0          & 0          & ~0                      & 0             & 0            &\\
0 & 2           & 1             & 1         & 1          & 1          & ~1                      & 1             & 1            & \vph\\
0 & 3           & 3             & 1         & 1          & 1          & ~1                      & 1             & 1            &\\
0 & 4           & 4             & 2         & 2          & 2          & ~2                      & 2             & 2            &\\[3pt]
\end{array}\right\}
$$

The recurrence relation for the coefficients of the fundamental period follows from the form of the differential equation.
\beq\begin{split}
9n^4 a_n \=
&  \hskip14pt 3 (n-1) \left(16 n^3-28 n^2+21 n-6\right) a_{n-1}\\
&  +\left(983 n^4-3764 n^3+5909 n^2-4392 n+1260\right) a_{n-2}\\
&  -2 \left(727 n^4-6384 n^3+20823 n^2-30294 n+16740\right) a_{n-3}\\
&  -4 \left(5849 n^4-46012 n^3+128695 n^2-148340 n+55848\right) a_{n-4}\\
& +8 (n-4) \left(3019 n^3-30072 n^2+93377 n-90756\right) a_{n-5}\\
& +288 (n-5) (n-4) \left(539 n^2-1503 n+624\right) a_{n-6} \\
& -3456 (n-6) (n-5) (n-4) (61 n-125)\, a_{n-7}\\
& +69120 (n-7) (n-6) (n-5) (n-4)\, a_{n-8}
\end{split}\notag\eeq
We do not give the explicit form of the recurrence relations for the coefficients $b_n,\,c_n,\,d_n$ of the functions $f_j(\vph)$ for $j\neq 0$. However these are obtained from the above relation by replacing $n$ by $n+\e$ and $a_n$ by $A_n(\e)$ and expanding in powers of $\e$.

\vskip30pt
\begin{figure}[H]
\begin{center}
\includegraphics[width=\textwidth]{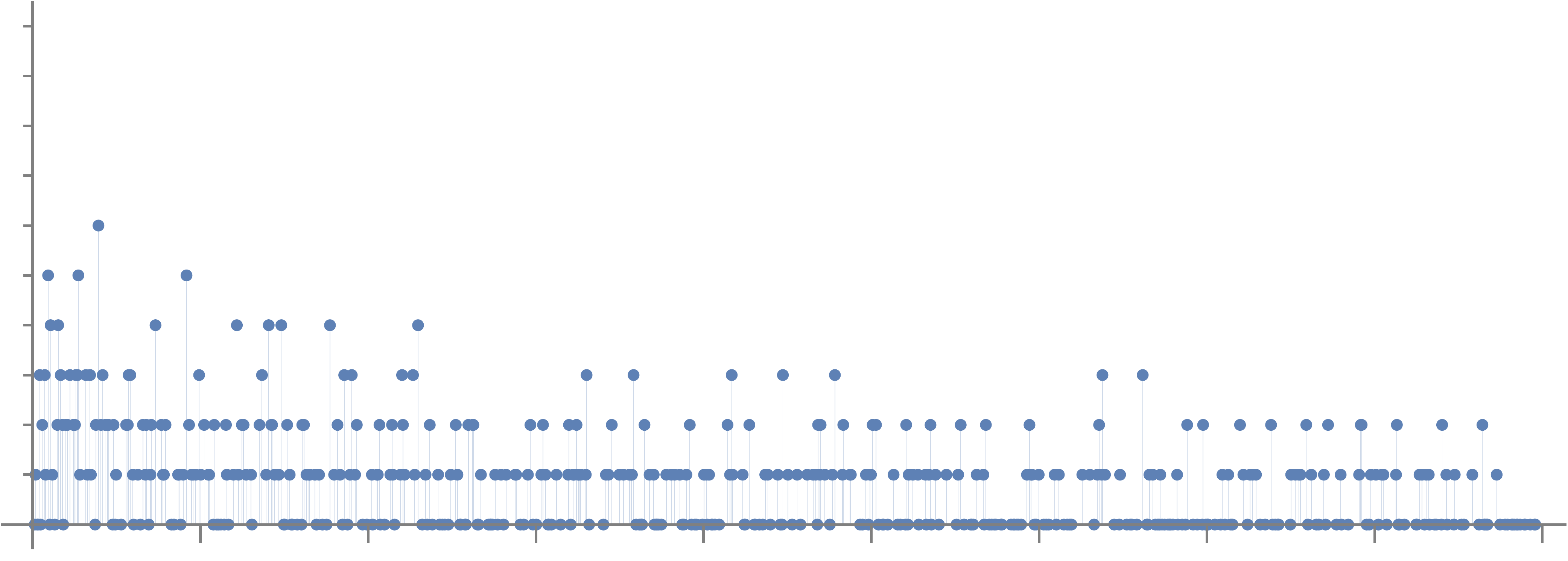}
\vskip0pt 
\place{-0.05}{0.54}{\scriptsize 1}
\place{-0.05}{0.75}{\scriptsize 2}
\place{-0.05}{0.96}{\scriptsize 3}
\place{-0.05}{1.16}{\scriptsize 4}
\place{-0.05}{1.37}{\scriptsize 5}
\place{-0.05}{1.57}{\scriptsize 6}
\place{-0.05}{1.78}{\scriptsize 7}
\place{-0.05}{1.99}{\scriptsize 8}
\place{-0.05}{2.20}{\scriptsize 9}
\place{-0.07}{2.41}{\scriptsize 10}
\place{0.73}{0.15}{\scriptsize 400}
\place{1.43}{0.15}{\scriptsize 800}
\place{2.10}{0.15}{\scriptsize 1200}
\place{2.80}{0.15}{\scriptsize 1600}
\place{3.49}{0.15}{\scriptsize 2000}
\place{4.19}{0.15}{\scriptsize 2400}
\place{4.89}{0.15}{\scriptsize 2800}
\place{5.59}{0.15}{\scriptsize 3200}
\place{6.29}{0.15}{\scriptsize 3600}
\capt{4.5in}{fig:Mnfld14}{The figure shows the numbers of factorisations into two quadrics for the manifold with Hodge numbers (1,4).}
\end{center}
\end{figure}
\vskip15pt
\begin{figure}[H] 
\begin{center}
\includegraphics[width=3.6in]{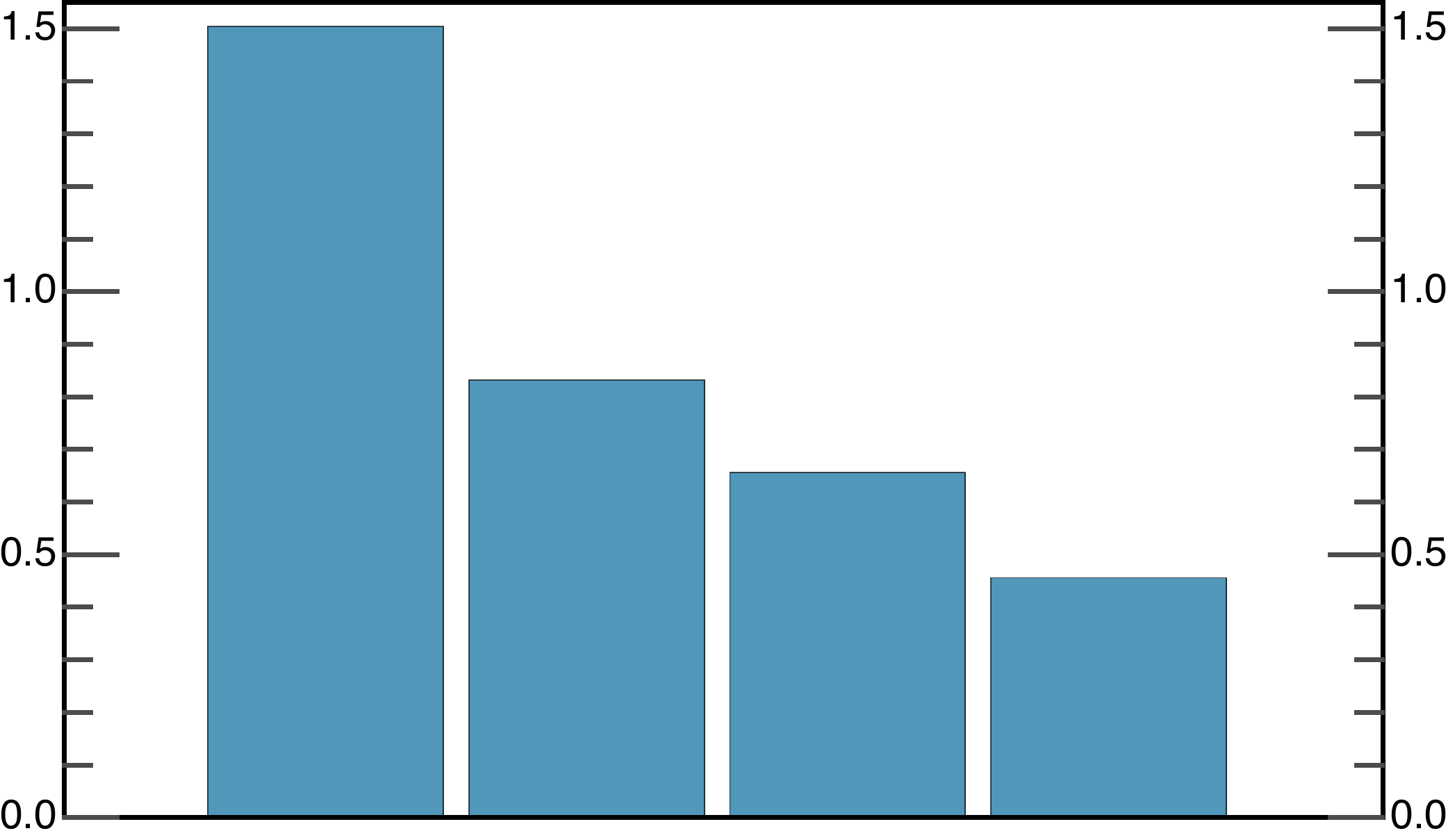}
\capt{5.0in}{fig:BarChart14}{Running averages for the data of the three generation manifold taken from 
     \fref{fig:Mnfld14}. The averages are taken for bins of 125 primes for the 500 primes $5\leqslant p\leqslant 3583$.}
\end{center} 
\end{figure}
\begin{table}[H]
\def\skip{\hskip10pt}
\renewcommand{\arraystretch}{1.1}
\begin{center}
\begin{tabular}[H]{|>{$}c<{$}|>{$}c<{$}|>{$}c<{$}|}
\hline
\vrule height15pt depth10pt width0pt \skip p \skip{} & \skip \frac32 \skip{} & \skip R \skip{} \\
\hline\hline
2   & 1 & (1 - \a pT + p^3T^2)(1 + 3T + p^3T^2) \\\hline
3   & 0 & $singular$ \\\hline
5   & 4 & $no unit root$ \\\hline
7   & 5 & $no unit root$ \\\hline
11 & 7 & (1 - \a pT + p^3T^2)(1 + 48T +p^3T^2)\\\hline
13 & 8 &  1+ 8T - 162 pT^2 + 8 p^3 T^3 + p^6 T^4 \\\hline
17 & 10 & $no unit root$ \\\hline
19 & 11 & 1 + 140 T + 690 pT^2 + 140 p^3T^3 + p^6T^4 \\\hline
23 & 13 & 1 - 6 T + 554 pT^2 - 6 p^3T^3 + p^6T^4 \\\hline
29 & 16 & 1 + 96 T + 926 pT^2 + 96 p^3T^3 + p^6T^4 \\\hline
31 & 17 & (1 - \a pT + p^3T^2)(1 - 56 T + p^3T^2) \\\hline
37 & 20 & 1 + 128 T + 2190 pT^2 + 128 p^3T^3 + p^6 T^4 \\\hline
\end{tabular}
\capt{5.0in}{tab:14UnitRoots}{The Frobenius polynomials for the first few primes at the apparent singularity $\vph{\;=\;}\frac{3}{2}$. These are calculated by the unit root method. For the cases that $R$ factorises, the coefficient $\a$ is left undetermined by this process.}
\end{center}
\end{table}

For this manifold, the hyperdiscriminants of the manifold and of the differential equation~are
\beq
\IDelta\= 2^{32}\, 3^{16}\, 5^2\, 7^2\, 17^2\, v^{30} \qquad \text{and} \qquad
\IS_4  \= 2^{42}\, 3^{16}\, 5^6\, 7^6\, 17^6\, v^{42}~.
\notag\eeq
On reducing $S_4$ modulo the bad primes, we find
\beq\begin{split}
S_4
&\;=\+ v^6(u-v)^2 \mod 2~,\\[2pt]
&\;= -u^2 v^3 (u-2v) (u-v)^2 \mod 3~,\\[2pt]
&\;= - v (u+v)^4 (u+2 v) (u+3 v) (u+4 v) \mod 5~,\\[2pt]
&\;=\+ 2(u + 2 v)^4 (u + 3 v) (u + 4 v) (u + 5 v) (u + 6 v) \mod 7~,\\[2pt]
&\;= -2 (u+3 v) (u+4 v) (u+7 v)^4 (u+11 v) (u+13 v) \mod 17~.\\
\end{split}\notag\eeq

Finally, we present \tref{(1,4) manifold} which presents the $\b_p$-coefficients for the values of $\vph$ for which the manifold is singular. For $p{\leqslant}41$ these were calculated using the unit root method (and, for the larger values of $p$, a supercomputer) and, for $p{\geqslant} 7$, also by the method of replacing the power series for $U(\vph)$ by a rational function. Certain entries to the Table are recorded in gray, such as the entry for $p{\=}31$ and $\vph{\=}1/12$ and those for $p{\=}17$ and $\vph{\=}-1/5$ and $\vph{\=}1/12$.

\begin{table}[H]
\begin{center}
\renewcommand{\arraystretch}{0.98}
\setlength{\extrarowheight}{1.0pt}
\renewcommand{\+}{\hspace*{10pt}}
\begin{tabular}{| >{$~} c <{~$} || >{$\,\bf} c <{\!$} | >{$\,\bf} c <{\!$} |>{$\,\bf} c <{\!$} | 
>{$\,\bf} c <{\!$} | >{$\,\bf} c <{\!$} | >{$\,\bf} c <{\!$} |} 
\hline
\text{Singularity} &  $$-\frac14 \+$$  & $$ -\frac15\+ $$ & $$ -\frac16\+ $$ & $$ \frac{1}{12 }$$ 
& $$ \frac14 $$ & $$ \frac13 $$ 
\tabularnewline[3pt]\hline
\text{Meyer label}         &  $$14/2$$ &  $$17/1$$  &  $$60/1$$  &  $$102/3$$ &  $$10/1$$ & $$21/2$$ 
\tabularnewline[0pt]\hline
\text{LMFDB} & \bf 14.4.a.b &\bf 17.4.a.a &\bf 60.4.a.a & \bf 102.4.a.c  & \bf 10.4.a.a & \bf 21.4.a.a
\\\hline\hline
p=2                    &\Gr   2        &  -3\+         &\Gr   0         &\Gr   2         &\Gr   2        &  -3\+           
\\\hline
p=3                    &     -2\+      &  -8\+         &\Gr   -3\+    &\Gr   -3\+    &   -8\+        &\Gr  -3\+       
\\\hline
p=5                    &   -12\+      &\Gr   6         &\Gr   -5\+   &  -12\+       &\Gr   5         & -18\+           
\\\hline
p=7                    &\Gr   7        &\Gr  -28\+   &\Gr  -28\+   &   -22\+       &   -4\+        &\Gr   7           
\\\hline
p=11                   &   48         &   -24\+       &  -24\+        &  -48\+         &  12            & -36\+          
\\\hline
p=13                  &   56          &    -58\+      &  -70\+        &      2            &-58\+         & -34\+           
\\\hline
p=17                  & -114\+     &\Gr   17       &  102           &\Gr   -17\+  & 66              & 42               
\\\hline
p=19                  &  2            &   116          &  20              & 20              &  -100\+      & -124\+          
\\\hline
p=23                  & -120\+     & -60\+        &  -72\+         & -54\+          & 132            &\Gr   0           
\\\hline
p=29                  & -54\+      & 30            &  306             & 84               & -90\+         &102                
\\\hline
p=31                  & 236         & -172\+      & -136\+         &\Gr   62       & 152           & -160\+          
\\\hline
p=37                  & 146         & -58\+        & -214\+         & 44              & -34\+        & 398             
\\\hline
p=41                 & 126         & -342\+        & -150\+         &-138\+       & -438\+        & -318\+        
\\\hline
p=43                 & -376\+    & -148\+        & -292\+         &428            & 32               & -268\+        
\\\hline
p=47                 & -12\+     &  288             & -72\+           &-516\+      & -204\+         &  240            
\\\hline
p=53                 & 174        &  318             & -414\+         &174           & 222              &  -498\+       
\\\hline
p=59                 & 138        &  252             & -744\+         & -852\+      & 420              &  -132\+       
\\\hline
p=61                 & 380        &  110             & -418\+         & 908           & 902              &  398       
\\\hline
p=67                 & -484\+   &  -484\+        & 188              & -508\+      & -1024\+       &  92       
\\\hline
p=71                 & 576        &  -708\+        & 480              & -426\+      & 432              &  -720\+       
\\\hline
p=73                 & -1150\+ &  362             & 434              & -574\+      & 362              &  -502\+       
\\\hline
p=79                 & 776        &  -484\+        & 1352            & 110           & -160\+         & -1024\+       
\\\hline
p=83                 & 378        &  756             & -612\+         & -1308\+    & 72                & -204\+       
\\\hline
p=89                 & -390\+   &  -774\+        & -30\+           & 798           & 810              &  354       
\\\hline
p=97                 & -1330\+ &  -382\+        & -286\+         & -1690\+     & 1106            &  -286\+       
\\\hline
p=101               & -1500\+ &  -210\+        & -1542\+       & -1890\+     & -258\+        &  414       
\\\hline
p=103               & 380        &  -232\+        & 1172           & -1900\+     & -988\+        &  56       
\\\hline
p=107               & 636        &  432             & 1956           & -480\+       & -24\+          &  12       
\\\hline
p=109               & 146        &  -1186\+      & -1858\+      & 1424          & 950              &  1478       
\\\hline
p=113               & 198        &  -366\+       & 174             & 402           & -1038\+         &  402       
\\\hline
p=127               & -376\+   &  -472\+       & -2068\+      & 2336         & -124\+           &  1280       
\\\hline
p=131              & 2130       &  2760          & 312             & 768           & 132                &  1764       
\\\hline
p=137              & -78\+      &  1098          & 2646           & 2106         & -1254\+         &  -2358\+       
\\\hline
\end{tabular}
\capt{5.5in}{(1,4) manifold}{\it For the (1,4)-manifold, the coefficient $\b_p$ for the characteristic polynomial of Frobenius for the cases that the manifold has (hyper-) conifold singularities.}
\end{center}
\end{table}
\newpage
\subsection{A quotient of the 24 cell with $\hodgenos {\=}(1,1)$, AESZ\hskip2pt 366}
\vskip-10pt
It was known already to Kreuzer and Skarke \cite{Kreuzer:2000xy} that the 24-cell is a reflexive polyhedron and occurs in their list of such polyhedra. It was observed by Braun \cite{Braun:2011hd} that the threefold that is thereby defined admits free quotients by three groups $G$, of order 24, where $G$ is homomorphic to $\text{SL}(2,3)$, $\IZ_3{\rtimes}\IZ_8$ or $\IZ_3{\times}\IQ_8$, and that these quotients are \cys with $\hodgenos{\=}(1,1)$. The Picard-Fuchs equation is common to these three manifolds and was already known, owing to the fact that the fundamental period has an interpretation as the generating function for lattice walks that return to the origin after $n$ steps. It had been observed by Guttmann \cite{Guttmann:2009, Guttmann:2010nj} and Broadhurst \cite{Broadhurst:2009} that, what is to us the fundamental period, but which they viewed as a lattice generating function, satisfies a differential equation of \cy type. This was recorded as operator $366$ of \cite{almkvist2005tables} and has designation 7.3 in the revised numbering. The manifold and the periods, for this case, are studied in somewhat greater detail in \cite{Braun:2015jdy}.

For the basic invariants of the manifold we have
\beq
\hodgenos \= (1,1)~;\quad\ch\= 0~;\quad c_2 H \= 4~;\quad H^3\= 4~.
\notag\eeq

The defining toric equation is
\beq
 1- \vph F(t) \= 0~,
\notag\eeq
where, in this case, the Laurent polynomial $F(t)$ takes the form
\beq
  \begin{split}
    F(t) ~&=~ t_1 + t_2 + t_3 + t_4 + \frac{1}{t_1} + \frac{1}{t_2} + \frac{1}{t_3} + \frac{1}{t_4}\\[5pt]
            &\hskip29pt +\frac{t_1}{t_2} + \frac{t_1}{t_3} +\frac{t_1}{t_4} + \frac{t_2}{t_1} + \frac{t_3}{t_1} 
                              + \frac{t_4}{t_1} + \frac{t_4}{t_2} + \frac{t_2}{t_4} + \frac{t_3}{t_4} +\frac{t_4}{t_3}\\[5pt]
            &\hskip29pt + \frac{t_1}{t_2 t_3} + \frac{t_2 t_3}{t_1}+\frac{t_2 t_3}{t_4}+\frac{t_4}{t_2 t_3} + \frac{t_1 t_4}{t_2 t_3}+ \frac{t_2 t_3}{t_1 t_4}\\[10pt]
            &=~ W(t_1,t_2) + W(t_3,t_4) +  W\left(\frac{t_1}{t_3},\frac{t_2}{t_4}\right) + W\left(\frac{t_1}{t_4}, \frac{t_2 t_3}{t_4}\right)~,
  \end{split}
\notag\eeq
where we used the same expression for $W$ as  in eq.~\eqref{eq:Wdef}.
The fundamental period has again an expansion with coefficients $\left[ F(t)^n \right]_0$ and one can use the last expression above to generate a sufficient number of the coefficients $a_n$. 

Explicit expressions may also be given in several ways and~Refs.\hskip2pt\cite{Broadhurst:2009, Guttmann:2010nj}
give, for example, the convenient series
\beq\begin{split}
  \varpi_0(\vph) ~&= \hskip-5pt\sum_{i,j,k,l,m=0}^\infty  
  \binom{2i}{i} \binom{2j}{j} \binom{2k}{k} \binom{l+m}{m} \binom{2(l+m)}{l+m}^2\times \\[7pt]
  &\hskip30pt
  \times\binom{i+j+k+l+m}{2(l+m)} \binom{i+j+k-l-m}{-i+j+k}\binom{2i-l-m}{i-k-l}\, \vph^{i+j+k+l+m}.
\end{split}\notag\eeq
The differential operator that annihilates this function has coefficient functions
\beq\begin{split}
 S_4&= 8957952 \left(\vph + \smallfrac{1}{18}\right)^2 \left(\vph + \smallfrac13\right)  \left(\vph + \smallfrac14\right) \left(\vph + \smallfrac18\right)  \left(\vph + \smallfrac{1}{12}\right)  
             \left(\vph - \smallfrac{1}{24}\right)   \\[10pt]
 S_3&= 36\vph \left(\vph + \smallfrac{1}{18}\right) \left(1990656 \vph^5+1257984 \vph^4+264384 \vph^3+22320\vph^2 +800 \vph +15\right) \\[15pt]
 S_2&= \vph  \big( 206032896 \vph^6+118195200 \vph^5+24103872 \vph^4+2276640 \vph^3+105552 \vph^2+2114 \vph +19 \big) \\[15pt]
 S_1&= 72\vph \left(\vph + \smallfrac{1}{18}\right) \left(3483648 \vph^5+1548288 \vph^4+225072 \vph^3+13572\vph^2+320 \vph +1\right) \\[15pt]
 S_0&= 96 \vph^2 \left(1119744 \vph^5+508032 \vph^4+82512 \vph^3+6318 \vph^2+237\vph +4\right)~. \\
\end{split}\notag\raisetag{18pt}\eeq
The singularities of the Picard-Fuchs equation are evident from the factorisation of $S_4$. For the present manifold $\vph{\;=}-1/18$ is an apparent singularity.

The Riemann symbol for this manifold is
\beq
  \renewcommand{\arraystretch}{1.1}
  \mathcal{P}
  \left\{
    \begin{array}{rrrrrrrrc}
      ~0 & 
      \hskip7pt\infty\hskip-2pt\null & 
      -\tfrac{1}{3} &
      -\tfrac{1}{4} &
      -\tfrac{1}{8} &
      \hskip-2pt-\tfrac{1}{12}\hskip-2pt\null &
      -\tfrac{1}{18}\hskip-2pt\null &
      ~\hskip3pt\tfrac{1}{24}\hskip-2pt\null 
      \\[2pt]
      \noalign{\hrule height1pt width7.3cm}\\[-17pt]
      0 &     
      1 &     
      0 &     
      0 &     
      0 &     
      0 &     
      0 &     
      0 &     
      \\
      0 &     
      2 &     
      1 &     
      1 &     
      1 &     
      1 &     
      1 &     
      1 &     
      \\
      0 &     
      2 &     
      1 &     
      1 &     
      1 &     
      1 &     
      3 &     
      1 &     
      \\
      0 &     
      3 &     
      2 &     
      2 &     
      2 &     
      2 &     
      4 &     
      2 &     
      \\[3pt] 
    \end{array}
    ~\vph\right\},
\label{eq:RiemannP}\eeq

From the differential equation the recurrence relation for the $a_n$ follows
\beq\begin{split}
n^4 a_n \=
& -(n-1) \left(39 n^3-147 n^2+158 n-54\right)a_{n-1}\\[3pt]
& -2 \left(16 n^4-1198 n^3+5747 n^2-9800 n+5748\right) a_{n-2}\\[3pt]
& +36\, (3 n-7) \left(171 n^3-973 n^2+1821 n-1007\right) a_{n-3}\\[3pt]
& +864 \left(384 n^4-4602 n^3+20995 n^2-43195 n+33786\right) a_{n-4}\\[3pt]
& +1728\, (n-4) \left(1393 n^3-15324 n^2+57143 n-72156\right) a_{n-5}\\[3pt]
& +248832\, (n-4)(n-5) \left(31 n^2-267 n+584\right) a_{n-6}\\[3pt]
& +8957952\, (n-4) (n-5)^2 (n-6)\, a_{n-7}~.
\end{split}\notag\eeq
Again, the Frobenius period may be obtained from this recurrence by replacing $n$ by $n+\e$ and $a_n$ by $A_n(\e)$.

If we now return to the Riemann symbol \eqref{eq:RiemannP} and make the change of variable
\beq
\vph\= \frac{\widetilde\vph}{1-18\,\widetilde\vph}~,~~~\text{so}~~~\widetilde\vph\= \frac{\vph}{1+18\,\vph}~.
\notag\eeq
Then, using the properties of the Riemann symbol we see that \eqref{eq:RiemannP} is equivalent to the scheme
\begin{equation}  \renewcommand{\arraystretch}{1.1}
 \frac{1}{1+18\,\vph}~ \mathcal{P}
  \left\{
    \begin{array}{rrrrrrrrc}
      ~0 & \infty\hskip-3pt{} &\;\tfrac{1}{42}\hskip-2pt{} & \;\tfrac{1}{18}\hskip-2pt{} 
      & \;\tfrac{1}{15}\hskip-2pt{} & \;\tfrac{1}{14}\hskip-2pt{} & \;\tfrac{1}{10}\hskip-2pt{} 
      &\;\,\tfrac{1}{6} 
      \\[2pt]
      \noalign{\hrule height1pt width 6.1cm}\\[-17pt]
      0 &     1 &     0 &     0 &     0 &     0 &     0 &     0 &     
      \\
      0 &     2 &     1 &     1 &     1 &     1 &     1 &     1 &     
      \\
      0 &     4 &     1 &     1 &     1 &     1 &     1 &     1 &     
      \\  
      0 &     5 &     2 &     2 &     2 &     2 &     2 &     2 &
      \\[3pt]        
    \end{array}
    ~\frac{\vph}{1+18\,\vph}
  \right\}~.
\notag\end{equation}

The change of variables has interchanged the roles of the singularities at $\vph{\,=\,}\infty$ and $\vph{\,=\,}\tfrac{1}{18}$ with those at $\widetilde\vph{\,=\,}\tfrac{1}{18}$ and 
$\widetilde\vph{\,=\,}\infty$. The six conifold points now appear on an equal footing in the symbol. Based on this new symbol we can choose new period vector
\beq
\widetilde\vp_j(\tilde\vph)~=~(1+18\,\vph)\, \vp_j(\vph)
\notag\eeq
This satisfies a somewhat simpler Picard-Fuchs equation with coefficient functions that are sixth order polynomials rather than seventh order, for detail see \cite[\SS4.3]{Broadhurst:2009}. The simpler differential operator may also be found in the AESZ list as operator 6.18.

The properly normalized Yukawa coupling is
\beq
    y_{\vph\vph\vph} \=
    \frac{1}{\vph^3}\,\exp\left(\! -\frac{1}{2}\int \frac{d\vph}{\vph}\,\frac{S_3(\vph)}{S_4(\vph)}\right)
    \= \! -\frac{1}{384}\; \frac{\vph + \smallfrac{1}{18}}{\vph^3\left(\vph + \smallfrac13\right) \left(\vph + \smallfrac14\right) \left(\vph + \smallfrac18\right) 
      \left(\vph + \smallfrac{1}{12}\right) \left(\vph - \smallfrac{1}{24}\right)}
\notag\end{equation}
where we have fixed the integration constant to match the classical intersection number (times $\vph^{-3}$) at the large complex structure limit $\vph=0$. 

For this manifold, the hyperdiscriminants of the manifold and of the differential equation~are
\beq
\IDelta\= 2^{32}\, 3^{14}\, 5^2\, 7^2\, v^{20} \qquad \text{and} \qquad
\IS_4  \= 2^{40}\, 3^{20}\, 5^6\, 7^6\, v^{30}~.
\notag\eeq
\vskip30pt
\begin{figure}[H]
\begin{center}
\includegraphics[width=\textwidth]{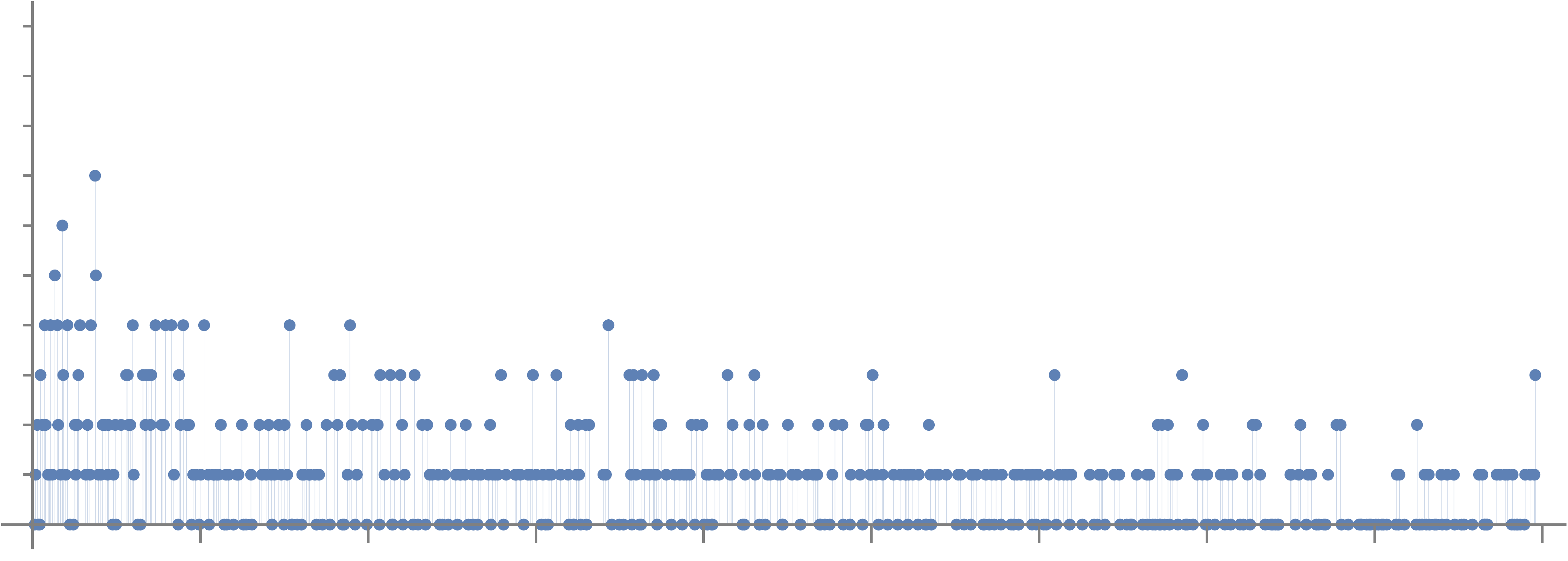}
\vskip0pt 
\place{-0.05}{0.54}{\scriptsize 1}
\place{-0.05}{0.75}{\scriptsize 2}
\place{-0.05}{0.96}{\scriptsize 3}
\place{-0.05}{1.16}{\scriptsize 4}
\place{-0.05}{1.37}{\scriptsize 5}
\place{-0.05}{1.57}{\scriptsize 6}
\place{-0.05}{1.78}{\scriptsize 7}
\place{-0.05}{1.99}{\scriptsize 8}
\place{-0.05}{2.20}{\scriptsize 9}
\place{-0.07}{2.41}{\scriptsize 10}
\place{0.73}{0.15}{\scriptsize 400}
\place{1.43}{0.15}{\scriptsize 800}
\place{2.10}{0.15}{\scriptsize 1200}
\place{2.80}{0.15}{\scriptsize 1600}
\place{3.49}{0.15}{\scriptsize 2000}
\place{4.19}{0.15}{\scriptsize 2400}
\place{4.89}{0.15}{\scriptsize 2800}
\place{5.59}{0.15}{\scriptsize 3200}
\place{6.29}{0.15}{\scriptsize 3600}
\capt{5in}{fig:Mnfld11}{The figure shows the number of factorisations into two quadrics for the manifold with Hodge numbers (1,1).}
\end{center}
\end{figure}
\begin{figure}[!t] 
\begin{center}
\includegraphics[width=4.0in]{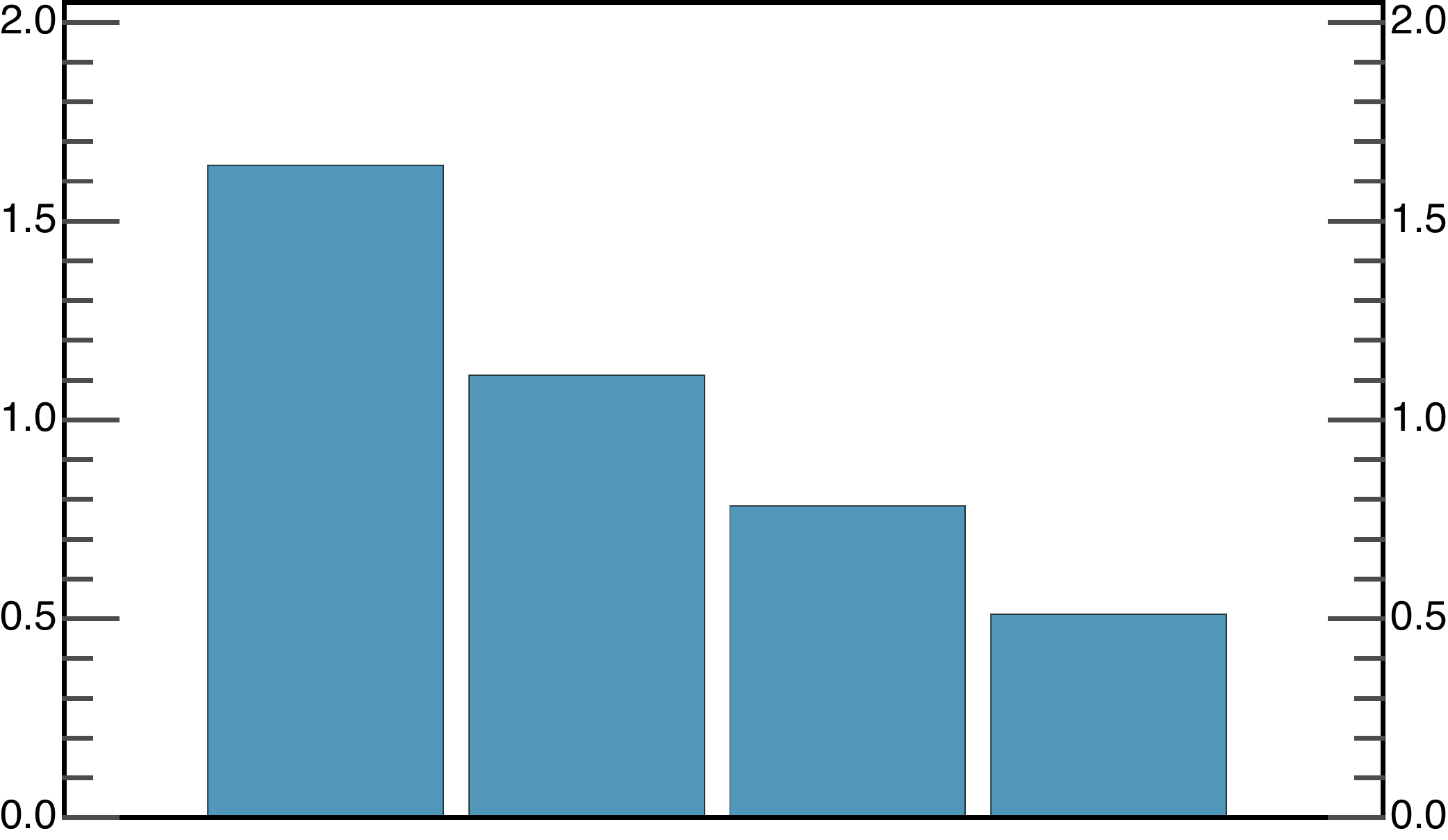}
\capt{5.0in}{fig:BarChart11}{Running averages for the data from 
     \fref{fig:Mnfld11}. The averages are taken for bins of 125 primes for the 500 primes $5\leqslant p\leqslant 3583$.}
\end{center} 
\end{figure}
\vskip20pt
\begin{table}[H]
\def\skip{\hskip10pt}
\renewcommand{\arraystretch}{1.1}
\begin{center}
\begin{tabular}[H]{|>{$}c<{$}|>{$}c<{$}|>{$}c<{$}|}
\hline
\vrule height15pt depth10pt width0pt \skip p \skip{} &  -\frac{1}{18}\+  & \skip R \skip{} \\
\hline\hline
2   & \infty & \\\hline
3   & \infty & \\\hline
5   & 3 & $no unit root$ \\\hline
7   & 5 & $no unit root$ \\\hline
11 & 3 & (1 - \a pT + p^3T^2)(1 - 12T +p^3T^2)\\\hline
13 & 8 &  $no unit root$ \\\hline
17 & 16 &(1 - \a pT + p^3T^2)(1 + 1187 T +p^3T^2) \\\hline
19 & 1   & (1 - \a pT + p^3T^2)(1 + 52 T +p^3T^2)\\\hline
23 & 14 & 1 - 132 T + 578 pT^2 - 132 p^3T^3 + p^6T^4 \\\hline
29 & 8   & 1 - 16 T + 446 pT^2 - 16 p^3T^3 + p^6T^4 \\\hline
31 & 12 & 1 + 76 T - 766 pT^2 + 76 p^3T^3 + p^6T^4\\\hline 
37 &   2 & 1 - 44 T - 730 pT^2 - 44 p^3T^3 + p^6 T^4 \\\hline
\end{tabular}
\capt{5.0in}{tab:11UnitRoots}{The Frobenius polynomials for the first few primes at the apparent singularity $\vph{\;=}-\frac{1}{18}$. These are calculated by the unit root method. For the cases that $R$ factorises, the coefficient $\a$ is left undetermined by this process.}
\end{center}
\end{table}

On reducing $S_4$ modulo the bad primes, we find
\beq\begin{split}
S_4
&\;=\+ v^6 (u+v) \mod 2~,\\[3pt]
&\;=\+ v^5 (u+v) (u+2 v) \mod 3~,\\[3pt]
&\;=\+ 2 (u+v) (u+2 v)^4 (u+3 v) (u+4 v) \mod 5~,\\[3pt]
&\;=\+ 3 (u+v) (u+2 v)^4 (u+3 v) (u+5 v) \mod 7~.\\[3pt]
\end{split}\notag\eeq

\newpage
\begin{table}[H]
\renewcommand\arraystretch{1}
\renewcommand{\+}{\hspace*{15pt}}
\begin{center}
\begin{tabular}{| >{$~} c <{~$} || >{$~} c <{~$} | >{$~} c <{~$}|
 >{$~} c <{~$}| >{$~} c <{~$}| >{$~} c <{~$}|}
\hline
\vrule height14pt width0pt\text{Singularity} & -\frac13\+ & -\frac14\+ & -\frac18\+ & -\frac{1}{12}\+  & \frac{1}{24}\\[3pt]\hline
\vrule height14pt width0pt\text{Meyer label}& 15/2         &     28/2      &   10/1       &      12/1               &     42/2 \\[3pt]\hline
\vrule height14pt width0pt\text{LMFDB}&\bf 15.4.a.a &\bf 28.4.a.b &\bf 10.4.a.a &\bf 12.4.a.a &\bf 42.4.a.b\\
\hline\hline
    p=2                &\bf 1           &\bf\Gr   0     &\bf\Gr 2      &\bf\Gr   0             &\bf\Gr   2 
\\\hline
    p=3                &\bf\Gr   3    &\bf 4           &\bf -8\+       &\bf\Gr   3             &\bf\Gr   3
\\\hline
    p=5                &\bf\Gr  5     &\bf	 6       &\bf\Gr   5     &\bf  -18\+            &\bf  2 
\\\hline 
    p=7                &\bf -24\+    &\bf\Gr  7     &\bf -4\+       &\bf   8                   &\bf\Gr   -7\+
\\\hline
    p=11              &\bf  52        &\bf -12\+    &\bf 12           &\bf   36                &\bf -8\+
\\\hline
    p=13              &\bf 22         &\bf -82\+    &\bf -58\+      &\bf  -10\+            &\bf -42\+
\\\hline
    p=17              &\bf -14\+    &\bf -30\+    &\bf 66           &\bf   18                &\bf -2\+
\\\hline
    p=19              &\bf -20\+    &\bf 68         &\bf -100\+    &\bf -100\+           &\bf -124\+
\\\hline 
    p=23              &\bf -168\+  &\bf 216       &\bf 132          &\bf 72                 &\bf 76
\\\hline 
    p=29              &\bf 230       &\bf 246       &\bf -90\+       &\bf -234\+          &\bf 254
\\\hline 
    p=31              &\bf -288\+   &\bf -112\+ &\bf 152          &\bf -16\+            &\bf -72\+
\\\hline 
    p=37              &\bf -34\+     &\bf 110      &\bf -34\+       &\bf -226\+          &\bf  398
\\\hline 
   p=41              &\bf 122         &\bf -246\+ &\bf -438\+     &\bf 90                 &\bf  462
\\\hline 
   p=43              &\bf -188\+   &\bf -172\+ &\bf 32             &\bf 452               &\bf  212
\\\hline
   p=47              &\bf 256        &\bf 192      &\bf -204\+      &\bf 432               &\bf  -264\+
\\\hline 
   p=53              &\bf -338\+   &\bf 558      &\bf 222           &\bf 414               &\bf  -162\+
\\\hline 
   p=59              &\bf 100        &\bf 540      &\bf 420           &\bf -684\+          &\bf  -772\+
\\\hline 
   p=61              &\bf 742        &\bf 110      &\bf 902           &\bf 422               &\bf  30
\\\hline 
   p=67              &\bf -84\+     &\bf 140      &\bf -1024\+    &\bf 332               &\bf  -764\+
\\\hline 
   p=71              &\bf -328\+   &\bf -840\+  &\bf 432          &\bf -360\+          &\bf  -236\+
\\\hline
   p=73              &\bf -38\+     &\bf -550\+  &\bf 362          &\bf 26                 &\bf  418
\\\hline 
   p=79              &\bf -240\+   &\bf -208\+  &\bf -160\+      &\bf 512               &\bf  552
\\\hline 
   p=83              &\bf 1212      &\bf 516       &\bf 72             &\bf -1188\+        &\bf  1036
\\\hline 
   p=89              &\bf 330        &\bf -1398\+ &\bf 810          &\bf -630\+          &\bf  30
\\\hline 
   p=97              &\bf 866        &\bf 1586      &\bf 1106        &\bf -1054\+         &\bf  -1190\+
\\\hline 
   p=101            &\bf -1218\+  &\bf -1242\+ &\bf -258\+     &\bf 558               &\bf  1370
\\\hline 
   p=103            &\bf -88\+      &\bf 680       &\bf -988\+     &\bf 8                  &\bf  464
\\\hline
   p=107            &\bf 36           &\bf 996       &\bf -24\+       &\bf 1764             &\bf  -2136\+
\\\hline 
   p=109            &\bf -970\+   &\bf 1382       &\bf 950          &\bf 1622             &\bf  -1226\+
\\\hline 
   p=113            &\bf 1042      &\bf -750\+    &\bf -1038\+   &\bf -1134\+         &\bf  338
\\\hline 
   p=127            &\bf 1936       &\bf 176        &\bf -124\+     &\bf -592\+          &\bf  2088
\\\hline
   p=131            &\bf 732        &\bf -1548\+  &\bf 132          &\bf -1908\+        &\bf  -292\+
\\\hline 
   p=137            &\bf -2214\+  &\bf 378        &\bf -1254\+   &\bf 954               &\bf  818
\\\hline
\end{tabular}
\capt{5.5in}{(1,1) manifold}{For the (1,1)-manifold, the coefficient $a_p$ for the characteristic polynomial of Frobenius for the cases that the manifold has (hyper-) conifold singularities.}
\end{center}
\end{table}
\newpage
\appendix
\section*{Acknowledgements}
\vskip-8pt
It is a pleasure to acknowledge fruitful conversations with Kilian B\"{o}nich, Mohamed Elmi, Minhyong Kim and Albrecht Klemm. We are especially grateful to John Voight for an extended discussion regarding the nature of the modular forms that are relevant to our results in \sref{sec:G25mfld} and \sref{sec:Rmfld}, and to Alan Lauder whose insights changed the course of this research. The authors wish to thank the MITP, Mainz for hospitality and support during the 2018 Workshop on String Theory, Geometry and String Model Building,  also  KIAS for hospitality and support associated with the 2019 International Conference on Arithmetic Geometry and Quantum Field Theory, PC and XD are grateful also for support and hospitality of KIAS in~2018.
\newpage
\addcontentsline{toc}{section}{Appendices}
\section{A Corollary to the Frobenius-Chebotar\"ev Theorem}\label{sec:Chebotarev}
\vskip-10pt
It follows from the celebrated Frobenius-Chebotar\"ev theorem~\cite{Stevenhagen1996} that if $f$ is a polynomial with rational coefficients, that is irreducible over $\IQ$, then the number of roots of $f$ in $\IF_{\! p}$, averaged over~$p$, is one.

For the proof of this corollary, which must be quite well known, we first recall the Cauchy-Frobenius-Burnside lemma. This is often referred to as Burnside's lemma, but was earlier known to both Cauchy and Frobenius~\cite{Neumann:1979}. This lemma states that if $X$ is a set on which a group $G$ acts, and $X^g$ denotes the set of elements of $X$ that are fixed by $g{\;\in\;}G$. Then
\beq
\frac{1}{|G|}\sum_{g\in G} |X^g| \= \left| X/G \right|~.
\label{CFBlemma}\eeq
Thus the average of the size of the fixed point sets, over the group, is given by the number of group orbits.

For the application to the Frobenius-Cebotar\"ev theorem, take a polynomial $f$, as above, $X$ to be the set whose elements are the roots of $f$ and $G$ to be the Galois group of $f$. Each $g{\;\in\;}G$ permutes the roots. Let us suppose that $g$ has cycle type $(n_1,n_2,\ldots,n_r)$, with $n_1{\;+\:}\ldots{\;+\;}n_r{\=}n$, the degree of $f$. Since the roots of an irreducible polynomial are permuted transitively by the Galois group, there is only one orbit in $X/G$ and the right hand side of \eqref{CFBlemma} is one. If we reduce $f$ mod $p$ (and avoid a finite number of bad primes) then $f$ will factor into irreducible factors mod $p$ of degree $(m_1,m_2,\ldots,m_r)$ with
      $m_1{\;+\:}\ldots{\;+\;}m_r{\=}n$. If $f$ mod $p$ has $k$ linear factors, exactly $k$ of the $m_j$ are equal to one, and $f$ has exactly $k$ roots in $\IF_{\! p}$. In \eqref{CFBlemma} we have $\sum_{g\in G} |X^g|{\=}k$.
      
The Frobenius-Chebotar\"ev theorem states that the density of primes $p$ where we have factorization type $(m_1,m_2,\ldots,m_r)$ is equal to the proportion of elements $g{\;\in\;}G$ which have cycle type $(m_1,m_2,\ldots,m_r)$. So the average number of roots of $f$ in $\IF_p$ is indeed one.

The statement of the corollary has been proved for the case that $f$ is irreducible over $\IQ$. For the case that a polynomial $F$, with rational coefficients, is reducible, we consider the factorisation of $F$ into irreducible factors. Suppose there are $k$ such factors $f_j$, so
\beq
F \= \prod_{j=1}^k f_j~.
\notag\eeq
Applying the corollary to each factor, we see that the number of roots of $F$ in $\IF_{\! p}$, averaged over $p$, is $k$.
\newpage
\section{The p-adic $\G$ and $\z$ functions}\label{sec:AppendixZeta}
\vskip-10pt
We set out here a brief account of some results that relate to the p-adic $\G$ and $\z$ functions. Excellent textbook accounts can be found in \cite{KoblitzPadicNumbers,Robert2013course,Villegas2007Experimental,Cohen2007NumberTheoryII}. There are however inequivalent definitions of $\z_p(3)$ in the literature, so, since this quantity is important to us, we recall the salient points of its definition here.

We begin with the p-adic $\G$-function. First note that as a p-adic function the factorial function $n!$ tends rapidly to zero as $n$ becomes large, since $n!$ then contains many factors of~$p$. A better behaved function is obtained by omitting the terms divisible by~$p$. Following from this observation, the Morita p-adic $\G$-function is defined initially for $n{\,\in\,}\IZ_{\geqslant 0}$ by
\beq
\Gp(n)\= (-1)^n \prod^{n-1}_{\substack{k=1\\[2pt] p\notdiv k}} k~.
\notag\eeq 
The definition can be extended to $\Gp(z)$ for $z{\,\in\,}\IZ_p$ by choosing a sequence of integers $\{n_j\}$ that converge to $z$ and noting that $\Gp$ is p-adically smooth, that is
\beq
\Gp(n+a p^m) \to \Gp(n)~~\text{as}~~m\to\infty~.
\notag\eeq
The process of extending the definition from a subset of $\IZ$ to $\IZ_p$, as above, is known as p-adic interpolation.

The p-adic function satisfies a recurrence relation that is an analogue of the familiar relation:
\beq
\Gp(z+1) \= 
\begin{cases} 
-z\Gp(z) & \text{if}~z\in\IZ_p^*~,\\ 
-\phantom{z}\Gp(z) & \text{if}~z\in p\IZ_p~.
\end{cases}
\notag\eeq

While $\Gp(z)$ can be computed from the definition, the following procedure is more convenient. Define coefficients $c_n$ via the expansion
\beq
\exp\left(x + \frac{x^p}{p}\right)\= \sum_{n=0}^\infty c_n x^n~.
\notag\eeq
By differentiating this expression, it is easy to show that the $c_n$ satisfy the recursion relation
\beq
nc_n \= c_{n-1} + c_{n-p}~,~~~c_0\= 1~,~~c_n\=0~~\text{for}~~n<0~.
\notag\eeq
It was shown by Dwork that if $0\leqslant a\leqslant p-1$ then
\beq
\Gp(-a + pz) \= \sum_{k=0}^\infty p^k c_{a+kp}(z)_k~,
\notag\eeq
where
\beq
(z)_k \= z(z+1)\ldots (z+k-1)~;~~(z)_0\=1
\notag\eeq
is the Pochammer symbol.

The function $\Gp$ is locally analytic, that is it can be expanded as a power series about every $z{\,\in\,\IZ_p}$. This being so, it makes sense to define the p-adic digamma function on $p\IZ_p$
\beq
\ps_p(z)\= \frac{\Gp'(z)}{\Gp(z)}
\notag\eeq
which satisfies the relation
\beq
\ps_p(z+1) - \ps_p(z) \= 
\begin{cases}
\frac{\displaystyle 1}{\displaystyle z} & \text{for}~z\in\IZ_p^* ~,\\
0 & \text{for}~z\in p\IZ_p~.
\end{cases}
\notag\eeq
The digamma function is also locally analytic on $p\IZ_p$
\beq
\ps_p(z)\= b_0 - \sum_{n=1}^\infty \frac{b_n}{n}\,z^n~;~~\norm{z}_p<1~.
\label{eq:DigammaExpansion}\eeq
The coefficients $b_n$ that appear in this relation can be expressed as integrals
\beq
b_0\= \int_{\IZ_p^*} \log z\, \dd z~~;\quad b_n\=  \int_{\IZ_p^*} z^{-n}\, \dd z~,~~n\geqslant 1~.
\label{eq:bdef}\eeq
We note in passing that the coefficient $b_0{\=}\ps_p(0){\=}\ps_p(1)$ that appears in the expansion above is the p-adic analogue of Euler's constant $\g_E{\=}-\G'(1)$.

The integral that appears here is the Volkenborn integral~\cite{Robert2013course}. It is somewhat analogous to a Riemann sum and has the definition
\beq
\int_{\IZ_p} f(z)\,\dd z\= \lim_{n\to\infty} \frac{1}{p^n} \sum_{k=0}^{p^n - 1} f(k)~.
\notag\eeq
and for compact subsets $Y{\subset}\IZ_p$ the integral is defined by
\beq
\int_Y  f(z)\,\dd z\= \int_{\IZ_p} \Th_Y(z) f(z)\,\dd z~,
\notag\eeq
where $\Th_Y$ is the characteristic function of $Y$. 

It is also straightforward to show that, for $k{\,\in\,}\IZ_{\geqslant 0}$
\beq
\int_{\IZ_p^*} z^k\, \dd z \= (1 - p^{k-1})B_k~,
\label{eq:Bernoulli}\eeq
where the $B_k$ are the Bernoulli numbers
\beq
B_0\=1~,\quad B_1\;=-\frac12~,\quad B_2\=\frac16~,\quad B_3\=0~,\quad B_4\;=-\frac{1}{30}~,\ldots
\notag\eeq
and have generating function
\beq
\frac{t}{\ee^t - 1}\= \sum_{k=0}^\infty B_k \frac{t^k}{k!}~.
\notag\eeq
The Bernoulli numbers $B_{2k+1}$ vanish for $k\geqslant1$, and so the quantities $(1 - p^{k-1})B_k$ vanish for $k$ odd, including $k{\=}1$. The numbers $b_n$ also vanish for odd $n$. 

Our interest in \eqref{eq:Bernoulli} and \eqref{eq:bdef} is that $\z(s)$ can be related to the Bernoulli numbers and these are, in a sense that we shall now examine, related to the constants $b_n$. The question arises as to what extent the constants $b_n$ and $B_n$ can be interpolated as functions of $n$. In order to examine this,
      we  enquire to what extent the function $z^s$, for $z{\,\in\,}\IZ_p^*$ can be defined and is smooth, as a function of p-adic $s$.       

Note first that it is not a priori obvious that this can be done. For suppose $z{\=}n$, a rational integer not divisible by $p$, then, if $n^s$ were continuous in $s$, the power $n^{1+p^m}$ would tend to $n$ as $m{\,\to\,}\infty$. However,
\beq
n^{1+p^m}\= n\, n^{p^m}\= n\big(n+\cO(p)\big)\= n^2 + \cO(p)~,
\notag\eeq
so, if $n\neq 1$ mod $p$, this is not close to $n$.

On the other hand, there are two cases that can be interpolated. If $z{\=}1{\,+\,}\cO(p)$, that is $z{\=}1{\,+\,}p z_1$, with $z_1{\,\in\,}\IZ_p$ then
\beq
z^s \= (1 + pz_1)^s
\notag\eeq
and the expression on the right is defined through the binomial expansion and is continuous as a function of $s$.
If the leading digit of $z$ is not 1, we can still proceed by considering a set of integers that are equal mod $p{\,-\,}1$
\beq
s\= s_0 + (p-1)s_1~,
\label{eq:ssequence}\eeq
We have
\beq
z^s \= z^{s_0} \left( z^{p-1}\right)^{s_1}
\notag\eeq
and $z^{p-1}{\=}1+\cO(p)$. This returns us to the previous case and we see that $z^s$ is $p$-adically smooth since $z^{s + a(p-1)p^m}{\,\to\,}z^s$ as $m{\,\to\,}\infty$.

In virtue of the above we can write
\beq
b_{2n}\= \lim_{m\to\infty} \int_{\IZ_p^*} z^{(p-1)p^m - 2n}\,\dd z
\=\lim_{m\to\infty}B_{(p-1)p^m - 2n}~.
\label{eq:bFromB}\eeq
The numbers $(p-1)p^m - 2n$ are, for $p{\,>\,}2$, a sequence of positive even integers that tend $p$-adically to $-2n$.

We can now apply these considerations to the definition of a $p$-adic $\z$-function.

The Riemann $z$-function is defined, for $\hbox{Re}(s){\,>\,}1$, by either of the relations
\beq
\z(s) \= \sum_{n=1}^\infty \frac{1}{n^s} \= \prod_{q~\text{prime}}\frac{1}{1-q^{-s}}~.
\notag\eeq
Consider first the representation as a sum. The sum contains terms for which $p|n$ and it seems best to remove these terms. So we define an apocopated function
\beq
\z^*(s) \= \sum^\infty_{\substack{n=1\\ p\notdiv n}}\frac{1}{n^s} 
\= \sum^\infty_{n=1}\frac{1}{n^s} -  \sum^\infty_{n=1}\frac{1}{(pn)^s}
\= (1 - p^{-s}) \z(s)~.
\notag\eeq
Note that
\beq
\z^*(s) \= \prod_{\substack{q~\text{prime}\\ q\neq p}}\frac{1}{1-q^{-s}}~.
\notag\eeq
For this reason, the process of passing from $\z(s)$ to $\z^*(s)$ is known as ``removing the Euler factor''.

For $s$ a positive even integer, we have the formula, due to Euler,
\beq
\z(2n)\= \frac{2^{2n-1} \p^{2n}}{(2n)!}| B_{2n} |~,
\notag\eeq
with $B_{2n}$ a Bernoulli number. For $s{\,>\,}1$, a positive odd number, we have the values $\z(3)$, $\z(5)$, $\ldots$, which are believed to be algebraically independent transcendents. In neither case can the $\z(n)$, for $n$ a positive integer, be understood as $p$-adic numbers.

The situation is better for the negative integers. For the negative even integers, we have
\beq
\z(-2k)\= \begin{cases} -\frac12 &\text{for}~k=0~,\\ \hfill 0 &\text{for}~k\geqslant 1~. \end{cases}
\notag\eeq
While for the negative odd integers we have
\beq
\z(1-2k) \= - \frac{B_{2k}}{2k}~,
\notag\eeq
and so
\beq
\z^*(1-2k) \= - \frac{1}{2k} (1- p^{2k-1}) B_{2k} \= - \frac{b_{-2k}}{2k}~.
\notag\eeq
We define $\z_p(s)$ by interpolating these values. Writing $1{\,-\,}2k{\=}s$, we have
\beq
\z_p(s) \=\frac{b_{s-1}}{s-1}~.
\notag\eeq
and, in particular, $\z_p(3) {\=} b_2/2$. The coefficient $b_2$ may be computed from \eqref{eq:bFromB} or, more efficiently, by noting that
\beq
b_1\=0 \quad \text{and} \quad b_2\;= -\ps_p''(0)
\notag\eeq
from which we learn that 
\beq
\Gp''(0) \= \Gp'(0)^2
\label{eq:FirstGammaId}\eeq 
and
\beq
\z_p(3)\=\frac12 b_2 \;= -\frac12 \left(\Gp'''(0) - \Gp'(0)^3\right)~.
\notag\eeq

\begin{table}[htb]
\renewcommand\arraystretch{1.2}
\begin{center}
\begin{tabular}{| c || l |}
\hline
\vrule height14pt depth 8pt width0pt $p$ & \hfil $\z_p(3)$\\
\hline\hline
 5 &\quad 5072570838186014721957419926005297 \\\hline
      &\quad\{2,4,1,2,3,1,4,2,0,2,0,0,2,3,3,0,0,4,2,4,0,1,4,0,2,3,2,2,3,2\} \\\hline\hline
 7 &\quad 117061310239336308334025231434747323634935 \\\hline
      &\quad\{1,3,5,4,1,0,2,0,2,4,1,2,5,0,4,6,2,4,4,5,6,3,2,2,5,3,2,1,5,2\} \\\hline\hline
 11 &\quad 149134818187057310308096045871911437019561347245302 \\\hline
        &\quad\{5,4,1,0,5,8,5,2,6,8,3,1,4,3,4,7,0,10,10,3,9,5,10,7,0,9,7,5,7\} \\\hline\hline
 13 &\quad 1959147439192239725032744234064691597958143832554774385 \\\hline
        &\quad\{6,6,5,2,11,12,5,1,5,2,7,5,10,10,6,7,12,0,10,2,12,0,8,3,1,8,12,7,0,2\}\\\hline\hline
 17 &\quad 1047241053891811275086066324080237679295844245162126884817922 \\\hline
        &\quad\{7,10,6,1,1,1,12,12,1,15,10,4,1,0,3,1,14,3,7,13,8,12,0,7,10,16,13,16,8,9\} \\\hline\hline
 19 &\quad 366705732565363086376907501556670998446992440018555462542006132\quad{} \\\hline 
        &\quad\{5,16,17,8,0,5,13,12,3,14,7,9,8,13,17,18,10,10,2,2,11,9,10,15,12,14,3,11,3\}\quad{} \\\hline
 \end{tabular}
 \capt{4.5in}{tab:zeta3}{The values of $\z_p(3)$ for $5\leqslant p\leqslant 19$. These are given as integers mod $p^{50}$ and also, mod $p^{30}$, as a list of $p$-adic digits .}
 \end{center}
 \end{table} 
 
Although the definition that we have used for $\z_p(3)$ is widely used, other choices are possible. In a celebrated paper, Kubota and Leopoldt~\cite{Kubota1964} seek a $p$-adic analogue $L_p(s,\chi)$ of the \hbox{$L$-function }
\beq
L(s,\chi) \= \sum_{n=1}^\infty\frac{\chi(n)}{n^s} \= \prod_{q~\text{prime}} \frac{1}{1 - \chi(q)\, q^{-s}}~,
\notag\eeq
where $\chi$ is a Dirichlet character. Clearly $L(s,1){\=}\z(s)$, however the procedure of $p$-adic interpolation followed by Kubota and Leopoldt leads to the relation
\[ 
\z_p(n) \= L_p(n,\omega^{1-n})~,
\]
where $\omega$ is the Teichm\"uller character. In concrete terms, this means
\beq
\z_p(n) \= \sum_{a=1}^{p-1} \omega(a)^{1-n} H_p(n,a,p)~,
\label{eq:KLrelation}\eeq
with 
\[ 
H_p(n,a,p)\= \frac{1}{(n-1)p}\, \langle a\rangle^{1-n} \sum_{j=0}^{\infty}\binom{1-n}{j} \left(\frac{p}{a}\right)^j B_j~,
\]
where~\cite{Washington1997CyclotomicFields,Beukers2008} 
\[ 
\langle a\rangle \= \frac{a}{\omega(a)}~.
\]

In fact one can replace the Teichm\"{u}ller character $\o(a)$ in \eqref{eq:KLrelation} by $\o(a)^k$ for the $p{-}1$ values $0\leqslant k\leqslant p{-}2$ and these have an equal right to be called the $p$-adic zeta value. The element of choice goes back to the choice made in \eqref{eq:ssequence} of interpolating the zeta function through a particular sequence of integers. Other choices are possible and this leads to a situation whereby $\z_p$ is a function with $p{-}1$ `branches'.

It has been shown that that for $p=2,3$ the number $\z_p(3)$ is irrational,
\cite{Calegari2005,Beukers2008} (in the case $p{\=}2$, there is only one branch and, in the case $p{\=}3$, the two possible definitions
coincide!). However, it is currently not even known if $\z_p(3)$ is nonzero for all $p$.  

Let us return to the $p$-adic gamma function. Even the briefest account must make mention of two fundamental identities, presented here without proof, that are very close to the corresponding identities for the classical gamma function. The first is the analogue of Euler's reflection formula. For $z{\,\in\,}\IZ_p$ write
\beq
z \= z_0 + p z_1
\notag\eeq
where $z_0{\,\in\,}\{1,2,\ldots,p\}$, in other words $z_0$ is the first $p$-adic digit of $z$, unless $z{\,\in\,}p\IZ_p$, in which case $z_0{\=}p$. Then we have
\beq
\Gp(z)\Gp(1-z) \= (-1)^{z_0}~.
\notag\eeq
The second identity is the analogue of the multiplication formula, which can be written in the form
\beq
\frac{\displaystyle \prod_{r=0}^{m-1}\Gp\left(\frac{z+r}{m}\right)}{\displaystyle \Gp(z)\,
\prod_{r=1}^{m-1}\Gp\left(\frac{r}{m}\right)} \= 
m^{1-z_0}\left(m^{-(p-1)}\right)^{z_1}~,
\notag\eeq
with $z_0$ and $z_1$ as above. Remarkably, the classical analogue of this formula can be written by replacing the $p$-adic gamma functions by the classical functions and the right hand side of the expression by~$m^{(1-z)}$.

One use of the reflection formula is the following. If $z{\,\in\,}p\IZ_p$, then the reflection formula becomes
\beq
\Gp(z)\Gp(-z) \= 1~.
\notag\eeq
On expanding the gamma functions on the left as Taylor series, we discover many identities relating the derivatives $\Gp^{(n)}(0)$; of which \eqref{eq:FirstGammaId} is but the first.

Finally, let us return to the expansion for the digamma function \eqref{eq:FirstGammaId}, substitute what we know for the coefficients and integrate. We come to the useful expression
\beq
\Gp(z) \= \exp\left( \Gp'(0)\, z - \sum_{k=1}^{\infty}\frac{\z_p(2k+1)}{2k+1}\,z^{2k+1}\right)~.
\label{eq:LogGpExpansion}\eeq
\newpage
\iftables
\renewcommand{\arraystretch}{0.9}

\newcommand{\tablepreamble}[1]{
\vspace{-1.7cm}
\begin{center}
\begin{longtable}{| >{\footnotesize$~} c <{~$} | >{~\footnotesize} l <{~} 
| >{\centering\footnotesize$}p{1in}<{$} |>{\centering\footnotesize $}p{3in}<{$} |}\hline
\multicolumn{4}{|c|}{\vrule height 13pt depth8pt width 0pt \small $p=#1$}\tabularnewline[0.5pt] \hline
\vph & smooth/sing. & $singularity$ & R(T)\tabularnewline[0.5pt] \hline\hline
\endfirsthead
\hline
\multicolumn{4}{|l|}{\footnotesize\sl $p=#1$, continued}\tabularnewline[0.5pt] \hline
\vph & smooth/sing. & $singularity$ & R(T)\tabularnewline[0.5pt] \hline\hline
\endhead
\hline\hline 
\multicolumn{4}{|r|}{{\footnotesize\sl Continued on the following page}}\tabularnewline[0.5pt] \hline
\endfoot
\hline
\endlastfoot}

\newcommand{\tablepostamble}{\end{longtable}\end{center}}

\pagestyle{fancy}
\setlength{\headheight}{14.5pt}
\renewcommand{\headrulewidth}{0pt}
\fancyfoot{}
\lhead{\ifthenelse{\isodd{\value{page}}}{\thepage}{\sl The $\z$-function for the mirror of the quintic threefold, AESZ\hskip2pt 1}}
\rhead{\ifthenelse{\isodd{\value{page}}}{\sl The $\z$-function for the mirror of the quintic threefold, AESZ\hskip2pt 1}{\thepage}}
\null\vskip10pt
\rightline{\<k*:t\,ob .hAzwoN w*bA'er `al--hal*u.hOt l:ma`an yArw*.s qwore' bwo;>}
\vskip10pt
\rightline{\it Write the vision, and make it plain in tables,}
\rightline{\it so that he that runs past may read it.}
\vskip5pt
\rightline{Habakkuk 2.2}
\vskip1.5cm
\section{$\z$-Function Tables}\label{sec:tables}
\vskip-12pt
\subsection{The $\z$-function for the mirror of the quintic threefold, AESZ\hskip2pt 1}
\vspace{1.5cm}
\tablepreamble{7}				
1	 &	smooth    &	          &	1+5 T+55 p T^2+5 p^3 T^3+p^6 T^4
\tabularnewline[0.5pt]\hline				
2	 &	smooth    &	          &	1+25 T+50 p T^2+25 p^3 T^3+p^6 T^4
\tabularnewline[0.5pt]\hline				
3	 &	smooth    &	          &	1+10 T+60 p T^2+10 p^3 T^3+p^6 T^4
\tabularnewline[0.5pt]\hline				
4	 &	smooth    &	          &	1-5 T-30 p T^2-5 p^3 T^3+p^6 T^4
\tabularnewline[0.5pt]\hline				
5	 &	singular  &	5^{-5}&	(1+p T) (1-6 T+p^3 T^2)
\tabularnewline[0.5pt]\hline				
6	 &	smooth    &	          &	1-5 p T+115 p T^2-5 p^4 T^3+p^6 T^4
\tabularnewline[0.5pt]\hline				
\tablepostamble				
\tablepreamble{11}				
1	 &	singular  &	5^{-5}&	(1-p T) (1+43 T+p^3 T^2)
\tabularnewline[0.5pt]\hline				
2	 &	smooth    &	          &	1-14 T-74 p T^2-14 p^3 T^3+p^6 T^4
\tabularnewline[0.5pt]\hline				
3	 &	smooth    &	          &	1-29 T+6 p^2 T^2-29 p^3 T^3+p^6 T^4
\tabularnewline[0.5pt]\hline				
4	 &	smooth    &	          &	1+31 T+131 p T^2+31 p^3 T^3+p^6 T^4
\tabularnewline[0.5pt]\hline				
5	 &	smooth    &	          &	1+T-14 p T^2+p^3 T^3+p^6 T^4
\tabularnewline[0.5pt]\hline				
6	 &	smooth    &	          &	1-9 T-4 p T^2-9 p^3 T^3+p^6 T^4
\tabularnewline[0.5pt]\hline				
7	 &	smooth    &	          &	1-54 T+266 p T^2-54 p^3 T^3+p^6 T^4
\tabularnewline[0.5pt]\hline				
8	 &	smooth    &	          &	1+31 T+206 p T^2+31 p^3 T^3+p^6 T^4
\tabularnewline[0.5pt]\hline				
9	 &	smooth    &	          &	1+26 T+236 p T^2+26 p^3 T^3+p^6 T^4
\tabularnewline[0.5pt]\hline				
10	 &	smooth    &	          &	1-14 T+76 p T^2-14 p^3 T^3+p^6 T^4
\tabularnewline[0.5pt]\hline				
\tablepostamble				
\tablepreamble{13}				
1	 &	smooth    &	          &	1-5 T+160 p T^2-5 p^3 T^3+p^6 T^4
\tabularnewline[0.5pt]\hline				
2	 &	smooth    &	          &	1+15 T+120 p T^2+15 p^3 T^3+p^6 T^4
\tabularnewline[0.5pt]\hline				
3	 &	smooth    &	          &	1+85 T+405 p T^2+85 p^3 T^3+p^6 T^4
\tabularnewline[0.5pt]\hline				
4	 &	smooth    &	          &	(1+6 p T+p^3 T^2)(1-68 T+p^3 T^2)
\tabularnewline[0.5pt]\hline				
5	 &	smooth    &	          &	1-15 T-20 p T^2-15 p^3 T^3+p^6 T^4
\tabularnewline[0.5pt]\hline				
6	 &	smooth    &	          &	(1-6 p T+p^3 T^2)(1-42 T+p^3 T^2)
\tabularnewline[0.5pt]\hline				
7	 &	smooth    &	          &	1+20 T-90 p T^2+20 p^3 T^3+p^6 T^4
\tabularnewline[0.5pt]\hline				
8	 &	singular  &	5^{-5}&	(1+p T) (1+28 T+p^3 T^2)
\tabularnewline[0.5pt]\hline				
9	 &	smooth    &	          &	1-5 T+20 p^2 T^2-5 p^3 T^3+p^6 T^4
\tabularnewline[0.5pt]\hline				
10	 &	smooth    &	          &	1+25 T+100 p T^2+25 p^3 T^3+p^6 T^4
\tabularnewline[0.5pt]\hline				
11	 &	smooth    &	          &	1-25 T+275 p T^2-25 p^3 T^3+p^6 T^4
\tabularnewline[0.5pt]\hline				
12	 &	smooth    &	          &	1-25 T+100 p T^2-25 p^3 T^3+p^6 T^4
\tabularnewline[0.5pt]\hline				
\tablepostamble				
\tablepreamble{17}				
1	 &	smooth    &	          &	1+5 p T+660 p T^2+5 p^4 T^3+p^6 T^4
\tabularnewline[0.5pt]\hline				
2	 &	smooth    &	          &	1+70 T+270 p T^2+70 p^3 T^3+p^6 T^4
\tabularnewline[0.5pt]\hline				
3	 &	smooth    &	          &	1-25 T-25 p T^2-25 p^3 T^3+p^6 T^4
\tabularnewline[0.5pt]\hline				
4	 &	smooth    &	          &	1+25 T+225 p T^2+25 p^3 T^3+p^6 T^4
\tabularnewline[0.5pt]\hline				
5	 &	smooth    &	          &	1+20 T+120 p T^2+20 p^3 T^3+p^6 T^4
\tabularnewline[0.5pt]\hline				
6	 &	smooth    &	          &	1+15 T+90 p T^2+15 p^3 T^3+p^6 T^4
\tabularnewline[0.5pt]\hline				
7	 &	smooth    &	          &	1-20 T+130 p T^2-20 p^3 T^3+p^6 T^4
\tabularnewline[0.5pt]\hline				
8	 &	smooth    &	          &	1+55 T+280 p T^2+55 p^3 T^3+p^6 T^4
\tabularnewline[0.5pt]\hline				
9	 &	smooth    &	          &	1-45 T+530 p T^2-45 p^3 T^3+p^6 T^4
\tabularnewline[0.5pt]\hline				
10	 &	smooth    &	          &	1+75 T+100 p T^2+75 p^3 T^3+p^6 T^4
\tabularnewline[0.5pt]\hline				
11	 &	singular  &	5^{-5}&	(1+p T) (1-91 T+p^3 T^2)
\tabularnewline[0.5pt]\hline				
12	 &	smooth    &	          &	1+5 T+5 p T^2+5 p^3 T^3+p^6 T^4
\tabularnewline[0.5pt]\hline				
13	 &	smooth    &	          &	1-75 T+150 p T^2-75 p^3 T^3+p^6 T^4
\tabularnewline[0.5pt]\hline				
14	 &	smooth    &	          &	1-140 T+710 p T^2-140 p^3 T^3+p^6 T^4
\tabularnewline[0.5pt]\hline				
15	 &	smooth    &	          &	1+120 T+570 p T^2+120 p^3 T^3+p^6 T^4
\tabularnewline[0.5pt]\hline				
16	 &	smooth    &	          &	1-90 T+610 p T^2-90 p^3 T^3+p^6 T^4
\tabularnewline[0.5pt]\hline				
\tablepostamble				
\tablepreamble{19}				
1	 &	smooth    &	          &	1-60 T+122 p T^2-60 p^3 T^3+p^6 T^4
\tabularnewline[0.5pt]\hline				
2	 &	smooth    &	          &	1-120 T+522 p T^2-120 p^3 T^3+p^6 T^4
\tabularnewline[0.5pt]\hline				
3	 &	smooth    &	          &	1-90 T+372 p T^2-90 p^3 T^3+p^6 T^4
\tabularnewline[0.5pt]\hline				
4	 &	smooth    &	          &	1+45 T-153 p T^2+45 p^3 T^3+p^6 T^4
\tabularnewline[0.5pt]\hline				
5	 &	smooth    &	          &	1+15 T-378 p T^2+15 p^3 T^3+p^6 T^4
\tabularnewline[0.5pt]\hline				
6	 &	smooth    &	          &	1+25 T-12 p^2 T^2+25 p^3 T^3+p^6 T^4
\tabularnewline[0.5pt]\hline				
7	 &	smooth    &	          &	1+185 T+1097 p T^2+185 p^3 T^3+p^6 T^4
\tabularnewline[0.5pt]\hline				
8	 &	smooth    &	          &	1+130 T+822 p T^2+130 p^3 T^3+p^6 T^4
\tabularnewline[0.5pt]\hline				
9	 &	smooth    &	          &	1+65 T+597 p T^2+65 p^3 T^3+p^6 T^4
\tabularnewline[0.5pt]\hline				
10	 &	smooth    &	          &	1+5 T+197 p T^2+5 p^3 T^3+p^6 T^4
\tabularnewline[0.5pt]\hline				
11	 &	smooth    &	          &	1-130 T+572 p T^2-130 p^3 T^3+p^6 T^4
\tabularnewline[0.5pt]\hline				
12	 &	smooth    &	          &	1+55 T+597 p T^2+55 p^3 T^3+p^6 T^4
\tabularnewline[0.5pt]\hline				
13	 &	smooth    &	          &	(1-5 p T+p^3 T^2)(1+100 T+p^3 T^2)
\tabularnewline[0.5pt]\hline				
14	 &	smooth    &	          &	1-75 T+422 p T^2-75 p^3 T^3+p^6 T^4
\tabularnewline[0.5pt]\hline				
15	 &	smooth    &	          &	1+20 T+22 p T^2+20 p^3 T^3+p^6 T^4
\tabularnewline[0.5pt]\hline				
16	 &	smooth    &	          &	1+20 T+222 p T^2+20 p^3 T^3+p^6 T^4
\tabularnewline[0.5pt]\hline				
17	 &	singular  &	5^{-5}&	(1-p T) (1+35 T+p^3 T^2)
\tabularnewline[0.5pt]\hline				
18	 &	smooth    &	          &	1-110 T+422 p T^2-110 p^3 T^3+p^6 T^4
\tabularnewline[0.5pt]\hline				
\tablepostamble				
\tablepreamble{23}				
1	 &	smooth    &	          &	1+130 T+240 p T^2+130 p^3 T^3+p^6 T^4
\tabularnewline[0.5pt]\hline				
2	 &	smooth    &	          &	1+40 T+670 p T^2+40 p^3 T^3+p^6 T^4
\tabularnewline[0.5pt]\hline				
3	 &	smooth    &	          &	1+160 T+930 p T^2+160 p^3 T^3+p^6 T^4
\tabularnewline[0.5pt]\hline				
4	 &	smooth    &	          &	1+105 T+90 p T^2+105 p^3 T^3+p^6 T^4
\tabularnewline[0.5pt]\hline				
5	 &	smooth    &	          &	1-15 T+730 p T^2-15 p^3 T^3+p^6 T^4
\tabularnewline[0.5pt]\hline				
6	 &	smooth    &	          &	1-55 T-90 p T^2-55 p^3 T^3+p^6 T^4
\tabularnewline[0.5pt]\hline				
7	 &	smooth    &	          &	1+50 T+450 p T^2+50 p^3 T^3+p^6 T^4
\tabularnewline[0.5pt]\hline				
8	 &	smooth    &	          &	1-60 T+220 p T^2-60 p^3 T^3+p^6 T^4
\tabularnewline[0.5pt]\hline				
9	 &	smooth    &	          &	1-135 T+395 p T^2-135 p^3 T^3+p^6 T^4
\tabularnewline[0.5pt]\hline				
10	 &	smooth    &	          &	1-5 T+210 p T^2-5 p^3 T^3+p^6 T^4
\tabularnewline[0.5pt]\hline				
11	 &	smooth    &	          &	1-40 T-520 p T^2-40 p^3 T^3+p^6 T^4
\tabularnewline[0.5pt]\hline				
12	 &	smooth    &	          &	1-5 p T+605 p T^2-5 p^4 T^3+p^6 T^4
\tabularnewline[0.5pt]\hline				
13	 &	smooth    &	          &	1+45 T+335 p T^2+45 p^3 T^3+p^6 T^4
\tabularnewline[0.5pt]\hline				
14	 &	smooth    &	          &	1+175 T+1100 p T^2+175 p^3 T^3+p^6 T^4
\tabularnewline[0.5pt]\hline				
15	 &	singular  &	5^{-5}&	(1+p T) (1-162 T+p^3 T^2)
\tabularnewline[0.5pt]\hline				
16	 &	smooth    &	          &	1+150 T+850 p T^2+150 p^3 T^3+p^6 T^4
\tabularnewline[0.5pt]\hline				
17	 &	smooth    &	          &	1-245 T+1665 p T^2-245 p^3 T^3+p^6 T^4
\tabularnewline[0.5pt]\hline				
18	 &	smooth    &	          &	1+350 p T^2+p^6 T^4
\tabularnewline[0.5pt]\hline				
19	 &	smooth    &	          &	1+105 T+365 p T^2+105 p^3 T^3+p^6 T^4
\tabularnewline[0.5pt]\hline				
20	 &	smooth    &	          &	1-100 T+850 p T^2-100 p^3 T^3+p^6 T^4
\tabularnewline[0.5pt]\hline				
21	 &	smooth    &	          &	1+55 T+865 p T^2+55 p^3 T^3+p^6 T^4
\tabularnewline[0.5pt]\hline				
22	 &	smooth    &	          &	1-105 T+960 p T^2-105 p^3 T^3+p^6 T^4
\tabularnewline[0.5pt]\hline				
\tablepostamble				
\tablepreamble{29}				
1	 &	smooth    &	          &	1+240 T+1582 p T^2+240 p^3 T^3+p^6 T^4
\tabularnewline[0.5pt]\hline				
2	 &	smooth    &	          &	1-245 T+1732 p T^2-245 p^3 T^3+p^6 T^4
\tabularnewline[0.5pt]\hline				
3	 &	smooth    &	          &	1-105 T+182 p T^2-105 p^3 T^3+p^6 T^4
\tabularnewline[0.5pt]\hline				
4	 &	singular  &	5^{-5}&	(1-p T) (1-160 T+p^3 T^2)
\tabularnewline[0.5pt]\hline				
5	 &	smooth    &	          &	1+80 T+782 p T^2+80 p^3 T^3+p^6 T^4
\tabularnewline[0.5pt]\hline				
6	 &	smooth    &	          &	1+125 T+8 p^2 T^2+125 p^3 T^3+p^6 T^4
\tabularnewline[0.5pt]\hline				
7	 &	smooth    &	          &	1+420 T+3032 p T^2+420 p^3 T^3+p^6 T^4
\tabularnewline[0.5pt]\hline				
8	 &	smooth    &	          &	1-90 T-118 p T^2-90 p^3 T^3+p^6 T^4
\tabularnewline[0.5pt]\hline				
9	 &	smooth    &	          &	1-15 T-68 p T^2-15 p^3 T^3+p^6 T^4
\tabularnewline[0.5pt]\hline				
10	 &	smooth    &	          &	1+15 T-318 p T^2+15 p^3 T^3+p^6 T^4
\tabularnewline[0.5pt]\hline				
11	 &	smooth    &	          &	1-10 T+932 p T^2-10 p^3 T^3+p^6 T^4
\tabularnewline[0.5pt]\hline				
12	 &	smooth    &	          &	1-175 T+1607 p T^2-175 p^3 T^3+p^6 T^4
\tabularnewline[0.5pt]\hline				
13	 &	smooth    &	          &	1-175 T+1607 p T^2-175 p^3 T^3+p^6 T^4
\tabularnewline[0.5pt]\hline				
14	 &	smooth    &	          &	1+135 T+557 p T^2+135 p^3 T^3+p^6 T^4
\tabularnewline[0.5pt]\hline				
15	 &	smooth    &	          &	1+45 T+932 p T^2+45 p^3 T^3+p^6 T^4
\tabularnewline[0.5pt]\hline				
16	 &	smooth    &	          &	1+50 T+1132 p T^2+50 p^3 T^3+p^6 T^4
\tabularnewline[0.5pt]\hline				
17	 &	smooth    &	          &	1+140 T+582 p T^2+140 p^3 T^3+p^6 T^4
\tabularnewline[0.5pt]\hline				
18	 &	smooth    &	          &	1-25 T+757 p T^2-25 p^3 T^3+p^6 T^4
\tabularnewline[0.5pt]\hline				
19	 &	smooth    &	          &	1+75 T+1357 p T^2+75 p^3 T^3+p^6 T^4
\tabularnewline[0.5pt]\hline				
20	 &	smooth    &	          &	1-35 T+432 p T^2-35 p^3 T^3+p^6 T^4
\tabularnewline[0.5pt]\hline				
21	 &	smooth    &	          &	1+90 T+682 p T^2+90 p^3 T^3+p^6 T^4
\tabularnewline[0.5pt]\hline				
22	 &	smooth    &	          &	1-185 T+682 p T^2-185 p^3 T^3+p^6 T^4
\tabularnewline[0.5pt]\hline				
23	 &	smooth    &	          &	1-80 T-718 p T^2-80 p^3 T^3+p^6 T^4
\tabularnewline[0.5pt]\hline				
24	 &	smooth    &	          &	1-125 T+107 p T^2-125 p^3 T^3+p^6 T^4
\tabularnewline[0.5pt]\hline				
25	 &	smooth    &	          &	1+45 T+482 p T^2+45 p^3 T^3+p^6 T^4
\tabularnewline[0.5pt]\hline				
26	 &	smooth    &	          &	1-20 T-718 p T^2-20 p^3 T^3+p^6 T^4
\tabularnewline[0.5pt]\hline				
27	 &	smooth    &	          &	1-250 T+1482 p T^2-250 p^3 T^3+p^6 T^4
\tabularnewline[0.5pt]\hline				
28	 &	smooth    &	          &	1+265 T+1932 p T^2+265 p^3 T^3+p^6 T^4
\tabularnewline[0.5pt]\hline				
\tablepostamble				
\tablepreamble{31}				
1	 &	smooth    &	          &	(1-7 p T+p^3 T^2)(1+108 T+p^3 T^2)
\tabularnewline[0.5pt]\hline				
2	 &	smooth    &	          &	1-114 T+846 p T^2-114 p^3 T^3+p^6 T^4
\tabularnewline[0.5pt]\hline				
3	 &	smooth    &	          &	1+111 T+946 p T^2+111 p^3 T^3+p^6 T^4
\tabularnewline[0.5pt]\hline				
4	 &	smooth    &	          &	1+366 T+2966 p T^2+366 p^3 T^3+p^6 T^4
\tabularnewline[0.5pt]\hline				
5	 &	singular  &	5^{-5}&	(1-p T) (1-42 T+p^3 T^2)
\tabularnewline[0.5pt]\hline				
6	 &	smooth    &	          &	1+66 T+566 p T^2+66 p^3 T^3+p^6 T^4
\tabularnewline[0.5pt]\hline				
7	 &	smooth    &	          &	1+81 T+1401 p T^2+81 p^3 T^3+p^6 T^4
\tabularnewline[0.5pt]\hline				
8	 &	smooth    &	          &	1-144 T+1026 p T^2-144 p^3 T^3+p^6 T^4
\tabularnewline[0.5pt]\hline				
9	 &	smooth    &	          &	1-39 T+471 p T^2-39 p^3 T^3+p^6 T^4
\tabularnewline[0.5pt]\hline				
10	 &	smooth    &	          &	1+191 T+1091 p T^2+191 p^3 T^3+p^6 T^4
\tabularnewline[0.5pt]\hline				
11	 &	smooth    &	          &	1+16 T-14 p^2 T^2+16 p^3 T^3+p^6 T^4
\tabularnewline[0.5pt]\hline				
12	 &	smooth    &	          &	1-174 T+1906 p T^2-174 p^3 T^3+p^6 T^4
\tabularnewline[0.5pt]\hline				
13	 &	smooth    &	          &	1+26 T+156 p T^2+26 p^3 T^3+p^6 T^4
\tabularnewline[0.5pt]\hline				
14	 &	smooth    &	          &	1-4 T-214 p T^2-4 p^3 T^3+p^6 T^4
\tabularnewline[0.5pt]\hline				
15	 &	smooth    &	          &	1-139 T+796 p T^2-139 p^3 T^3+p^6 T^4
\tabularnewline[0.5pt]\hline				
16	 &	smooth    &	          &	1-44 T-774 p T^2-44 p^3 T^3+p^6 T^4
\tabularnewline[0.5pt]\hline				
17	 &	smooth    &	          &	1+T+206 p T^2+p^3 T^3+p^6 T^4
\tabularnewline[0.5pt]\hline				
18	 &	smooth    &	          &	1-154 T+6 p^2 T^2-154 p^3 T^3+p^6 T^4
\tabularnewline[0.5pt]\hline				
19	 &	smooth    &	          &	1+81 T-99 p T^2+81 p^3 T^3+p^6 T^4
\tabularnewline[0.5pt]\hline				
20	 &	smooth    &	          &	1-4 p T+1056 p T^2-4 p^4 T^3+p^6 T^4
\tabularnewline[0.5pt]\hline				
21	 &	smooth    &	          &	1+111 T+846 p T^2+111 p^3 T^3+p^6 T^4
\tabularnewline[0.5pt]\hline				
22	 &	smooth    &	          &	1+171 T+1611 p T^2+171 p^3 T^3+p^6 T^4
\tabularnewline[0.5pt]\hline				
23	 &	smooth    &	          &	1+p T+1576 p T^2+p^4 T^3+p^6 T^4
\tabularnewline[0.5pt]\hline				
24	 &	smooth    &	          &	1-119 T+76 p T^2-119 p^3 T^3+p^6 T^4
\tabularnewline[0.5pt]\hline				
25	 &	smooth    &	          &	1+391 T+86 p^2 T^2+391 p^3 T^3+p^6 T^4
\tabularnewline[0.5pt]\hline				
26	 &	smooth    &	          &	1-259 T+1591 p T^2-259 p^3 T^3+p^6 T^4
\tabularnewline[0.5pt]\hline				
27	 &	smooth    &	          &	1-339 T+2746 p T^2-339 p^3 T^3+p^6 T^4
\tabularnewline[0.5pt]\hline				
28	 &	smooth    &	          &	1+176 T+1406 p T^2+176 p^3 T^3+p^6 T^4
\tabularnewline[0.5pt]\hline				
29	 &	smooth    &	          &	1-49 T+26 p^2 T^2-49 p^3 T^3+p^6 T^4
\tabularnewline[0.5pt]\hline				
30	 &	smooth    &	          &	1+66 T-684 p T^2+66 p^3 T^3+p^6 T^4
\tabularnewline[0.5pt]\hline				
\tablepostamble				
\tablepreamble{37}				
1	 &	smooth    &	          &	1-265 T+1885 p T^2-265 p^3 T^3+p^6 T^4
\tabularnewline[0.5pt]\hline				
2	 &	smooth    &	          &	1-540 T+4210 p T^2-540 p^3 T^3+p^6 T^4
\tabularnewline[0.5pt]\hline				
3	 &	smooth    &	          &	1+180 T+1080 p T^2+180 p^3 T^3+p^6 T^4
\tabularnewline[0.5pt]\hline				
4	 &	smooth    &	          &	1+5 p T+1585 p T^2+5 p^4 T^3+p^6 T^4
\tabularnewline[0.5pt]\hline				
5	 &	smooth    &	          &	1+160 T+510 p T^2+160 p^3 T^3+p^6 T^4
\tabularnewline[0.5pt]\hline				
6	 &	smooth    &	          &	1-205 T+1470 p T^2-205 p^3 T^3+p^6 T^4
\tabularnewline[0.5pt]\hline				
7	 &	smooth    &	          &	1-315 T+1760 p T^2-315 p^3 T^3+p^6 T^4
\tabularnewline[0.5pt]\hline				
8	 &	smooth    &	          &	1+5 p T+960 p T^2+5 p^4 T^3+p^6 T^4
\tabularnewline[0.5pt]\hline				
9	 &	smooth    &	          &	1-15 T-1265 p T^2-15 p^3 T^3+p^6 T^4
\tabularnewline[0.5pt]\hline				
10	 &	smooth    &	          &	1+130 T+2130 p T^2+130 p^3 T^3+p^6 T^4
\tabularnewline[0.5pt]\hline				
11	 &	smooth    &	          &	1+155 T-245 p T^2+155 p^3 T^3+p^6 T^4
\tabularnewline[0.5pt]\hline				
12	 &	smooth    &	          &	1+385 T+2910 p T^2+385 p^3 T^3+p^6 T^4
\tabularnewline[0.5pt]\hline				
13	 &	smooth    &	          &	1-155 T+2645 p T^2-155 p^3 T^3+p^6 T^4
\tabularnewline[0.5pt]\hline				
14	 &	smooth    &	          &	1-55 T+60 p^2 T^2-55 p^3 T^3+p^6 T^4
\tabularnewline[0.5pt]\hline				
15	 &	smooth    &	          &	1-130 T+2170 p T^2-130 p^3 T^3+p^6 T^4
\tabularnewline[0.5pt]\hline				
16	 &	smooth    &	          &	1-55 T+845 p T^2-55 p^3 T^3+p^6 T^4
\tabularnewline[0.5pt]\hline				
17	 &	smooth    &	          &	1+15 T+2040 p T^2+15 p^3 T^3+p^6 T^4
\tabularnewline[0.5pt]\hline				
18	 &	smooth    &	          &	1-325 T+1900 p T^2-325 p^3 T^3+p^6 T^4
\tabularnewline[0.5pt]\hline				
19	 &	smooth    &	          &	1+155 T+1080 p T^2+155 p^3 T^3+p^6 T^4
\tabularnewline[0.5pt]\hline				
20	 &	smooth    &	          &	1-105 T-705 p T^2-105 p^3 T^3+p^6 T^4
\tabularnewline[0.5pt]\hline				
21	 &	smooth    &	          &	1+80 T+1430 p T^2+80 p^3 T^3+p^6 T^4
\tabularnewline[0.5pt]\hline				
22	 &	smooth    &	          &	1+145 T+70 p T^2+145 p^3 T^3+p^6 T^4
\tabularnewline[0.5pt]\hline				
23	 &	smooth    &	          &	1+75 T-800 p T^2+75 p^3 T^3+p^6 T^4
\tabularnewline[0.5pt]\hline				
24	 &	singular  &	5^{-5}&	(1+p T) (1+314 T+p^3 T^2)
\tabularnewline[0.5pt]\hline				
25	 &	smooth    &	          &	1-320 T+1430 p T^2-320 p^3 T^3+p^6 T^4
\tabularnewline[0.5pt]\hline				
26	 &	smooth    &	          &	1-140 T+1910 p T^2-140 p^3 T^3+p^6 T^4
\tabularnewline[0.5pt]\hline				
27	 &	smooth    &	          &	1+380 T+3580 p T^2+380 p^3 T^3+p^6 T^4
\tabularnewline[0.5pt]\hline				
28	 &	smooth    &	          &	1+135 T+1310 p T^2+135 p^3 T^3+p^6 T^4
\tabularnewline[0.5pt]\hline				
29	 &	smooth    &	          &	1+30 T-1020 p T^2+30 p^3 T^3+p^6 T^4
\tabularnewline[0.5pt]\hline				
30	 &	smooth    &	          &	1+25 T+500 p T^2+25 p^3 T^3+p^6 T^4
\tabularnewline[0.5pt]\hline				
31	 &	smooth    &	          &	1+120 T+170 p T^2+120 p^3 T^3+p^6 T^4
\tabularnewline[0.5pt]\hline				
32	 &	smooth    &	          &	1+90 T+1590 p T^2+90 p^3 T^3+p^6 T^4
\tabularnewline[0.5pt]\hline				
33	 &	smooth    &	          &	1+180 T+1780 p T^2+180 p^3 T^3+p^6 T^4
\tabularnewline[0.5pt]\hline				
34	 &	smooth    &	          &	1-210 T+1290 p T^2-210 p^3 T^3+p^6 T^4
\tabularnewline[0.5pt]\hline				
35	 &	smooth    &	          &	1-495 T+3855 p T^2-495 p^3 T^3+p^6 T^4
\tabularnewline[0.5pt]\hline				
36	 &	smooth    &	          &	1+170 T+1320 p T^2+170 p^3 T^3+p^6 T^4
\tabularnewline[0.5pt]\hline				
\tablepostamble				
\tablepreamble{41}				
1	 &	smooth    &	          &	(1+8 p T+p^3 T^2)(1-372 T+p^3 T^2)
\tabularnewline[0.5pt]\hline				
2	 &	smooth    &	          &	1-154 T-204 p T^2-154 p^3 T^3+p^6 T^4
\tabularnewline[0.5pt]\hline				
3	 &	smooth    &	          &	1-219 T-389 p T^2-219 p^3 T^3+p^6 T^4
\tabularnewline[0.5pt]\hline				
4	 &	smooth    &	          &	1+371 T+2646 p T^2+371 p^3 T^3+p^6 T^4
\tabularnewline[0.5pt]\hline				
5	 &	smooth    &	          &	1-44 T-914 p T^2-44 p^3 T^3+p^6 T^4
\tabularnewline[0.5pt]\hline				
6	 &	smooth    &	          &	1-509 T+4676 p T^2-509 p^3 T^3+p^6 T^4
\tabularnewline[0.5pt]\hline				
7	 &	smooth    &	          &	1+111 T+2906 p T^2+111 p^3 T^3+p^6 T^4
\tabularnewline[0.5pt]\hline				
8	 &	smooth    &	          &	1-4 p T+1206 p T^2-4 p^4 T^3+p^6 T^4
\tabularnewline[0.5pt]\hline				
9	 &	smooth    &	          &	1+581 T+4761 p T^2+581 p^3 T^3+p^6 T^4
\tabularnewline[0.5pt]\hline				
10	 &	smooth    &	          &	1+546 T+4196 p T^2+546 p^3 T^3+p^6 T^4
\tabularnewline[0.5pt]\hline				
11	 &	smooth    &	          &	1+146 T-854 p T^2+146 p^3 T^3+p^6 T^4
\tabularnewline[0.5pt]\hline				
12	 &	smooth    &	          &	1-184 T-124 p T^2-184 p^3 T^3+p^6 T^4
\tabularnewline[0.5pt]\hline				
13	 &	smooth    &	          &	1+201 T+716 p T^2+201 p^3 T^3+p^6 T^4
\tabularnewline[0.5pt]\hline				
14	 &	smooth    &	          &	1-144 T+2086 p T^2-144 p^3 T^3+p^6 T^4
\tabularnewline[0.5pt]\hline				
15	 &	smooth    &	          &	1+126 T-134 p T^2+126 p^3 T^3+p^6 T^4
\tabularnewline[0.5pt]\hline				
16	 &	smooth    &	          &	1+11 T+1956 p T^2+11 p^3 T^3+p^6 T^4
\tabularnewline[0.5pt]\hline				
17	 &	smooth    &	          &	1+301 T+2291 p T^2+301 p^3 T^3+p^6 T^4
\tabularnewline[0.5pt]\hline				
18	 &	smooth    &	          &	1+381 T+2186 p T^2+381 p^3 T^3+p^6 T^4
\tabularnewline[0.5pt]\hline				
19	 &	smooth    &	          &	1+T-2284 p T^2+p^3 T^3+p^6 T^4
\tabularnewline[0.5pt]\hline				
20	 &	smooth    &	          &	1-154 T+846 p T^2-154 p^3 T^3+p^6 T^4
\tabularnewline[0.5pt]\hline				
21	 &	smooth    &	          &	1+266 T+1576 p T^2+266 p^3 T^3+p^6 T^4
\tabularnewline[0.5pt]\hline				
22	 &	smooth    &	          &	1-434 T+4226 p T^2-434 p^3 T^3+p^6 T^4
\tabularnewline[0.5pt]\hline				
23	 &	smooth    &	          &	1-194 T+1586 p T^2-194 p^3 T^3+p^6 T^4
\tabularnewline[0.5pt]\hline				
24	 &	smooth    &	          &	1-239 T+1906 p T^2-239 p^3 T^3+p^6 T^4
\tabularnewline[0.5pt]\hline				
25	 &	smooth    &	          &	1-14 T+406 p T^2-14 p^3 T^3+p^6 T^4
\tabularnewline[0.5pt]\hline				
26	 &	smooth    &	          &	1-179 T+371 p T^2-179 p^3 T^3+p^6 T^4
\tabularnewline[0.5pt]\hline				
27	 &	smooth    &	          &	1-219 T+2736 p T^2-219 p^3 T^3+p^6 T^4
\tabularnewline[0.5pt]\hline				
28	 &	smooth    &	          &	1+216 T+2276 p T^2+216 p^3 T^3+p^6 T^4
\tabularnewline[0.5pt]\hline				
29	 &	smooth    &	          &	1+31 T-989 p T^2+31 p^3 T^3+p^6 T^4
\tabularnewline[0.5pt]\hline				
30	 &	smooth    &	          &	1+66 T+1976 p T^2+66 p^3 T^3+p^6 T^4
\tabularnewline[0.5pt]\hline				
31	 &	smooth    &	          &	1-469 T+3611 p T^2-469 p^3 T^3+p^6 T^4
\tabularnewline[0.5pt]\hline				
32	 &	singular  &	5^{-5}&	(1-p T) (1+203 T+p^3 T^2)
\tabularnewline[0.5pt]\hline				
33	 &	smooth    &	          &	1-169 T+936 p T^2-169 p^3 T^3+p^6 T^4
\tabularnewline[0.5pt]\hline				
34	 &	smooth    &	          &	1+301 T+1966 p T^2+301 p^3 T^3+p^6 T^4
\tabularnewline[0.5pt]\hline				
35	 &	smooth    &	          &	1+56 T+2386 p T^2+56 p^3 T^3+p^6 T^4
\tabularnewline[0.5pt]\hline				
36	 &	smooth    &	          &	1-89 T+2781 p T^2-89 p^3 T^3+p^6 T^4
\tabularnewline[0.5pt]\hline				
37	 &	smooth    &	          &	1+61 T+2256 p T^2+61 p^3 T^3+p^6 T^4
\tabularnewline[0.5pt]\hline				
38	 &	smooth    &	          &	1-269 T+2336 p T^2-269 p^3 T^3+p^6 T^4
\tabularnewline[0.5pt]\hline				
39	 &	smooth    &	          &	1+T+1216 p T^2+p^3 T^3+p^6 T^4
\tabularnewline[0.5pt]\hline				
40	 &	smooth    &	          &	1-44 T+2886 p T^2-44 p^3 T^3+p^6 T^4
\tabularnewline[0.5pt]\hline				
\tablepostamble				
\tablepreamble{43}				
1	 &	smooth    &	          &	1-135 T+670 p T^2-135 p^3 T^3+p^6 T^4
\tabularnewline[0.5pt]\hline				
2	 &	smooth    &	          &	1-540 T+4330 p T^2-540 p^3 T^3+p^6 T^4
\tabularnewline[0.5pt]\hline				
3	 &	singular  &	5^{-5}&	(1+p T) (1-92 T+p^3 T^2)
\tabularnewline[0.5pt]\hline				
4	 &	smooth    &	          &	1+180 T+1590 p T^2+180 p^3 T^3+p^6 T^4
\tabularnewline[0.5pt]\hline				
5	 &	smooth    &	          &	1-80 T-390 p T^2-80 p^3 T^3+p^6 T^4
\tabularnewline[0.5pt]\hline				
6	 &	smooth    &	          &	1-140 T+10 p^2 T^2-140 p^3 T^3+p^6 T^4
\tabularnewline[0.5pt]\hline				
7	 &	smooth    &	          &	1-875 T+8125 p T^2-875 p^3 T^3+p^6 T^4
\tabularnewline[0.5pt]\hline				
8	 &	smooth    &	          &	1+100 T+2800 p T^2+100 p^3 T^3+p^6 T^4
\tabularnewline[0.5pt]\hline				
9	 &	smooth    &	          &	1+160 T-570 p T^2+160 p^3 T^3+p^6 T^4
\tabularnewline[0.5pt]\hline				
10	 &	smooth    &	          &	1-365 T+4430 p T^2-365 p^3 T^3+p^6 T^4
\tabularnewline[0.5pt]\hline				
11	 &	smooth    &	          &	1-370 T+3790 p T^2-370 p^3 T^3+p^6 T^4
\tabularnewline[0.5pt]\hline				
12	 &	smooth    &	          &	1-25 T+2250 p T^2-25 p^3 T^3+p^6 T^4
\tabularnewline[0.5pt]\hline				
13	 &	smooth    &	          &	1-55 T+510 p T^2-55 p^3 T^3+p^6 T^4
\tabularnewline[0.5pt]\hline				
14	 &	smooth    &	          &	1-20 T+2290 p T^2-20 p^3 T^3+p^6 T^4
\tabularnewline[0.5pt]\hline				
15	 &	smooth    &	          &	1+375 T+2450 p T^2+375 p^3 T^3+p^6 T^4
\tabularnewline[0.5pt]\hline				
16	 &	smooth    &	          &	1-105 T+1410 p T^2-105 p^3 T^3+p^6 T^4
\tabularnewline[0.5pt]\hline				
17	 &	smooth    &	          &	1+600 T+4450 p T^2+600 p^3 T^3+p^6 T^4
\tabularnewline[0.5pt]\hline				
18	 &	smooth    &	          &	1+115 T+195 p T^2+115 p^3 T^3+p^6 T^4
\tabularnewline[0.5pt]\hline				
19	 &	smooth    &	          &	1-445 T+2890 p T^2-445 p^3 T^3+p^6 T^4
\tabularnewline[0.5pt]\hline				
20	 &	smooth    &	          &	1-265 T+2155 p T^2-265 p^3 T^3+p^6 T^4
\tabularnewline[0.5pt]\hline				
21	 &	smooth    &	          &	1-105 T+3110 p T^2-105 p^3 T^3+p^6 T^4
\tabularnewline[0.5pt]\hline				
22	 &	smooth    &	          &	1+160 T+330 p T^2+160 p^3 T^3+p^6 T^4
\tabularnewline[0.5pt]\hline				
23	 &	smooth    &	          &	1-185 T+670 p T^2-185 p^3 T^3+p^6 T^4
\tabularnewline[0.5pt]\hline				
24	 &	smooth    &	          &	1+565 T+5370 p T^2+565 p^3 T^3+p^6 T^4
\tabularnewline[0.5pt]\hline				
25	 &	smooth    &	          &	1+405 T+3765 p T^2+405 p^3 T^3+p^6 T^4
\tabularnewline[0.5pt]\hline				
26	 &	smooth    &	          &	1+165 T+3445 p T^2+165 p^3 T^3+p^6 T^4
\tabularnewline[0.5pt]\hline				
27	 &	smooth    &	          &	1+180 T+390 p T^2+180 p^3 T^3+p^6 T^4
\tabularnewline[0.5pt]\hline				
28	 &	smooth    &	          &	1-100 T-50 p T^2-100 p^3 T^3+p^6 T^4
\tabularnewline[0.5pt]\hline				
29	 &	smooth    &	          &	1+245 T+960 p T^2+245 p^3 T^3+p^6 T^4
\tabularnewline[0.5pt]\hline				
30	 &	smooth    &	          &	1+230 T+1440 p T^2+230 p^3 T^3+p^6 T^4
\tabularnewline[0.5pt]\hline				
31	 &	smooth    &	          &	1-85 T-1605 p T^2-85 p^3 T^3+p^6 T^4
\tabularnewline[0.5pt]\hline				
32	 &	smooth    &	          &	1+125 T+2500 p T^2+125 p^3 T^3+p^6 T^4
\tabularnewline[0.5pt]\hline				
33	 &	smooth    &	          &	1+25 T-1850 p T^2+25 p^3 T^3+p^6 T^4
\tabularnewline[0.5pt]\hline				
34	 &	smooth    &	          &	1-25 T+2350 p T^2-25 p^3 T^3+p^6 T^4
\tabularnewline[0.5pt]\hline				
35	 &	smooth    &	          &	1+425 T+3975 p T^2+425 p^3 T^3+p^6 T^4
\tabularnewline[0.5pt]\hline				
36	 &	smooth    &	          &	1-30 T+3310 p T^2-30 p^3 T^3+p^6 T^4
\tabularnewline[0.5pt]\hline				
37	 &	smooth    &	          &	1+90 T+40 p^2 T^2+90 p^3 T^3+p^6 T^4
\tabularnewline[0.5pt]\hline				
38	 &	smooth    &	          &	1-90 T-40 p^2 T^2-90 p^3 T^3+p^6 T^4
\tabularnewline[0.5pt]\hline				
39	 &	smooth    &	          &	1-105 T+1185 p T^2-105 p^3 T^3+p^6 T^4
\tabularnewline[0.5pt]\hline				
40	 &	smooth    &	          &	1+175 T+1350 p T^2+175 p^3 T^3+p^6 T^4
\tabularnewline[0.5pt]\hline				
41	 &	smooth    &	          &	1-275 T+700 p T^2-275 p^3 T^3+p^6 T^4
\tabularnewline[0.5pt]\hline				
42	 &	smooth    &	          &	1+150 T+750 p T^2+150 p^3 T^3+p^6 T^4
\tabularnewline[0.5pt]\hline				
\tablepostamble				
\tablepreamble{47}				
1	 &	smooth    &	          &	1+65 T+1790 p T^2+65 p^3 T^3+p^6 T^4
\tabularnewline[0.5pt]\hline				
2	 &	smooth    &	          &	1-190 T+3410 p T^2-190 p^3 T^3+p^6 T^4
\tabularnewline[0.5pt]\hline				
3	 &	smooth    &	          &	1+110 T+1110 p T^2+110 p^3 T^3+p^6 T^4
\tabularnewline[0.5pt]\hline				
4	 &	smooth    &	          &	1+330 T+4530 p T^2+330 p^3 T^3+p^6 T^4
\tabularnewline[0.5pt]\hline				
5	 &	smooth    &	          &	1+385 T+2610 p T^2+385 p^3 T^3+p^6 T^4
\tabularnewline[0.5pt]\hline				
6	 &	smooth    &	          &	1+390 T+4040 p T^2+390 p^3 T^3+p^6 T^4
\tabularnewline[0.5pt]\hline				
7	 &	smooth    &	          &	1+450 T+4450 p T^2+450 p^3 T^3+p^6 T^4
\tabularnewline[0.5pt]\hline				
8	 &	smooth    &	          &	1-325 T+3825 p T^2-325 p^3 T^3+p^6 T^4
\tabularnewline[0.5pt]\hline				
9	 &	smooth    &	          &	1+555 T+3930 p T^2+555 p^3 T^3+p^6 T^4
\tabularnewline[0.5pt]\hline				
10	 &	smooth    &	          &	1-45 T-1145 p T^2-45 p^3 T^3+p^6 T^4
\tabularnewline[0.5pt]\hline				
11	 &	smooth    &	          &	1+230 T-1070 p T^2+230 p^3 T^3+p^6 T^4
\tabularnewline[0.5pt]\hline				
12	 &	smooth    &	          &	1+40 T-30 p^2 T^2+40 p^3 T^3+p^6 T^4
\tabularnewline[0.5pt]\hline				
13	 &	smooth    &	          &	1-450 T+3500 p T^2-450 p^3 T^3+p^6 T^4
\tabularnewline[0.5pt]\hline				
14	 &	smooth    &	          &	1+120 T+2420 p T^2+120 p^3 T^3+p^6 T^4
\tabularnewline[0.5pt]\hline				
15	 &	smooth    &	          &	1-285 T+3065 p T^2-285 p^3 T^3+p^6 T^4
\tabularnewline[0.5pt]\hline				
16	 &	smooth    &	          &	1+80 T+480 p T^2+80 p^3 T^3+p^6 T^4
\tabularnewline[0.5pt]\hline				
17	 &	smooth    &	          &	1+410 T+1910 p T^2+410 p^3 T^3+p^6 T^4
\tabularnewline[0.5pt]\hline				
18	 &	smooth    &	          &	1-205 T+770 p T^2-205 p^3 T^3+p^6 T^4
\tabularnewline[0.5pt]\hline				
19	 &	smooth    &	          &	1+285 T+3410 p T^2+285 p^3 T^3+p^6 T^4
\tabularnewline[0.5pt]\hline				
20	 &	smooth    &	          &	1-15 T-640 p T^2-15 p^3 T^3+p^6 T^4
\tabularnewline[0.5pt]\hline				
21	 &	smooth    &	          &	1-385 T+3165 p T^2-385 p^3 T^3+p^6 T^4
\tabularnewline[0.5pt]\hline				
22	 &	smooth    &	          &	1-105 T+770 p T^2-105 p^3 T^3+p^6 T^4
\tabularnewline[0.5pt]\hline				
23	 &	smooth    &	          &	1-435 T+3065 p T^2-435 p^3 T^3+p^6 T^4
\tabularnewline[0.5pt]\hline				
24	 &	smooth    &	          &	1+145 T+1345 p T^2+145 p^3 T^3+p^6 T^4
\tabularnewline[0.5pt]\hline				
25	 &	smooth    &	          &	1+365 T+2340 p T^2+365 p^3 T^3+p^6 T^4
\tabularnewline[0.5pt]\hline				
26	 &	smooth    &	          &	1+215 T+2065 p T^2+215 p^3 T^3+p^6 T^4
\tabularnewline[0.5pt]\hline				
27	 &	smooth    &	          &	1+415 T+4065 p T^2+415 p^3 T^3+p^6 T^4
\tabularnewline[0.5pt]\hline				
28	 &	smooth    &	          &	1-720 T+5530 p T^2-720 p^3 T^3+p^6 T^4
\tabularnewline[0.5pt]\hline				
29	 &	smooth    &	          &	1-240 T+610 p T^2-240 p^3 T^3+p^6 T^4
\tabularnewline[0.5pt]\hline				
30	 &	smooth    &	          &	1-15 T-1765 p T^2-15 p^3 T^3+p^6 T^4
\tabularnewline[0.5pt]\hline				
31	 &	smooth    &	          &	1-305 T+3170 p T^2-305 p^3 T^3+p^6 T^4
\tabularnewline[0.5pt]\hline				
32	 &	smooth    &	          &	1+280 T+2780 p T^2+280 p^3 T^3+p^6 T^4
\tabularnewline[0.5pt]\hline				
33	 &	smooth    &	          &	1-210 T+2090 p T^2-210 p^3 T^3+p^6 T^4
\tabularnewline[0.5pt]\hline				
34	 &	smooth    &	          &	1-45 T+1780 p T^2-45 p^3 T^3+p^6 T^4
\tabularnewline[0.5pt]\hline				
35	 &	smooth    &	          &	1-75 T-275 p T^2-75 p^3 T^3+p^6 T^4
\tabularnewline[0.5pt]\hline				
36	 &	smooth    &	          &	1-145 T+2755 p T^2-145 p^3 T^3+p^6 T^4
\tabularnewline[0.5pt]\hline				
37	 &	smooth    &	          &	1-335 T+1690 p T^2-335 p^3 T^3+p^6 T^4
\tabularnewline[0.5pt]\hline				
38	 &	smooth    &	          &	1-5 p T+2190 p T^2-5 p^4 T^3+p^6 T^4
\tabularnewline[0.5pt]\hline				
39	 &	smooth    &	          &	1-870 T+7730 p T^2-870 p^3 T^3+p^6 T^4
\tabularnewline[0.5pt]\hline				
40	 &	smooth    &	          &	1+355 T+755 p T^2+355 p^3 T^3+p^6 T^4
\tabularnewline[0.5pt]\hline				
41	 &	smooth    &	          &	1+480 T+4430 p T^2+480 p^3 T^3+p^6 T^4
\tabularnewline[0.5pt]\hline				
42	 &	smooth    &	          &	(1-12 p T+p^3 T^2)(1+264 T+p^3 T^2)
\tabularnewline[0.5pt]\hline				
43	 &	smooth    &	          &	1-70 T+2080 p T^2-70 p^3 T^3+p^6 T^4
\tabularnewline[0.5pt]\hline				
44	 &	smooth    &	          &	1+260 T+1010 p T^2+260 p^3 T^3+p^6 T^4
\tabularnewline[0.5pt]\hline				
45	 &	singular  &	5^{-5}&	(1+p T) (1-196 T+p^3 T^2)
\tabularnewline[0.5pt]\hline				
46	 &	smooth    &	          &	1+190 T+3990 p T^2+190 p^3 T^3+p^6 T^4
\tabularnewline[0.5pt]\hline				
\tablepostamble				
\tablepreamble{53}				
1	 &	smooth    &	          &	1-270 T+2890 p T^2-270 p^3 T^3+p^6 T^4
\tabularnewline[0.5pt]\hline				
2	 &	smooth    &	          &	1+10 T-270 p T^2+10 p^3 T^3+p^6 T^4
\tabularnewline[0.5pt]\hline				
3	 &	smooth    &	          &	1+125 T-3200 p T^2+125 p^3 T^3+p^6 T^4
\tabularnewline[0.5pt]\hline				
4	 &	smooth    &	          &	1-405 T+4435 p T^2-405 p^3 T^3+p^6 T^4
\tabularnewline[0.5pt]\hline				
5	 &	smooth    &	          &	1-710 T+5670 p T^2-710 p^3 T^3+p^6 T^4
\tabularnewline[0.5pt]\hline				
6	 &	smooth    &	          &	1+210 T+2030 p T^2+210 p^3 T^3+p^6 T^4
\tabularnewline[0.5pt]\hline				
7	 &	smooth    &	          &	1+465 T+1645 p T^2+465 p^3 T^3+p^6 T^4
\tabularnewline[0.5pt]\hline				
8	 &	smooth    &	          &	1+245 T-590 p T^2+245 p^3 T^3+p^6 T^4
\tabularnewline[0.5pt]\hline				
9	 &	smooth    &	          &	1+310 T+3130 p T^2+310 p^3 T^3+p^6 T^4
\tabularnewline[0.5pt]\hline				
10	 &	smooth    &	          &	1+140 T+3570 p T^2+140 p^3 T^3+p^6 T^4
\tabularnewline[0.5pt]\hline				
11	 &	smooth    &	          &	1-425 T+4100 p T^2-425 p^3 T^3+p^6 T^4
\tabularnewline[0.5pt]\hline				
12	 &	smooth    &	          &	1-180 T+1010 p T^2-180 p^3 T^3+p^6 T^4
\tabularnewline[0.5pt]\hline				
13	 &	smooth    &	          &	1+185 T+4530 p T^2+185 p^3 T^3+p^6 T^4
\tabularnewline[0.5pt]\hline				
14	 &	smooth    &	          &	1+430 T+4590 p T^2+430 p^3 T^3+p^6 T^4
\tabularnewline[0.5pt]\hline				
15	 &	smooth    &	          &	1+90 T+2570 p T^2+90 p^3 T^3+p^6 T^4
\tabularnewline[0.5pt]\hline				
16	 &	smooth    &	          &	1+640 T+4870 p T^2+640 p^3 T^3+p^6 T^4
\tabularnewline[0.5pt]\hline				
17	 &	smooth    &	          &	1+10 T+2530 p T^2+10 p^3 T^3+p^6 T^4
\tabularnewline[0.5pt]\hline				
18	 &	smooth    &	          &	1+10 p T+4890 p T^2+10 p^4 T^3+p^6 T^4
\tabularnewline[0.5pt]\hline				
19	 &	smooth    &	          &	1+15 T+2395 p T^2+15 p^3 T^3+p^6 T^4
\tabularnewline[0.5pt]\hline				
20	 &	smooth    &	          &	1-75 T+5475 p T^2-75 p^3 T^3+p^6 T^4
\tabularnewline[0.5pt]\hline				
21	 &	smooth    &	          &	1-985 T+9970 p T^2-985 p^3 T^3+p^6 T^4
\tabularnewline[0.5pt]\hline				
22	 &	smooth    &	          &	1+340 T+5770 p T^2+340 p^3 T^3+p^6 T^4
\tabularnewline[0.5pt]\hline				
23	 &	smooth    &	          &	1+150 T+3900 p T^2+150 p^3 T^3+p^6 T^4
\tabularnewline[0.5pt]\hline				
24	 &	smooth    &	          &	1-160 T+4470 p T^2-160 p^3 T^3+p^6 T^4
\tabularnewline[0.5pt]\hline				
25	 &	smooth    &	          &	1+30 T+2140 p T^2+30 p^3 T^3+p^6 T^4
\tabularnewline[0.5pt]\hline				
26	 &	singular  &	5^{-5}&	(1+p T) (1-82 T+p^3 T^2)
\tabularnewline[0.5pt]\hline				
27	 &	smooth    &	          &	1+5 T+2540 p T^2+5 p^3 T^3+p^6 T^4
\tabularnewline[0.5pt]\hline				
28	 &	smooth    &	          &	1-225 T-475 p T^2-225 p^3 T^3+p^6 T^4
\tabularnewline[0.5pt]\hline				
29	 &	smooth    &	          &	1+70 T-1090 p T^2+70 p^3 T^3+p^6 T^4
\tabularnewline[0.5pt]\hline				
30	 &	smooth    &	          &	1-555 T+4060 p T^2-555 p^3 T^3+p^6 T^4
\tabularnewline[0.5pt]\hline				
31	 &	smooth    &	          &	1+260 T-1170 p T^2+260 p^3 T^3+p^6 T^4
\tabularnewline[0.5pt]\hline				
32	 &	smooth    &	          &	1-235 T-55 p T^2-235 p^3 T^3+p^6 T^4
\tabularnewline[0.5pt]\hline				
33	 &	smooth    &	          &	1+695 T+5435 p T^2+695 p^3 T^3+p^6 T^4
\tabularnewline[0.5pt]\hline				
34	 &	smooth    &	          &	1-90 T+1430 p T^2-90 p^3 T^3+p^6 T^4
\tabularnewline[0.5pt]\hline				
35	 &	smooth    &	          &	1+35 T-1595 p T^2+35 p^3 T^3+p^6 T^4
\tabularnewline[0.5pt]\hline				
36	 &	smooth    &	          &	1+370 T+5610 p T^2+370 p^3 T^3+p^6 T^4
\tabularnewline[0.5pt]\hline				
37	 &	smooth    &	          &	1-405 T+2360 p T^2-405 p^3 T^3+p^6 T^4
\tabularnewline[0.5pt]\hline				
38	 &	smooth    &	          &	1-305 T+3760 p T^2-305 p^3 T^3+p^6 T^4
\tabularnewline[0.5pt]\hline				
39	 &	smooth    &	          &	1-225 T+5375 p T^2-225 p^3 T^3+p^6 T^4
\tabularnewline[0.5pt]\hline				
40	 &	smooth    &	          &	1+5 T+340 p T^2+5 p^3 T^3+p^6 T^4
\tabularnewline[0.5pt]\hline				
41	 &	smooth    &	          &	1-555 T+4360 p T^2-555 p^3 T^3+p^6 T^4
\tabularnewline[0.5pt]\hline				
42	 &	smooth    &	          &	1+825 T+8000 p T^2+825 p^3 T^3+p^6 T^4
\tabularnewline[0.5pt]\hline				
43	 &	smooth    &	          &	1-100 T-3700 p T^2-100 p^3 T^3+p^6 T^4
\tabularnewline[0.5pt]\hline				
44	 &	smooth    &	          &	1+55 T-3785 p T^2+55 p^3 T^3+p^6 T^4
\tabularnewline[0.5pt]\hline				
45	 &	smooth    &	          &	1+50 T+3050 p T^2+50 p^3 T^3+p^6 T^4
\tabularnewline[0.5pt]\hline				
46	 &	smooth    &	          &	1+575 T+5600 p T^2+575 p^3 T^3+p^6 T^4
\tabularnewline[0.5pt]\hline				
47	 &	smooth    &	          &	1-165 T+1380 p T^2-165 p^3 T^3+p^6 T^4
\tabularnewline[0.5pt]\hline				
48	 &	smooth    &	          &	1-5 p T+4230 p T^2-5 p^4 T^3+p^6 T^4
\tabularnewline[0.5pt]\hline				
49	 &	smooth    &	          &	1-165 T+355 p T^2-165 p^3 T^3+p^6 T^4
\tabularnewline[0.5pt]\hline				
50	 &	smooth    &	          &	1-1005 T+9860 p T^2-1005 p^3 T^3+p^6 T^4
\tabularnewline[0.5pt]\hline				
51	 &	smooth    &	          &	1+615 T+3695 p T^2+615 p^3 T^3+p^6 T^4
\tabularnewline[0.5pt]\hline				
52	 &	smooth    &	          &	1+50 T+1250 p T^2+50 p^3 T^3+p^6 T^4
\tabularnewline[0.5pt]\hline				
\tablepostamble				
\tablepreamble{59}				
1	 &	smooth    &	          &	1-70 T+2812 p T^2-70 p^3 T^3+p^6 T^4
\tabularnewline[0.5pt]\hline				
2	 &	smooth    &	          &	1+70 T-4938 p T^2+70 p^3 T^3+p^6 T^4
\tabularnewline[0.5pt]\hline				
3	 &	smooth    &	          &	1-195 T+3212 p T^2-195 p^3 T^3+p^6 T^4
\tabularnewline[0.5pt]\hline				
4	 &	smooth    &	          &	1+1185 T+12712 p T^2+1185 p^3 T^3+p^6 T^4
\tabularnewline[0.5pt]\hline				
5	 &	smooth    &	          &	1-375 T+662 p T^2-375 p^3 T^3+p^6 T^4
\tabularnewline[0.5pt]\hline				
6	 &	smooth    &	          &	1+4162 p T^2+p^6 T^4
\tabularnewline[0.5pt]\hline				
7	 &	smooth    &	          &	1+380 T-538 p T^2+380 p^3 T^3+p^6 T^4
\tabularnewline[0.5pt]\hline				
8	 &	smooth    &	          &	1-385 T+1987 p T^2-385 p^3 T^3+p^6 T^4
\tabularnewline[0.5pt]\hline				
9	 &	smooth    &	          &	1-200 T+4162 p T^2-200 p^3 T^3+p^6 T^4
\tabularnewline[0.5pt]\hline				
10	 &	smooth    &	          &	1-355 T+2837 p T^2-355 p^3 T^3+p^6 T^4
\tabularnewline[0.5pt]\hline				
11	 &	smooth    &	          &	1+1160 T+11962 p T^2+1160 p^3 T^3+p^6 T^4
\tabularnewline[0.5pt]\hline				
12	 &	smooth    &	          &	1-65 T+2362 p T^2-65 p^3 T^3+p^6 T^4
\tabularnewline[0.5pt]\hline				
13	 &	smooth    &	          &	1+145 T-3538 p T^2+145 p^3 T^3+p^6 T^4
\tabularnewline[0.5pt]\hline				
14	 &	smooth    &	          &	1-765 T+8962 p T^2-765 p^3 T^3+p^6 T^4
\tabularnewline[0.5pt]\hline				
15	 &	smooth    &	          &	1-475 T+3387 p T^2-475 p^3 T^3+p^6 T^4
\tabularnewline[0.5pt]\hline				
16	 &	smooth    &	          &	1-455 T+2837 p T^2-455 p^3 T^3+p^6 T^4
\tabularnewline[0.5pt]\hline				
17	 &	smooth    &	          &	1+510 T+4862 p T^2+510 p^3 T^3+p^6 T^4
\tabularnewline[0.5pt]\hline				
18	 &	smooth    &	          &	1-220 T+5812 p T^2-220 p^3 T^3+p^6 T^4
\tabularnewline[0.5pt]\hline				
19	 &	smooth    &	          &	1+225 T+2787 p T^2+225 p^3 T^3+p^6 T^4
\tabularnewline[0.5pt]\hline				
20	 &	smooth    &	          &	1+30 T+3212 p T^2+30 p^3 T^3+p^6 T^4
\tabularnewline[0.5pt]\hline				
21	 &	smooth    &	          &	1+705 T+6087 p T^2+705 p^3 T^3+p^6 T^4
\tabularnewline[0.5pt]\hline				
22	 &	smooth    &	          &	1-370 T+2212 p T^2-370 p^3 T^3+p^6 T^4
\tabularnewline[0.5pt]\hline				
23	 &	smooth    &	          &	1-315 T+1862 p T^2-315 p^3 T^3+p^6 T^4
\tabularnewline[0.5pt]\hline				
24	 &	smooth    &	          &	1-510 T+6362 p T^2-510 p^3 T^3+p^6 T^4
\tabularnewline[0.5pt]\hline				
25	 &	smooth    &	          &	1-130 T+1812 p T^2-130 p^3 T^3+p^6 T^4
\tabularnewline[0.5pt]\hline				
26	 &	smooth    &	          &	1-250 T+5162 p T^2-250 p^3 T^3+p^6 T^4
\tabularnewline[0.5pt]\hline				
27	 &	smooth    &	          &	(1+10 p T+p^3 T^2)(1+150 T+p^3 T^2)
\tabularnewline[0.5pt]\hline				
28	 &	smooth    &	          &	1+230 T+6562 p T^2+230 p^3 T^3+p^6 T^4
\tabularnewline[0.5pt]\hline				
29	 &	singular  &	5^{-5}&	(1-p T) (1+280 T+p^3 T^2)
\tabularnewline[0.5pt]\hline				
30	 &	smooth    &	          &	1-60 T-4588 p T^2-60 p^3 T^3+p^6 T^4
\tabularnewline[0.5pt]\hline				
31	 &	smooth    &	          &	1+260 T-138 p T^2+260 p^3 T^3+p^6 T^4
\tabularnewline[0.5pt]\hline				
32	 &	smooth    &	          &	(1-5 p T+p^3 T^2)(1-50 T+p^3 T^2)
\tabularnewline[0.5pt]\hline				
33	 &	smooth    &	          &	1+450 T+5762 p T^2+450 p^3 T^3+p^6 T^4
\tabularnewline[0.5pt]\hline				
34	 &	smooth    &	          &	1+645 T+5212 p T^2+645 p^3 T^3+p^6 T^4
\tabularnewline[0.5pt]\hline				
35	 &	smooth    &	          &	1+195 T+562 p T^2+195 p^3 T^3+p^6 T^4
\tabularnewline[0.5pt]\hline				
36	 &	smooth    &	          &	1+35 T+587 p T^2+35 p^3 T^3+p^6 T^4
\tabularnewline[0.5pt]\hline				
37	 &	smooth    &	          &	1-165 T+6587 p T^2-165 p^3 T^3+p^6 T^4
\tabularnewline[0.5pt]\hline				
38	 &	smooth    &	          &	1+105 T+18 p^2 T^2+105 p^3 T^3+p^6 T^4
\tabularnewline[0.5pt]\hline				
39	 &	smooth    &	          &	1-530 T+3562 p T^2-530 p^3 T^3+p^6 T^4
\tabularnewline[0.5pt]\hline				
40	 &	smooth    &	          &	1-475 T+2262 p T^2-475 p^3 T^3+p^6 T^4
\tabularnewline[0.5pt]\hline				
41	 &	smooth    &	          &	1-195 T+1337 p T^2-195 p^3 T^3+p^6 T^4
\tabularnewline[0.5pt]\hline				
42	 &	smooth    &	          &	1-565 T+4862 p T^2-565 p^3 T^3+p^6 T^4
\tabularnewline[0.5pt]\hline				
43	 &	smooth    &	          &	1-995 T+9962 p T^2-995 p^3 T^3+p^6 T^4
\tabularnewline[0.5pt]\hline				
44	 &	smooth    &	          &	1+185 T+1412 p T^2+185 p^3 T^3+p^6 T^4
\tabularnewline[0.5pt]\hline				
45	 &	smooth    &	          &	1-75 T+6662 p T^2-75 p^3 T^3+p^6 T^4
\tabularnewline[0.5pt]\hline				
46	 &	smooth    &	          &	(1+p^3 T^2)(1+770 T+p^3 T^2)
\tabularnewline[0.5pt]\hline				
47	 &	smooth    &	          &	1+175 T-213 p T^2+175 p^3 T^3+p^6 T^4
\tabularnewline[0.5pt]\hline				
48	 &	smooth    &	          &	1-355 T-563 p T^2-355 p^3 T^3+p^6 T^4
\tabularnewline[0.5pt]\hline				
49	 &	smooth    &	          &	1-825 T+7262 p T^2-825 p^3 T^3+p^6 T^4
\tabularnewline[0.5pt]\hline				
50	 &	smooth    &	          &	1+350 T+4262 p T^2+350 p^3 T^3+p^6 T^4
\tabularnewline[0.5pt]\hline				
51	 &	smooth    &	          &	1-415 T+3462 p T^2-415 p^3 T^3+p^6 T^4
\tabularnewline[0.5pt]\hline				
52	 &	smooth    &	          &	1+455 T+2062 p T^2+455 p^3 T^3+p^6 T^4
\tabularnewline[0.5pt]\hline				
53	 &	smooth    &	          &	1+540 T+3212 p T^2+540 p^3 T^3+p^6 T^4
\tabularnewline[0.5pt]\hline				
54	 &	smooth    &	          &	1-200 T+4162 p T^2-200 p^3 T^3+p^6 T^4
\tabularnewline[0.5pt]\hline				
55	 &	smooth    &	          &	1-105 T+312 p T^2-105 p^3 T^3+p^6 T^4
\tabularnewline[0.5pt]\hline				
56	 &	smooth    &	          &	1-120 T+962 p T^2-120 p^3 T^3+p^6 T^4
\tabularnewline[0.5pt]\hline				
57	 &	smooth    &	          &	1+425 T+287 p T^2+425 p^3 T^3+p^6 T^4
\tabularnewline[0.5pt]\hline				
58	 &	smooth    &	          &	1+370 T-988 p T^2+370 p^3 T^3+p^6 T^4
\tabularnewline[0.5pt]\hline				
\tablepostamble				
\tablepreamble{61}				
1	 &	smooth    &	          &	1+11 T+2276 p T^2+11 p^3 T^3+p^6 T^4
\tabularnewline[0.5pt]\hline				
2	 &	smooth    &	          &	1-124 T+4236 p T^2-124 p^3 T^3+p^6 T^4
\tabularnewline[0.5pt]\hline				
3	 &	smooth    &	          &	1-869 T+6356 p T^2-869 p^3 T^3+p^6 T^4
\tabularnewline[0.5pt]\hline				
4	 &	smooth    &	          &	1-384 T+4796 p T^2-384 p^3 T^3+p^6 T^4
\tabularnewline[0.5pt]\hline				
5	 &	smooth    &	          &	1-344 T+4206 p T^2-344 p^3 T^3+p^6 T^4
\tabularnewline[0.5pt]\hline				
6	 &	smooth    &	          &	1+401 T-664 p T^2+401 p^3 T^3+p^6 T^4
\tabularnewline[0.5pt]\hline				
7	 &	smooth    &	          &	1+216 T+3996 p T^2+216 p^3 T^3+p^6 T^4
\tabularnewline[0.5pt]\hline				
8	 &	smooth    &	          &	1-64 T+1926 p T^2-64 p^3 T^3+p^6 T^4
\tabularnewline[0.5pt]\hline				
9	 &	smooth    &	          &	1-599 T+1936 p T^2-599 p^3 T^3+p^6 T^4
\tabularnewline[0.5pt]\hline				
10	 &	smooth    &	          &	1-419 T+96 p^2 T^2-419 p^3 T^3+p^6 T^4
\tabularnewline[0.5pt]\hline				
11	 &	smooth    &	          &	1+636 T+8526 p T^2+636 p^3 T^3+p^6 T^4
\tabularnewline[0.5pt]\hline				
12	 &	smooth    &	          &	1-249 T-1489 p T^2-249 p^3 T^3+p^6 T^4
\tabularnewline[0.5pt]\hline				
13	 &	smooth    &	          &	1-214 T+3726 p T^2-214 p^3 T^3+p^6 T^4
\tabularnewline[0.5pt]\hline				
14	 &	smooth    &	          &	1+486 T+326 p T^2+486 p^3 T^3+p^6 T^4
\tabularnewline[0.5pt]\hline				
15	 &	smooth    &	          &	1-429 T+1666 p T^2-429 p^3 T^3+p^6 T^4
\tabularnewline[0.5pt]\hline				
16	 &	smooth    &	          &	1+736 T+7126 p T^2+736 p^3 T^3+p^6 T^4
\tabularnewline[0.5pt]\hline				
17	 &	smooth    &	          &	1+21 T+2916 p T^2+21 p^3 T^3+p^6 T^4
\tabularnewline[0.5pt]\hline				
18	 &	smooth    &	          &	1+351 T+1436 p T^2+351 p^3 T^3+p^6 T^4
\tabularnewline[0.5pt]\hline				
19	 &	smooth    &	          &	1+836 T+9126 p T^2+836 p^3 T^3+p^6 T^4
\tabularnewline[0.5pt]\hline				
20	 &	smooth    &	          &	1-494 T+8156 p T^2-494 p^3 T^3+p^6 T^4
\tabularnewline[0.5pt]\hline				
21	 &	smooth    &	          &	1-364 T+526 p T^2-364 p^3 T^3+p^6 T^4
\tabularnewline[0.5pt]\hline				
22	 &	smooth    &	          &	1-134 T-1554 p T^2-134 p^3 T^3+p^6 T^4
\tabularnewline[0.5pt]\hline				
23	 &	smooth    &	          &	1-529 T+1966 p T^2-529 p^3 T^3+p^6 T^4
\tabularnewline[0.5pt]\hline				
24	 &	smooth    &	          &	1-509 T+5796 p T^2-509 p^3 T^3+p^6 T^4
\tabularnewline[0.5pt]\hline				
25	 &	smooth    &	          &	1+501 T+786 p T^2+501 p^3 T^3+p^6 T^4
\tabularnewline[0.5pt]\hline				
26	 &	smooth    &	          &	1+116 T+3696 p T^2+116 p^3 T^3+p^6 T^4
\tabularnewline[0.5pt]\hline				
27	 &	smooth    &	          &	1+1036 T+11326 p T^2+1036 p^3 T^3+p^6 T^4
\tabularnewline[0.5pt]\hline				
28	 &	smooth    &	          &	1-594 T+3356 p T^2-594 p^3 T^3+p^6 T^4
\tabularnewline[0.5pt]\hline				
29	 &	smooth    &	          &	1+561 T+8176 p T^2+561 p^3 T^3+p^6 T^4
\tabularnewline[0.5pt]\hline				
30	 &	smooth    &	          &	1+346 T+4866 p T^2+346 p^3 T^3+p^6 T^4
\tabularnewline[0.5pt]\hline				
31	 &	smooth    &	          &	1-9 p T+7961 p T^2-9 p^4 T^3+p^6 T^4
\tabularnewline[0.5pt]\hline				
32	 &	smooth    &	          &	1-489 T+3276 p T^2-489 p^3 T^3+p^6 T^4
\tabularnewline[0.5pt]\hline				
33	 &	smooth    &	          &	1-1024 T+9786 p T^2-1024 p^3 T^3+p^6 T^4
\tabularnewline[0.5pt]\hline				
34	 &	smooth    &	          &	(1+13 p T+p^3 T^2)(1-182 T+p^3 T^2)
\tabularnewline[0.5pt]\hline				
35	 &	smooth    &	          &	1+486 T+6676 p T^2+486 p^3 T^3+p^6 T^4
\tabularnewline[0.5pt]\hline				
36	 &	smooth    &	          &	1-279 T+1091 p T^2-279 p^3 T^3+p^6 T^4
\tabularnewline[0.5pt]\hline				
37	 &	smooth    &	          &	1-64 T+926 p T^2-64 p^3 T^3+p^6 T^4
\tabularnewline[0.5pt]\hline				
38	 &	smooth    &	          &	1+381 T+3756 p T^2+381 p^3 T^3+p^6 T^4
\tabularnewline[0.5pt]\hline				
39	 &	smooth    &	          &	1-29 T-2309 p T^2-29 p^3 T^3+p^6 T^4
\tabularnewline[0.5pt]\hline				
40	 &	smooth    &	          &	1-489 T+2651 p T^2-489 p^3 T^3+p^6 T^4
\tabularnewline[0.5pt]\hline				
41	 &	smooth    &	          &	1-34 T+3996 p T^2-34 p^3 T^3+p^6 T^4
\tabularnewline[0.5pt]\hline				
42	 &	smooth    &	          &	1-79 T+5766 p T^2-79 p^3 T^3+p^6 T^4
\tabularnewline[0.5pt]\hline				
43	 &	smooth    &	          &	1+31 T-494 p T^2+31 p^3 T^3+p^6 T^4
\tabularnewline[0.5pt]\hline				
44	 &	smooth    &	          &	1+541 T+3746 p T^2+541 p^3 T^3+p^6 T^4
\tabularnewline[0.5pt]\hline				
45	 &	smooth    &	          &	1+206 T+2306 p T^2+206 p^3 T^3+p^6 T^4
\tabularnewline[0.5pt]\hline				
46	 &	smooth    &	          &	1+21 T-6684 p T^2+21 p^3 T^3+p^6 T^4
\tabularnewline[0.5pt]\hline				
47	 &	smooth    &	          &	1+11 T-2099 p T^2+11 p^3 T^3+p^6 T^4
\tabularnewline[0.5pt]\hline				
48	 &	singular  &	5^{-5}&	(1-p T) (1+518 T+p^3 T^2)
\tabularnewline[0.5pt]\hline				
49	 &	smooth    &	          &	1+111 T-1299 p T^2+111 p^3 T^3+p^6 T^4
\tabularnewline[0.5pt]\hline				
50	 &	smooth    &	          &	1-689 T+8176 p T^2-689 p^3 T^3+p^6 T^4
\tabularnewline[0.5pt]\hline				
51	 &	smooth    &	          &	1-154 T+4266 p T^2-154 p^3 T^3+p^6 T^4
\tabularnewline[0.5pt]\hline				
52	 &	smooth    &	          &	1+191 T+2171 p T^2+191 p^3 T^3+p^6 T^4
\tabularnewline[0.5pt]\hline				
53	 &	smooth    &	          &	1-594 T+5156 p T^2-594 p^3 T^3+p^6 T^4
\tabularnewline[0.5pt]\hline				
54	 &	smooth    &	          &	1-274 T+4786 p T^2-274 p^3 T^3+p^6 T^4
\tabularnewline[0.5pt]\hline				
55	 &	smooth    &	          &	1+271 T+3891 p T^2+271 p^3 T^3+p^6 T^4
\tabularnewline[0.5pt]\hline				
56	 &	smooth    &	          &	1-44 T+7056 p T^2-44 p^3 T^3+p^6 T^4
\tabularnewline[0.5pt]\hline				
57	 &	smooth    &	          &	1+91 T+2771 p T^2+91 p^3 T^3+p^6 T^4
\tabularnewline[0.5pt]\hline				
58	 &	smooth    &	          &	1+876 T+10386 p T^2+876 p^3 T^3+p^6 T^4
\tabularnewline[0.5pt]\hline				
59	 &	smooth    &	          &	1+711 T+4751 p T^2+711 p^3 T^3+p^6 T^4
\tabularnewline[0.5pt]\hline				
60	 &	smooth    &	          &	1-139 T-299 p T^2-139 p^3 T^3+p^6 T^4
\tabularnewline[0.5pt]\hline				
\tablepostamble				
\tablepreamble{67}				
1	 &	smooth    &	          &	1+550 T+4650 p T^2+550 p^3 T^3+p^6 T^4
\tabularnewline[0.5pt]\hline				
2	 &	smooth    &	          &	1-260 T+5490 p T^2-260 p^3 T^3+p^6 T^4
\tabularnewline[0.5pt]\hline				
3	 &	smooth    &	          &	1+435 T+7885 p T^2+435 p^3 T^3+p^6 T^4
\tabularnewline[0.5pt]\hline				
4	 &	smooth    &	          &	1+160 T-5890 p T^2+160 p^3 T^3+p^6 T^4
\tabularnewline[0.5pt]\hline				
5	 &	smooth    &	          &	1+40 T+4490 p T^2+40 p^3 T^3+p^6 T^4
\tabularnewline[0.5pt]\hline				
6	 &	smooth    &	          &	1+315 T+7615 p T^2+315 p^3 T^3+p^6 T^4
\tabularnewline[0.5pt]\hline				
7	 &	smooth    &	          &	1-185 T-1810 p T^2-185 p^3 T^3+p^6 T^4
\tabularnewline[0.5pt]\hline				
8	 &	smooth    &	          &	1+60 T+4610 p T^2+60 p^3 T^3+p^6 T^4
\tabularnewline[0.5pt]\hline				
9	 &	smooth    &	          &	1-190 T-1190 p T^2-190 p^3 T^3+p^6 T^4
\tabularnewline[0.5pt]\hline				
10	 &	smooth    &	          &	1+390 T-30 p^2 T^2+390 p^3 T^3+p^6 T^4
\tabularnewline[0.5pt]\hline				
11	 &	smooth    &	          &	1+5450 p T^2+p^6 T^4
\tabularnewline[0.5pt]\hline				
12	 &	smooth    &	          &	1-135 T+915 p T^2-135 p^3 T^3+p^6 T^4
\tabularnewline[0.5pt]\hline				
13	 &	smooth    &	          &	1+145 T+1370 p T^2+145 p^3 T^3+p^6 T^4
\tabularnewline[0.5pt]\hline				
14	 &	smooth    &	          &	1-435 T+115 p T^2-435 p^3 T^3+p^6 T^4
\tabularnewline[0.5pt]\hline				
15	 &	smooth    &	          &	1+610 T+3910 p T^2+610 p^3 T^3+p^6 T^4
\tabularnewline[0.5pt]\hline				
16	 &	smooth    &	          &	1-555 T+2370 p T^2-555 p^3 T^3+p^6 T^4
\tabularnewline[0.5pt]\hline				
17	 &	smooth    &	          &	1-265 T-2015 p T^2-265 p^3 T^3+p^6 T^4
\tabularnewline[0.5pt]\hline				
18	 &	smooth    &	          &	1+280 T+2830 p T^2+280 p^3 T^3+p^6 T^4
\tabularnewline[0.5pt]\hline				
19	 &	smooth    &	          &	1-210 T+4090 p T^2-210 p^3 T^3+p^6 T^4
\tabularnewline[0.5pt]\hline				
20	 &	smooth    &	          &	1+115 T+7390 p T^2+115 p^3 T^3+p^6 T^4
\tabularnewline[0.5pt]\hline				
21	 &	smooth    &	          &	1+470 T+3370 p T^2+470 p^3 T^3+p^6 T^4
\tabularnewline[0.5pt]\hline				
22	 &	smooth    &	          &	1-130 T+5920 p T^2-130 p^3 T^3+p^6 T^4
\tabularnewline[0.5pt]\hline				
23	 &	smooth    &	          &	1-850 T+7800 p T^2-850 p^3 T^3+p^6 T^4
\tabularnewline[0.5pt]\hline				
24	 &	smooth    &	          &	1-185 T-1635 p T^2-185 p^3 T^3+p^6 T^4
\tabularnewline[0.5pt]\hline				
25	 &	smooth    &	          &	1-955 T+9220 p T^2-955 p^3 T^3+p^6 T^4
\tabularnewline[0.5pt]\hline				
26	 &	smooth    &	          &	1-45 T-2270 p T^2-45 p^3 T^3+p^6 T^4
\tabularnewline[0.5pt]\hline				
27	 &	smooth    &	          &	1+580 T+9880 p T^2+580 p^3 T^3+p^6 T^4
\tabularnewline[0.5pt]\hline				
28	 &	smooth    &	          &	1+790 T+5290 p T^2+790 p^3 T^3+p^6 T^4
\tabularnewline[0.5pt]\hline				
29	 &	smooth    &	          &	1+35 T+710 p T^2+35 p^3 T^3+p^6 T^4
\tabularnewline[0.5pt]\hline				
30	 &	smooth    &	          &	1-170 T+1030 p T^2-170 p^3 T^3+p^6 T^4
\tabularnewline[0.5pt]\hline				
31	 &	smooth    &	          &	1+15 T+6140 p T^2+15 p^3 T^3+p^6 T^4
\tabularnewline[0.5pt]\hline				
32	 &	smooth    &	          &	1+6900 p T^2+p^6 T^4
\tabularnewline[0.5pt]\hline				
33	 &	smooth    &	          &	1+365 T+1590 p T^2+365 p^3 T^3+p^6 T^4
\tabularnewline[0.5pt]\hline				
34	 &	smooth    &	          &	1-10 T-3060 p T^2-10 p^3 T^3+p^6 T^4
\tabularnewline[0.5pt]\hline				
35	 &	smooth    &	          &	1+265 T+4440 p T^2+265 p^3 T^3+p^6 T^4
\tabularnewline[0.5pt]\hline				
36	 &	smooth    &	          &	1-45 T+5530 p T^2-45 p^3 T^3+p^6 T^4
\tabularnewline[0.5pt]\hline				
37	 &	smooth    &	          &	1-605 T+1445 p T^2-605 p^3 T^3+p^6 T^4
\tabularnewline[0.5pt]\hline				
38	 &	smooth    &	          &	1-905 T+9320 p T^2-905 p^3 T^3+p^6 T^4
\tabularnewline[0.5pt]\hline				
39	 &	smooth    &	          &	1+310 T+3610 p T^2+310 p^3 T^3+p^6 T^4
\tabularnewline[0.5pt]\hline				
40	 &	smooth    &	          &	1+1660 T+18460 p T^2+1660 p^3 T^3+p^6 T^4
\tabularnewline[0.5pt]\hline				
41	 &	smooth    &	          &	1-325 T+4750 p T^2-325 p^3 T^3+p^6 T^4
\tabularnewline[0.5pt]\hline				
42	 &	smooth    &	          &	1-425 T+7750 p T^2-425 p^3 T^3+p^6 T^4
\tabularnewline[0.5pt]\hline				
43	 &	smooth    &	          &	1+220 T-1580 p T^2+220 p^3 T^3+p^6 T^4
\tabularnewline[0.5pt]\hline				
44	 &	smooth    &	          &	1+30 T+4230 p T^2+30 p^3 T^3+p^6 T^4
\tabularnewline[0.5pt]\hline				
45	 &	smooth    &	          &	1-245 T+4405 p T^2-245 p^3 T^3+p^6 T^4
\tabularnewline[0.5pt]\hline				
46	 &	smooth    &	          &	1+1035 T+8510 p T^2+1035 p^3 T^3+p^6 T^4
\tabularnewline[0.5pt]\hline				
47	 &	smooth    &	          &	1+995 T+9870 p T^2+995 p^3 T^3+p^6 T^4
\tabularnewline[0.5pt]\hline				
48	 &	smooth    &	          &	1+255 T+5530 p T^2+255 p^3 T^3+p^6 T^4
\tabularnewline[0.5pt]\hline				
49	 &	smooth    &	          &	1+645 T+5095 p T^2+645 p^3 T^3+p^6 T^4
\tabularnewline[0.5pt]\hline				
50	 &	smooth    &	          &	1+615 T+7015 p T^2+615 p^3 T^3+p^6 T^4
\tabularnewline[0.5pt]\hline				
51	 &	smooth    &	          &	1+310 T+3910 p T^2+310 p^3 T^3+p^6 T^4
\tabularnewline[0.5pt]\hline				
52	 &	smooth    &	          &	1-850 T+150 p^2 T^2-850 p^3 T^3+p^6 T^4
\tabularnewline[0.5pt]\hline				
53	 &	singular  &	5^{-5}&	(1+p T) (1-141 T+p^3 T^2)
\tabularnewline[0.5pt]\hline				
54	 &	smooth    &	          &	1+35 T+4810 p T^2+35 p^3 T^3+p^6 T^4
\tabularnewline[0.5pt]\hline				
55	 &	smooth    &	          &	1+390 T+3190 p T^2+390 p^3 T^3+p^6 T^4
\tabularnewline[0.5pt]\hline				
56	 &	smooth    &	          &	1+60 T+2610 p T^2+60 p^3 T^3+p^6 T^4
\tabularnewline[0.5pt]\hline				
57	 &	smooth    &	          &	1-35 T+2790 p T^2-35 p^3 T^3+p^6 T^4
\tabularnewline[0.5pt]\hline				
58	 &	smooth    &	          &	1-755 T+7920 p T^2-755 p^3 T^3+p^6 T^4
\tabularnewline[0.5pt]\hline				
59	 &	smooth    &	          &	1+320 T+820 p T^2+320 p^3 T^3+p^6 T^4
\tabularnewline[0.5pt]\hline				
60	 &	smooth    &	          &	1+445 T+2870 p T^2+445 p^3 T^3+p^6 T^4
\tabularnewline[0.5pt]\hline				
61	 &	smooth    &	          &	1-1385 T+14040 p T^2-1385 p^3 T^3+p^6 T^4
\tabularnewline[0.5pt]\hline				
62	 &	smooth    &	          &	1+5 T+2480 p T^2+5 p^3 T^3+p^6 T^4
\tabularnewline[0.5pt]\hline				
63	 &	smooth    &	          &	1+25 T+2450 p T^2+25 p^3 T^3+p^6 T^4
\tabularnewline[0.5pt]\hline				
64	 &	smooth    &	          &	1-1235 T+12190 p T^2-1235 p^3 T^3+p^6 T^4
\tabularnewline[0.5pt]\hline				
65	 &	smooth    &	          &	1-75 T+6125 p T^2-75 p^3 T^3+p^6 T^4
\tabularnewline[0.5pt]\hline				
66	 &	smooth    &	          &	1-1435 T+16140 p T^2-1435 p^3 T^3+p^6 T^4
\tabularnewline[0.5pt]\hline				
\tablepostamble				
\tablepreamble{71}				
1	 &	singular  &	5^{-5}&	(1-p T) (1-412 T+p^3 T^2)
\tabularnewline[0.5pt]\hline				
2	 &	smooth    &	          &	1+726 T+10846 p T^2+726 p^3 T^3+p^6 T^4
\tabularnewline[0.5pt]\hline				
3	 &	smooth    &	          &	1-569 T+10666 p T^2-569 p^3 T^3+p^6 T^4
\tabularnewline[0.5pt]\hline				
4	 &	smooth    &	          &	1-109 T+2131 p T^2-109 p^3 T^3+p^6 T^4
\tabularnewline[0.5pt]\hline				
5	 &	smooth    &	          &	1-279 T-3374 p T^2-279 p^3 T^3+p^6 T^4
\tabularnewline[0.5pt]\hline				
6	 &	smooth    &	          &	1-379 T+9751 p T^2-379 p^3 T^3+p^6 T^4
\tabularnewline[0.5pt]\hline				
7	 &	smooth    &	          &	1-689 T+7686 p T^2-689 p^3 T^3+p^6 T^4
\tabularnewline[0.5pt]\hline				
8	 &	smooth    &	          &	1+1186 T+14536 p T^2+1186 p^3 T^3+p^6 T^4
\tabularnewline[0.5pt]\hline				
9	 &	smooth    &	          &	1-204 T+8526 p T^2-204 p^3 T^3+p^6 T^4
\tabularnewline[0.5pt]\hline				
10	 &	smooth    &	          &	1-434 T-694 p T^2-434 p^3 T^3+p^6 T^4
\tabularnewline[0.5pt]\hline				
11	 &	smooth    &	          &	1+121 T+6526 p T^2+121 p^3 T^3+p^6 T^4
\tabularnewline[0.5pt]\hline				
12	 &	smooth    &	          &	1+186 T+1586 p T^2+186 p^3 T^3+p^6 T^4
\tabularnewline[0.5pt]\hline				
13	 &	smooth    &	          &	1+86 T-2814 p T^2+86 p^3 T^3+p^6 T^4
\tabularnewline[0.5pt]\hline				
14	 &	smooth    &	          &	1-789 T+3761 p T^2-789 p^3 T^3+p^6 T^4
\tabularnewline[0.5pt]\hline				
15	 &	smooth    &	          &	1+116 T-894 p T^2+116 p^3 T^3+p^6 T^4
\tabularnewline[0.5pt]\hline				
16	 &	smooth    &	          &	1-19 T+3591 p T^2-19 p^3 T^3+p^6 T^4
\tabularnewline[0.5pt]\hline				
17	 &	smooth    &	          &	1+256 T+2866 p T^2+256 p^3 T^3+p^6 T^4
\tabularnewline[0.5pt]\hline				
18	 &	smooth    &	          &	1+1666 T+19406 p T^2+1666 p^3 T^3+p^6 T^4
\tabularnewline[0.5pt]\hline				
19	 &	smooth    &	          &	1+491 T+1081 p T^2+491 p^3 T^3+p^6 T^4
\tabularnewline[0.5pt]\hline				
20	 &	smooth    &	          &	1+601 T+3121 p T^2+601 p^3 T^3+p^6 T^4
\tabularnewline[0.5pt]\hline				
21	 &	smooth    &	          &	1-224 T+5946 p T^2-224 p^3 T^3+p^6 T^4
\tabularnewline[0.5pt]\hline				
22	 &	smooth    &	          &	1-169 T+7741 p T^2-169 p^3 T^3+p^6 T^4
\tabularnewline[0.5pt]\hline				
23	 &	smooth    &	          &	1+176 T+6846 p T^2+176 p^3 T^3+p^6 T^4
\tabularnewline[0.5pt]\hline				
24	 &	smooth    &	          &	1+1076 T+11346 p T^2+1076 p^3 T^3+p^6 T^4
\tabularnewline[0.5pt]\hline				
25	 &	smooth    &	          &	1-379 T+10526 p T^2-379 p^3 T^3+p^6 T^4
\tabularnewline[0.5pt]\hline				
26	 &	smooth    &	          &	1-624 T+1946 p T^2-624 p^3 T^3+p^6 T^4
\tabularnewline[0.5pt]\hline				
27	 &	smooth    &	          &	(1-12 p T+p^3 T^2)(1+153 T+p^3 T^2)
\tabularnewline[0.5pt]\hline				
28	 &	smooth    &	          &	1-744 T+3966 p T^2-744 p^3 T^3+p^6 T^4
\tabularnewline[0.5pt]\hline				
29	 &	smooth    &	          &	1-594 T+4466 p T^2-594 p^3 T^3+p^6 T^4
\tabularnewline[0.5pt]\hline				
30	 &	smooth    &	          &	1-549 T+1546 p T^2-549 p^3 T^3+p^6 T^4
\tabularnewline[0.5pt]\hline				
31	 &	smooth    &	          &	1+76 T+126 p^2 T^2+76 p^3 T^3+p^6 T^4
\tabularnewline[0.5pt]\hline				
32	 &	smooth    &	          &	1+726 T+5996 p T^2+726 p^3 T^3+p^6 T^4
\tabularnewline[0.5pt]\hline				
33	 &	smooth    &	          &	1-699 T+3296 p T^2-699 p^3 T^3+p^6 T^4
\tabularnewline[0.5pt]\hline				
34	 &	smooth    &	          &	1+301 T-279 p T^2+301 p^3 T^3+p^6 T^4
\tabularnewline[0.5pt]\hline				
35	 &	smooth    &	          &	1-1449 T+16121 p T^2-1449 p^3 T^3+p^6 T^4
\tabularnewline[0.5pt]\hline				
36	 &	smooth    &	          &	1+406 T+6266 p T^2+406 p^3 T^3+p^6 T^4
\tabularnewline[0.5pt]\hline				
37	 &	smooth    &	          &	1-299 T+421 p T^2-299 p^3 T^3+p^6 T^4
\tabularnewline[0.5pt]\hline				
38	 &	smooth    &	          &	1+176 T+7446 p T^2+176 p^3 T^3+p^6 T^4
\tabularnewline[0.5pt]\hline				
39	 &	smooth    &	          &	1-524 T+5996 p T^2-524 p^3 T^3+p^6 T^4
\tabularnewline[0.5pt]\hline				
40	 &	smooth    &	          &	1-149 T-3754 p T^2-149 p^3 T^3+p^6 T^4
\tabularnewline[0.5pt]\hline				
41	 &	smooth    &	          &	1+626 T+8196 p T^2+626 p^3 T^3+p^6 T^4
\tabularnewline[0.5pt]\hline				
42	 &	smooth    &	          &	1+151 T+7771 p T^2+151 p^3 T^3+p^6 T^4
\tabularnewline[0.5pt]\hline				
43	 &	smooth    &	          &	1+306 T+8316 p T^2+306 p^3 T^3+p^6 T^4
\tabularnewline[0.5pt]\hline				
44	 &	smooth    &	          &	1+331 T+6266 p T^2+331 p^3 T^3+p^6 T^4
\tabularnewline[0.5pt]\hline				
45	 &	smooth    &	          &	1-1049 T+10046 p T^2-1049 p^3 T^3+p^6 T^4
\tabularnewline[0.5pt]\hline				
46	 &	smooth    &	          &	1+126 T-3604 p T^2+126 p^3 T^3+p^6 T^4
\tabularnewline[0.5pt]\hline				
47	 &	smooth    &	          &	1+321 T-3899 p T^2+321 p^3 T^3+p^6 T^4
\tabularnewline[0.5pt]\hline				
48	 &	smooth    &	          &	1-299 T+7296 p T^2-299 p^3 T^3+p^6 T^4
\tabularnewline[0.5pt]\hline				
49	 &	smooth    &	          &	1-349 T-3154 p T^2-349 p^3 T^3+p^6 T^4
\tabularnewline[0.5pt]\hline				
50	 &	smooth    &	          &	1+731 T+5966 p T^2+731 p^3 T^3+p^6 T^4
\tabularnewline[0.5pt]\hline				
51	 &	smooth    &	          &	1+726 T+4746 p T^2+726 p^3 T^3+p^6 T^4
\tabularnewline[0.5pt]\hline				
52	 &	smooth    &	          &	1+151 T+1321 p T^2+151 p^3 T^3+p^6 T^4
\tabularnewline[0.5pt]\hline				
53	 &	smooth    &	          &	1-189 T+6386 p T^2-189 p^3 T^3+p^6 T^4
\tabularnewline[0.5pt]\hline				
54	 &	smooth    &	          &	1-464 T+2236 p T^2-464 p^3 T^3+p^6 T^4
\tabularnewline[0.5pt]\hline				
55	 &	smooth    &	          &	1+56 T+96 p^2 T^2+56 p^3 T^3+p^6 T^4
\tabularnewline[0.5pt]\hline				
56	 &	smooth    &	          &	1-119 T-7309 p T^2-119 p^3 T^3+p^6 T^4
\tabularnewline[0.5pt]\hline				
57	 &	smooth    &	          &	1+561 T+7086 p T^2+561 p^3 T^3+p^6 T^4
\tabularnewline[0.5pt]\hline				
58	 &	smooth    &	          &	1+266 T+1906 p T^2+266 p^3 T^3+p^6 T^4
\tabularnewline[0.5pt]\hline				
59	 &	smooth    &	          &	1-34 T-894 p T^2-34 p^3 T^3+p^6 T^4
\tabularnewline[0.5pt]\hline				
60	 &	smooth    &	          &	1+616 T+3156 p T^2+616 p^3 T^3+p^6 T^4
\tabularnewline[0.5pt]\hline				
61	 &	smooth    &	          &	1-404 T+5076 p T^2-404 p^3 T^3+p^6 T^4
\tabularnewline[0.5pt]\hline				
62	 &	smooth    &	          &	1+1261 T+14161 p T^2+1261 p^3 T^3+p^6 T^4
\tabularnewline[0.5pt]\hline				
63	 &	smooth    &	          &	1-29 T+7151 p T^2-29 p^3 T^3+p^6 T^4
\tabularnewline[0.5pt]\hline				
64	 &	smooth    &	          &	1-9 T+4106 p T^2-9 p^3 T^3+p^6 T^4
\tabularnewline[0.5pt]\hline				
65	 &	smooth    &	          &	1-234 T-2494 p T^2-234 p^3 T^3+p^6 T^4
\tabularnewline[0.5pt]\hline				
66	 &	smooth    &	          &	1-904 T+7826 p T^2-904 p^3 T^3+p^6 T^4
\tabularnewline[0.5pt]\hline				
67	 &	smooth    &	          &	1+596 T+7826 p T^2+596 p^3 T^3+p^6 T^4
\tabularnewline[0.5pt]\hline				
68	 &	smooth    &	          &	1-1039 T+8786 p T^2-1039 p^3 T^3+p^6 T^4
\tabularnewline[0.5pt]\hline				
69	 &	smooth    &	          &	1+331 T+2266 p T^2+331 p^3 T^3+p^6 T^4
\tabularnewline[0.5pt]\hline				
70	 &	smooth    &	          &	1+651 T+2646 p T^2+651 p^3 T^3+p^6 T^4
\tabularnewline[0.5pt]\hline				
\tablepostamble				
\tablepreamble{73}				
1	 &	smooth    &	          &	1+395 T+8410 p T^2+395 p^3 T^3+p^6 T^4
\tabularnewline[0.5pt]\hline				
2	 &	smooth    &	          &	1+1005 T+11890 p T^2+1005 p^3 T^3+p^6 T^4
\tabularnewline[0.5pt]\hline				
3	 &	smooth    &	          &	1-560 T+5270 p T^2-560 p^3 T^3+p^6 T^4
\tabularnewline[0.5pt]\hline				
4	 &	smooth    &	          &	1+175 T+2200 p T^2+175 p^3 T^3+p^6 T^4
\tabularnewline[0.5pt]\hline				
5	 &	smooth    &	          &	1+250 T+4050 p T^2+250 p^3 T^3+p^6 T^4
\tabularnewline[0.5pt]\hline				
6	 &	smooth    &	          &	1+555 T+4240 p T^2+555 p^3 T^3+p^6 T^4
\tabularnewline[0.5pt]\hline				
7	 &	smooth    &	          &	1+445 T+6735 p T^2+445 p^3 T^3+p^6 T^4
\tabularnewline[0.5pt]\hline				
8	 &	smooth    &	          &	1-140 T+4730 p T^2-140 p^3 T^3+p^6 T^4
\tabularnewline[0.5pt]\hline				
9	 &	smooth    &	          &	1+945 T+11035 p T^2+945 p^3 T^3+p^6 T^4
\tabularnewline[0.5pt]\hline				
10	 &	smooth    &	          &	1+345 T+4985 p T^2+345 p^3 T^3+p^6 T^4
\tabularnewline[0.5pt]\hline				
11	 &	smooth    &	          &	1-275 T+7700 p T^2-275 p^3 T^3+p^6 T^4
\tabularnewline[0.5pt]\hline				
12	 &	smooth    &	          &	1+115 T+4045 p T^2+115 p^3 T^3+p^6 T^4
\tabularnewline[0.5pt]\hline				
13	 &	smooth    &	          &	1+535 T+5680 p T^2+535 p^3 T^3+p^6 T^4
\tabularnewline[0.5pt]\hline				
14	 &	smooth    &	          &	1-10 p T+70 p^2 T^2-10 p^4 T^3+p^6 T^4
\tabularnewline[0.5pt]\hline				
15	 &	smooth    &	          &	1-420 T-1360 p T^2-420 p^3 T^3+p^6 T^4
\tabularnewline[0.5pt]\hline				
16	 &	smooth    &	          &	1-275 T+7275 p T^2-275 p^3 T^3+p^6 T^4
\tabularnewline[0.5pt]\hline				
17	 &	smooth    &	          &	1-65 T-2895 p T^2-65 p^3 T^3+p^6 T^4
\tabularnewline[0.5pt]\hline				
18	 &	smooth    &	          &	1-570 T+1890 p T^2-570 p^3 T^3+p^6 T^4
\tabularnewline[0.5pt]\hline				
19	 &	smooth    &	          &	1+1185 T+13905 p T^2+1185 p^3 T^3+p^6 T^4
\tabularnewline[0.5pt]\hline				
20	 &	smooth    &	          &	1+120 T-3490 p T^2+120 p^3 T^3+p^6 T^4
\tabularnewline[0.5pt]\hline				
21	 &	smooth    &	          &	1-110 T+1320 p T^2-110 p^3 T^3+p^6 T^4
\tabularnewline[0.5pt]\hline				
22	 &	smooth    &	          &	1+555 T+6415 p T^2+555 p^3 T^3+p^6 T^4
\tabularnewline[0.5pt]\hline				
23	 &	smooth    &	          &	1+280 T+390 p T^2+280 p^3 T^3+p^6 T^4
\tabularnewline[0.5pt]\hline				
24	 &	smooth    &	          &	1-620 T+2740 p T^2-620 p^3 T^3+p^6 T^4
\tabularnewline[0.5pt]\hline				
25	 &	smooth    &	          &	1+795 T+6760 p T^2+795 p^3 T^3+p^6 T^4
\tabularnewline[0.5pt]\hline				
26	 &	singular  &	5^{-5}&	(1+p T) (1+763 T+p^3 T^2)
\tabularnewline[0.5pt]\hline				
27	 &	smooth    &	          &	1+490 T+9520 p T^2+490 p^3 T^3+p^6 T^4
\tabularnewline[0.5pt]\hline				
28	 &	smooth    &	          &	1+655 T+1665 p T^2+655 p^3 T^3+p^6 T^4
\tabularnewline[0.5pt]\hline				
29	 &	smooth    &	          &	1-1035 T+10120 p T^2-1035 p^3 T^3+p^6 T^4
\tabularnewline[0.5pt]\hline				
30	 &	smooth    &	          &	1+555 T-685 p T^2+555 p^3 T^3+p^6 T^4
\tabularnewline[0.5pt]\hline				
31	 &	smooth    &	          &	1-20 T-3210 p T^2-20 p^3 T^3+p^6 T^4
\tabularnewline[0.5pt]\hline				
32	 &	smooth    &	          &	1-810 T+10770 p T^2-810 p^3 T^3+p^6 T^4
\tabularnewline[0.5pt]\hline				
33	 &	smooth    &	          &	1+105 T+2740 p T^2+105 p^3 T^3+p^6 T^4
\tabularnewline[0.5pt]\hline				
34	 &	smooth    &	          &	1-1415 T+14280 p T^2-1415 p^3 T^3+p^6 T^4
\tabularnewline[0.5pt]\hline				
35	 &	smooth    &	          &	1+990 T+11470 p T^2+990 p^3 T^3+p^6 T^4
\tabularnewline[0.5pt]\hline				
36	 &	smooth    &	          &	1+190 T+4070 p T^2+190 p^3 T^3+p^6 T^4
\tabularnewline[0.5pt]\hline				
37	 &	smooth    &	          &	1+220 T+6510 p T^2+220 p^3 T^3+p^6 T^4
\tabularnewline[0.5pt]\hline				
38	 &	smooth    &	          &	(1-9 p T+p^3 T^2)(1-508 T+p^3 T^2)
\tabularnewline[0.5pt]\hline				
39	 &	smooth    &	          &	1+550 T+10450 p T^2+550 p^3 T^3+p^6 T^4
\tabularnewline[0.5pt]\hline				
40	 &	smooth    &	          &	1+470 T-440 p T^2+470 p^3 T^3+p^6 T^4
\tabularnewline[0.5pt]\hline				
41	 &	smooth    &	          &	1+285 T-4120 p T^2+285 p^3 T^3+p^6 T^4
\tabularnewline[0.5pt]\hline				
42	 &	smooth    &	          &	1+400 T+5350 p T^2+400 p^3 T^3+p^6 T^4
\tabularnewline[0.5pt]\hline				
43	 &	smooth    &	          &	1-850 T+11050 p T^2-850 p^3 T^3+p^6 T^4
\tabularnewline[0.5pt]\hline				
44	 &	smooth    &	          &	(1+14 p T+p^3 T^2)(1-872 T+p^3 T^2)
\tabularnewline[0.5pt]\hline				
45	 &	smooth    &	          &	1-75 T+1100 p T^2-75 p^3 T^3+p^6 T^4
\tabularnewline[0.5pt]\hline				
46	 &	smooth    &	          &	1+425 T+1025 p T^2+425 p^3 T^3+p^6 T^4
\tabularnewline[0.5pt]\hline				
47	 &	smooth    &	          &	1-260 T-5230 p T^2-260 p^3 T^3+p^6 T^4
\tabularnewline[0.5pt]\hline				
48	 &	smooth    &	          &	1+595 T+11435 p T^2+595 p^3 T^3+p^6 T^4
\tabularnewline[0.5pt]\hline				
49	 &	smooth    &	          &	1-230 T+860 p T^2-230 p^3 T^3+p^6 T^4
\tabularnewline[0.5pt]\hline				
50	 &	smooth    &	          &	1-90 T+1930 p T^2-90 p^3 T^3+p^6 T^4
\tabularnewline[0.5pt]\hline				
51	 &	smooth    &	          &	1-1010 T+10570 p T^2-1010 p^3 T^3+p^6 T^4
\tabularnewline[0.5pt]\hline				
52	 &	smooth    &	          &	1-1430 T+16410 p T^2-1430 p^3 T^3+p^6 T^4
\tabularnewline[0.5pt]\hline				
53	 &	smooth    &	          &	1+535 T+10055 p T^2+535 p^3 T^3+p^6 T^4
\tabularnewline[0.5pt]\hline				
54	 &	smooth    &	          &	1+495 T+3960 p T^2+495 p^3 T^3+p^6 T^4
\tabularnewline[0.5pt]\hline				
55	 &	smooth    &	          &	1-335 T+3945 p T^2-335 p^3 T^3+p^6 T^4
\tabularnewline[0.5pt]\hline				
56	 &	smooth    &	          &	1-860 T+6270 p T^2-860 p^3 T^3+p^6 T^4
\tabularnewline[0.5pt]\hline				
57	 &	smooth    &	          &	1-790 T+7930 p T^2-790 p^3 T^3+p^6 T^4
\tabularnewline[0.5pt]\hline				
58	 &	smooth    &	          &	1-1315 T+12180 p T^2-1315 p^3 T^3+p^6 T^4
\tabularnewline[0.5pt]\hline				
59	 &	smooth    &	          &	1-10 T-3430 p T^2-10 p^3 T^3+p^6 T^4
\tabularnewline[0.5pt]\hline				
60	 &	smooth    &	          &	1+140 T+3320 p T^2+140 p^3 T^3+p^6 T^4
\tabularnewline[0.5pt]\hline				
61	 &	smooth    &	          &	1-430 T+10860 p T^2-430 p^3 T^3+p^6 T^4
\tabularnewline[0.5pt]\hline				
62	 &	smooth    &	          &	1+470 T+2410 p T^2+470 p^3 T^3+p^6 T^4
\tabularnewline[0.5pt]\hline				
63	 &	smooth    &	          &	1+435 T+8705 p T^2+435 p^3 T^3+p^6 T^4
\tabularnewline[0.5pt]\hline				
64	 &	smooth    &	          &	1-250 T-2550 p T^2-250 p^3 T^3+p^6 T^4
\tabularnewline[0.5pt]\hline				
65	 &	smooth    &	          &	1-215 T+7855 p T^2-215 p^3 T^3+p^6 T^4
\tabularnewline[0.5pt]\hline				
66	 &	smooth    &	          &	1+130 T+6940 p T^2+130 p^3 T^3+p^6 T^4
\tabularnewline[0.5pt]\hline				
67	 &	smooth    &	          &	1+485 T+6030 p T^2+485 p^3 T^3+p^6 T^4
\tabularnewline[0.5pt]\hline				
68	 &	smooth    &	          &	1-465 T+10380 p T^2-465 p^3 T^3+p^6 T^4
\tabularnewline[0.5pt]\hline				
69	 &	smooth    &	          &	1-855 T+7885 p T^2-855 p^3 T^3+p^6 T^4
\tabularnewline[0.5pt]\hline				
70	 &	smooth    &	          &	1+25 T-4700 p T^2+25 p^3 T^3+p^6 T^4
\tabularnewline[0.5pt]\hline				
71	 &	smooth    &	          &	1+1115 T+11020 p T^2+1115 p^3 T^3+p^6 T^4
\tabularnewline[0.5pt]\hline				
72	 &	smooth    &	          &	1-765 T+8180 p T^2-765 p^3 T^3+p^6 T^4
\tabularnewline[0.5pt]\hline				
\tablepostamble				
\tablepreamble{79}				
1	 &	smooth    &	          &	1+845 T+6982 p T^2+845 p^3 T^3+p^6 T^4
\tabularnewline[0.5pt]\hline				
2	 &	smooth    &	          &	1+420 T+4482 p T^2+420 p^3 T^3+p^6 T^4
\tabularnewline[0.5pt]\hline				
3	 &	smooth    &	          &	1+20 T+8982 p T^2+20 p^3 T^3+p^6 T^4
\tabularnewline[0.5pt]\hline				
4	 &	smooth    &	          &	1-425 T-4068 p T^2-425 p^3 T^3+p^6 T^4
\tabularnewline[0.5pt]\hline				
5	 &	smooth    &	          &	1+210 T+6132 p T^2+210 p^3 T^3+p^6 T^4
\tabularnewline[0.5pt]\hline				
6	 &	smooth    &	          &	1-190 T+4632 p T^2-190 p^3 T^3+p^6 T^4
\tabularnewline[0.5pt]\hline				
7	 &	smooth    &	          &	1-460 T+7882 p T^2-460 p^3 T^3+p^6 T^4
\tabularnewline[0.5pt]\hline				
8	 &	smooth    &	          &	1+975 T+8932 p T^2+975 p^3 T^3+p^6 T^4
\tabularnewline[0.5pt]\hline				
9	 &	singular  &	5^{-5}&	(1-p T) (1-510 T+p^3 T^2)
\tabularnewline[0.5pt]\hline				
10	 &	smooth    &	          &	1+75 T+182 p T^2+75 p^3 T^3+p^6 T^4
\tabularnewline[0.5pt]\hline				
11	 &	smooth    &	          &	1-415 T+9882 p T^2-415 p^3 T^3+p^6 T^4
\tabularnewline[0.5pt]\hline				
12	 &	smooth    &	          &	1-105 T+5957 p T^2-105 p^3 T^3+p^6 T^4
\tabularnewline[0.5pt]\hline				
13	 &	smooth    &	          &	1-845 T+13482 p T^2-845 p^3 T^3+p^6 T^4
\tabularnewline[0.5pt]\hline				
14	 &	smooth    &	          &	1+200 T+8032 p T^2+200 p^3 T^3+p^6 T^4
\tabularnewline[0.5pt]\hline				
15	 &	smooth    &	          &	1+115 T-5018 p T^2+115 p^3 T^3+p^6 T^4
\tabularnewline[0.5pt]\hline				
16	 &	smooth    &	          &	1-885 T+13382 p T^2-885 p^3 T^3+p^6 T^4
\tabularnewline[0.5pt]\hline				
17	 &	smooth    &	          &	1-5 p T+2082 p T^2-5 p^4 T^3+p^6 T^4
\tabularnewline[0.5pt]\hline				
18	 &	smooth    &	          &	1+2482 p T^2+p^6 T^4
\tabularnewline[0.5pt]\hline				
19	 &	smooth    &	          &	1-715 T+4557 p T^2-715 p^3 T^3+p^6 T^4
\tabularnewline[0.5pt]\hline				
20	 &	smooth    &	          &	1+515 T+5007 p T^2+515 p^3 T^3+p^6 T^4
\tabularnewline[0.5pt]\hline				
21	 &	smooth    &	          &	1-385 T+9382 p T^2-385 p^3 T^3+p^6 T^4
\tabularnewline[0.5pt]\hline				
22	 &	smooth    &	          &	1+435 T+4107 p T^2+435 p^3 T^3+p^6 T^4
\tabularnewline[0.5pt]\hline				
23	 &	smooth    &	          &	1+1260 T+15382 p T^2+1260 p^3 T^3+p^6 T^4
\tabularnewline[0.5pt]\hline				
24	 &	smooth    &	          &	1-745 T+2957 p T^2-745 p^3 T^3+p^6 T^4
\tabularnewline[0.5pt]\hline				
25	 &	smooth    &	          &	1-425 T+8107 p T^2-425 p^3 T^3+p^6 T^4
\tabularnewline[0.5pt]\hline				
26	 &	smooth    &	          &	1+450 T+12532 p T^2+450 p^3 T^3+p^6 T^4
\tabularnewline[0.5pt]\hline				
27	 &	smooth    &	          &	1-955 T+11232 p T^2-955 p^3 T^3+p^6 T^4
\tabularnewline[0.5pt]\hline				
28	 &	smooth    &	          &	(1+p^3 T^2)(1+435 T+p^3 T^2)
\tabularnewline[0.5pt]\hline				
29	 &	smooth    &	          &	1-1470 T+14832 p T^2-1470 p^3 T^3+p^6 T^4
\tabularnewline[0.5pt]\hline				
30	 &	smooth    &	          &	1+380 T+2232 p T^2+380 p^3 T^3+p^6 T^4
\tabularnewline[0.5pt]\hline				
31	 &	smooth    &	          &	1-795 T+6482 p T^2-795 p^3 T^3+p^6 T^4
\tabularnewline[0.5pt]\hline				
32	 &	smooth    &	          &	1-495 T+8482 p T^2-495 p^3 T^3+p^6 T^4
\tabularnewline[0.5pt]\hline				
33	 &	smooth    &	          &	1-565 T+2482 p T^2-565 p^3 T^3+p^6 T^4
\tabularnewline[0.5pt]\hline				
34	 &	smooth    &	          &	1+1140 T+11982 p T^2+1140 p^3 T^3+p^6 T^4
\tabularnewline[0.5pt]\hline				
35	 &	smooth    &	          &	1+80 T+6732 p T^2+80 p^3 T^3+p^6 T^4
\tabularnewline[0.5pt]\hline				
36	 &	smooth    &	          &	1-1020 T+6082 p T^2-1020 p^3 T^3+p^6 T^4
\tabularnewline[0.5pt]\hline				
37	 &	smooth    &	          &	1+225 T+8657 p T^2+225 p^3 T^3+p^6 T^4
\tabularnewline[0.5pt]\hline				
38	 &	smooth    &	          &	1-510 T+7182 p T^2-510 p^3 T^3+p^6 T^4
\tabularnewline[0.5pt]\hline				
39	 &	smooth    &	          &	(1+p^3 T^2)(1+1055 T+p^3 T^2)
\tabularnewline[0.5pt]\hline				
40	 &	smooth    &	          &	1+190 T-2618 p T^2+190 p^3 T^3+p^6 T^4
\tabularnewline[0.5pt]\hline				
41	 &	smooth    &	          &	1+1765 T+19732 p T^2+1765 p^3 T^3+p^6 T^4
\tabularnewline[0.5pt]\hline				
42	 &	smooth    &	          &	1+430 T-2918 p T^2+430 p^3 T^3+p^6 T^4
\tabularnewline[0.5pt]\hline				
43	 &	smooth    &	          &	1+1290 T+14882 p T^2+1290 p^3 T^3+p^6 T^4
\tabularnewline[0.5pt]\hline				
44	 &	smooth    &	          &	1-135 T+1432 p T^2-135 p^3 T^3+p^6 T^4
\tabularnewline[0.5pt]\hline				
45	 &	smooth    &	          &	1+80 T+7482 p T^2+80 p^3 T^3+p^6 T^4
\tabularnewline[0.5pt]\hline				
46	 &	smooth    &	          &	1-350 T+1282 p T^2-350 p^3 T^3+p^6 T^4
\tabularnewline[0.5pt]\hline				
47	 &	smooth    &	          &	1-30 T-10418 p T^2-30 p^3 T^3+p^6 T^4
\tabularnewline[0.5pt]\hline				
48	 &	smooth    &	          &	1-1135 T+13757 p T^2-1135 p^3 T^3+p^6 T^4
\tabularnewline[0.5pt]\hline				
49	 &	smooth    &	          &	1+1525 T+16307 p T^2+1525 p^3 T^3+p^6 T^4
\tabularnewline[0.5pt]\hline				
50	 &	smooth    &	          &	1+180 T+5782 p T^2+180 p^3 T^3+p^6 T^4
\tabularnewline[0.5pt]\hline				
51	 &	smooth    &	          &	1+30 T-5918 p T^2+30 p^3 T^3+p^6 T^4
\tabularnewline[0.5pt]\hline				
52	 &	smooth    &	          &	1-80 T+58 p^2 T^2-80 p^3 T^3+p^6 T^4
\tabularnewline[0.5pt]\hline				
53	 &	smooth    &	          &	1-95 T+6732 p T^2-95 p^3 T^3+p^6 T^4
\tabularnewline[0.5pt]\hline				
54	 &	smooth    &	          &	1-400 T+1282 p T^2-400 p^3 T^3+p^6 T^4
\tabularnewline[0.5pt]\hline				
55	 &	smooth    &	          &	1+515 T+11982 p T^2+515 p^3 T^3+p^6 T^4
\tabularnewline[0.5pt]\hline				
56	 &	smooth    &	          &	1+225 T-6818 p T^2+225 p^3 T^3+p^6 T^4
\tabularnewline[0.5pt]\hline				
57	 &	smooth    &	          &	1+1025 T+8682 p T^2+1025 p^3 T^3+p^6 T^4
\tabularnewline[0.5pt]\hline				
58	 &	smooth    &	          &	1-140 T-2118 p T^2-140 p^3 T^3+p^6 T^4
\tabularnewline[0.5pt]\hline				
59	 &	smooth    &	          &	1-645 T+11582 p T^2-645 p^3 T^3+p^6 T^4
\tabularnewline[0.5pt]\hline				
60	 &	smooth    &	          &	1+335 T+6232 p T^2+335 p^3 T^3+p^6 T^4
\tabularnewline[0.5pt]\hline				
61	 &	smooth    &	          &	1+160 T+382 p T^2+160 p^3 T^3+p^6 T^4
\tabularnewline[0.5pt]\hline				
62	 &	smooth    &	          &	1+485 T+4232 p T^2+485 p^3 T^3+p^6 T^4
\tabularnewline[0.5pt]\hline				
63	 &	smooth    &	          &	1-265 T+1557 p T^2-265 p^3 T^3+p^6 T^4
\tabularnewline[0.5pt]\hline				
64	 &	smooth    &	          &	1+1335 T+14982 p T^2+1335 p^3 T^3+p^6 T^4
\tabularnewline[0.5pt]\hline				
65	 &	smooth    &	          &	1+110 T-618 p T^2+110 p^3 T^3+p^6 T^4
\tabularnewline[0.5pt]\hline				
66	 &	smooth    &	          &	1-1105 T+10957 p T^2-1105 p^3 T^3+p^6 T^4
\tabularnewline[0.5pt]\hline				
67	 &	smooth    &	          &	1+610 T+6882 p T^2+610 p^3 T^3+p^6 T^4
\tabularnewline[0.5pt]\hline				
68	 &	smooth    &	          &	1-875 T+6682 p T^2-875 p^3 T^3+p^6 T^4
\tabularnewline[0.5pt]\hline				
69	 &	smooth    &	          &	1-120 T+9282 p T^2-120 p^3 T^3+p^6 T^4
\tabularnewline[0.5pt]\hline				
70	 &	smooth    &	          &	1-105 T-4718 p T^2-105 p^3 T^3+p^6 T^4
\tabularnewline[0.5pt]\hline				
71	 &	smooth    &	          &	1-1015 T+8732 p T^2-1015 p^3 T^3+p^6 T^4
\tabularnewline[0.5pt]\hline				
72	 &	smooth    &	          &	1+280 T+3232 p T^2+280 p^3 T^3+p^6 T^4
\tabularnewline[0.5pt]\hline				
73	 &	smooth    &	          &	1+1315 T+14607 p T^2+1315 p^3 T^3+p^6 T^4
\tabularnewline[0.5pt]\hline				
74	 &	smooth    &	          &	1-1380 T+12832 p T^2-1380 p^3 T^3+p^6 T^4
\tabularnewline[0.5pt]\hline				
75	 &	smooth    &	          &	1-65 T-5993 p T^2-65 p^3 T^3+p^6 T^4
\tabularnewline[0.5pt]\hline				
76	 &	smooth    &	          &	1-1080 T+14082 p T^2-1080 p^3 T^3+p^6 T^4
\tabularnewline[0.5pt]\hline				
77	 &	smooth    &	          &	1+755 T+9857 p T^2+755 p^3 T^3+p^6 T^4
\tabularnewline[0.5pt]\hline				
78	 &	smooth    &	          &	1-65 T-8118 p T^2-65 p^3 T^3+p^6 T^4
\tabularnewline[0.5pt]\hline				
\tablepostamble				
\tablepreamble{83}				
1	 &	smooth    &	          &	1-275 T+12350 p T^2-275 p^3 T^3+p^6 T^4
\tabularnewline[0.5pt]\hline				
2	 &	smooth    &	          &	1-1075 T+10225 p T^2-1075 p^3 T^3+p^6 T^4
\tabularnewline[0.5pt]\hline				
3	 &	smooth    &	          &	1+295 T-5815 p T^2+295 p^3 T^3+p^6 T^4
\tabularnewline[0.5pt]\hline				
4	 &	smooth    &	          &	1-525 T+175 p T^2-525 p^3 T^3+p^6 T^4
\tabularnewline[0.5pt]\hline				
5	 &	smooth    &	          &	1+1465 T+15170 p T^2+1465 p^3 T^3+p^6 T^4
\tabularnewline[0.5pt]\hline				
6	 &	smooth    &	          &	1-980 T+15010 p T^2-980 p^3 T^3+p^6 T^4
\tabularnewline[0.5pt]\hline				
7	 &	smooth    &	          &	1-195 T-260 p T^2-195 p^3 T^3+p^6 T^4
\tabularnewline[0.5pt]\hline				
8	 &	smooth    &	          &	1-345 T+11340 p T^2-345 p^3 T^3+p^6 T^4
\tabularnewline[0.5pt]\hline				
9	 &	smooth    &	          &	1-895 T+3515 p T^2-895 p^3 T^3+p^6 T^4
\tabularnewline[0.5pt]\hline				
10	 &	smooth    &	          &	1-825 T+11150 p T^2-825 p^3 T^3+p^6 T^4
\tabularnewline[0.5pt]\hline				
11	 &	smooth    &	          &	1+45 T-6590 p T^2+45 p^3 T^3+p^6 T^4
\tabularnewline[0.5pt]\hline				
12	 &	smooth    &	          &	1-855 T+13910 p T^2-855 p^3 T^3+p^6 T^4
\tabularnewline[0.5pt]\hline				
13	 &	smooth    &	          &	1+35 T+4230 p T^2+35 p^3 T^3+p^6 T^4
\tabularnewline[0.5pt]\hline				
14	 &	smooth    &	          &	1-40 T+13330 p T^2-40 p^3 T^3+p^6 T^4
\tabularnewline[0.5pt]\hline				
15	 &	smooth    &	          &	1+465 T-205 p T^2+465 p^3 T^3+p^6 T^4
\tabularnewline[0.5pt]\hline				
16	 &	smooth    &	          &	1+575 T+3225 p T^2+575 p^3 T^3+p^6 T^4
\tabularnewline[0.5pt]\hline				
17	 &	smooth    &	          &	1+115 T+5470 p T^2+115 p^3 T^3+p^6 T^4
\tabularnewline[0.5pt]\hline				
18	 &	smooth    &	          &	1-270 T+13190 p T^2-270 p^3 T^3+p^6 T^4
\tabularnewline[0.5pt]\hline				
19	 &	smooth    &	          &	1+225 T-3250 p T^2+225 p^3 T^3+p^6 T^4
\tabularnewline[0.5pt]\hline				
20	 &	singular  &	5^{-5}&	(1+p T) (1-777 T+p^3 T^2)
\tabularnewline[0.5pt]\hline				
21	 &	smooth    &	          &	1+1070 T+14210 p T^2+1070 p^3 T^3+p^6 T^4
\tabularnewline[0.5pt]\hline				
22	 &	smooth    &	          &	1+1180 T+15340 p T^2+1180 p^3 T^3+p^6 T^4
\tabularnewline[0.5pt]\hline				
23	 &	smooth    &	          &	1-105 T+8060 p T^2-105 p^3 T^3+p^6 T^4
\tabularnewline[0.5pt]\hline				
24	 &	smooth    &	          &	1-15 T-10570 p T^2-15 p^3 T^3+p^6 T^4
\tabularnewline[0.5pt]\hline				
25	 &	smooth    &	          &	1-710 T+4370 p T^2-710 p^3 T^3+p^6 T^4
\tabularnewline[0.5pt]\hline				
26	 &	smooth    &	          &	1-365 T+12230 p T^2-365 p^3 T^3+p^6 T^4
\tabularnewline[0.5pt]\hline				
27	 &	smooth    &	          &	1-120 T+6590 p T^2-120 p^3 T^3+p^6 T^4
\tabularnewline[0.5pt]\hline				
28	 &	smooth    &	          &	1+495 T+4785 p T^2+495 p^3 T^3+p^6 T^4
\tabularnewline[0.5pt]\hline				
29	 &	smooth    &	          &	1+870 T+5010 p T^2+870 p^3 T^3+p^6 T^4
\tabularnewline[0.5pt]\hline				
30	 &	smooth    &	          &	1-570 T+13440 p T^2-570 p^3 T^3+p^6 T^4
\tabularnewline[0.5pt]\hline				
31	 &	smooth    &	          &	1+310 T+530 p T^2+310 p^3 T^3+p^6 T^4
\tabularnewline[0.5pt]\hline				
32	 &	smooth    &	          &	1-435 T-2330 p T^2-435 p^3 T^3+p^6 T^4
\tabularnewline[0.5pt]\hline				
33	 &	smooth    &	          &	1+325 T-5275 p T^2+325 p^3 T^3+p^6 T^4
\tabularnewline[0.5pt]\hline				
34	 &	smooth    &	          &	1+365 T-4080 p T^2+365 p^3 T^3+p^6 T^4
\tabularnewline[0.5pt]\hline				
35	 &	smooth    &	          &	1-750 T+10350 p T^2-750 p^3 T^3+p^6 T^4
\tabularnewline[0.5pt]\hline				
36	 &	smooth    &	          &	1+1390 T+13170 p T^2+1390 p^3 T^3+p^6 T^4
\tabularnewline[0.5pt]\hline				
37	 &	smooth    &	          &	1-1020 T+13040 p T^2-1020 p^3 T^3+p^6 T^4
\tabularnewline[0.5pt]\hline				
38	 &	smooth    &	          &	1-155 T+10085 p T^2-155 p^3 T^3+p^6 T^4
\tabularnewline[0.5pt]\hline				
39	 &	smooth    &	          &	1-1730 T+20760 p T^2-1730 p^3 T^3+p^6 T^4
\tabularnewline[0.5pt]\hline				
40	 &	smooth    &	          &	1-635 T+4595 p T^2-635 p^3 T^3+p^6 T^4
\tabularnewline[0.5pt]\hline				
41	 &	smooth    &	          &	1+1210 T+16280 p T^2+1210 p^3 T^3+p^6 T^4
\tabularnewline[0.5pt]\hline				
42	 &	smooth    &	          &	1-1270 T+15940 p T^2-1270 p^3 T^3+p^6 T^4
\tabularnewline[0.5pt]\hline				
43	 &	smooth    &	          &	1-1290 T+15630 p T^2-1290 p^3 T^3+p^6 T^4
\tabularnewline[0.5pt]\hline				
44	 &	smooth    &	          &	1-190 T+5430 p T^2-190 p^3 T^3+p^6 T^4
\tabularnewline[0.5pt]\hline				
45	 &	smooth    &	          &	1-55 T+2985 p T^2-55 p^3 T^3+p^6 T^4
\tabularnewline[0.5pt]\hline				
46	 &	smooth    &	          &	1-925 T+11750 p T^2-925 p^3 T^3+p^6 T^4
\tabularnewline[0.5pt]\hline				
47	 &	smooth    &	          &	1-1160 T+10370 p T^2-1160 p^3 T^3+p^6 T^4
\tabularnewline[0.5pt]\hline				
48	 &	smooth    &	          &	1-600 T+9950 p T^2-600 p^3 T^3+p^6 T^4
\tabularnewline[0.5pt]\hline				
49	 &	smooth    &	          &	1-205 T-6940 p T^2-205 p^3 T^3+p^6 T^4
\tabularnewline[0.5pt]\hline				
50	 &	smooth    &	          &	1+835 T+12005 p T^2+835 p^3 T^3+p^6 T^4
\tabularnewline[0.5pt]\hline				
51	 &	smooth    &	          &	1+1170 T+10410 p T^2+1170 p^3 T^3+p^6 T^4
\tabularnewline[0.5pt]\hline				
52	 &	smooth    &	          &	1-435 T+12970 p T^2-435 p^3 T^3+p^6 T^4
\tabularnewline[0.5pt]\hline				
53	 &	smooth    &	          &	1+310 T+4430 p T^2+310 p^3 T^3+p^6 T^4
\tabularnewline[0.5pt]\hline				
54	 &	smooth    &	          &	1+350 T+400 p T^2+350 p^3 T^3+p^6 T^4
\tabularnewline[0.5pt]\hline				
55	 &	smooth    &	          &	1-1205 T+14160 p T^2-1205 p^3 T^3+p^6 T^4
\tabularnewline[0.5pt]\hline				
56	 &	smooth    &	          &	1-185 T+1645 p T^2-185 p^3 T^3+p^6 T^4
\tabularnewline[0.5pt]\hline				
57	 &	smooth    &	          &	1-1335 T+17870 p T^2-1335 p^3 T^3+p^6 T^4
\tabularnewline[0.5pt]\hline				
58	 &	smooth    &	          &	1+815 T+8970 p T^2+815 p^3 T^3+p^6 T^4
\tabularnewline[0.5pt]\hline				
59	 &	smooth    &	          &	1-555 T+9985 p T^2-555 p^3 T^3+p^6 T^4
\tabularnewline[0.5pt]\hline				
60	 &	smooth    &	          &	1+275 T+5075 p T^2+275 p^3 T^3+p^6 T^4
\tabularnewline[0.5pt]\hline				
61	 &	smooth    &	          &	1+570 T+8810 p T^2+570 p^3 T^3+p^6 T^4
\tabularnewline[0.5pt]\hline				
62	 &	smooth    &	          &	1+590 T+12270 p T^2+590 p^3 T^3+p^6 T^4
\tabularnewline[0.5pt]\hline				
63	 &	smooth    &	          &	1-605 T+1110 p T^2-605 p^3 T^3+p^6 T^4
\tabularnewline[0.5pt]\hline				
64	 &	smooth    &	          &	1+135 T-1020 p T^2+135 p^3 T^3+p^6 T^4
\tabularnewline[0.5pt]\hline				
65	 &	smooth    &	          &	1+1565 T+17070 p T^2+1565 p^3 T^3+p^6 T^4
\tabularnewline[0.5pt]\hline				
66	 &	smooth    &	          &	1+410 T-2170 p T^2+410 p^3 T^3+p^6 T^4
\tabularnewline[0.5pt]\hline				
67	 &	smooth    &	          &	1+550 T+13050 p T^2+550 p^3 T^3+p^6 T^4
\tabularnewline[0.5pt]\hline				
68	 &	smooth    &	          &	1-510 T+7020 p T^2-510 p^3 T^3+p^6 T^4
\tabularnewline[0.5pt]\hline				
69	 &	smooth    &	          &	1+335 T-920 p T^2+335 p^3 T^3+p^6 T^4
\tabularnewline[0.5pt]\hline				
70	 &	smooth    &	          &	1-950 T+13650 p T^2-950 p^3 T^3+p^6 T^4
\tabularnewline[0.5pt]\hline				
71	 &	smooth    &	          &	1+110 T-4870 p T^2+110 p^3 T^3+p^6 T^4
\tabularnewline[0.5pt]\hline				
72	 &	smooth    &	          &	1+630 T+2240 p T^2+630 p^3 T^3+p^6 T^4
\tabularnewline[0.5pt]\hline				
73	 &	smooth    &	          &	1+630 T+1140 p T^2+630 p^3 T^3+p^6 T^4
\tabularnewline[0.5pt]\hline				
74	 &	smooth    &	          &	1+100 T+3300 p T^2+100 p^3 T^3+p^6 T^4
\tabularnewline[0.5pt]\hline				
75	 &	smooth    &	          &	1+5 p T+10870 p T^2+5 p^4 T^3+p^6 T^4
\tabularnewline[0.5pt]\hline				
76	 &	smooth    &	          &	1+580 T+940 p T^2+580 p^3 T^3+p^6 T^4
\tabularnewline[0.5pt]\hline				
77	 &	smooth    &	          &	1+1255 T+17390 p T^2+1255 p^3 T^3+p^6 T^4
\tabularnewline[0.5pt]\hline				
78	 &	smooth    &	          &	1+1820 T+21010 p T^2+1820 p^3 T^3+p^6 T^4
\tabularnewline[0.5pt]\hline				
79	 &	smooth    &	          &	1+805 T+3890 p T^2+805 p^3 T^3+p^6 T^4
\tabularnewline[0.5pt]\hline				
80	 &	smooth    &	          &	1+350 T+6950 p T^2+350 p^3 T^3+p^6 T^4
\tabularnewline[0.5pt]\hline				
81	 &	smooth    &	          &	1+345 T-4540 p T^2+345 p^3 T^3+p^6 T^4
\tabularnewline[0.5pt]\hline				
82	 &	smooth    &	          &	1-325 T+1800 p T^2-325 p^3 T^3+p^6 T^4
\tabularnewline[0.5pt]\hline				
\tablepostamble				
\tablepreamble{89}				
1	 &	smooth    &	          &	1+355 T-6033 p T^2+355 p^3 T^3+p^6 T^4
\tabularnewline[0.5pt]\hline				
2	 &	smooth    &	          &	1+200 T-4858 p T^2+200 p^3 T^3+p^6 T^4
\tabularnewline[0.5pt]\hline				
3	 &	smooth    &	          &	1-785 T+5492 p T^2-785 p^3 T^3+p^6 T^4
\tabularnewline[0.5pt]\hline				
4	 &	smooth    &	          &	1+320 T+11642 p T^2+320 p^3 T^3+p^6 T^4
\tabularnewline[0.5pt]\hline				
5	 &	smooth    &	          &	(1+15 p T+p^3 T^2)(1-570 T+p^3 T^2)
\tabularnewline[0.5pt]\hline				
6	 &	smooth    &	          &	1-185 T-97 p^2 T^2-185 p^3 T^3+p^6 T^4
\tabularnewline[0.5pt]\hline				
7	 &	smooth    &	          &	1+1570 T+17442 p T^2+1570 p^3 T^3+p^6 T^4
\tabularnewline[0.5pt]\hline				
8	 &	smooth    &	          &	1+275 T-3333 p T^2+275 p^3 T^3+p^6 T^4
\tabularnewline[0.5pt]\hline				
9	 &	singular  &	5^{-5}&	(1-p T) (1+945 T+p^3 T^2)
\tabularnewline[0.5pt]\hline				
10	 &	smooth    &	          &	1-675 T+10792 p T^2-675 p^3 T^3+p^6 T^4
\tabularnewline[0.5pt]\hline				
11	 &	smooth    &	          &	1+80 T+13642 p T^2+80 p^3 T^3+p^6 T^4
\tabularnewline[0.5pt]\hline				
12	 &	smooth    &	          &	1-1690 T+20342 p T^2-1690 p^3 T^3+p^6 T^4
\tabularnewline[0.5pt]\hline				
13	 &	smooth    &	          &	1-450 T+14142 p T^2-450 p^3 T^3+p^6 T^4
\tabularnewline[0.5pt]\hline				
14	 &	smooth    &	          &	1+285 T+3742 p T^2+285 p^3 T^3+p^6 T^4
\tabularnewline[0.5pt]\hline				
15	 &	smooth    &	          &	1+1295 T+17217 p T^2+1295 p^3 T^3+p^6 T^4
\tabularnewline[0.5pt]\hline				
16	 &	smooth    &	          &	1-870 T+2942 p T^2-870 p^3 T^3+p^6 T^4
\tabularnewline[0.5pt]\hline				
17	 &	smooth    &	          &	1+895 T+13717 p T^2+895 p^3 T^3+p^6 T^4
\tabularnewline[0.5pt]\hline				
18	 &	smooth    &	          &	1-655 T+7267 p T^2-655 p^3 T^3+p^6 T^4
\tabularnewline[0.5pt]\hline				
19	 &	smooth    &	          &	1-865 T+14342 p T^2-865 p^3 T^3+p^6 T^4
\tabularnewline[0.5pt]\hline				
20	 &	smooth    &	          &	1-180 T-9358 p T^2-180 p^3 T^3+p^6 T^4
\tabularnewline[0.5pt]\hline				
21	 &	smooth    &	          &	1+860 T+8742 p T^2+860 p^3 T^3+p^6 T^4
\tabularnewline[0.5pt]\hline				
22	 &	smooth    &	          &	1-600 T+14792 p T^2-600 p^3 T^3+p^6 T^4
\tabularnewline[0.5pt]\hline				
23	 &	smooth    &	          &	1+310 T+1842 p T^2+310 p^3 T^3+p^6 T^4
\tabularnewline[0.5pt]\hline				
24	 &	smooth    &	          &	1-1655 T+22217 p T^2-1655 p^3 T^3+p^6 T^4
\tabularnewline[0.5pt]\hline				
25	 &	smooth    &	          &	1+700 T+8342 p T^2+700 p^3 T^3+p^6 T^4
\tabularnewline[0.5pt]\hline				
26	 &	smooth    &	          &	1+330 T-5558 p T^2+330 p^3 T^3+p^6 T^4
\tabularnewline[0.5pt]\hline				
27	 &	smooth    &	          &	1+130 T+5842 p T^2+130 p^3 T^3+p^6 T^4
\tabularnewline[0.5pt]\hline				
28	 &	smooth    &	          &	1-2070 T+26342 p T^2-2070 p^3 T^3+p^6 T^4
\tabularnewline[0.5pt]\hline				
29	 &	smooth    &	          &	1+1060 T+11592 p T^2+1060 p^3 T^3+p^6 T^4
\tabularnewline[0.5pt]\hline				
30	 &	smooth    &	          &	1-1115 T+12242 p T^2-1115 p^3 T^3+p^6 T^4
\tabularnewline[0.5pt]\hline				
31	 &	smooth    &	          &	1-180 T+10342 p T^2-180 p^3 T^3+p^6 T^4
\tabularnewline[0.5pt]\hline				
32	 &	smooth    &	          &	1-555 T+7842 p T^2-555 p^3 T^3+p^6 T^4
\tabularnewline[0.5pt]\hline				
33	 &	smooth    &	          &	1+655 T+7092 p T^2+655 p^3 T^3+p^6 T^4
\tabularnewline[0.5pt]\hline				
34	 &	smooth    &	          &	1+1755 T+23017 p T^2+1755 p^3 T^3+p^6 T^4
\tabularnewline[0.5pt]\hline				
35	 &	smooth    &	          &	1+1645 T+20092 p T^2+1645 p^3 T^3+p^6 T^4
\tabularnewline[0.5pt]\hline				
36	 &	smooth    &	          &	1-1105 T+12192 p T^2-1105 p^3 T^3+p^6 T^4
\tabularnewline[0.5pt]\hline				
37	 &	smooth    &	          &	1-775 T+11517 p T^2-775 p^3 T^3+p^6 T^4
\tabularnewline[0.5pt]\hline				
38	 &	smooth    &	          &	1-815 T+1867 p T^2-815 p^3 T^3+p^6 T^4
\tabularnewline[0.5pt]\hline				
39	 &	smooth    &	          &	1+1140 T+10492 p T^2+1140 p^3 T^3+p^6 T^4
\tabularnewline[0.5pt]\hline				
40	 &	smooth    &	          &	1+850 T+17042 p T^2+850 p^3 T^3+p^6 T^4
\tabularnewline[0.5pt]\hline				
41	 &	smooth    &	          &	1-150 T+7292 p T^2-150 p^3 T^3+p^6 T^4
\tabularnewline[0.5pt]\hline				
42	 &	smooth    &	          &	1-290 T-5508 p T^2-290 p^3 T^3+p^6 T^4
\tabularnewline[0.5pt]\hline				
43	 &	smooth    &	          &	1+250 T-3958 p T^2+250 p^3 T^3+p^6 T^4
\tabularnewline[0.5pt]\hline				
44	 &	smooth    &	          &	1-885 T+2342 p T^2-885 p^3 T^3+p^6 T^4
\tabularnewline[0.5pt]\hline				
45	 &	smooth    &	          &	1+270 T+8642 p T^2+270 p^3 T^3+p^6 T^4
\tabularnewline[0.5pt]\hline				
46	 &	smooth    &	          &	1-755 T+8317 p T^2-755 p^3 T^3+p^6 T^4
\tabularnewline[0.5pt]\hline				
47	 &	smooth    &	          &	1+1080 T+15342 p T^2+1080 p^3 T^3+p^6 T^4
\tabularnewline[0.5pt]\hline				
48	 &	smooth    &	          &	1-360 T+12742 p T^2-360 p^3 T^3+p^6 T^4
\tabularnewline[0.5pt]\hline				
49	 &	smooth    &	          &	1-425 T+4592 p T^2-425 p^3 T^3+p^6 T^4
\tabularnewline[0.5pt]\hline				
50	 &	smooth    &	          &	1-1070 T+4842 p T^2-1070 p^3 T^3+p^6 T^4
\tabularnewline[0.5pt]\hline				
51	 &	smooth    &	          &	1+640 T+8492 p T^2+640 p^3 T^3+p^6 T^4
\tabularnewline[0.5pt]\hline				
52	 &	smooth    &	          &	1+175 T-9108 p T^2+175 p^3 T^3+p^6 T^4
\tabularnewline[0.5pt]\hline				
53	 &	smooth    &	          &	1-325 T-7283 p T^2-325 p^3 T^3+p^6 T^4
\tabularnewline[0.5pt]\hline				
54	 &	smooth    &	          &	1-700 T+13142 p T^2-700 p^3 T^3+p^6 T^4
\tabularnewline[0.5pt]\hline				
55	 &	smooth    &	          &	1-590 T+242 p T^2-590 p^3 T^3+p^6 T^4
\tabularnewline[0.5pt]\hline				
56	 &	smooth    &	          &	1-490 T+3742 p T^2-490 p^3 T^3+p^6 T^4
\tabularnewline[0.5pt]\hline				
57	 &	smooth    &	          &	1-390 T+9242 p T^2-390 p^3 T^3+p^6 T^4
\tabularnewline[0.5pt]\hline				
58	 &	smooth    &	          &	1+1230 T+8342 p T^2+1230 p^3 T^3+p^6 T^4
\tabularnewline[0.5pt]\hline				
59	 &	smooth    &	          &	1+300 T+10292 p T^2+300 p^3 T^3+p^6 T^4
\tabularnewline[0.5pt]\hline				
60	 &	smooth    &	          &	1-555 T-1033 p T^2-555 p^3 T^3+p^6 T^4
\tabularnewline[0.5pt]\hline				
61	 &	smooth    &	          &	1-55 T+2317 p T^2-55 p^3 T^3+p^6 T^4
\tabularnewline[0.5pt]\hline				
62	 &	smooth    &	          &	1-525 T+16142 p T^2-525 p^3 T^3+p^6 T^4
\tabularnewline[0.5pt]\hline				
63	 &	smooth    &	          &	1+285 T-4633 p T^2+285 p^3 T^3+p^6 T^4
\tabularnewline[0.5pt]\hline				
64	 &	smooth    &	          &	1+465 T+5417 p T^2+465 p^3 T^3+p^6 T^4
\tabularnewline[0.5pt]\hline				
65	 &	smooth    &	          &	1-15 T+1867 p T^2-15 p^3 T^3+p^6 T^4
\tabularnewline[0.5pt]\hline				
66	 &	smooth    &	          &	1+1320 T+15342 p T^2+1320 p^3 T^3+p^6 T^4
\tabularnewline[0.5pt]\hline				
67	 &	smooth    &	          &	1+510 T+7492 p T^2+510 p^3 T^3+p^6 T^4
\tabularnewline[0.5pt]\hline				
68	 &	smooth    &	          &	1+265 T+6242 p T^2+265 p^3 T^3+p^6 T^4
\tabularnewline[0.5pt]\hline				
69	 &	smooth    &	          &	1+750 T+8392 p T^2+750 p^3 T^3+p^6 T^4
\tabularnewline[0.5pt]\hline				
70	 &	smooth    &	          &	1-1175 T+16917 p T^2-1175 p^3 T^3+p^6 T^4
\tabularnewline[0.5pt]\hline				
71	 &	smooth    &	          &	1-505 T+3092 p T^2-505 p^3 T^3+p^6 T^4
\tabularnewline[0.5pt]\hline				
72	 &	smooth    &	          &	1+1125 T+15142 p T^2+1125 p^3 T^3+p^6 T^4
\tabularnewline[0.5pt]\hline				
73	 &	smooth    &	          &	1-875 T+8967 p T^2-875 p^3 T^3+p^6 T^4
\tabularnewline[0.5pt]\hline				
74	 &	smooth    &	          &	1+1755 T+23592 p T^2+1755 p^3 T^3+p^6 T^4
\tabularnewline[0.5pt]\hline				
75	 &	smooth    &	          &	1+1265 T+17967 p T^2+1265 p^3 T^3+p^6 T^4
\tabularnewline[0.5pt]\hline				
76	 &	smooth    &	          &	1-110 T+7242 p T^2-110 p^3 T^3+p^6 T^4
\tabularnewline[0.5pt]\hline				
77	 &	smooth    &	          &	1+225 T+292 p T^2+225 p^3 T^3+p^6 T^4
\tabularnewline[0.5pt]\hline				
78	 &	smooth    &	          &	1-705 T+11517 p T^2-705 p^3 T^3+p^6 T^4
\tabularnewline[0.5pt]\hline				
79	 &	smooth    &	          &	1-380 T+2942 p T^2-380 p^3 T^3+p^6 T^4
\tabularnewline[0.5pt]\hline				
80	 &	smooth    &	          &	1-15 T+10292 p T^2-15 p^3 T^3+p^6 T^4
\tabularnewline[0.5pt]\hline				
81	 &	smooth    &	          &	1+1340 T+14242 p T^2+1340 p^3 T^3+p^6 T^4
\tabularnewline[0.5pt]\hline				
82	 &	smooth    &	          &	1-1005 T+17192 p T^2-1005 p^3 T^3+p^6 T^4
\tabularnewline[0.5pt]\hline				
83	 &	smooth    &	          &	1-1625 T+17792 p T^2-1625 p^3 T^3+p^6 T^4
\tabularnewline[0.5pt]\hline				
84	 &	smooth    &	          &	1-340 T-3508 p T^2-340 p^3 T^3+p^6 T^4
\tabularnewline[0.5pt]\hline				
85	 &	smooth    &	          &	1-330 T+1442 p T^2-330 p^3 T^3+p^6 T^4
\tabularnewline[0.5pt]\hline				
86	 &	smooth    &	          &	1-580 T-4558 p T^2-580 p^3 T^3+p^6 T^4
\tabularnewline[0.5pt]\hline				
87	 &	smooth    &	          &	1+875 T+12092 p T^2+875 p^3 T^3+p^6 T^4
\tabularnewline[0.5pt]\hline				
88	 &	smooth    &	          &	1-5 T-3408 p T^2-5 p^3 T^3+p^6 T^4
\tabularnewline[0.5pt]\hline				
\tablepostamble				
\tablepreamble{97}				
1	 &	smooth    &	          &	1+15 p T+21730 p T^2+15 p^4 T^3+p^6 T^4
\tabularnewline[0.5pt]\hline				
2	 &	smooth    &	          &	1-370 T+13230 p T^2-370 p^3 T^3+p^6 T^4
\tabularnewline[0.5pt]\hline				
3	 &	smooth    &	          &	1-555 T+14795 p T^2-555 p^3 T^3+p^6 T^4
\tabularnewline[0.5pt]\hline				
4	 &	smooth    &	          &	1+15 p T+22730 p T^2+15 p^4 T^3+p^6 T^4
\tabularnewline[0.5pt]\hline				
5	 &	smooth    &	          &	1-285 T-460 p T^2-285 p^3 T^3+p^6 T^4
\tabularnewline[0.5pt]\hline				
6	 &	smooth    &	          &	1-2440 T+29610 p T^2-2440 p^3 T^3+p^6 T^4
\tabularnewline[0.5pt]\hline				
7	 &	smooth    &	          &	1-1470 T+21580 p T^2-1470 p^3 T^3+p^6 T^4
\tabularnewline[0.5pt]\hline				
8	 &	smooth    &	          &	1+1255 T+13505 p T^2+1255 p^3 T^3+p^6 T^4
\tabularnewline[0.5pt]\hline				
9	 &	smooth    &	          &	1+175 T+7225 p T^2+175 p^3 T^3+p^6 T^4
\tabularnewline[0.5pt]\hline				
10	 &	smooth    &	          &	1-300 T+12950 p T^2-300 p^3 T^3+p^6 T^4
\tabularnewline[0.5pt]\hline				
11	 &	smooth    &	          &	1-385 T+14415 p T^2-385 p^3 T^3+p^6 T^4
\tabularnewline[0.5pt]\hline				
12	 &	smooth    &	          &	1-305 T+17920 p T^2-305 p^3 T^3+p^6 T^4
\tabularnewline[0.5pt]\hline				
13	 &	smooth    &	          &	1-275 T-1000 p T^2-275 p^3 T^3+p^6 T^4
\tabularnewline[0.5pt]\hline				
14	 &	smooth    &	          &	1+890 T+5590 p T^2+890 p^3 T^3+p^6 T^4
\tabularnewline[0.5pt]\hline				
15	 &	smooth    &	          &	1+1495 T+16895 p T^2+1495 p^3 T^3+p^6 T^4
\tabularnewline[0.5pt]\hline				
16	 &	smooth    &	          &	1-775 T+19500 p T^2-775 p^3 T^3+p^6 T^4
\tabularnewline[0.5pt]\hline				
17	 &	smooth    &	          &	1+115 T+12840 p T^2+115 p^3 T^3+p^6 T^4
\tabularnewline[0.5pt]\hline				
18	 &	smooth    &	          &	1+1185 T+17360 p T^2+1185 p^3 T^3+p^6 T^4
\tabularnewline[0.5pt]\hline				
19	 &	smooth    &	          &	1-400 T+350 p T^2-400 p^3 T^3+p^6 T^4
\tabularnewline[0.5pt]\hline				
20	 &	smooth    &	          &	1+120 T+9270 p T^2+120 p^3 T^3+p^6 T^4
\tabularnewline[0.5pt]\hline				
21	 &	smooth    &	          &	1+870 T+1170 p T^2+870 p^3 T^3+p^6 T^4
\tabularnewline[0.5pt]\hline				
22	 &	smooth    &	          &	1+1605 T+21105 p T^2+1605 p^3 T^3+p^6 T^4
\tabularnewline[0.5pt]\hline				
23	 &	smooth    &	          &	1+1180 T+20230 p T^2+1180 p^3 T^3+p^6 T^4
\tabularnewline[0.5pt]\hline				
24	 &	smooth    &	          &	1-55 T+3120 p T^2-55 p^3 T^3+p^6 T^4
\tabularnewline[0.5pt]\hline				
25	 &	smooth    &	          &	1-160 T+9490 p T^2-160 p^3 T^3+p^6 T^4
\tabularnewline[0.5pt]\hline				
26	 &	smooth    &	          &	1+475 T-3400 p T^2+475 p^3 T^3+p^6 T^4
\tabularnewline[0.5pt]\hline				
27	 &	smooth    &	          &	1-2410 T+29390 p T^2-2410 p^3 T^3+p^6 T^4
\tabularnewline[0.5pt]\hline				
28	 &	smooth    &	          &	1+1410 T+12910 p T^2+1410 p^3 T^3+p^6 T^4
\tabularnewline[0.5pt]\hline				
29	 &	smooth    &	          &	1+165 T+11415 p T^2+165 p^3 T^3+p^6 T^4
\tabularnewline[0.5pt]\hline				
30	 &	smooth    &	          &	1+1390 T+22590 p T^2+1390 p^3 T^3+p^6 T^4
\tabularnewline[0.5pt]\hline				
31	 &	smooth    &	          &	1+1345 T+12420 p T^2+1345 p^3 T^3+p^6 T^4
\tabularnewline[0.5pt]\hline				
32	 &	smooth    &	          &	1-795 T+14005 p T^2-795 p^3 T^3+p^6 T^4
\tabularnewline[0.5pt]\hline				
33	 &	smooth    &	          &	1+50 T+8850 p T^2+50 p^3 T^3+p^6 T^4
\tabularnewline[0.5pt]\hline				
34	 &	smooth    &	          &	1-55 T-9880 p T^2-55 p^3 T^3+p^6 T^4
\tabularnewline[0.5pt]\hline				
35	 &	smooth    &	          &	1+175 T+900 p T^2+175 p^3 T^3+p^6 T^4
\tabularnewline[0.5pt]\hline				
36	 &	smooth    &	          &	1-450 T+5600 p T^2-450 p^3 T^3+p^6 T^4
\tabularnewline[0.5pt]\hline				
37	 &	singular  &	5^{-5}&	(1+p T) (1-1246 T+p^3 T^2)
\tabularnewline[0.5pt]\hline				
38	 &	smooth    &	          &	1-505 T+6020 p T^2-505 p^3 T^3+p^6 T^4
\tabularnewline[0.5pt]\hline				
39	 &	smooth    &	          &	1-105 T-1755 p T^2-105 p^3 T^3+p^6 T^4
\tabularnewline[0.5pt]\hline				
40	 &	smooth    &	          &	1-1250 T+17550 p T^2-1250 p^3 T^3+p^6 T^4
\tabularnewline[0.5pt]\hline				
41	 &	smooth    &	          &	1-320 T+13230 p T^2-320 p^3 T^3+p^6 T^4
\tabularnewline[0.5pt]\hline				
42	 &	smooth    &	          &	1+20 T+17920 p T^2+20 p^3 T^3+p^6 T^4
\tabularnewline[0.5pt]\hline				
43	 &	smooth    &	          &	1-460 T+17740 p T^2-460 p^3 T^3+p^6 T^4
\tabularnewline[0.5pt]\hline				
44	 &	smooth    &	          &	1-175 T-13825 p T^2-175 p^3 T^3+p^6 T^4
\tabularnewline[0.5pt]\hline				
45	 &	smooth    &	          &	1-265 T+4835 p T^2-265 p^3 T^3+p^6 T^4
\tabularnewline[0.5pt]\hline				
46	 &	smooth    &	          &	1-300 T+150 p^2 T^2-300 p^3 T^3+p^6 T^4
\tabularnewline[0.5pt]\hline				
47	 &	smooth    &	          &	1-595 T+280 p T^2-595 p^3 T^3+p^6 T^4
\tabularnewline[0.5pt]\hline				
48	 &	smooth    &	          &	1+1340 T+15590 p T^2+1340 p^3 T^3+p^6 T^4
\tabularnewline[0.5pt]\hline				
49	 &	smooth    &	          &	1-685 T+590 p T^2-685 p^3 T^3+p^6 T^4
\tabularnewline[0.5pt]\hline				
50	 &	smooth    &	          &	1+1355 T+14655 p T^2+1355 p^3 T^3+p^6 T^4
\tabularnewline[0.5pt]\hline				
51	 &	smooth    &	          &	1+225 T-3250 p T^2+225 p^3 T^3+p^6 T^4
\tabularnewline[0.5pt]\hline				
52	 &	smooth    &	          &	1+430 T+2880 p T^2+430 p^3 T^3+p^6 T^4
\tabularnewline[0.5pt]\hline				
53	 &	smooth    &	          &	1+1025 T+13900 p T^2+1025 p^3 T^3+p^6 T^4
\tabularnewline[0.5pt]\hline				
54	 &	smooth    &	          &	1+965 T+17340 p T^2+965 p^3 T^3+p^6 T^4
\tabularnewline[0.5pt]\hline				
55	 &	smooth    &	          &	1+575 T-2375 p T^2+575 p^3 T^3+p^6 T^4
\tabularnewline[0.5pt]\hline				
56	 &	smooth    &	          &	1-1540 T+19060 p T^2-1540 p^3 T^3+p^6 T^4
\tabularnewline[0.5pt]\hline				
57	 &	smooth    &	          &	1-1235 T+10840 p T^2-1235 p^3 T^3+p^6 T^4
\tabularnewline[0.5pt]\hline				
58	 &	smooth    &	          &	1-325 T-1100 p T^2-325 p^3 T^3+p^6 T^4
\tabularnewline[0.5pt]\hline				
59	 &	smooth    &	          &	1+675 T-725 p T^2+675 p^3 T^3+p^6 T^4
\tabularnewline[0.5pt]\hline				
60	 &	smooth    &	          &	1-1565 T+21510 p T^2-1565 p^3 T^3+p^6 T^4
\tabularnewline[0.5pt]\hline				
61	 &	smooth    &	          &	1-730 T+2770 p T^2-730 p^3 T^3+p^6 T^4
\tabularnewline[0.5pt]\hline				
62	 &	smooth    &	          &	1-1460 T+15690 p T^2-1460 p^3 T^3+p^6 T^4
\tabularnewline[0.5pt]\hline				
63	 &	smooth    &	          &	1-1320 T+11980 p T^2-1320 p^3 T^3+p^6 T^4
\tabularnewline[0.5pt]\hline				
64	 &	smooth    &	          &	1+790 T+7490 p T^2+790 p^3 T^3+p^6 T^4
\tabularnewline[0.5pt]\hline				
65	 &	smooth    &	          &	1+195 T+9795 p T^2+195 p^3 T^3+p^6 T^4
\tabularnewline[0.5pt]\hline				
66	 &	smooth    &	          &	1+1360 T+15410 p T^2+1360 p^3 T^3+p^6 T^4
\tabularnewline[0.5pt]\hline				
67	 &	smooth    &	          &	1-2050 T+27200 p T^2-2050 p^3 T^3+p^6 T^4
\tabularnewline[0.5pt]\hline				
68	 &	smooth    &	          &	1+2045 T+26120 p T^2+2045 p^3 T^3+p^6 T^4
\tabularnewline[0.5pt]\hline				
69	 &	smooth    &	          &	1+95 T-3855 p T^2+95 p^3 T^3+p^6 T^4
\tabularnewline[0.5pt]\hline				
70	 &	smooth    &	          &	1-170 T-6120 p T^2-170 p^3 T^3+p^6 T^4
\tabularnewline[0.5pt]\hline				
71	 &	smooth    &	          &	1+340 T-7710 p T^2+340 p^3 T^3+p^6 T^4
\tabularnewline[0.5pt]\hline				
72	 &	smooth    &	          &	1+90 T+1290 p T^2+90 p^3 T^3+p^6 T^4
\tabularnewline[0.5pt]\hline				
73	 &	smooth    &	          &	1+1420 T+10070 p T^2+1420 p^3 T^3+p^6 T^4
\tabularnewline[0.5pt]\hline				
74	 &	smooth    &	          &	1-1580 T+16470 p T^2-1580 p^3 T^3+p^6 T^4
\tabularnewline[0.5pt]\hline				
75	 &	smooth    &	          &	1+705 T+3805 p T^2+705 p^3 T^3+p^6 T^4
\tabularnewline[0.5pt]\hline				
76	 &	smooth    &	          &	1+355 T+6330 p T^2+355 p^3 T^3+p^6 T^4
\tabularnewline[0.5pt]\hline				
77	 &	smooth    &	          &	1-845 T+11330 p T^2-845 p^3 T^3+p^6 T^4
\tabularnewline[0.5pt]\hline				
78	 &	smooth    &	          &	1+590 T+4090 p T^2+590 p^3 T^3+p^6 T^4
\tabularnewline[0.5pt]\hline				
79	 &	smooth    &	          &	1-455 T+10320 p T^2-455 p^3 T^3+p^6 T^4
\tabularnewline[0.5pt]\hline				
80	 &	smooth    &	          &	1+410 T+6510 p T^2+410 p^3 T^3+p^6 T^4
\tabularnewline[0.5pt]\hline				
81	 &	smooth    &	          &	1-465 T+8460 p T^2-465 p^3 T^3+p^6 T^4
\tabularnewline[0.5pt]\hline				
82	 &	smooth    &	          &	1-780 T-1330 p T^2-780 p^3 T^3+p^6 T^4
\tabularnewline[0.5pt]\hline				
83	 &	smooth    &	          &	1-300 T+550 p T^2-300 p^3 T^3+p^6 T^4
\tabularnewline[0.5pt]\hline				
84	 &	smooth    &	          &	1+205 T+13180 p T^2+205 p^3 T^3+p^6 T^4
\tabularnewline[0.5pt]\hline				
85	 &	smooth    &	          &	1+1060 T+13510 p T^2+1060 p^3 T^3+p^6 T^4
\tabularnewline[0.5pt]\hline				
86	 &	smooth    &	          &	1-1440 T+20410 p T^2-1440 p^3 T^3+p^6 T^4
\tabularnewline[0.5pt]\hline				
87	 &	smooth    &	          &	1-1140 T+9110 p T^2-1140 p^3 T^3+p^6 T^4
\tabularnewline[0.5pt]\hline				
88	 &	smooth    &	          &	1+1585 T+20285 p T^2+1585 p^3 T^3+p^6 T^4
\tabularnewline[0.5pt]\hline				
89	 &	smooth    &	          &	1-130 T-7730 p T^2-130 p^3 T^3+p^6 T^4
\tabularnewline[0.5pt]\hline				
90	 &	smooth    &	          &	1+345 T-180 p T^2+345 p^3 T^3+p^6 T^4
\tabularnewline[0.5pt]\hline				
91	 &	smooth    &	          &	1+450 T+12050 p T^2+450 p^3 T^3+p^6 T^4
\tabularnewline[0.5pt]\hline				
92	 &	smooth    &	          &	1+235 T+7210 p T^2+235 p^3 T^3+p^6 T^4
\tabularnewline[0.5pt]\hline				
93	 &	smooth    &	          &	1+365 T+8315 p T^2+365 p^3 T^3+p^6 T^4
\tabularnewline[0.5pt]\hline				
94	 &	smooth    &	          &	1-1325 T+18500 p T^2-1325 p^3 T^3+p^6 T^4
\tabularnewline[0.5pt]\hline				
95	 &	smooth    &	          &	1-295 T-6620 p T^2-295 p^3 T^3+p^6 T^4
\tabularnewline[0.5pt]\hline				
96	 &	smooth    &	          &	1+385 T-5015 p T^2+385 p^3 T^3+p^6 T^4
\tabularnewline[0.5pt]\hline				
\tablepostamble				
\newpage
\lhead{\ifthenelse{\isodd{\value{page}}}{\thepage}{\sl The $\z$-function for the mirror of a hypersurface in G(2,5), AESZ\hskip2pt 25}}
\rhead{\ifthenelse{\isodd{\value{page}}}{\sl The $\z$-function for the mirror of a hypersurface in G(2,5), AESZ\hskip2pt 25}{\thepage}}
\subsection{The $\z$-function for the mirror of a hypersurface in G(2,5), AESZ\hskip2pt 25}
\vspace{1.5cm}				
\tablepreamble{5}				
1	 &	smooth    &	          &	1+13 T+16 p T^2+13 p^3 T^3+p^6 T^4
\tabularnewline[0.5pt]\hline				
2	 &	singular   &	        2&	1+14T+p^3 T^2
\tabularnewline[0.5pt]\hline				
3	 &	smooth    &	          &	(1-2 p T+p^3 T^2)(1+12 T+p^3 T^2)
\tabularnewline[0.5pt]\hline				
4	 &	smooth    &	          &	1-3 T+16 p T^2-3 p^3 T^3+p^6 T^4
\tabularnewline[0.5pt]\hline				
\tablepostamble				
\tablepreamble{7}				
1	 &	smooth    &	          &	(1-2 p T+p^3 T^2)(1+24 T+p^3 T^2)
\tabularnewline[0.5pt]\hline				
2	 &	smooth    &	          &	1+25 T+74 p T^2+25 p^3 T^3+p^6 T^4
\tabularnewline[0.5pt]\hline				
3	 &	smooth    &	          &	(1+p^3 T^2)(1+30 T+p^3 T^2)
\tabularnewline[0.5pt]\hline				
4	 &	smooth    &	          &	1-10 T+82 p T^2-10 p^3 T^3+p^6 T^4
\tabularnewline[0.5pt]\hline				
5	 &	smooth    &	          &	1+10 T-30 p T^2+10 p^3 T^3+p^6 T^4
\tabularnewline[0.5pt]\hline				
6	 &	smooth    &	          &	1-15 T+26 p T^2-15 p^3 T^3+p^6 T^4
\tabularnewline[0.5pt]\hline				
\tablepostamble				
\tablepreamble{11}				
1	 &	smooth    &	          &	1+39 T+262 p T^2+39 p^3 T^3+p^6 T^4
\tabularnewline[0.5pt]\hline				
2	 &	singular  &	2&	(1+p T) (1-20 T+p^3 T^2)
\tabularnewline[0.5pt]\hline				
3	 &	smooth    &	          &	(1-4 p T+p^3 T^2)(1+46 T+p^3 T^2)
\tabularnewline[0.5pt]\hline				
4	 &	smooth    &	          &	1-26 T+42 p T^2-26 p^3 T^3+p^6 T^4
\tabularnewline[0.5pt]\hline				
5	 &	smooth    &	          &	1+15 T+166 p T^2+15 p^3 T^3+p^6 T^4
\tabularnewline[0.5pt]\hline				
6	 &	smooth    &	          &	1+10 T-134 p T^2+10 p^3 T^3+p^6 T^4
\tabularnewline[0.5pt]\hline				
7	 &	smooth    &	          &	1+39 T+142 p T^2+39 p^3 T^3+p^6 T^4
\tabularnewline[0.5pt]\hline				
8	 &	smooth    &	          &	1+47 T+78 p T^2+47 p^3 T^3+p^6 T^4
\tabularnewline[0.5pt]\hline				
9	 &	singular  &	9&	(1+p T) (1+20 T+p^3 T^2)
\tabularnewline[0.5pt]\hline				
10	 &	smooth    &	          &	1-26 T+122 p T^2-26 p^3 T^3+p^6 T^4
\tabularnewline[0.5pt]\hline				
\tablepostamble				
\tablepreamble{13}				
1	 &	smooth    &	          &	(1+2 p T+p^3 T^2)(1-86 T+p^3 T^2)
\tabularnewline[0.5pt]\hline				
2	 &	smooth    &	          &	(1+p^3 T^2)(1-50 T+p^3 T^2)
\tabularnewline[0.5pt]\hline				
3	 &	smooth    &	          &	1-5 p T+332 p T^2-5 p^4 T^3+p^6 T^4
\tabularnewline[0.5pt]\hline				
4	 &	smooth    &	          &	(1+p^3 T^2)(1+90 T+p^3 T^2)
\tabularnewline[0.5pt]\hline				
5	 &	smooth    &	          &	(1+p^3 T^2)(1+70 T+p^3 T^2)
\tabularnewline[0.5pt]\hline				
6	 &	smooth    &	          &	1-30 T+90 p T^2-30 p^3 T^3+p^6 T^4
\tabularnewline[0.5pt]\hline				
7	 &	smooth    &	          &	1+20 T+214 p T^2+20 p^3 T^3+p^6 T^4
\tabularnewline[0.5pt]\hline				
8	 &	smooth    &	          &	1+35 T+120 p T^2+35 p^3 T^3+p^6 T^4
\tabularnewline[0.5pt]\hline				
9	 &	smooth    &	          &	1+50 T+98 p T^2+50 p^3 T^3+p^6 T^4
\tabularnewline[0.5pt]\hline				
10	 &	smooth    &	          &	1+60 T+246 p T^2+60 p^3 T^3+p^6 T^4
\tabularnewline[0.5pt]\hline				
11	 &	smooth    &	          &	1+35 T-40 p T^2+35 p^3 T^3+p^6 T^4
\tabularnewline[0.5pt]\hline				
12	 &	smooth    &	          &	1+15 T+12 p T^2+15 p^3 T^3+p^6 T^4
\tabularnewline[0.5pt]\hline				
\tablepostamble				
\tablepreamble{17}				
1	 &	smooth    &	          &	1-60 T+22 p^2 T^2-60 p^3 T^3+p^6 T^4
\tabularnewline[0.5pt]\hline				
2	 &	smooth    &	          &	1+115 T+744 p T^2+115 p^3 T^3+p^6 T^4
\tabularnewline[0.5pt]\hline				
3	 &	smooth    &	          &	(1+6 p T+p^3 T^2)(1-82 T+p^3 T^2)
\tabularnewline[0.5pt]\hline				
4	 &	smooth    &	          &	1-10 T+50 p T^2-10 p^3 T^3+p^6 T^4
\tabularnewline[0.5pt]\hline				
5	 &	smooth    &	          &	1-15 T+368 p T^2-15 p^3 T^3+p^6 T^4
\tabularnewline[0.5pt]\hline				
6	 &	smooth    &	          &	1-25 T+60 p T^2-25 p^3 T^3+p^6 T^4
\tabularnewline[0.5pt]\hline				
7	 &	smooth    &	          &	(1+2 p T+p^3 T^2)(1+106 T+p^3 T^2)
\tabularnewline[0.5pt]\hline				
8	 &	smooth    &	          &	1+35 T-56 p T^2+35 p^3 T^3+p^6 T^4
\tabularnewline[0.5pt]\hline				
9	 &	smooth    &	          &	1-5 p T+632 p T^2-5 p^4 T^3+p^6 T^4
\tabularnewline[0.5pt]\hline				
10	 &	smooth    &	          &	1-15 T+208 p T^2-15 p^3 T^3+p^6 T^4
\tabularnewline[0.5pt]\hline				
11	 &	smooth    &	          &	1+20 T-394 p T^2+20 p^3 T^3+p^6 T^4
\tabularnewline[0.5pt]\hline				
12	 &	smooth    &	          &	1-20 T-330 p T^2-20 p^3 T^3+p^6 T^4
\tabularnewline[0.5pt]\hline				
13	 &	smooth    &	          &	1+446 p T^2+p^6 T^4
\tabularnewline[0.5pt]\hline				
14	 &	smooth    &	          &	1+55 T+540 p T^2+55 p^3 T^3+p^6 T^4
\tabularnewline[0.5pt]\hline				
15	 &	smooth    &	          &	1+75 T+632 p T^2+75 p^3 T^3+p^6 T^4
\tabularnewline[0.5pt]\hline				
16	 &	smooth    &	          &	1+60 T+134 p T^2+60 p^3 T^3+p^6 T^4
\tabularnewline[0.5pt]\hline				
\tablepostamble				
\tablepreamble{19}				
1	 &	smooth    &	          &	1+145 T+838 p T^2+145 p^3 T^3+p^6 T^4
\tabularnewline[0.5pt]\hline				
2	 &	smooth    &	          &	1+33 T-6 p^2 T^2+33 p^3 T^3+p^6 T^4
\tabularnewline[0.5pt]\hline				
3	 &	singular  &	3&	(1-p T) (1+84 T+p^3 T^2)
\tabularnewline[0.5pt]\hline				
4	 &	smooth    &	          &	1-62 p T^2+p^6 T^4
\tabularnewline[0.5pt]\hline				
5	 &	smooth    &	          &	1+113 T+710 p T^2+113 p^3 T^3+p^6 T^4
\tabularnewline[0.5pt]\hline				
6	 &	smooth    &	          &	1+56 T+322 p T^2+56 p^3 T^3+p^6 T^4
\tabularnewline[0.5pt]\hline				
7	 &	singular  &	7&	(1-p T) (1-84 T+p^3 T^2)
\tabularnewline[0.5pt]\hline				
8	 &	smooth    &	          &	1-31 T+654 p T^2-31 p^3 T^3+p^6 T^4
\tabularnewline[0.5pt]\hline				
9	 &	smooth    &	          &	1-17 T+286 p T^2-17 p^3 T^3+p^6 T^4
\tabularnewline[0.5pt]\hline				
10	 &	smooth    &	          &	(1+4 p T+p^3 T^2)(1-124 T+p^3 T^2)
\tabularnewline[0.5pt]\hline				
11	 &	smooth    &	          &	1+4 T+562 p T^2+4 p^3 T^3+p^6 T^4
\tabularnewline[0.5pt]\hline				
12	 &	smooth    &	          &	1+79 T+662 p T^2+79 p^3 T^3+p^6 T^4
\tabularnewline[0.5pt]\hline				
13	 &	smooth    &	          &	1-24 T-238 p T^2-24 p^3 T^3+p^6 T^4
\tabularnewline[0.5pt]\hline				
14	 &	smooth    &	          &	(1+p^3 T^2)(1+124 T+p^3 T^2)
\tabularnewline[0.5pt]\hline				
15	 &	smooth    &	          &	1+15 T+438 p T^2+15 p^3 T^3+p^6 T^4
\tabularnewline[0.5pt]\hline				
16	 &	smooth    &	          &	1-T+382 p T^2-p^3 T^3+p^6 T^4
\tabularnewline[0.5pt]\hline				
17	 &	smooth    &	          &	1-16 T+34 p T^2-16 p^3 T^3+p^6 T^4
\tabularnewline[0.5pt]\hline				
18	 &	smooth    &	          &	1-32 T-350 p T^2-32 p^3 T^3+p^6 T^4
\tabularnewline[0.5pt]\hline				
\tablepostamble				
\tablepreamble{23}				
1	 &	smooth    &	          &	1+35 T+842 p T^2+35 p^3 T^3+p^6 T^4
\tabularnewline[0.5pt]\hline				
2	 &	smooth    &	          &	1-40 T+130 p T^2-40 p^3 T^3+p^6 T^4
\tabularnewline[0.5pt]\hline				
3	 &	smooth    &	          &	1-30 T+626 p T^2-30 p^3 T^3+p^6 T^4
\tabularnewline[0.5pt]\hline				
4	 &	smooth    &	          &	1+60 T+258 p T^2+60 p^3 T^3+p^6 T^4
\tabularnewline[0.5pt]\hline				
5	 &	smooth    &	          &	1+170 T+1074 p T^2+170 p^3 T^3+p^6 T^4
\tabularnewline[0.5pt]\hline				
6	 &	smooth    &	          &	1-100 T+738 p T^2-100 p^3 T^3+p^6 T^4
\tabularnewline[0.5pt]\hline				
7	 &	smooth    &	          &	1+30 T+290 p T^2+30 p^3 T^3+p^6 T^4
\tabularnewline[0.5pt]\hline				
8	 &	smooth    &	          &	1-50 T-334 p T^2-50 p^3 T^3+p^6 T^4
\tabularnewline[0.5pt]\hline				
9	 &	smooth    &	          &	1+100 T+418 p T^2+100 p^3 T^3+p^6 T^4
\tabularnewline[0.5pt]\hline				
10	 &	smooth    &	          &	1-35 T-26 p^2 T^2-35 p^3 T^3+p^6 T^4
\tabularnewline[0.5pt]\hline				
11	 &	smooth    &	          &	1+50 T+578 p T^2+50 p^3 T^3+p^6 T^4
\tabularnewline[0.5pt]\hline				
12	 &	smooth    &	          &	1+30 T+370 p T^2+30 p^3 T^3+p^6 T^4
\tabularnewline[0.5pt]\hline				
13	 &	smooth    &	          &	(1-5 p T+p^3 T^2)(1+200 T+p^3 T^2)
\tabularnewline[0.5pt]\hline				
14	 &	smooth    &	          &	1-180 T+722 p T^2-180 p^3 T^3+p^6 T^4
\tabularnewline[0.5pt]\hline				
15	 &	smooth    &	          &	1-85 T+218 p T^2-85 p^3 T^3+p^6 T^4
\tabularnewline[0.5pt]\hline				
16	 &	smooth    &	          &	1-70 T+82 p T^2-70 p^3 T^3+p^6 T^4
\tabularnewline[0.5pt]\hline				
17	 &	smooth    &	          &	1+100 T+562 p T^2+100 p^3 T^3+p^6 T^4
\tabularnewline[0.5pt]\hline				
18	 &	smooth    &	          &	1+245 T+1530 p T^2+245 p^3 T^3+p^6 T^4
\tabularnewline[0.5pt]\hline				
19	 &	smooth    &	          &	1+240 T+1570 p T^2+240 p^3 T^3+p^6 T^4
\tabularnewline[0.5pt]\hline				
20	 &	smooth    &	          &	1+205 T+1450 p T^2+205 p^3 T^3+p^6 T^4
\tabularnewline[0.5pt]\hline				
21	 &	smooth    &	          &	1-180 T+802 p T^2-180 p^3 T^3+p^6 T^4
\tabularnewline[0.5pt]\hline				
22	 &	smooth    &	          &	1-50 T+274 p T^2-50 p^3 T^3+p^6 T^4
\tabularnewline[0.5pt]\hline				
\tablepostamble				
\tablepreamble{29}				
1	 &	smooth    &	          &	1+49 T-832 p T^2+49 p^3 T^3+p^6 T^4
\tabularnewline[0.5pt]\hline				
2	 &	smooth    &	          &	1-195 T+1748 p T^2-195 p^3 T^3+p^6 T^4
\tabularnewline[0.5pt]\hline				
3	 &	smooth    &	          &	1-35 T-492 p T^2-35 p^3 T^3+p^6 T^4
\tabularnewline[0.5pt]\hline				
4	 &	smooth    &	          &	1-214 T+914 p T^2-214 p^3 T^3+p^6 T^4
\tabularnewline[0.5pt]\hline				
5	 &	singular  &	5&	(1+p T) (1-6 T+p^3 T^2)
\tabularnewline[0.5pt]\hline				
6	 &	smooth    &	          &	1+209 T+1248 p T^2+209 p^3 T^3+p^6 T^4
\tabularnewline[0.5pt]\hline				
7	 &	singular  &	7&	(1+p T) (1-6 T+p^3 T^2)
\tabularnewline[0.5pt]\hline				
8	 &	smooth    &	          &	(1+6 p T+p^3 T^2)(1-214 T+p^3 T^2)
\tabularnewline[0.5pt]\hline				
9	 &	smooth    &	          &	(1+8 p T+p^3 T^2)(1-166 T+p^3 T^2)
\tabularnewline[0.5pt]\hline				
10	 &	smooth    &	          &	1+71 T+832 p T^2+71 p^3 T^3+p^6 T^4
\tabularnewline[0.5pt]\hline				
11	 &	smooth    &	          &	1+26 T+394 p T^2+26 p^3 T^3+p^6 T^4
\tabularnewline[0.5pt]\hline				
12	 &	smooth    &	          &	1-66 T+898 p T^2-66 p^3 T^3+p^6 T^4
\tabularnewline[0.5pt]\hline				
13	 &	smooth    &	          &	1+94 T+1178 p T^2+94 p^3 T^3+p^6 T^4
\tabularnewline[0.5pt]\hline				
14	 &	smooth    &	          &	1-306 T+2402 p T^2-306 p^3 T^3+p^6 T^4
\tabularnewline[0.5pt]\hline				
15	 &	smooth    &	          &	1+286 T+1538 p T^2+286 p^3 T^3+p^6 T^4
\tabularnewline[0.5pt]\hline				
16	 &	smooth    &	          &	1+266 T+1394 p T^2+266 p^3 T^3+p^6 T^4
\tabularnewline[0.5pt]\hline				
17	 &	smooth    &	          &	1+126 T+322 p T^2+126 p^3 T^3+p^6 T^4
\tabularnewline[0.5pt]\hline				
18	 &	smooth    &	          &	1-41 T-448 p T^2-41 p^3 T^3+p^6 T^4
\tabularnewline[0.5pt]\hline				
19	 &	smooth    &	          &	1+114 T+714 p T^2+114 p^3 T^3+p^6 T^4
\tabularnewline[0.5pt]\hline				
20	 &	smooth    &	          &	1-94 T-286 p T^2-94 p^3 T^3+p^6 T^4
\tabularnewline[0.5pt]\hline				
21	 &	smooth    &	          &	1+260 T+2006 p T^2+260 p^3 T^3+p^6 T^4
\tabularnewline[0.5pt]\hline				
22	 &	smooth    &	          &	1-133 T+372 p T^2-133 p^3 T^3+p^6 T^4
\tabularnewline[0.5pt]\hline				
23	 &	smooth    &	          &	1+16 T+1214 p T^2+16 p^3 T^3+p^6 T^4
\tabularnewline[0.5pt]\hline				
24	 &	smooth    &	          &	1+267 T+1652 p T^2+267 p^3 T^3+p^6 T^4
\tabularnewline[0.5pt]\hline				
25	 &	smooth    &	          &	1-106 T+778 p T^2-106 p^3 T^3+p^6 T^4
\tabularnewline[0.5pt]\hline				
26	 &	smooth    &	          &	1+12 T+1366 p T^2+12 p^3 T^3+p^6 T^4
\tabularnewline[0.5pt]\hline				
27	 &	smooth    &	          &	1+188 T+566 p T^2+188 p^3 T^3+p^6 T^4
\tabularnewline[0.5pt]\hline				
28	 &	smooth    &	          &	1-24 T-146 p T^2-24 p^3 T^3+p^6 T^4
\tabularnewline[0.5pt]\hline				
\tablepostamble				
\tablepreamble{31}				
1	 &	smooth    &	          &	1+156 T+962 p T^2+156 p^3 T^3+p^6 T^4
\tabularnewline[0.5pt]\hline				
2	 &	smooth    &	          &	1-19 T-350 p T^2-19 p^3 T^3+p^6 T^4
\tabularnewline[0.5pt]\hline				
3	 &	smooth    &	          &	1+245 T+2018 p T^2+245 p^3 T^3+p^6 T^4
\tabularnewline[0.5pt]\hline				
4	 &	smooth    &	          &	1-22 T+802 p T^2-22 p^3 T^3+p^6 T^4
\tabularnewline[0.5pt]\hline				
5	 &	smooth    &	          &	1+54 T-510 p T^2+54 p^3 T^3+p^6 T^4
\tabularnewline[0.5pt]\hline				
6	 &	smooth    &	          &	1+21 T+1282 p T^2+21 p^3 T^3+p^6 T^4
\tabularnewline[0.5pt]\hline				
7	 &	smooth    &	          &	1-16 T+322 p T^2-16 p^3 T^3+p^6 T^4
\tabularnewline[0.5pt]\hline				
8	 &	smooth    &	          &	1+298 T+2242 p T^2+298 p^3 T^3+p^6 T^4
\tabularnewline[0.5pt]\hline				
9	 &	smooth    &	          &	1+221 T+1506 p T^2+221 p^3 T^3+p^6 T^4
\tabularnewline[0.5pt]\hline				
10	 &	smooth    &	          &	1+157 T+802 p T^2+157 p^3 T^3+p^6 T^4
\tabularnewline[0.5pt]\hline				
11	 &	smooth    &	          &	(1+p^3 T^2)(1-185 T+p^3 T^2)
\tabularnewline[0.5pt]\hline				
12	 &	smooth    &	          &	1+357 T+2242 p T^2+357 p^3 T^3+p^6 T^4
\tabularnewline[0.5pt]\hline				
13	 &	smooth    &	          &	1+86 T-510 p T^2+86 p^3 T^3+p^6 T^4
\tabularnewline[0.5pt]\hline				
14	 &	smooth    &	          &	1-122 T+834 p T^2-122 p^3 T^3+p^6 T^4
\tabularnewline[0.5pt]\hline				
15	 &	smooth    &	          &	1+82 T-78 p T^2+82 p^3 T^3+p^6 T^4
\tabularnewline[0.5pt]\hline				
16	 &	smooth    &	          &	1+127 T+1474 p T^2+127 p^3 T^3+p^6 T^4
\tabularnewline[0.5pt]\hline				
17	 &	smooth    &	          &	1+48 T+2 p T^2+48 p^3 T^3+p^6 T^4
\tabularnewline[0.5pt]\hline				
18	 &	smooth    &	          &	1+112 T+322 p T^2+112 p^3 T^3+p^6 T^4
\tabularnewline[0.5pt]\hline				
19	 &	singular  &	19&	(1+p T) (1-224 T+p^3 T^2)
\tabularnewline[0.5pt]\hline				
20	 &	smooth    &	          &	1-211 T+962 p T^2-211 p^3 T^3+p^6 T^4
\tabularnewline[0.5pt]\hline				
21	 &	singular  &	21&	(1+p T) (1+224 T+p^3 T^2)
\tabularnewline[0.5pt]\hline				
22	 &	smooth    &	          &	1-340 T+2578 p T^2-340 p^3 T^3+p^6 T^4
\tabularnewline[0.5pt]\hline				
23	 &	smooth    &	          &	1-361 T+2530 p T^2-361 p^3 T^3+p^6 T^4
\tabularnewline[0.5pt]\hline				
24	 &	smooth    &	          &	1+22 T-206 p T^2+22 p^3 T^3+p^6 T^4
\tabularnewline[0.5pt]\hline				
25	 &	smooth    &	          &	(1+p^3 T^2)(1-65 T+p^3 T^2)
\tabularnewline[0.5pt]\hline				
26	 &	smooth    &	          &	1+176 T+962 p T^2+176 p^3 T^3+p^6 T^4
\tabularnewline[0.5pt]\hline				
27	 &	smooth    &	          &	1+4 T+882 p T^2+4 p^3 T^3+p^6 T^4
\tabularnewline[0.5pt]\hline				
28	 &	smooth    &	          &	1+4 T-894 p T^2+4 p^3 T^3+p^6 T^4
\tabularnewline[0.5pt]\hline				
29	 &	smooth    &	          &	1+213 T+674 p T^2+213 p^3 T^3+p^6 T^4
\tabularnewline[0.5pt]\hline				
30	 &	smooth    &	          &	1-142 T+322 p T^2-142 p^3 T^3+p^6 T^4
\tabularnewline[0.5pt]\hline				
\tablepostamble				
\tablepreamble{37}				
1	 &	smooth    &	          &	1-50 T-86 p T^2-50 p^3 T^3+p^6 T^4
\tabularnewline[0.5pt]\hline				
2	 &	smooth    &	          &	1+425 T+3332 p T^2+425 p^3 T^3+p^6 T^4
\tabularnewline[0.5pt]\hline				
3	 &	smooth    &	          &	1+80 T-1250 p T^2+80 p^3 T^3+p^6 T^4
\tabularnewline[0.5pt]\hline				
4	 &	smooth    &	          &	1+35 T+1676 p T^2+35 p^3 T^3+p^6 T^4
\tabularnewline[0.5pt]\hline				
5	 &	smooth    &	          &	1+10 T+490 p T^2+10 p^3 T^3+p^6 T^4
\tabularnewline[0.5pt]\hline				
6	 &	smooth    &	          &	1+55 T+1400 p T^2+55 p^3 T^3+p^6 T^4
\tabularnewline[0.5pt]\hline				
7	 &	smooth    &	          &	1+195 T+812 p T^2+195 p^3 T^3+p^6 T^4
\tabularnewline[0.5pt]\hline				
8	 &	smooth    &	          &	1+5 p T+1604 p T^2+5 p^4 T^3+p^6 T^4
\tabularnewline[0.5pt]\hline				
9	 &	smooth    &	          &	1+90 T+2018 p T^2+90 p^3 T^3+p^6 T^4
\tabularnewline[0.5pt]\hline				
10	 &	smooth    &	          &	1+125 T+1872 p T^2+125 p^3 T^3+p^6 T^4
\tabularnewline[0.5pt]\hline				
11	 &	smooth    &	          &	1-150 T-702 p T^2-150 p^3 T^3+p^6 T^4
\tabularnewline[0.5pt]\hline				
12	 &	smooth    &	          &	1-60 T+358 p T^2-60 p^3 T^3+p^6 T^4
\tabularnewline[0.5pt]\hline				
13	 &	smooth    &	          &	(1-10 p T+p^3 T^2)(1+90 T+p^3 T^2)
\tabularnewline[0.5pt]\hline				
14	 &	smooth    &	          &	1+360 T+3134 p T^2+360 p^3 T^3+p^6 T^4
\tabularnewline[0.5pt]\hline				
15	 &	smooth    &	          &	1+40 T+894 p T^2+40 p^3 T^3+p^6 T^4
\tabularnewline[0.5pt]\hline				
16	 &	smooth    &	          &	1-220 T+46 p^2 T^2-220 p^3 T^3+p^6 T^4
\tabularnewline[0.5pt]\hline				
17	 &	smooth    &	          &	1-55 T+164 p T^2-55 p^3 T^3+p^6 T^4
\tabularnewline[0.5pt]\hline				
18	 &	smooth    &	          &	1+180 T+2534 p T^2+180 p^3 T^3+p^6 T^4
\tabularnewline[0.5pt]\hline				
19	 &	smooth    &	          &	1+280 T+1038 p T^2+280 p^3 T^3+p^6 T^4
\tabularnewline[0.5pt]\hline				
20	 &	smooth    &	          &	1-310 T+2842 p T^2-310 p^3 T^3+p^6 T^4
\tabularnewline[0.5pt]\hline				
21	 &	smooth    &	          &	1+45 T+1232 p T^2+45 p^3 T^3+p^6 T^4
\tabularnewline[0.5pt]\hline				
22	 &	smooth    &	          &	1-160 T+1246 p T^2-160 p^3 T^3+p^6 T^4
\tabularnewline[0.5pt]\hline				
23	 &	smooth    &	          &	1-160 T+2046 p T^2-160 p^3 T^3+p^6 T^4
\tabularnewline[0.5pt]\hline				
24	 &	smooth    &	          &	1-20 T-506 p T^2-20 p^3 T^3+p^6 T^4
\tabularnewline[0.5pt]\hline				
25	 &	smooth    &	          &	1+230 T+2234 p T^2+230 p^3 T^3+p^6 T^4
\tabularnewline[0.5pt]\hline				
26	 &	smooth    &	          &	1+210 T+2514 p T^2+210 p^3 T^3+p^6 T^4
\tabularnewline[0.5pt]\hline				
27	 &	smooth    &	          &	1-140 T+2182 p T^2-140 p^3 T^3+p^6 T^4
\tabularnewline[0.5pt]\hline				
28	 &	smooth    &	          &	1-190 T+914 p T^2-190 p^3 T^3+p^6 T^4
\tabularnewline[0.5pt]\hline				
29	 &	smooth    &	          &	1-30 T+762 p T^2-30 p^3 T^3+p^6 T^4
\tabularnewline[0.5pt]\hline				
30	 &	smooth    &	          &	1+35 T+972 p T^2+35 p^3 T^3+p^6 T^4
\tabularnewline[0.5pt]\hline				
31	 &	smooth    &	          &	1-135 T-348 p T^2-135 p^3 T^3+p^6 T^4
\tabularnewline[0.5pt]\hline				
32	 &	smooth    &	          &	1+200 T+1262 p T^2+200 p^3 T^3+p^6 T^4
\tabularnewline[0.5pt]\hline				
33	 &	smooth    &	          &	1-40 T+1710 p T^2-40 p^3 T^3+p^6 T^4
\tabularnewline[0.5pt]\hline				
34	 &	smooth    &	          &	1-45 T+396 p T^2-45 p^3 T^3+p^6 T^4
\tabularnewline[0.5pt]\hline				
35	 &	smooth    &	          &	1+135 T+760 p T^2+135 p^3 T^3+p^6 T^4
\tabularnewline[0.5pt]\hline				
36	 &	smooth    &	          &	(1+10 p T+p^3 T^2)(1+130 T+p^3 T^2)
\tabularnewline[0.5pt]\hline				
\tablepostamble				
\tablepreamble{41}				
1	 &	smooth    &	          &	1+284 T+1030 p T^2+284 p^3 T^3+p^6 T^4
\tabularnewline[0.5pt]\hline				
2	 &	smooth    &	          &	1+30 T-334 p T^2+30 p^3 T^3+p^6 T^4
\tabularnewline[0.5pt]\hline				
3	 &	smooth    &	          &	1-232 T-178 p T^2-232 p^3 T^3+p^6 T^4
\tabularnewline[0.5pt]\hline				
4	 &	smooth    &	          &	1+76 T-730 p T^2+76 p^3 T^3+p^6 T^4
\tabularnewline[0.5pt]\hline				
5	 &	smooth    &	          &	1+242 T+1514 p T^2+242 p^3 T^3+p^6 T^4
\tabularnewline[0.5pt]\hline				
6	 &	smooth    &	          &	1+206 T+1242 p T^2+206 p^3 T^3+p^6 T^4
\tabularnewline[0.5pt]\hline				
7	 &	smooth    &	          &	1+142 T+2538 p T^2+142 p^3 T^3+p^6 T^4
\tabularnewline[0.5pt]\hline				
8	 &	smooth    &	          &	1+534 T+4482 p T^2+534 p^3 T^3+p^6 T^4
\tabularnewline[0.5pt]\hline				
9	 &	smooth    &	          &	1-182 T+474 p T^2-182 p^3 T^3+p^6 T^4
\tabularnewline[0.5pt]\hline				
10	 &	smooth    &	          &	1-60 T+3062 p T^2-60 p^3 T^3+p^6 T^4
\tabularnewline[0.5pt]\hline				
11	 &	smooth    &	          &	1+157 T+3328 p T^2+157 p^3 T^3+p^6 T^4
\tabularnewline[0.5pt]\hline				
12	 &	smooth    &	          &	1-166 T+1890 p T^2-166 p^3 T^3+p^6 T^4
\tabularnewline[0.5pt]\hline				
13	 &	singular  &	13&	(1-p T) (1+266 T+p^3 T^2)
\tabularnewline[0.5pt]\hline				
14	 &	smooth    &	          &	1-46 T+50 p^2 T^2-46 p^3 T^3+p^6 T^4
\tabularnewline[0.5pt]\hline				
15	 &	smooth    &	          &	1-472 T+3502 p T^2-472 p^3 T^3+p^6 T^4
\tabularnewline[0.5pt]\hline				
16	 &	smooth    &	          &	1+7 p T+3208 p T^2+7 p^4 T^3+p^6 T^4
\tabularnewline[0.5pt]\hline				
17	 &	smooth    &	          &	1-116 T+1750 p T^2-116 p^3 T^3+p^6 T^4
\tabularnewline[0.5pt]\hline				
18	 &	smooth    &	          &	1-342 T+2234 p T^2-342 p^3 T^3+p^6 T^4
\tabularnewline[0.5pt]\hline				
19	 &	smooth    &	          &	1+237 T+608 p T^2+237 p^3 T^3+p^6 T^4
\tabularnewline[0.5pt]\hline				
20	 &	smooth    &	          &	1+401 T+3092 p T^2+401 p^3 T^3+p^6 T^4
\tabularnewline[0.5pt]\hline				
21	 &	smooth    &	          &	1+126 T+2962 p T^2+126 p^3 T^3+p^6 T^4
\tabularnewline[0.5pt]\hline				
22	 &	smooth    &	          &	1+304 T+2110 p T^2+304 p^3 T^3+p^6 T^4
\tabularnewline[0.5pt]\hline				
23	 &	smooth    &	          &	1-78 T-1366 p T^2-78 p^3 T^3+p^6 T^4
\tabularnewline[0.5pt]\hline				
24	 &	smooth    &	          &	1+102 T-502 p T^2+102 p^3 T^3+p^6 T^4
\tabularnewline[0.5pt]\hline				
25	 &	smooth    &	          &	1+420 T+3062 p T^2+420 p^3 T^3+p^6 T^4
\tabularnewline[0.5pt]\hline				
26	 &	smooth    &	          &	1-59 T-1168 p T^2-59 p^3 T^3+p^6 T^4
\tabularnewline[0.5pt]\hline				
27	 &	smooth    &	          &	(1+10 p T+p^3 T^2)(1+225 T+p^3 T^2)
\tabularnewline[0.5pt]\hline				
28	 &	smooth    &	          &	1+326 T+1562 p T^2+326 p^3 T^3+p^6 T^4
\tabularnewline[0.5pt]\hline				
29	 &	smooth    &	          &	1-165 T+2412 p T^2-165 p^3 T^3+p^6 T^4
\tabularnewline[0.5pt]\hline				
30	 &	smooth    &	          &	1-256 T+190 p T^2-256 p^3 T^3+p^6 T^4
\tabularnewline[0.5pt]\hline				
31	 &	smooth    &	          &	1+63 T+1608 p T^2+63 p^3 T^3+p^6 T^4
\tabularnewline[0.5pt]\hline				
32	 &	smooth    &	          &	1-359 T+2116 p T^2-359 p^3 T^3+p^6 T^4
\tabularnewline[0.5pt]\hline				
33	 &	smooth    &	          &	1+129 T+1332 p T^2+129 p^3 T^3+p^6 T^4
\tabularnewline[0.5pt]\hline				
34	 &	smooth    &	          &	1+4 T+70 p^2 T^2+4 p^3 T^3+p^6 T^4
\tabularnewline[0.5pt]\hline				
35	 &	singular  &	35&	(1-p T) (1+266 T+p^3 T^2)
\tabularnewline[0.5pt]\hline				
36	 &	smooth    &	          &	1-151 T-604 p T^2-151 p^3 T^3+p^6 T^4
\tabularnewline[0.5pt]\hline				
37	 &	smooth    &	          &	1-120 T+2638 p T^2-120 p^3 T^3+p^6 T^4
\tabularnewline[0.5pt]\hline				
38	 &	smooth    &	          &	1-219 T+592 p T^2-219 p^3 T^3+p^6 T^4
\tabularnewline[0.5pt]\hline				
39	 &	smooth    &	          &	1-210 T+1298 p T^2-210 p^3 T^3+p^6 T^4
\tabularnewline[0.5pt]\hline				
40	 &	smooth    &	          &	1-240 T+2558 p T^2-240 p^3 T^3+p^6 T^4
\tabularnewline[0.5pt]\hline				
\tablepostamble				
\tablepreamble{43}				
1	 &	smooth    &	          &	1+70 T+1706 p T^2+70 p^3 T^3+p^6 T^4
\tabularnewline[0.5pt]\hline				
2	 &	smooth    &	          &	1-10 T-1750 p T^2-10 p^3 T^3+p^6 T^4
\tabularnewline[0.5pt]\hline				
3	 &	smooth    &	          &	1-240 T+3650 p T^2-240 p^3 T^3+p^6 T^4
\tabularnewline[0.5pt]\hline				
4	 &	smooth    &	          &	1+80 T+898 p T^2+80 p^3 T^3+p^6 T^4
\tabularnewline[0.5pt]\hline				
5	 &	smooth    &	          &	1-110 T-1878 p T^2-110 p^3 T^3+p^6 T^4
\tabularnewline[0.5pt]\hline				
6	 &	smooth    &	          &	1-355 T+1454 p T^2-355 p^3 T^3+p^6 T^4
\tabularnewline[0.5pt]\hline				
7	 &	smooth    &	          &	1-125 T-98 p T^2-125 p^3 T^3+p^6 T^4
\tabularnewline[0.5pt]\hline				
8	 &	smooth    &	          &	1-150 T+3210 p T^2-150 p^3 T^3+p^6 T^4
\tabularnewline[0.5pt]\hline				
9	 &	smooth    &	          &	1-75 T+2286 p T^2-75 p^3 T^3+p^6 T^4
\tabularnewline[0.5pt]\hline				
10	 &	smooth    &	          &	1+100 T+2898 p T^2+100 p^3 T^3+p^6 T^4
\tabularnewline[0.5pt]\hline				
11	 &	smooth    &	          &	(1+6 p T+p^3 T^2)(1-268 T+p^3 T^2)
\tabularnewline[0.5pt]\hline				
12	 &	smooth    &	          &	1-235 T+2374 p T^2-235 p^3 T^3+p^6 T^4
\tabularnewline[0.5pt]\hline				
13	 &	smooth    &	          &	1+250 T-102 p T^2+250 p^3 T^3+p^6 T^4
\tabularnewline[0.5pt]\hline				
14	 &	smooth    &	          &	1+75 T-522 p T^2+75 p^3 T^3+p^6 T^4
\tabularnewline[0.5pt]\hline				
15	 &	smooth    &	          &	1+160 T+1922 p T^2+160 p^3 T^3+p^6 T^4
\tabularnewline[0.5pt]\hline				
16	 &	smooth    &	          &	1-160 T+1122 p T^2-160 p^3 T^3+p^6 T^4
\tabularnewline[0.5pt]\hline				
17	 &	smooth    &	          &	1+205 T+10 p^2 T^2+205 p^3 T^3+p^6 T^4
\tabularnewline[0.5pt]\hline				
18	 &	smooth    &	          &	1+125 T+2886 p T^2+125 p^3 T^3+p^6 T^4
\tabularnewline[0.5pt]\hline				
19	 &	smooth    &	          &	1+205 T+710 p T^2+205 p^3 T^3+p^6 T^4
\tabularnewline[0.5pt]\hline				
20	 &	smooth    &	          &	1-325 T+1662 p T^2-325 p^3 T^3+p^6 T^4
\tabularnewline[0.5pt]\hline				
21	 &	smooth    &	          &	1+360 T+3042 p T^2+360 p^3 T^3+p^6 T^4
\tabularnewline[0.5pt]\hline				
22	 &	smooth    &	          &	1+70 T+2154 p T^2+70 p^3 T^3+p^6 T^4
\tabularnewline[0.5pt]\hline				
23	 &	smooth    &	          &	1-435 T+3694 p T^2-435 p^3 T^3+p^6 T^4
\tabularnewline[0.5pt]\hline				
24	 &	smooth    &	          &	1+565 T+4398 p T^2+565 p^3 T^3+p^6 T^4
\tabularnewline[0.5pt]\hline				
25	 &	smooth    &	          &	1+200 T-30 p T^2+200 p^3 T^3+p^6 T^4
\tabularnewline[0.5pt]\hline				
26	 &	smooth    &	          &	1+165 T+2822 p T^2+165 p^3 T^3+p^6 T^4
\tabularnewline[0.5pt]\hline				
27	 &	smooth    &	          &	1+40 T-318 p T^2+40 p^3 T^3+p^6 T^4
\tabularnewline[0.5pt]\hline				
28	 &	smooth    &	          &	1+340 T+1682 p T^2+340 p^3 T^3+p^6 T^4
\tabularnewline[0.5pt]\hline				
29	 &	smooth    &	          &	1-275 T+1126 p T^2-275 p^3 T^3+p^6 T^4
\tabularnewline[0.5pt]\hline				
30	 &	smooth    &	          &	1-20 T+1986 p T^2-20 p^3 T^3+p^6 T^4
\tabularnewline[0.5pt]\hline				
31	 &	smooth    &	          &	1-265 T+3334 p T^2-265 p^3 T^3+p^6 T^4
\tabularnewline[0.5pt]\hline				
32	 &	smooth    &	          &	1-25 T+2510 p T^2-25 p^3 T^3+p^6 T^4
\tabularnewline[0.5pt]\hline				
33	 &	smooth    &	          &	1+440 T+4578 p T^2+440 p^3 T^3+p^6 T^4
\tabularnewline[0.5pt]\hline				
34	 &	smooth    &	          &	1+615 T+4846 p T^2+615 p^3 T^3+p^6 T^4
\tabularnewline[0.5pt]\hline				
35	 &	smooth    &	          &	1-350 T+3450 p T^2-350 p^3 T^3+p^6 T^4
\tabularnewline[0.5pt]\hline				
36	 &	smooth    &	          &	1+360 T+3170 p T^2+360 p^3 T^3+p^6 T^4
\tabularnewline[0.5pt]\hline				
37	 &	smooth    &	          &	(1+4 p T+p^3 T^2)(1-4 p T+p^3 T^2)
\tabularnewline[0.5pt]\hline				
38	 &	smooth    &	          &	1+140 T-1006 p T^2+140 p^3 T^3+p^6 T^4
\tabularnewline[0.5pt]\hline				
39	 &	smooth    &	          &	1+120 T+3218 p T^2+120 p^3 T^3+p^6 T^4
\tabularnewline[0.5pt]\hline				
40	 &	smooth    &	          &	1+115 T+438 p T^2+115 p^3 T^3+p^6 T^4
\tabularnewline[0.5pt]\hline				
41	 &	smooth    &	          &	1-105 T+1510 p T^2-105 p^3 T^3+p^6 T^4
\tabularnewline[0.5pt]\hline				
42	 &	smooth    &	          &	1+320 T+3138 p T^2+320 p^3 T^3+p^6 T^4
\tabularnewline[0.5pt]\hline				
\tablepostamble				
\tablepreamble{47}				
1	 &	smooth    &	          &	1+30 T+2594 p T^2+30 p^3 T^3+p^6 T^4
\tabularnewline[0.5pt]\hline				
2	 &	smooth    &	          &	1-480 T+3394 p T^2-480 p^3 T^3+p^6 T^4
\tabularnewline[0.5pt]\hline				
3	 &	smooth    &	          &	1-160 T-318 p T^2-160 p^3 T^3+p^6 T^4
\tabularnewline[0.5pt]\hline				
4	 &	smooth    &	          &	1+395 T+2930 p T^2+395 p^3 T^3+p^6 T^4
\tabularnewline[0.5pt]\hline				
5	 &	smooth    &	          &	1-10 p T+2578 p T^2-10 p^4 T^3+p^6 T^4
\tabularnewline[0.5pt]\hline				
6	 &	smooth    &	          &	1-360 T+3522 p T^2-360 p^3 T^3+p^6 T^4
\tabularnewline[0.5pt]\hline				
7	 &	smooth    &	          &	1-420 T+3330 p T^2-420 p^3 T^3+p^6 T^4
\tabularnewline[0.5pt]\hline				
8	 &	smooth    &	          &	1+260 T+898 p T^2+260 p^3 T^3+p^6 T^4
\tabularnewline[0.5pt]\hline				
9	 &	smooth    &	          &	1+305 T+178 p T^2+305 p^3 T^3+p^6 T^4
\tabularnewline[0.5pt]\hline				
10	 &	smooth    &	          &	1+530 T+5538 p T^2+530 p^3 T^3+p^6 T^4
\tabularnewline[0.5pt]\hline				
11	 &	smooth    &	          &	1+110 T+1058 p T^2+110 p^3 T^3+p^6 T^4
\tabularnewline[0.5pt]\hline				
12	 &	smooth    &	          &	1+230 T+2242 p T^2+230 p^3 T^3+p^6 T^4
\tabularnewline[0.5pt]\hline				
13	 &	smooth    &	          &	1+625 T+4818 p T^2+625 p^3 T^3+p^6 T^4
\tabularnewline[0.5pt]\hline				
14	 &	smooth    &	          &	1+320 T+1730 p T^2+320 p^3 T^3+p^6 T^4
\tabularnewline[0.5pt]\hline				
15	 &	smooth    &	          &	1-540 T+5026 p T^2-540 p^3 T^3+p^6 T^4
\tabularnewline[0.5pt]\hline				
16	 &	smooth    &	          &	1+585 T+4594 p T^2+585 p^3 T^3+p^6 T^4
\tabularnewline[0.5pt]\hline				
17	 &	smooth    &	          &	1-310 T+4642 p T^2-310 p^3 T^3+p^6 T^4
\tabularnewline[0.5pt]\hline				
18	 &	smooth    &	          &	1+220 T+2178 p T^2+220 p^3 T^3+p^6 T^4
\tabularnewline[0.5pt]\hline				
19	 &	smooth    &	          &	1-320 T+1986 p T^2-320 p^3 T^3+p^6 T^4
\tabularnewline[0.5pt]\hline				
20	 &	smooth    &	          &	1-15 T+146 p T^2-15 p^3 T^3+p^6 T^4
\tabularnewline[0.5pt]\hline				
21	 &	smooth    &	          &	1-615 T+5682 p T^2-615 p^3 T^3+p^6 T^4
\tabularnewline[0.5pt]\hline				
22	 &	smooth    &	          &	1-160 T-462 p T^2-160 p^3 T^3+p^6 T^4
\tabularnewline[0.5pt]\hline				
23	 &	smooth    &	          &	1-580 T+4482 p T^2-580 p^3 T^3+p^6 T^4
\tabularnewline[0.5pt]\hline				
24	 &	smooth    &	          &	1+75 T-2606 p T^2+75 p^3 T^3+p^6 T^4
\tabularnewline[0.5pt]\hline				
25	 &	smooth    &	          &	1+280 T+2498 p T^2+280 p^3 T^3+p^6 T^4
\tabularnewline[0.5pt]\hline				
26	 &	smooth    &	          &	1+635 T+4530 p T^2+635 p^3 T^3+p^6 T^4
\tabularnewline[0.5pt]\hline				
27	 &	smooth    &	          &	1+145 T-1614 p T^2+145 p^3 T^3+p^6 T^4
\tabularnewline[0.5pt]\hline				
28	 &	smooth    &	          &	1-300 T+2850 p T^2-300 p^3 T^3+p^6 T^4
\tabularnewline[0.5pt]\hline				
29	 &	smooth    &	          &	1+355 T+1106 p T^2+355 p^3 T^3+p^6 T^4
\tabularnewline[0.5pt]\hline				
30	 &	smooth    &	          &	1-175 T+722 p T^2-175 p^3 T^3+p^6 T^4
\tabularnewline[0.5pt]\hline				
31	 &	smooth    &	          &	1-135 T+1874 p T^2-135 p^3 T^3+p^6 T^4
\tabularnewline[0.5pt]\hline				
32	 &	smooth    &	          &	1+50 T-478 p T^2+50 p^3 T^3+p^6 T^4
\tabularnewline[0.5pt]\hline				
33	 &	smooth    &	          &	1-45 T-1870 p T^2-45 p^3 T^3+p^6 T^4
\tabularnewline[0.5pt]\hline				
34	 &	smooth    &	          &	1-205 T+690 p T^2-205 p^3 T^3+p^6 T^4
\tabularnewline[0.5pt]\hline				
35	 &	smooth    &	          &	1+575 T+5522 p T^2+575 p^3 T^3+p^6 T^4
\tabularnewline[0.5pt]\hline				
36	 &	smooth    &	          &	1+135 T-142 p T^2+135 p^3 T^3+p^6 T^4
\tabularnewline[0.5pt]\hline				
37	 &	smooth    &	          &	1+100 T-1406 p T^2+100 p^3 T^3+p^6 T^4
\tabularnewline[0.5pt]\hline				
38	 &	smooth    &	          &	1+310 T+3346 p T^2+310 p^3 T^3+p^6 T^4
\tabularnewline[0.5pt]\hline				
39	 &	smooth    &	          &	1-60 T-190 p T^2-60 p^3 T^3+p^6 T^4
\tabularnewline[0.5pt]\hline				
40	 &	smooth    &	          &	1-20 T+2562 p T^2-20 p^3 T^3+p^6 T^4
\tabularnewline[0.5pt]\hline				
41	 &	smooth    &	          &	1-440 T+4322 p T^2-440 p^3 T^3+p^6 T^4
\tabularnewline[0.5pt]\hline				
42	 &	smooth    &	          &	1+825 T+7378 p T^2+825 p^3 T^3+p^6 T^4
\tabularnewline[0.5pt]\hline				
43	 &	smooth    &	          &	1+480 T+4450 p T^2+480 p^3 T^3+p^6 T^4
\tabularnewline[0.5pt]\hline				
44	 &	smooth    &	          &	1+160 T+1058 p T^2+160 p^3 T^3+p^6 T^4
\tabularnewline[0.5pt]\hline				
45	 &	smooth    &	          &	1+180 T+690 p T^2+180 p^3 T^3+p^6 T^4
\tabularnewline[0.5pt]\hline				
46	 &	smooth    &	          &	1+105 T+3602 p T^2+105 p^3 T^3+p^6 T^4
\tabularnewline[0.5pt]\hline				
\tablepostamble				
\tablepreamble{53}				
1	 &	smooth    &	          &	1-25 T-924 p T^2-25 p^3 T^3+p^6 T^4
\tabularnewline[0.5pt]\hline				
2	 &	smooth    &	          &	1+665 T+5012 p T^2+665 p^3 T^3+p^6 T^4
\tabularnewline[0.5pt]\hline				
3	 &	smooth    &	          &	1-615 T+4532 p T^2-615 p^3 T^3+p^6 T^4
\tabularnewline[0.5pt]\hline				
4	 &	smooth    &	          &	1+185 T+632 p T^2+185 p^3 T^3+p^6 T^4
\tabularnewline[0.5pt]\hline				
5	 &	smooth    &	          &	1+35 T+1488 p T^2+35 p^3 T^3+p^6 T^4
\tabularnewline[0.5pt]\hline				
6	 &	smooth    &	          &	1+455 T+2116 p T^2+455 p^3 T^3+p^6 T^4
\tabularnewline[0.5pt]\hline				
7	 &	smooth    &	          &	1-120 T-2674 p T^2-120 p^3 T^3+p^6 T^4
\tabularnewline[0.5pt]\hline				
8	 &	smooth    &	          &	1-510 T+6330 p T^2-510 p^3 T^3+p^6 T^4
\tabularnewline[0.5pt]\hline				
9	 &	smooth    &	          &	1-955 T+8432 p T^2-955 p^3 T^3+p^6 T^4
\tabularnewline[0.5pt]\hline				
10	 &	smooth    &	          &	1-280 T+1998 p T^2-280 p^3 T^3+p^6 T^4
\tabularnewline[0.5pt]\hline				
11	 &	smooth    &	          &	1-20 T-1770 p T^2-20 p^3 T^3+p^6 T^4
\tabularnewline[0.5pt]\hline				
12	 &	smooth    &	          &	1-135 T-844 p T^2-135 p^3 T^3+p^6 T^4
\tabularnewline[0.5pt]\hline				
13	 &	smooth    &	          &	1-15 T+3336 p T^2-15 p^3 T^3+p^6 T^4
\tabularnewline[0.5pt]\hline				
14	 &	smooth    &	          &	1+10 p T+3610 p T^2+10 p^4 T^3+p^6 T^4
\tabularnewline[0.5pt]\hline				
15	 &	smooth    &	          &	1+100 T+4390 p T^2+100 p^3 T^3+p^6 T^4
\tabularnewline[0.5pt]\hline				
16	 &	smooth    &	          &	1+520 T+5646 p T^2+520 p^3 T^3+p^6 T^4
\tabularnewline[0.5pt]\hline				
17	 &	smooth    &	          &	1+900 T+8006 p T^2+900 p^3 T^3+p^6 T^4
\tabularnewline[0.5pt]\hline				
18	 &	smooth    &	          &	1-360 T+4334 p T^2-360 p^3 T^3+p^6 T^4
\tabularnewline[0.5pt]\hline				
19	 &	smooth    &	          &	1+35 T+80 p^2 T^2+35 p^3 T^3+p^6 T^4
\tabularnewline[0.5pt]\hline				
20	 &	smooth    &	          &	1+120 T-866 p T^2+120 p^3 T^3+p^6 T^4
\tabularnewline[0.5pt]\hline				
21	 &	smooth    &	          &	1-170 T+3122 p T^2-170 p^3 T^3+p^6 T^4
\tabularnewline[0.5pt]\hline				
22	 &	smooth    &	          &	1+140 T+966 p T^2+140 p^3 T^3+p^6 T^4
\tabularnewline[0.5pt]\hline				
23	 &	smooth    &	          &	1-210 T+2962 p T^2-210 p^3 T^3+p^6 T^4
\tabularnewline[0.5pt]\hline				
24	 &	smooth    &	          &	1+565 T+6256 p T^2+565 p^3 T^3+p^6 T^4
\tabularnewline[0.5pt]\hline				
25	 &	smooth    &	          &	1-280 T+526 p T^2-280 p^3 T^3+p^6 T^4
\tabularnewline[0.5pt]\hline				
26	 &	smooth    &	          &	1+40 T-1122 p T^2+40 p^3 T^3+p^6 T^4
\tabularnewline[0.5pt]\hline				
27	 &	smooth    &	          &	1-135 T+916 p T^2-135 p^3 T^3+p^6 T^4
\tabularnewline[0.5pt]\hline				
28	 &	smooth    &	          &	1-215 T-8 p T^2-215 p^3 T^3+p^6 T^4
\tabularnewline[0.5pt]\hline				
29	 &	smooth    &	          &	1+625 T+6376 p T^2+625 p^3 T^3+p^6 T^4
\tabularnewline[0.5pt]\hline				
30	 &	smooth    &	          &	1-260 T+5222 p T^2-260 p^3 T^3+p^6 T^4
\tabularnewline[0.5pt]\hline				
31	 &	smooth    &	          &	1-365 T+2000 p T^2-365 p^3 T^3+p^6 T^4
\tabularnewline[0.5pt]\hline				
32	 &	smooth    &	          &	1+460 T+3782 p T^2+460 p^3 T^3+p^6 T^4
\tabularnewline[0.5pt]\hline				
33	 &	smooth    &	          &	1-125 T+2768 p T^2-125 p^3 T^3+p^6 T^4
\tabularnewline[0.5pt]\hline				
34	 &	smooth    &	          &	(1-2 p T+p^3 T^2)(1+486 T+p^3 T^2)
\tabularnewline[0.5pt]\hline				
35	 &	smooth    &	          &	1+490 T+3066 p T^2+490 p^3 T^3+p^6 T^4
\tabularnewline[0.5pt]\hline				
36	 &	smooth    &	          &	1-75 T+2352 p T^2-75 p^3 T^3+p^6 T^4
\tabularnewline[0.5pt]\hline				
37	 &	smooth    &	          &	1+360 T+2414 p T^2+360 p^3 T^3+p^6 T^4
\tabularnewline[0.5pt]\hline				
38	 &	smooth    &	          &	1-280 T+5614 p T^2-280 p^3 T^3+p^6 T^4
\tabularnewline[0.5pt]\hline				
39	 &	smooth    &	          &	1+370 T+2042 p T^2+370 p^3 T^3+p^6 T^4
\tabularnewline[0.5pt]\hline				
40	 &	smooth    &	          &	1+5 T-464 p T^2+5 p^3 T^3+p^6 T^4
\tabularnewline[0.5pt]\hline				
41	 &	smooth    &	          &	1+240 T+158 p T^2+240 p^3 T^3+p^6 T^4
\tabularnewline[0.5pt]\hline				
42	 &	smooth    &	          &	1+120 T+4654 p T^2+120 p^3 T^3+p^6 T^4
\tabularnewline[0.5pt]\hline				
43	 &	smooth    &	          &	1-360 T+5998 p T^2-360 p^3 T^3+p^6 T^4
\tabularnewline[0.5pt]\hline				
44	 &	smooth    &	          &	1+620 T+4726 p T^2+620 p^3 T^3+p^6 T^4
\tabularnewline[0.5pt]\hline				
45	 &	smooth    &	          &	1+610 T+5082 p T^2+610 p^3 T^3+p^6 T^4
\tabularnewline[0.5pt]\hline				
46	 &	smooth    &	          &	1+80 T-1186 p T^2+80 p^3 T^3+p^6 T^4
\tabularnewline[0.5pt]\hline				
47	 &	smooth    &	          &	1+50 T+2610 p T^2+50 p^3 T^3+p^6 T^4
\tabularnewline[0.5pt]\hline				
48	 &	smooth    &	          &	1+240 T+5534 p T^2+240 p^3 T^3+p^6 T^4
\tabularnewline[0.5pt]\hline				
49	 &	smooth    &	          &	1+120 T+5486 p T^2+120 p^3 T^3+p^6 T^4
\tabularnewline[0.5pt]\hline				
50	 &	smooth    &	          &	1+65 T-2780 p T^2+65 p^3 T^3+p^6 T^4
\tabularnewline[0.5pt]\hline				
51	 &	smooth    &	          &	1-175 T-1340 p T^2-175 p^3 T^3+p^6 T^4
\tabularnewline[0.5pt]\hline				
52	 &	smooth    &	          &	1-630 T+4130 p T^2-630 p^3 T^3+p^6 T^4
\tabularnewline[0.5pt]\hline				
\tablepostamble				
\tablepreamble{59}				
1	 &	smooth    &	          &	1-557 T+7910 p T^2-557 p^3 T^3+p^6 T^4
\tabularnewline[0.5pt]\hline				
2	 &	smooth    &	          &	1+352 T+5042 p T^2+352 p^3 T^3+p^6 T^4
\tabularnewline[0.5pt]\hline				
3	 &	smooth    &	          &	1+647 T+6806 p T^2+647 p^3 T^3+p^6 T^4
\tabularnewline[0.5pt]\hline				
4	 &	smooth    &	          &	1+140 T+4818 p T^2+140 p^3 T^3+p^6 T^4
\tabularnewline[0.5pt]\hline				
5	 &	smooth    &	          &	1+132 T+3730 p T^2+132 p^3 T^3+p^6 T^4
\tabularnewline[0.5pt]\hline				
6	 &	smooth    &	          &	1+244 T-686 p T^2+244 p^3 T^3+p^6 T^4
\tabularnewline[0.5pt]\hline				
7	 &	smooth    &	          &	1+26 T-422 p T^2+26 p^3 T^3+p^6 T^4
\tabularnewline[0.5pt]\hline				
8	 &	smooth    &	          &	1-30 T+1146 p T^2-30 p^3 T^3+p^6 T^4
\tabularnewline[0.5pt]\hline				
9	 &	smooth    &	          &	1-82 T+3530 p T^2-82 p^3 T^3+p^6 T^4
\tabularnewline[0.5pt]\hline				
10	 &	smooth    &	          &	1-210 T-1222 p T^2-210 p^3 T^3+p^6 T^4
\tabularnewline[0.5pt]\hline				
11	 &	smooth    &	          &	1+7 T-1954 p T^2+7 p^3 T^3+p^6 T^4
\tabularnewline[0.5pt]\hline				
12	 &	smooth    &	          &	1+164 T+882 p T^2+164 p^3 T^3+p^6 T^4
\tabularnewline[0.5pt]\hline				
13	 &	smooth    &	          &	1+337 T+4582 p T^2+337 p^3 T^3+p^6 T^4
\tabularnewline[0.5pt]\hline				
14	 &	smooth    &	          &	1-1052 T+11026 p T^2-1052 p^3 T^3+p^6 T^4
\tabularnewline[0.5pt]\hline				
15	 &	singular  &	15&	(1-p T) (1-28 T+p^3 T^2)
\tabularnewline[0.5pt]\hline				
16	 &	smooth    &	          &	1-430 T+4058 p T^2-430 p^3 T^3+p^6 T^4
\tabularnewline[0.5pt]\hline				
17	 &	smooth    &	          &	1-599 T+5134 p T^2-599 p^3 T^3+p^6 T^4
\tabularnewline[0.5pt]\hline				
18	 &	smooth    &	          &	1+40 T-1774 p T^2+40 p^3 T^3+p^6 T^4
\tabularnewline[0.5pt]\hline				
19	 &	smooth    &	          &	1+223 T+3542 p T^2+223 p^3 T^3+p^6 T^4
\tabularnewline[0.5pt]\hline				
20	 &	smooth    &	          &	1-356 T+5522 p T^2-356 p^3 T^3+p^6 T^4
\tabularnewline[0.5pt]\hline				
21	 &	smooth    &	          &	1+274 T+378 p T^2+274 p^3 T^3+p^6 T^4
\tabularnewline[0.5pt]\hline				
22	 &	smooth    &	          &	1+61 T-1378 p T^2+61 p^3 T^3+p^6 T^4
\tabularnewline[0.5pt]\hline				
23	 &	smooth    &	          &	1-4 T+5634 p T^2-4 p^3 T^3+p^6 T^4
\tabularnewline[0.5pt]\hline				
24	 &	smooth    &	          &	1+381 T+1142 p T^2+381 p^3 T^3+p^6 T^4
\tabularnewline[0.5pt]\hline				
25	 &	smooth    &	          &	1+432 T+38 p^2 T^2+432 p^3 T^3+p^6 T^4
\tabularnewline[0.5pt]\hline				
26	 &	smooth    &	          &	1-8 T+5250 p T^2-8 p^3 T^3+p^6 T^4
\tabularnewline[0.5pt]\hline				
27	 &	smooth    &	          &	1-63 T-2578 p T^2-63 p^3 T^3+p^6 T^4
\tabularnewline[0.5pt]\hline				
28	 &	smooth    &	          &	1+592 T+4802 p T^2+592 p^3 T^3+p^6 T^4
\tabularnewline[0.5pt]\hline				
29	 &	smooth    &	          &	1-452 T+6610 p T^2-452 p^3 T^3+p^6 T^4
\tabularnewline[0.5pt]\hline				
30	 &	smooth    &	          &	1-702 T+7562 p T^2-702 p^3 T^3+p^6 T^4
\tabularnewline[0.5pt]\hline				
31	 &	smooth    &	          &	1-276 T+2434 p T^2-276 p^3 T^3+p^6 T^4
\tabularnewline[0.5pt]\hline				
32	 &	smooth    &	          &	1+680 T+4626 p T^2+680 p^3 T^3+p^6 T^4
\tabularnewline[0.5pt]\hline				
33	 &	smooth    &	          &	1-78 T+2906 p T^2-78 p^3 T^3+p^6 T^4
\tabularnewline[0.5pt]\hline				
34	 &	smooth    &	          &	1-497 T+2142 p T^2-497 p^3 T^3+p^6 T^4
\tabularnewline[0.5pt]\hline				
35	 &	smooth    &	          &	1-804 T+7762 p T^2-804 p^3 T^3+p^6 T^4
\tabularnewline[0.5pt]\hline				
36	 &	smooth    &	          &	1+846 T+126 p^2 T^2+846 p^3 T^3+p^6 T^4
\tabularnewline[0.5pt]\hline				
37	 &	smooth    &	          &	1+116 T+2642 p T^2+116 p^3 T^3+p^6 T^4
\tabularnewline[0.5pt]\hline				
38	 &	smooth    &	          &	1-263 T+2630 p T^2-263 p^3 T^3+p^6 T^4
\tabularnewline[0.5pt]\hline				
39	 &	smooth    &	          &	1+86 T+5194 p T^2+86 p^3 T^3+p^6 T^4
\tabularnewline[0.5pt]\hline				
40	 &	smooth    &	          &	1+192 T+4402 p T^2+192 p^3 T^3+p^6 T^4
\tabularnewline[0.5pt]\hline				
41	 &	smooth    &	          &	1-82 T+1546 p T^2-82 p^3 T^3+p^6 T^4
\tabularnewline[0.5pt]\hline				
42	 &	smooth    &	          &	1+10 T-3014 p T^2+10 p^3 T^3+p^6 T^4
\tabularnewline[0.5pt]\hline				
43	 &	smooth    &	          &	1+936 T+8034 p T^2+936 p^3 T^3+p^6 T^4
\tabularnewline[0.5pt]\hline				
44	 &	smooth    &	          &	1+124 T+642 p T^2+124 p^3 T^3+p^6 T^4
\tabularnewline[0.5pt]\hline				
45	 &	smooth    &	          &	1+777 T+7310 p T^2+777 p^3 T^3+p^6 T^4
\tabularnewline[0.5pt]\hline				
46	 &	smooth    &	          &	1-264 T+1698 p T^2-264 p^3 T^3+p^6 T^4
\tabularnewline[0.5pt]\hline				
47	 &	singular  &	47&	(1-p T) (1+28 T+p^3 T^2)
\tabularnewline[0.5pt]\hline				
48	 &	smooth    &	          &	1+375 T+4278 p T^2+375 p^3 T^3+p^6 T^4
\tabularnewline[0.5pt]\hline				
49	 &	smooth    &	          &	(1+12 p T+p^3 T^2)(1-92 T+p^3 T^2)
\tabularnewline[0.5pt]\hline				
50	 &	smooth    &	          &	(1-4 p T+p^3 T^2)(1+560 T+p^3 T^2)
\tabularnewline[0.5pt]\hline				
51	 &	smooth    &	          &	1+54 T+1034 p T^2+54 p^3 T^3+p^6 T^4
\tabularnewline[0.5pt]\hline				
52	 &	smooth    &	          &	1-258 T+1850 p T^2-258 p^3 T^3+p^6 T^4
\tabularnewline[0.5pt]\hline				
53	 &	smooth    &	          &	1+18 T+2522 p T^2+18 p^3 T^3+p^6 T^4
\tabularnewline[0.5pt]\hline				
54	 &	smooth    &	          &	1+718 T+7610 p T^2+718 p^3 T^3+p^6 T^4
\tabularnewline[0.5pt]\hline				
55	 &	smooth    &	          &	1-39 T+5894 p T^2-39 p^3 T^3+p^6 T^4
\tabularnewline[0.5pt]\hline				
56	 &	smooth    &	          &	1+659 T+4014 p T^2+659 p^3 T^3+p^6 T^4
\tabularnewline[0.5pt]\hline				
57	 &	smooth    &	          &	1+148 T+370 p T^2+148 p^3 T^3+p^6 T^4
\tabularnewline[0.5pt]\hline				
58	 &	smooth    &	          &	1-25 T+158 p T^2-25 p^3 T^3+p^6 T^4
\tabularnewline[0.5pt]\hline				
\tablepostamble				
\tablepreamble{61}				
1	 &	smooth    &	          &	1-416 T+6750 p T^2-416 p^3 T^3+p^6 T^4
\tabularnewline[0.5pt]\hline				
2	 &	smooth    &	          &	1-89 T-3968 p T^2-89 p^3 T^3+p^6 T^4
\tabularnewline[0.5pt]\hline				
3	 &	smooth    &	          &	1+1067 T+11444 p T^2+1067 p^3 T^3+p^6 T^4
\tabularnewline[0.5pt]\hline				
4	 &	smooth    &	          &	1-151 T-1808 p T^2-151 p^3 T^3+p^6 T^4
\tabularnewline[0.5pt]\hline				
5	 &	smooth    &	          &	1-64 T-1890 p T^2-64 p^3 T^3+p^6 T^4
\tabularnewline[0.5pt]\hline				
6	 &	smooth    &	          &	1-772 T+7014 p T^2-772 p^3 T^3+p^6 T^4
\tabularnewline[0.5pt]\hline				
7	 &	smooth    &	          &	1+147 T+2952 p T^2+147 p^3 T^3+p^6 T^4
\tabularnewline[0.5pt]\hline				
8	 &	smooth    &	          &	1+1269 T+13636 p T^2+1269 p^3 T^3+p^6 T^4
\tabularnewline[0.5pt]\hline				
9	 &	smooth    &	          &	1+58 T-3502 p T^2+58 p^3 T^3+p^6 T^4
\tabularnewline[0.5pt]\hline				
10	 &	smooth    &	          &	1-41 T+1216 p T^2-41 p^3 T^3+p^6 T^4
\tabularnewline[0.5pt]\hline				
11	 &	smooth    &	          &	1-292 T+6694 p T^2-292 p^3 T^3+p^6 T^4
\tabularnewline[0.5pt]\hline				
12	 &	smooth    &	          &	1+240 T+2942 p T^2+240 p^3 T^3+p^6 T^4
\tabularnewline[0.5pt]\hline				
13	 &	smooth    &	          &	1+523 T+5972 p T^2+523 p^3 T^3+p^6 T^4
\tabularnewline[0.5pt]\hline				
14	 &	smooth    &	          &	1+111 T-1988 p T^2+111 p^3 T^3+p^6 T^4
\tabularnewline[0.5pt]\hline				
15	 &	smooth    &	          &	1+834 T+8482 p T^2+834 p^3 T^3+p^6 T^4
\tabularnewline[0.5pt]\hline				
16	 &	smooth    &	          &	1-58 T-1558 p T^2-58 p^3 T^3+p^6 T^4
\tabularnewline[0.5pt]\hline				
17	 &	smooth    &	          &	1-277 T+5144 p T^2-277 p^3 T^3+p^6 T^4
\tabularnewline[0.5pt]\hline				
18	 &	smooth    &	          &	1-708 T+8902 p T^2-708 p^3 T^3+p^6 T^4
\tabularnewline[0.5pt]\hline				
19	 &	smooth    &	          &	1-602 T+7658 p T^2-602 p^3 T^3+p^6 T^4
\tabularnewline[0.5pt]\hline				
20	 &	smooth    &	          &	1+550 T+1578 p T^2+550 p^3 T^3+p^6 T^4
\tabularnewline[0.5pt]\hline				
21	 &	smooth    &	          &	1+84 T+3766 p T^2+84 p^3 T^3+p^6 T^4
\tabularnewline[0.5pt]\hline				
22	 &	smooth    &	          &	1-597 T+5844 p T^2-597 p^3 T^3+p^6 T^4
\tabularnewline[0.5pt]\hline				
23	 &	smooth    &	          &	1+444 T+3222 p T^2+444 p^3 T^3+p^6 T^4
\tabularnewline[0.5pt]\hline				
24	 &	smooth    &	          &	1+92 T-3578 p T^2+92 p^3 T^3+p^6 T^4
\tabularnewline[0.5pt]\hline				
25	 &	smooth    &	          &	1-137 T-2068 p T^2-137 p^3 T^3+p^6 T^4
\tabularnewline[0.5pt]\hline				
26	 &	smooth    &	          &	1+442 T+18 p^2 T^2+442 p^3 T^3+p^6 T^4
\tabularnewline[0.5pt]\hline				
27	 &	smooth    &	          &	1-682 T+4202 p T^2-682 p^3 T^3+p^6 T^4
\tabularnewline[0.5pt]\hline				
28	 &	singular  &	28&	(1-p T) (1+182 T+p^3 T^2)
\tabularnewline[0.5pt]\hline				
29	 &	smooth    &	          &	1+763 T+8184 p T^2+763 p^3 T^3+p^6 T^4
\tabularnewline[0.5pt]\hline				
30	 &	smooth    &	          &	1+144 T-1714 p T^2+144 p^3 T^3+p^6 T^4
\tabularnewline[0.5pt]\hline				
31	 &	smooth    &	          &	1+119 T+5536 p T^2+119 p^3 T^3+p^6 T^4
\tabularnewline[0.5pt]\hline				
32	 &	smooth    &	          &	1-886 T+7802 p T^2-886 p^3 T^3+p^6 T^4
\tabularnewline[0.5pt]\hline				
33	 &	smooth    &	          &	1+151 T+5472 p T^2+151 p^3 T^3+p^6 T^4
\tabularnewline[0.5pt]\hline				
34	 &	smooth    &	          &	1+54 T-118 p T^2+54 p^3 T^3+p^6 T^4
\tabularnewline[0.5pt]\hline				
35	 &	smooth    &	          &	1-326 T+2746 p T^2-326 p^3 T^3+p^6 T^4
\tabularnewline[0.5pt]\hline				
36	 &	smooth    &	          &	1-80 T-258 p T^2-80 p^3 T^3+p^6 T^4
\tabularnewline[0.5pt]\hline				
37	 &	smooth    &	          &	1-96 T-594 p T^2-96 p^3 T^3+p^6 T^4
\tabularnewline[0.5pt]\hline				
38	 &	smooth    &	          &	1+404 T+6006 p T^2+404 p^3 T^3+p^6 T^4
\tabularnewline[0.5pt]\hline				
39	 &	smooth    &	          &	1+942 T+9434 p T^2+942 p^3 T^3+p^6 T^4
\tabularnewline[0.5pt]\hline				
40	 &	smooth    &	          &	1+764 T+7062 p T^2+764 p^3 T^3+p^6 T^4
\tabularnewline[0.5pt]\hline				
41	 &	smooth    &	          &	1-254 T+4162 p T^2-254 p^3 T^3+p^6 T^4
\tabularnewline[0.5pt]\hline				
42	 &	smooth    &	          &	1-778 T+8778 p T^2-778 p^3 T^3+p^6 T^4
\tabularnewline[0.5pt]\hline				
43	 &	smooth    &	          &	1+504 T+4206 p T^2+504 p^3 T^3+p^6 T^4
\tabularnewline[0.5pt]\hline				
44	 &	smooth    &	          &	1+834 T+8442 p T^2+834 p^3 T^3+p^6 T^4
\tabularnewline[0.5pt]\hline				
45	 &	smooth    &	          &	1-194 T+2842 p T^2-194 p^3 T^3+p^6 T^4
\tabularnewline[0.5pt]\hline				
46	 &	smooth    &	          &	1-130 T+378 p T^2-130 p^3 T^3+p^6 T^4
\tabularnewline[0.5pt]\hline				
47	 &	smooth    &	          &	1+761 T+8112 p T^2+761 p^3 T^3+p^6 T^4
\tabularnewline[0.5pt]\hline				
48	 &	smooth    &	          &	1-38 T+1618 p T^2-38 p^3 T^3+p^6 T^4
\tabularnewline[0.5pt]\hline				
49	 &	smooth    &	          &	1-393 T+1292 p T^2-393 p^3 T^3+p^6 T^4
\tabularnewline[0.5pt]\hline				
50	 &	smooth    &	          &	1+578 T+3978 p T^2+578 p^3 T^3+p^6 T^4
\tabularnewline[0.5pt]\hline				
51	 &	smooth    &	          &	1-296 T+5806 p T^2-296 p^3 T^3+p^6 T^4
\tabularnewline[0.5pt]\hline				
52	 &	smooth    &	          &	1+747 T+6132 p T^2+747 p^3 T^3+p^6 T^4
\tabularnewline[0.5pt]\hline				
53	 &	smooth    &	          &	1-13 T-4728 p T^2-13 p^3 T^3+p^6 T^4
\tabularnewline[0.5pt]\hline				
54	 &	smooth    &	          &	1-491 T+6116 p T^2-491 p^3 T^3+p^6 T^4
\tabularnewline[0.5pt]\hline				
55	 &	smooth    &	          &	1-822 T+7978 p T^2-822 p^3 T^3+p^6 T^4
\tabularnewline[0.5pt]\hline				
56	 &	smooth    &	          &	1+94 T+4122 p T^2+94 p^3 T^3+p^6 T^4
\tabularnewline[0.5pt]\hline				
57	 &	smooth    &	          &	1-81 T-3108 p T^2-81 p^3 T^3+p^6 T^4
\tabularnewline[0.5pt]\hline				
58	 &	smooth    &	          &	1+718 T+6394 p T^2+718 p^3 T^3+p^6 T^4
\tabularnewline[0.5pt]\hline				
59	 &	singular  &	59&	(1-p T) (1+182 T+p^3 T^2)
\tabularnewline[0.5pt]\hline				
60	 &	smooth    &	          &	1-194 T+5402 p T^2-194 p^3 T^3+p^6 T^4
\tabularnewline[0.5pt]\hline				
\tablepostamble				
\tablepreamble{67}				
1	 &	smooth    &	          &	1-1290 T+13242 p T^2-1290 p^3 T^3+p^6 T^4
\tabularnewline[0.5pt]\hline				
2	 &	smooth    &	          &	1+880 T+7650 p T^2+880 p^3 T^3+p^6 T^4
\tabularnewline[0.5pt]\hline				
3	 &	smooth    &	          &	1-935 T+9470 p T^2-935 p^3 T^3+p^6 T^4
\tabularnewline[0.5pt]\hline				
4	 &	smooth    &	          &	1-215 T+5942 p T^2-215 p^3 T^3+p^6 T^4
\tabularnewline[0.5pt]\hline				
5	 &	smooth    &	          &	1+40 T+2 p T^2+40 p^3 T^3+p^6 T^4
\tabularnewline[0.5pt]\hline				
6	 &	smooth    &	          &	1+600 T+6210 p T^2+600 p^3 T^3+p^6 T^4
\tabularnewline[0.5pt]\hline				
7	 &	smooth    &	          &	1+505 T+3678 p T^2+505 p^3 T^3+p^6 T^4
\tabularnewline[0.5pt]\hline				
8	 &	smooth    &	          &	1+675 T+5622 p T^2+675 p^3 T^3+p^6 T^4
\tabularnewline[0.5pt]\hline				
9	 &	smooth    &	          &	1+155 T+2846 p T^2+155 p^3 T^3+p^6 T^4
\tabularnewline[0.5pt]\hline				
10	 &	smooth    &	          &	1-400 T+4482 p T^2-400 p^3 T^3+p^6 T^4
\tabularnewline[0.5pt]\hline				
11	 &	smooth    &	          &	1-65 T-7706 p T^2-65 p^3 T^3+p^6 T^4
\tabularnewline[0.5pt]\hline				
12	 &	smooth    &	          &	1-120 T+5906 p T^2-120 p^3 T^3+p^6 T^4
\tabularnewline[0.5pt]\hline				
13	 &	smooth    &	          &	1+130 T+3370 p T^2+130 p^3 T^3+p^6 T^4
\tabularnewline[0.5pt]\hline				
14	 &	smooth    &	          &	1+200 T+1410 p T^2+200 p^3 T^3+p^6 T^4
\tabularnewline[0.5pt]\hline				
15	 &	smooth    &	          &	1+400 T+1698 p T^2+400 p^3 T^3+p^6 T^4
\tabularnewline[0.5pt]\hline				
16	 &	smooth    &	          &	1+520 T+7810 p T^2+520 p^3 T^3+p^6 T^4
\tabularnewline[0.5pt]\hline				
17	 &	smooth    &	          &	1-610 T+2682 p T^2-610 p^3 T^3+p^6 T^4
\tabularnewline[0.5pt]\hline				
18	 &	smooth    &	          &	1-725 T+7734 p T^2-725 p^3 T^3+p^6 T^4
\tabularnewline[0.5pt]\hline				
19	 &	smooth    &	          &	1-600 T+5378 p T^2-600 p^3 T^3+p^6 T^4
\tabularnewline[0.5pt]\hline				
20	 &	smooth    &	          &	1-295 T+6526 p T^2-295 p^3 T^3+p^6 T^4
\tabularnewline[0.5pt]\hline				
21	 &	smooth    &	          &	1+115 T-930 p T^2+115 p^3 T^3+p^6 T^4
\tabularnewline[0.5pt]\hline				
22	 &	smooth    &	          &	1-880 T+8482 p T^2-880 p^3 T^3+p^6 T^4
\tabularnewline[0.5pt]\hline				
23	 &	smooth    &	          &	1+255 T+4270 p T^2+255 p^3 T^3+p^6 T^4
\tabularnewline[0.5pt]\hline				
24	 &	smooth    &	          &	1+1740 T+20178 p T^2+1740 p^3 T^3+p^6 T^4
\tabularnewline[0.5pt]\hline				
25	 &	smooth    &	          &	1-175 T+2870 p T^2-175 p^3 T^3+p^6 T^4
\tabularnewline[0.5pt]\hline				
26	 &	smooth    &	          &	1-40 T+482 p T^2-40 p^3 T^3+p^6 T^4
\tabularnewline[0.5pt]\hline				
27	 &	smooth    &	          &	1+320 T+4290 p T^2+320 p^3 T^3+p^6 T^4
\tabularnewline[0.5pt]\hline				
28	 &	smooth    &	          &	1+30 T+8218 p T^2+30 p^3 T^3+p^6 T^4
\tabularnewline[0.5pt]\hline				
29	 &	smooth    &	          &	1+400 T+9122 p T^2+400 p^3 T^3+p^6 T^4
\tabularnewline[0.5pt]\hline				
30	 &	smooth    &	          &	1+495 T+3430 p T^2+495 p^3 T^3+p^6 T^4
\tabularnewline[0.5pt]\hline				
31	 &	smooth    &	          &	1+160 T+8178 p T^2+160 p^3 T^3+p^6 T^4
\tabularnewline[0.5pt]\hline				
32	 &	smooth    &	          &	1+480 T+4338 p T^2+480 p^3 T^3+p^6 T^4
\tabularnewline[0.5pt]\hline				
33	 &	smooth    &	          &	1+5 p T-594 p T^2+5 p^4 T^3+p^6 T^4
\tabularnewline[0.5pt]\hline				
34	 &	smooth    &	          &	1-565 T+18 p^2 T^2-565 p^3 T^3+p^6 T^4
\tabularnewline[0.5pt]\hline				
35	 &	smooth    &	          &	1+610 T+2058 p T^2+610 p^3 T^3+p^6 T^4
\tabularnewline[0.5pt]\hline				
36	 &	smooth    &	          &	1-190 T+4682 p T^2-190 p^3 T^3+p^6 T^4
\tabularnewline[0.5pt]\hline				
37	 &	smooth    &	          &	1+355 T+3678 p T^2+355 p^3 T^3+p^6 T^4
\tabularnewline[0.5pt]\hline				
38	 &	smooth    &	          &	1-510 T+94 p^2 T^2-510 p^3 T^3+p^6 T^4
\tabularnewline[0.5pt]\hline				
39	 &	smooth    &	          &	(1-8 p T+p^3 T^2)(1+636 T+p^3 T^2)
\tabularnewline[0.5pt]\hline				
40	 &	smooth    &	          &	1+190 T+826 p T^2+190 p^3 T^3+p^6 T^4
\tabularnewline[0.5pt]\hline				
41	 &	smooth    &	          &	1+930 T+10202 p T^2+930 p^3 T^3+p^6 T^4
\tabularnewline[0.5pt]\hline				
42	 &	smooth    &	          &	1-385 T+7270 p T^2-385 p^3 T^3+p^6 T^4
\tabularnewline[0.5pt]\hline				
43	 &	smooth    &	          &	1-160 T-4702 p T^2-160 p^3 T^3+p^6 T^4
\tabularnewline[0.5pt]\hline				
44	 &	smooth    &	          &	1+465 T+4350 p T^2+465 p^3 T^3+p^6 T^4
\tabularnewline[0.5pt]\hline				
45	 &	smooth    &	          &	1+150 T-1862 p T^2+150 p^3 T^3+p^6 T^4
\tabularnewline[0.5pt]\hline				
46	 &	smooth    &	          &	1+435 T+310 p T^2+435 p^3 T^3+p^6 T^4
\tabularnewline[0.5pt]\hline				
47	 &	smooth    &	          &	1-280 T-3646 p T^2-280 p^3 T^3+p^6 T^4
\tabularnewline[0.5pt]\hline				
48	 &	smooth    &	          &	1-270 T+5786 p T^2-270 p^3 T^3+p^6 T^4
\tabularnewline[0.5pt]\hline				
49	 &	smooth    &	          &	1-100 T+7890 p T^2-100 p^3 T^3+p^6 T^4
\tabularnewline[0.5pt]\hline				
50	 &	smooth    &	          &	1+440 T+7938 p T^2+440 p^3 T^3+p^6 T^4
\tabularnewline[0.5pt]\hline				
51	 &	smooth    &	          &	1+360 T+722 p T^2+360 p^3 T^3+p^6 T^4
\tabularnewline[0.5pt]\hline				
52	 &	smooth    &	          &	1+160 T+4450 p T^2+160 p^3 T^3+p^6 T^4
\tabularnewline[0.5pt]\hline				
53	 &	smooth    &	          &	1-1100 T+9490 p T^2-1100 p^3 T^3+p^6 T^4
\tabularnewline[0.5pt]\hline				
54	 &	smooth    &	          &	1-375 T+4150 p T^2-375 p^3 T^3+p^6 T^4
\tabularnewline[0.5pt]\hline				
55	 &	smooth    &	          &	1-615 T+4854 p T^2-615 p^3 T^3+p^6 T^4
\tabularnewline[0.5pt]\hline				
56	 &	smooth    &	          &	1-485 T+2494 p T^2-485 p^3 T^3+p^6 T^4
\tabularnewline[0.5pt]\hline				
57	 &	smooth    &	          &	1+100 T-2318 p T^2+100 p^3 T^3+p^6 T^4
\tabularnewline[0.5pt]\hline				
58	 &	smooth    &	          &	1+660 T+3570 p T^2+660 p^3 T^3+p^6 T^4
\tabularnewline[0.5pt]\hline				
59	 &	smooth    &	          &	1+815 T+9870 p T^2+815 p^3 T^3+p^6 T^4
\tabularnewline[0.5pt]\hline				
60	 &	smooth    &	          &	1+110 T+634 p T^2+110 p^3 T^3+p^6 T^4
\tabularnewline[0.5pt]\hline				
61	 &	smooth    &	          &	1-10 T+2874 p T^2-10 p^3 T^3+p^6 T^4
\tabularnewline[0.5pt]\hline				
62	 &	smooth    &	          &	1+450 T+5034 p T^2+450 p^3 T^3+p^6 T^4
\tabularnewline[0.5pt]\hline				
63	 &	smooth    &	          &	1+390 T+3146 p T^2+390 p^3 T^3+p^6 T^4
\tabularnewline[0.5pt]\hline				
64	 &	smooth    &	          &	1+1250 T+14730 p T^2+1250 p^3 T^3+p^6 T^4
\tabularnewline[0.5pt]\hline				
65	 &	smooth    &	          &	1-100 T+2130 p T^2-100 p^3 T^3+p^6 T^4
\tabularnewline[0.5pt]\hline				
66	 &	smooth    &	          &	(1-8 p T+p^3 T^2)(1+116 T+p^3 T^2)
\tabularnewline[0.5pt]\hline				
\tablepostamble				
\tablepreamble{71}				
1	 &	smooth    &	          &	1+54 T-8494 p T^2+54 p^3 T^3+p^6 T^4
\tabularnewline[0.5pt]\hline				
2	 &	smooth    &	          &	1-1185 T+13082 p T^2-1185 p^3 T^3+p^6 T^4
\tabularnewline[0.5pt]\hline				
3	 &	smooth    &	          &	1-1353 T+13722 p T^2-1353 p^3 T^3+p^6 T^4
\tabularnewline[0.5pt]\hline				
4	 &	smooth    &	          &	1+309 T+2602 p T^2+309 p^3 T^3+p^6 T^4
\tabularnewline[0.5pt]\hline				
5	 &	smooth    &	          &	1+348 T-1918 p T^2+348 p^3 T^3+p^6 T^4
\tabularnewline[0.5pt]\hline				
6	 &	smooth    &	          &	1+356 T-510 p T^2+356 p^3 T^3+p^6 T^4
\tabularnewline[0.5pt]\hline				
7	 &	smooth    &	          &	1+63 T+7082 p T^2+63 p^3 T^3+p^6 T^4
\tabularnewline[0.5pt]\hline				
8	 &	smooth    &	          &	1-713 T+5242 p T^2-713 p^3 T^3+p^6 T^4
\tabularnewline[0.5pt]\hline				
9	 &	smooth    &	          &	(1+8 p T+p^3 T^2)(1-672 T+p^3 T^2)
\tabularnewline[0.5pt]\hline				
10	 &	smooth    &	          &	1-394 T+7186 p T^2-394 p^3 T^3+p^6 T^4
\tabularnewline[0.5pt]\hline				
11	 &	smooth    &	          &	1-513 T+3242 p T^2-513 p^3 T^3+p^6 T^4
\tabularnewline[0.5pt]\hline				
12	 &	smooth    &	          &	1+505 T+10282 p T^2+505 p^3 T^3+p^6 T^4
\tabularnewline[0.5pt]\hline				
13	 &	smooth    &	          &	1+223 T+1002 p T^2+223 p^3 T^3+p^6 T^4
\tabularnewline[0.5pt]\hline				
14	 &	smooth    &	          &	1-844 T+11602 p T^2-844 p^3 T^3+p^6 T^4
\tabularnewline[0.5pt]\hline				
15	 &	smooth    &	          &	1+184 T+8866 p T^2+184 p^3 T^3+p^6 T^4
\tabularnewline[0.5pt]\hline				
16	 &	smooth    &	          &	1-240 T+2082 p T^2-240 p^3 T^3+p^6 T^4
\tabularnewline[0.5pt]\hline				
17	 &	smooth    &	          &	1+234 T+8466 p T^2+234 p^3 T^3+p^6 T^4
\tabularnewline[0.5pt]\hline				
18	 &	smooth    &	          &	1-1052 T+9602 p T^2-1052 p^3 T^3+p^6 T^4
\tabularnewline[0.5pt]\hline				
19	 &	smooth    &	          &	1+303 T-3942 p T^2+303 p^3 T^3+p^6 T^4
\tabularnewline[0.5pt]\hline				
20	 &	smooth    &	          &	1+402 T+8338 p T^2+402 p^3 T^3+p^6 T^4
\tabularnewline[0.5pt]\hline				
21	 &	smooth    &	          &	1+544 T+9986 p T^2+544 p^3 T^3+p^6 T^4
\tabularnewline[0.5pt]\hline				
22	 &	smooth    &	          &	1+972 T+11714 p T^2+972 p^3 T^3+p^6 T^4
\tabularnewline[0.5pt]\hline				
23	 &	smooth    &	          &	1-705 T+8298 p T^2-705 p^3 T^3+p^6 T^4
\tabularnewline[0.5pt]\hline				
24	 &	smooth    &	          &	1+117 T+4202 p T^2+117 p^3 T^3+p^6 T^4
\tabularnewline[0.5pt]\hline				
25	 &	smooth    &	          &	1+647 T+3802 p T^2+647 p^3 T^3+p^6 T^4
\tabularnewline[0.5pt]\hline				
26	 &	smooth    &	          &	1-339 T+4442 p T^2-339 p^3 T^3+p^6 T^4
\tabularnewline[0.5pt]\hline				
27	 &	smooth    &	          &	1-350 T+3858 p T^2-350 p^3 T^3+p^6 T^4
\tabularnewline[0.5pt]\hline				
28	 &	smooth    &	          &	1-67 T+2842 p T^2-67 p^3 T^3+p^6 T^4
\tabularnewline[0.5pt]\hline				
29	 &	smooth    &	          &	1+988 T+7842 p T^2+988 p^3 T^3+p^6 T^4
\tabularnewline[0.5pt]\hline				
30	 &	smooth    &	          &	1-12 T+6722 p T^2-12 p^3 T^3+p^6 T^4
\tabularnewline[0.5pt]\hline				
31	 &	smooth    &	          &	(1-15 p T+p^3 T^2)(1+744 T+p^3 T^2)
\tabularnewline[0.5pt]\hline				
32	 &	smooth    &	          &	1+480 T+8738 p T^2+480 p^3 T^3+p^6 T^4
\tabularnewline[0.5pt]\hline				
33	 &	smooth    &	          &	1+310 T-2702 p T^2+310 p^3 T^3+p^6 T^4
\tabularnewline[0.5pt]\hline				
34	 &	smooth    &	          &	1+359 T+4362 p T^2+359 p^3 T^3+p^6 T^4
\tabularnewline[0.5pt]\hline				
35	 &	smooth    &	          &	1+428 T-2238 p T^2+428 p^3 T^3+p^6 T^4
\tabularnewline[0.5pt]\hline				
36	 &	smooth    &	          &	1+1032 T+11362 p T^2+1032 p^3 T^3+p^6 T^4
\tabularnewline[0.5pt]\hline				
37	 &	smooth    &	          &	1-956 T+11490 p T^2-956 p^3 T^3+p^6 T^4
\tabularnewline[0.5pt]\hline				
38	 &	smooth    &	          &	1+233 T+1514 p T^2+233 p^3 T^3+p^6 T^4
\tabularnewline[0.5pt]\hline				
39	 &	smooth    &	          &	1+700 T+7218 p T^2+700 p^3 T^3+p^6 T^4
\tabularnewline[0.5pt]\hline				
40	 &	smooth    &	          &	1+439 T+6522 p T^2+439 p^3 T^3+p^6 T^4
\tabularnewline[0.5pt]\hline				
41	 &	smooth    &	          &	1-12 T+2274 p T^2-12 p^3 T^3+p^6 T^4
\tabularnewline[0.5pt]\hline				
42	 &	smooth    &	          &	(1+p^3 T^2)(1+880 T+p^3 T^2)
\tabularnewline[0.5pt]\hline				
43	 &	smooth    &	          &	1+1896 T+22306 p T^2+1896 p^3 T^3+p^6 T^4
\tabularnewline[0.5pt]\hline				
44	 &	smooth    &	          &	1+23 T+2602 p T^2+23 p^3 T^3+p^6 T^4
\tabularnewline[0.5pt]\hline				
45	 &	smooth    &	          &	1+400 T+6818 p T^2+400 p^3 T^3+p^6 T^4
\tabularnewline[0.5pt]\hline				
46	 &	smooth    &	          &	1+111 T+8810 p T^2+111 p^3 T^3+p^6 T^4
\tabularnewline[0.5pt]\hline				
47	 &	smooth    &	          &	1+760 T+8738 p T^2+760 p^3 T^3+p^6 T^4
\tabularnewline[0.5pt]\hline				
48	 &	smooth    &	          &	1-156 T+1442 p T^2-156 p^3 T^3+p^6 T^4
\tabularnewline[0.5pt]\hline				
49	 &	smooth    &	          &	1-129 T+6682 p T^2-129 p^3 T^3+p^6 T^4
\tabularnewline[0.5pt]\hline				
50	 &	smooth    &	          &	1+884 T+6466 p T^2+884 p^3 T^3+p^6 T^4
\tabularnewline[0.5pt]\hline				
51	 &	smooth    &	          &	1+560 T+5938 p T^2+560 p^3 T^3+p^6 T^4
\tabularnewline[0.5pt]\hline				
52	 &	smooth    &	          &	1+455 T-118 p T^2+455 p^3 T^3+p^6 T^4
\tabularnewline[0.5pt]\hline				
53	 &	smooth    &	          &	1-490 T+498 p T^2-490 p^3 T^3+p^6 T^4
\tabularnewline[0.5pt]\hline				
54	 &	smooth    &	          &	1-657 T+9722 p T^2-657 p^3 T^3+p^6 T^4
\tabularnewline[0.5pt]\hline				
55	 &	smooth    &	          &	1-548 T+2802 p T^2-548 p^3 T^3+p^6 T^4
\tabularnewline[0.5pt]\hline				
56	 &	smooth    &	          &	1-175 T-1318 p T^2-175 p^3 T^3+p^6 T^4
\tabularnewline[0.5pt]\hline				
57	 &	smooth    &	          &	1-489 T+7226 p T^2-489 p^3 T^3+p^6 T^4
\tabularnewline[0.5pt]\hline				
58	 &	smooth    &	          &	1+152 T+4002 p T^2+152 p^3 T^3+p^6 T^4
\tabularnewline[0.5pt]\hline				
59	 &	smooth    &	          &	1-120 T+2178 p T^2-120 p^3 T^3+p^6 T^4
\tabularnewline[0.5pt]\hline				
60	 &	singular  &	60&	(1+p T) (1+408 T+p^3 T^2)
\tabularnewline[0.5pt]\hline				
61	 &	smooth    &	          &	1+52 T+8162 p T^2+52 p^3 T^3+p^6 T^4
\tabularnewline[0.5pt]\hline				
62	 &	smooth    &	          &	1+170 T-814 p T^2+170 p^3 T^3+p^6 T^4
\tabularnewline[0.5pt]\hline				
63	 &	smooth    &	          &	1-312 T+6962 p T^2-312 p^3 T^3+p^6 T^4
\tabularnewline[0.5pt]\hline				
64	 &	smooth    &	          &	1+103 T-1286 p T^2+103 p^3 T^3+p^6 T^4
\tabularnewline[0.5pt]\hline				
65	 &	smooth    &	          &	1-188 T-318 p T^2-188 p^3 T^3+p^6 T^4
\tabularnewline[0.5pt]\hline				
66	 &	smooth    &	          &	1-32 T+2322 p T^2-32 p^3 T^3+p^6 T^4
\tabularnewline[0.5pt]\hline				
67	 &	smooth    &	          &	1+656 T+2786 p T^2+656 p^3 T^3+p^6 T^4
\tabularnewline[0.5pt]\hline				
68	 &	singular  &	68&	(1+p T) (1-408 T+p^3 T^2)
\tabularnewline[0.5pt]\hline				
69	 &	smooth    &	          &	1+135 T+1066 p T^2+135 p^3 T^3+p^6 T^4
\tabularnewline[0.5pt]\hline				
70	 &	smooth    &	          &	1-111 T-4390 p T^2-111 p^3 T^3+p^6 T^4
\tabularnewline[0.5pt]\hline				
\tablepostamble				
\tablepreamble{73}				
1	 &	smooth    &	          &	1+960 T+11038 p T^2+960 p^3 T^3+p^6 T^4
\tabularnewline[0.5pt]\hline				
2	 &	smooth    &	          &	1+420 T+1718 p T^2+420 p^3 T^3+p^6 T^4
\tabularnewline[0.5pt]\hline				
3	 &	smooth    &	          &	1-1475 T+16252 p T^2-1475 p^3 T^3+p^6 T^4
\tabularnewline[0.5pt]\hline				
4	 &	smooth    &	          &	1-655 T+8820 p T^2-655 p^3 T^3+p^6 T^4
\tabularnewline[0.5pt]\hline				
5	 &	smooth    &	          &	1+480 T+3678 p T^2+480 p^3 T^3+p^6 T^4
\tabularnewline[0.5pt]\hline				
6	 &	smooth    &	          &	1+170 T+7930 p T^2+170 p^3 T^3+p^6 T^4
\tabularnewline[0.5pt]\hline				
7	 &	smooth    &	          &	1+355 T+5084 p T^2+355 p^3 T^3+p^6 T^4
\tabularnewline[0.5pt]\hline				
8	 &	smooth    &	          &	1+90 T-7142 p T^2+90 p^3 T^3+p^6 T^4
\tabularnewline[0.5pt]\hline				
9	 &	smooth    &	          &	1+1085 T+156 p^2 T^2+1085 p^3 T^3+p^6 T^4
\tabularnewline[0.5pt]\hline				
10	 &	smooth    &	          &	1-400 T+30 p^2 T^2-400 p^3 T^3+p^6 T^4
\tabularnewline[0.5pt]\hline				
11	 &	smooth    &	          &	1+690 T+114 p^2 T^2+690 p^3 T^3+p^6 T^4
\tabularnewline[0.5pt]\hline				
12	 &	smooth    &	          &	1+710 T+7874 p T^2+710 p^3 T^3+p^6 T^4
\tabularnewline[0.5pt]\hline				
13	 &	smooth    &	          &	1-390 T+3490 p T^2-390 p^3 T^3+p^6 T^4
\tabularnewline[0.5pt]\hline				
14	 &	smooth    &	          &	1+320 T+7326 p T^2+320 p^3 T^3+p^6 T^4
\tabularnewline[0.5pt]\hline				
15	 &	smooth    &	          &	1+675 T+1212 p T^2+675 p^3 T^3+p^6 T^4
\tabularnewline[0.5pt]\hline				
16	 &	smooth    &	          &	1-1155 T+14588 p T^2-1155 p^3 T^3+p^6 T^4
\tabularnewline[0.5pt]\hline				
17	 &	smooth    &	          &	1-1155 T+10176 p T^2-1155 p^3 T^3+p^6 T^4
\tabularnewline[0.5pt]\hline				
18	 &	smooth    &	          &	1+330 T-3142 p T^2+330 p^3 T^3+p^6 T^4
\tabularnewline[0.5pt]\hline				
19	 &	smooth    &	          &	1-390 T+58 p T^2-390 p^3 T^3+p^6 T^4
\tabularnewline[0.5pt]\hline				
20	 &	smooth    &	          &	1+40 T-7250 p T^2+40 p^3 T^3+p^6 T^4
\tabularnewline[0.5pt]\hline				
21	 &	smooth    &	          &	1-605 T+8988 p T^2-605 p^3 T^3+p^6 T^4
\tabularnewline[0.5pt]\hline				
22	 &	smooth    &	          &	1-420 T+3702 p T^2-420 p^3 T^3+p^6 T^4
\tabularnewline[0.5pt]\hline				
23	 &	smooth    &	          &	(1-10 p T+p^3 T^2)(1+995 T+p^3 T^2)
\tabularnewline[0.5pt]\hline				
24	 &	smooth    &	          &	1-70 T+8410 p T^2-70 p^3 T^3+p^6 T^4
\tabularnewline[0.5pt]\hline				
25	 &	smooth    &	          &	1+55 T+5624 p T^2+55 p^3 T^3+p^6 T^4
\tabularnewline[0.5pt]\hline				
26	 &	smooth    &	          &	1-415 T-2584 p T^2-415 p^3 T^3+p^6 T^4
\tabularnewline[0.5pt]\hline				
27	 &	smooth    &	          &	1-540 T+1078 p T^2-540 p^3 T^3+p^6 T^4
\tabularnewline[0.5pt]\hline				
28	 &	smooth    &	          &	1-575 T+4776 p T^2-575 p^3 T^3+p^6 T^4
\tabularnewline[0.5pt]\hline				
29	 &	smooth    &	          &	1-80 T-7250 p T^2-80 p^3 T^3+p^6 T^4
\tabularnewline[0.5pt]\hline				
30	 &	smooth    &	          &	1-100 T+4918 p T^2-100 p^3 T^3+p^6 T^4
\tabularnewline[0.5pt]\hline				
31	 &	smooth    &	          &	1-125 T-1316 p T^2-125 p^3 T^3+p^6 T^4
\tabularnewline[0.5pt]\hline				
32	 &	smooth    &	          &	1+270 T+10770 p T^2+270 p^3 T^3+p^6 T^4
\tabularnewline[0.5pt]\hline				
33	 &	smooth    &	          &	1+610 T+11842 p T^2+610 p^3 T^3+p^6 T^4
\tabularnewline[0.5pt]\hline				
34	 &	smooth    &	          &	1+620 T+10838 p T^2+620 p^3 T^3+p^6 T^4
\tabularnewline[0.5pt]\hline				
35	 &	smooth    &	          &	1-3234 p T^2+p^6 T^4
\tabularnewline[0.5pt]\hline				
36	 &	smooth    &	          &	1+705 T+2260 p T^2+705 p^3 T^3+p^6 T^4
\tabularnewline[0.5pt]\hline				
37	 &	smooth    &	          &	1-230 T-1254 p T^2-230 p^3 T^3+p^6 T^4
\tabularnewline[0.5pt]\hline				
38	 &	smooth    &	          &	1-1015 T+8708 p T^2-1015 p^3 T^3+p^6 T^4
\tabularnewline[0.5pt]\hline				
39	 &	smooth    &	          &	1+80 T+1566 p T^2+80 p^3 T^3+p^6 T^4
\tabularnewline[0.5pt]\hline				
40	 &	smooth    &	          &	1+645 T+5904 p T^2+645 p^3 T^3+p^6 T^4
\tabularnewline[0.5pt]\hline				
41	 &	smooth    &	          &	1+790 T+10370 p T^2+790 p^3 T^3+p^6 T^4
\tabularnewline[0.5pt]\hline				
42	 &	smooth    &	          &	1+1210 T+15394 p T^2+1210 p^3 T^3+p^6 T^4
\tabularnewline[0.5pt]\hline				
43	 &	smooth    &	          &	1+240 T-5474 p T^2+240 p^3 T^3+p^6 T^4
\tabularnewline[0.5pt]\hline				
44	 &	smooth    &	          &	1+130 T-2638 p T^2+130 p^3 T^3+p^6 T^4
\tabularnewline[0.5pt]\hline				
45	 &	smooth    &	          &	1+20 T+2438 p T^2+20 p^3 T^3+p^6 T^4
\tabularnewline[0.5pt]\hline				
46	 &	smooth    &	          &	1+1130 T+14298 p T^2+1130 p^3 T^3+p^6 T^4
\tabularnewline[0.5pt]\hline				
47	 &	smooth    &	          &	1-1660 T+19398 p T^2-1660 p^3 T^3+p^6 T^4
\tabularnewline[0.5pt]\hline				
48	 &	smooth    &	          &	1+45 T+5852 p T^2+45 p^3 T^3+p^6 T^4
\tabularnewline[0.5pt]\hline				
49	 &	smooth    &	          &	1+100 T+6070 p T^2+100 p^3 T^3+p^6 T^4
\tabularnewline[0.5pt]\hline				
50	 &	smooth    &	          &	1+1040 T+11006 p T^2+1040 p^3 T^3+p^6 T^4
\tabularnewline[0.5pt]\hline				
51	 &	smooth    &	          &	1+120 T+590 p T^2+120 p^3 T^3+p^6 T^4
\tabularnewline[0.5pt]\hline				
52	 &	smooth    &	          &	1+1080 T+10846 p T^2+1080 p^3 T^3+p^6 T^4
\tabularnewline[0.5pt]\hline				
53	 &	smooth    &	          &	1+900 T+8022 p T^2+900 p^3 T^3+p^6 T^4
\tabularnewline[0.5pt]\hline				
54	 &	smooth    &	          &	1+660 T+10198 p T^2+660 p^3 T^3+p^6 T^4
\tabularnewline[0.5pt]\hline				
55	 &	smooth    &	          &	1+870 T+9122 p T^2+870 p^3 T^3+p^6 T^4
\tabularnewline[0.5pt]\hline				
56	 &	smooth    &	          &	1-30 T+1890 p T^2-30 p^3 T^3+p^6 T^4
\tabularnewline[0.5pt]\hline				
57	 &	smooth    &	          &	1-735 T+5652 p T^2-735 p^3 T^3+p^6 T^4
\tabularnewline[0.5pt]\hline				
58	 &	smooth    &	          &	1-80 T+9598 p T^2-80 p^3 T^3+p^6 T^4
\tabularnewline[0.5pt]\hline				
59	 &	smooth    &	          &	1+595 T+3068 p T^2+595 p^3 T^3+p^6 T^4
\tabularnewline[0.5pt]\hline				
60	 &	smooth    &	          &	1+45 T+6336 p T^2+45 p^3 T^3+p^6 T^4
\tabularnewline[0.5pt]\hline				
61	 &	smooth    &	          &	1-1300 T+13670 p T^2-1300 p^3 T^3+p^6 T^4
\tabularnewline[0.5pt]\hline				
62	 &	smooth    &	          &	1+805 T+10224 p T^2+805 p^3 T^3+p^6 T^4
\tabularnewline[0.5pt]\hline				
63	 &	smooth    &	          &	1+250 T+10162 p T^2+250 p^3 T^3+p^6 T^4
\tabularnewline[0.5pt]\hline				
64	 &	smooth    &	          &	1-175 T-2508 p T^2-175 p^3 T^3+p^6 T^4
\tabularnewline[0.5pt]\hline				
65	 &	smooth    &	          &	1-370 T+722 p T^2-370 p^3 T^3+p^6 T^4
\tabularnewline[0.5pt]\hline				
66	 &	smooth    &	          &	1-790 T+5954 p T^2-790 p^3 T^3+p^6 T^4
\tabularnewline[0.5pt]\hline				
67	 &	smooth    &	          &	1+1340 T+13670 p T^2+1340 p^3 T^3+p^6 T^4
\tabularnewline[0.5pt]\hline				
68	 &	smooth    &	          &	1-5 p T-3908 p T^2-5 p^4 T^3+p^6 T^4
\tabularnewline[0.5pt]\hline				
69	 &	smooth    &	          &	1-630 T+3706 p T^2-630 p^3 T^3+p^6 T^4
\tabularnewline[0.5pt]\hline				
70	 &	smooth    &	          &	1+855 T+4984 p T^2+855 p^3 T^3+p^6 T^4
\tabularnewline[0.5pt]\hline				
71	 &	smooth    &	          &	1-80 T-2 p T^2-80 p^3 T^3+p^6 T^4
\tabularnewline[0.5pt]\hline				
72	 &	smooth    &	          &	1-460 T+5398 p T^2-460 p^3 T^3+p^6 T^4
\tabularnewline[0.5pt]\hline				
\tablepostamble				
\tablepreamble{79}				
1	 &	smooth    &	          &	1+792 T+13602 p T^2+792 p^3 T^3+p^6 T^4
\tabularnewline[0.5pt]\hline				
2	 &	smooth    &	          &	1-404 T+3714 p T^2-404 p^3 T^3+p^6 T^4
\tabularnewline[0.5pt]\hline				
3	 &	smooth    &	          &	1-326 T+18 p T^2-326 p^3 T^3+p^6 T^4
\tabularnewline[0.5pt]\hline				
4	 &	smooth    &	          &	1+1097 T+13042 p T^2+1097 p^3 T^3+p^6 T^4
\tabularnewline[0.5pt]\hline				
5	 &	smooth    &	          &	1-198 T-2718 p T^2-198 p^3 T^3+p^6 T^4
\tabularnewline[0.5pt]\hline				
6	 &	smooth    &	          &	1-21 T+12402 p T^2-21 p^3 T^3+p^6 T^4
\tabularnewline[0.5pt]\hline				
7	 &	smooth    &	          &	1-226 T+9874 p T^2-226 p^3 T^3+p^6 T^4
\tabularnewline[0.5pt]\hline				
8	 &	smooth    &	          &	1-582 T+6242 p T^2-582 p^3 T^3+p^6 T^4
\tabularnewline[0.5pt]\hline				
9	 &	smooth    &	          &	1-349 T+10642 p T^2-349 p^3 T^3+p^6 T^4
\tabularnewline[0.5pt]\hline				
10	 &	smooth    &	          &	1+1290 T+13986 p T^2+1290 p^3 T^3+p^6 T^4
\tabularnewline[0.5pt]\hline				
11	 &	smooth    &	          &	1+1167 T+14386 p T^2+1167 p^3 T^3+p^6 T^4
\tabularnewline[0.5pt]\hline				
12	 &	smooth    &	          &	1-184 T+5202 p T^2-184 p^3 T^3+p^6 T^4
\tabularnewline[0.5pt]\hline				
13	 &	smooth    &	          &	1-504 T+7234 p T^2-504 p^3 T^3+p^6 T^4
\tabularnewline[0.5pt]\hline				
14	 &	smooth    &	          &	1+821 T+5714 p T^2+821 p^3 T^3+p^6 T^4
\tabularnewline[0.5pt]\hline				
15	 &	smooth    &	          &	1-858 T+11042 p T^2-858 p^3 T^3+p^6 T^4
\tabularnewline[0.5pt]\hline				
16	 &	smooth    &	          &	1+258 T+5922 p T^2+258 p^3 T^3+p^6 T^4
\tabularnewline[0.5pt]\hline				
17	 &	smooth    &	          &	1-198 T+2642 p T^2-198 p^3 T^3+p^6 T^4
\tabularnewline[0.5pt]\hline				
18	 &	smooth    &	          &	1-950 T+9026 p T^2-950 p^3 T^3+p^6 T^4
\tabularnewline[0.5pt]\hline				
19	 &	smooth    &	          &	1+478 T+10530 p T^2+478 p^3 T^3+p^6 T^4
\tabularnewline[0.5pt]\hline				
20	 &	smooth    &	          &	1+455 T-718 p T^2+455 p^3 T^3+p^6 T^4
\tabularnewline[0.5pt]\hline				
21	 &	smooth    &	          &	1-355 T+1586 p T^2-355 p^3 T^3+p^6 T^4
\tabularnewline[0.5pt]\hline				
22	 &	smooth    &	          &	1+1350 T+12706 p T^2+1350 p^3 T^3+p^6 T^4
\tabularnewline[0.5pt]\hline				
23	 &	smooth    &	          &	1-206 T+4994 p T^2-206 p^3 T^3+p^6 T^4
\tabularnewline[0.5pt]\hline				
24	 &	smooth    &	          &	1+1886 T+22738 p T^2+1886 p^3 T^3+p^6 T^4
\tabularnewline[0.5pt]\hline				
25	 &	singular  &	25&	(1-p T) (1+48 T+p^3 T^2)
\tabularnewline[0.5pt]\hline				
26	 &	smooth    &	          &	1-1143 T+12370 p T^2-1143 p^3 T^3+p^6 T^4
\tabularnewline[0.5pt]\hline				
27	 &	smooth    &	          &	1+708 T+11970 p T^2+708 p^3 T^3+p^6 T^4
\tabularnewline[0.5pt]\hline				
28	 &	smooth    &	          &	1+1088 T+11330 p T^2+1088 p^3 T^3+p^6 T^4
\tabularnewline[0.5pt]\hline				
29	 &	smooth    &	          &	1-162 T+2466 p T^2-162 p^3 T^3+p^6 T^4
\tabularnewline[0.5pt]\hline				
30	 &	smooth    &	          &	1-809 T+10994 p T^2-809 p^3 T^3+p^6 T^4
\tabularnewline[0.5pt]\hline				
31	 &	smooth    &	          &	1+240 T-1214 p T^2+240 p^3 T^3+p^6 T^4
\tabularnewline[0.5pt]\hline				
32	 &	smooth    &	          &	1-558 T+9890 p T^2-558 p^3 T^3+p^6 T^4
\tabularnewline[0.5pt]\hline				
33	 &	smooth    &	          &	1+482 T+722 p T^2+482 p^3 T^3+p^6 T^4
\tabularnewline[0.5pt]\hline				
34	 &	smooth    &	          &	1+204 T+3954 p T^2+204 p^3 T^3+p^6 T^4
\tabularnewline[0.5pt]\hline				
35	 &	smooth    &	          &	1-7 T+6130 p T^2-7 p^3 T^3+p^6 T^4
\tabularnewline[0.5pt]\hline				
36	 &	smooth    &	          &	1-304 T-6846 p T^2-304 p^3 T^3+p^6 T^4
\tabularnewline[0.5pt]\hline				
37	 &	smooth    &	          &	1+506 T-622 p T^2+506 p^3 T^3+p^6 T^4
\tabularnewline[0.5pt]\hline				
38	 &	smooth    &	          &	1+727 T+3250 p T^2+727 p^3 T^3+p^6 T^4
\tabularnewline[0.5pt]\hline				
39	 &	smooth    &	          &	1-354 T+3442 p T^2-354 p^3 T^3+p^6 T^4
\tabularnewline[0.5pt]\hline				
40	 &	smooth    &	          &	1+110 T-4702 p T^2+110 p^3 T^3+p^6 T^4
\tabularnewline[0.5pt]\hline				
41	 &	smooth    &	          &	1+448 T+770 p T^2+448 p^3 T^3+p^6 T^4
\tabularnewline[0.5pt]\hline				
42	 &	smooth    &	          &	1+1110 T+11778 p T^2+1110 p^3 T^3+p^6 T^4
\tabularnewline[0.5pt]\hline				
43	 &	smooth    &	          &	1+319 T-3086 p T^2+319 p^3 T^3+p^6 T^4
\tabularnewline[0.5pt]\hline				
44	 &	smooth    &	          &	1+264 T+4802 p T^2+264 p^3 T^3+p^6 T^4
\tabularnewline[0.5pt]\hline				
45	 &	smooth    &	          &	1+402 T+3362 p T^2+402 p^3 T^3+p^6 T^4
\tabularnewline[0.5pt]\hline				
46	 &	smooth    &	          &	1-664 T+3202 p T^2-664 p^3 T^3+p^6 T^4
\tabularnewline[0.5pt]\hline				
47	 &	smooth    &	          &	1+122 T+10530 p T^2+122 p^3 T^3+p^6 T^4
\tabularnewline[0.5pt]\hline				
48	 &	smooth    &	          &	1-1385 T+16178 p T^2-1385 p^3 T^3+p^6 T^4
\tabularnewline[0.5pt]\hline				
49	 &	smooth    &	          &	1+173 T+2610 p T^2+173 p^3 T^3+p^6 T^4
\tabularnewline[0.5pt]\hline				
50	 &	smooth    &	          &	1-346 T-4318 p T^2-346 p^3 T^3+p^6 T^4
\tabularnewline[0.5pt]\hline				
51	 &	smooth    &	          &	1+278 T+7522 p T^2+278 p^3 T^3+p^6 T^4
\tabularnewline[0.5pt]\hline				
52	 &	smooth    &	          &	1-64 T-3678 p T^2-64 p^3 T^3+p^6 T^4
\tabularnewline[0.5pt]\hline				
53	 &	smooth    &	          &	1-143 T+8082 p T^2-143 p^3 T^3+p^6 T^4
\tabularnewline[0.5pt]\hline				
54	 &	smooth    &	          &	1-488 T+2962 p T^2-488 p^3 T^3+p^6 T^4
\tabularnewline[0.5pt]\hline				
55	 &	smooth    &	          &	1-338 T+6370 p T^2-338 p^3 T^3+p^6 T^4
\tabularnewline[0.5pt]\hline				
56	 &	smooth    &	          &	1+394 T-286 p T^2+394 p^3 T^3+p^6 T^4
\tabularnewline[0.5pt]\hline				
57	 &	smooth    &	          &	1+294 T+2994 p T^2+294 p^3 T^3+p^6 T^4
\tabularnewline[0.5pt]\hline				
58	 &	smooth    &	          &	1+735 T+10482 p T^2+735 p^3 T^3+p^6 T^4
\tabularnewline[0.5pt]\hline				
59	 &	smooth    &	          &	1-379 T+3058 p T^2-379 p^3 T^3+p^6 T^4
\tabularnewline[0.5pt]\hline				
60	 &	smooth    &	          &	1-373 T+7602 p T^2-373 p^3 T^3+p^6 T^4
\tabularnewline[0.5pt]\hline				
61	 &	smooth    &	          &	1+137 T+690 p T^2+137 p^3 T^3+p^6 T^4
\tabularnewline[0.5pt]\hline				
62	 &	smooth    &	          &	1+228 T+3010 p T^2+228 p^3 T^3+p^6 T^4
\tabularnewline[0.5pt]\hline				
63	 &	smooth    &	          &	1+744 T+4482 p T^2+744 p^3 T^3+p^6 T^4
\tabularnewline[0.5pt]\hline				
64	 &	smooth    &	          &	1+464 T+6594 p T^2+464 p^3 T^3+p^6 T^4
\tabularnewline[0.5pt]\hline				
65	 &	smooth    &	          &	1-193 T+11890 p T^2-193 p^3 T^3+p^6 T^4
\tabularnewline[0.5pt]\hline				
66	 &	smooth    &	          &	1-1118 T+12962 p T^2-1118 p^3 T^3+p^6 T^4
\tabularnewline[0.5pt]\hline				
67	 &	smooth    &	          &	1-879 T+12914 p T^2-879 p^3 T^3+p^6 T^4
\tabularnewline[0.5pt]\hline				
68	 &	smooth    &	          &	(1-8 p T+p^3 T^2)(1+1368 T+p^3 T^2)
\tabularnewline[0.5pt]\hline				
69	 &	smooth    &	          &	1+238 T-2334 p T^2+238 p^3 T^3+p^6 T^4
\tabularnewline[0.5pt]\hline				
70	 &	smooth    &	          &	1-466 T+9698 p T^2-466 p^3 T^3+p^6 T^4
\tabularnewline[0.5pt]\hline				
71	 &	smooth    &	          &	1-968 T+10050 p T^2-968 p^3 T^3+p^6 T^4
\tabularnewline[0.5pt]\hline				
72	 &	smooth    &	          &	1+467 T+6322 p T^2+467 p^3 T^3+p^6 T^4
\tabularnewline[0.5pt]\hline				
73	 &	smooth    &	          &	1+30 T+6626 p T^2+30 p^3 T^3+p^6 T^4
\tabularnewline[0.5pt]\hline				
74	 &	smooth    &	          &	1+1578 T+17122 p T^2+1578 p^3 T^3+p^6 T^4
\tabularnewline[0.5pt]\hline				
75	 &	smooth    &	          &	1+929 T+11154 p T^2+929 p^3 T^3+p^6 T^4
\tabularnewline[0.5pt]\hline				
76	 &	smooth    &	          &	1-191 T-3566 p T^2-191 p^3 T^3+p^6 T^4
\tabularnewline[0.5pt]\hline				
77	 &	smooth    &	          &	1-656 T+4882 p T^2-656 p^3 T^3+p^6 T^4
\tabularnewline[0.5pt]\hline				
78	 &	singular  &	78&	(1-p T) (1-48 T+p^3 T^2)
\tabularnewline[0.5pt]\hline				
\tablepostamble				
\tablepreamble{83}				
1	 &	smooth    &	          &	1+960 T+4514 p T^2+960 p^3 T^3+p^6 T^4
\tabularnewline[0.5pt]\hline				
2	 &	smooth    &	          &	1+300 T-4030 p T^2+300 p^3 T^3+p^6 T^4
\tabularnewline[0.5pt]\hline				
3	 &	smooth    &	          &	1-570 T+3706 p T^2-570 p^3 T^3+p^6 T^4
\tabularnewline[0.5pt]\hline				
4	 &	smooth    &	          &	1-320 T-2238 p T^2-320 p^3 T^3+p^6 T^4
\tabularnewline[0.5pt]\hline				
5	 &	smooth    &	          &	1-610 T+3642 p T^2-610 p^3 T^3+p^6 T^4
\tabularnewline[0.5pt]\hline				
6	 &	smooth    &	          &	1+170 T+8010 p T^2+170 p^3 T^3+p^6 T^4
\tabularnewline[0.5pt]\hline				
7	 &	smooth    &	          &	1+180 T+1586 p T^2+180 p^3 T^3+p^6 T^4
\tabularnewline[0.5pt]\hline				
8	 &	smooth    &	          &	1+1515 T+17638 p T^2+1515 p^3 T^3+p^6 T^4
\tabularnewline[0.5pt]\hline				
9	 &	smooth    &	          &	1+610 T+3690 p T^2+610 p^3 T^3+p^6 T^4
\tabularnewline[0.5pt]\hline				
10	 &	smooth    &	          &	1-235 T-7050 p T^2-235 p^3 T^3+p^6 T^4
\tabularnewline[0.5pt]\hline				
11	 &	smooth    &	          &	1+1575 T+16574 p T^2+1575 p^3 T^3+p^6 T^4
\tabularnewline[0.5pt]\hline				
12	 &	smooth    &	          &	1-110 T+3498 p T^2-110 p^3 T^3+p^6 T^4
\tabularnewline[0.5pt]\hline				
13	 &	smooth    &	          &	1+50 T+12970 p T^2+50 p^3 T^3+p^6 T^4
\tabularnewline[0.5pt]\hline				
14	 &	smooth    &	          &	1+295 T+4566 p T^2+295 p^3 T^3+p^6 T^4
\tabularnewline[0.5pt]\hline				
15	 &	smooth    &	          &	1+1720 T+18978 p T^2+1720 p^3 T^3+p^6 T^4
\tabularnewline[0.5pt]\hline				
16	 &	smooth    &	          &	1+520 T+9922 p T^2+520 p^3 T^3+p^6 T^4
\tabularnewline[0.5pt]\hline				
17	 &	smooth    &	          &	1-605 T+14062 p T^2-605 p^3 T^3+p^6 T^4
\tabularnewline[0.5pt]\hline				
18	 &	smooth    &	          &	1+275 T+8326 p T^2+275 p^3 T^3+p^6 T^4
\tabularnewline[0.5pt]\hline				
19	 &	smooth    &	          &	1+340 T+13410 p T^2+340 p^3 T^3+p^6 T^4
\tabularnewline[0.5pt]\hline				
20	 &	smooth    &	          &	1+1058 p T^2+p^6 T^4
\tabularnewline[0.5pt]\hline				
21	 &	smooth    &	          &	1-590 T+10666 p T^2-590 p^3 T^3+p^6 T^4
\tabularnewline[0.5pt]\hline				
22	 &	smooth    &	          &	1+540 T+7154 p T^2+540 p^3 T^3+p^6 T^4
\tabularnewline[0.5pt]\hline				
23	 &	smooth    &	          &	1+860 T+6930 p T^2+860 p^3 T^3+p^6 T^4
\tabularnewline[0.5pt]\hline				
24	 &	smooth    &	          &	1+675 T+8166 p T^2+675 p^3 T^3+p^6 T^4
\tabularnewline[0.5pt]\hline				
25	 &	smooth    &	          &	1-10 T-1286 p T^2-10 p^3 T^3+p^6 T^4
\tabularnewline[0.5pt]\hline				
26	 &	smooth    &	          &	1-590 T+5162 p T^2-590 p^3 T^3+p^6 T^4
\tabularnewline[0.5pt]\hline				
27	 &	smooth    &	          &	1+35 T+11054 p T^2+35 p^3 T^3+p^6 T^4
\tabularnewline[0.5pt]\hline				
28	 &	smooth    &	          &	1+600 T+1794 p T^2+600 p^3 T^3+p^6 T^4
\tabularnewline[0.5pt]\hline				
29	 &	smooth    &	          &	1-1525 T+16334 p T^2-1525 p^3 T^3+p^6 T^4
\tabularnewline[0.5pt]\hline				
30	 &	smooth    &	          &	1-1360 T+14562 p T^2-1360 p^3 T^3+p^6 T^4
\tabularnewline[0.5pt]\hline				
31	 &	smooth    &	          &	1-310 T+2250 p T^2-310 p^3 T^3+p^6 T^4
\tabularnewline[0.5pt]\hline				
32	 &	smooth    &	          &	1-555 T+3070 p T^2-555 p^3 T^3+p^6 T^4
\tabularnewline[0.5pt]\hline				
33	 &	smooth    &	          &	1+235 T+3374 p T^2+235 p^3 T^3+p^6 T^4
\tabularnewline[0.5pt]\hline				
34	 &	smooth    &	          &	1-475 T+11902 p T^2-475 p^3 T^3+p^6 T^4
\tabularnewline[0.5pt]\hline				
35	 &	smooth    &	          &	1+930 T+14714 p T^2+930 p^3 T^3+p^6 T^4
\tabularnewline[0.5pt]\hline				
36	 &	smooth    &	          &	1+175 T+4542 p T^2+175 p^3 T^3+p^6 T^4
\tabularnewline[0.5pt]\hline				
37	 &	smooth    &	          &	1+1430 T+13946 p T^2+1430 p^3 T^3+p^6 T^4
\tabularnewline[0.5pt]\hline				
38	 &	smooth    &	          &	1+770 T+6954 p T^2+770 p^3 T^3+p^6 T^4
\tabularnewline[0.5pt]\hline				
39	 &	smooth    &	          &	1+730 T+8026 p T^2+730 p^3 T^3+p^6 T^4
\tabularnewline[0.5pt]\hline				
40	 &	smooth    &	          &	1+955 T+14542 p T^2+955 p^3 T^3+p^6 T^4
\tabularnewline[0.5pt]\hline				
41	 &	smooth    &	          &	1-45 T+11054 p T^2-45 p^3 T^3+p^6 T^4
\tabularnewline[0.5pt]\hline				
42	 &	smooth    &	          &	1+15 T+9846 p T^2+15 p^3 T^3+p^6 T^4
\tabularnewline[0.5pt]\hline				
43	 &	smooth    &	          &	1+160 T+5282 p T^2+160 p^3 T^3+p^6 T^4
\tabularnewline[0.5pt]\hline				
44	 &	smooth    &	          &	1-355 T-5482 p T^2-355 p^3 T^3+p^6 T^4
\tabularnewline[0.5pt]\hline				
45	 &	smooth    &	          &	1+1405 T+13342 p T^2+1405 p^3 T^3+p^6 T^4
\tabularnewline[0.5pt]\hline				
46	 &	smooth    &	          &	1-1290 T+16346 p T^2-1290 p^3 T^3+p^6 T^4
\tabularnewline[0.5pt]\hline				
47	 &	smooth    &	          &	1-160 T+226 p T^2-160 p^3 T^3+p^6 T^4
\tabularnewline[0.5pt]\hline				
48	 &	smooth    &	          &	1+840 T+13442 p T^2+840 p^3 T^3+p^6 T^4
\tabularnewline[0.5pt]\hline				
49	 &	smooth    &	          &	1+4642 p T^2+p^6 T^4
\tabularnewline[0.5pt]\hline				
50	 &	smooth    &	          &	1-180 T+1250 p T^2-180 p^3 T^3+p^6 T^4
\tabularnewline[0.5pt]\hline				
51	 &	smooth    &	          &	1+510 T+10330 p T^2+510 p^3 T^3+p^6 T^4
\tabularnewline[0.5pt]\hline				
52	 &	smooth    &	          &	1-245 T-3866 p T^2-245 p^3 T^3+p^6 T^4
\tabularnewline[0.5pt]\hline				
53	 &	smooth    &	          &	1+1395 T+18278 p T^2+1395 p^3 T^3+p^6 T^4
\tabularnewline[0.5pt]\hline				
54	 &	smooth    &	          &	1-1190 T+9450 p T^2-1190 p^3 T^3+p^6 T^4
\tabularnewline[0.5pt]\hline				
55	 &	smooth    &	          &	1-225 T+598 p T^2-225 p^3 T^3+p^6 T^4
\tabularnewline[0.5pt]\hline				
56	 &	smooth    &	          &	1-500 T+2466 p T^2-500 p^3 T^3+p^6 T^4
\tabularnewline[0.5pt]\hline				
57	 &	smooth    &	          &	1-1990 T+24218 p T^2-1990 p^3 T^3+p^6 T^4
\tabularnewline[0.5pt]\hline				
58	 &	smooth    &	          &	1-530 T+9034 p T^2-530 p^3 T^3+p^6 T^4
\tabularnewline[0.5pt]\hline				
59	 &	smooth    &	          &	1-65 T-12066 p T^2-65 p^3 T^3+p^6 T^4
\tabularnewline[0.5pt]\hline				
60	 &	smooth    &	          &	1+1035 T+11206 p T^2+1035 p^3 T^3+p^6 T^4
\tabularnewline[0.5pt]\hline				
61	 &	smooth    &	          &	1+200 T+5378 p T^2+200 p^3 T^3+p^6 T^4
\tabularnewline[0.5pt]\hline				
62	 &	smooth    &	          &	1-80 T+7378 p T^2-80 p^3 T^3+p^6 T^4
\tabularnewline[0.5pt]\hline				
63	 &	smooth    &	          &	1-330 T+8090 p T^2-330 p^3 T^3+p^6 T^4
\tabularnewline[0.5pt]\hline				
64	 &	smooth    &	          &	1+110 T+2458 p T^2+110 p^3 T^3+p^6 T^4
\tabularnewline[0.5pt]\hline				
65	 &	smooth    &	          &	1-140 T+9554 p T^2-140 p^3 T^3+p^6 T^4
\tabularnewline[0.5pt]\hline				
66	 &	smooth    &	          &	1-1400 T+17378 p T^2-1400 p^3 T^3+p^6 T^4
\tabularnewline[0.5pt]\hline				
67	 &	smooth    &	          &	1+1310 T+13690 p T^2+1310 p^3 T^3+p^6 T^4
\tabularnewline[0.5pt]\hline				
68	 &	smooth    &	          &	1-75 T+10518 p T^2-75 p^3 T^3+p^6 T^4
\tabularnewline[0.5pt]\hline				
69	 &	smooth    &	          &	1+350 T+7162 p T^2+350 p^3 T^3+p^6 T^4
\tabularnewline[0.5pt]\hline				
70	 &	smooth    &	          &	1-965 T+9806 p T^2-965 p^3 T^3+p^6 T^4
\tabularnewline[0.5pt]\hline				
71	 &	smooth    &	          &	1+160 T+4146 p T^2+160 p^3 T^3+p^6 T^4
\tabularnewline[0.5pt]\hline				
72	 &	smooth    &	          &	1-1330 T+12298 p T^2-1330 p^3 T^3+p^6 T^4
\tabularnewline[0.5pt]\hline				
73	 &	smooth    &	          &	1+510 T+10010 p T^2+510 p^3 T^3+p^6 T^4
\tabularnewline[0.5pt]\hline				
74	 &	smooth    &	          &	1-5 T-474 p T^2-5 p^3 T^3+p^6 T^4
\tabularnewline[0.5pt]\hline				
75	 &	smooth    &	          &	1-120 T-4862 p T^2-120 p^3 T^3+p^6 T^4
\tabularnewline[0.5pt]\hline				
76	 &	smooth    &	          &	1+400 T-798 p T^2+400 p^3 T^3+p^6 T^4
\tabularnewline[0.5pt]\hline				
77	 &	smooth    &	          &	1-340 T-3310 p T^2-340 p^3 T^3+p^6 T^4
\tabularnewline[0.5pt]\hline				
78	 &	smooth    &	          &	1+780 T+5330 p T^2+780 p^3 T^3+p^6 T^4
\tabularnewline[0.5pt]\hline				
79	 &	smooth    &	          &	1-250 T+8234 p T^2-250 p^3 T^3+p^6 T^4
\tabularnewline[0.5pt]\hline				
80	 &	smooth    &	          &	1+440 T+4178 p T^2+440 p^3 T^3+p^6 T^4
\tabularnewline[0.5pt]\hline				
81	 &	smooth    &	          &	1+410 T-2870 p T^2+410 p^3 T^3+p^6 T^4
\tabularnewline[0.5pt]\hline				
82	 &	smooth    &	          &	1+90 T+5402 p T^2+90 p^3 T^3+p^6 T^4
\tabularnewline[0.5pt]\hline				
\tablepostamble				
\tablepreamble{89}				
1	 &	smooth    &	          &	1-352 T+3742 p T^2-352 p^3 T^3+p^6 T^4
\tabularnewline[0.5pt]\hline				
2	 &	smooth    &	          &	1+37 T+1116 p T^2+37 p^3 T^3+p^6 T^4
\tabularnewline[0.5pt]\hline				
3	 &	smooth    &	          &	1+1668 T+19446 p T^2+1668 p^3 T^3+p^6 T^4
\tabularnewline[0.5pt]\hline				
4	 &	smooth    &	          &	1+290 T+5738 p T^2+290 p^3 T^3+p^6 T^4
\tabularnewline[0.5pt]\hline				
5	 &	smooth    &	          &	1-724 T+10982 p T^2-724 p^3 T^3+p^6 T^4
\tabularnewline[0.5pt]\hline				
6	 &	smooth    &	          &	1+278 T+10890 p T^2+278 p^3 T^3+p^6 T^4
\tabularnewline[0.5pt]\hline				
7	 &	smooth    &	          &	1+40 T-34 p T^2+40 p^3 T^3+p^6 T^4
\tabularnewline[0.5pt]\hline				
8	 &	smooth    &	          &	1+541 T+12204 p T^2+541 p^3 T^3+p^6 T^4
\tabularnewline[0.5pt]\hline				
9	 &	smooth    &	          &	1+172 T+5606 p T^2+172 p^3 T^3+p^6 T^4
\tabularnewline[0.5pt]\hline				
10	 &	smooth    &	          &	1-360 T+12142 p T^2-360 p^3 T^3+p^6 T^4
\tabularnewline[0.5pt]\hline				
11	 &	smooth    &	          &	1+30 T+8978 p T^2+30 p^3 T^3+p^6 T^4
\tabularnewline[0.5pt]\hline				
12	 &	smooth    &	          &	1+2813 T+37712 p T^2+2813 p^3 T^3+p^6 T^4
\tabularnewline[0.5pt]\hline				
13	 &	smooth    &	          &	1-6 T+2114 p T^2-6 p^3 T^3+p^6 T^4
\tabularnewline[0.5pt]\hline				
14	 &	smooth    &	          &	(1+2 p T+p^3 T^2)(1-1054 T+p^3 T^2)
\tabularnewline[0.5pt]\hline				
15	 &	smooth    &	          &	1+455 T-172 p T^2+455 p^3 T^3+p^6 T^4
\tabularnewline[0.5pt]\hline				
16	 &	smooth    &	          &	1-1145 T+18792 p T^2-1145 p^3 T^3+p^6 T^4
\tabularnewline[0.5pt]\hline				
17	 &	smooth    &	          &	(1-10 p T+p^3 T^2)(1+1410 T+p^3 T^2)
\tabularnewline[0.5pt]\hline				
18	 &	smooth    &	          &	1+1566 T+21362 p T^2+1566 p^3 T^3+p^6 T^4
\tabularnewline[0.5pt]\hline				
19	 &	smooth    &	          &	1+562 T+5282 p T^2+562 p^3 T^3+p^6 T^4
\tabularnewline[0.5pt]\hline				
20	 &	smooth    &	          &	1+335 T-2408 p T^2+335 p^3 T^3+p^6 T^4
\tabularnewline[0.5pt]\hline				
21	 &	smooth    &	          &	1-523 T+2876 p T^2-523 p^3 T^3+p^6 T^4
\tabularnewline[0.5pt]\hline				
22	 &	smooth    &	          &	1+664 T+15534 p T^2+664 p^3 T^3+p^6 T^4
\tabularnewline[0.5pt]\hline				
23	 &	smooth    &	          &	1-1022 T+17762 p T^2-1022 p^3 T^3+p^6 T^4
\tabularnewline[0.5pt]\hline				
24	 &	smooth    &	          &	1+304 T-6658 p T^2+304 p^3 T^3+p^6 T^4
\tabularnewline[0.5pt]\hline				
25	 &	smooth    &	          &	1-289 T+1272 p T^2-289 p^3 T^3+p^6 T^4
\tabularnewline[0.5pt]\hline				
26	 &	smooth    &	          &	1-123 T-10528 p T^2-123 p^3 T^3+p^6 T^4
\tabularnewline[0.5pt]\hline				
27	 &	smooth    &	          &	1-188 T+7766 p T^2-188 p^3 T^3+p^6 T^4
\tabularnewline[0.5pt]\hline				
28	 &	smooth    &	          &	1+68 T-10554 p T^2+68 p^3 T^3+p^6 T^4
\tabularnewline[0.5pt]\hline				
29	 &	smooth    &	          &	1-1443 T+18192 p T^2-1443 p^3 T^3+p^6 T^4
\tabularnewline[0.5pt]\hline				
30	 &	smooth    &	          &	1-1601 T+18884 p T^2-1601 p^3 T^3+p^6 T^4
\tabularnewline[0.5pt]\hline				
31	 &	smooth    &	          &	1-421 T+11452 p T^2-421 p^3 T^3+p^6 T^4
\tabularnewline[0.5pt]\hline				
32	 &	smooth    &	          &	1+904 T+12814 p T^2+904 p^3 T^3+p^6 T^4
\tabularnewline[0.5pt]\hline				
33	 &	smooth    &	          &	1+640 T+2078 p T^2+640 p^3 T^3+p^6 T^4
\tabularnewline[0.5pt]\hline				
34	 &	smooth    &	          &	1+156 T+262 p T^2+156 p^3 T^3+p^6 T^4
\tabularnewline[0.5pt]\hline				
35	 &	smooth    &	          &	1-402 T-518 p T^2-402 p^3 T^3+p^6 T^4
\tabularnewline[0.5pt]\hline				
36	 &	singular  &	36&	(1+p T) (1+1526 T+p^3 T^2)
\tabularnewline[0.5pt]\hline				
37	 &	smooth    &	          &	1+9886 p T^2+p^6 T^4
\tabularnewline[0.5pt]\hline				
38	 &	smooth    &	          &	1+1493 T+20096 p T^2+1493 p^3 T^3+p^6 T^4
\tabularnewline[0.5pt]\hline				
39	 &	smooth    &	          &	1+1415 T+17896 p T^2+1415 p^3 T^3+p^6 T^4
\tabularnewline[0.5pt]\hline				
40	 &	smooth    &	          &	1+66 T+2762 p T^2+66 p^3 T^3+p^6 T^4
\tabularnewline[0.5pt]\hline				
41	 &	smooth    &	          &	1-505 T-3692 p T^2-505 p^3 T^3+p^6 T^4
\tabularnewline[0.5pt]\hline				
42	 &	smooth    &	          &	1+997 T+11356 p T^2+997 p^3 T^3+p^6 T^4
\tabularnewline[0.5pt]\hline				
43	 &	smooth    &	          &	1+342 T+11162 p T^2+342 p^3 T^3+p^6 T^4
\tabularnewline[0.5pt]\hline				
44	 &	smooth    &	          &	1-265 T+3592 p T^2-265 p^3 T^3+p^6 T^4
\tabularnewline[0.5pt]\hline				
45	 &	smooth    &	          &	1+1010 T+14538 p T^2+1010 p^3 T^3+p^6 T^4
\tabularnewline[0.5pt]\hline				
46	 &	smooth    &	          &	1-44 T+4422 p T^2-44 p^3 T^3+p^6 T^4
\tabularnewline[0.5pt]\hline				
47	 &	smooth    &	          &	1-924 T+3062 p T^2-924 p^3 T^3+p^6 T^4
\tabularnewline[0.5pt]\hline				
48	 &	smooth    &	          &	1+1135 T+17508 p T^2+1135 p^3 T^3+p^6 T^4
\tabularnewline[0.5pt]\hline				
49	 &	smooth    &	          &	1-114 T+5522 p T^2-114 p^3 T^3+p^6 T^4
\tabularnewline[0.5pt]\hline				
50	 &	smooth    &	          &	1+135 T-664 p T^2+135 p^3 T^3+p^6 T^4
\tabularnewline[0.5pt]\hline				
51	 &	smooth    &	          &	1+1280 T+142 p^2 T^2+1280 p^3 T^3+p^6 T^4
\tabularnewline[0.5pt]\hline				
52	 &	smooth    &	          &	1-426 T+4762 p T^2-426 p^3 T^3+p^6 T^4
\tabularnewline[0.5pt]\hline				
53	 &	smooth    &	          &	1-945 T+13592 p T^2-945 p^3 T^3+p^6 T^4
\tabularnewline[0.5pt]\hline				
54	 &	smooth    &	          &	1+84 T+9222 p T^2+84 p^3 T^3+p^6 T^4
\tabularnewline[0.5pt]\hline				
55	 &	smooth    &	          &	1-1683 T+17036 p T^2-1683 p^3 T^3+p^6 T^4
\tabularnewline[0.5pt]\hline				
56	 &	smooth    &	          &	1+197 T+9056 p T^2+197 p^3 T^3+p^6 T^4
\tabularnewline[0.5pt]\hline				
57	 &	smooth    &	          &	1-126 T+3114 p T^2-126 p^3 T^3+p^6 T^4
\tabularnewline[0.5pt]\hline				
58	 &	smooth    &	          &	1+142 T+12170 p T^2+142 p^3 T^3+p^6 T^4
\tabularnewline[0.5pt]\hline				
59	 &	smooth    &	          &	1+576 T+5342 p T^2+576 p^3 T^3+p^6 T^4
\tabularnewline[0.5pt]\hline				
60	 &	smooth    &	          &	1+953 T+8360 p T^2+953 p^3 T^3+p^6 T^4
\tabularnewline[0.5pt]\hline				
61	 &	smooth    &	          &	1+826 T+10114 p T^2+826 p^3 T^3+p^6 T^4
\tabularnewline[0.5pt]\hline				
62	 &	smooth    &	          &	1+62 T-2710 p T^2+62 p^3 T^3+p^6 T^4
\tabularnewline[0.5pt]\hline				
63	 &	smooth    &	          &	1-124 T+5254 p T^2-124 p^3 T^3+p^6 T^4
\tabularnewline[0.5pt]\hline				
64	 &	smooth    &	          &	1+590 T+7538 p T^2+590 p^3 T^3+p^6 T^4
\tabularnewline[0.5pt]\hline				
65	 &	smooth    &	          &	1-369 T+14084 p T^2-369 p^3 T^3+p^6 T^4
\tabularnewline[0.5pt]\hline				
66	 &	smooth    &	          &	1+460 T+358 p T^2+460 p^3 T^3+p^6 T^4
\tabularnewline[0.5pt]\hline				
67	 &	smooth    &	          &	1-886 T+11994 p T^2-886 p^3 T^3+p^6 T^4
\tabularnewline[0.5pt]\hline				
68	 &	smooth    &	          &	1-364 T+4822 p T^2-364 p^3 T^3+p^6 T^4
\tabularnewline[0.5pt]\hline				
69	 &	singular  &	69&	(1+p T) (1+1526 T+p^3 T^2)
\tabularnewline[0.5pt]\hline				
70	 &	smooth    &	          &	1+260 T-442 p T^2+260 p^3 T^3+p^6 T^4
\tabularnewline[0.5pt]\hline				
71	 &	smooth    &	          &	1+1306 T+12602 p T^2+1306 p^3 T^3+p^6 T^4
\tabularnewline[0.5pt]\hline				
72	 &	smooth    &	          &	1-344 T-4018 p T^2-344 p^3 T^3+p^6 T^4
\tabularnewline[0.5pt]\hline				
73	 &	smooth    &	          &	1+330 T+42 p^2 T^2+330 p^3 T^3+p^6 T^4
\tabularnewline[0.5pt]\hline				
74	 &	smooth    &	          &	1+95 T+11588 p T^2+95 p^3 T^3+p^6 T^4
\tabularnewline[0.5pt]\hline				
75	 &	smooth    &	          &	1-3 p T-448 p T^2-3 p^4 T^3+p^6 T^4
\tabularnewline[0.5pt]\hline				
76	 &	smooth    &	          &	1-108 T+11686 p T^2-108 p^3 T^3+p^6 T^4
\tabularnewline[0.5pt]\hline				
77	 &	smooth    &	          &	1-1383 T+11720 p T^2-1383 p^3 T^3+p^6 T^4
\tabularnewline[0.5pt]\hline				
78	 &	smooth    &	          &	1+591 T-488 p T^2+591 p^3 T^3+p^6 T^4
\tabularnewline[0.5pt]\hline				
79	 &	smooth    &	          &	1+336 T+13342 p T^2+336 p^3 T^3+p^6 T^4
\tabularnewline[0.5pt]\hline				
80	 &	smooth    &	          &	1-928 T+11806 p T^2-928 p^3 T^3+p^6 T^4
\tabularnewline[0.5pt]\hline				
81	 &	smooth    &	          &	1-72 T+14702 p T^2-72 p^3 T^3+p^6 T^4
\tabularnewline[0.5pt]\hline				
82	 &	smooth    &	          &	1-114 T-13158 p T^2-114 p^3 T^3+p^6 T^4
\tabularnewline[0.5pt]\hline				
83	 &	smooth    &	          &	1-229 T+2172 p T^2-229 p^3 T^3+p^6 T^4
\tabularnewline[0.5pt]\hline				
84	 &	smooth    &	          &	1-243 T-1204 p T^2-243 p^3 T^3+p^6 T^4
\tabularnewline[0.5pt]\hline				
85	 &	smooth    &	          &	1-2203 T+27356 p T^2-2203 p^3 T^3+p^6 T^4
\tabularnewline[0.5pt]\hline				
86	 &	smooth    &	          &	1-1082 T+12650 p T^2-1082 p^3 T^3+p^6 T^4
\tabularnewline[0.5pt]\hline				
87	 &	smooth    &	          &	1-70 T+13178 p T^2-70 p^3 T^3+p^6 T^4
\tabularnewline[0.5pt]\hline				
88	 &	smooth    &	          &	1+1261 T+11404 p T^2+1261 p^3 T^3+p^6 T^4
\tabularnewline[0.5pt]\hline				
\tablepostamble				
\tablepreamble{97}				
1	 &	smooth    &	          &	1-285 T+15000 p T^2-285 p^3 T^3+p^6 T^4
\tabularnewline[0.5pt]\hline				
2	 &	smooth    &	          &	1-770 T+8546 p T^2-770 p^3 T^3+p^6 T^4
\tabularnewline[0.5pt]\hline				
3	 &	smooth    &	          &	1+120 T+7630 p T^2+120 p^3 T^3+p^6 T^4
\tabularnewline[0.5pt]\hline				
4	 &	smooth    &	          &	1+120 T+3918 p T^2+120 p^3 T^3+p^6 T^4
\tabularnewline[0.5pt]\hline				
5	 &	smooth    &	          &	1-450 T+9978 p T^2-450 p^3 T^3+p^6 T^4
\tabularnewline[0.5pt]\hline				
6	 &	smooth    &	          &	1+1285 T+20372 p T^2+1285 p^3 T^3+p^6 T^4
\tabularnewline[0.5pt]\hline				
7	 &	smooth    &	          &	(1+4 p T+p^3 T^2)(1+1002 T+p^3 T^2)
\tabularnewline[0.5pt]\hline				
8	 &	smooth    &	          &	1-570 T+11314 p T^2-570 p^3 T^3+p^6 T^4
\tabularnewline[0.5pt]\hline				
9	 &	smooth    &	          &	1+175 T+14208 p T^2+175 p^3 T^3+p^6 T^4
\tabularnewline[0.5pt]\hline				
10	 &	smooth    &	          &	1+420 T-2906 p T^2+420 p^3 T^3+p^6 T^4
\tabularnewline[0.5pt]\hline				
11	 &	smooth    &	          &	1-250 T+6450 p T^2-250 p^3 T^3+p^6 T^4
\tabularnewline[0.5pt]\hline				
12	 &	smooth    &	          &	1-380 T+7158 p T^2-380 p^3 T^3+p^6 T^4
\tabularnewline[0.5pt]\hline				
13	 &	smooth    &	          &	1-905 T+10988 p T^2-905 p^3 T^3+p^6 T^4
\tabularnewline[0.5pt]\hline				
14	 &	smooth    &	          &	1+1650 T+15522 p T^2+1650 p^3 T^3+p^6 T^4
\tabularnewline[0.5pt]\hline				
15	 &	smooth    &	          &	1+1260 T+22742 p T^2+1260 p^3 T^3+p^6 T^4
\tabularnewline[0.5pt]\hline				
16	 &	smooth    &	          &	1+2490 T+33610 p T^2+2490 p^3 T^3+p^6 T^4
\tabularnewline[0.5pt]\hline				
17	 &	smooth    &	          &	1+140 T+11862 p T^2+140 p^3 T^3+p^6 T^4
\tabularnewline[0.5pt]\hline				
18	 &	smooth    &	          &	1-865 T+7968 p T^2-865 p^3 T^3+p^6 T^4
\tabularnewline[0.5pt]\hline				
19	 &	smooth    &	          &	1+500 T+1318 p T^2+500 p^3 T^3+p^6 T^4
\tabularnewline[0.5pt]\hline				
20	 &	smooth    &	          &	1-760 T+11630 p T^2-760 p^3 T^3+p^6 T^4
\tabularnewline[0.5pt]\hline				
21	 &	smooth    &	          &	1-390 T+9378 p T^2-390 p^3 T^3+p^6 T^4
\tabularnewline[0.5pt]\hline				
22	 &	smooth    &	          &	1+50 T+7770 p T^2+50 p^3 T^3+p^6 T^4
\tabularnewline[0.5pt]\hline				
23	 &	smooth    &	          &	1-1860 T+19638 p T^2-1860 p^3 T^3+p^6 T^4
\tabularnewline[0.5pt]\hline				
24	 &	smooth    &	          &	1+600 T+1422 p T^2+600 p^3 T^3+p^6 T^4
\tabularnewline[0.5pt]\hline				
25	 &	smooth    &	          &	1+1085 T+15076 p T^2+1085 p^3 T^3+p^6 T^4
\tabularnewline[0.5pt]\hline				
26	 &	smooth    &	          &	1+1050 T+9858 p T^2+1050 p^3 T^3+p^6 T^4
\tabularnewline[0.5pt]\hline				
27	 &	smooth    &	          &	1+1280 T+12670 p T^2+1280 p^3 T^3+p^6 T^4
\tabularnewline[0.5pt]\hline				
28	 &	smooth    &	          &	1+1840 T+19694 p T^2+1840 p^3 T^3+p^6 T^4
\tabularnewline[0.5pt]\hline				
29	 &	smooth    &	          &	1+1575 T+23244 p T^2+1575 p^3 T^3+p^6 T^4
\tabularnewline[0.5pt]\hline				
30	 &	smooth    &	          &	1+1200 T+8030 p T^2+1200 p^3 T^3+p^6 T^4
\tabularnewline[0.5pt]\hline				
31	 &	smooth    &	          &	1+1600 T+15870 p T^2+1600 p^3 T^3+p^6 T^4
\tabularnewline[0.5pt]\hline				
32	 &	smooth    &	          &	1+230 T-4110 p T^2+230 p^3 T^3+p^6 T^4
\tabularnewline[0.5pt]\hline				
33	 &	smooth    &	          &	(1-2 p T+p^3 T^2)(1-796 T+p^3 T^2)
\tabularnewline[0.5pt]\hline				
34	 &	smooth    &	          &	1+1365 T+11816 p T^2+1365 p^3 T^3+p^6 T^4
\tabularnewline[0.5pt]\hline				
35	 &	smooth    &	          &	1-740 T+5830 p T^2-740 p^3 T^3+p^6 T^4
\tabularnewline[0.5pt]\hline				
36	 &	smooth    &	          &	1-775 T+17964 p T^2-775 p^3 T^3+p^6 T^4
\tabularnewline[0.5pt]\hline				
37	 &	smooth    &	          &	1+600 T+2990 p T^2+600 p^3 T^3+p^6 T^4
\tabularnewline[0.5pt]\hline				
38	 &	smooth    &	          &	1-10 T-2822 p T^2-10 p^3 T^3+p^6 T^4
\tabularnewline[0.5pt]\hline				
39	 &	smooth    &	          &	1-190 T+3522 p T^2-190 p^3 T^3+p^6 T^4
\tabularnewline[0.5pt]\hline				
40	 &	smooth    &	          &	1+1310 T+20810 p T^2+1310 p^3 T^3+p^6 T^4
\tabularnewline[0.5pt]\hline				
41	 &	smooth    &	          &	1+240 T+2270 p T^2+240 p^3 T^3+p^6 T^4
\tabularnewline[0.5pt]\hline				
42	 &	smooth    &	          &	1+55 T+2668 p T^2+55 p^3 T^3+p^6 T^4
\tabularnewline[0.5pt]\hline				
43	 &	smooth    &	          &	1-750 T+2650 p T^2-750 p^3 T^3+p^6 T^4
\tabularnewline[0.5pt]\hline				
44	 &	smooth    &	          &	1+950 T+10834 p T^2+950 p^3 T^3+p^6 T^4
\tabularnewline[0.5pt]\hline				
45	 &	smooth    &	          &	1-1530 T+23418 p T^2-1530 p^3 T^3+p^6 T^4
\tabularnewline[0.5pt]\hline				
46	 &	smooth    &	          &	1-2245 T+26548 p T^2-2245 p^3 T^3+p^6 T^4
\tabularnewline[0.5pt]\hline				
47	 &	smooth    &	          &	1-1240 T+22382 p T^2-1240 p^3 T^3+p^6 T^4
\tabularnewline[0.5pt]\hline				
48	 &	smooth    &	          &	1-80 T-8546 p T^2-80 p^3 T^3+p^6 T^4
\tabularnewline[0.5pt]\hline				
49	 &	smooth    &	          &	1+605 T+16356 p T^2+605 p^3 T^3+p^6 T^4
\tabularnewline[0.5pt]\hline				
50	 &	smooth    &	          &	1-270 T+6554 p T^2-270 p^3 T^3+p^6 T^4
\tabularnewline[0.5pt]\hline				
51	 &	smooth    &	          &	1+30 T+2298 p T^2+30 p^3 T^3+p^6 T^4
\tabularnewline[0.5pt]\hline				
52	 &	smooth    &	          &	1-920 T+5310 p T^2-920 p^3 T^3+p^6 T^4
\tabularnewline[0.5pt]\hline				
53	 &	smooth    &	          &	1-270 T-4806 p T^2-270 p^3 T^3+p^6 T^4
\tabularnewline[0.5pt]\hline				
54	 &	smooth    &	          &	1-1140 T+19494 p T^2-1140 p^3 T^3+p^6 T^4
\tabularnewline[0.5pt]\hline				
55	 &	smooth    &	          &	1+665 T+3696 p T^2+665 p^3 T^3+p^6 T^4
\tabularnewline[0.5pt]\hline				
56	 &	smooth    &	          &	1+325 T-10904 p T^2+325 p^3 T^3+p^6 T^4
\tabularnewline[0.5pt]\hline				
57	 &	smooth    &	          &	1-240 T+30 p^2 T^2-240 p^3 T^3+p^6 T^4
\tabularnewline[0.5pt]\hline				
58	 &	smooth    &	          &	1+925 T+11608 p T^2+925 p^3 T^3+p^6 T^4
\tabularnewline[0.5pt]\hline				
59	 &	smooth    &	          &	1-480 T+6270 p T^2-480 p^3 T^3+p^6 T^4
\tabularnewline[0.5pt]\hline				
60	 &	smooth    &	          &	1-270 T+82 p T^2-270 p^3 T^3+p^6 T^4
\tabularnewline[0.5pt]\hline				
61	 &	smooth    &	          &	1+595 T+18360 p T^2+595 p^3 T^3+p^6 T^4
\tabularnewline[0.5pt]\hline				
62	 &	smooth    &	          &	1+550 T+7282 p T^2+550 p^3 T^3+p^6 T^4
\tabularnewline[0.5pt]\hline				
63	 &	smooth    &	          &	1+680 T-3330 p T^2+680 p^3 T^3+p^6 T^4
\tabularnewline[0.5pt]\hline				
64	 &	smooth    &	          &	1-220 T+14774 p T^2-220 p^3 T^3+p^6 T^4
\tabularnewline[0.5pt]\hline				
65	 &	smooth    &	          &	1-355 T+7780 p T^2-355 p^3 T^3+p^6 T^4
\tabularnewline[0.5pt]\hline				
66	 &	smooth    &	          &	1+805 T+16660 p T^2+805 p^3 T^3+p^6 T^4
\tabularnewline[0.5pt]\hline				
67	 &	smooth    &	          &	1+470 T+6890 p T^2+470 p^3 T^3+p^6 T^4
\tabularnewline[0.5pt]\hline				
68	 &	smooth    &	          &	1+520 T+5134 p T^2+520 p^3 T^3+p^6 T^4
\tabularnewline[0.5pt]\hline				
69	 &	smooth    &	          &	1+455 T-756 p T^2+455 p^3 T^3+p^6 T^4
\tabularnewline[0.5pt]\hline				
70	 &	smooth    &	          &	1+85 T+6260 p T^2+85 p^3 T^3+p^6 T^4
\tabularnewline[0.5pt]\hline				
71	 &	smooth    &	          &	1-1015 T+19088 p T^2-1015 p^3 T^3+p^6 T^4
\tabularnewline[0.5pt]\hline				
72	 &	smooth    &	          &	1+1200 T+13854 p T^2+1200 p^3 T^3+p^6 T^4
\tabularnewline[0.5pt]\hline				
73	 &	smooth    &	          &	1+1130 T+42 p^2 T^2+1130 p^3 T^3+p^6 T^4
\tabularnewline[0.5pt]\hline				
74	 &	smooth    &	          &	1+420 T+11494 p T^2+420 p^3 T^3+p^6 T^4
\tabularnewline[0.5pt]\hline				
75	 &	smooth    &	          &	1+540 T-58 p T^2+540 p^3 T^3+p^6 T^4
\tabularnewline[0.5pt]\hline				
76	 &	smooth    &	          &	1-1335 T+12848 p T^2-1335 p^3 T^3+p^6 T^4
\tabularnewline[0.5pt]\hline				
77	 &	smooth    &	          &	1-630 T-1198 p T^2-630 p^3 T^3+p^6 T^4
\tabularnewline[0.5pt]\hline				
78	 &	smooth    &	          &	1-1290 T+10986 p T^2-1290 p^3 T^3+p^6 T^4
\tabularnewline[0.5pt]\hline				
79	 &	smooth    &	          &	1+1990 T+25266 p T^2+1990 p^3 T^3+p^6 T^4
\tabularnewline[0.5pt]\hline				
80	 &	smooth    &	          &	1+205 T+9624 p T^2+205 p^3 T^3+p^6 T^4
\tabularnewline[0.5pt]\hline				
81	 &	smooth    &	          &	1-140 T+10262 p T^2-140 p^3 T^3+p^6 T^4
\tabularnewline[0.5pt]\hline				
82	 &	smooth    &	          &	1+205 T+15864 p T^2+205 p^3 T^3+p^6 T^4
\tabularnewline[0.5pt]\hline				
83	 &	smooth    &	          &	1-595 T+4888 p T^2-595 p^3 T^3+p^6 T^4
\tabularnewline[0.5pt]\hline				
84	 &	smooth    &	          &	1+825 T+9296 p T^2+825 p^3 T^3+p^6 T^4
\tabularnewline[0.5pt]\hline				
85	 &	smooth    &	          &	1+1640 T+15790 p T^2+1640 p^3 T^3+p^6 T^4
\tabularnewline[0.5pt]\hline				
86	 &	smooth    &	          &	1-675 T+6180 p T^2-675 p^3 T^3+p^6 T^4
\tabularnewline[0.5pt]\hline				
87	 &	smooth    &	          &	1+620 T+9718 p T^2+620 p^3 T^3+p^6 T^4
\tabularnewline[0.5pt]\hline				
88	 &	smooth    &	          &	1-320 T+1278 p T^2-320 p^3 T^3+p^6 T^4
\tabularnewline[0.5pt]\hline				
89	 &	smooth    &	          &	1-350 T-10086 p T^2-350 p^3 T^3+p^6 T^4
\tabularnewline[0.5pt]\hline				
90	 &	smooth    &	          &	1-700 T+3718 p T^2-700 p^3 T^3+p^6 T^4
\tabularnewline[0.5pt]\hline				
91	 &	smooth    &	          &	1-1275 T+16532 p T^2-1275 p^3 T^3+p^6 T^4
\tabularnewline[0.5pt]\hline				
92	 &	smooth    &	          &	1-5 p T-2732 p T^2-5 p^4 T^3+p^6 T^4
\tabularnewline[0.5pt]\hline				
93	 &	smooth    &	          &	1+15 T+1088 p T^2+15 p^3 T^3+p^6 T^4
\tabularnewline[0.5pt]\hline				
94	 &	smooth    &	          &	1-500 T-5562 p T^2-500 p^3 T^3+p^6 T^4
\tabularnewline[0.5pt]\hline				
95	 &	smooth    &	          &	1+735 T+16928 p T^2+735 p^3 T^3+p^6 T^4
\tabularnewline[0.5pt]\hline				
96	 &	smooth    &	          &	1-1895 T+26604 p T^2-1895 p^3 T^3+p^6 T^4
\tabularnewline[0.5pt]\hline				
\tablepostamble	
\newpage
\lhead{\ifthenelse{\isodd{\value{page}}}{\thepage}{\sl The $\z$-function for a Hulek--Verrill manifold, AESZ\hskip2pt 34}}
\rhead{\ifthenelse{\isodd{\value{page}}}{\sl The $\z$-function for a Hulek--Verrill manifold, AESZ\hskip2pt 34}{\thepage}}
\subsection{The $\z$-function for a Hulek--Verrill manifold, AESZ\hskip2pt 34}
\vspace{1.5cm}
\tablepreamble{5}				
1	 &	singular  &	1&	(1-p T) (1-6 T+p^3 T^2)
\tabularnewline[0.5pt]\hline				
2	 &	smooth    &	          &	(1+p^3 T^2)(1+14 T+p^3 T^2)
\tabularnewline[0.5pt]\hline				
3	 &	smooth    &	          &	1+4 T-2 p^2 T^2+4 p^3 T^3+p^6 T^4
\tabularnewline[0.5pt]\hline				
4	 &	singular  &	\frac{1}{9}&	(1+p T) (1-6 T+p^3 T^2)
\tabularnewline[0.5pt]\hline				
\tablepostamble				
\tablepreamble{7}				
1	 &	singular  &	1&	(1-p T) (1+16 T+p^3 T^2)
\tabularnewline[0.5pt]\hline				
2	 &	singular  &	\frac{1}{25}&	(1+p T) (1-32 T+p^3 T^2)
\tabularnewline[0.5pt]\hline				
3	 &	smooth    &	          &	1+2 T-54 p T^2+2 p^3 T^3+p^6 T^4
\tabularnewline[0.5pt]\hline				
4	 &	singular  &	\frac{1}{9}&	(1-p T) (1+16 T+p^3 T^2)
\tabularnewline[0.5pt]\hline				
5	 &	smooth    &	          &	(1+4 p T+p^3 T^2)(1-34 T+p^3 T^2)
\tabularnewline[0.5pt]\hline				
6	 &	smooth    &	          &	1+12 T-2 p^2 T^2+12 p^3 T^3+p^6 T^4
\tabularnewline[0.5pt]\hline				
\tablepostamble				
\tablepreamble{11}				
1	 &	singular  &	1&	(1-p T) (1-12 T+p^3 T^2)
\tabularnewline[0.5pt]\hline				
2	 &	smooth    &	          &	1+42 T+194 p T^2+42 p^3 T^3+p^6 T^4
\tabularnewline[0.5pt]\hline				
3	 &	smooth    &	          &	(1+p^3 T^2)(1+28 T+p^3 T^2)
\tabularnewline[0.5pt]\hline				
4	 &	singular  &	\frac{1}{25}&	(1-p T) (1+60 T+p^3 T^2)
\tabularnewline[0.5pt]\hline				
5	 &	singular  &	\frac{1}{9}&	(1+p T) (1-12 T+p^3 T^2)
\tabularnewline[0.5pt]\hline				
6	 &	smooth    &	          &	1-2 T+2 p T^2-2 p^3 T^3+p^6 T^4
\tabularnewline[0.5pt]\hline				
7	 &	smooth    &	          &	1-2 p T+122 p T^2-2 p^4 T^3+p^6 T^4
\tabularnewline[0.5pt]\hline				
8	 &	smooth    &	          &	1-16 T+50 p T^2-16 p^3 T^3+p^6 T^4
\tabularnewline[0.5pt]\hline				
9	 &	smooth    &	          &	1-32 T+2 p T^2-32 p^3 T^3+p^6 T^4
\tabularnewline[0.5pt]\hline				
10	 &	smooth    &	          &	1-2 p T+2 p T^2-2 p^4 T^3+p^6 T^4
\tabularnewline[0.5pt]\hline				
\tablepostamble				
\tablepreamble{13}				
1	 &	singular  &	1&	(1-p T) (1-38 T+p^3 T^2)
\tabularnewline[0.5pt]\hline				
2	 &	smooth    &	          &	(1-6 p T+p^3 T^2)(1+34 T+p^3 T^2)
\tabularnewline[0.5pt]\hline				
3	 &	singular  &	\frac{1}{9}&	(1-p T) (1-38 T+p^3 T^2)
\tabularnewline[0.5pt]\hline				
4	 &	smooth    &	          &	(1-2 p T+p^3 T^2)(1+42 T+p^3 T^2)
\tabularnewline[0.5pt]\hline				
5	 &	smooth    &	          &	1-36 T+166 p T^2-36 p^3 T^3+p^6 T^4
\tabularnewline[0.5pt]\hline				
6	 &	smooth    &	          &	1-2 p T+266 p T^2-2 p^4 T^3+p^6 T^4
\tabularnewline[0.5pt]\hline				
7	 &	smooth    &	          &	1+14 T-54 p T^2+14 p^3 T^3+p^6 T^4
\tabularnewline[0.5pt]\hline				
8	 &	smooth    &	          &	(1+4 p T+p^3 T^2)(1-6 T+p^3 T^2)
\tabularnewline[0.5pt]\hline				
9	 &	smooth    &	          &	1+36 T-2 p^2 T^2+36 p^3 T^3+p^6 T^4
\tabularnewline[0.5pt]\hline				
10	 &	smooth    &	          &	(1-2 p T+p^3 T^2)(1+42 T+p^3 T^2)
\tabularnewline[0.5pt]\hline				
11	 &	smooth    &	          &	(1+4 p T+p^3 T^2)(1-18 T+p^3 T^2)
\tabularnewline[0.5pt]\hline				
12	 &	singular  &	\frac{1}{25}&	(1+p T) (1+34 T+p^3 T^2)
\tabularnewline[0.5pt]\hline				
\tablepostamble				
\tablepreamble{17}				
1	 &	singular  &	1&	(1-p T) (1+126 T+p^3 T^2)
\tabularnewline[0.5pt]\hline				
2	 &	singular  &	\frac{1}{9}&	(1+p T) (1+126 T+p^3 T^2)
\tabularnewline[0.5pt]\hline				
3	 &	smooth    &	          &	1+98 T+674 p T^2+98 p^3 T^3+p^6 T^4
\tabularnewline[0.5pt]\hline				
4	 &	smooth    &	          &	(1+6 p T+p^3 T^2)(1-114 T+p^3 T^2)
\tabularnewline[0.5pt]\hline				
5	 &	smooth    &	          &	1-4 T-202 p T^2-4 p^3 T^3+p^6 T^4
\tabularnewline[0.5pt]\hline				
6	 &	smooth    &	          &	1-44 T+38 p T^2-44 p^3 T^3+p^6 T^4
\tabularnewline[0.5pt]\hline				
7	 &	smooth    &	          &	1-2 T+194 p T^2-2 p^3 T^3+p^6 T^4
\tabularnewline[0.5pt]\hline				
8	 &	smooth    &	          &	(1+6 p T+p^3 T^2)(1-94 T+p^3 T^2)
\tabularnewline[0.5pt]\hline				
9	 &	smooth    &	          &	1+4 T+182 p T^2+4 p^3 T^3+p^6 T^4
\tabularnewline[0.5pt]\hline				
10	 &	smooth    &	          &	1+2 p T+2 p T^2+2 p^4 T^3+p^6 T^4
\tabularnewline[0.5pt]\hline				
11	 &	smooth    &	          &	1-124 T+518 p T^2-124 p^3 T^3+p^6 T^4
\tabularnewline[0.5pt]\hline				
12	 &	smooth    &	          &	(1-6 p T+p^3 T^2)(1-74 T+p^3 T^2)
\tabularnewline[0.5pt]\hline				
13	 &	smooth    &	          &	(1-2 p T+p^3 T^2)(1+78 T+p^3 T^2)
\tabularnewline[0.5pt]\hline				
14	 &	smooth    &	          &	(1+p^3 T^2)(1-134 T+p^3 T^2)
\tabularnewline[0.5pt]\hline				
15	 &	singular  &	\frac{1}{25}&	(1+p T) (1-42 T+p^3 T^2)
\tabularnewline[0.5pt]\hline				
16	 &	smooth    &	          &	(1+6 p T+p^3 T^2)(1-2 p T+p^3 T^2)
\tabularnewline[0.5pt]\hline				
\tablepostamble				
\tablepreamble{19}				
1	 &	singular  &	1&	(1-p T) (1-20 T+p^3 T^2)
\tabularnewline[0.5pt]\hline				
2	 &	smooth    &	          &	1+4 p T+2 p T^2+4 p^4 T^3+p^6 T^4
\tabularnewline[0.5pt]\hline				
3	 &	smooth    &	          &	1-8 T+242 p T^2-8 p^3 T^3+p^6 T^4
\tabularnewline[0.5pt]\hline				
4	 &	smooth    &	          &	(1+4 p T+p^3 T^2)(1-60 T+p^3 T^2)
\tabularnewline[0.5pt]\hline				
5	 &	smooth    &	          &	(1+4 p T+p^3 T^2)(1-60 T+p^3 T^2)
\tabularnewline[0.5pt]\hline				
6	 &	smooth    &	          &	1+8 T-318 p T^2+8 p^3 T^3+p^6 T^4
\tabularnewline[0.5pt]\hline				
7	 &	smooth    &	          &	1-44 T-238 p T^2-44 p^3 T^3+p^6 T^4
\tabularnewline[0.5pt]\hline				
8	 &	smooth    &	          &	(1-2 p T+p^3 T^2)(1-80 T+p^3 T^2)
\tabularnewline[0.5pt]\hline				
9	 &	smooth    &	          &	(1+4 p T+p^3 T^2)(1-160 T+p^3 T^2)
\tabularnewline[0.5pt]\hline				
10	 &	smooth    &	          &	1+12 T+562 p T^2+12 p^3 T^3+p^6 T^4
\tabularnewline[0.5pt]\hline				
11	 &	smooth    &	          &	(1+4 p T+p^3 T^2)(1-140 T+p^3 T^2)
\tabularnewline[0.5pt]\hline				
12	 &	smooth    &	          &	1+12 T+82 p T^2+12 p^3 T^3+p^6 T^4
\tabularnewline[0.5pt]\hline				
13	 &	smooth    &	          &	1+178 T+1082 p T^2+178 p^3 T^3+p^6 T^4
\tabularnewline[0.5pt]\hline				
14	 &	smooth    &	          &	1+12 T-158 p T^2+12 p^3 T^3+p^6 T^4
\tabularnewline[0.5pt]\hline				
15	 &	smooth    &	          &	1+42 T-2 p^2 T^2+42 p^3 T^3+p^6 T^4
\tabularnewline[0.5pt]\hline				
16	 &	singular  &	\frac{1}{25}&	(1-p T) (1+76 T+p^3 T^2)
\tabularnewline[0.5pt]\hline				
17	 &	singular  &	\frac{1}{9}&	(1-p T) (1-20 T+p^3 T^2)
\tabularnewline[0.5pt]\hline				
18	 &	smooth    &	          &	1-54 T+322 p T^2-54 p^3 T^3+p^6 T^4
\tabularnewline[0.5pt]\hline				
\tablepostamble				
\tablepreamble{23}				
1	 &	singular  &	1&	(1-p T) (1-168 T+p^3 T^2)
\tabularnewline[0.5pt]\hline				
2	 &	smooth    &	          &	1-28 T+98 p T^2-28 p^3 T^3+p^6 T^4
\tabularnewline[0.5pt]\hline				
3	 &	smooth    &	          &	1-68 T+98 p T^2-68 p^3 T^3+p^6 T^4
\tabularnewline[0.5pt]\hline				
4	 &	smooth    &	          &	(1+4 p T+p^3 T^2)(1-120 T+p^3 T^2)
\tabularnewline[0.5pt]\hline				
5	 &	smooth    &	          &	1-90 T+890 p T^2-90 p^3 T^3+p^6 T^4
\tabularnewline[0.5pt]\hline				
6	 &	smooth    &	          &	1+72 T+98 p T^2+72 p^3 T^3+p^6 T^4
\tabularnewline[0.5pt]\hline				
7	 &	smooth    &	          &	1+52 T+338 p T^2+52 p^3 T^3+p^6 T^4
\tabularnewline[0.5pt]\hline				
8	 &	smooth    &	          &	1+12 T+98 p T^2+12 p^3 T^3+p^6 T^4
\tabularnewline[0.5pt]\hline				
9	 &	smooth    &	          &	1+32 T+98 p T^2+32 p^3 T^3+p^6 T^4
\tabularnewline[0.5pt]\hline				
10	 &	smooth    &	          &	(1+4 p T+p^3 T^2)(1-132 T+p^3 T^2)
\tabularnewline[0.5pt]\hline				
11	 &	smooth    &	          &	1+94 T-94 p T^2+94 p^3 T^3+p^6 T^4
\tabularnewline[0.5pt]\hline				
12	 &	singular  &	\frac{1}{25}&	(1+p T) (1+p^3 T^2)
\tabularnewline[0.5pt]\hline				
13	 &	smooth    &	          &	(1+p^3 T^2)(1+112 T+p^3 T^2)
\tabularnewline[0.5pt]\hline				
14	 &	smooth    &	          &	1+94 T+266 p T^2+94 p^3 T^3+p^6 T^4
\tabularnewline[0.5pt]\hline				
15	 &	smooth    &	          &	1+2 T-142 p T^2+2 p^3 T^3+p^6 T^4
\tabularnewline[0.5pt]\hline				
16	 &	smooth    &	          &	1-28 T+98 p T^2-28 p^3 T^3+p^6 T^4
\tabularnewline[0.5pt]\hline				
17	 &	smooth    &	          &	1+144 T+626 p T^2+144 p^3 T^3+p^6 T^4
\tabularnewline[0.5pt]\hline				
18	 &	singular  &	\frac{1}{9}&	(1+p T) (1-168 T+p^3 T^2)
\tabularnewline[0.5pt]\hline				
19	 &	smooth    &	          &	1+162 T+1178 p T^2+162 p^3 T^3+p^6 T^4
\tabularnewline[0.5pt]\hline				
20	 &	smooth    &	          &	1-120 T+530 p T^2-120 p^3 T^3+p^6 T^4
\tabularnewline[0.5pt]\hline				
21	 &	smooth    &	          &	1-120 T+530 p T^2-120 p^3 T^3+p^6 T^4
\tabularnewline[0.5pt]\hline				
22	 &	smooth    &	          &	1+94 T+866 p T^2+94 p^3 T^3+p^6 T^4
\tabularnewline[0.5pt]\hline				
\tablepostamble				
\tablepreamble{29}				
1	 &	singular  &	1&	(1-p T) (1-30 T+p^3 T^2)
\tabularnewline[0.5pt]\hline				
2	 &	smooth    &	          &	1-150 T+362 p T^2-150 p^3 T^3+p^6 T^4
\tabularnewline[0.5pt]\hline				
3	 &	smooth    &	          &	1-6 T+722 p T^2-6 p^3 T^3+p^6 T^4
\tabularnewline[0.5pt]\hline				
4	 &	smooth    &	          &	(1+6 p T+p^3 T^2)(1-190 T+p^3 T^2)
\tabularnewline[0.5pt]\hline				
5	 &	smooth    &	          &	1+44 T+182 p T^2+44 p^3 T^3+p^6 T^4
\tabularnewline[0.5pt]\hline				
6	 &	smooth    &	          &	1+96 T-418 p T^2+96 p^3 T^3+p^6 T^4
\tabularnewline[0.5pt]\hline				
7	 &	singular  &	\frac{1}{25}&	(1-p T) (1-6 T+p^3 T^2)
\tabularnewline[0.5pt]\hline				
8	 &	smooth    &	          &	1+24 T+542 p T^2+24 p^3 T^3+p^6 T^4
\tabularnewline[0.5pt]\hline				
9	 &	smooth    &	          &	1-16 T-418 p T^2-16 p^3 T^3+p^6 T^4
\tabularnewline[0.5pt]\hline				
10	 &	smooth    &	          &	1-80 T+542 p T^2-80 p^3 T^3+p^6 T^4
\tabularnewline[0.5pt]\hline				
11	 &	smooth    &	          &	1-126 T+242 p T^2-126 p^3 T^3+p^6 T^4
\tabularnewline[0.5pt]\hline				
12	 &	smooth    &	          &	1+100 T+182 p T^2+100 p^3 T^3+p^6 T^4
\tabularnewline[0.5pt]\hline				
13	 &	singular  &	\frac{1}{9}&	(1+p T) (1-30 T+p^3 T^2)
\tabularnewline[0.5pt]\hline				
14	 &	smooth    &	          &	1-24 T+542 p T^2-24 p^3 T^3+p^6 T^4
\tabularnewline[0.5pt]\hline				
15	 &	smooth    &	          &	1-156 T+1382 p T^2-156 p^3 T^3+p^6 T^4
\tabularnewline[0.5pt]\hline				
16	 &	smooth    &	          &	1-396 T+2822 p T^2-396 p^3 T^3+p^6 T^4
\tabularnewline[0.5pt]\hline				
17	 &	smooth    &	          &	1+6 T-238 p T^2+6 p^3 T^3+p^6 T^4
\tabularnewline[0.5pt]\hline				
18	 &	smooth    &	          &	1+10 T-1438 p T^2+10 p^3 T^3+p^6 T^4
\tabularnewline[0.5pt]\hline				
19	 &	smooth    &	          &	1+130 T+722 p T^2+130 p^3 T^3+p^6 T^4
\tabularnewline[0.5pt]\hline				
20	 &	smooth    &	          &	1+276 T+1862 p T^2+276 p^3 T^3+p^6 T^4
\tabularnewline[0.5pt]\hline				
21	 &	smooth    &	          &	1+1502 p T^2+p^6 T^4
\tabularnewline[0.5pt]\hline				
22	 &	smooth    &	          &	1+24 T+1262 p T^2+24 p^3 T^3+p^6 T^4
\tabularnewline[0.5pt]\hline				
23	 &	smooth    &	          &	(1+10 p T+p^3 T^2)(1-294 T+p^3 T^2)
\tabularnewline[0.5pt]\hline				
24	 &	smooth    &	          &	1+56 T+782 p T^2+56 p^3 T^3+p^6 T^4
\tabularnewline[0.5pt]\hline				
25	 &	smooth    &	          &	1-44 T-2 p^2 T^2-44 p^3 T^3+p^6 T^4
\tabularnewline[0.5pt]\hline				
26	 &	smooth    &	          &	1+206 T+1922 p T^2+206 p^3 T^3+p^6 T^4
\tabularnewline[0.5pt]\hline				
27	 &	smooth    &	          &	1+6 T-118 p T^2+6 p^3 T^3+p^6 T^4
\tabularnewline[0.5pt]\hline				
28	 &	smooth    &	          &	1+136 T+782 p T^2+136 p^3 T^3+p^6 T^4
\tabularnewline[0.5pt]\hline				
\tablepostamble				
\tablepreamble{31}				
1	 &	singular  &	1&	(1-p T) (1+88 T+p^3 T^2)
\tabularnewline[0.5pt]\hline				
2	 &	smooth    &	          &	1-288 T+2434 p T^2-288 p^3 T^3+p^6 T^4
\tabularnewline[0.5pt]\hline				
3	 &	smooth    &	          &	1+256 T+1106 p T^2+256 p^3 T^3+p^6 T^4
\tabularnewline[0.5pt]\hline				
4	 &	smooth    &	          &	1+352 T+2114 p T^2+352 p^3 T^3+p^6 T^4
\tabularnewline[0.5pt]\hline				
5	 &	singular  &	\frac{1}{25}&	(1-p T) (1+232 T+p^3 T^2)
\tabularnewline[0.5pt]\hline				
6	 &	smooth    &	          &	1+258 T+2122 p T^2+258 p^3 T^3+p^6 T^4
\tabularnewline[0.5pt]\hline				
7	 &	singular  &	\frac{1}{9}&	(1-p T) (1+88 T+p^3 T^2)
\tabularnewline[0.5pt]\hline				
8	 &	smooth    &	          &	(1-8 p T+p^3 T^2)(1+268 T+p^3 T^2)
\tabularnewline[0.5pt]\hline				
9	 &	smooth    &	          &	(1-8 p T+p^3 T^2)(1+168 T+p^3 T^2)
\tabularnewline[0.5pt]\hline				
10	 &	smooth    &	          &	1+52 T-286 p T^2+52 p^3 T^3+p^6 T^4
\tabularnewline[0.5pt]\hline				
11	 &	smooth    &	          &	1-98 T+1514 p T^2-98 p^3 T^3+p^6 T^4
\tabularnewline[0.5pt]\hline				
12	 &	smooth    &	          &	1-264 T+1906 p T^2-264 p^3 T^3+p^6 T^4
\tabularnewline[0.5pt]\hline				
13	 &	smooth    &	          &	1-98 T+674 p T^2-98 p^3 T^3+p^6 T^4
\tabularnewline[0.5pt]\hline				
14	 &	smooth    &	          &	1-8 T+1154 p T^2-8 p^3 T^3+p^6 T^4
\tabularnewline[0.5pt]\hline				
15	 &	smooth    &	          &	1-200 T+1058 p T^2-200 p^3 T^3+p^6 T^4
\tabularnewline[0.5pt]\hline				
16	 &	smooth    &	          &	(1+4 p T+p^3 T^2)(1-192 T+p^3 T^2)
\tabularnewline[0.5pt]\hline				
17	 &	smooth    &	          &	1-90 T-422 p T^2-90 p^3 T^3+p^6 T^4
\tabularnewline[0.5pt]\hline				
18	 &	smooth    &	          &	(1-8 p T+p^3 T^2)(1+128 T+p^3 T^2)
\tabularnewline[0.5pt]\hline				
19	 &	smooth    &	          &	1-148 T+1794 p T^2-148 p^3 T^3+p^6 T^4
\tabularnewline[0.5pt]\hline				
20	 &	smooth    &	          &	(1+8 p T+p^3 T^2)(1-8 p T+p^3 T^2)
\tabularnewline[0.5pt]\hline				
21	 &	smooth    &	          &	1+12 T+754 p T^2+12 p^3 T^3+p^6 T^4
\tabularnewline[0.5pt]\hline				
22	 &	smooth    &	          &	(1+4 p T+p^3 T^2)(1-72 T+p^3 T^2)
\tabularnewline[0.5pt]\hline				
23	 &	smooth    &	          &	1-4 T+1266 p T^2-4 p^3 T^3+p^6 T^4
\tabularnewline[0.5pt]\hline				
24	 &	smooth    &	          &	1+82 T+1394 p T^2+82 p^3 T^3+p^6 T^4
\tabularnewline[0.5pt]\hline				
25	 &	smooth    &	          &	1+72 T-446 p T^2+72 p^3 T^3+p^6 T^4
\tabularnewline[0.5pt]\hline				
26	 &	smooth    &	          &	(1+4 p T+p^3 T^2)(1-202 T+p^3 T^2)
\tabularnewline[0.5pt]\hline				
27	 &	smooth    &	          &	1+198 T+562 p T^2+198 p^3 T^3+p^6 T^4
\tabularnewline[0.5pt]\hline				
28	 &	smooth    &	          &	1+1858 p T^2+p^6 T^4
\tabularnewline[0.5pt]\hline				
29	 &	smooth    &	          &	1+196 T+1346 p T^2+196 p^3 T^3+p^6 T^4
\tabularnewline[0.5pt]\hline				
30	 &	smooth    &	          &	1-318 T+1954 p T^2-318 p^3 T^3+p^6 T^4
\tabularnewline[0.5pt]\hline				
\tablepostamble				
\tablepreamble{37}				
1	 &	singular  &	1&	(1-p T) (1-254 T+p^3 T^2)
\tabularnewline[0.5pt]\hline				
2	 &	smooth    &	          &	1+54 T+322 p T^2+54 p^3 T^3+p^6 T^4
\tabularnewline[0.5pt]\hline				
3	 &	singular  &	\frac{1}{25}&	(1+p T) (1-134 T+p^3 T^2)
\tabularnewline[0.5pt]\hline				
4	 &	smooth    &	          &	1-48 T+2206 p T^2-48 p^3 T^3+p^6 T^4
\tabularnewline[0.5pt]\hline				
5	 &	smooth    &	          &	(1-2 p T+p^3 T^2)(1-404 T+p^3 T^2)
\tabularnewline[0.5pt]\hline				
6	 &	smooth    &	          &	1+64 T+302 p T^2+64 p^3 T^3+p^6 T^4
\tabularnewline[0.5pt]\hline				
7	 &	smooth    &	          &	1-204 T+838 p T^2-204 p^3 T^3+p^6 T^4
\tabularnewline[0.5pt]\hline				
8	 &	smooth    &	          &	1-274 T+2418 p T^2-274 p^3 T^3+p^6 T^4
\tabularnewline[0.5pt]\hline				
9	 &	smooth    &	          &	1+232 T+1166 p T^2+232 p^3 T^3+p^6 T^4
\tabularnewline[0.5pt]\hline				
10	 &	smooth    &	          &	1-244 T+2358 p T^2-244 p^3 T^3+p^6 T^4
\tabularnewline[0.5pt]\hline				
11	 &	smooth    &	          &	1+252 T+1366 p T^2+252 p^3 T^3+p^6 T^4
\tabularnewline[0.5pt]\hline				
12	 &	smooth    &	          &	1+176 T+1758 p T^2+176 p^3 T^3+p^6 T^4
\tabularnewline[0.5pt]\hline				
13	 &	smooth    &	          &	1-10 T+930 p T^2-10 p^3 T^3+p^6 T^4
\tabularnewline[0.5pt]\hline				
14	 &	smooth    &	          &	1-18 T+826 p T^2-18 p^3 T^3+p^6 T^4
\tabularnewline[0.5pt]\hline				
15	 &	smooth    &	          &	1+182 T+186 p T^2+182 p^3 T^3+p^6 T^4
\tabularnewline[0.5pt]\hline				
16	 &	smooth    &	          &	1-28 T-474 p T^2-28 p^3 T^3+p^6 T^4
\tabularnewline[0.5pt]\hline				
17	 &	smooth    &	          &	1-18 T-1454 p T^2-18 p^3 T^3+p^6 T^4
\tabularnewline[0.5pt]\hline				
18	 &	smooth    &	          &	1+316 T+3158 p T^2+316 p^3 T^3+p^6 T^4
\tabularnewline[0.5pt]\hline				
19	 &	smooth    &	          &	1+162 T+1186 p T^2+162 p^3 T^3+p^6 T^4
\tabularnewline[0.5pt]\hline				
20	 &	smooth    &	          &	1-188 T+1766 p T^2-188 p^3 T^3+p^6 T^4
\tabularnewline[0.5pt]\hline				
21	 &	smooth    &	          &	(1-2 p T+p^3 T^2)(1+346 T+p^3 T^2)
\tabularnewline[0.5pt]\hline				
22	 &	smooth    &	          &	1-258 T+58 p^2 T^2-258 p^3 T^3+p^6 T^4
\tabularnewline[0.5pt]\hline				
23	 &	smooth    &	          &	1+170 T+1890 p T^2+170 p^3 T^3+p^6 T^4
\tabularnewline[0.5pt]\hline				
24	 &	smooth    &	          &	(1-2 p T+p^3 T^2)(1+358 T+p^3 T^2)
\tabularnewline[0.5pt]\hline				
25	 &	smooth    &	          &	1-24 T+238 p T^2-24 p^3 T^3+p^6 T^4
\tabularnewline[0.5pt]\hline				
26	 &	smooth    &	          &	1+12 T+1846 p T^2+12 p^3 T^3+p^6 T^4
\tabularnewline[0.5pt]\hline				
27	 &	smooth    &	          &	1+432 T+3166 p T^2+432 p^3 T^3+p^6 T^4
\tabularnewline[0.5pt]\hline				
28	 &	smooth    &	          &	1-88 T-1074 p T^2-88 p^3 T^3+p^6 T^4
\tabularnewline[0.5pt]\hline				
29	 &	smooth    &	          &	1+432 T+3886 p T^2+432 p^3 T^3+p^6 T^4
\tabularnewline[0.5pt]\hline				
30	 &	smooth    &	          &	1-8 T+2126 p T^2-8 p^3 T^3+p^6 T^4
\tabularnewline[0.5pt]\hline				
31	 &	smooth    &	          &	1+30 T+1090 p T^2+30 p^3 T^3+p^6 T^4
\tabularnewline[0.5pt]\hline				
32	 &	smooth    &	          &	1+30 T+2410 p T^2+30 p^3 T^3+p^6 T^4
\tabularnewline[0.5pt]\hline				
33	 &	singular  &	\frac{1}{9}&	(1-p T) (1-254 T+p^3 T^2)
\tabularnewline[0.5pt]\hline				
34	 &	smooth    &	          &	1-28 T+1446 p T^2-28 p^3 T^3+p^6 T^4
\tabularnewline[0.5pt]\hline				
35	 &	smooth    &	          &	1-196 T+1062 p T^2-196 p^3 T^3+p^6 T^4
\tabularnewline[0.5pt]\hline				
36	 &	smooth    &	          &	1-308 T+2486 p T^2-308 p^3 T^3+p^6 T^4
\tabularnewline[0.5pt]\hline				
\tablepostamble				
\tablepreamble{41}				
1	 &	singular  &	1&	(1-p T) (1-42 T+p^3 T^2)
\tabularnewline[0.5pt]\hline				
2	 &	smooth    &	          &	1-48 T+1214 p T^2-48 p^3 T^3+p^6 T^4
\tabularnewline[0.5pt]\hline				
3	 &	smooth    &	          &	1-230 T+2018 p T^2-230 p^3 T^3+p^6 T^4
\tabularnewline[0.5pt]\hline				
4	 &	smooth    &	          &	1+112 T+2174 p T^2+112 p^3 T^3+p^6 T^4
\tabularnewline[0.5pt]\hline				
5	 &	smooth    &	          &	1+192 T-226 p T^2+192 p^3 T^3+p^6 T^4
\tabularnewline[0.5pt]\hline				
6	 &	smooth    &	          &	1+396 T+2966 p T^2+396 p^3 T^3+p^6 T^4
\tabularnewline[0.5pt]\hline				
7	 &	smooth    &	          &	1-12 T+2342 p T^2-12 p^3 T^3+p^6 T^4
\tabularnewline[0.5pt]\hline				
8	 &	smooth    &	          &	1-396 T+2390 p T^2-396 p^3 T^3+p^6 T^4
\tabularnewline[0.5pt]\hline				
9	 &	smooth    &	          &	1+24 T-850 p T^2+24 p^3 T^3+p^6 T^4
\tabularnewline[0.5pt]\hline				
10	 &	smooth    &	          &	(1+6 p T+p^3 T^2)(1-422 T+p^3 T^2)
\tabularnewline[0.5pt]\hline				
11	 &	smooth    &	          &	1-6 p T+1970 p T^2-6 p^4 T^3+p^6 T^4
\tabularnewline[0.5pt]\hline				
12	 &	smooth    &	          &	1-282 T+482 p T^2-282 p^3 T^3+p^6 T^4
\tabularnewline[0.5pt]\hline				
13	 &	smooth    &	          &	1-106 T+290 p T^2-106 p^3 T^3+p^6 T^4
\tabularnewline[0.5pt]\hline				
14	 &	smooth    &	          &	1+8 T-2338 p T^2+8 p^3 T^3+p^6 T^4
\tabularnewline[0.5pt]\hline				
15	 &	smooth    &	          &	1+182 T+1154 p T^2+182 p^3 T^3+p^6 T^4
\tabularnewline[0.5pt]\hline				
16	 &	smooth    &	          &	1-36 T-2170 p T^2-36 p^3 T^3+p^6 T^4
\tabularnewline[0.5pt]\hline				
17	 &	smooth    &	          &	1+148 T+422 p T^2+148 p^3 T^3+p^6 T^4
\tabularnewline[0.5pt]\hline				
18	 &	smooth    &	          &	1-76 T-490 p T^2-76 p^3 T^3+p^6 T^4
\tabularnewline[0.5pt]\hline				
19	 &	smooth    &	          &	1-30 T+578 p T^2-30 p^3 T^3+p^6 T^4
\tabularnewline[0.5pt]\hline				
20	 &	smooth    &	          &	1-108 T-106 p T^2-108 p^3 T^3+p^6 T^4
\tabularnewline[0.5pt]\hline				
21	 &	smooth    &	          &	1-128 T-1186 p T^2-128 p^3 T^3+p^6 T^4
\tabularnewline[0.5pt]\hline				
22	 &	smooth    &	          &	1-336 T+2030 p T^2-336 p^3 T^3+p^6 T^4
\tabularnewline[0.5pt]\hline				
23	 &	singular  &	\frac{1}{25}&	(1-p T) (1-234 T+p^3 T^2)
\tabularnewline[0.5pt]\hline				
24	 &	smooth    &	          &	1+4 T+470 p T^2+4 p^3 T^3+p^6 T^4
\tabularnewline[0.5pt]\hline				
25	 &	smooth    &	          &	1-36 T-250 p T^2-36 p^3 T^3+p^6 T^4
\tabularnewline[0.5pt]\hline				
26	 &	smooth    &	          &	1-312 T+1742 p T^2-312 p^3 T^3+p^6 T^4
\tabularnewline[0.5pt]\hline				
27	 &	smooth    &	          &	1+168 T-898 p T^2+168 p^3 T^3+p^6 T^4
\tabularnewline[0.5pt]\hline				
28	 &	smooth    &	          &	1+48 T-1138 p T^2+48 p^3 T^3+p^6 T^4
\tabularnewline[0.5pt]\hline				
29	 &	smooth    &	          &	1+172 T+614 p T^2+172 p^3 T^3+p^6 T^4
\tabularnewline[0.5pt]\hline				
30	 &	smooth    &	          &	1-2 p T+2642 p T^2-2 p^4 T^3+p^6 T^4
\tabularnewline[0.5pt]\hline				
31	 &	smooth    &	          &	1+104 T+590 p T^2+104 p^3 T^3+p^6 T^4
\tabularnewline[0.5pt]\hline				
32	 &	singular  &	\frac{1}{9}&	(1+p T) (1-42 T+p^3 T^2)
\tabularnewline[0.5pt]\hline				
33	 &	smooth    &	          &	1+292 T+1334 p T^2+292 p^3 T^3+p^6 T^4
\tabularnewline[0.5pt]\hline				
34	 &	smooth    &	          &	1+282 T+3554 p T^2+282 p^3 T^3+p^6 T^4
\tabularnewline[0.5pt]\hline				
35	 &	smooth    &	          &	(1-6 p T+p^3 T^2)(1-162 T+p^3 T^2)
\tabularnewline[0.5pt]\hline				
36	 &	smooth    &	          &	1+392 T+3374 p T^2+392 p^3 T^3+p^6 T^4
\tabularnewline[0.5pt]\hline				
37	 &	smooth    &	          &	1+284 T+1670 p T^2+284 p^3 T^3+p^6 T^4
\tabularnewline[0.5pt]\hline				
38	 &	smooth    &	          &	1-44 T-154 p T^2-44 p^3 T^3+p^6 T^4
\tabularnewline[0.5pt]\hline				
39	 &	smooth    &	          &	1+552 T+4334 p T^2+552 p^3 T^3+p^6 T^4
\tabularnewline[0.5pt]\hline				
40	 &	smooth    &	          &	1+92 T+2054 p T^2+92 p^3 T^3+p^6 T^4
\tabularnewline[0.5pt]\hline				
\tablepostamble				
\tablepreamble{43}				
1	 &	singular  &	1&	(1-p T) (1+52 T+p^3 T^2)
\tabularnewline[0.5pt]\hline				
2	 &	smooth    &	          &	1-212 T+482 p T^2-212 p^3 T^3+p^6 T^4
\tabularnewline[0.5pt]\hline				
3	 &	smooth    &	          &	1+94 T+2186 p T^2+94 p^3 T^3+p^6 T^4
\tabularnewline[0.5pt]\hline				
4	 &	smooth    &	          &	1+48 T+1282 p T^2+48 p^3 T^3+p^6 T^4
\tabularnewline[0.5pt]\hline				
5	 &	smooth    &	          &	1-72 T+1762 p T^2-72 p^3 T^3+p^6 T^4
\tabularnewline[0.5pt]\hline				
6	 &	smooth    &	          &	(1-8 p T+p^3 T^2)(1+412 T+p^3 T^2)
\tabularnewline[0.5pt]\hline				
7	 &	smooth    &	          &	1+356 T+2994 p T^2+356 p^3 T^3+p^6 T^4
\tabularnewline[0.5pt]\hline				
8	 &	smooth    &	          &	1-336 T+2626 p T^2-336 p^3 T^3+p^6 T^4
\tabularnewline[0.5pt]\hline				
9	 &	smooth    &	          &	1-196 T+3666 p T^2-196 p^3 T^3+p^6 T^4
\tabularnewline[0.5pt]\hline				
10	 &	smooth    &	          &	(1+4 p T+p^3 T^2)(1-508 T+p^3 T^2)
\tabularnewline[0.5pt]\hline				
11	 &	smooth    &	          &	1+84 T-1454 p T^2+84 p^3 T^3+p^6 T^4
\tabularnewline[0.5pt]\hline				
12	 &	smooth    &	          &	1+228 T+322 p T^2+228 p^3 T^3+p^6 T^4
\tabularnewline[0.5pt]\hline				
13	 &	smooth    &	          &	(1+4 p T+p^3 T^2)(1-508 T+p^3 T^2)
\tabularnewline[0.5pt]\hline				
14	 &	smooth    &	          &	1+368 T+2562 p T^2+368 p^3 T^3+p^6 T^4
\tabularnewline[0.5pt]\hline				
15	 &	smooth    &	          &	1+48 T+1282 p T^2+48 p^3 T^3+p^6 T^4
\tabularnewline[0.5pt]\hline				
16	 &	smooth    &	          &	1-196 T-654 p T^2-196 p^3 T^3+p^6 T^4
\tabularnewline[0.5pt]\hline				
17	 &	smooth    &	          &	1+224 T+1986 p T^2+224 p^3 T^3+p^6 T^4
\tabularnewline[0.5pt]\hline				
18	 &	smooth    &	          &	1+152 T+2418 p T^2+152 p^3 T^3+p^6 T^4
\tabularnewline[0.5pt]\hline				
19	 &	smooth    &	          &	1-128 T+818 p T^2-128 p^3 T^3+p^6 T^4
\tabularnewline[0.5pt]\hline				
20	 &	smooth    &	          &	1-12 T-1358 p T^2-12 p^3 T^3+p^6 T^4
\tabularnewline[0.5pt]\hline				
21	 &	smooth    &	          &	1+168 T-26 p^2 T^2+168 p^3 T^3+p^6 T^4
\tabularnewline[0.5pt]\hline				
22	 &	smooth    &	          &	1+38 T-2238 p T^2+38 p^3 T^3+p^6 T^4
\tabularnewline[0.5pt]\hline				
23	 &	smooth    &	          &	1+44 T+2226 p T^2+44 p^3 T^3+p^6 T^4
\tabularnewline[0.5pt]\hline				
24	 &	singular  &	\frac{1}{9}&	(1-p T) (1+52 T+p^3 T^2)
\tabularnewline[0.5pt]\hline				
25	 &	smooth    &	          &	1+124 T+1586 p T^2+124 p^3 T^3+p^6 T^4
\tabularnewline[0.5pt]\hline				
26	 &	smooth    &	          &	1-178 T+1938 p T^2-178 p^3 T^3+p^6 T^4
\tabularnewline[0.5pt]\hline				
27	 &	smooth    &	          &	(1-8 p T+p^3 T^2)(1+292 T+p^3 T^2)
\tabularnewline[0.5pt]\hline				
28	 &	smooth    &	          &	1+96 T-1166 p T^2+96 p^3 T^3+p^6 T^4
\tabularnewline[0.5pt]\hline				
29	 &	smooth    &	          &	1-162 T+1882 p T^2-162 p^3 T^3+p^6 T^4
\tabularnewline[0.5pt]\hline				
30	 &	smooth    &	          &	1-138 T+1738 p T^2-138 p^3 T^3+p^6 T^4
\tabularnewline[0.5pt]\hline				
31	 &	singular  &	\frac{1}{25}&	(1+p T) (1+412 T+p^3 T^2)
\tabularnewline[0.5pt]\hline				
32	 &	smooth    &	          &	1-214 T-366 p T^2-214 p^3 T^3+p^6 T^4
\tabularnewline[0.5pt]\hline				
33	 &	smooth    &	          &	1+326 T+2394 p T^2+326 p^3 T^3+p^6 T^4
\tabularnewline[0.5pt]\hline				
34	 &	smooth    &	          &	1-498 T+4618 p T^2-498 p^3 T^3+p^6 T^4
\tabularnewline[0.5pt]\hline				
35	 &	smooth    &	          &	1-176 T+1346 p T^2-176 p^3 T^3+p^6 T^4
\tabularnewline[0.5pt]\hline				
36	 &	smooth    &	          &	(1+4 p T+p^3 T^2)(1-244 T+p^3 T^2)
\tabularnewline[0.5pt]\hline				
37	 &	smooth    &	          &	1-136 T+3426 p T^2-136 p^3 T^3+p^6 T^4
\tabularnewline[0.5pt]\hline				
38	 &	smooth    &	          &	1+188 T+3282 p T^2+188 p^3 T^3+p^6 T^4
\tabularnewline[0.5pt]\hline				
39	 &	smooth    &	          &	1+274 T+3506 p T^2+274 p^3 T^3+p^6 T^4
\tabularnewline[0.5pt]\hline				
40	 &	smooth    &	          &	1+264 T+1186 p T^2+264 p^3 T^3+p^6 T^4
\tabularnewline[0.5pt]\hline				
41	 &	smooth    &	          &	(1+4 p T+p^3 T^2)(1-428 T+p^3 T^2)
\tabularnewline[0.5pt]\hline				
42	 &	smooth    &	          &	1+42 T-1982 p T^2+42 p^3 T^3+p^6 T^4
\tabularnewline[0.5pt]\hline				
\tablepostamble				
\tablepreamble{47}				
1	 &	singular  &	1&	(1-p T) (1+96 T+p^3 T^2)
\tabularnewline[0.5pt]\hline				
2	 &	smooth    &	          &	(1+p^3 T^2)(1-264 T+p^3 T^2)
\tabularnewline[0.5pt]\hline				
3	 &	smooth    &	          &	(1+p^3 T^2)(1+136 T+p^3 T^2)
\tabularnewline[0.5pt]\hline				
4	 &	smooth    &	          &	1-240 T+1730 p T^2-240 p^3 T^3+p^6 T^4
\tabularnewline[0.5pt]\hline				
5	 &	smooth    &	          &	1-216 T+1202 p T^2-216 p^3 T^3+p^6 T^4
\tabularnewline[0.5pt]\hline				
6	 &	smooth    &	          &	1-480 T+3650 p T^2-480 p^3 T^3+p^6 T^4
\tabularnewline[0.5pt]\hline				
7	 &	smooth    &	          &	1+8 p T+3458 p T^2+8 p^4 T^3+p^6 T^4
\tabularnewline[0.5pt]\hline				
8	 &	smooth    &	          &	1-144 T+1538 p T^2-144 p^3 T^3+p^6 T^4
\tabularnewline[0.5pt]\hline				
9	 &	smooth    &	          &	(1+p^3 T^2)(1-444 T+p^3 T^2)
\tabularnewline[0.5pt]\hline				
10	 &	smooth    &	          &	1+72 T+866 p T^2+72 p^3 T^3+p^6 T^4
\tabularnewline[0.5pt]\hline				
11	 &	smooth    &	          &	(1+8 p T+p^3 T^2)(1-402 T+p^3 T^2)
\tabularnewline[0.5pt]\hline				
12	 &	smooth    &	          &	1-124 T+578 p T^2-124 p^3 T^3+p^6 T^4
\tabularnewline[0.5pt]\hline				
13	 &	smooth    &	          &	1-2 T+74 p T^2-2 p^3 T^3+p^6 T^4
\tabularnewline[0.5pt]\hline				
14	 &	smooth    &	          &	1-264 T+578 p T^2-264 p^3 T^3+p^6 T^4
\tabularnewline[0.5pt]\hline				
15	 &	smooth    &	          &	1-276 T+3842 p T^2-276 p^3 T^3+p^6 T^4
\tabularnewline[0.5pt]\hline				
16	 &	smooth    &	          &	(1+p^3 T^2)(1+136 T+p^3 T^2)
\tabularnewline[0.5pt]\hline				
17	 &	smooth    &	          &	(1-12 p T+p^3 T^2)(1+416 T+p^3 T^2)
\tabularnewline[0.5pt]\hline				
18	 &	smooth    &	          &	1-64 T+3458 p T^2-64 p^3 T^3+p^6 T^4
\tabularnewline[0.5pt]\hline				
19	 &	smooth    &	          &	1+98 T+1754 p T^2+98 p^3 T^3+p^6 T^4
\tabularnewline[0.5pt]\hline				
20	 &	smooth    &	          &	(1+12 p T+p^3 T^2)(1-24 T+p^3 T^2)
\tabularnewline[0.5pt]\hline				
21	 &	singular  &	\frac{1}{9}&	(1+p T) (1+96 T+p^3 T^2)
\tabularnewline[0.5pt]\hline				
22	 &	smooth    &	          &	1+436 T+3218 p T^2+436 p^3 T^3+p^6 T^4
\tabularnewline[0.5pt]\hline				
23	 &	smooth    &	          &	1+18 T+434 p T^2+18 p^3 T^3+p^6 T^4
\tabularnewline[0.5pt]\hline				
24	 &	smooth    &	          &	1+440 T+3650 p T^2+440 p^3 T^3+p^6 T^4
\tabularnewline[0.5pt]\hline				
25	 &	smooth    &	          &	1-244 T+2018 p T^2-244 p^3 T^3+p^6 T^4
\tabularnewline[0.5pt]\hline				
26	 &	smooth    &	          &	(1-12 p T+p^3 T^2)(1-32 T+p^3 T^2)
\tabularnewline[0.5pt]\hline				
27	 &	smooth    &	          &	1-288 T+1346 p T^2-288 p^3 T^3+p^6 T^4
\tabularnewline[0.5pt]\hline				
28	 &	smooth    &	          &	1-304 T+3458 p T^2-304 p^3 T^3+p^6 T^4
\tabularnewline[0.5pt]\hline				
29	 &	smooth    &	          &	1+166 T+218 p T^2+166 p^3 T^3+p^6 T^4
\tabularnewline[0.5pt]\hline				
30	 &	smooth    &	          &	1-106 T+3602 p T^2-106 p^3 T^3+p^6 T^4
\tabularnewline[0.5pt]\hline				
31	 &	smooth    &	          &	1+156 T-382 p T^2+156 p^3 T^3+p^6 T^4
\tabularnewline[0.5pt]\hline				
32	 &	singular  &	\frac{1}{25}&	(1+p T) (1+360 T+p^3 T^2)
\tabularnewline[0.5pt]\hline				
33	 &	smooth    &	          &	1+536 T+4898 p T^2+536 p^3 T^3+p^6 T^4
\tabularnewline[0.5pt]\hline				
34	 &	smooth    &	          &	1+676 T+4898 p T^2+676 p^3 T^3+p^6 T^4
\tabularnewline[0.5pt]\hline				
35	 &	smooth    &	          &	1-560 T+5570 p T^2-560 p^3 T^3+p^6 T^4
\tabularnewline[0.5pt]\hline				
36	 &	smooth    &	          &	1-304 T+578 p T^2-304 p^3 T^3+p^6 T^4
\tabularnewline[0.5pt]\hline				
37	 &	smooth    &	          &	1+16 T-2302 p T^2+16 p^3 T^3+p^6 T^4
\tabularnewline[0.5pt]\hline				
38	 &	smooth    &	          &	1+296 T+2978 p T^2+296 p^3 T^3+p^6 T^4
\tabularnewline[0.5pt]\hline				
39	 &	smooth    &	          &	1+708 T+5714 p T^2+708 p^3 T^3+p^6 T^4
\tabularnewline[0.5pt]\hline				
40	 &	smooth    &	          &	1+270 T+2930 p T^2+270 p^3 T^3+p^6 T^4
\tabularnewline[0.5pt]\hline				
41	 &	smooth    &	          &	1+222 T+1826 p T^2+222 p^3 T^3+p^6 T^4
\tabularnewline[0.5pt]\hline				
42	 &	smooth    &	          &	1-168 T+3266 p T^2-168 p^3 T^3+p^6 T^4
\tabularnewline[0.5pt]\hline				
43	 &	smooth    &	          &	1+488 T+3314 p T^2+488 p^3 T^3+p^6 T^4
\tabularnewline[0.5pt]\hline				
44	 &	smooth    &	          &	1-158 T+1826 p T^2-158 p^3 T^3+p^6 T^4
\tabularnewline[0.5pt]\hline				
45	 &	smooth    &	          &	1-366 T+3122 p T^2-366 p^3 T^3+p^6 T^4
\tabularnewline[0.5pt]\hline				
46	 &	smooth    &	          &	1-612 T+5714 p T^2-612 p^3 T^3+p^6 T^4
\tabularnewline[0.5pt]\hline				
\tablepostamble				
\tablepreamble{53}				
1	 &	singular  &	1&	(1-p T) (1-198 T+p^3 T^2)
\tabularnewline[0.5pt]\hline				
2	 &	smooth    &	          &	1-560 T+6350 p T^2-560 p^3 T^3+p^6 T^4
\tabularnewline[0.5pt]\hline				
3	 &	smooth    &	          &	1+744 T+7886 p T^2+744 p^3 T^3+p^6 T^4
\tabularnewline[0.5pt]\hline				
4	 &	smooth    &	          &	1+984 T+8846 p T^2+984 p^3 T^3+p^6 T^4
\tabularnewline[0.5pt]\hline				
5	 &	smooth    &	          &	1+330 T+5450 p T^2+330 p^3 T^3+p^6 T^4
\tabularnewline[0.5pt]\hline				
6	 &	singular  &	\frac{1}{9}&	(1+p T) (1-198 T+p^3 T^2)
\tabularnewline[0.5pt]\hline				
7	 &	smooth    &	          &	1+444 T+4886 p T^2+444 p^3 T^3+p^6 T^4
\tabularnewline[0.5pt]\hline				
8	 &	smooth    &	          &	1+198 T+5042 p T^2+198 p^3 T^3+p^6 T^4
\tabularnewline[0.5pt]\hline				
9	 &	smooth    &	          &	1+280 T+4430 p T^2+280 p^3 T^3+p^6 T^4
\tabularnewline[0.5pt]\hline				
10	 &	smooth    &	          &	(1-6 p T+p^3 T^2)(1+498 T+p^3 T^2)
\tabularnewline[0.5pt]\hline				
11	 &	smooth    &	          &	1-200 T+2030 p T^2-200 p^3 T^3+p^6 T^4
\tabularnewline[0.5pt]\hline				
12	 &	smooth    &	          &	1-30 T+3290 p T^2-30 p^3 T^3+p^6 T^4
\tabularnewline[0.5pt]\hline				
13	 &	smooth    &	          &	1+64 T-4354 p T^2+64 p^3 T^3+p^6 T^4
\tabularnewline[0.5pt]\hline				
14	 &	smooth    &	          &	1+336 T+4574 p T^2+336 p^3 T^3+p^6 T^4
\tabularnewline[0.5pt]\hline				
15	 &	smooth    &	          &	(1-6 p T+p^3 T^2)2
\tabularnewline[0.5pt]\hline				
16	 &	smooth    &	          &	1-240 T+4190 p T^2-240 p^3 T^3+p^6 T^4
\tabularnewline[0.5pt]\hline				
17	 &	singular  &	\frac{1}{25}&	(1+p T) (1-222 T+p^3 T^2)
\tabularnewline[0.5pt]\hline				
18	 &	smooth    &	          &	1+442 T+4298 p T^2+442 p^3 T^3+p^6 T^4
\tabularnewline[0.5pt]\hline				
19	 &	smooth    &	          &	1+6 p T+1082 p T^2+6 p^4 T^3+p^6 T^4
\tabularnewline[0.5pt]\hline				
20	 &	smooth    &	          &	1-336 T+4766 p T^2-336 p^3 T^3+p^6 T^4
\tabularnewline[0.5pt]\hline				
21	 &	smooth    &	          &	1+108 T-58 p T^2+108 p^3 T^3+p^6 T^4
\tabularnewline[0.5pt]\hline				
22	 &	smooth    &	          &	1+182 T+698 p T^2+182 p^3 T^3+p^6 T^4
\tabularnewline[0.5pt]\hline				
23	 &	smooth    &	          &	1-8 p T+3614 p T^2-8 p^4 T^3+p^6 T^4
\tabularnewline[0.5pt]\hline				
24	 &	smooth    &	          &	(1-6 p T+p^3 T^2)2
\tabularnewline[0.5pt]\hline				
25	 &	smooth    &	          &	1-460 T+3590 p T^2-460 p^3 T^3+p^6 T^4
\tabularnewline[0.5pt]\hline				
26	 &	smooth    &	          &	1+182 T+338 p T^2+182 p^3 T^3+p^6 T^4
\tabularnewline[0.5pt]\hline				
27	 &	smooth    &	          &	1-66 T+2906 p T^2-66 p^3 T^3+p^6 T^4
\tabularnewline[0.5pt]\hline				
28	 &	smooth    &	          &	1+580 T+4070 p T^2+580 p^3 T^3+p^6 T^4
\tabularnewline[0.5pt]\hline				
29	 &	smooth    &	          &	1-480 T+1790 p T^2-480 p^3 T^3+p^6 T^4
\tabularnewline[0.5pt]\hline				
30	 &	smooth    &	          &	1-316 T-154 p T^2-316 p^3 T^3+p^6 T^4
\tabularnewline[0.5pt]\hline				
31	 &	smooth    &	          &	1-194 T+4394 p T^2-194 p^3 T^3+p^6 T^4
\tabularnewline[0.5pt]\hline				
32	 &	smooth    &	          &	1+270 T+1130 p T^2+270 p^3 T^3+p^6 T^4
\tabularnewline[0.5pt]\hline				
33	 &	smooth    &	          &	1+14 T+26 p T^2+14 p^3 T^3+p^6 T^4
\tabularnewline[0.5pt]\hline				
34	 &	smooth    &	          &	1-148 T+1478 p T^2-148 p^3 T^3+p^6 T^4
\tabularnewline[0.5pt]\hline				
35	 &	smooth    &	          &	1+702 T+5018 p T^2+702 p^3 T^3+p^6 T^4
\tabularnewline[0.5pt]\hline				
36	 &	smooth    &	          &	1-256 T+4286 p T^2-256 p^3 T^3+p^6 T^4
\tabularnewline[0.5pt]\hline				
37	 &	smooth    &	          &	1+300 T+2870 p T^2+300 p^3 T^3+p^6 T^4
\tabularnewline[0.5pt]\hline				
38	 &	smooth    &	          &	1+420 T+5990 p T^2+420 p^3 T^3+p^6 T^4
\tabularnewline[0.5pt]\hline				
39	 &	smooth    &	          &	1+16 T+254 p T^2+16 p^3 T^3+p^6 T^4
\tabularnewline[0.5pt]\hline				
40	 &	smooth    &	          &	1+100 T+1190 p T^2+100 p^3 T^3+p^6 T^4
\tabularnewline[0.5pt]\hline				
41	 &	smooth    &	          &	1-798 T+6938 p T^2-798 p^3 T^3+p^6 T^4
\tabularnewline[0.5pt]\hline				
42	 &	smooth    &	          &	(1-6 p T+p^3 T^2)2
\tabularnewline[0.5pt]\hline				
43	 &	smooth    &	          &	1-296 T+686 p T^2-296 p^3 T^3+p^6 T^4
\tabularnewline[0.5pt]\hline				
44	 &	smooth    &	          &	1+4 T-154 p T^2+4 p^3 T^3+p^6 T^4
\tabularnewline[0.5pt]\hline				
45	 &	smooth    &	          &	1+218 T-358 p T^2+218 p^3 T^3+p^6 T^4
\tabularnewline[0.5pt]\hline				
46	 &	smooth    &	          &	(1+14 p T+p^3 T^2)(1-6 p T+p^3 T^2)
\tabularnewline[0.5pt]\hline				
47	 &	smooth    &	          &	1+384 T+2366 p T^2+384 p^3 T^3+p^6 T^4
\tabularnewline[0.5pt]\hline				
48	 &	smooth    &	          &	1-550 T+5690 p T^2-550 p^3 T^3+p^6 T^4
\tabularnewline[0.5pt]\hline				
49	 &	smooth    &	          &	(1-6 p T+p^3 T^2)(1+42 T+p^3 T^2)
\tabularnewline[0.5pt]\hline				
50	 &	smooth    &	          &	1-310 T+4250 p T^2-310 p^3 T^3+p^6 T^4
\tabularnewline[0.5pt]\hline				
51	 &	smooth    &	          &	1+224 T+4286 p T^2+224 p^3 T^3+p^6 T^4
\tabularnewline[0.5pt]\hline				
52	 &	smooth    &	          &	1-120 T-2770 p T^2-120 p^3 T^3+p^6 T^4
\tabularnewline[0.5pt]\hline				
\tablepostamble				
\tablepreamble{59}				
1	 &	singular  &	1&	(1-p T) (1+660 T+p^3 T^2)
\tabularnewline[0.5pt]\hline				
2	 &	smooth    &	          &	1+200 T+482 p T^2+200 p^3 T^3+p^6 T^4
\tabularnewline[0.5pt]\hline				
3	 &	smooth    &	          &	1+72 T-1438 p T^2+72 p^3 T^3+p^6 T^4
\tabularnewline[0.5pt]\hline				
4	 &	smooth    &	          &	1+268 T+242 p T^2+268 p^3 T^3+p^6 T^4
\tabularnewline[0.5pt]\hline				
5	 &	smooth    &	          &	1-392 T-478 p T^2-392 p^3 T^3+p^6 T^4
\tabularnewline[0.5pt]\hline				
6	 &	smooth    &	          &	1+850 T+7442 p T^2+850 p^3 T^3+p^6 T^4
\tabularnewline[0.5pt]\hline				
7	 &	smooth    &	          &	1+300 T-2158 p T^2+300 p^3 T^3+p^6 T^4
\tabularnewline[0.5pt]\hline				
8	 &	smooth    &	          &	1+648 T+4322 p T^2+648 p^3 T^3+p^6 T^4
\tabularnewline[0.5pt]\hline				
9	 &	smooth    &	          &	1-840 T+9122 p T^2-840 p^3 T^3+p^6 T^4
\tabularnewline[0.5pt]\hline				
10	 &	smooth    &	          &	1+96 T-4558 p T^2+96 p^3 T^3+p^6 T^4
\tabularnewline[0.5pt]\hline				
11	 &	smooth    &	          &	1+240 T-958 p T^2+240 p^3 T^3+p^6 T^4
\tabularnewline[0.5pt]\hline				
12	 &	smooth    &	          &	1+512 T+3842 p T^2+512 p^3 T^3+p^6 T^4
\tabularnewline[0.5pt]\hline				
13	 &	smooth    &	          &	1+108 T+2 p T^2+108 p^3 T^3+p^6 T^4
\tabularnewline[0.5pt]\hline				
14	 &	smooth    &	          &	1+86 T-2 p^2 T^2+86 p^3 T^3+p^6 T^4
\tabularnewline[0.5pt]\hline				
15	 &	smooth    &	          &	1+160 T+1922 p T^2+160 p^3 T^3+p^6 T^4
\tabularnewline[0.5pt]\hline				
16	 &	smooth    &	          &	1-448 T+962 p T^2-448 p^3 T^3+p^6 T^4
\tabularnewline[0.5pt]\hline				
17	 &	smooth    &	          &	1-80 T+962 p T^2-80 p^3 T^3+p^6 T^4
\tabularnewline[0.5pt]\hline				
18	 &	smooth    &	          &	1-54 T-958 p T^2-54 p^3 T^3+p^6 T^4
\tabularnewline[0.5pt]\hline				
19	 &	smooth    &	          &	1+288 T+2 p T^2+288 p^3 T^3+p^6 T^4
\tabularnewline[0.5pt]\hline				
20	 &	smooth    &	          &	1-480 T+5762 p T^2-480 p^3 T^3+p^6 T^4
\tabularnewline[0.5pt]\hline				
21	 &	smooth    &	          &	(1-12 p T+p^3 T^2)(1-300 T+p^3 T^2)
\tabularnewline[0.5pt]\hline				
22	 &	smooth    &	          &	1-292 T+1202 p T^2-292 p^3 T^3+p^6 T^4
\tabularnewline[0.5pt]\hline				
23	 &	smooth    &	          &	1-10 T-958 p T^2-10 p^3 T^3+p^6 T^4
\tabularnewline[0.5pt]\hline				
24	 &	smooth    &	          &	1-470 T+6122 p T^2-470 p^3 T^3+p^6 T^4
\tabularnewline[0.5pt]\hline				
25	 &	smooth    &	          &	1-72 T-478 p T^2-72 p^3 T^3+p^6 T^4
\tabularnewline[0.5pt]\hline				
26	 &	singular  &	\frac{1}{25}&	(1-p T) (1-660 T+p^3 T^2)
\tabularnewline[0.5pt]\hline				
27	 &	smooth    &	          &	1+100 T+1202 p T^2+100 p^3 T^3+p^6 T^4
\tabularnewline[0.5pt]\hline				
28	 &	smooth    &	          &	1+112 T+6722 p T^2+112 p^3 T^3+p^6 T^4
\tabularnewline[0.5pt]\hline				
29	 &	smooth    &	          &	1-180 T+242 p T^2-180 p^3 T^3+p^6 T^4
\tabularnewline[0.5pt]\hline				
30	 &	smooth    &	          &	1-786 T+7682 p T^2-786 p^3 T^3+p^6 T^4
\tabularnewline[0.5pt]\hline				
31	 &	smooth    &	          &	1-210 T+602 p T^2-210 p^3 T^3+p^6 T^4
\tabularnewline[0.5pt]\hline				
32	 &	smooth    &	          &	1-54 T+3002 p T^2-54 p^3 T^3+p^6 T^4
\tabularnewline[0.5pt]\hline				
33	 &	smooth    &	          &	1-464 T+5042 p T^2-464 p^3 T^3+p^6 T^4
\tabularnewline[0.5pt]\hline				
34	 &	smooth    &	          &	1+192 T+2 p T^2+192 p^3 T^3+p^6 T^4
\tabularnewline[0.5pt]\hline				
35	 &	smooth    &	          &	1-232 T+3362 p T^2-232 p^3 T^3+p^6 T^4
\tabularnewline[0.5pt]\hline				
36	 &	smooth    &	          &	1-188 T-238 p T^2-188 p^3 T^3+p^6 T^4
\tabularnewline[0.5pt]\hline				
37	 &	smooth    &	          &	1-56 T+242 p T^2-56 p^3 T^3+p^6 T^4
\tabularnewline[0.5pt]\hline				
38	 &	smooth    &	          &	1+776 T+6482 p T^2+776 p^3 T^3+p^6 T^4
\tabularnewline[0.5pt]\hline				
39	 &	smooth    &	          &	1-680 T+4322 p T^2-680 p^3 T^3+p^6 T^4
\tabularnewline[0.5pt]\hline				
40	 &	smooth    &	          &	1-1006 T+10202 p T^2-1006 p^3 T^3+p^6 T^4
\tabularnewline[0.5pt]\hline				
41	 &	smooth    &	          &	1+912 T+8642 p T^2+912 p^3 T^3+p^6 T^4
\tabularnewline[0.5pt]\hline				
42	 &	smooth    &	          &	(1+6 p T+p^3 T^2)(1+200 T+p^3 T^2)
\tabularnewline[0.5pt]\hline				
43	 &	smooth    &	          &	1+978 T+8282 p T^2+978 p^3 T^3+p^6 T^4
\tabularnewline[0.5pt]\hline				
44	 &	smooth    &	          &	1+726 T+7562 p T^2+726 p^3 T^3+p^6 T^4
\tabularnewline[0.5pt]\hline				
45	 &	smooth    &	          &	(1-12 p T+p^3 T^2)(1-300 T+p^3 T^2)
\tabularnewline[0.5pt]\hline				
46	 &	singular  &	\frac{1}{9}&	(1+p T) (1+660 T+p^3 T^2)
\tabularnewline[0.5pt]\hline				
47	 &	smooth    &	          &	1+380 T+1922 p T^2+380 p^3 T^3+p^6 T^4
\tabularnewline[0.5pt]\hline				
48	 &	smooth    &	          &	1-632 T+8162 p T^2-632 p^3 T^3+p^6 T^4
\tabularnewline[0.5pt]\hline				
49	 &	smooth    &	          &	1-148 T+1202 p T^2-148 p^3 T^3+p^6 T^4
\tabularnewline[0.5pt]\hline				
50	 &	smooth    &	          &	1+2 T+1202 p T^2+2 p^3 T^3+p^6 T^4
\tabularnewline[0.5pt]\hline				
51	 &	smooth    &	          &	1-280 T-478 p T^2-280 p^3 T^3+p^6 T^4
\tabularnewline[0.5pt]\hline				
52	 &	smooth    &	          &	1+54 T-1918 p T^2+54 p^3 T^3+p^6 T^4
\tabularnewline[0.5pt]\hline				
53	 &	smooth    &	          &	1+208 T-1918 p T^2+208 p^3 T^3+p^6 T^4
\tabularnewline[0.5pt]\hline				
54	 &	smooth    &	          &	1+6 p T-238 p T^2+6 p^4 T^3+p^6 T^4
\tabularnewline[0.5pt]\hline				
55	 &	smooth    &	          &	1+544 T+2162 p T^2+544 p^3 T^3+p^6 T^4
\tabularnewline[0.5pt]\hline				
56	 &	smooth    &	          &	1-240 T+5762 p T^2-240 p^3 T^3+p^6 T^4
\tabularnewline[0.5pt]\hline				
57	 &	smooth    &	          &	1-352 T+1922 p T^2-352 p^3 T^3+p^6 T^4
\tabularnewline[0.5pt]\hline				
58	 &	smooth    &	          &	1+342 T+3362 p T^2+342 p^3 T^3+p^6 T^4
\tabularnewline[0.5pt]\hline				
\tablepostamble				
\tablepreamble{61}				
1	 &	singular  &	1&	(1-p T) (1+538 T+p^3 T^2)
\tabularnewline[0.5pt]\hline				
2	 &	smooth    &	          &	1-278 T+3914 p T^2-278 p^3 T^3+p^6 T^4
\tabularnewline[0.5pt]\hline				
3	 &	smooth    &	          &	1+376 T+1646 p T^2+376 p^3 T^3+p^6 T^4
\tabularnewline[0.5pt]\hline				
4	 &	smooth    &	          &	1-1264 T+13566 p T^2-1264 p^3 T^3+p^6 T^4
\tabularnewline[0.5pt]\hline				
5	 &	smooth    &	          &	1-564 T+46 p^2 T^2-564 p^3 T^3+p^6 T^4
\tabularnewline[0.5pt]\hline				
6	 &	smooth    &	          &	1+330 T+7258 p T^2+330 p^3 T^3+p^6 T^4
\tabularnewline[0.5pt]\hline				
7	 &	smooth    &	          &	1-60 T+118 p T^2-60 p^3 T^3+p^6 T^4
\tabularnewline[0.5pt]\hline				
8	 &	smooth    &	          &	1-274 T-534 p T^2-274 p^3 T^3+p^6 T^4
\tabularnewline[0.5pt]\hline				
9	 &	smooth    &	          &	1+8 p T+1902 p T^2+8 p^4 T^3+p^6 T^4
\tabularnewline[0.5pt]\hline				
10	 &	smooth    &	          &	1-1410 T+15298 p T^2-1410 p^3 T^3+p^6 T^4
\tabularnewline[0.5pt]\hline				
11	 &	smooth    &	          &	1+168 T+4462 p T^2+168 p^3 T^3+p^6 T^4
\tabularnewline[0.5pt]\hline				
12	 &	smooth    &	          &	(1-14 p T+p^3 T^2)(1-102 T+p^3 T^2)
\tabularnewline[0.5pt]\hline				
13	 &	smooth    &	          &	1-276 T+5110 p T^2-276 p^3 T^3+p^6 T^4
\tabularnewline[0.5pt]\hline				
14	 &	smooth    &	          &	1-272 T+1982 p T^2-272 p^3 T^3+p^6 T^4
\tabularnewline[0.5pt]\hline				
15	 &	smooth    &	          &	1+556 T+1526 p T^2+556 p^3 T^3+p^6 T^4
\tabularnewline[0.5pt]\hline				
16	 &	smooth    &	          &	1+628 T+2822 p T^2+628 p^3 T^3+p^6 T^4
\tabularnewline[0.5pt]\hline				
17	 &	smooth    &	          &	(1-2 p T+p^3 T^2)(1+824 T+p^3 T^2)
\tabularnewline[0.5pt]\hline				
18	 &	smooth    &	          &	1-388 T+2214 p T^2-388 p^3 T^3+p^6 T^4
\tabularnewline[0.5pt]\hline				
19	 &	smooth    &	          &	1+116 T+966 p T^2+116 p^3 T^3+p^6 T^4
\tabularnewline[0.5pt]\hline				
20	 &	smooth    &	          &	1+844 T+8630 p T^2+844 p^3 T^3+p^6 T^4
\tabularnewline[0.5pt]\hline				
21	 &	smooth    &	          &	1+276 T+5446 p T^2+276 p^3 T^3+p^6 T^4
\tabularnewline[0.5pt]\hline				
22	 &	singular  &	\frac{1}{25}&	(1-p T) (1+490 T+p^3 T^2)
\tabularnewline[0.5pt]\hline				
23	 &	smooth    &	          &	1-470 T+6218 p T^2-470 p^3 T^3+p^6 T^4
\tabularnewline[0.5pt]\hline				
24	 &	smooth    &	          &	1-532 T+7542 p T^2-532 p^3 T^3+p^6 T^4
\tabularnewline[0.5pt]\hline				
25	 &	smooth    &	          &	1+648 T+6382 p T^2+648 p^3 T^3+p^6 T^4
\tabularnewline[0.5pt]\hline				
26	 &	smooth    &	          &	(1-8 p T+p^3 T^2)(1+198 T+p^3 T^2)
\tabularnewline[0.5pt]\hline				
27	 &	smooth    &	          &	1-744 T+4846 p T^2-744 p^3 T^3+p^6 T^4
\tabularnewline[0.5pt]\hline				
28	 &	smooth    &	          &	1-60 T+3478 p T^2-60 p^3 T^3+p^6 T^4
\tabularnewline[0.5pt]\hline				
29	 &	smooth    &	          &	1+42 T+2194 p T^2+42 p^3 T^3+p^6 T^4
\tabularnewline[0.5pt]\hline				
30	 &	smooth    &	          &	1-30 T-3182 p T^2-30 p^3 T^3+p^6 T^4
\tabularnewline[0.5pt]\hline				
31	 &	smooth    &	          &	1-1000 T+9438 p T^2-1000 p^3 T^3+p^6 T^4
\tabularnewline[0.5pt]\hline				
32	 &	smooth    &	          &	1+188 T+2262 p T^2+188 p^3 T^3+p^6 T^4
\tabularnewline[0.5pt]\hline				
33	 &	smooth    &	          &	1+110 T+858 p T^2+110 p^3 T^3+p^6 T^4
\tabularnewline[0.5pt]\hline				
34	 &	singular  &	\frac{1}{9}&	(1-p T) (1+538 T+p^3 T^2)
\tabularnewline[0.5pt]\hline				
35	 &	smooth    &	          &	1-332 T+62 p^2 T^2-332 p^3 T^3+p^6 T^4
\tabularnewline[0.5pt]\hline				
36	 &	smooth    &	          &	(1-2 p T+p^3 T^2)(1-590 T+p^3 T^2)
\tabularnewline[0.5pt]\hline				
37	 &	smooth    &	          &	1+408 T+382 p T^2+408 p^3 T^3+p^6 T^4
\tabularnewline[0.5pt]\hline				
38	 &	smooth    &	          &	1-180 T+838 p T^2-180 p^3 T^3+p^6 T^4
\tabularnewline[0.5pt]\hline				
39	 &	smooth    &	          &	1-424 T+2286 p T^2-424 p^3 T^3+p^6 T^4
\tabularnewline[0.5pt]\hline				
40	 &	smooth    &	          &	1-700 T+5718 p T^2-700 p^3 T^3+p^6 T^4
\tabularnewline[0.5pt]\hline				
41	 &	smooth    &	          &	1-96 T+2110 p T^2-96 p^3 T^3+p^6 T^4
\tabularnewline[0.5pt]\hline				
42	 &	smooth    &	          &	1-252 T+1702 p T^2-252 p^3 T^3+p^6 T^4
\tabularnewline[0.5pt]\hline				
43	 &	smooth    &	          &	1+498 T+2362 p T^2+498 p^3 T^3+p^6 T^4
\tabularnewline[0.5pt]\hline				
44	 &	smooth    &	          &	1-246 T-1790 p T^2-246 p^3 T^3+p^6 T^4
\tabularnewline[0.5pt]\hline				
45	 &	smooth    &	          &	1+248 T-18 p T^2+248 p^3 T^3+p^6 T^4
\tabularnewline[0.5pt]\hline				
46	 &	smooth    &	          &	1-192 T+862 p T^2-192 p^3 T^3+p^6 T^4
\tabularnewline[0.5pt]\hline				
47	 &	smooth    &	          &	1-264 T+526 p T^2-264 p^3 T^3+p^6 T^4
\tabularnewline[0.5pt]\hline				
48	 &	smooth    &	          &	1-264 T+5326 p T^2-264 p^3 T^3+p^6 T^4
\tabularnewline[0.5pt]\hline				
49	 &	smooth    &	          &	1+476 T+4566 p T^2+476 p^3 T^3+p^6 T^4
\tabularnewline[0.5pt]\hline				
50	 &	smooth    &	          &	(1+8 p T+p^3 T^2)(1+718 T+p^3 T^2)
\tabularnewline[0.5pt]\hline				
51	 &	smooth    &	          &	1+636 T+9046 p T^2+636 p^3 T^3+p^6 T^4
\tabularnewline[0.5pt]\hline				
52	 &	smooth    &	          &	1+996 T+10726 p T^2+996 p^3 T^3+p^6 T^4
\tabularnewline[0.5pt]\hline				
53	 &	smooth    &	          &	1+378 T+4762 p T^2+378 p^3 T^3+p^6 T^4
\tabularnewline[0.5pt]\hline				
54	 &	smooth    &	          &	1+570 T+6658 p T^2+570 p^3 T^3+p^6 T^4
\tabularnewline[0.5pt]\hline				
55	 &	smooth    &	          &	1+128 T-2178 p T^2+128 p^3 T^3+p^6 T^4
\tabularnewline[0.5pt]\hline				
56	 &	smooth    &	          &	1+168 T-1298 p T^2+168 p^3 T^3+p^6 T^4
\tabularnewline[0.5pt]\hline				
57	 &	smooth    &	          &	1+168 T+2542 p T^2+168 p^3 T^3+p^6 T^4
\tabularnewline[0.5pt]\hline				
58	 &	smooth    &	          &	1+268 T-1738 p T^2+268 p^3 T^3+p^6 T^4
\tabularnewline[0.5pt]\hline				
59	 &	smooth    &	          &	1-326 T+7370 p T^2-326 p^3 T^3+p^6 T^4
\tabularnewline[0.5pt]\hline				
60	 &	smooth    &	          &	1-264 T+1006 p T^2-264 p^3 T^3+p^6 T^4
\tabularnewline[0.5pt]\hline				
\tablepostamble				
\tablepreamble{67}				
1	 &	singular  &	1&	(1-p T) (1-884 T+p^3 T^2)
\tabularnewline[0.5pt]\hline				
2	 &	smooth    &	          &	1+460 T+1730 p T^2+460 p^3 T^3+p^6 T^4
\tabularnewline[0.5pt]\hline				
3	 &	smooth    &	          &	1-594 T+9418 p T^2-594 p^3 T^3+p^6 T^4
\tabularnewline[0.5pt]\hline				
4	 &	smooth    &	          &	(1+12 p T+p^3 T^2)(1-1004 T+p^3 T^2)
\tabularnewline[0.5pt]\hline				
5	 &	smooth    &	          &	1-364 T+3858 p T^2-364 p^3 T^3+p^6 T^4
\tabularnewline[0.5pt]\hline				
6	 &	smooth    &	          &	1-476 T+6482 p T^2-476 p^3 T^3+p^6 T^4
\tabularnewline[0.5pt]\hline				
7	 &	smooth    &	          &	1-156 T-1118 p T^2-156 p^3 T^3+p^6 T^4
\tabularnewline[0.5pt]\hline				
8	 &	smooth    &	          &	1-388 T+8226 p T^2-388 p^3 T^3+p^6 T^4
\tabularnewline[0.5pt]\hline				
9	 &	smooth    &	          &	1+364 T+3602 p T^2+364 p^3 T^3+p^6 T^4
\tabularnewline[0.5pt]\hline				
10	 &	smooth    &	          &	1-556 T+8562 p T^2-556 p^3 T^3+p^6 T^4
\tabularnewline[0.5pt]\hline				
11	 &	smooth    &	          &	1+202 T+6386 p T^2+202 p^3 T^3+p^6 T^4
\tabularnewline[0.5pt]\hline				
12	 &	smooth    &	          &	1-102 T+2074 p T^2-102 p^3 T^3+p^6 T^4
\tabularnewline[0.5pt]\hline				
13	 &	smooth    &	          &	1+112 T+4946 p T^2+112 p^3 T^3+p^6 T^4
\tabularnewline[0.5pt]\hline				
14	 &	smooth    &	          &	1-16 T+2082 p T^2-16 p^3 T^3+p^6 T^4
\tabularnewline[0.5pt]\hline				
15	 &	singular  &	\frac{1}{9}&	(1-p T) (1-884 T+p^3 T^2)
\tabularnewline[0.5pt]\hline				
16	 &	smooth    &	          &	1+304 T+2402 p T^2+304 p^3 T^3+p^6 T^4
\tabularnewline[0.5pt]\hline				
17	 &	smooth    &	          &	1-836 T+10802 p T^2-836 p^3 T^3+p^6 T^4
\tabularnewline[0.5pt]\hline				
18	 &	smooth    &	          &	1+278 T+6354 p T^2+278 p^3 T^3+p^6 T^4
\tabularnewline[0.5pt]\hline				
19	 &	smooth    &	          &	(1+4 p T+p^3 T^2)(1+716 T+p^3 T^2)
\tabularnewline[0.5pt]\hline				
20	 &	smooth    &	          &	1+848 T+8274 p T^2+848 p^3 T^3+p^6 T^4
\tabularnewline[0.5pt]\hline				
21	 &	smooth    &	          &	1+64 T+7202 p T^2+64 p^3 T^3+p^6 T^4
\tabularnewline[0.5pt]\hline				
22	 &	smooth    &	          &	1-1032 T+10114 p T^2-1032 p^3 T^3+p^6 T^4
\tabularnewline[0.5pt]\hline				
23	 &	smooth    &	          &	1+144 T-1118 p T^2+144 p^3 T^3+p^6 T^4
\tabularnewline[0.5pt]\hline				
24	 &	smooth    &	          &	1-56 T+1922 p T^2-56 p^3 T^3+p^6 T^4
\tabularnewline[0.5pt]\hline				
25	 &	smooth    &	          &	1-136 T-1278 p T^2-136 p^3 T^3+p^6 T^4
\tabularnewline[0.5pt]\hline				
26	 &	smooth    &	          &	(1+12 p T+p^3 T^2)(1-524 T+p^3 T^2)
\tabularnewline[0.5pt]\hline				
27	 &	smooth    &	          &	1-1158 T+13426 p T^2-1158 p^3 T^3+p^6 T^4
\tabularnewline[0.5pt]\hline				
28	 &	smooth    &	          &	1-630 T+1930 p T^2-630 p^3 T^3+p^6 T^4
\tabularnewline[0.5pt]\hline				
29	 &	smooth    &	          &	(1+4 p T+p^3 T^2)(1+676 T+p^3 T^2)
\tabularnewline[0.5pt]\hline				
30	 &	smooth    &	          &	1+294 T-158 p T^2+294 p^3 T^3+p^6 T^4
\tabularnewline[0.5pt]\hline				
31	 &	smooth    &	          &	1-84 T+8578 p T^2-84 p^3 T^3+p^6 T^4
\tabularnewline[0.5pt]\hline				
32	 &	smooth    &	          &	1+248 T+102 p^2 T^2+248 p^3 T^3+p^6 T^4
\tabularnewline[0.5pt]\hline				
33	 &	smooth    &	          &	1-456 T+5122 p T^2-456 p^3 T^3+p^6 T^4
\tabularnewline[0.5pt]\hline				
34	 &	smooth    &	          &	1+614 T+1602 p T^2+614 p^3 T^3+p^6 T^4
\tabularnewline[0.5pt]\hline				
35	 &	smooth    &	          &	(1+4 p T+p^3 T^2)(1+36 T+p^3 T^2)
\tabularnewline[0.5pt]\hline				
36	 &	smooth    &	          &	1-216 T+4162 p T^2-216 p^3 T^3+p^6 T^4
\tabularnewline[0.5pt]\hline				
37	 &	smooth    &	          &	(1+4 p T+p^3 T^2)(1+676 T+p^3 T^2)
\tabularnewline[0.5pt]\hline				
38	 &	smooth    &	          &	1-210 T+2050 p T^2-210 p^3 T^3+p^6 T^4
\tabularnewline[0.5pt]\hline				
39	 &	smooth    &	          &	1-936 T+8002 p T^2-936 p^3 T^3+p^6 T^4
\tabularnewline[0.5pt]\hline				
40	 &	smooth    &	          &	1+164 T-78 p T^2+164 p^3 T^3+p^6 T^4
\tabularnewline[0.5pt]\hline				
41	 &	smooth    &	          &	1+920 T+7170 p T^2+920 p^3 T^3+p^6 T^4
\tabularnewline[0.5pt]\hline				
42	 &	smooth    &	          &	1+124 T+1202 p T^2+124 p^3 T^3+p^6 T^4
\tabularnewline[0.5pt]\hline				
43	 &	smooth    &	          &	1+344 T-2238 p T^2+344 p^3 T^3+p^6 T^4
\tabularnewline[0.5pt]\hline				
44	 &	smooth    &	          &	1-298 T-294 p T^2-298 p^3 T^3+p^6 T^4
\tabularnewline[0.5pt]\hline				
45	 &	smooth    &	          &	1-92 T+2354 p T^2-92 p^3 T^3+p^6 T^4
\tabularnewline[0.5pt]\hline				
46	 &	smooth    &	          &	1-90 T-3110 p T^2-90 p^3 T^3+p^6 T^4
\tabularnewline[0.5pt]\hline				
47	 &	smooth    &	          &	1+44 T+1362 p T^2+44 p^3 T^3+p^6 T^4
\tabularnewline[0.5pt]\hline				
48	 &	smooth    &	          &	1+840 T+10690 p T^2+840 p^3 T^3+p^6 T^4
\tabularnewline[0.5pt]\hline				
49	 &	smooth    &	          &	1+664 T+8642 p T^2+664 p^3 T^3+p^6 T^4
\tabularnewline[0.5pt]\hline				
50	 &	smooth    &	          &	1-6 T+6802 p T^2-6 p^3 T^3+p^6 T^4
\tabularnewline[0.5pt]\hline				
51	 &	smooth    &	          &	1+188 T+5634 p T^2+188 p^3 T^3+p^6 T^4
\tabularnewline[0.5pt]\hline				
52	 &	smooth    &	          &	1+132 T+5986 p T^2+132 p^3 T^3+p^6 T^4
\tabularnewline[0.5pt]\hline				
53	 &	smooth    &	          &	1-76 T+4722 p T^2-76 p^3 T^3+p^6 T^4
\tabularnewline[0.5pt]\hline				
54	 &	smooth    &	          &	1+960 T+10210 p T^2+960 p^3 T^3+p^6 T^4
\tabularnewline[0.5pt]\hline				
55	 &	smooth    &	          &	1+4 T-3598 p T^2+4 p^3 T^3+p^6 T^4
\tabularnewline[0.5pt]\hline				
56	 &	smooth    &	          &	1-356 T+2162 p T^2-356 p^3 T^3+p^6 T^4
\tabularnewline[0.5pt]\hline				
57	 &	smooth    &	          &	1-122 T+1394 p T^2-122 p^3 T^3+p^6 T^4
\tabularnewline[0.5pt]\hline				
58	 &	smooth    &	          &	1+164 T-2958 p T^2+164 p^3 T^3+p^6 T^4
\tabularnewline[0.5pt]\hline				
59	 &	singular  &	\frac{1}{25}&	(1+p T) (1-812 T+p^3 T^2)
\tabularnewline[0.5pt]\hline				
60	 &	smooth    &	          &	1-156 T+5362 p T^2-156 p^3 T^3+p^6 T^4
\tabularnewline[0.5pt]\hline				
61	 &	smooth    &	          &	1+416 T+7698 p T^2+416 p^3 T^3+p^6 T^4
\tabularnewline[0.5pt]\hline				
62	 &	smooth    &	          &	1+80 T+4770 p T^2+80 p^3 T^3+p^6 T^4
\tabularnewline[0.5pt]\hline				
63	 &	smooth    &	          &	1-498 T+8626 p T^2-498 p^3 T^3+p^6 T^4
\tabularnewline[0.5pt]\hline				
64	 &	smooth    &	          &	1+20 T-1230 p T^2+20 p^3 T^3+p^6 T^4
\tabularnewline[0.5pt]\hline				
65	 &	smooth    &	          &	1-180 T+2290 p T^2-180 p^3 T^3+p^6 T^4
\tabularnewline[0.5pt]\hline				
66	 &	smooth    &	          &	1+672 T+5266 p T^2+672 p^3 T^3+p^6 T^4
\tabularnewline[0.5pt]\hline				
\tablepostamble				
\tablepreamble{71}				
1	 &	singular  &	1&	(1-p T) (1-792 T+p^3 T^2)
\tabularnewline[0.5pt]\hline				
2	 &	smooth    &	          &	1+1068 T+9122 p T^2+1068 p^3 T^3+p^6 T^4
\tabularnewline[0.5pt]\hline				
3	 &	smooth    &	          &	1+436 T+6626 p T^2+436 p^3 T^3+p^6 T^4
\tabularnewline[0.5pt]\hline				
4	 &	smooth    &	          &	1+248 T-478 p T^2+248 p^3 T^3+p^6 T^4
\tabularnewline[0.5pt]\hline				
5	 &	smooth    &	          &	1+240 T-1822 p T^2+240 p^3 T^3+p^6 T^4
\tabularnewline[0.5pt]\hline				
6	 &	smooth    &	          &	1-800 T+6818 p T^2-800 p^3 T^3+p^6 T^4
\tabularnewline[0.5pt]\hline				
7	 &	smooth    &	          &	1+654 T+3530 p T^2+654 p^3 T^3+p^6 T^4
\tabularnewline[0.5pt]\hline				
8	 &	singular  &	\frac{1}{9}&	(1+p T) (1-792 T+p^3 T^2)
\tabularnewline[0.5pt]\hline				
9	 &	smooth    &	          &	1-532 T+2882 p T^2-532 p^3 T^3+p^6 T^4
\tabularnewline[0.5pt]\hline				
10	 &	smooth    &	          &	(1+p^3 T^2)(1-392 T+p^3 T^2)
\tabularnewline[0.5pt]\hline				
11	 &	smooth    &	          &	1+526 T+6506 p T^2+526 p^3 T^3+p^6 T^4
\tabularnewline[0.5pt]\hline				
12	 &	smooth    &	          &	1+160 T-862 p T^2+160 p^3 T^3+p^6 T^4
\tabularnewline[0.5pt]\hline				
13	 &	smooth    &	          &	1+260 T-2 p^2 T^2+260 p^3 T^3+p^6 T^4
\tabularnewline[0.5pt]\hline				
14	 &	smooth    &	          &	1-1218 T+11474 p T^2-1218 p^3 T^3+p^6 T^4
\tabularnewline[0.5pt]\hline				
15	 &	smooth    &	          &	(1+p^3 T^2)(1-912 T+p^3 T^2)
\tabularnewline[0.5pt]\hline				
16	 &	smooth    &	          &	1-132 T-2878 p T^2-132 p^3 T^3+p^6 T^4
\tabularnewline[0.5pt]\hline				
17	 &	smooth    &	          &	1-654 T+6266 p T^2-654 p^3 T^3+p^6 T^4
\tabularnewline[0.5pt]\hline				
18	 &	smooth    &	          &	1-112 T-4318 p T^2-112 p^3 T^3+p^6 T^4
\tabularnewline[0.5pt]\hline				
19	 &	smooth    &	          &	1-320 T+5858 p T^2-320 p^3 T^3+p^6 T^4
\tabularnewline[0.5pt]\hline				
20	 &	smooth    &	          &	1+168 T+4322 p T^2+168 p^3 T^3+p^6 T^4
\tabularnewline[0.5pt]\hline				
21	 &	smooth    &	          &	(1-12 p T+p^3 T^2)(1+126 T+p^3 T^2)
\tabularnewline[0.5pt]\hline				
22	 &	smooth    &	          &	1-420 T+7778 p T^2-420 p^3 T^3+p^6 T^4
\tabularnewline[0.5pt]\hline				
23	 &	smooth    &	          &	1+644 T+7730 p T^2+644 p^3 T^3+p^6 T^4
\tabularnewline[0.5pt]\hline				
24	 &	smooth    &	          &	1-232 T-1438 p T^2-232 p^3 T^3+p^6 T^4
\tabularnewline[0.5pt]\hline				
25	 &	smooth    &	          &	1+1176 T+12386 p T^2+1176 p^3 T^3+p^6 T^4
\tabularnewline[0.5pt]\hline				
26	 &	smooth    &	          &	1-140 T+5138 p T^2-140 p^3 T^3+p^6 T^4
\tabularnewline[0.5pt]\hline				
27	 &	smooth    &	          &	1+28 T+3362 p T^2+28 p^3 T^3+p^6 T^4
\tabularnewline[0.5pt]\hline				
28	 &	smooth    &	          &	1+936 T+6146 p T^2+936 p^3 T^3+p^6 T^4
\tabularnewline[0.5pt]\hline				
29	 &	smooth    &	          &	1-112 T-6238 p T^2-112 p^3 T^3+p^6 T^4
\tabularnewline[0.5pt]\hline				
30	 &	smooth    &	          &	(1+8 p T+p^3 T^2)(1+768 T+p^3 T^2)
\tabularnewline[0.5pt]\hline				
31	 &	smooth    &	          &	1-706 T+9770 p T^2-706 p^3 T^3+p^6 T^4
\tabularnewline[0.5pt]\hline				
32	 &	smooth    &	          &	1-512 T+6242 p T^2-512 p^3 T^3+p^6 T^4
\tabularnewline[0.5pt]\hline				
33	 &	smooth    &	          &	1-660 T+4178 p T^2-660 p^3 T^3+p^6 T^4
\tabularnewline[0.5pt]\hline				
34	 &	smooth    &	          &	1+808 T+7202 p T^2+808 p^3 T^3+p^6 T^4
\tabularnewline[0.5pt]\hline				
35	 &	smooth    &	          &	1+102 T+1154 p T^2+102 p^3 T^3+p^6 T^4
\tabularnewline[0.5pt]\hline				
36	 &	smooth    &	          &	1-152 T-1438 p T^2-152 p^3 T^3+p^6 T^4
\tabularnewline[0.5pt]\hline				
37	 &	smooth    &	          &	1-672 T+2402 p T^2-672 p^3 T^3+p^6 T^4
\tabularnewline[0.5pt]\hline				
38	 &	smooth    &	          &	1+1368 T+13922 p T^2+1368 p^3 T^3+p^6 T^4
\tabularnewline[0.5pt]\hline				
39	 &	smooth    &	          &	1-248 T+94 p^2 T^2-248 p^3 T^3+p^6 T^4
\tabularnewline[0.5pt]\hline				
40	 &	smooth    &	          &	1-184 T+8546 p T^2-184 p^3 T^3+p^6 T^4
\tabularnewline[0.5pt]\hline				
41	 &	smooth    &	          &	1-1318 T+13514 p T^2-1318 p^3 T^3+p^6 T^4
\tabularnewline[0.5pt]\hline				
42	 &	smooth    &	          &	1-252 T-1438 p T^2-252 p^3 T^3+p^6 T^4
\tabularnewline[0.5pt]\hline				
43	 &	smooth    &	          &	1+368 T+2402 p T^2+368 p^3 T^3+p^6 T^4
\tabularnewline[0.5pt]\hline				
44	 &	smooth    &	          &	1+46 T+7706 p T^2+46 p^3 T^3+p^6 T^4
\tabularnewline[0.5pt]\hline				
45	 &	smooth    &	          &	1+8 p T+4322 p T^2+8 p^4 T^3+p^6 T^4
\tabularnewline[0.5pt]\hline				
46	 &	smooth    &	          &	1-282 T+3002 p T^2-282 p^3 T^3+p^6 T^4
\tabularnewline[0.5pt]\hline				
47	 &	smooth    &	          &	1-442 T+4082 p T^2-442 p^3 T^3+p^6 T^4
\tabularnewline[0.5pt]\hline				
48	 &	smooth    &	          &	1+796 T+6626 p T^2+796 p^3 T^3+p^6 T^4
\tabularnewline[0.5pt]\hline				
49	 &	smooth    &	          &	1+216 T+2786 p T^2+216 p^3 T^3+p^6 T^4
\tabularnewline[0.5pt]\hline				
50	 &	smooth    &	          &	1+800 T+6818 p T^2+800 p^3 T^3+p^6 T^4
\tabularnewline[0.5pt]\hline				
51	 &	smooth    &	          &	1+210 T-4822 p T^2+210 p^3 T^3+p^6 T^4
\tabularnewline[0.5pt]\hline				
52	 &	smooth    &	          &	1-144 T-2014 p T^2-144 p^3 T^3+p^6 T^4
\tabularnewline[0.5pt]\hline				
53	 &	smooth    &	          &	1+14 T+7490 p T^2+14 p^3 T^3+p^6 T^4
\tabularnewline[0.5pt]\hline				
54	 &	singular  &	\frac{1}{25}&	(1-p T) (1-120 T+p^3 T^2)
\tabularnewline[0.5pt]\hline				
55	 &	smooth    &	          &	1-66 T+4250 p T^2-66 p^3 T^3+p^6 T^4
\tabularnewline[0.5pt]\hline				
56	 &	smooth    &	          &	1+564 T+4370 p T^2+564 p^3 T^3+p^6 T^4
\tabularnewline[0.5pt]\hline				
57	 &	smooth    &	          &	1-840 T+10658 p T^2-840 p^3 T^3+p^6 T^4
\tabularnewline[0.5pt]\hline				
58	 &	smooth    &	          &	1-472 T+7202 p T^2-472 p^3 T^3+p^6 T^4
\tabularnewline[0.5pt]\hline				
59	 &	smooth    &	          &	1-318 T+7034 p T^2-318 p^3 T^3+p^6 T^4
\tabularnewline[0.5pt]\hline				
60	 &	smooth    &	          &	1+28 T+3842 p T^2+28 p^3 T^3+p^6 T^4
\tabularnewline[0.5pt]\hline				
61	 &	smooth    &	          &	1+1112 T+11474 p T^2+1112 p^3 T^3+p^6 T^4
\tabularnewline[0.5pt]\hline				
62	 &	smooth    &	          &	1+492 T+4754 p T^2+492 p^3 T^3+p^6 T^4
\tabularnewline[0.5pt]\hline				
63	 &	smooth    &	          &	1+366 T+5906 p T^2+366 p^3 T^3+p^6 T^4
\tabularnewline[0.5pt]\hline				
64	 &	smooth    &	          &	1+228 T-4798 p T^2+228 p^3 T^3+p^6 T^4
\tabularnewline[0.5pt]\hline				
65	 &	smooth    &	          &	1-82 T+3962 p T^2-82 p^3 T^3+p^6 T^4
\tabularnewline[0.5pt]\hline				
66	 &	smooth    &	          &	1-216 T+1490 p T^2-216 p^3 T^3+p^6 T^4
\tabularnewline[0.5pt]\hline				
67	 &	smooth    &	          &	(1+p^3 T^2)(1+488 T+p^3 T^2)
\tabularnewline[0.5pt]\hline				
68	 &	smooth    &	          &	1+162 T+4754 p T^2+162 p^3 T^3+p^6 T^4
\tabularnewline[0.5pt]\hline				
69	 &	smooth    &	          &	1-182 T+1922 p T^2-182 p^3 T^3+p^6 T^4
\tabularnewline[0.5pt]\hline				
70	 &	smooth    &	          &	1-390 T-2302 p T^2-390 p^3 T^3+p^6 T^4
\tabularnewline[0.5pt]\hline				
\tablepostamble				
\tablepreamble{73}				
1	 &	singular  &	1&	(1-p T) (1-218 T+p^3 T^2)
\tabularnewline[0.5pt]\hline				
2	 &	smooth    &	          &	1-44 T-4906 p T^2-44 p^3 T^3+p^6 T^4
\tabularnewline[0.5pt]\hline				
3	 &	smooth    &	          &	1-640 T+6750 p T^2-640 p^3 T^3+p^6 T^4
\tabularnewline[0.5pt]\hline				
4	 &	smooth    &	          &	(1+6 p T+p^3 T^2)(1+22 T+p^3 T^2)
\tabularnewline[0.5pt]\hline				
5	 &	smooth    &	          &	1+560 T+5790 p T^2+560 p^3 T^3+p^6 T^4
\tabularnewline[0.5pt]\hline				
6	 &	smooth    &	          &	1-448 T+7998 p T^2-448 p^3 T^3+p^6 T^4
\tabularnewline[0.5pt]\hline				
7	 &	smooth    &	          &	1-1236 T+13846 p T^2-1236 p^3 T^3+p^6 T^4
\tabularnewline[0.5pt]\hline				
8	 &	smooth    &	          &	1+76 T+7334 p T^2+76 p^3 T^3+p^6 T^4
\tabularnewline[0.5pt]\hline				
9	 &	smooth    &	          &	1+156 T+7174 p T^2+156 p^3 T^3+p^6 T^4
\tabularnewline[0.5pt]\hline				
10	 &	smooth    &	          &	1-274 T+4794 p T^2-274 p^3 T^3+p^6 T^4
\tabularnewline[0.5pt]\hline				
11	 &	smooth    &	          &	1-212 T+4982 p T^2-212 p^3 T^3+p^6 T^4
\tabularnewline[0.5pt]\hline				
12	 &	smooth    &	          &	1+536 T+7374 p T^2+536 p^3 T^3+p^6 T^4
\tabularnewline[0.5pt]\hline				
13	 &	smooth    &	          &	1-270 T+7570 p T^2-270 p^3 T^3+p^6 T^4
\tabularnewline[0.5pt]\hline				
14	 &	smooth    &	          &	1-54 T+1354 p T^2-54 p^3 T^3+p^6 T^4
\tabularnewline[0.5pt]\hline				
15	 &	smooth    &	          &	1+348 T+2902 p T^2+348 p^3 T^3+p^6 T^4
\tabularnewline[0.5pt]\hline				
16	 &	smooth    &	          &	1-284 T-2506 p T^2-284 p^3 T^3+p^6 T^4
\tabularnewline[0.5pt]\hline				
17	 &	smooth    &	          &	1+462 T+1018 p T^2+462 p^3 T^3+p^6 T^4
\tabularnewline[0.5pt]\hline				
18	 &	smooth    &	          &	1-68 T-6202 p T^2-68 p^3 T^3+p^6 T^4
\tabularnewline[0.5pt]\hline				
19	 &	smooth    &	          &	1+256 T-226 p T^2+256 p^3 T^3+p^6 T^4
\tabularnewline[0.5pt]\hline				
20	 &	smooth    &	          &	1+1110 T+14170 p T^2+1110 p^3 T^3+p^6 T^4
\tabularnewline[0.5pt]\hline				
21	 &	smooth    &	          &	1+492 T+9958 p T^2+492 p^3 T^3+p^6 T^4
\tabularnewline[0.5pt]\hline				
22	 &	smooth    &	          &	1+306 T-3926 p T^2+306 p^3 T^3+p^6 T^4
\tabularnewline[0.5pt]\hline				
23	 &	smooth    &	          &	1-648 T+9358 p T^2-648 p^3 T^3+p^6 T^4
\tabularnewline[0.5pt]\hline				
24	 &	smooth    &	          &	1+552 T+2638 p T^2+552 p^3 T^3+p^6 T^4
\tabularnewline[0.5pt]\hline				
25	 &	smooth    &	          &	1-348 T+2038 p T^2-348 p^3 T^3+p^6 T^4
\tabularnewline[0.5pt]\hline				
26	 &	smooth    &	          &	1-26 T+7706 p T^2-26 p^3 T^3+p^6 T^4
\tabularnewline[0.5pt]\hline				
27	 &	smooth    &	          &	1-488 T+9038 p T^2-488 p^3 T^3+p^6 T^4
\tabularnewline[0.5pt]\hline				
28	 &	smooth    &	          &	1+230 T+5850 p T^2+230 p^3 T^3+p^6 T^4
\tabularnewline[0.5pt]\hline				
29	 &	smooth    &	          &	1+164 T+8646 p T^2+164 p^3 T^3+p^6 T^4
\tabularnewline[0.5pt]\hline				
30	 &	smooth    &	          &	1-296 T-3874 p T^2-296 p^3 T^3+p^6 T^4
\tabularnewline[0.5pt]\hline				
31	 &	smooth    &	          &	1+520 T+6350 p T^2+520 p^3 T^3+p^6 T^4
\tabularnewline[0.5pt]\hline				
32	 &	smooth    &	          &	1+252 T+8518 p T^2+252 p^3 T^3+p^6 T^4
\tabularnewline[0.5pt]\hline				
33	 &	smooth    &	          &	1+166 T+7754 p T^2+166 p^3 T^3+p^6 T^4
\tabularnewline[0.5pt]\hline				
34	 &	smooth    &	          &	1-618 T+6178 p T^2-618 p^3 T^3+p^6 T^4
\tabularnewline[0.5pt]\hline				
35	 &	smooth    &	          &	1+776 T+5934 p T^2+776 p^3 T^3+p^6 T^4
\tabularnewline[0.5pt]\hline				
36	 &	smooth    &	          &	1+916 T+10454 p T^2+916 p^3 T^3+p^6 T^4
\tabularnewline[0.5pt]\hline				
37	 &	smooth    &	          &	1+636 T+6214 p T^2+636 p^3 T^3+p^6 T^4
\tabularnewline[0.5pt]\hline				
38	 &	singular  &	\frac{1}{25}&	(1+p T) (1-746 T+p^3 T^2)
\tabularnewline[0.5pt]\hline				
39	 &	smooth    &	          &	1-406 T-1854 p T^2-406 p^3 T^3+p^6 T^4
\tabularnewline[0.5pt]\hline				
40	 &	smooth    &	          &	1-936 T+7006 p T^2-936 p^3 T^3+p^6 T^4
\tabularnewline[0.5pt]\hline				
41	 &	smooth    &	          &	1-224 T+5534 p T^2-224 p^3 T^3+p^6 T^4
\tabularnewline[0.5pt]\hline				
42	 &	smooth    &	          &	1-540 T+3670 p T^2-540 p^3 T^3+p^6 T^4
\tabularnewline[0.5pt]\hline				
43	 &	smooth    &	          &	1+116 T+7254 p T^2+116 p^3 T^3+p^6 T^4
\tabularnewline[0.5pt]\hline				
44	 &	smooth    &	          &	1-994 T+10194 p T^2-994 p^3 T^3+p^6 T^4
\tabularnewline[0.5pt]\hline				
45	 &	smooth    &	          &	1-760 T+9390 p T^2-760 p^3 T^3+p^6 T^4
\tabularnewline[0.5pt]\hline				
46	 &	smooth    &	          &	1+260 T+6390 p T^2+260 p^3 T^3+p^6 T^4
\tabularnewline[0.5pt]\hline				
47	 &	smooth    &	          &	1+272 T+5358 p T^2+272 p^3 T^3+p^6 T^4
\tabularnewline[0.5pt]\hline				
48	 &	smooth    &	          &	1-1304 T+14894 p T^2-1304 p^3 T^3+p^6 T^4
\tabularnewline[0.5pt]\hline				
49	 &	smooth    &	          &	1-368 T+1598 p T^2-368 p^3 T^3+p^6 T^4
\tabularnewline[0.5pt]\hline				
50	 &	smooth    &	          &	1+200 T+30 p^2 T^2+200 p^3 T^3+p^6 T^4
\tabularnewline[0.5pt]\hline				
51	 &	smooth    &	          &	1+224 T+4446 p T^2+224 p^3 T^3+p^6 T^4
\tabularnewline[0.5pt]\hline				
52	 &	smooth    &	          &	(1-2 p T+p^3 T^2)(1-538 T+p^3 T^2)
\tabularnewline[0.5pt]\hline				
53	 &	smooth    &	          &	1+1218 T+11482 p T^2+1218 p^3 T^3+p^6 T^4
\tabularnewline[0.5pt]\hline				
54	 &	smooth    &	          &	1-304 T+4254 p T^2-304 p^3 T^3+p^6 T^4
\tabularnewline[0.5pt]\hline				
55	 &	smooth    &	          &	1-184 T+1614 p T^2-184 p^3 T^3+p^6 T^4
\tabularnewline[0.5pt]\hline				
56	 &	smooth    &	          &	1+16 T-4786 p T^2+16 p^3 T^3+p^6 T^4
\tabularnewline[0.5pt]\hline				
57	 &	smooth    &	          &	1+636 T+10054 p T^2+636 p^3 T^3+p^6 T^4
\tabularnewline[0.5pt]\hline				
58	 &	smooth    &	          &	1+340 T-1690 p T^2+340 p^3 T^3+p^6 T^4
\tabularnewline[0.5pt]\hline				
59	 &	smooth    &	          &	(1+14 p T+p^3 T^2)(1-764 T+p^3 T^2)
\tabularnewline[0.5pt]\hline				
60	 &	smooth    &	          &	1-284 T+9014 p T^2-284 p^3 T^3+p^6 T^4
\tabularnewline[0.5pt]\hline				
61	 &	smooth    &	          &	(1+10 p T+p^3 T^2)(1+142 T+p^3 T^2)
\tabularnewline[0.5pt]\hline				
62	 &	smooth    &	          &	1-504 T+2014 p T^2-504 p^3 T^3+p^6 T^4
\tabularnewline[0.5pt]\hline				
63	 &	smooth    &	          &	1-776 T+7166 p T^2-776 p^3 T^3+p^6 T^4
\tabularnewline[0.5pt]\hline				
64	 &	smooth    &	          &	(1-2 p T+p^3 T^2)(1+1062 T+p^3 T^2)
\tabularnewline[0.5pt]\hline				
65	 &	singular  &	\frac{1}{9}&	(1-p T) (1-218 T+p^3 T^2)
\tabularnewline[0.5pt]\hline				
66	 &	smooth    &	          &	1+406 T+7754 p T^2+406 p^3 T^3+p^6 T^4
\tabularnewline[0.5pt]\hline				
67	 &	smooth    &	          &	1-160 T+9150 p T^2-160 p^3 T^3+p^6 T^4
\tabularnewline[0.5pt]\hline				
68	 &	smooth    &	          &	1+566 T+7194 p T^2+566 p^3 T^3+p^6 T^4
\tabularnewline[0.5pt]\hline				
69	 &	smooth    &	          &	1-4 T+3654 p T^2-4 p^3 T^3+p^6 T^4
\tabularnewline[0.5pt]\hline				
70	 &	smooth    &	          &	1+112 T+9278 p T^2+112 p^3 T^3+p^6 T^4
\tabularnewline[0.5pt]\hline				
71	 &	smooth    &	          &	1+660 T+1750 p T^2+660 p^3 T^3+p^6 T^4
\tabularnewline[0.5pt]\hline				
72	 &	smooth    &	          &	1-404 T+2534 p T^2-404 p^3 T^3+p^6 T^4
\tabularnewline[0.5pt]\hline				
\tablepostamble				
\tablepreamble{79}				
1	 &	singular  &	1&	(1-p T) (1+520 T+p^3 T^2)
\tabularnewline[0.5pt]\hline				
2	 &	smooth    &	          &	1+1308 T+13762 p T^2+1308 p^3 T^3+p^6 T^4
\tabularnewline[0.5pt]\hline				
3	 &	smooth    &	          &	1-82 T-7158 p T^2-82 p^3 T^3+p^6 T^4
\tabularnewline[0.5pt]\hline				
4	 &	smooth    &	          &	1+404 T+2082 p T^2+404 p^3 T^3+p^6 T^4
\tabularnewline[0.5pt]\hline				
5	 &	smooth    &	          &	1-752 T+4802 p T^2-752 p^3 T^3+p^6 T^4
\tabularnewline[0.5pt]\hline				
6	 &	smooth    &	          &	1-620 T+7682 p T^2-620 p^3 T^3+p^6 T^4
\tabularnewline[0.5pt]\hline				
7	 &	smooth    &	          &	1+440 T+4722 p T^2+440 p^3 T^3+p^6 T^4
\tabularnewline[0.5pt]\hline				
8	 &	smooth    &	          &	1+1048 T+10562 p T^2+1048 p^3 T^3+p^6 T^4
\tabularnewline[0.5pt]\hline				
9	 &	smooth    &	          &	1+288 T-2558 p T^2+288 p^3 T^3+p^6 T^4
\tabularnewline[0.5pt]\hline				
10	 &	smooth    &	          &	1-92 T-7678 p T^2-92 p^3 T^3+p^6 T^4
\tabularnewline[0.5pt]\hline				
11	 &	smooth    &	          &	1-296 T+2882 p T^2-296 p^3 T^3+p^6 T^4
\tabularnewline[0.5pt]\hline				
12	 &	smooth    &	          &	1+1202 T+14442 p T^2+1202 p^3 T^3+p^6 T^4
\tabularnewline[0.5pt]\hline				
13	 &	smooth    &	          &	(1+p^3 T^2)(1-776 T+p^3 T^2)
\tabularnewline[0.5pt]\hline				
14	 &	smooth    &	          &	1-342 T+6922 p T^2-342 p^3 T^3+p^6 T^4
\tabularnewline[0.5pt]\hline				
15	 &	smooth    &	          &	1-710 T+5042 p T^2-710 p^3 T^3+p^6 T^4
\tabularnewline[0.5pt]\hline				
16	 &	smooth    &	          &	1-272 T-2878 p T^2-272 p^3 T^3+p^6 T^4
\tabularnewline[0.5pt]\hline				
17	 &	smooth    &	          &	1-706 T+8682 p T^2-706 p^3 T^3+p^6 T^4
\tabularnewline[0.5pt]\hline				
18	 &	smooth    &	          &	1+528 T+322 p T^2+528 p^3 T^3+p^6 T^4
\tabularnewline[0.5pt]\hline				
19	 &	singular  &	\frac{1}{25}&	(1-p T) (1-152 T+p^3 T^2)
\tabularnewline[0.5pt]\hline				
20	 &	smooth    &	          &	1-912 T+13762 p T^2-912 p^3 T^3+p^6 T^4
\tabularnewline[0.5pt]\hline				
21	 &	smooth    &	          &	1-104 T-6718 p T^2-104 p^3 T^3+p^6 T^4
\tabularnewline[0.5pt]\hline				
22	 &	smooth    &	          &	(1-8 p T+p^3 T^2)(1+880 T+p^3 T^2)
\tabularnewline[0.5pt]\hline				
23	 &	smooth    &	          &	1-156 T+2722 p T^2-156 p^3 T^3+p^6 T^4
\tabularnewline[0.5pt]\hline				
24	 &	smooth    &	          &	1+190 T-2278 p T^2+190 p^3 T^3+p^6 T^4
\tabularnewline[0.5pt]\hline				
25	 &	smooth    &	          &	1-1512 T+15682 p T^2-1512 p^3 T^3+p^6 T^4
\tabularnewline[0.5pt]\hline				
26	 &	smooth    &	          &	1+736 T+962 p T^2+736 p^3 T^3+p^6 T^4
\tabularnewline[0.5pt]\hline				
27	 &	smooth    &	          &	1-64 T-1278 p T^2-64 p^3 T^3+p^6 T^4
\tabularnewline[0.5pt]\hline				
28	 &	smooth    &	          &	1+398 T-3918 p T^2+398 p^3 T^3+p^6 T^4
\tabularnewline[0.5pt]\hline				
29	 &	smooth    &	          &	1+1164 T+15442 p T^2+1164 p^3 T^3+p^6 T^4
\tabularnewline[0.5pt]\hline				
30	 &	smooth    &	          &	1+972 T+9682 p T^2+972 p^3 T^3+p^6 T^4
\tabularnewline[0.5pt]\hline				
31	 &	smooth    &	          &	1-32 T-6718 p T^2-32 p^3 T^3+p^6 T^4
\tabularnewline[0.5pt]\hline				
32	 &	smooth    &	          &	1+1368 T+13762 p T^2+1368 p^3 T^3+p^6 T^4
\tabularnewline[0.5pt]\hline				
33	 &	smooth    &	          &	1-1008 T+8722 p T^2-1008 p^3 T^3+p^6 T^4
\tabularnewline[0.5pt]\hline				
34	 &	smooth    &	          &	1-224 T+1922 p T^2-224 p^3 T^3+p^6 T^4
\tabularnewline[0.5pt]\hline				
35	 &	smooth    &	          &	1-362 T+11042 p T^2-362 p^3 T^3+p^6 T^4
\tabularnewline[0.5pt]\hline				
36	 &	smooth    &	          &	1-232 T-318 p T^2-232 p^3 T^3+p^6 T^4
\tabularnewline[0.5pt]\hline				
37	 &	smooth    &	          &	1-1312 T+13122 p T^2-1312 p^3 T^3+p^6 T^4
\tabularnewline[0.5pt]\hline				
38	 &	smooth    &	          &	1-904 T+5442 p T^2-904 p^3 T^3+p^6 T^4
\tabularnewline[0.5pt]\hline				
39	 &	smooth    &	          &	1+254 T+2922 p T^2+254 p^3 T^3+p^6 T^4
\tabularnewline[0.5pt]\hline				
40	 &	smooth    &	          &	1+888 T+9922 p T^2+888 p^3 T^3+p^6 T^4
\tabularnewline[0.5pt]\hline				
41	 &	smooth    &	          &	1-952 T+7842 p T^2-952 p^3 T^3+p^6 T^4
\tabularnewline[0.5pt]\hline				
42	 &	smooth    &	          &	(1-8 p T+p^3 T^2)(1+320 T+p^3 T^2)
\tabularnewline[0.5pt]\hline				
43	 &	smooth    &	          &	1-420 T+6562 p T^2-420 p^3 T^3+p^6 T^4
\tabularnewline[0.5pt]\hline				
44	 &	singular  &	\frac{1}{9}&	(1-p T) (1+520 T+p^3 T^2)
\tabularnewline[0.5pt]\hline				
45	 &	smooth    &	          &	(1-8 p T+p^3 T^2)(1-240 T+p^3 T^2)
\tabularnewline[0.5pt]\hline				
46	 &	smooth    &	          &	1+808 T+6722 p T^2+808 p^3 T^3+p^6 T^4
\tabularnewline[0.5pt]\hline				
47	 &	smooth    &	          &	1-20 T+1442 p T^2-20 p^3 T^3+p^6 T^4
\tabularnewline[0.5pt]\hline				
48	 &	smooth    &	          &	1+830 T+9042 p T^2+830 p^3 T^3+p^6 T^4
\tabularnewline[0.5pt]\hline				
49	 &	smooth    &	          &	1+776 T+9282 p T^2+776 p^3 T^3+p^6 T^4
\tabularnewline[0.5pt]\hline				
50	 &	smooth    &	          &	1-576 T+4162 p T^2-576 p^3 T^3+p^6 T^4
\tabularnewline[0.5pt]\hline				
51	 &	smooth    &	          &	1-964 T+10242 p T^2-964 p^3 T^3+p^6 T^4
\tabularnewline[0.5pt]\hline				
52	 &	smooth    &	          &	1-756 T+6562 p T^2-756 p^3 T^3+p^6 T^4
\tabularnewline[0.5pt]\hline				
53	 &	smooth    &	          &	1-294 T-5558 p T^2-294 p^3 T^3+p^6 T^4
\tabularnewline[0.5pt]\hline				
54	 &	smooth    &	          &	1+582 T-878 p T^2+582 p^3 T^3+p^6 T^4
\tabularnewline[0.5pt]\hline				
55	 &	smooth    &	          &	1+88 T+4802 p T^2+88 p^3 T^3+p^6 T^4
\tabularnewline[0.5pt]\hline				
56	 &	smooth    &	          &	1+104 T-7758 p T^2+104 p^3 T^3+p^6 T^4
\tabularnewline[0.5pt]\hline				
57	 &	smooth    &	          &	1+500 T+1842 p T^2+500 p^3 T^3+p^6 T^4
\tabularnewline[0.5pt]\hline				
58	 &	smooth    &	          &	1+408 T+1282 p T^2+408 p^3 T^3+p^6 T^4
\tabularnewline[0.5pt]\hline				
59	 &	smooth    &	          &	1-214 T+8802 p T^2-214 p^3 T^3+p^6 T^4
\tabularnewline[0.5pt]\hline				
60	 &	smooth    &	          &	1+1330 T+15122 p T^2+1330 p^3 T^3+p^6 T^4
\tabularnewline[0.5pt]\hline				
61	 &	smooth    &	          &	1+406 T+3602 p T^2+406 p^3 T^3+p^6 T^4
\tabularnewline[0.5pt]\hline				
62	 &	smooth    &	          &	(1-8 p T+p^3 T^2)(1-200 T+p^3 T^2)
\tabularnewline[0.5pt]\hline				
63	 &	smooth    &	          &	1-196 T+3762 p T^2-196 p^3 T^3+p^6 T^4
\tabularnewline[0.5pt]\hline				
64	 &	smooth    &	          &	1-552 T+3202 p T^2-552 p^3 T^3+p^6 T^4
\tabularnewline[0.5pt]\hline				
65	 &	smooth    &	          &	(1-8 p T+p^3 T^2)(1+1080 T+p^3 T^2)
\tabularnewline[0.5pt]\hline				
66	 &	smooth    &	          &	1+266 T+7962 p T^2+266 p^3 T^3+p^6 T^4
\tabularnewline[0.5pt]\hline				
67	 &	smooth    &	          &	1+56 T-7998 p T^2+56 p^3 T^3+p^6 T^4
\tabularnewline[0.5pt]\hline				
68	 &	smooth    &	          &	1+10 T+3482 p T^2+10 p^3 T^3+p^6 T^4
\tabularnewline[0.5pt]\hline				
69	 &	smooth    &	          &	1-620 T+7922 p T^2-620 p^3 T^3+p^6 T^4
\tabularnewline[0.5pt]\hline				
70	 &	smooth    &	          &	1+564 T+2962 p T^2+564 p^3 T^3+p^6 T^4
\tabularnewline[0.5pt]\hline				
71	 &	smooth    &	          &	1-1294 T+14322 p T^2-1294 p^3 T^3+p^6 T^4
\tabularnewline[0.5pt]\hline				
72	 &	smooth    &	          &	1+848 T+13122 p T^2+848 p^3 T^3+p^6 T^4
\tabularnewline[0.5pt]\hline				
73	 &	smooth    &	          &	1-232 T+7362 p T^2-232 p^3 T^3+p^6 T^4
\tabularnewline[0.5pt]\hline				
74	 &	smooth    &	          &	1+750 T+5962 p T^2+750 p^3 T^3+p^6 T^4
\tabularnewline[0.5pt]\hline				
75	 &	smooth    &	          &	1+90 T-8798 p T^2+90 p^3 T^3+p^6 T^4
\tabularnewline[0.5pt]\hline				
76	 &	smooth    &	          &	1+616 T+2882 p T^2+616 p^3 T^3+p^6 T^4
\tabularnewline[0.5pt]\hline				
77	 &	smooth    &	          &	1-40 T+882 p T^2-40 p^3 T^3+p^6 T^4
\tabularnewline[0.5pt]\hline				
78	 &	smooth    &	          &	1-852 T+12802 p T^2-852 p^3 T^3+p^6 T^4
\tabularnewline[0.5pt]\hline				
\tablepostamble				
\tablepreamble{83}				
1	 &	singular  &	1&	(1-p T) (1+492 T+p^3 T^2)
\tabularnewline[0.5pt]\hline				
2	 &	smooth    &	          &	1-688 T+14018 p T^2-688 p^3 T^3+p^6 T^4
\tabularnewline[0.5pt]\hline				
3	 &	smooth    &	          &	1-1964 T+22994 p T^2-1964 p^3 T^3+p^6 T^4
\tabularnewline[0.5pt]\hline				
4	 &	smooth    &	          &	1+36 T+11474 p T^2+36 p^3 T^3+p^6 T^4
\tabularnewline[0.5pt]\hline				
5	 &	smooth    &	          &	1-1728 T+19778 p T^2-1728 p^3 T^3+p^6 T^4
\tabularnewline[0.5pt]\hline				
6	 &	smooth    &	          &	1-1048 T+15938 p T^2-1048 p^3 T^3+p^6 T^4
\tabularnewline[0.5pt]\hline				
7	 &	smooth    &	          &	1+296 T-2686 p T^2+296 p^3 T^3+p^6 T^4
\tabularnewline[0.5pt]\hline				
8	 &	smooth    &	          &	1-446 T-2734 p T^2-446 p^3 T^3+p^6 T^4
\tabularnewline[0.5pt]\hline				
9	 &	smooth    &	          &	1+216 T+7874 p T^2+216 p^3 T^3+p^6 T^4
\tabularnewline[0.5pt]\hline				
10	 &	singular  &	\frac{1}{25}&	(1+p T) (1+804 T+p^3 T^2)
\tabularnewline[0.5pt]\hline				
11	 &	smooth    &	          &	1+868 T+9842 p T^2+868 p^3 T^3+p^6 T^4
\tabularnewline[0.5pt]\hline				
12	 &	smooth    &	          &	1-1072 T+6242 p T^2-1072 p^3 T^3+p^6 T^4
\tabularnewline[0.5pt]\hline				
13	 &	smooth    &	          &	1+76 T-5086 p T^2+76 p^3 T^3+p^6 T^4
\tabularnewline[0.5pt]\hline				
14	 &	smooth    &	          &	1-514 T+1754 p T^2-514 p^3 T^3+p^6 T^4
\tabularnewline[0.5pt]\hline				
15	 &	smooth    &	          &	1-10 p T+14810 p T^2-10 p^4 T^3+p^6 T^4
\tabularnewline[0.5pt]\hline				
16	 &	smooth    &	          &	1+176 T+7394 p T^2+176 p^3 T^3+p^6 T^4
\tabularnewline[0.5pt]\hline				
17	 &	smooth    &	          &	1+48 T-1438 p T^2+48 p^3 T^3+p^6 T^4
\tabularnewline[0.5pt]\hline				
18	 &	smooth    &	          &	1+302 T+4898 p T^2+302 p^3 T^3+p^6 T^4
\tabularnewline[0.5pt]\hline				
19	 &	smooth    &	          &	1+420 T-2110 p T^2+420 p^3 T^3+p^6 T^4
\tabularnewline[0.5pt]\hline				
20	 &	smooth    &	          &	1+382 T+9098 p T^2+382 p^3 T^3+p^6 T^4
\tabularnewline[0.5pt]\hline				
21	 &	smooth    &	          &	1-8 p T+6914 p T^2-8 p^4 T^3+p^6 T^4
\tabularnewline[0.5pt]\hline				
22	 &	smooth    &	          &	1+408 T+1922 p T^2+408 p^3 T^3+p^6 T^4
\tabularnewline[0.5pt]\hline				
23	 &	smooth    &	          &	1+776 T+7874 p T^2+776 p^3 T^3+p^6 T^4
\tabularnewline[0.5pt]\hline				
24	 &	smooth    &	          &	1-1038 T+8738 p T^2-1038 p^3 T^3+p^6 T^4
\tabularnewline[0.5pt]\hline				
25	 &	smooth    &	          &	1+376 T+6914 p T^2+376 p^3 T^3+p^6 T^4
\tabularnewline[0.5pt]\hline				
26	 &	smooth    &	          &	1+96 T-8926 p T^2+96 p^3 T^3+p^6 T^4
\tabularnewline[0.5pt]\hline				
27	 &	smooth    &	          &	1+32 T-862 p T^2+32 p^3 T^3+p^6 T^4
\tabularnewline[0.5pt]\hline				
28	 &	smooth    &	          &	1-8 T+8258 p T^2-8 p^3 T^3+p^6 T^4
\tabularnewline[0.5pt]\hline				
29	 &	smooth    &	          &	1-552 T+3842 p T^2-552 p^3 T^3+p^6 T^4
\tabularnewline[0.5pt]\hline				
30	 &	smooth    &	          &	(1+12 p T+p^3 T^2)(1+1132 T+p^3 T^2)
\tabularnewline[0.5pt]\hline				
31	 &	smooth    &	          &	1+536 T+118 p^2 T^2+536 p^3 T^3+p^6 T^4
\tabularnewline[0.5pt]\hline				
32	 &	smooth    &	          &	1+126 T-7726 p T^2+126 p^3 T^3+p^6 T^4
\tabularnewline[0.5pt]\hline				
33	 &	smooth    &	          &	1-1452 T+18962 p T^2-1452 p^3 T^3+p^6 T^4
\tabularnewline[0.5pt]\hline				
34	 &	smooth    &	          &	1-476 T+1586 p T^2-476 p^3 T^3+p^6 T^4
\tabularnewline[0.5pt]\hline				
35	 &	smooth    &	          &	1-244 T+1154 p T^2-244 p^3 T^3+p^6 T^4
\tabularnewline[0.5pt]\hline				
36	 &	smooth    &	          &	(1+12 p T+p^3 T^2)(1+1132 T+p^3 T^2)
\tabularnewline[0.5pt]\hline				
37	 &	singular  &	\frac{1}{9}&	(1+p T) (1+492 T+p^3 T^2)
\tabularnewline[0.5pt]\hline				
38	 &	smooth    &	          &	(1+4 p T+p^3 T^2)(1+1116 T+p^3 T^2)
\tabularnewline[0.5pt]\hline				
39	 &	smooth    &	          &	1+98 T+962 p T^2+98 p^3 T^3+p^6 T^4
\tabularnewline[0.5pt]\hline				
40	 &	smooth    &	          &	1+168 T+6722 p T^2+168 p^3 T^3+p^6 T^4
\tabularnewline[0.5pt]\hline				
41	 &	smooth    &	          &	1-1112 T+10562 p T^2-1112 p^3 T^3+p^6 T^4
\tabularnewline[0.5pt]\hline				
42	 &	smooth    &	          &	1+780 T+10610 p T^2+780 p^3 T^3+p^6 T^4
\tabularnewline[0.5pt]\hline				
43	 &	smooth    &	          &	(1-12 p T+p^3 T^2)(1+1362 T+p^3 T^2)
\tabularnewline[0.5pt]\hline				
44	 &	smooth    &	          &	1-828 T+6578 p T^2-828 p^3 T^3+p^6 T^4
\tabularnewline[0.5pt]\hline				
45	 &	smooth    &	          &	1+740 T+70 p^2 T^2+740 p^3 T^3+p^6 T^4
\tabularnewline[0.5pt]\hline				
46	 &	smooth    &	          &	1+240 T-2350 p T^2+240 p^3 T^3+p^6 T^4
\tabularnewline[0.5pt]\hline				
47	 &	smooth    &	          &	1+124 T+9506 p T^2+124 p^3 T^3+p^6 T^4
\tabularnewline[0.5pt]\hline				
48	 &	smooth    &	          &	1-888 T+11138 p T^2-888 p^3 T^3+p^6 T^4
\tabularnewline[0.5pt]\hline				
49	 &	smooth    &	          &	1-32 T-478 p T^2-32 p^3 T^3+p^6 T^4
\tabularnewline[0.5pt]\hline				
50	 &	smooth    &	          &	1-1586 T+19106 p T^2-1586 p^3 T^3+p^6 T^4
\tabularnewline[0.5pt]\hline				
51	 &	smooth    &	          &	1+876 T+15314 p T^2+876 p^3 T^3+p^6 T^4
\tabularnewline[0.5pt]\hline				
52	 &	smooth    &	          &	1+240 T-9070 p T^2+240 p^3 T^3+p^6 T^4
\tabularnewline[0.5pt]\hline				
53	 &	smooth    &	          &	1-206 T+26 p T^2-206 p^3 T^3+p^6 T^4
\tabularnewline[0.5pt]\hline				
54	 &	smooth    &	          &	1+864 T+6866 p T^2+864 p^3 T^3+p^6 T^4
\tabularnewline[0.5pt]\hline				
55	 &	smooth    &	          &	1-388 T-1582 p T^2-388 p^3 T^3+p^6 T^4
\tabularnewline[0.5pt]\hline				
56	 &	smooth    &	          &	1-812 T+12242 p T^2-812 p^3 T^3+p^6 T^4
\tabularnewline[0.5pt]\hline				
57	 &	smooth    &	          &	1+196 T-2926 p T^2+196 p^3 T^3+p^6 T^4
\tabularnewline[0.5pt]\hline				
58	 &	smooth    &	          &	1-12 p T+12866 p T^2-12 p^4 T^3+p^6 T^4
\tabularnewline[0.5pt]\hline				
59	 &	smooth    &	          &	1-1044 T+8594 p T^2-1044 p^3 T^3+p^6 T^4
\tabularnewline[0.5pt]\hline				
60	 &	smooth    &	          &	1+222 T+2978 p T^2+222 p^3 T^3+p^6 T^4
\tabularnewline[0.5pt]\hline				
61	 &	smooth    &	          &	1+776 T+8834 p T^2+776 p^3 T^3+p^6 T^4
\tabularnewline[0.5pt]\hline				
62	 &	smooth    &	          &	1+342 T+10178 p T^2+342 p^3 T^3+p^6 T^4
\tabularnewline[0.5pt]\hline				
63	 &	smooth    &	          &	1-408 T+2498 p T^2-408 p^3 T^3+p^6 T^4
\tabularnewline[0.5pt]\hline				
64	 &	smooth    &	          &	1-304 T+7394 p T^2-304 p^3 T^3+p^6 T^4
\tabularnewline[0.5pt]\hline				
65	 &	smooth    &	          &	1+948 T+8882 p T^2+948 p^3 T^3+p^6 T^4
\tabularnewline[0.5pt]\hline				
66	 &	smooth    &	          &	1+198 T+1922 p T^2+198 p^3 T^3+p^6 T^4
\tabularnewline[0.5pt]\hline				
67	 &	smooth    &	          &	1-70 T-3670 p T^2-70 p^3 T^3+p^6 T^4
\tabularnewline[0.5pt]\hline				
68	 &	smooth    &	          &	1+252 T+338 p T^2+252 p^3 T^3+p^6 T^4
\tabularnewline[0.5pt]\hline				
69	 &	smooth    &	          &	1-1048 T+13058 p T^2-1048 p^3 T^3+p^6 T^4
\tabularnewline[0.5pt]\hline				
70	 &	smooth    &	          &	1-632 T+4802 p T^2-632 p^3 T^3+p^6 T^4
\tabularnewline[0.5pt]\hline				
71	 &	smooth    &	          &	(1+6 p T+p^3 T^2)(1+1072 T+p^3 T^2)
\tabularnewline[0.5pt]\hline				
72	 &	smooth    &	          &	1+28 T+6482 p T^2+28 p^3 T^3+p^6 T^4
\tabularnewline[0.5pt]\hline				
73	 &	smooth    &	          &	1-56 T+2546 p T^2-56 p^3 T^3+p^6 T^4
\tabularnewline[0.5pt]\hline				
74	 &	smooth    &	          &	1-196 T-3454 p T^2-196 p^3 T^3+p^6 T^4
\tabularnewline[0.5pt]\hline				
75	 &	smooth    &	          &	1+1392 T+11618 p T^2+1392 p^3 T^3+p^6 T^4
\tabularnewline[0.5pt]\hline				
76	 &	smooth    &	          &	1+140 T-1390 p T^2+140 p^3 T^3+p^6 T^4
\tabularnewline[0.5pt]\hline				
77	 &	smooth    &	          &	1-1168 T+9698 p T^2-1168 p^3 T^3+p^6 T^4
\tabularnewline[0.5pt]\hline				
78	 &	smooth    &	          &	1+208 T-3358 p T^2+208 p^3 T^3+p^6 T^4
\tabularnewline[0.5pt]\hline				
79	 &	smooth    &	          &	1-550 T+3170 p T^2-550 p^3 T^3+p^6 T^4
\tabularnewline[0.5pt]\hline				
80	 &	smooth    &	          &	1+1078 T+10442 p T^2+1078 p^3 T^3+p^6 T^4
\tabularnewline[0.5pt]\hline				
81	 &	smooth    &	          &	1-552 T+11522 p T^2-552 p^3 T^3+p^6 T^4
\tabularnewline[0.5pt]\hline				
82	 &	smooth    &	          &	1+1164 T+13106 p T^2+1164 p^3 T^3+p^6 T^4
\tabularnewline[0.5pt]\hline				
\tablepostamble				
\tablepreamble{89}				
1	 &	singular  &	1&	(1-p T) (1-810 T+p^3 T^2)
\tabularnewline[0.5pt]\hline				
2	 &	smooth    &	          &	1+124 T+9542 p T^2+124 p^3 T^3+p^6 T^4
\tabularnewline[0.5pt]\hline				
3	 &	smooth    &	          &	1-600 T+12302 p T^2-600 p^3 T^3+p^6 T^4
\tabularnewline[0.5pt]\hline				
4	 &	smooth    &	          &	1+1236 T+9302 p T^2+1236 p^3 T^3+p^6 T^4
\tabularnewline[0.5pt]\hline				
5	 &	smooth    &	          &	1-264 T-3538 p T^2-264 p^3 T^3+p^6 T^4
\tabularnewline[0.5pt]\hline				
6	 &	smooth    &	          &	1+1114 T+6242 p T^2+1114 p^3 T^3+p^6 T^4
\tabularnewline[0.5pt]\hline				
7	 &	smooth    &	          &	1-294 T+15002 p T^2-294 p^3 T^3+p^6 T^4
\tabularnewline[0.5pt]\hline				
8	 &	smooth    &	          &	1-1256 T+16622 p T^2-1256 p^3 T^3+p^6 T^4
\tabularnewline[0.5pt]\hline				
9	 &	smooth    &	          &	1+968 T+5582 p T^2+968 p^3 T^3+p^6 T^4
\tabularnewline[0.5pt]\hline				
10	 &	singular  &	\frac{1}{9}&	(1+p T) (1-810 T+p^3 T^2)
\tabularnewline[0.5pt]\hline				
11	 &	smooth    &	          &	1-68 T+13382 p T^2-68 p^3 T^3+p^6 T^4
\tabularnewline[0.5pt]\hline				
12	 &	smooth    &	          &	1-1384 T+17822 p T^2-1384 p^3 T^3+p^6 T^4
\tabularnewline[0.5pt]\hline				
13	 &	smooth    &	          &	1-538 T-3598 p T^2-538 p^3 T^3+p^6 T^4
\tabularnewline[0.5pt]\hline				
14	 &	smooth    &	          &	1-1990 T+21962 p T^2-1990 p^3 T^3+p^6 T^4
\tabularnewline[0.5pt]\hline				
15	 &	smooth    &	          &	1-348 T+4742 p T^2-348 p^3 T^3+p^6 T^4
\tabularnewline[0.5pt]\hline				
16	 &	smooth    &	          &	1+764 T+9542 p T^2+764 p^3 T^3+p^6 T^4
\tabularnewline[0.5pt]\hline				
17	 &	smooth    &	          &	1+376 T+6062 p T^2+376 p^3 T^3+p^6 T^4
\tabularnewline[0.5pt]\hline				
18	 &	smooth    &	          &	1+64 T-6178 p T^2+64 p^3 T^3+p^6 T^4
\tabularnewline[0.5pt]\hline				
19	 &	smooth    &	          &	1+666 T+16322 p T^2+666 p^3 T^3+p^6 T^4
\tabularnewline[0.5pt]\hline				
20	 &	smooth    &	          &	1+856 T+5582 p T^2+856 p^3 T^3+p^6 T^4
\tabularnewline[0.5pt]\hline				
21	 &	smooth    &	          &	(1+10 p T+p^3 T^2)(1-1146 T+p^3 T^2)
\tabularnewline[0.5pt]\hline				
22	 &	smooth    &	          &	1-656 T-2818 p T^2-656 p^3 T^3+p^6 T^4
\tabularnewline[0.5pt]\hline				
23	 &	smooth    &	          &	1+782 T+5402 p T^2+782 p^3 T^3+p^6 T^4
\tabularnewline[0.5pt]\hline				
24	 &	smooth    &	          &	1+576 T+7262 p T^2+576 p^3 T^3+p^6 T^4
\tabularnewline[0.5pt]\hline				
25	 &	smooth    &	          &	1+1364 T+13142 p T^2+1364 p^3 T^3+p^6 T^4
\tabularnewline[0.5pt]\hline				
26	 &	smooth    &	          &	1-434 T+10682 p T^2-434 p^3 T^3+p^6 T^4
\tabularnewline[0.5pt]\hline				
27	 &	smooth    &	          &	1-410 T-478 p T^2-410 p^3 T^3+p^6 T^4
\tabularnewline[0.5pt]\hline				
28	 &	smooth    &	          &	1+1002 T+11882 p T^2+1002 p^3 T^3+p^6 T^4
\tabularnewline[0.5pt]\hline				
29	 &	smooth    &	          &	1-1116 T+12662 p T^2-1116 p^3 T^3+p^6 T^4
\tabularnewline[0.5pt]\hline				
30	 &	smooth    &	          &	1+310 T-5758 p T^2+310 p^3 T^3+p^6 T^4
\tabularnewline[0.5pt]\hline				
31	 &	smooth    &	          &	1-330 T+13922 p T^2-330 p^3 T^3+p^6 T^4
\tabularnewline[0.5pt]\hline				
32	 &	smooth    &	          &	1+64 T+9182 p T^2+64 p^3 T^3+p^6 T^4
\tabularnewline[0.5pt]\hline				
33	 &	smooth    &	          &	1-20 T-6058 p T^2-20 p^3 T^3+p^6 T^4
\tabularnewline[0.5pt]\hline				
34	 &	smooth    &	          &	1+1236 T+16022 p T^2+1236 p^3 T^3+p^6 T^4
\tabularnewline[0.5pt]\hline				
35	 &	smooth    &	          &	1-992 T+14222 p T^2-992 p^3 T^3+p^6 T^4
\tabularnewline[0.5pt]\hline				
36	 &	smooth    &	          &	1-504 T+302 p T^2-504 p^3 T^3+p^6 T^4
\tabularnewline[0.5pt]\hline				
37	 &	smooth    &	          &	1+258 T+9362 p T^2+258 p^3 T^3+p^6 T^4
\tabularnewline[0.5pt]\hline				
38	 &	smooth    &	          &	(1+6 p T+p^3 T^2)(1-810 T+p^3 T^2)
\tabularnewline[0.5pt]\hline				
39	 &	smooth    &	          &	(1+6 p T+p^3 T^2)(1-610 T+p^3 T^2)
\tabularnewline[0.5pt]\hline				
40	 &	smooth    &	          &	1-88 T+3662 p T^2-88 p^3 T^3+p^6 T^4
\tabularnewline[0.5pt]\hline				
41	 &	smooth    &	          &	1-790 T+8642 p T^2-790 p^3 T^3+p^6 T^4
\tabularnewline[0.5pt]\hline				
42	 &	smooth    &	          &	1+896 T+5822 p T^2+896 p^3 T^3+p^6 T^4
\tabularnewline[0.5pt]\hline				
43	 &	smooth    &	          &	1-1772 T+21542 p T^2-1772 p^3 T^3+p^6 T^4
\tabularnewline[0.5pt]\hline				
44	 &	smooth    &	          &	(1-10 p T+p^3 T^2)(1+654 T+p^3 T^2)
\tabularnewline[0.5pt]\hline				
45	 &	smooth    &	          &	1+144 T+10622 p T^2+144 p^3 T^3+p^6 T^4
\tabularnewline[0.5pt]\hline				
46	 &	smooth    &	          &	1-292 T+13382 p T^2-292 p^3 T^3+p^6 T^4
\tabularnewline[0.5pt]\hline				
47	 &	smooth    &	          &	1-136 T+11342 p T^2-136 p^3 T^3+p^6 T^4
\tabularnewline[0.5pt]\hline				
48	 &	smooth    &	          &	1-330 T+2042 p T^2-330 p^3 T^3+p^6 T^4
\tabularnewline[0.5pt]\hline				
49	 &	smooth    &	          &	1-884 T+12902 p T^2-884 p^3 T^3+p^6 T^4
\tabularnewline[0.5pt]\hline				
50	 &	smooth    &	          &	1-84 T-2458 p T^2-84 p^3 T^3+p^6 T^4
\tabularnewline[0.5pt]\hline				
51	 &	smooth    &	          &	1+12 T+662 p T^2+12 p^3 T^3+p^6 T^4
\tabularnewline[0.5pt]\hline				
52	 &	smooth    &	          &	1-300 T+3302 p T^2-300 p^3 T^3+p^6 T^4
\tabularnewline[0.5pt]\hline				
53	 &	smooth    &	          &	1-136 T+1742 p T^2-136 p^3 T^3+p^6 T^4
\tabularnewline[0.5pt]\hline				
54	 &	smooth    &	          &	1+634 T+2882 p T^2+634 p^3 T^3+p^6 T^4
\tabularnewline[0.5pt]\hline				
55	 &	smooth    &	          &	(1+14 p T+p^3 T^2)(1+510 T+p^3 T^2)
\tabularnewline[0.5pt]\hline				
56	 &	smooth    &	          &	1-322 T+9482 p T^2-322 p^3 T^3+p^6 T^4
\tabularnewline[0.5pt]\hline				
57	 &	singular  &	\frac{1}{25}&	(1-p T) (1+678 T+p^3 T^2)
\tabularnewline[0.5pt]\hline				
58	 &	smooth    &	          &	1-318 T-2878 p T^2-318 p^3 T^3+p^6 T^4
\tabularnewline[0.5pt]\hline				
59	 &	smooth    &	          &	1-480 T+13262 p T^2-480 p^3 T^3+p^6 T^4
\tabularnewline[0.5pt]\hline				
60	 &	smooth    &	          &	1+518 T-958 p T^2+518 p^3 T^3+p^6 T^4
\tabularnewline[0.5pt]\hline				
61	 &	smooth    &	          &	1+1206 T+9842 p T^2+1206 p^3 T^3+p^6 T^4
\tabularnewline[0.5pt]\hline				
62	 &	smooth    &	          &	1-66 T+6722 p T^2-66 p^3 T^3+p^6 T^4
\tabularnewline[0.5pt]\hline				
63	 &	smooth    &	          &	1-110 T+3722 p T^2-110 p^3 T^3+p^6 T^4
\tabularnewline[0.5pt]\hline				
64	 &	smooth    &	          &	1+76 T+2342 p T^2+76 p^3 T^3+p^6 T^4
\tabularnewline[0.5pt]\hline				
65	 &	smooth    &	          &	1+44 T+5222 p T^2+44 p^3 T^3+p^6 T^4
\tabularnewline[0.5pt]\hline				
66	 &	smooth    &	          &	1+662 T-1918 p T^2+662 p^3 T^3+p^6 T^4
\tabularnewline[0.5pt]\hline				
67	 &	smooth    &	          &	1+88 T+7022 p T^2+88 p^3 T^3+p^6 T^4
\tabularnewline[0.5pt]\hline				
68	 &	smooth    &	          &	1+816 T+13502 p T^2+816 p^3 T^3+p^6 T^4
\tabularnewline[0.5pt]\hline				
69	 &	smooth    &	          &	1-464 T+8702 p T^2-464 p^3 T^3+p^6 T^4
\tabularnewline[0.5pt]\hline				
70	 &	smooth    &	          &	1+148 T+10022 p T^2+148 p^3 T^3+p^6 T^4
\tabularnewline[0.5pt]\hline				
71	 &	smooth    &	          &	(1-6 p T+p^3 T^2)(1+350 T+p^3 T^2)
\tabularnewline[0.5pt]\hline				
72	 &	smooth    &	          &	1-708 T+16262 p T^2-708 p^3 T^3+p^6 T^4
\tabularnewline[0.5pt]\hline				
73	 &	smooth    &	          &	1+1484 T+17702 p T^2+1484 p^3 T^3+p^6 T^4
\tabularnewline[0.5pt]\hline				
74	 &	smooth    &	          &	1+76 T+7142 p T^2+76 p^3 T^3+p^6 T^4
\tabularnewline[0.5pt]\hline				
75	 &	smooth    &	          &	1+774 T+13202 p T^2+774 p^3 T^3+p^6 T^4
\tabularnewline[0.5pt]\hline				
76	 &	smooth    &	          &	1-244 T+422 p T^2-244 p^3 T^3+p^6 T^4
\tabularnewline[0.5pt]\hline				
77	 &	smooth    &	          &	1-40 T-7618 p T^2-40 p^3 T^3+p^6 T^4
\tabularnewline[0.5pt]\hline				
78	 &	smooth    &	          &	(1+6 p T+p^3 T^2)(1-1530 T+p^3 T^2)
\tabularnewline[0.5pt]\hline				
79	 &	smooth    &	          &	1+396 T+422 p T^2+396 p^3 T^3+p^6 T^4
\tabularnewline[0.5pt]\hline				
80	 &	smooth    &	          &	1+276 T-1258 p T^2+276 p^3 T^3+p^6 T^4
\tabularnewline[0.5pt]\hline				
81	 &	smooth    &	          &	1-1644 T+16982 p T^2-1644 p^3 T^3+p^6 T^4
\tabularnewline[0.5pt]\hline				
82	 &	smooth    &	          &	1+1244 T+18182 p T^2+1244 p^3 T^3+p^6 T^4
\tabularnewline[0.5pt]\hline				
83	 &	smooth    &	          &	1-202 T+9602 p T^2-202 p^3 T^3+p^6 T^4
\tabularnewline[0.5pt]\hline				
84	 &	smooth    &	          &	(1-6 p T+p^3 T^2)(1+350 T+p^3 T^2)
\tabularnewline[0.5pt]\hline				
85	 &	smooth    &	          &	1-696 T+4622 p T^2-696 p^3 T^3+p^6 T^4
\tabularnewline[0.5pt]\hline				
86	 &	smooth    &	          &	1+172 T+6422 p T^2+172 p^3 T^3+p^6 T^4
\tabularnewline[0.5pt]\hline				
87	 &	smooth    &	          &	1-676 T+16262 p T^2-676 p^3 T^3+p^6 T^4
\tabularnewline[0.5pt]\hline				
88	 &	smooth    &	          &	(1+6 p T+p^3 T^2)(1+1510 T+p^3 T^2)
\tabularnewline[0.5pt]\hline				
\tablepostamble				
\tablepreamble{97}				
1	 &	singular  &	1&	(1-p T) (1-1154 T+p^3 T^2)
\tabularnewline[0.5pt]\hline				
2	 &	smooth    &	          &	1-232 T+654 p T^2-232 p^3 T^3+p^6 T^4
\tabularnewline[0.5pt]\hline				
3	 &	smooth    &	          &	(1-10 p T+p^3 T^2)(1+406 T+p^3 T^2)
\tabularnewline[0.5pt]\hline				
4	 &	smooth    &	          &	1+848 T+9054 p T^2+848 p^3 T^3+p^6 T^4
\tabularnewline[0.5pt]\hline				
5	 &	smooth    &	          &	1-172 T-1866 p T^2-172 p^3 T^3+p^6 T^4
\tabularnewline[0.5pt]\hline				
6	 &	smooth    &	          &	1+1172 T+8406 p T^2+1172 p^3 T^3+p^6 T^4
\tabularnewline[0.5pt]\hline				
7	 &	smooth    &	          &	1+182 T-4614 p T^2+182 p^3 T^3+p^6 T^4
\tabularnewline[0.5pt]\hline				
8	 &	smooth    &	          &	1+1352 T+22926 p T^2+1352 p^3 T^3+p^6 T^4
\tabularnewline[0.5pt]\hline				
9	 &	smooth    &	          &	1+92 T-4794 p T^2+92 p^3 T^3+p^6 T^4
\tabularnewline[0.5pt]\hline				
10	 &	smooth    &	          &	1+46 T+11498 p T^2+46 p^3 T^3+p^6 T^4
\tabularnewline[0.5pt]\hline				
11	 &	smooth    &	          &	1+216 T+3118 p T^2+216 p^3 T^3+p^6 T^4
\tabularnewline[0.5pt]\hline				
12	 &	smooth    &	          &	1+248 T+9294 p T^2+248 p^3 T^3+p^6 T^4
\tabularnewline[0.5pt]\hline				
13	 &	smooth    &	          &	1-150 T+4210 p T^2-150 p^3 T^3+p^6 T^4
\tabularnewline[0.5pt]\hline				
14	 &	smooth    &	          &	1+156 T-1082 p T^2+156 p^3 T^3+p^6 T^4
\tabularnewline[0.5pt]\hline				
15	 &	smooth    &	          &	1-1864 T+24558 p T^2-1864 p^3 T^3+p^6 T^4
\tabularnewline[0.5pt]\hline				
16	 &	smooth    &	          &	1-124 T+54 p^2 T^2-124 p^3 T^3+p^6 T^4
\tabularnewline[0.5pt]\hline				
17	 &	smooth    &	          &	1+430 T-6070 p T^2+430 p^3 T^3+p^6 T^4
\tabularnewline[0.5pt]\hline				
18	 &	smooth    &	          &	1-684 T+10198 p T^2-684 p^3 T^3+p^6 T^4
\tabularnewline[0.5pt]\hline				
19	 &	smooth    &	          &	1-784 T+7758 p T^2-784 p^3 T^3+p^6 T^4
\tabularnewline[0.5pt]\hline				
20	 &	smooth    &	          &	1+216 T-8882 p T^2+216 p^3 T^3+p^6 T^4
\tabularnewline[0.5pt]\hline				
21	 &	smooth    &	          &	1-1462 T+14154 p T^2-1462 p^3 T^3+p^6 T^4
\tabularnewline[0.5pt]\hline				
22	 &	smooth    &	          &	1+432 T+13726 p T^2+432 p^3 T^3+p^6 T^4
\tabularnewline[0.5pt]\hline				
23	 &	smooth    &	          &	1-604 T+16278 p T^2-604 p^3 T^3+p^6 T^4
\tabularnewline[0.5pt]\hline				
24	 &	smooth    &	          &	(1+18 p T+p^3 T^2)(1+46 T+p^3 T^2)
\tabularnewline[0.5pt]\hline				
25	 &	smooth    &	          &	1-484 T+6918 p T^2-484 p^3 T^3+p^6 T^4
\tabularnewline[0.5pt]\hline				
26	 &	smooth    &	          &	1+1220 T+11910 p T^2+1220 p^3 T^3+p^6 T^4
\tabularnewline[0.5pt]\hline				
27	 &	smooth    &	          &	1+888 T+17134 p T^2+888 p^3 T^3+p^6 T^4
\tabularnewline[0.5pt]\hline				
28	 &	smooth    &	          &	1-1220 T+12470 p T^2-1220 p^3 T^3+p^6 T^4
\tabularnewline[0.5pt]\hline				
29	 &	smooth    &	          &	1-1382 T+14234 p T^2-1382 p^3 T^3+p^6 T^4
\tabularnewline[0.5pt]\hline				
30	 &	smooth    &	          &	1+260 T+1350 p T^2+260 p^3 T^3+p^6 T^4
\tabularnewline[0.5pt]\hline				
31	 &	smooth    &	          &	1-808 T+17646 p T^2-808 p^3 T^3+p^6 T^4
\tabularnewline[0.5pt]\hline				
32	 &	smooth    &	          &	1-192 T-10946 p T^2-192 p^3 T^3+p^6 T^4
\tabularnewline[0.5pt]\hline				
33	 &	smooth    &	          &	1+972 T+1126 p T^2+972 p^3 T^3+p^6 T^4
\tabularnewline[0.5pt]\hline				
34	 &	smooth    &	          &	1-1970 T+22250 p T^2-1970 p^3 T^3+p^6 T^4
\tabularnewline[0.5pt]\hline				
35	 &	smooth    &	          &	1+972 T+12646 p T^2+972 p^3 T^3+p^6 T^4
\tabularnewline[0.5pt]\hline				
36	 &	smooth    &	          &	(1-2 p T+p^3 T^2)(1+1206 T+p^3 T^2)
\tabularnewline[0.5pt]\hline				
37	 &	smooth    &	          &	1-100 T-10890 p T^2-100 p^3 T^3+p^6 T^4
\tabularnewline[0.5pt]\hline				
38	 &	smooth    &	          &	1+1530 T+20650 p T^2+1530 p^3 T^3+p^6 T^4
\tabularnewline[0.5pt]\hline				
39	 &	smooth    &	          &	1-258 T-1694 p T^2-258 p^3 T^3+p^6 T^4
\tabularnewline[0.5pt]\hline				
40	 &	smooth    &	          &	1+508 T+8774 p T^2+508 p^3 T^3+p^6 T^4
\tabularnewline[0.5pt]\hline				
41	 &	smooth    &	          &	1-1394 T+10418 p T^2-1394 p^3 T^3+p^6 T^4
\tabularnewline[0.5pt]\hline				
42	 &	smooth    &	          &	1-32 T+14 p T^2-32 p^3 T^3+p^6 T^4
\tabularnewline[0.5pt]\hline				
43	 &	smooth    &	          &	1-468 T+13606 p T^2-468 p^3 T^3+p^6 T^4
\tabularnewline[0.5pt]\hline				
44	 &	smooth    &	          &	1-584 T+2318 p T^2-584 p^3 T^3+p^6 T^4
\tabularnewline[0.5pt]\hline				
45	 &	smooth    &	          &	1-82 T+9594 p T^2-82 p^3 T^3+p^6 T^4
\tabularnewline[0.5pt]\hline				
46	 &	smooth    &	          &	1+810 T+6490 p T^2+810 p^3 T^3+p^6 T^4
\tabularnewline[0.5pt]\hline				
47	 &	smooth    &	          &	1+316 T-1402 p T^2+316 p^3 T^3+p^6 T^4
\tabularnewline[0.5pt]\hline				
48	 &	smooth    &	          &	1+1488 T+14494 p T^2+1488 p^3 T^3+p^6 T^4
\tabularnewline[0.5pt]\hline				
49	 &	smooth    &	          &	1-624 T+15838 p T^2-624 p^3 T^3+p^6 T^4
\tabularnewline[0.5pt]\hline				
50	 &	smooth    &	          &	1+12 T+14566 p T^2+12 p^3 T^3+p^6 T^4
\tabularnewline[0.5pt]\hline				
51	 &	smooth    &	          &	1-894 T+16018 p T^2-894 p^3 T^3+p^6 T^4
\tabularnewline[0.5pt]\hline				
52	 &	smooth    &	          &	1-210 T-110 p T^2-210 p^3 T^3+p^6 T^4
\tabularnewline[0.5pt]\hline				
53	 &	smooth    &	          &	1+848 T+9534 p T^2+848 p^3 T^3+p^6 T^4
\tabularnewline[0.5pt]\hline				
54	 &	singular  &	\frac{1}{9}&	(1-p T) (1-1154 T+p^3 T^2)
\tabularnewline[0.5pt]\hline				
55	 &	smooth    &	          &	1-708 T+18886 p T^2-708 p^3 T^3+p^6 T^4
\tabularnewline[0.5pt]\hline				
56	 &	smooth    &	          &	1+170 T+90 p T^2+170 p^3 T^3+p^6 T^4
\tabularnewline[0.5pt]\hline				
57	 &	smooth    &	          &	1+40 T-13570 p T^2+40 p^3 T^3+p^6 T^4
\tabularnewline[0.5pt]\hline				
58	 &	smooth    &	          &	1-632 T+5534 p T^2-632 p^3 T^3+p^6 T^4
\tabularnewline[0.5pt]\hline				
59	 &	smooth    &	          &	1-524 T+10838 p T^2-524 p^3 T^3+p^6 T^4
\tabularnewline[0.5pt]\hline				
60	 &	smooth    &	          &	1+1500 T+19990 p T^2+1500 p^3 T^3+p^6 T^4
\tabularnewline[0.5pt]\hline				
61	 &	smooth    &	          &	1-1024 T+12798 p T^2-1024 p^3 T^3+p^6 T^4
\tabularnewline[0.5pt]\hline				
62	 &	smooth    &	          &	1+348 T-1466 p T^2+348 p^3 T^3+p^6 T^4
\tabularnewline[0.5pt]\hline				
63	 &	smooth    &	          &	1+572 T+6 p T^2+572 p^3 T^3+p^6 T^4
\tabularnewline[0.5pt]\hline				
64	 &	smooth    &	          &	1-324 T+3718 p T^2-324 p^3 T^3+p^6 T^4
\tabularnewline[0.5pt]\hline				
65	 &	smooth    &	          &	1+1308 T+19654 p T^2+1308 p^3 T^3+p^6 T^4
\tabularnewline[0.5pt]\hline				
66	 &	singular  &	\frac{1}{25}&	(1+p T) (1-194 T+p^3 T^2)
\tabularnewline[0.5pt]\hline				
67	 &	smooth    &	          &	1+710 T+10890 p T^2+710 p^3 T^3+p^6 T^4
\tabularnewline[0.5pt]\hline				
68	 &	smooth    &	          &	1+182 T+13746 p T^2+182 p^3 T^3+p^6 T^4
\tabularnewline[0.5pt]\hline				
69	 &	smooth    &	          &	1+678 T+12514 p T^2+678 p^3 T^3+p^6 T^4
\tabularnewline[0.5pt]\hline				
70	 &	smooth    &	          &	1+56 T+8238 p T^2+56 p^3 T^3+p^6 T^4
\tabularnewline[0.5pt]\hline				
71	 &	smooth    &	          &	1+1324 T+18902 p T^2+1324 p^3 T^3+p^6 T^4
\tabularnewline[0.5pt]\hline				
72	 &	smooth    &	          &	1+1256 T+13518 p T^2+1256 p^3 T^3+p^6 T^4
\tabularnewline[0.5pt]\hline				
73	 &	smooth    &	          &	1-208 T-8034 p T^2-208 p^3 T^3+p^6 T^4
\tabularnewline[0.5pt]\hline				
74	 &	smooth    &	          &	1+834 T+6802 p T^2+834 p^3 T^3+p^6 T^4
\tabularnewline[0.5pt]\hline				
75	 &	smooth    &	          &	1-68 T-634 p T^2-68 p^3 T^3+p^6 T^4
\tabularnewline[0.5pt]\hline				
76	 &	smooth    &	          &	1+322 T-6574 p T^2+322 p^3 T^3+p^6 T^4
\tabularnewline[0.5pt]\hline				
77	 &	smooth    &	          &	1-778 T+8466 p T^2-778 p^3 T^3+p^6 T^4
\tabularnewline[0.5pt]\hline				
78	 &	smooth    &	          &	1+280 T-7810 p T^2+280 p^3 T^3+p^6 T^4
\tabularnewline[0.5pt]\hline				
79	 &	smooth    &	          &	1-104 T-4882 p T^2-104 p^3 T^3+p^6 T^4
\tabularnewline[0.5pt]\hline				
80	 &	smooth    &	          &	1+264 T-1778 p T^2+264 p^3 T^3+p^6 T^4
\tabularnewline[0.5pt]\hline				
81	 &	smooth    &	          &	1+692 T-2154 p T^2+692 p^3 T^3+p^6 T^4
\tabularnewline[0.5pt]\hline				
82	 &	smooth    &	          &	1-1346 T+10442 p T^2-1346 p^3 T^3+p^6 T^4
\tabularnewline[0.5pt]\hline				
83	 &	smooth    &	          &	(1+10 p T+p^3 T^2)(1-1354 T+p^3 T^2)
\tabularnewline[0.5pt]\hline				
84	 &	smooth    &	          &	1+1382 T+12786 p T^2+1382 p^3 T^3+p^6 T^4
\tabularnewline[0.5pt]\hline				
85	 &	smooth    &	          &	1-1144 T+13518 p T^2-1144 p^3 T^3+p^6 T^4
\tabularnewline[0.5pt]\hline				
86	 &	smooth    &	          &	1-592 T+1374 p T^2-592 p^3 T^3+p^6 T^4
\tabularnewline[0.5pt]\hline				
87	 &	smooth    &	          &	1+186 T+7858 p T^2+186 p^3 T^3+p^6 T^4
\tabularnewline[0.5pt]\hline				
88	 &	smooth    &	          &	1-44 T-6442 p T^2-44 p^3 T^3+p^6 T^4
\tabularnewline[0.5pt]\hline				
89	 &	smooth    &	          &	1-1448 T+22286 p T^2-1448 p^3 T^3+p^6 T^4
\tabularnewline[0.5pt]\hline				
90	 &	smooth    &	          &	1-2424 T+32638 p T^2-2424 p^3 T^3+p^6 T^4
\tabularnewline[0.5pt]\hline				
91	 &	smooth    &	          &	1+876 T+8998 p T^2+876 p^3 T^3+p^6 T^4
\tabularnewline[0.5pt]\hline				
92	 &	smooth    &	          &	1+1380 T+9670 p T^2+1380 p^3 T^3+p^6 T^4
\tabularnewline[0.5pt]\hline				
93	 &	smooth    &	          &	1-68 T-634 p T^2-68 p^3 T^3+p^6 T^4
\tabularnewline[0.5pt]\hline				
94	 &	smooth    &	          &	1-804 T+7558 p T^2-804 p^3 T^3+p^6 T^4
\tabularnewline[0.5pt]\hline				
95	 &	smooth    &	          &	1+476 T-4602 p T^2+476 p^3 T^3+p^6 T^4
\tabularnewline[0.5pt]\hline				
96	 &	smooth    &	          &	1-288 T+766 p T^2-288 p^3 T^3+p^6 T^4
\tabularnewline[0.5pt]\hline				
\tablepostamble											
\newpage
\lhead{\ifthenelse{\isodd{\value{page}}}{\thepage}{\sl The $\z$ function for the  \Rodland manifold, AESZ\hskip2pt 27}}
\rhead{\ifthenelse{\isodd{\value{page}}}{\sl The $\z$ function for the \Rodland manifold, AESZ\hskip2pt 27}{\thepage}}
\subsection{The $\z$ function for the \Rodland manifold, AESZ\hskip2pt 27}
\vspace{1.5cm}				
\tablepreamble{5}				
1	 &	smooth    &	          &	(1+p^3 T^2)(1+12 T+p^3 T^2)
\tabularnewline[0.5pt]\hline				
2	 &	smooth    &	          &	1+26 T+78 p T^2+26 p^3 T^3+p^6 T^4
\tabularnewline[0.5pt]\hline				
3	 &	smooth$^*$&	3&	  
\tabularnewline[0.5pt]\hline				
4	 &	smooth    &	          &	1-9 T+8 p T^2-9 p^3 T^3+p^6 T^4
\tabularnewline[0.5pt]\hline				
\tablepostamble				
\tablepreamble{7}				
1	 &	smooth    &	          &	1-8 T+8 p T^2-8 p^3 T^3+p^6 T^4
\tabularnewline[0.5pt]\hline				
2	 &	smooth    &	          &	1+20 T+64 p T^2+20 p^3 T^3+p^6 T^4
\tabularnewline[0.5pt]\hline				
3	 &	singular  &	\{3,3,3,3\}&	  
\tabularnewline[0.5pt]\hline				
4	 &	smooth    &	          &	1+6 T-6 p T^2+6 p^3 T^3+p^6 T^4
\tabularnewline[0.5pt]\hline				
5	 &	smooth    &	          &	1+20 T+50 p T^2+20 p^3 T^3+p^6 T^4
\tabularnewline[0.5pt]\hline				
6	 &	smooth    &	          &	1+6 T+22 p T^2+6 p^3 T^3+p^6 T^4
\tabularnewline[0.5pt]\hline				
\tablepostamble				
\tablepreamble{11}				
1	 &	smooth    &	          &	1-T+182 p T^2-p^3 T^3+p^6 T^4
\tabularnewline[0.5pt]\hline				
2	 &	smooth    &	          &	1+48 T+182 p T^2+48 p^3 T^3+p^6 T^4
\tabularnewline[0.5pt]\hline				
3	 &	smooth$^*$&	3&	  
\tabularnewline[0.5pt]\hline				
4	 &	smooth    &	          &	1+6 T+14 p^2 T^2+6 p^3 T^3+p^6 T^4
\tabularnewline[0.5pt]\hline				
5	 &	smooth    &	          &	1+6 T-42 p T^2+6 p^3 T^3+p^6 T^4
\tabularnewline[0.5pt]\hline				
6	 &	smooth    &	          &	1-64 T+238 p T^2-64 p^3 T^3+p^6 T^4
\tabularnewline[0.5pt]\hline				
7	 &	smooth    &	          &	1+41 T+210 p T^2+41 p^3 T^3+p^6 T^4
\tabularnewline[0.5pt]\hline				
8	 &	smooth    &	          &	1+13 T-70 p T^2+13 p^3 T^3+p^6 T^4
\tabularnewline[0.5pt]\hline				
9	 &	smooth    &	          &	1+69 T+294 p T^2+69 p^3 T^3+p^6 T^4
\tabularnewline[0.5pt]\hline				
10	 &	smooth    &	          &	1-29 T+98 p T^2-29 p^3 T^3+p^6 T^4
\tabularnewline[0.5pt]\hline				
\tablepostamble				
\tablepreamble{13}				
1	 &	smooth    &	          &	1-35 T+44 p T^2-35 p^3 T^3+p^6 T^4
\tabularnewline[0.5pt]\hline				
2	 &	singular  &	2&	(1-p T) (1+42 T+p^3 T^2)
\tabularnewline[0.5pt]\hline				
3	 &	smooth$^*$&	3&	  
\tabularnewline[0.5pt]\hline				
4	 &	smooth    &	          &	(1+p^3 T^2)(1+84 T+p^3 T^2)
\tabularnewline[0.5pt]\hline				
5	 &	singular  &	5&	(1-p T) (1-42 T+p^3 T^2)
\tabularnewline[0.5pt]\hline				
6	 &	smooth    &	          &	1+28 T-54 p T^2+28 p^3 T^3+p^6 T^4
\tabularnewline[0.5pt]\hline				
7	 &	smooth    &	          &	(1+p^3 T^2)(1+63 T+p^3 T^2)
\tabularnewline[0.5pt]\hline				
8	 &	smooth    &	          &	1+49 T+142 p T^2+49 p^3 T^3+p^6 T^4
\tabularnewline[0.5pt]\hline				
9	 &	singular  &	        9& 	(1-p T) (1+14 T+p^3 T^2)
\tabularnewline[0.5pt]\hline				
10	 &	smooth    &	          &	1-14 T+142 p T^2-14 p^3 T^3+p^6 T^4
\tabularnewline[0.5pt]\hline				
11	 &	smooth    &	          &	1-42 T+240 p T^2-42 p^3 T^3+p^6 T^4
\tabularnewline[0.5pt]\hline				
12	 &	smooth    &	          &	(1+7 p T+p^3 T^2)(1-42 T+p^3 T^2)
\tabularnewline[0.5pt]\hline				
\tablepostamble				
\tablepreamble{17}				
1	 &	smooth    &	          &	1+3 p T+452 p T^2+3 p^4 T^3+p^6 T^4
\tabularnewline[0.5pt]\hline				
2	 &	smooth    &	          &	(1+p^3 T^2)(1+58 T+p^3 T^2)
\tabularnewline[0.5pt]\hline				
3	 &	smooth$^*$&	3&	  
\tabularnewline[0.5pt]\hline				
4	 &	smooth    &	          &	1-96 T+438 p T^2-96 p^3 T^3+p^6 T^4
\tabularnewline[0.5pt]\hline				
5	 &	smooth    &	          &	1+100 T+620 p T^2+100 p^3 T^3+p^6 T^4
\tabularnewline[0.5pt]\hline				
6	 &	smooth    &	          &	1+44 T-66 p T^2+44 p^3 T^3+p^6 T^4
\tabularnewline[0.5pt]\hline				
7	 &	smooth    &	          &	1-33 T-10 p T^2-33 p^3 T^3+p^6 T^4
\tabularnewline[0.5pt]\hline				
8	 &	smooth    &	          &	1-54 T+466 p T^2-54 p^3 T^3+p^6 T^4
\tabularnewline[0.5pt]\hline				
9	 &	smooth    &	          &	1+72 T+270 p T^2+72 p^3 T^3+p^6 T^4
\tabularnewline[0.5pt]\hline				
10	 &	smooth    &	          &	1+2 T+466 p T^2+2 p^3 T^3+p^6 T^4
\tabularnewline[0.5pt]\hline				
11	 &	smooth    &	          &	(1-4 p T+p^3 T^2)(1+63 T+p^3 T^2)
\tabularnewline[0.5pt]\hline				
12	 &	smooth    &	          &	1-5 T-178 p T^2-5 p^3 T^3+p^6 T^4
\tabularnewline[0.5pt]\hline				
13	 &	smooth    &	          &	1+142 T+634 p T^2+142 p^3 T^3+p^6 T^4
\tabularnewline[0.5pt]\hline				
14	 &	smooth    &	          &	1-33 T+382 p T^2-33 p^3 T^3+p^6 T^4
\tabularnewline[0.5pt]\hline				
15	 &	smooth    &	          &	1+16 T+158 p T^2+16 p^3 T^3+p^6 T^4
\tabularnewline[0.5pt]\hline				
16	 &	smooth    &	          &	1-26 T+242 p T^2-26 p^3 T^3+p^6 T^4
\tabularnewline[0.5pt]\hline				
\tablepostamble				
\tablepreamble{19}				
1	 &	smooth    &	          &	1-72 T+610 p T^2-72 p^3 T^3+p^6 T^4
\tabularnewline[0.5pt]\hline				
2	 &	smooth    &	          &	1+12 T-258 p T^2+12 p^3 T^3+p^6 T^4
\tabularnewline[0.5pt]\hline				
3	 &	smooth$^*$&	3&	  
\tabularnewline[0.5pt]\hline				
4	 &	smooth    &	          &	1+54 T+610 p T^2+54 p^3 T^3+p^6 T^4
\tabularnewline[0.5pt]\hline				
5	 &	smooth    &	          &	1+12 T-20 p T^2+12 p^3 T^3+p^6 T^4
\tabularnewline[0.5pt]\hline				
6	 &	smooth    &	          &	1+166 T+876 p T^2+166 p^3 T^3+p^6 T^4
\tabularnewline[0.5pt]\hline				
7	 &	smooth    &	          &	1+33 T-62 p T^2+33 p^3 T^3+p^6 T^4
\tabularnewline[0.5pt]\hline				
8	 &	smooth    &	          &	1+5 T+330 p T^2+5 p^3 T^3+p^6 T^4
\tabularnewline[0.5pt]\hline				
9	 &	smooth    &	          &	1+110 T+484 p T^2+110 p^3 T^3+p^6 T^4
\tabularnewline[0.5pt]\hline				
10	 &	smooth    &	          &	(1+8 p T+p^3 T^2)(1-35 T+p^3 T^2)
\tabularnewline[0.5pt]\hline				
11	 &	smooth    &	          &	1+p T-426 p T^2+p^4 T^3+p^6 T^4
\tabularnewline[0.5pt]\hline				
12	 &	smooth    &	          &	1-9 T-230 p T^2-9 p^3 T^3+p^6 T^4
\tabularnewline[0.5pt]\hline				
13	 &	smooth    &	          &	1+40 T-118 p T^2+40 p^3 T^3+p^6 T^4
\tabularnewline[0.5pt]\hline				
14	 &	smooth    &	          &	1+26 T+358 p T^2+26 p^3 T^3+p^6 T^4
\tabularnewline[0.5pt]\hline				
15	 &	smooth    &	          &	(1-6 p T+p^3 T^2)(1+98 T+p^3 T^2)
\tabularnewline[0.5pt]\hline				
16	 &	smooth    &	          &	1-163 T+1030 p T^2-163 p^3 T^3+p^6 T^4
\tabularnewline[0.5pt]\hline				
17	 &	smooth    &	          &	1+131 T+610 p T^2+131 p^3 T^3+p^6 T^4
\tabularnewline[0.5pt]\hline				
18	 &	smooth    &	          &	1+12 T+106 p T^2+12 p^3 T^3+p^6 T^4
\tabularnewline[0.5pt]\hline				
\tablepostamble				
\tablepreamble{23}				
1	 &	smooth    &	          &	1-71 T+14 p T^2-71 p^3 T^3+p^6 T^4
\tabularnewline[0.5pt]\hline				
2	 &	smooth    &	          &	1-99 T+630 p T^2-99 p^3 T^3+p^6 T^4
\tabularnewline[0.5pt]\hline				
3	 &	smooth$^*$&	3&	  
\tabularnewline[0.5pt]\hline				
4	 &	smooth    &	          &	1+20 T-616 p T^2+20 p^3 T^3+p^6 T^4
\tabularnewline[0.5pt]\hline				
5	 &	smooth    &	          &	1+132 T+350 p T^2+132 p^3 T^3+p^6 T^4
\tabularnewline[0.5pt]\hline				
6	 &	smooth    &	          &	1-4 p T+84 p T^2-4 p^4 T^3+p^6 T^4
\tabularnewline[0.5pt]\hline				
7	 &	smooth    &	          &	1+48 T-350 p T^2+48 p^3 T^3+p^6 T^4
\tabularnewline[0.5pt]\hline				
8	 &	smooth    &	          &	1-176 T+1050 p T^2-176 p^3 T^3+p^6 T^4
\tabularnewline[0.5pt]\hline				
9	 &	smooth    &	          &	1+90 T+980 p T^2+90 p^3 T^3+p^6 T^4
\tabularnewline[0.5pt]\hline				
10	 &	smooth    &	          &	1+216 T+1442 p T^2+216 p^3 T^3+p^6 T^4
\tabularnewline[0.5pt]\hline				
11	 &	smooth    &	          &	1+20 T+70 p T^2+20 p^3 T^3+p^6 T^4
\tabularnewline[0.5pt]\hline				
12	 &	smooth    &	          &	1-43 T-210 p T^2-43 p^3 T^3+p^6 T^4
\tabularnewline[0.5pt]\hline				
13	 &	smooth    &	          &	1+48 T+728 p T^2+48 p^3 T^3+p^6 T^4
\tabularnewline[0.5pt]\hline				
14	 &	smooth    &	          &	1-8 T-686 p T^2-8 p^3 T^3+p^6 T^4
\tabularnewline[0.5pt]\hline				
15	 &	smooth    &	          &	1-78 T+658 p T^2-78 p^3 T^3+p^6 T^4
\tabularnewline[0.5pt]\hline				
16	 &	smooth    &	          &	1+3 p T+854 p T^2+3 p^4 T^3+p^6 T^4
\tabularnewline[0.5pt]\hline				
17	 &	smooth    &	          &	1+41 T+490 p T^2+41 p^3 T^3+p^6 T^4
\tabularnewline[0.5pt]\hline				
18	 &	smooth    &	          &	1-78 T+658 p T^2-78 p^3 T^3+p^6 T^4
\tabularnewline[0.5pt]\hline				
19	 &	smooth    &	          &	1+125 T+602 p T^2+125 p^3 T^3+p^6 T^4
\tabularnewline[0.5pt]\hline				
20	 &	smooth    &	          &	1+160 T+714 p T^2+160 p^3 T^3+p^6 T^4
\tabularnewline[0.5pt]\hline				
21	 &	smooth    &	          &	1+97 T+826 p T^2+97 p^3 T^3+p^6 T^4
\tabularnewline[0.5pt]\hline				
22	 &	smooth    &	          &	1+48 T+1022 p T^2+48 p^3 T^3+p^6 T^4
\tabularnewline[0.5pt]\hline				
\tablepostamble				
\tablepreamble{29}				
1	 &	smooth    &	          &	1+122 T+734 p T^2+122 p^3 T^3+p^6 T^4
\tabularnewline[0.5pt]\hline				
2	 &	smooth    &	          &	1-123 T-190 p T^2-123 p^3 T^3+p^6 T^4
\tabularnewline[0.5pt]\hline				
3	 &	smooth$^*$&	3&	  
\tabularnewline[0.5pt]\hline				
4	 &	smooth    &	          &	1+31 T+1056 p T^2+31 p^3 T^3+p^6 T^4
\tabularnewline[0.5pt]\hline				
5	 &	smooth    &	          &	1-95 T+524 p T^2-95 p^3 T^3+p^6 T^4
\tabularnewline[0.5pt]\hline				
6	 &	smooth    &	          &	1-95 T+720 p T^2-95 p^3 T^3+p^6 T^4
\tabularnewline[0.5pt]\hline				
7	 &	smooth    &	          &	1+38 T+1658 p T^2+38 p^3 T^3+p^6 T^4
\tabularnewline[0.5pt]\hline				
8	 &	smooth    &	          &	1-88 T-568 p T^2-88 p^3 T^3+p^6 T^4
\tabularnewline[0.5pt]\hline				
9	 &	smooth    &	          &	1-11 T+1280 p T^2-11 p^3 T^3+p^6 T^4
\tabularnewline[0.5pt]\hline				
10	 &	smooth    &	          &	1+80 T+1462 p T^2+80 p^3 T^3+p^6 T^4
\tabularnewline[0.5pt]\hline				
11	 &	smooth    &	          &	1+437 T+3058 p T^2+437 p^3 T^3+p^6 T^4
\tabularnewline[0.5pt]\hline				
12	 &	singular  &	12&	(1-p T) (1+306 T+p^3 T^2)
\tabularnewline[0.5pt]\hline				
13	 &	smooth    &	          &	1+45 T-876 p T^2+45 p^3 T^3+p^6 T^4
\tabularnewline[0.5pt]\hline				
14	 &	smooth    &	          &	(1-2 p T+p^3 T^2)(1+194 T+p^3 T^2)
\tabularnewline[0.5pt]\hline				
15	 &	smooth    &	          &	1+38 T+1210 p T^2+38 p^3 T^3+p^6 T^4
\tabularnewline[0.5pt]\hline				
16	 &	smooth    &	          &	1+192 T+1658 p T^2+192 p^3 T^3+p^6 T^4
\tabularnewline[0.5pt]\hline				
17	 &	singular  &	17&	(1-p T) (1+110 T+p^3 T^2)
\tabularnewline[0.5pt]\hline				
18	 &	smooth    &	          &	1-137 T+958 p T^2-137 p^3 T^3+p^6 T^4
\tabularnewline[0.5pt]\hline				
19	 &	smooth    &	          &	1-123 T-134 p T^2-123 p^3 T^3+p^6 T^4
\tabularnewline[0.5pt]\hline				
20	 &	smooth    &	          &	1+17 T+384 p T^2+17 p^3 T^3+p^6 T^4
\tabularnewline[0.5pt]\hline				
21	 &	smooth    &	          &	1+276 T+1644 p T^2+276 p^3 T^3+p^6 T^4
\tabularnewline[0.5pt]\hline				
22	 &	smooth    &	          &	1+115 T+356 p T^2+115 p^3 T^3+p^6 T^4
\tabularnewline[0.5pt]\hline				
23	 &	smooth    &	          &	1+360 T+2554 p T^2+360 p^3 T^3+p^6 T^4
\tabularnewline[0.5pt]\hline				
24	 &	smooth    &	          &	1-25 T+804 p T^2-25 p^3 T^3+p^6 T^4
\tabularnewline[0.5pt]\hline				
25	 &	smooth    &	          &	1-284 T+1434 p T^2-284 p^3 T^3+p^6 T^4
\tabularnewline[0.5pt]\hline				
26	 &	smooth    &	          &	1+24 T+944 p T^2+24 p^3 T^3+p^6 T^4
\tabularnewline[0.5pt]\hline				
27	 &	smooth    &	          &	1-4 T+524 p T^2-4 p^3 T^3+p^6 T^4
\tabularnewline[0.5pt]\hline				
28	 &	singular  &	28&	(1-p T) (1-282 T+p^3 T^2)
\tabularnewline[0.5pt]\hline				
\tablepostamble				
\tablepreamble{31}				
1	 &	smooth    &	          &	1-5 T+578 p T^2-5 p^3 T^3+p^6 T^4
\tabularnewline[0.5pt]\hline				
2	 &	smooth    &	          &	1+219 T+2202 p T^2+219 p^3 T^3+p^6 T^4
\tabularnewline[0.5pt]\hline				
3	 &	smooth$^*$&	3&	  
\tabularnewline[0.5pt]\hline				
4	 &	smooth    &	          &	1+135 T+1138 p T^2+135 p^3 T^3+p^6 T^4
\tabularnewline[0.5pt]\hline				
5	 &	smooth    &	          &	1+247 T+1894 p T^2+247 p^3 T^3+p^6 T^4
\tabularnewline[0.5pt]\hline				
6	 &	smooth    &	          &	1+44 T+522 p T^2+44 p^3 T^3+p^6 T^4
\tabularnewline[0.5pt]\hline				
7	 &	smooth    &	          &	1-110 T+1124 p T^2-110 p^3 T^3+p^6 T^4
\tabularnewline[0.5pt]\hline				
8	 &	smooth    &	          &	1+16 T+18 p T^2+16 p^3 T^3+p^6 T^4
\tabularnewline[0.5pt]\hline				
9	 &	smooth    &	          &	1+170 T+368 p T^2+170 p^3 T^3+p^6 T^4
\tabularnewline[0.5pt]\hline				
10	 &	smooth    &	          &	1-61 T+914 p T^2-61 p^3 T^3+p^6 T^4
\tabularnewline[0.5pt]\hline				
11	 &	smooth    &	          &	1+114 T+74 p T^2+114 p^3 T^3+p^6 T^4
\tabularnewline[0.5pt]\hline				
12	 &	smooth    &	          &	1+324 T+1810 p T^2+324 p^3 T^3+p^6 T^4
\tabularnewline[0.5pt]\hline				
13	 &	smooth    &	          &	1-96 T+130 p T^2-96 p^3 T^3+p^6 T^4
\tabularnewline[0.5pt]\hline				
14	 &	smooth    &	          &	1+142 T+634 p T^2+142 p^3 T^3+p^6 T^4
\tabularnewline[0.5pt]\hline				
15	 &	smooth    &	          &	1-166 T+438 p T^2-166 p^3 T^3+p^6 T^4
\tabularnewline[0.5pt]\hline				
16	 &	smooth    &	          &	1+331 T+2398 p T^2+331 p^3 T^3+p^6 T^4
\tabularnewline[0.5pt]\hline				
17	 &	smooth    &	          &	1+296 T+1978 p T^2+296 p^3 T^3+p^6 T^4
\tabularnewline[0.5pt]\hline				
18	 &	smooth    &	          &	1-12 T-724 p T^2-12 p^3 T^3+p^6 T^4
\tabularnewline[0.5pt]\hline				
19	 &	smooth    &	          &	1-243 T+1446 p T^2-243 p^3 T^3+p^6 T^4
\tabularnewline[0.5pt]\hline				
20	 &	smooth    &	          &	1-131 T+2034 p T^2-131 p^3 T^3+p^6 T^4
\tabularnewline[0.5pt]\hline				
21	 &	smooth    &	          &	1-40 T-542 p T^2-40 p^3 T^3+p^6 T^4
\tabularnewline[0.5pt]\hline				
22	 &	smooth    &	          &	1+128 T-430 p T^2+128 p^3 T^3+p^6 T^4
\tabularnewline[0.5pt]\hline				
23	 &	smooth    &	          &	1+9 T+1754 p T^2+9 p^3 T^3+p^6 T^4
\tabularnewline[0.5pt]\hline				
24	 &	smooth    &	          &	1-180 T+494 p T^2-180 p^3 T^3+p^6 T^4
\tabularnewline[0.5pt]\hline				
25	 &	smooth    &	          &	1-208 T+1348 p T^2-208 p^3 T^3+p^6 T^4
\tabularnewline[0.5pt]\hline				
26	 &	smooth    &	          &	1-131 T+130 p T^2-131 p^3 T^3+p^6 T^4
\tabularnewline[0.5pt]\hline				
27	 &	smooth    &	          &	1+233 T+2090 p T^2+233 p^3 T^3+p^6 T^4
\tabularnewline[0.5pt]\hline				
28	 &	smooth    &	          &	1-40 T-598 p T^2-40 p^3 T^3+p^6 T^4
\tabularnewline[0.5pt]\hline				
29	 &	smooth    &	          &	1-33 T-38 p T^2-33 p^3 T^3+p^6 T^4
\tabularnewline[0.5pt]\hline				
30	 &	smooth    &	          &	1+37 T-766 p T^2+37 p^3 T^3+p^6 T^4
\tabularnewline[0.5pt]\hline				
\tablepostamble				
\tablepreamble{37}				
1	 &	smooth    &	          &	1+76 T+1694 p T^2+76 p^3 T^3+p^6 T^4
\tabularnewline[0.5pt]\hline				
2	 &	smooth    &	          &	1+62 T+1904 p T^2+62 p^3 T^3+p^6 T^4
\tabularnewline[0.5pt]\hline				
3	 &	smooth$^*$&	3&	  
\tabularnewline[0.5pt]\hline				
4	 &	smooth    &	          &	1+244 T+742 p T^2+244 p^3 T^3+p^6 T^4
\tabularnewline[0.5pt]\hline				
5	 &	smooth    &	          &	1-92 T+2156 p T^2-92 p^3 T^3+p^6 T^4
\tabularnewline[0.5pt]\hline				
6	 &	smooth    &	          &	1+202 T+1470 p T^2+202 p^3 T^3+p^6 T^4
\tabularnewline[0.5pt]\hline				
7	 &	smooth    &	          &	1+272 T+3066 p T^2+272 p^3 T^3+p^6 T^4
\tabularnewline[0.5pt]\hline				
8	 &	smooth    &	          &	1+447 T+2842 p T^2+447 p^3 T^3+p^6 T^4
\tabularnewline[0.5pt]\hline				
9	 &	smooth    &	          &	1-414 T+2478 p T^2-414 p^3 T^3+p^6 T^4
\tabularnewline[0.5pt]\hline				
10	 &	smooth    &	          &	1-36 T+2198 p T^2-36 p^3 T^3+p^6 T^4
\tabularnewline[0.5pt]\hline				
11	 &	smooth    &	          &	1+181 T+1344 p T^2+181 p^3 T^3+p^6 T^4
\tabularnewline[0.5pt]\hline				
12	 &	smooth    &	          &	1-99 T+2016 p T^2-99 p^3 T^3+p^6 T^4
\tabularnewline[0.5pt]\hline				
13	 &	smooth    &	          &	1+13 T-1134 p T^2+13 p^3 T^3+p^6 T^4
\tabularnewline[0.5pt]\hline				
14	 &	smooth    &	          &	1-120 T+714 p T^2-120 p^3 T^3+p^6 T^4
\tabularnewline[0.5pt]\hline				
15	 &	smooth    &	          &	1-22 T-168 p T^2-22 p^3 T^3+p^6 T^4
\tabularnewline[0.5pt]\hline				
16	 &	smooth    &	          &	1+545 T+4508 p T^2+545 p^3 T^3+p^6 T^4
\tabularnewline[0.5pt]\hline				
17	 &	smooth    &	          &	1-8 T-1064 p T^2-8 p^3 T^3+p^6 T^4
\tabularnewline[0.5pt]\hline				
18	 &	smooth    &	          &	1-120 T+910 p T^2-120 p^3 T^3+p^6 T^4
\tabularnewline[0.5pt]\hline				
19	 &	smooth    &	          &	1-50 T+938 p T^2-50 p^3 T^3+p^6 T^4
\tabularnewline[0.5pt]\hline				
20	 &	smooth    &	          &	1+328 T+84 p^2 T^2+328 p^3 T^3+p^6 T^4
\tabularnewline[0.5pt]\hline				
21	 &	smooth    &	          &	1-57 T+1876 p T^2-57 p^3 T^3+p^6 T^4
\tabularnewline[0.5pt]\hline				
22	 &	smooth    &	          &	1+230 T+1442 p T^2+230 p^3 T^3+p^6 T^4
\tabularnewline[0.5pt]\hline				
23	 &	smooth    &	          &	1+83 T-714 p T^2+83 p^3 T^3+p^6 T^4
\tabularnewline[0.5pt]\hline				
24	 &	smooth    &	          &	1-57 T+2170 p T^2-57 p^3 T^3+p^6 T^4
\tabularnewline[0.5pt]\hline				
25	 &	smooth    &	          &	1+104 T+686 p T^2+104 p^3 T^3+p^6 T^4
\tabularnewline[0.5pt]\hline				
26	 &	smooth    &	          &	1-106 T+798 p T^2-106 p^3 T^3+p^6 T^4
\tabularnewline[0.5pt]\hline				
27	 &	smooth    &	          &	1-484 T+4018 p T^2-484 p^3 T^3+p^6 T^4
\tabularnewline[0.5pt]\hline				
28	 &	smooth    &	          &	1-29 T+1848 p T^2-29 p^3 T^3+p^6 T^4
\tabularnewline[0.5pt]\hline				
29	 &	smooth    &	          &	1+55 T+1666 p T^2+55 p^3 T^3+p^6 T^4
\tabularnewline[0.5pt]\hline				
30	 &	smooth    &	          &	1+230 T+462 p T^2+230 p^3 T^3+p^6 T^4
\tabularnewline[0.5pt]\hline				
31	 &	smooth    &	          &	1-316 T+2478 p T^2-316 p^3 T^3+p^6 T^4
\tabularnewline[0.5pt]\hline				
32	 &	smooth    &	          &	1-8 T+1092 p T^2-8 p^3 T^3+p^6 T^4
\tabularnewline[0.5pt]\hline				
33	 &	smooth    &	          &	1+181 T+952 p T^2+181 p^3 T^3+p^6 T^4
\tabularnewline[0.5pt]\hline				
34	 &	smooth    &	          &	(1-p T+p^3 T^2)(1-202 T+p^3 T^2)
\tabularnewline[0.5pt]\hline				
35	 &	smooth    &	          &	1+293 T+3094 p T^2+293 p^3 T^3+p^6 T^4
\tabularnewline[0.5pt]\hline				
36	 &	smooth    &	          &	1+90 T-182 p T^2+90 p^3 T^3+p^6 T^4
\tabularnewline[0.5pt]\hline				
\tablepostamble				
\tablepreamble{41}				
1	 &	smooth    &	          &	1-21 T-1636 p T^2-21 p^3 T^3+p^6 T^4
\tabularnewline[0.5pt]\hline				
2	 &	smooth    &	          &	1+175 T+2088 p T^2+175 p^3 T^3+p^6 T^4
\tabularnewline[0.5pt]\hline				
3	 &	smooth$^*$&	3&	  
\tabularnewline[0.5pt]\hline				
4	 &	smooth    &	          &	1-448 T+3166 p T^2-448 p^3 T^3+p^6 T^4
\tabularnewline[0.5pt]\hline				
5	 &	smooth    &	          &	1+469 T+4048 p T^2+469 p^3 T^3+p^6 T^4
\tabularnewline[0.5pt]\hline				
6	 &	smooth    &	          &	1+217 T+1402 p T^2+217 p^3 T^3+p^6 T^4
\tabularnewline[0.5pt]\hline				
7	 &	smooth    &	          &	1+476 T+3558 p T^2+476 p^3 T^3+p^6 T^4
\tabularnewline[0.5pt]\hline				
8	 &	smooth    &	          &	1+266 T+2578 p T^2+266 p^3 T^3+p^6 T^4
\tabularnewline[0.5pt]\hline				
9	 &	smooth    &	          &	1+28 T+30 p T^2+28 p^3 T^3+p^6 T^4
\tabularnewline[0.5pt]\hline				
10	 &	smooth    &	          &	1-70 T+1794 p T^2-70 p^3 T^3+p^6 T^4
\tabularnewline[0.5pt]\hline				
11	 &	smooth    &	          &	1-336 T+2480 p T^2-336 p^3 T^3+p^6 T^4
\tabularnewline[0.5pt]\hline				
12	 &	smooth    &	          &	1+406 T+1892 p T^2+406 p^3 T^3+p^6 T^4
\tabularnewline[0.5pt]\hline				
13	 &	smooth    &	          &	1+70 T-950 p T^2+70 p^3 T^3+p^6 T^4
\tabularnewline[0.5pt]\hline				
14	 &	smooth    &	          &	1-70 T+422 p T^2-70 p^3 T^3+p^6 T^4
\tabularnewline[0.5pt]\hline				
15	 &	smooth    &	          &	1+35 T-754 p T^2+35 p^3 T^3+p^6 T^4
\tabularnewline[0.5pt]\hline				
16	 &	smooth    &	          &	1-623 T+5224 p T^2-623 p^3 T^3+p^6 T^4
\tabularnewline[0.5pt]\hline				
17	 &	smooth    &	          &	1-133 T+1402 p T^2-133 p^3 T^3+p^6 T^4
\tabularnewline[0.5pt]\hline				
18	 &	smooth    &	          &	1-21 T-68 p T^2-21 p^3 T^3+p^6 T^4
\tabularnewline[0.5pt]\hline				
19	 &	smooth    &	          &	1-518 T+4146 p T^2-518 p^3 T^3+p^6 T^4
\tabularnewline[0.5pt]\hline				
20	 &	smooth    &	          &	1-210 T+226 p T^2-210 p^3 T^3+p^6 T^4
\tabularnewline[0.5pt]\hline				
21	 &	smooth    &	          &	1+203 T+3068 p T^2+203 p^3 T^3+p^6 T^4
\tabularnewline[0.5pt]\hline				
22	 &	smooth    &	          &	1+182 T+2480 p T^2+182 p^3 T^3+p^6 T^4
\tabularnewline[0.5pt]\hline				
23	 &	singular  &	23&	(1-p T) (1-70 T+p^3 T^2)
\tabularnewline[0.5pt]\hline				
24	 &	smooth    &	          &	1-140 T-460 p T^2-140 p^3 T^3+p^6 T^4
\tabularnewline[0.5pt]\hline				
25	 &	smooth    &	          &	1+140 T-166 p T^2+140 p^3 T^3+p^6 T^4
\tabularnewline[0.5pt]\hline				
26	 &	smooth    &	          &	1+252 T+1304 p T^2+252 p^3 T^3+p^6 T^4
\tabularnewline[0.5pt]\hline				
27	 &	singular  &	27&	(1-p T) (1+70 T+p^3 T^2)
\tabularnewline[0.5pt]\hline				
28	 &	smooth    &	          &	1+182 T+1598 p T^2+182 p^3 T^3+p^6 T^4
\tabularnewline[0.5pt]\hline				
29	 &	smooth    &	          &	1+182 T+3068 p T^2+182 p^3 T^3+p^6 T^4
\tabularnewline[0.5pt]\hline				
30	 &	smooth    &	          &	1+266 T+618 p T^2+266 p^3 T^3+p^6 T^4
\tabularnewline[0.5pt]\hline				
31	 &	smooth    &	          &	1+14 T-2910 p T^2+14 p^3 T^3+p^6 T^4
\tabularnewline[0.5pt]\hline				
32	 &	smooth    &	          &	1-539 T+3852 p T^2-539 p^3 T^3+p^6 T^4
\tabularnewline[0.5pt]\hline				
33	 &	smooth    &	          &	1+196 T+2382 p T^2+196 p^3 T^3+p^6 T^4
\tabularnewline[0.5pt]\hline				
34	 &	singular  &	34&	(1-p T) (1+434 T+p^3 T^2)
\tabularnewline[0.5pt]\hline				
35	 &	smooth    &	          &	1-70 T+128 p T^2-70 p^3 T^3+p^6 T^4
\tabularnewline[0.5pt]\hline				
36	 &	smooth    &	          &	1+532 T+4734 p T^2+532 p^3 T^3+p^6 T^4
\tabularnewline[0.5pt]\hline				
37	 &	smooth    &	          &	1+35 T+2284 p T^2+35 p^3 T^3+p^6 T^4
\tabularnewline[0.5pt]\hline				
38	 &	smooth    &	          &	1-301 T+2578 p T^2-301 p^3 T^3+p^6 T^4
\tabularnewline[0.5pt]\hline				
39	 &	smooth    &	          &	1-224 T+1598 p T^2-224 p^3 T^3+p^6 T^4
\tabularnewline[0.5pt]\hline				
40	 &	smooth    &	          &	1+308 T+2774 p T^2+308 p^3 T^3+p^6 T^4
\tabularnewline[0.5pt]\hline				
\tablepostamble				
\tablepreamble{43}				
1	 &	singular  &	1&	(1-p T) (1-128 T+p^3 T^2)
\tabularnewline[0.5pt]\hline				
2	 &	smooth    &	          &	1-67 T-638 p T^2-67 p^3 T^3+p^6 T^4
\tabularnewline[0.5pt]\hline				
3	 &	smooth$^*$&	3&	  
\tabularnewline[0.5pt]\hline				
4	 &	smooth    &	          &	1+283 T+2330 p T^2+283 p^3 T^3+p^6 T^4
\tabularnewline[0.5pt]\hline				
5	 &	smooth    &	          &	1-11 T+1602 p T^2-11 p^3 T^3+p^6 T^4
\tabularnewline[0.5pt]\hline				
6	 &	smooth    &	          &	1+234 T+1392 p T^2+234 p^3 T^3+p^6 T^4
\tabularnewline[0.5pt]\hline				
7	 &	smooth    &	          &	1-354 T+3254 p T^2-354 p^3 T^3+p^6 T^4
\tabularnewline[0.5pt]\hline				
8	 &	smooth    &	          &	1+185 T+2834 p T^2+185 p^3 T^3+p^6 T^4
\tabularnewline[0.5pt]\hline				
9	 &	smooth    &	          &	1+787 T+7062 p T^2+787 p^3 T^3+p^6 T^4
\tabularnewline[0.5pt]\hline				
10	 &	smooth    &	          &	1+24 T-2360 p T^2+24 p^3 T^3+p^6 T^4
\tabularnewline[0.5pt]\hline				
11	 &	smooth    &	          &	1-74 T-974 p T^2-74 p^3 T^3+p^6 T^4
\tabularnewline[0.5pt]\hline				
12	 &	smooth    &	          &	1+150 T+2414 p T^2+150 p^3 T^3+p^6 T^4
\tabularnewline[0.5pt]\hline				
13	 &	smooth    &	          &	1-81 T+1266 p T^2-81 p^3 T^3+p^6 T^4
\tabularnewline[0.5pt]\hline				
14	 &	smooth    &	          &	1+395 T+4290 p T^2+395 p^3 T^3+p^6 T^4
\tabularnewline[0.5pt]\hline				
15	 &	smooth    &	          &	1-459 T+3842 p T^2-459 p^3 T^3+p^6 T^4
\tabularnewline[0.5pt]\hline				
16	 &	smooth    &	          &	1-529 T+4542 p T^2-529 p^3 T^3+p^6 T^4
\tabularnewline[0.5pt]\hline				
17	 &	smooth    &	          &	1-494 T+4752 p T^2-494 p^3 T^3+p^6 T^4
\tabularnewline[0.5pt]\hline				
18	 &	smooth    &	          &	1-60 T+1910 p T^2-60 p^3 T^3+p^6 T^4
\tabularnewline[0.5pt]\hline				
19	 &	smooth    &	          &	1-564 T+4514 p T^2-564 p^3 T^3+p^6 T^4
\tabularnewline[0.5pt]\hline				
20	 &	smooth    &	          &	1+31 T-78 p T^2+31 p^3 T^3+p^6 T^4
\tabularnewline[0.5pt]\hline				
21	 &	smooth    &	          &	1+10 T-1786 p T^2+10 p^3 T^3+p^6 T^4
\tabularnewline[0.5pt]\hline				
22	 &	smooth    &	          &	1+122 T+1714 p T^2+122 p^3 T^3+p^6 T^4
\tabularnewline[0.5pt]\hline				
23	 &	smooth    &	          &	1+311 T+3142 p T^2+311 p^3 T^3+p^6 T^4
\tabularnewline[0.5pt]\hline				
24	 &	smooth    &	          &	1-130 T-1170 p T^2-130 p^3 T^3+p^6 T^4
\tabularnewline[0.5pt]\hline				
25	 &	smooth    &	          &	1+101 T+1042 p T^2+101 p^3 T^3+p^6 T^4
\tabularnewline[0.5pt]\hline				
26	 &	smooth    &	          &	1+136 T+34 p^2 T^2+136 p^3 T^3+p^6 T^4
\tabularnewline[0.5pt]\hline				
27	 &	smooth    &	          &	1+661 T+5410 p T^2+661 p^3 T^3+p^6 T^4
\tabularnewline[0.5pt]\hline				
28	 &	smooth    &	          &	1-242 T-386 p T^2-242 p^3 T^3+p^6 T^4
\tabularnewline[0.5pt]\hline				
29	 &	smooth    &	          &	1+192 T+1714 p T^2+192 p^3 T^3+p^6 T^4
\tabularnewline[0.5pt]\hline				
30	 &	smooth    &	          &	1-144 T-1226 p T^2-144 p^3 T^3+p^6 T^4
\tabularnewline[0.5pt]\hline				
31	 &	smooth    &	          &	1+10 T-2038 p T^2+10 p^3 T^3+p^6 T^4
\tabularnewline[0.5pt]\hline				
32	 &	smooth    &	          &	1+185 T+762 p T^2+185 p^3 T^3+p^6 T^4
\tabularnewline[0.5pt]\hline				
33	 &	smooth    &	          &	1-158 T-554 p T^2-158 p^3 T^3+p^6 T^4
\tabularnewline[0.5pt]\hline				
34	 &	smooth    &	          &	1-375 T+3730 p T^2-375 p^3 T^3+p^6 T^4
\tabularnewline[0.5pt]\hline				
35	 &	smooth    &	          &	(1+12 p T+p^3 T^2)(1-44 T+p^3 T^2)
\tabularnewline[0.5pt]\hline				
36	 &	singular  &	36&	(1-p T) (1+264 T+p^3 T^2)
\tabularnewline[0.5pt]\hline				
37	 &	singular  &	37&	(1-p T) (1+68 T+p^3 T^2)
\tabularnewline[0.5pt]\hline				
38	 &	smooth    &	          &	1+248 T+1490 p T^2+248 p^3 T^3+p^6 T^4
\tabularnewline[0.5pt]\hline				
39	 &	smooth    &	          &	1+262 T+398 p T^2+262 p^3 T^3+p^6 T^4
\tabularnewline[0.5pt]\hline				
40	 &	smooth    &	          &	1+66 T-36 p^2 T^2+66 p^3 T^3+p^6 T^4
\tabularnewline[0.5pt]\hline				
41	 &	smooth    &	          &	1+234 T+2218 p T^2+234 p^3 T^3+p^6 T^4
\tabularnewline[0.5pt]\hline				
42	 &	smooth    &	          &	(1+5 p T+p^3 T^2)(1-128 T+p^3 T^2)
\tabularnewline[0.5pt]\hline				
\tablepostamble				
\tablepreamble{47}				
1	 &	smooth    &	          &	1-30 T-2890 p T^2-30 p^3 T^3+p^6 T^4
\tabularnewline[0.5pt]\hline				
2	 &	smooth    &	          &	1+222 T+3802 p T^2+222 p^3 T^3+p^6 T^4
\tabularnewline[0.5pt]\hline				
3	 &	smooth$^*$&	3&	  
\tabularnewline[0.5pt]\hline				
4	 &	smooth    &	          &	1-23 T+3214 p T^2-23 p^3 T^3+p^6 T^4
\tabularnewline[0.5pt]\hline				
5	 &	smooth    &	          &	1-13 p T+4754 p T^2-13 p^4 T^3+p^6 T^4
\tabularnewline[0.5pt]\hline				
6	 &	smooth    &	          &	1+166 T+1520 p T^2+166 p^3 T^3+p^6 T^4
\tabularnewline[0.5pt]\hline				
7	 &	smooth    &	          &	1-205 T+2234 p T^2-205 p^3 T^3+p^6 T^4
\tabularnewline[0.5pt]\hline				
8	 &	smooth    &	          &	1-23 T+2430 p T^2-23 p^3 T^3+p^6 T^4
\tabularnewline[0.5pt]\hline				
9	 &	smooth    &	          &	1+68 T+1338 p T^2+68 p^3 T^3+p^6 T^4
\tabularnewline[0.5pt]\hline				
10	 &	smooth    &	          &	1-58 T+750 p T^2-58 p^3 T^3+p^6 T^4
\tabularnewline[0.5pt]\hline				
11	 &	smooth    &	          &	1-296 T+1730 p T^2-296 p^3 T^3+p^6 T^4
\tabularnewline[0.5pt]\hline				
12	 &	smooth    &	          &	1+446 T+4782 p T^2+446 p^3 T^3+p^6 T^4
\tabularnewline[0.5pt]\hline				
13	 &	smooth    &	          &	1+166 T+3074 p T^2+166 p^3 T^3+p^6 T^4
\tabularnewline[0.5pt]\hline				
14	 &	smooth    &	          &	1-58 T-3198 p T^2-58 p^3 T^3+p^6 T^4
\tabularnewline[0.5pt]\hline				
15	 &	smooth    &	          &	1+5 T-1238 p T^2+5 p^3 T^3+p^6 T^4
\tabularnewline[0.5pt]\hline				
16	 &	smooth    &	          &	1+866 T+8114 p T^2+866 p^3 T^3+p^6 T^4
\tabularnewline[0.5pt]\hline				
17	 &	smooth    &	          &	1-65 T-482 p T^2-65 p^3 T^3+p^6 T^4
\tabularnewline[0.5pt]\hline				
18	 &	smooth    &	          &	1+p T+862 p T^2+p^4 T^3+p^6 T^4
\tabularnewline[0.5pt]\hline				
19	 &	smooth    &	          &	1+159 T-678 p T^2+159 p^3 T^3+p^6 T^4
\tabularnewline[0.5pt]\hline				
20	 &	smooth    &	          &	1+82 T-622 p T^2+82 p^3 T^3+p^6 T^4
\tabularnewline[0.5pt]\hline				
21	 &	smooth    &	          &	1-58 T-1168 p T^2-58 p^3 T^3+p^6 T^4
\tabularnewline[0.5pt]\hline				
22	 &	smooth    &	          &	1+404 T+3466 p T^2+404 p^3 T^3+p^6 T^4
\tabularnewline[0.5pt]\hline				
23	 &	smooth    &	          &	1-170 T+2710 p T^2-170 p^3 T^3+p^6 T^4
\tabularnewline[0.5pt]\hline				
24	 &	smooth    &	          &	1+467 T+2934 p T^2+467 p^3 T^3+p^6 T^4
\tabularnewline[0.5pt]\hline				
25	 &	smooth    &	          &	1+383 T+4978 p T^2+383 p^3 T^3+p^6 T^4
\tabularnewline[0.5pt]\hline				
26	 &	smooth    &	          &	1-44 T+2654 p T^2-44 p^3 T^3+p^6 T^4
\tabularnewline[0.5pt]\hline				
27	 &	smooth    &	          &	1-114 T+1268 p T^2-114 p^3 T^3+p^6 T^4
\tabularnewline[0.5pt]\hline				
28	 &	smooth    &	          &	1+166 T+2346 p T^2+166 p^3 T^3+p^6 T^4
\tabularnewline[0.5pt]\hline				
29	 &	smooth    &	          &	1-716 T+6322 p T^2-716 p^3 T^3+p^6 T^4
\tabularnewline[0.5pt]\hline				
30	 &	smooth    &	          &	1+68 T+3410 p T^2+68 p^3 T^3+p^6 T^4
\tabularnewline[0.5pt]\hline				
31	 &	smooth    &	          &	1-9 T+3522 p T^2-9 p^3 T^3+p^6 T^4
\tabularnewline[0.5pt]\hline				
32	 &	smooth    &	          &	1+138 T+4222 p T^2+138 p^3 T^3+p^6 T^4
\tabularnewline[0.5pt]\hline				
33	 &	smooth    &	          &	1-100 T+274 p T^2-100 p^3 T^3+p^6 T^4
\tabularnewline[0.5pt]\hline				
34	 &	smooth    &	          &	1+12 T+1450 p T^2+12 p^3 T^3+p^6 T^4
\tabularnewline[0.5pt]\hline				
35	 &	smooth    &	          &	1+677 T+6210 p T^2+677 p^3 T^3+p^6 T^4
\tabularnewline[0.5pt]\hline				
36	 &	smooth    &	          &	1+467 T+2934 p T^2+467 p^3 T^3+p^6 T^4
\tabularnewline[0.5pt]\hline				
37	 &	smooth    &	          &	1+96 T+3536 p T^2+96 p^3 T^3+p^6 T^4
\tabularnewline[0.5pt]\hline				
38	 &	smooth    &	          &	1+117 T+3914 p T^2+117 p^3 T^3+p^6 T^4
\tabularnewline[0.5pt]\hline				
39	 &	smooth    &	          &	1+271 T+2234 p T^2+271 p^3 T^3+p^6 T^4
\tabularnewline[0.5pt]\hline				
40	 &	smooth    &	          &	1+222 T+2570 p T^2+222 p^3 T^3+p^6 T^4
\tabularnewline[0.5pt]\hline				
41	 &	smooth    &	          &	1+103 T-1294 p T^2+103 p^3 T^3+p^6 T^4
\tabularnewline[0.5pt]\hline				
42	 &	smooth    &	          &	1-492 T+4810 p T^2-492 p^3 T^3+p^6 T^4
\tabularnewline[0.5pt]\hline				
43	 &	smooth    &	          &	1+285 T+50 p T^2+285 p^3 T^3+p^6 T^4
\tabularnewline[0.5pt]\hline				
44	 &	smooth    &	          &	1+12 T-1574 p T^2+12 p^3 T^3+p^6 T^4
\tabularnewline[0.5pt]\hline				
45	 &	smooth    &	          &	1-247 T+1842 p T^2-247 p^3 T^3+p^6 T^4
\tabularnewline[0.5pt]\hline				
46	 &	smooth    &	          &	1-191 T+890 p T^2-191 p^3 T^3+p^6 T^4
\tabularnewline[0.5pt]\hline				
\tablepostamble				
\tablepreamble{53}				
1	 &	smooth    &	          &	1-267 T+4172 p T^2-267 p^3 T^3+p^6 T^4
\tabularnewline[0.5pt]\hline				
2	 &	smooth    &	          &	1-22 T-532 p T^2-22 p^3 T^3+p^6 T^4
\tabularnewline[0.5pt]\hline				
3	 &	smooth$^*$&	3&	  
\tabularnewline[0.5pt]\hline				
4	 &	smooth    &	          &	1+552 T+4130 p T^2+552 p^3 T^3+p^6 T^4
\tabularnewline[0.5pt]\hline				
5	 &	smooth    &	          &	1-232 T+5012 p T^2-232 p^3 T^3+p^6 T^4
\tabularnewline[0.5pt]\hline				
6	 &	smooth    &	          &	1+545 T+4256 p T^2+545 p^3 T^3+p^6 T^4
\tabularnewline[0.5pt]\hline				
7	 &	smooth    &	          &	1-533 T+4844 p T^2-533 p^3 T^3+p^6 T^4
\tabularnewline[0.5pt]\hline				
8	 &	smooth    &	          &	1-57 T+3038 p T^2-57 p^3 T^3+p^6 T^4
\tabularnewline[0.5pt]\hline				
9	 &	smooth    &	          &	1+412 T+5082 p T^2+412 p^3 T^3+p^6 T^4
\tabularnewline[0.5pt]\hline				
10	 &	smooth    &	          &	1+27 T+3976 p T^2+27 p^3 T^3+p^6 T^4
\tabularnewline[0.5pt]\hline				
11	 &	smooth    &	          &	1+111 T-280 p T^2+111 p^3 T^3+p^6 T^4
\tabularnewline[0.5pt]\hline				
12	 &	smooth    &	          &	1+1105 T+10542 p T^2+1105 p^3 T^3+p^6 T^4
\tabularnewline[0.5pt]\hline				
13	 &	smooth    &	          &	1-36 T-2338 p T^2-36 p^3 T^3+p^6 T^4
\tabularnewline[0.5pt]\hline				
14	 &	smooth    &	          &	1-148 T+4088 p T^2-148 p^3 T^3+p^6 T^4
\tabularnewline[0.5pt]\hline				
15	 &	smooth    &	          &	1-50 T+1050 p T^2-50 p^3 T^3+p^6 T^4
\tabularnewline[0.5pt]\hline				
16	 &	smooth    &	          &	1+48 T+4382 p T^2+48 p^3 T^3+p^6 T^4
\tabularnewline[0.5pt]\hline				
17	 &	smooth    &	          &	1+5 p T+2828 p T^2+5 p^4 T^3+p^6 T^4
\tabularnewline[0.5pt]\hline				
18	 &	smooth    &	          &	1-204 T-1666 p T^2-204 p^3 T^3+p^6 T^4
\tabularnewline[0.5pt]\hline				
19	 &	smooth    &	          &	1+76 T+2506 p T^2+76 p^3 T^3+p^6 T^4
\tabularnewline[0.5pt]\hline				
20	 &	smooth    &	          &	1+468 T+3290 p T^2+468 p^3 T^3+p^6 T^4
\tabularnewline[0.5pt]\hline				
21	 &	smooth    &	          &	(1-4 p T+p^3 T^2)(1+498 T+p^3 T^2)
\tabularnewline[0.5pt]\hline				
22	 &	smooth    &	          &	1-743 T+5586 p T^2-743 p^3 T^3+p^6 T^4
\tabularnewline[0.5pt]\hline				
23	 &	smooth    &	          &	1+230 T+4732 p T^2+230 p^3 T^3+p^6 T^4
\tabularnewline[0.5pt]\hline				
24	 &	smooth    &	          &	1+314 T+5866 p T^2+314 p^3 T^3+p^6 T^4
\tabularnewline[0.5pt]\hline				
25	 &	smooth    &	          &	(1-4 p T+p^3 T^2)(1+624 T+p^3 T^2)
\tabularnewline[0.5pt]\hline				
26	 &	smooth    &	          &	1+244 T+1246 p T^2+244 p^3 T^3+p^6 T^4
\tabularnewline[0.5pt]\hline				
27	 &	smooth    &	          &	1+118 T+2828 p T^2+118 p^3 T^3+p^6 T^4
\tabularnewline[0.5pt]\hline				
28	 &	smooth    &	          &	1-2 p T+4998 p T^2-2 p^4 T^3+p^6 T^4
\tabularnewline[0.5pt]\hline				
29	 &	smooth    &	          &	1+202 T-1330 p T^2+202 p^3 T^3+p^6 T^4
\tabularnewline[0.5pt]\hline				
30	 &	smooth    &	          &	1+195 T-1106 p T^2+195 p^3 T^3+p^6 T^4
\tabularnewline[0.5pt]\hline				
31	 &	smooth    &	          &	1+251 T+2982 p T^2+251 p^3 T^3+p^6 T^4
\tabularnewline[0.5pt]\hline				
32	 &	smooth    &	          &	1+384 T+3724 p T^2+384 p^3 T^3+p^6 T^4
\tabularnewline[0.5pt]\hline				
33	 &	smooth    &	          &	1+272 T+4270 p T^2+272 p^3 T^3+p^6 T^4
\tabularnewline[0.5pt]\hline				
34	 &	smooth    &	          &	1-813 T+7630 p T^2-813 p^3 T^3+p^6 T^4
\tabularnewline[0.5pt]\hline				
35	 &	smooth    &	          &	1+202 T+3570 p T^2+202 p^3 T^3+p^6 T^4
\tabularnewline[0.5pt]\hline				
36	 &	smooth    &	          &	1-176 T-2366 p T^2-176 p^3 T^3+p^6 T^4
\tabularnewline[0.5pt]\hline				
37	 &	smooth    &	          &	1-8 T+882 p T^2-8 p^3 T^3+p^6 T^4
\tabularnewline[0.5pt]\hline				
38	 &	smooth    &	          &	1-407 T+812 p T^2-407 p^3 T^3+p^6 T^4
\tabularnewline[0.5pt]\hline				
39	 &	smooth    &	          &	1+48 T+3500 p T^2+48 p^3 T^3+p^6 T^4
\tabularnewline[0.5pt]\hline				
40	 &	smooth    &	          &	1+608 T+4886 p T^2+608 p^3 T^3+p^6 T^4
\tabularnewline[0.5pt]\hline				
41	 &	smooth    &	          &	1-596 T+5978 p T^2-596 p^3 T^3+p^6 T^4
\tabularnewline[0.5pt]\hline				
42	 &	smooth    &	          &	1+13 T-1652 p T^2+13 p^3 T^3+p^6 T^4
\tabularnewline[0.5pt]\hline				
43	 &	smooth    &	          &	1+321 T+2016 p T^2+321 p^3 T^3+p^6 T^4
\tabularnewline[0.5pt]\hline				
44	 &	smooth    &	          &	1+321 T+4172 p T^2+321 p^3 T^3+p^6 T^4
\tabularnewline[0.5pt]\hline				
45	 &	smooth    &	          &	1-190 T+2002 p T^2-190 p^3 T^3+p^6 T^4
\tabularnewline[0.5pt]\hline				
46	 &	smooth    &	          &	1-64 T+2478 p T^2-64 p^3 T^3+p^6 T^4
\tabularnewline[0.5pt]\hline				
47	 &	smooth    &	          &	1-316 T+3878 p T^2-316 p^3 T^3+p^6 T^4
\tabularnewline[0.5pt]\hline				
48	 &	smooth    &	          &	1-218 T+252 p T^2-218 p^3 T^3+p^6 T^4
\tabularnewline[0.5pt]\hline				
49	 &	smooth    &	          &	1-344 T+658 p T^2-344 p^3 T^3+p^6 T^4
\tabularnewline[0.5pt]\hline				
50	 &	smooth    &	          &	1-64 T-1834 p T^2-64 p^3 T^3+p^6 T^4
\tabularnewline[0.5pt]\hline				
51	 &	smooth    &	          &	1+496 T+3374 p T^2+496 p^3 T^3+p^6 T^4
\tabularnewline[0.5pt]\hline				
52	 &	smooth    &	          &	1+335 T+5096 p T^2+335 p^3 T^3+p^6 T^4
\tabularnewline[0.5pt]\hline				
\tablepostamble				
\tablepreamble{59}				
1	 &	smooth    &	          &	1-446 T+3434 p T^2-446 p^3 T^3+p^6 T^4
\tabularnewline[0.5pt]\hline				
2	 &	smooth    &	          &	1+569 T+2762 p T^2+569 p^3 T^3+p^6 T^4
\tabularnewline[0.5pt]\hline				
3	 &	smooth$^*$&	3&	  
\tabularnewline[0.5pt]\hline				
4	 &	smooth    &	          &	1-334 T+200 p T^2-334 p^3 T^3+p^6 T^4
\tabularnewline[0.5pt]\hline				
5	 &	smooth    &	          &	1-579 T+3406 p T^2-579 p^3 T^3+p^6 T^4
\tabularnewline[0.5pt]\hline				
6	 &	smooth    &	          &	1+3 p T+1306 p T^2+3 p^4 T^3+p^6 T^4
\tabularnewline[0.5pt]\hline				
7	 &	smooth    &	          &	1+394 T+2804 p T^2+394 p^3 T^3+p^6 T^4
\tabularnewline[0.5pt]\hline				
8	 &	smooth    &	          &	1-96 T-2698 p T^2-96 p^3 T^3+p^6 T^4
\tabularnewline[0.5pt]\hline				
9	 &	smooth    &	          &	1+373 T+5450 p T^2+373 p^3 T^3+p^6 T^4
\tabularnewline[0.5pt]\hline				
10	 &	smooth    &	          &	1+2 T+2370 p T^2+2 p^3 T^3+p^6 T^4
\tabularnewline[0.5pt]\hline				
11	 &	smooth    &	          &	1+534 T+3042 p T^2+534 p^3 T^3+p^6 T^4
\tabularnewline[0.5pt]\hline				
12	 &	smooth    &	          &	1+800 T+6612 p T^2+800 p^3 T^3+p^6 T^4
\tabularnewline[0.5pt]\hline				
13	 &	smooth    &	          &	1+44 T+3518 p T^2+44 p^3 T^3+p^6 T^4
\tabularnewline[0.5pt]\hline				
14	 &	smooth    &	          &	1+639 T+4442 p T^2+639 p^3 T^3+p^6 T^4
\tabularnewline[0.5pt]\hline				
15	 &	smooth    &	          &	1+324 T+6066 p T^2+324 p^3 T^3+p^6 T^4
\tabularnewline[0.5pt]\hline				
16	 &	smooth    &	          &	1+44 T+4386 p T^2+44 p^3 T^3+p^6 T^4
\tabularnewline[0.5pt]\hline				
17	 &	smooth    &	          &	1+436 T+5170 p T^2+436 p^3 T^3+p^6 T^4
\tabularnewline[0.5pt]\hline				
18	 &	smooth    &	          &	1-1118 T+10854 p T^2-1118 p^3 T^3+p^6 T^4
\tabularnewline[0.5pt]\hline				
19	 &	smooth    &	          &	1+247 T+3378 p T^2+247 p^3 T^3+p^6 T^4
\tabularnewline[0.5pt]\hline				
20	 &	smooth    &	          &	1-642 T+4330 p T^2-642 p^3 T^3+p^6 T^4
\tabularnewline[0.5pt]\hline				
21	 &	smooth    &	          &	1+310 T-402 p T^2+310 p^3 T^3+p^6 T^4
\tabularnewline[0.5pt]\hline				
22	 &	smooth    &	          &	1+128 T-4294 p T^2+128 p^3 T^3+p^6 T^4
\tabularnewline[0.5pt]\hline				
23	 &	smooth    &	          &	1+30 T-850 p T^2+30 p^3 T^3+p^6 T^4
\tabularnewline[0.5pt]\hline				
24	 &	smooth    &	          &	1-320 T+1026 p T^2-320 p^3 T^3+p^6 T^4
\tabularnewline[0.5pt]\hline				
25	 &	smooth    &	          &	1+478 T+4204 p T^2+478 p^3 T^3+p^6 T^4
\tabularnewline[0.5pt]\hline				
26	 &	smooth    &	          &	1-523 T+3126 p T^2-523 p^3 T^3+p^6 T^4
\tabularnewline[0.5pt]\hline				
27	 &	smooth    &	          &	1+366 T+2398 p T^2+366 p^3 T^3+p^6 T^4
\tabularnewline[0.5pt]\hline				
28	 &	smooth    &	          &	1-250 T+4890 p T^2-250 p^3 T^3+p^6 T^4
\tabularnewline[0.5pt]\hline				
29	 &	smooth    &	          &	1+674 T+5100 p T^2+674 p^3 T^3+p^6 T^4
\tabularnewline[0.5pt]\hline				
30	 &	smooth    &	          &	1+891 T+9818 p T^2+891 p^3 T^3+p^6 T^4
\tabularnewline[0.5pt]\hline				
31	 &	smooth    &	          &	1-803 T+7914 p T^2-803 p^3 T^3+p^6 T^4
\tabularnewline[0.5pt]\hline				
32	 &	smooth    &	          &	1+723 T+6458 p T^2+723 p^3 T^3+p^6 T^4
\tabularnewline[0.5pt]\hline				
33	 &	smooth    &	          &	1+93 T-1942 p T^2+93 p^3 T^3+p^6 T^4
\tabularnewline[0.5pt]\hline				
34	 &	smooth    &	          &	1+646 T+6206 p T^2+646 p^3 T^3+p^6 T^4
\tabularnewline[0.5pt]\hline				
35	 &	smooth    &	          &	1-12 T+410 p T^2-12 p^3 T^3+p^6 T^4
\tabularnewline[0.5pt]\hline				
36	 &	smooth    &	          &	1+506 T+2762 p T^2+506 p^3 T^3+p^6 T^4
\tabularnewline[0.5pt]\hline				
37	 &	smooth    &	          &	1-824 T+7214 p T^2-824 p^3 T^3+p^6 T^4
\tabularnewline[0.5pt]\hline				
38	 &	smooth    &	          &	1-320 T+1810 p T^2-320 p^3 T^3+p^6 T^4
\tabularnewline[0.5pt]\hline				
39	 &	smooth    &	          &	1+366 T+1222 p T^2+366 p^3 T^3+p^6 T^4
\tabularnewline[0.5pt]\hline				
40	 &	smooth    &	          &	1+338 T+6318 p T^2+338 p^3 T^3+p^6 T^4
\tabularnewline[0.5pt]\hline				
41	 &	smooth    &	          &	1-159 T+158 p T^2-159 p^3 T^3+p^6 T^4
\tabularnewline[0.5pt]\hline				
42	 &	smooth    &	          &	1+548 T+6094 p T^2+548 p^3 T^3+p^6 T^4
\tabularnewline[0.5pt]\hline				
43	 &	smooth    &	          &	1-572 T+7270 p T^2-572 p^3 T^3+p^6 T^4
\tabularnewline[0.5pt]\hline				
44	 &	smooth    &	          &	1+247 T+186 p T^2+247 p^3 T^3+p^6 T^4
\tabularnewline[0.5pt]\hline				
45	 &	smooth    &	          &	1+254 T+2272 p T^2+254 p^3 T^3+p^6 T^4
\tabularnewline[0.5pt]\hline				
46	 &	smooth    &	          &	1-47 T-318 p T^2-47 p^3 T^3+p^6 T^4
\tabularnewline[0.5pt]\hline				
47	 &	smooth    &	          &	1-124 T-962 p T^2-124 p^3 T^3+p^6 T^4
\tabularnewline[0.5pt]\hline				
48	 &	smooth    &	          &	1-250 T-962 p T^2-250 p^3 T^3+p^6 T^4
\tabularnewline[0.5pt]\hline				
49	 &	smooth    &	          &	1+478 T+6108 p T^2+478 p^3 T^3+p^6 T^4
\tabularnewline[0.5pt]\hline				
50	 &	smooth    &	          &	1-320 T+1726 p T^2-320 p^3 T^3+p^6 T^4
\tabularnewline[0.5pt]\hline				
51	 &	smooth    &	          &	1+555 T+5702 p T^2+555 p^3 T^3+p^6 T^4
\tabularnewline[0.5pt]\hline				
52	 &	smooth    &	          &	1-47 T+4106 p T^2-47 p^3 T^3+p^6 T^4
\tabularnewline[0.5pt]\hline				
53	 &	smooth    &	          &	1-950 T+9538 p T^2-950 p^3 T^3+p^6 T^4
\tabularnewline[0.5pt]\hline				
54	 &	smooth    &	          &	1+324 T+2622 p T^2+324 p^3 T^3+p^6 T^4
\tabularnewline[0.5pt]\hline				
55	 &	smooth    &	          &	1+436 T+1110 p T^2+436 p^3 T^3+p^6 T^4
\tabularnewline[0.5pt]\hline				
56	 &	smooth    &	          &	1-138 T+3014 p T^2-138 p^3 T^3+p^6 T^4
\tabularnewline[0.5pt]\hline				
57	 &	smooth    &	          &	1+121 T+5086 p T^2+121 p^3 T^3+p^6 T^4
\tabularnewline[0.5pt]\hline				
58	 &	smooth    &	          &	1-572 T+7046 p T^2-572 p^3 T^3+p^6 T^4
\tabularnewline[0.5pt]\hline				
\tablepostamble				
\tablepreamble{61}				
1	 &	smooth    &	          &	1-422 T+134 p T^2-422 p^3 T^3+p^6 T^4
\tabularnewline[0.5pt]\hline				
2	 &	smooth    &	          &	1+8 p T+3914 p T^2+8 p^4 T^3+p^6 T^4
\tabularnewline[0.5pt]\hline				
3	 &	smooth$^*$&	3&	  
\tabularnewline[0.5pt]\hline				
4	 &	smooth    &	          &	1+439 T+3060 p T^2+439 p^3 T^3+p^6 T^4
\tabularnewline[0.5pt]\hline				
5	 &	smooth    &	          &	1+516 T+6070 p T^2+516 p^3 T^3+p^6 T^4
\tabularnewline[0.5pt]\hline				
6	 &	smooth    &	          &	1+1181 T+12090 p T^2+1181 p^3 T^3+p^6 T^4
\tabularnewline[0.5pt]\hline				
7	 &	smooth    &	          &	1+194 T-160 p T^2+194 p^3 T^3+p^6 T^4
\tabularnewline[0.5pt]\hline				
8	 &	smooth    &	          &	1+537 T+3074 p T^2+537 p^3 T^3+p^6 T^4
\tabularnewline[0.5pt]\hline				
9	 &	smooth    &	          &	1-170 T-678 p T^2-170 p^3 T^3+p^6 T^4
\tabularnewline[0.5pt]\hline				
10	 &	smooth    &	          &	1-121 T+2990 p T^2-121 p^3 T^3+p^6 T^4
\tabularnewline[0.5pt]\hline				
11	 &	smooth    &	          &	1+397 T+4670 p T^2+397 p^3 T^3+p^6 T^4
\tabularnewline[0.5pt]\hline				
12	 &	smooth    &	          &	1+656 T+4558 p T^2+656 p^3 T^3+p^6 T^4
\tabularnewline[0.5pt]\hline				
13	 &	smooth    &	          &	1+250 T+4194 p T^2+250 p^3 T^3+p^6 T^4
\tabularnewline[0.5pt]\hline				
14	 &	smooth    &	          &	1+390 T+1674 p T^2+390 p^3 T^3+p^6 T^4
\tabularnewline[0.5pt]\hline				
15	 &	smooth    &	          &	1+47 T+1828 p T^2+47 p^3 T^3+p^6 T^4
\tabularnewline[0.5pt]\hline				
16	 &	smooth    &	          &	1+334 T+1702 p T^2+334 p^3 T^3+p^6 T^4
\tabularnewline[0.5pt]\hline				
17	 &	smooth    &	          &	1-828 T+8142 p T^2-828 p^3 T^3+p^6 T^4
\tabularnewline[0.5pt]\hline				
18	 &	smooth    &	          &	1+110 T+4292 p T^2+110 p^3 T^3+p^6 T^4
\tabularnewline[0.5pt]\hline				
19	 &	smooth    &	          &	1+565 T+8100 p T^2+565 p^3 T^3+p^6 T^4
\tabularnewline[0.5pt]\hline				
20	 &	smooth    &	          &	1+663 T+7288 p T^2+663 p^3 T^3+p^6 T^4
\tabularnewline[0.5pt]\hline				
21	 &	smooth    &	          &	1+439 T+5202 p T^2+439 p^3 T^3+p^6 T^4
\tabularnewline[0.5pt]\hline				
22	 &	smooth    &	          &	1+775 T+8884 p T^2+775 p^3 T^3+p^6 T^4
\tabularnewline[0.5pt]\hline				
23	 &	smooth    &	          &	1+362 T+2108 p T^2+362 p^3 T^3+p^6 T^4
\tabularnewline[0.5pt]\hline				
24	 &	smooth    &	          &	1-324 T+3354 p T^2-324 p^3 T^3+p^6 T^4
\tabularnewline[0.5pt]\hline				
25	 &	smooth    &	          &	1-296 T+5202 p T^2-296 p^3 T^3+p^6 T^4
\tabularnewline[0.5pt]\hline				
26	 &	smooth    &	          &	1+159 T+2654 p T^2+159 p^3 T^3+p^6 T^4
\tabularnewline[0.5pt]\hline				
27	 &	smooth    &	          &	1+334 T+2822 p T^2+334 p^3 T^3+p^6 T^4
\tabularnewline[0.5pt]\hline				
28	 &	smooth    &	          &	1-576 T+4726 p T^2-576 p^3 T^3+p^6 T^4
\tabularnewline[0.5pt]\hline				
29	 &	smooth    &	          &	1-548 T+6364 p T^2-548 p^3 T^3+p^6 T^4
\tabularnewline[0.5pt]\hline				
30	 &	smooth    &	          &	1-6 p T+1758 p T^2-6 p^4 T^3+p^6 T^4
\tabularnewline[0.5pt]\hline				
31	 &	smooth    &	          &	1-205 T+7190 p T^2-205 p^3 T^3+p^6 T^4
\tabularnewline[0.5pt]\hline				
32	 &	smooth    &	          &	1+236 T+526 p T^2+236 p^3 T^3+p^6 T^4
\tabularnewline[0.5pt]\hline				
33	 &	smooth    &	          &	1-282 T+358 p T^2-282 p^3 T^3+p^6 T^4
\tabularnewline[0.5pt]\hline				
34	 &	smooth    &	          &	1-688 T+7106 p T^2-688 p^3 T^3+p^6 T^4
\tabularnewline[0.5pt]\hline				
35	 &	smooth    &	          &	1+33 T+3690 p T^2+33 p^3 T^3+p^6 T^4
\tabularnewline[0.5pt]\hline				
36	 &	smooth    &	          &	1-877 T+7232 p T^2-877 p^3 T^3+p^6 T^4
\tabularnewline[0.5pt]\hline				
37	 &	smooth    &	          &	1-359 T+3718 p T^2-359 p^3 T^3+p^6 T^4
\tabularnewline[0.5pt]\hline				
38	 &	smooth    &	          &	1-380 T+2934 p T^2-380 p^3 T^3+p^6 T^4
\tabularnewline[0.5pt]\hline				
39	 &	smooth    &	          &	1-352 T+778 p T^2-352 p^3 T^3+p^6 T^4
\tabularnewline[0.5pt]\hline				
40	 &	smooth    &	          &	1+1013 T+9570 p T^2+1013 p^3 T^3+p^6 T^4
\tabularnewline[0.5pt]\hline				
41	 &	smooth    &	          &	1+320 T+4894 p T^2+320 p^3 T^3+p^6 T^4
\tabularnewline[0.5pt]\hline				
42	 &	smooth    &	          &	1+68 T-650 p T^2+68 p^3 T^3+p^6 T^4
\tabularnewline[0.5pt]\hline				
43	 &	smooth    &	          &	1-170 T+890 p T^2-170 p^3 T^3+p^6 T^4
\tabularnewline[0.5pt]\hline				
44	 &	smooth    &	          &	1+712 T+4278 p T^2+712 p^3 T^3+p^6 T^4
\tabularnewline[0.5pt]\hline				
45	 &	smooth    &	          &	1-149 T+3088 p T^2-149 p^3 T^3+p^6 T^4
\tabularnewline[0.5pt]\hline				
46	 &	smooth    &	          &	1+537 T+4292 p T^2+537 p^3 T^3+p^6 T^4
\tabularnewline[0.5pt]\hline				
47	 &	smooth    &	          &	1-604 T+2346 p T^2-604 p^3 T^3+p^6 T^4
\tabularnewline[0.5pt]\hline				
48	 &	smooth    &	          &	1+208 T+4726 p T^2+208 p^3 T^3+p^6 T^4
\tabularnewline[0.5pt]\hline				
49	 &	smooth    &	          &	1+40 T-426 p T^2+40 p^3 T^3+p^6 T^4
\tabularnewline[0.5pt]\hline				
50	 &	smooth    &	          &	1-86 T+7120 p T^2-86 p^3 T^3+p^6 T^4
\tabularnewline[0.5pt]\hline				
51	 &	smooth    &	          &	1-352 T+1310 p T^2-352 p^3 T^3+p^6 T^4
\tabularnewline[0.5pt]\hline				
52	 &	smooth    &	          &	1-184 T+946 p T^2-184 p^3 T^3+p^6 T^4
\tabularnewline[0.5pt]\hline				
53	 &	smooth    &	          &	1+117 T+1170 p T^2+117 p^3 T^3+p^6 T^4
\tabularnewline[0.5pt]\hline				
54	 &	smooth    &	          &	1+82 T+4614 p T^2+82 p^3 T^3+p^6 T^4
\tabularnewline[0.5pt]\hline				
55	 &	smooth    &	          &	1-275 T+1030 p T^2-275 p^3 T^3+p^6 T^4
\tabularnewline[0.5pt]\hline				
56	 &	smooth    &	          &	1+68 T-2050 p T^2+68 p^3 T^3+p^6 T^4
\tabularnewline[0.5pt]\hline				
57	 &	smooth    &	          &	1+831 T+8800 p T^2+831 p^3 T^3+p^6 T^4
\tabularnewline[0.5pt]\hline				
58	 &	smooth    &	          &	1-380 T-1070 p T^2-380 p^3 T^3+p^6 T^4
\tabularnewline[0.5pt]\hline				
59	 &	smooth    &	          &	1-828 T+9010 p T^2-828 p^3 T^3+p^6 T^4
\tabularnewline[0.5pt]\hline				
60	 &	smooth    &	          &	1+222 T+5986 p T^2+222 p^3 T^3+p^6 T^4
\tabularnewline[0.5pt]\hline				
\tablepostamble				
\tablepreamble{67}				
1	 &	smooth    &	          &	1-435 T+70 p T^2-435 p^3 T^3+p^6 T^4
\tabularnewline[0.5pt]\hline				
2	 &	smooth    &	          &	1+55 T-7182 p T^2+55 p^3 T^3+p^6 T^4
\tabularnewline[0.5pt]\hline				
3	 &	smooth$^*$&	3&	  
\tabularnewline[0.5pt]\hline				
4	 &	smooth    &	          &	1-148 T-148 p^3 T^3+p^6 T^4
\tabularnewline[0.5pt]\hline				
5	 &	smooth    &	          &	1+825 T+7378 p T^2+825 p^3 T^3+p^6 T^4
\tabularnewline[0.5pt]\hline				
6	 &	smooth    &	          &	1+286 T+6986 p T^2+286 p^3 T^3+p^6 T^4
\tabularnewline[0.5pt]\hline				
7	 &	smooth    &	          &	1-50 T+7742 p T^2-50 p^3 T^3+p^6 T^4
\tabularnewline[0.5pt]\hline				
8	 &	smooth    &	          &	1+811 T+7042 p T^2+811 p^3 T^3+p^6 T^4
\tabularnewline[0.5pt]\hline				
9	 &	smooth    &	          &	1+118 T-5474 p T^2+118 p^3 T^3+p^6 T^4
\tabularnewline[0.5pt]\hline				
10	 &	smooth    &	          &	1-92 T+6636 p T^2-92 p^3 T^3+p^6 T^4
\tabularnewline[0.5pt]\hline				
11	 &	smooth    &	          &	1+132 T+1918 p T^2+132 p^3 T^3+p^6 T^4
\tabularnewline[0.5pt]\hline				
12	 &	smooth    &	          &	1+223 T+6258 p T^2+223 p^3 T^3+p^6 T^4
\tabularnewline[0.5pt]\hline				
13	 &	smooth    &	          &	1-533 T-126 p T^2-533 p^3 T^3+p^6 T^4
\tabularnewline[0.5pt]\hline				
14	 &	smooth    &	          &	1+321 T+2926 p T^2+321 p^3 T^3+p^6 T^4
\tabularnewline[0.5pt]\hline				
15	 &	smooth    &	          &	1+587 T+3234 p T^2+587 p^3 T^3+p^6 T^4
\tabularnewline[0.5pt]\hline				
16	 &	smooth    &	          &	1+636 T+3234 p T^2+636 p^3 T^3+p^6 T^4
\tabularnewline[0.5pt]\hline				
17	 &	smooth    &	          &	1-169 T+966 p T^2-169 p^3 T^3+p^6 T^4
\tabularnewline[0.5pt]\hline				
18	 &	smooth    &	          &	1+377 T+3290 p T^2+377 p^3 T^3+p^6 T^4
\tabularnewline[0.5pt]\hline				
19	 &	smooth    &	          &	1+433 T+7378 p T^2+433 p^3 T^3+p^6 T^4
\tabularnewline[0.5pt]\hline				
20	 &	smooth    &	          &	1+552 T+4746 p T^2+552 p^3 T^3+p^6 T^4
\tabularnewline[0.5pt]\hline				
21	 &	smooth    &	          &	1-162 T+1232 p T^2-162 p^3 T^3+p^6 T^4
\tabularnewline[0.5pt]\hline				
22	 &	smooth    &	          &	1-400 T+5418 p T^2-400 p^3 T^3+p^6 T^4
\tabularnewline[0.5pt]\hline				
23	 &	smooth    &	          &	1+258 T-2016 p T^2+258 p^3 T^3+p^6 T^4
\tabularnewline[0.5pt]\hline				
24	 &	smooth    &	          &	1+538 T+5978 p T^2+538 p^3 T^3+p^6 T^4
\tabularnewline[0.5pt]\hline				
25	 &	smooth    &	          &	1-596 T+6202 p T^2-596 p^3 T^3+p^6 T^4
\tabularnewline[0.5pt]\hline				
26	 &	smooth    &	          &	1-204 T-854 p T^2-204 p^3 T^3+p^6 T^4
\tabularnewline[0.5pt]\hline				
27	 &	smooth    &	          &	1-106 T+1694 p T^2-106 p^3 T^3+p^6 T^4
\tabularnewline[0.5pt]\hline				
28	 &	smooth    &	          &	1+97 T+882 p T^2+97 p^3 T^3+p^6 T^4
\tabularnewline[0.5pt]\hline				
29	 &	smooth    &	          &	1+69 T+6286 p T^2+69 p^3 T^3+p^6 T^4
\tabularnewline[0.5pt]\hline				
30	 &	smooth    &	          &	1-106 T-3598 p T^2-106 p^3 T^3+p^6 T^4
\tabularnewline[0.5pt]\hline				
31	 &	smooth    &	          &	1+433 T+8162 p T^2+433 p^3 T^3+p^6 T^4
\tabularnewline[0.5pt]\hline				
32	 &	smooth    &	          &	1-519 T+5306 p T^2-519 p^3 T^3+p^6 T^4
\tabularnewline[0.5pt]\hline				
33	 &	smooth    &	          &	1-484 T+658 p T^2-484 p^3 T^3+p^6 T^4
\tabularnewline[0.5pt]\hline				
34	 &	smooth    &	          &	1-218 T+8414 p T^2-218 p^3 T^3+p^6 T^4
\tabularnewline[0.5pt]\hline				
35	 &	smooth    &	          &	1+678 T+4928 p T^2+678 p^3 T^3+p^6 T^4
\tabularnewline[0.5pt]\hline				
36	 &	smooth    &	          &	1+622 T+6426 p T^2+622 p^3 T^3+p^6 T^4
\tabularnewline[0.5pt]\hline				
37	 &	smooth    &	          &	1-421 T+3542 p T^2-421 p^3 T^3+p^6 T^4
\tabularnewline[0.5pt]\hline				
38	 &	smooth    &	          &	1+727 T+8946 p T^2+727 p^3 T^3+p^6 T^4
\tabularnewline[0.5pt]\hline				
39	 &	smooth    &	          &	1-974 T+10850 p T^2-974 p^3 T^3+p^6 T^4
\tabularnewline[0.5pt]\hline				
40	 &	smooth    &	          &	1+62 T-6818 p T^2+62 p^3 T^3+p^6 T^4
\tabularnewline[0.5pt]\hline				
41	 &	smooth    &	          &	1-708 T+9198 p T^2-708 p^3 T^3+p^6 T^4
\tabularnewline[0.5pt]\hline				
42	 &	smooth    &	          &	1-624 T+4158 p T^2-624 p^3 T^3+p^6 T^4
\tabularnewline[0.5pt]\hline				
43	 &	smooth    &	          &	1+377 T+2506 p T^2+377 p^3 T^3+p^6 T^4
\tabularnewline[0.5pt]\hline				
44	 &	smooth    &	          &	1+419 T+6258 p T^2+419 p^3 T^3+p^6 T^4
\tabularnewline[0.5pt]\hline				
45	 &	smooth    &	          &	1-36 T-6230 p T^2-36 p^3 T^3+p^6 T^4
\tabularnewline[0.5pt]\hline				
46	 &	smooth    &	          &	(1+2 p T+p^3 T^2)(1-688 T+p^3 T^2)
\tabularnewline[0.5pt]\hline				
47	 &	smooth    &	          &	1+1532 T+17388 p T^2+1532 p^3 T^3+p^6 T^4
\tabularnewline[0.5pt]\hline				
48	 &	smooth    &	          &	1+356 T+1806 p T^2+356 p^3 T^3+p^6 T^4
\tabularnewline[0.5pt]\hline				
49	 &	smooth    &	          &	1+1042 T+10234 p T^2+1042 p^3 T^3+p^6 T^4
\tabularnewline[0.5pt]\hline				
50	 &	smooth    &	          &	1-484 T+8498 p T^2-484 p^3 T^3+p^6 T^4
\tabularnewline[0.5pt]\hline				
51	 &	smooth    &	          &	1-316 T+2338 p T^2-316 p^3 T^3+p^6 T^4
\tabularnewline[0.5pt]\hline				
52	 &	smooth    &	          &	1+188 T+4242 p T^2+188 p^3 T^3+p^6 T^4
\tabularnewline[0.5pt]\hline				
53	 &	smooth    &	          &	1-554 T+2702 p T^2-554 p^3 T^3+p^6 T^4
\tabularnewline[0.5pt]\hline				
54	 &	smooth    &	          &	1-358 T+4172 p T^2-358 p^3 T^3+p^6 T^4
\tabularnewline[0.5pt]\hline				
55	 &	smooth    &	          &	1+13 T+7686 p T^2+13 p^3 T^3+p^6 T^4
\tabularnewline[0.5pt]\hline				
56	 &	smooth    &	          &	1-330 T+2688 p T^2-330 p^3 T^3+p^6 T^4
\tabularnewline[0.5pt]\hline				
57	 &	smooth    &	          &	1-204 T+8750 p T^2-204 p^3 T^3+p^6 T^4
\tabularnewline[0.5pt]\hline				
58	 &	smooth    &	          &	1-183 T-2702 p T^2-183 p^3 T^3+p^6 T^4
\tabularnewline[0.5pt]\hline				
59	 &	smooth    &	          &	1+503 T+7294 p T^2+503 p^3 T^3+p^6 T^4
\tabularnewline[0.5pt]\hline				
60	 &	smooth    &	          &	1+258 T-2310 p T^2+258 p^3 T^3+p^6 T^4
\tabularnewline[0.5pt]\hline				
61	 &	smooth    &	          &	1+223 T+7042 p T^2+223 p^3 T^3+p^6 T^4
\tabularnewline[0.5pt]\hline				
62	 &	smooth    &	          &	1+83 T+3486 p T^2+83 p^3 T^3+p^6 T^4
\tabularnewline[0.5pt]\hline				
63	 &	smooth    &	          &	1+69 T-966 p T^2+69 p^3 T^3+p^6 T^4
\tabularnewline[0.5pt]\hline				
64	 &	smooth    &	          &	1+944 T+9352 p T^2+944 p^3 T^3+p^6 T^4
\tabularnewline[0.5pt]\hline				
65	 &	smooth    &	          &	1+272 T+1652 p T^2+272 p^3 T^3+p^6 T^4
\tabularnewline[0.5pt]\hline				
66	 &	smooth    &	          &	1-652 T+9954 p T^2-652 p^3 T^3+p^6 T^4
\tabularnewline[0.5pt]\hline				
\tablepostamble				
\tablepreamble{71}				
1	 &	smooth    &	          &	1+556 T+5354 p T^2+556 p^3 T^3+p^6 T^4
\tabularnewline[0.5pt]\hline				
2	 &	smooth    &	          &	1-578 T+1868 p T^2-578 p^3 T^3+p^6 T^4
\tabularnewline[0.5pt]\hline				
3	 &	smooth$^*$&	3&	  
\tabularnewline[0.5pt]\hline				
4	 &	smooth    &	          &	1+150 T+4920 p T^2+150 p^3 T^3+p^6 T^4
\tabularnewline[0.5pt]\hline				
5	 &	singular  &	5&	(1-p T) (1-604 T+p^3 T^2)
\tabularnewline[0.5pt]\hline				
6	 &	smooth    &	          &	1+654 T+8938 p T^2+654 p^3 T^3+p^6 T^4
\tabularnewline[0.5pt]\hline				
7	 &	smooth    &	          &	1+220 T+706 p T^2+220 p^3 T^3+p^6 T^4
\tabularnewline[0.5pt]\hline				
8	 &	smooth    &	          &	1-606 T+4136 p T^2-606 p^3 T^3+p^6 T^4
\tabularnewline[0.5pt]\hline				
9	 &	smooth    &	          &	1+479 T+4822 p T^2+479 p^3 T^3+p^6 T^4
\tabularnewline[0.5pt]\hline				
10	 &	smooth    &	          &	1+353 T+4486 p T^2+353 p^3 T^3+p^6 T^4
\tabularnewline[0.5pt]\hline				
11	 &	smooth    &	          &	1+871 T+8994 p T^2+871 p^3 T^3+p^6 T^4
\tabularnewline[0.5pt]\hline				
12	 &	smooth    &	          &	1+52 T+7426 p T^2+52 p^3 T^3+p^6 T^4
\tabularnewline[0.5pt]\hline				
13	 &	smooth    &	          &	1+570 T+4738 p T^2+570 p^3 T^3+p^6 T^4
\tabularnewline[0.5pt]\hline				
14	 &	smooth    &	          &	1+752 T+10450 p T^2+752 p^3 T^3+p^6 T^4
\tabularnewline[0.5pt]\hline				
15	 &	smooth    &	          &	1+472 T-2080 p T^2+472 p^3 T^3+p^6 T^4
\tabularnewline[0.5pt]\hline				
16	 &	smooth    &	          &	1-704 T+10352 p T^2-704 p^3 T^3+p^6 T^4
\tabularnewline[0.5pt]\hline				
17	 &	smooth    &	          &	1-1159 T+13586 p T^2-1159 p^3 T^3+p^6 T^4
\tabularnewline[0.5pt]\hline				
18	 &	smooth    &	          &	1-326 T+1028 p T^2-326 p^3 T^3+p^6 T^4
\tabularnewline[0.5pt]\hline				
19	 &	smooth    &	          &	1-88 T+5186 p T^2-88 p^3 T^3+p^6 T^4
\tabularnewline[0.5pt]\hline				
20	 &	smooth    &	          &	1+332 T+580 p T^2+332 p^3 T^3+p^6 T^4
\tabularnewline[0.5pt]\hline				
21	 &	smooth    &	          &	1+108 T-3214 p T^2+108 p^3 T^3+p^6 T^4
\tabularnewline[0.5pt]\hline				
22	 &	smooth    &	          &	1+136 T-1478 p T^2+136 p^3 T^3+p^6 T^4
\tabularnewline[0.5pt]\hline				
23	 &	smooth    &	          &	1-4 T+3898 p T^2-4 p^3 T^3+p^6 T^4
\tabularnewline[0.5pt]\hline				
24	 &	smooth    &	          &	1+318 T-4334 p T^2+318 p^3 T^3+p^6 T^4
\tabularnewline[0.5pt]\hline				
25	 &	singular  &	25&	(1-p T) (1+180 T+p^3 T^2)
\tabularnewline[0.5pt]\hline				
26	 &	smooth    &	          &	1+857 T+5130 p T^2+857 p^3 T^3+p^6 T^4
\tabularnewline[0.5pt]\hline				
27	 &	smooth    &	          &	1+577 T+10058 p T^2+577 p^3 T^3+p^6 T^4
\tabularnewline[0.5pt]\hline				
28	 &	smooth    &	          &	1-389 T+7426 p T^2-389 p^3 T^3+p^6 T^4
\tabularnewline[0.5pt]\hline				
29	 &	smooth    &	          &	1+311 T-3774 p T^2+311 p^3 T^3+p^6 T^4
\tabularnewline[0.5pt]\hline				
30	 &	smooth    &	          &	1+318 T+9274 p T^2+318 p^3 T^3+p^6 T^4
\tabularnewline[0.5pt]\hline				
31	 &	smooth    &	          &	1-179 T+2442 p T^2-179 p^3 T^3+p^6 T^4
\tabularnewline[0.5pt]\hline				
32	 &	smooth    &	          &	1+500 T+3044 p T^2+500 p^3 T^3+p^6 T^4
\tabularnewline[0.5pt]\hline				
33	 &	smooth    &	          &	1-984 T+7930 p T^2-984 p^3 T^3+p^6 T^4
\tabularnewline[0.5pt]\hline				
34	 &	smooth    &	          &	1-452 T+9638 p T^2-452 p^3 T^3+p^6 T^4
\tabularnewline[0.5pt]\hline				
35	 &	smooth    &	          &	1+500 T+2134 p T^2+500 p^3 T^3+p^6 T^4
\tabularnewline[0.5pt]\hline				
36	 &	smooth    &	          &	1+696 T+8756 p T^2+696 p^3 T^3+p^6 T^4
\tabularnewline[0.5pt]\hline				
37	 &	smooth    &	          &	1+1060 T+9778 p T^2+1060 p^3 T^3+p^6 T^4
\tabularnewline[0.5pt]\hline				
38	 &	smooth    &	          &	1-1054 T+13250 p T^2-1054 p^3 T^3+p^6 T^4
\tabularnewline[0.5pt]\hline				
39	 &	smooth    &	          &	1-389 T+6194 p T^2-389 p^3 T^3+p^6 T^4
\tabularnewline[0.5pt]\hline				
40	 &	smooth    &	          &	1+381 T+8210 p T^2+381 p^3 T^3+p^6 T^4
\tabularnewline[0.5pt]\hline				
41	 &	smooth    &	          &	1+332 T+3058 p T^2+332 p^3 T^3+p^6 T^4
\tabularnewline[0.5pt]\hline				
42	 &	smooth    &	          &	1+395 T+4290 p T^2+395 p^3 T^3+p^6 T^4
\tabularnewline[0.5pt]\hline				
43	 &	smooth    &	          &	1+248 T+1098 p T^2+248 p^3 T^3+p^6 T^4
\tabularnewline[0.5pt]\hline				
44	 &	smooth    &	          &	1+528 T+1378 p T^2+528 p^3 T^3+p^6 T^4
\tabularnewline[0.5pt]\hline				
45	 &	smooth    &	          &	1-606 T+4444 p T^2-606 p^3 T^3+p^6 T^4
\tabularnewline[0.5pt]\hline				
46	 &	singular  &	46&	(1-p T) (1+376 T+p^3 T^2)
\tabularnewline[0.5pt]\hline				
47	 &	smooth    &	          &	1+17 T-2206 p T^2+17 p^3 T^3+p^6 T^4
\tabularnewline[0.5pt]\hline				
48	 &	smooth    &	          &	1-536 T+9568 p T^2-536 p^3 T^3+p^6 T^4
\tabularnewline[0.5pt]\hline				
49	 &	smooth    &	          &	1-228 T+6404 p T^2-228 p^3 T^3+p^6 T^4
\tabularnewline[0.5pt]\hline				
50	 &	smooth    &	          &	1-725 T+9890 p T^2-725 p^3 T^3+p^6 T^4
\tabularnewline[0.5pt]\hline				
51	 &	smooth    &	          &	1+199 T-1534 p T^2+199 p^3 T^3+p^6 T^4
\tabularnewline[0.5pt]\hline				
52	 &	smooth    &	          &	1-186 T+7874 p T^2-186 p^3 T^3+p^6 T^4
\tabularnewline[0.5pt]\hline				
53	 &	smooth    &	          &	1+367 T+4682 p T^2+367 p^3 T^3+p^6 T^4
\tabularnewline[0.5pt]\hline				
54	 &	smooth    &	          &	1+66 T+2218 p T^2+66 p^3 T^3+p^6 T^4
\tabularnewline[0.5pt]\hline				
55	 &	smooth    &	          &	1+325 T+5802 p T^2+325 p^3 T^3+p^6 T^4
\tabularnewline[0.5pt]\hline				
56	 &	smooth    &	          &	1-508 T+4066 p T^2-508 p^3 T^3+p^6 T^4
\tabularnewline[0.5pt]\hline				
57	 &	smooth    &	          &	1+262 T+7314 p T^2+262 p^3 T^3+p^6 T^4
\tabularnewline[0.5pt]\hline				
58	 &	smooth    &	          &	1+171 T+4346 p T^2+171 p^3 T^3+p^6 T^4
\tabularnewline[0.5pt]\hline				
59	 &	smooth    &	          &	1-340 T+2834 p T^2-340 p^3 T^3+p^6 T^4
\tabularnewline[0.5pt]\hline				
60	 &	smooth    &	          &	1+80 T+5018 p T^2+80 p^3 T^3+p^6 T^4
\tabularnewline[0.5pt]\hline				
61	 &	smooth    &	          &	1+332 T+1042 p T^2+332 p^3 T^3+p^6 T^4
\tabularnewline[0.5pt]\hline				
62	 &	smooth    &	          &	1-599 T+2442 p T^2-599 p^3 T^3+p^6 T^4
\tabularnewline[0.5pt]\hline				
63	 &	smooth    &	          &	1+920 T+6838 p T^2+920 p^3 T^3+p^6 T^4
\tabularnewline[0.5pt]\hline				
64	 &	smooth    &	          &	1-67 T-4530 p T^2-67 p^3 T^3+p^6 T^4
\tabularnewline[0.5pt]\hline				
65	 &	smooth    &	          &	1+941 T+8714 p T^2+941 p^3 T^3+p^6 T^4
\tabularnewline[0.5pt]\hline				
66	 &	smooth    &	          &	1+17 T+5746 p T^2+17 p^3 T^3+p^6 T^4
\tabularnewline[0.5pt]\hline				
67	 &	smooth    &	          &	1-438 T+7650 p T^2-438 p^3 T^3+p^6 T^4
\tabularnewline[0.5pt]\hline				
68	 &	smooth    &	          &	1-599 T+4794 p T^2-599 p^3 T^3+p^6 T^4
\tabularnewline[0.5pt]\hline				
69	 &	smooth    &	          &	1-767 T+4962 p T^2-767 p^3 T^3+p^6 T^4
\tabularnewline[0.5pt]\hline				
70	 &	smooth    &	          &	1-130 T-2878 p T^2-130 p^3 T^3+p^6 T^4
\tabularnewline[0.5pt]\hline				
\tablepostamble				
\tablepreamble{73}				
1	 &	smooth    &	          &	1+2 T-990 p T^2+2 p^3 T^3+p^6 T^4
\tabularnewline[0.5pt]\hline				
2	 &	smooth    &	          &	1+576 T+7214 p T^2+576 p^3 T^3+p^6 T^4
\tabularnewline[0.5pt]\hline				
3	 &	smooth$^*$&	3&	  
\tabularnewline[0.5pt]\hline				
4	 &	smooth    &	          &	1+1206 T+10042 p T^2+1206 p^3 T^3+p^6 T^4
\tabularnewline[0.5pt]\hline				
5	 &	smooth    &	          &	1+72 T-5232 p T^2+72 p^3 T^3+p^6 T^4
\tabularnewline[0.5pt]\hline				
6	 &	smooth    &	          &	1-726 T+6178 p T^2-726 p^3 T^3+p^6 T^4
\tabularnewline[0.5pt]\hline				
7	 &	smooth    &	          &	1+121 T+7130 p T^2+121 p^3 T^3+p^6 T^4
\tabularnewline[0.5pt]\hline				
8	 &	smooth    &	          &	1-1118 T+13738 p T^2-1118 p^3 T^3+p^6 T^4
\tabularnewline[0.5pt]\hline				
9	 &	smooth    &	          &	1+450 T-1046 p T^2+450 p^3 T^3+p^6 T^4
\tabularnewline[0.5pt]\hline				
10	 &	smooth    &	          &	1+163 T+6906 p T^2+163 p^3 T^3+p^6 T^4
\tabularnewline[0.5pt]\hline				
11	 &	smooth    &	          &	1+149 T-2586 p T^2+149 p^3 T^3+p^6 T^4
\tabularnewline[0.5pt]\hline				
12	 &	smooth    &	          &	1+58 T+7298 p T^2+58 p^3 T^3+p^6 T^4
\tabularnewline[0.5pt]\hline				
13	 &	smooth    &	          &	1+324 T+4176 p T^2+324 p^3 T^3+p^6 T^4
\tabularnewline[0.5pt]\hline				
14	 &	smooth    &	          &	1+156 T+4036 p T^2+156 p^3 T^3+p^6 T^4
\tabularnewline[0.5pt]\hline				
15	 &	smooth    &	          &	1-40 T-2838 p T^2-40 p^3 T^3+p^6 T^4
\tabularnewline[0.5pt]\hline				
16	 &	smooth    &	          &	1-152 T+270 p T^2-152 p^3 T^3+p^6 T^4
\tabularnewline[0.5pt]\hline				
17	 &	smooth    &	          &	1+317 T+8418 p T^2+317 p^3 T^3+p^6 T^4
\tabularnewline[0.5pt]\hline				
18	 &	smooth    &	          &	1+737 T+8180 p T^2+737 p^3 T^3+p^6 T^4
\tabularnewline[0.5pt]\hline				
19	 &	smooth    &	          &	1-201 T-3916 p T^2-201 p^3 T^3+p^6 T^4
\tabularnewline[0.5pt]\hline				
20	 &	smooth    &	          &	1+1010 T+12954 p T^2+1010 p^3 T^3+p^6 T^4
\tabularnewline[0.5pt]\hline				
21	 &	smooth    &	          &	1+30 T-2348 p T^2+30 p^3 T^3+p^6 T^4
\tabularnewline[0.5pt]\hline				
22	 &	smooth    &	          &	1+1164 T+12044 p T^2+1164 p^3 T^3+p^6 T^4
\tabularnewline[0.5pt]\hline				
23	 &	smooth    &	          &	1-726 T+8698 p T^2-726 p^3 T^3+p^6 T^4
\tabularnewline[0.5pt]\hline				
24	 &	smooth    &	          &	1-621 T+10560 p T^2-621 p^3 T^3+p^6 T^4
\tabularnewline[0.5pt]\hline				
25	 &	smooth    &	          &	1-621 T+2076 p T^2-621 p^3 T^3+p^6 T^4
\tabularnewline[0.5pt]\hline				
26	 &	smooth    &	          &	1-873 T+11022 p T^2-873 p^3 T^3+p^6 T^4
\tabularnewline[0.5pt]\hline				
27	 &	smooth    &	          &	1+79 T+2776 p T^2+79 p^3 T^3+p^6 T^4
\tabularnewline[0.5pt]\hline				
28	 &	smooth    &	          &	1+289 T-22 p^2 T^2+289 p^3 T^3+p^6 T^4
\tabularnewline[0.5pt]\hline				
29	 &	smooth    &	          &	1-334 T+4274 p T^2-334 p^3 T^3+p^6 T^4
\tabularnewline[0.5pt]\hline				
30	 &	smooth    &	          &	1-684 T+8194 p T^2-684 p^3 T^3+p^6 T^4
\tabularnewline[0.5pt]\hline				
31	 &	smooth    &	          &	1-278 T+7620 p T^2-278 p^3 T^3+p^6 T^4
\tabularnewline[0.5pt]\hline				
32	 &	smooth    &	          &	1-313 T-1256 p T^2-313 p^3 T^3+p^6 T^4
\tabularnewline[0.5pt]\hline				
33	 &	smooth    &	          &	1-96 T+2048 p T^2-96 p^3 T^3+p^6 T^4
\tabularnewline[0.5pt]\hline				
34	 &	smooth    &	          &	1+114 T+8852 p T^2+114 p^3 T^3+p^6 T^4
\tabularnewline[0.5pt]\hline				
35	 &	smooth    &	          &	1-530 T+5058 p T^2-530 p^3 T^3+p^6 T^4
\tabularnewline[0.5pt]\hline				
36	 &	smooth    &	          &	1+674 T+3266 p T^2+674 p^3 T^3+p^6 T^4
\tabularnewline[0.5pt]\hline				
37	 &	smooth    &	          &	1+324 T+3910 p T^2+324 p^3 T^3+p^6 T^4
\tabularnewline[0.5pt]\hline				
38	 &	smooth    &	          &	1-348 T+6038 p T^2-348 p^3 T^3+p^6 T^4
\tabularnewline[0.5pt]\hline				
39	 &	smooth    &	          &	1+135 T+3938 p T^2+135 p^3 T^3+p^6 T^4
\tabularnewline[0.5pt]\hline				
40	 &	smooth    &	          &	1+163 T-7374 p T^2+163 p^3 T^3+p^6 T^4
\tabularnewline[0.5pt]\hline				
41	 &	smooth    &	          &	1+478 T+90 p^2 T^2+478 p^3 T^3+p^6 T^4
\tabularnewline[0.5pt]\hline				
42	 &	smooth    &	          &	1-264 T-2040 p T^2-264 p^3 T^3+p^6 T^4
\tabularnewline[0.5pt]\hline				
43	 &	smooth    &	          &	1+338 T-3286 p T^2+338 p^3 T^3+p^6 T^4
\tabularnewline[0.5pt]\hline				
44	 &	smooth    &	          &	1+10 p T+5730 p T^2+10 p^4 T^3+p^6 T^4
\tabularnewline[0.5pt]\hline				
45	 &	smooth    &	          &	1-320 T+3196 p T^2-320 p^3 T^3+p^6 T^4
\tabularnewline[0.5pt]\hline				
46	 &	smooth    &	          &	1+1066 T+13850 p T^2+1066 p^3 T^3+p^6 T^4
\tabularnewline[0.5pt]\hline				
47	 &	smooth    &	          &	1-649 T+90 p^2 T^2-649 p^3 T^3+p^6 T^4
\tabularnewline[0.5pt]\hline				
48	 &	smooth    &	          &	1+1262 T+13458 p T^2+1262 p^3 T^3+p^6 T^4
\tabularnewline[0.5pt]\hline				
49	 &	smooth    &	          &	1-180 T-6618 p T^2-180 p^3 T^3+p^6 T^4
\tabularnewline[0.5pt]\hline				
50	 &	smooth    &	          &	1+513 T+3336 p T^2+513 p^3 T^3+p^6 T^4
\tabularnewline[0.5pt]\hline				
51	 &	smooth    &	          &	1+282 T+5674 p T^2+282 p^3 T^3+p^6 T^4
\tabularnewline[0.5pt]\hline				
52	 &	smooth    &	          &	1+184 T-4854 p T^2+184 p^3 T^3+p^6 T^4
\tabularnewline[0.5pt]\hline				
53	 &	smooth    &	          &	1+1248 T+13038 p T^2+1248 p^3 T^3+p^6 T^4
\tabularnewline[0.5pt]\hline				
54	 &	smooth    &	          &	1-26 T+410 p T^2-26 p^3 T^3+p^6 T^4
\tabularnewline[0.5pt]\hline				
55	 &	smooth    &	          &	1-397 T+6864 p T^2-397 p^3 T^3+p^6 T^4
\tabularnewline[0.5pt]\hline				
56	 &	smooth    &	          &	1-488 T+7788 p T^2-488 p^3 T^3+p^6 T^4
\tabularnewline[0.5pt]\hline				
57	 &	smooth    &	          &	1-11 p T+7452 p T^2-11 p^4 T^3+p^6 T^4
\tabularnewline[0.5pt]\hline				
58	 &	smooth    &	          &	1-313 T+10630 p T^2-313 p^3 T^3+p^6 T^4
\tabularnewline[0.5pt]\hline				
59	 &	smooth    &	          &	1+10 p T+5534 p T^2+10 p^4 T^3+p^6 T^4
\tabularnewline[0.5pt]\hline				
60	 &	smooth    &	          &	1+51 T-4098 p T^2+51 p^3 T^3+p^6 T^4
\tabularnewline[0.5pt]\hline				
61	 &	smooth    &	          &	1-376 T-3986 p T^2-376 p^3 T^3+p^6 T^4
\tabularnewline[0.5pt]\hline				
62	 &	smooth    &	          &	1+79 T+2594 p T^2+79 p^3 T^3+p^6 T^4
\tabularnewline[0.5pt]\hline				
63	 &	smooth    &	          &	1+394 T+4890 p T^2+394 p^3 T^3+p^6 T^4
\tabularnewline[0.5pt]\hline				
64	 &	smooth    &	          &	1+828 T+4750 p T^2+828 p^3 T^3+p^6 T^4
\tabularnewline[0.5pt]\hline				
65	 &	smooth    &	          &	1-705 T+7480 p T^2-705 p^3 T^3+p^6 T^4
\tabularnewline[0.5pt]\hline				
66	 &	smooth    &	          &	1-4 p T+536 p T^2-4 p^4 T^3+p^6 T^4
\tabularnewline[0.5pt]\hline				
67	 &	smooth    &	          &	1+1045 T+7900 p T^2+1045 p^3 T^3+p^6 T^4
\tabularnewline[0.5pt]\hline				
68	 &	smooth    &	          &	1-460 T+11218 p T^2-460 p^3 T^3+p^6 T^4
\tabularnewline[0.5pt]\hline				
69	 &	smooth    &	          &	1+79 T+5156 p T^2+79 p^3 T^3+p^6 T^4
\tabularnewline[0.5pt]\hline				
70	 &	smooth    &	          &	1-110 T+3322 p T^2-110 p^3 T^3+p^6 T^4
\tabularnewline[0.5pt]\hline				
71	 &	smooth    &	          &	1-173 T-2292 p T^2-173 p^3 T^3+p^6 T^4
\tabularnewline[0.5pt]\hline				
72	 &	smooth    &	          &	1+226 T+1418 p T^2+226 p^3 T^3+p^6 T^4
\tabularnewline[0.5pt]\hline				
\tablepostamble				
\tablepreamble{79}				
1	 &	smooth    &	          &	1-246 T+490 p T^2-246 p^3 T^3+p^6 T^4
\tabularnewline[0.5pt]\hline				
2	 &	smooth    &	          &	1+468 T+1442 p T^2+468 p^3 T^3+p^6 T^4
\tabularnewline[0.5pt]\hline				
3	 &	smooth$^*$&	3&	  
\tabularnewline[0.5pt]\hline				
4	 &	smooth    &	          &	1-687 T+5586 p T^2-687 p^3 T^3+p^6 T^4
\tabularnewline[0.5pt]\hline				
5	 &	smooth    &	          &	1+769 T+9814 p T^2+769 p^3 T^3+p^6 T^4
\tabularnewline[0.5pt]\hline				
6	 &	smooth    &	          &	1+34 T+1386 p T^2+34 p^3 T^3+p^6 T^4
\tabularnewline[0.5pt]\hline				
7	 &	smooth    &	          &	1-197 T+5978 p T^2-197 p^3 T^3+p^6 T^4
\tabularnewline[0.5pt]\hline				
8	 &	smooth    &	          &	1+90 T-2590 p T^2+90 p^3 T^3+p^6 T^4
\tabularnewline[0.5pt]\hline				
9	 &	smooth    &	          &	1-288 T+6216 p T^2-288 p^3 T^3+p^6 T^4
\tabularnewline[0.5pt]\hline				
10	 &	smooth    &	          &	1+125 T+12222 p T^2+125 p^3 T^3+p^6 T^4
\tabularnewline[0.5pt]\hline				
11	 &	smooth    &	          &	1+713 T+2422 p T^2+713 p^3 T^3+p^6 T^4
\tabularnewline[0.5pt]\hline				
12	 &	smooth    &	          &	1+594 T+2002 p T^2+594 p^3 T^3+p^6 T^4
\tabularnewline[0.5pt]\hline				
13	 &	smooth    &	          &	1+433 T+3486 p T^2+433 p^3 T^3+p^6 T^4
\tabularnewline[0.5pt]\hline				
14	 &	smooth    &	          &	1+335 T+2898 p T^2+335 p^3 T^3+p^6 T^4
\tabularnewline[0.5pt]\hline				
15	 &	smooth    &	          &	1+1000 T+9730 p T^2+1000 p^3 T^3+p^6 T^4
\tabularnewline[0.5pt]\hline				
16	 &	smooth    &	          &	1+69 T+2282 p T^2+69 p^3 T^3+p^6 T^4
\tabularnewline[0.5pt]\hline				
17	 &	smooth    &	          &	1-498 T+8974 p T^2-498 p^3 T^3+p^6 T^4
\tabularnewline[0.5pt]\hline				
18	 &	smooth    &	          &	1+517 T+4186 p T^2+517 p^3 T^3+p^6 T^4
\tabularnewline[0.5pt]\hline				
19	 &	smooth    &	          &	1-379 T+9394 p T^2-379 p^3 T^3+p^6 T^4
\tabularnewline[0.5pt]\hline				
20	 &	smooth    &	          &	1-1513 T+17878 p T^2-1513 p^3 T^3+p^6 T^4
\tabularnewline[0.5pt]\hline				
21	 &	smooth    &	          &	1-267 T-1302 p T^2-267 p^3 T^3+p^6 T^4
\tabularnewline[0.5pt]\hline				
22	 &	smooth    &	          &	1-568 T+8162 p T^2-568 p^3 T^3+p^6 T^4
\tabularnewline[0.5pt]\hline				
23	 &	smooth    &	          &	1+1819 T+21602 p T^2+1819 p^3 T^3+p^6 T^4
\tabularnewline[0.5pt]\hline				
24	 &	smooth    &	          &	(1-10 p T+p^3 T^2)(1+1272 T+p^3 T^2)
\tabularnewline[0.5pt]\hline				
25	 &	smooth    &	          &	1+314 T+10220 p T^2+314 p^3 T^3+p^6 T^4
\tabularnewline[0.5pt]\hline				
26	 &	smooth    &	          &	1+286 T+2506 p T^2+286 p^3 T^3+p^6 T^4
\tabularnewline[0.5pt]\hline				
27	 &	smooth    &	          &	1+76 T-126 p T^2+76 p^3 T^3+p^6 T^4
\tabularnewline[0.5pt]\hline				
28	 &	smooth    &	          &	1-197 T-6958 p T^2-197 p^3 T^3+p^6 T^4
\tabularnewline[0.5pt]\hline				
29	 &	smooth    &	          &	1+97 T-3822 p T^2+97 p^3 T^3+p^6 T^4
\tabularnewline[0.5pt]\hline				
30	 &	smooth    &	          &	1-232 T-210 p T^2-232 p^3 T^3+p^6 T^4
\tabularnewline[0.5pt]\hline				
31	 &	smooth    &	          &	1+692 T+6706 p T^2+692 p^3 T^3+p^6 T^4
\tabularnewline[0.5pt]\hline				
32	 &	smooth    &	          &	1-78 T+1204 p T^2-78 p^3 T^3+p^6 T^4
\tabularnewline[0.5pt]\hline				
33	 &	smooth    &	          &	1+216 T+5810 p T^2+216 p^3 T^3+p^6 T^4
\tabularnewline[0.5pt]\hline				
34	 &	smooth    &	          &	1-1548 T+16786 p T^2-1548 p^3 T^3+p^6 T^4
\tabularnewline[0.5pt]\hline				
35	 &	smooth    &	          &	1-1156 T+9730 p T^2-1156 p^3 T^3+p^6 T^4
\tabularnewline[0.5pt]\hline				
36	 &	smooth    &	          &	1+762 T+11242 p T^2+762 p^3 T^3+p^6 T^4
\tabularnewline[0.5pt]\hline				
37	 &	smooth    &	          &	1-1646 T+19922 p T^2-1646 p^3 T^3+p^6 T^4
\tabularnewline[0.5pt]\hline				
38	 &	smooth    &	          &	1-1506 T+15862 p T^2-1506 p^3 T^3+p^6 T^4
\tabularnewline[0.5pt]\hline				
39	 &	smooth    &	          &	1+76 T+6342 p T^2+76 p^3 T^3+p^6 T^4
\tabularnewline[0.5pt]\hline				
40	 &	smooth    &	          &	1-85 T+2142 p T^2-85 p^3 T^3+p^6 T^4
\tabularnewline[0.5pt]\hline				
41	 &	smooth    &	          &	1-694 T+7014 p T^2-694 p^3 T^3+p^6 T^4
\tabularnewline[0.5pt]\hline				
42	 &	smooth    &	          &	1-302 T-1806 p T^2-302 p^3 T^3+p^6 T^4
\tabularnewline[0.5pt]\hline				
43	 &	smooth    &	          &	1+622 T+9030 p T^2+622 p^3 T^3+p^6 T^4
\tabularnewline[0.5pt]\hline				
44	 &	smooth    &	          &	1+664 T+10164 p T^2+664 p^3 T^3+p^6 T^4
\tabularnewline[0.5pt]\hline				
45	 &	smooth    &	          &	1+762 T+6734 p T^2+762 p^3 T^3+p^6 T^4
\tabularnewline[0.5pt]\hline				
46	 &	smooth    &	          &	1-393 T-98 p T^2-393 p^3 T^3+p^6 T^4
\tabularnewline[0.5pt]\hline				
47	 &	smooth    &	          &	1+979 T+12642 p T^2+979 p^3 T^3+p^6 T^4
\tabularnewline[0.5pt]\hline				
48	 &	smooth    &	          &	1-932 T+8330 p T^2-932 p^3 T^3+p^6 T^4
\tabularnewline[0.5pt]\hline				
49	 &	smooth    &	          &	1+1049 T+7966 p T^2+1049 p^3 T^3+p^6 T^4
\tabularnewline[0.5pt]\hline				
50	 &	smooth    &	          &	1-22 T+3206 p T^2-22 p^3 T^3+p^6 T^4
\tabularnewline[0.5pt]\hline				
51	 &	smooth    &	          &	1+1665 T+21070 p T^2+1665 p^3 T^3+p^6 T^4
\tabularnewline[0.5pt]\hline				
52	 &	smooth    &	          &	1+132 T-6062 p T^2+132 p^3 T^3+p^6 T^4
\tabularnewline[0.5pt]\hline				
53	 &	smooth    &	          &	1+272 T+8498 p T^2+272 p^3 T^3+p^6 T^4
\tabularnewline[0.5pt]\hline				
54	 &	smooth    &	          &	1+146 T-6762 p T^2+146 p^3 T^3+p^6 T^4
\tabularnewline[0.5pt]\hline				
55	 &	smooth    &	          &	1-631 T+6510 p T^2-631 p^3 T^3+p^6 T^4
\tabularnewline[0.5pt]\hline				
56	 &	smooth    &	          &	1+104 T-5054 p T^2+104 p^3 T^3+p^6 T^4
\tabularnewline[0.5pt]\hline				
57	 &	smooth    &	          &	1-1044 T+8442 p T^2-1044 p^3 T^3+p^6 T^4
\tabularnewline[0.5pt]\hline				
58	 &	smooth    &	          &	1+692 T+8274 p T^2+692 p^3 T^3+p^6 T^4
\tabularnewline[0.5pt]\hline				
59	 &	smooth    &	          &	1+20 T+6202 p T^2+20 p^3 T^3+p^6 T^4
\tabularnewline[0.5pt]\hline				
60	 &	smooth    &	          &	1-1485 T+14322 p T^2-1485 p^3 T^3+p^6 T^4
\tabularnewline[0.5pt]\hline				
61	 &	smooth    &	          &	1+1112 T+11578 p T^2+1112 p^3 T^3+p^6 T^4
\tabularnewline[0.5pt]\hline				
62	 &	smooth    &	          &	1-414 T+5852 p T^2-414 p^3 T^3+p^6 T^4
\tabularnewline[0.5pt]\hline				
63	 &	smooth    &	          &	1+48 T+5390 p T^2+48 p^3 T^3+p^6 T^4
\tabularnewline[0.5pt]\hline				
64	 &	smooth    &	          &	1-T+4214 p T^2-p^3 T^3+p^6 T^4
\tabularnewline[0.5pt]\hline				
65	 &	smooth    &	          &	1-64 T+5698 p T^2-64 p^3 T^3+p^6 T^4
\tabularnewline[0.5pt]\hline				
66	 &	smooth    &	          &	1+1098 T+13454 p T^2+1098 p^3 T^3+p^6 T^4
\tabularnewline[0.5pt]\hline				
67	 &	smooth    &	          &	1+1203 T+9086 p T^2+1203 p^3 T^3+p^6 T^4
\tabularnewline[0.5pt]\hline				
68	 &	smooth    &	          &	1+867 T+10794 p T^2+867 p^3 T^3+p^6 T^4
\tabularnewline[0.5pt]\hline				
69	 &	smooth    &	          &	1+1224 T+15974 p T^2+1224 p^3 T^3+p^6 T^4
\tabularnewline[0.5pt]\hline				
70	 &	smooth    &	          &	1-631 T+11018 p T^2-631 p^3 T^3+p^6 T^4
\tabularnewline[0.5pt]\hline				
71	 &	smooth    &	          &	1-806 T+7714 p T^2-806 p^3 T^3+p^6 T^4
\tabularnewline[0.5pt]\hline				
72	 &	smooth    &	          &	1+804 T+5712 p T^2+804 p^3 T^3+p^6 T^4
\tabularnewline[0.5pt]\hline				
73	 &	smooth    &	          &	1-477 T+3906 p T^2-477 p^3 T^3+p^6 T^4
\tabularnewline[0.5pt]\hline				
74	 &	smooth    &	          &	1+433 T+546 p T^2+433 p^3 T^3+p^6 T^4
\tabularnewline[0.5pt]\hline				
75	 &	smooth    &	          &	1+167 T+3066 p T^2+167 p^3 T^3+p^6 T^4
\tabularnewline[0.5pt]\hline				
76	 &	smooth    &	          &	1+699 T+9982 p T^2+699 p^3 T^3+p^6 T^4
\tabularnewline[0.5pt]\hline				
77	 &	smooth    &	          &	1+146 T-686 p T^2+146 p^3 T^3+p^6 T^4
\tabularnewline[0.5pt]\hline				
78	 &	smooth    &	          &	1-85 T+3906 p T^2-85 p^3 T^3+p^6 T^4
\tabularnewline[0.5pt]\hline				
\tablepostamble				
\tablepreamble{83}				
1	 &	smooth    &	          &	1-231 T+3194 p T^2-231 p^3 T^3+p^6 T^4
\tabularnewline[0.5pt]\hline				
2	 &	smooth    &	          &	1+259 T+58 p T^2+259 p^3 T^3+p^6 T^4
\tabularnewline[0.5pt]\hline				
3	 &	smooth$^*$&	3&	  
\tabularnewline[0.5pt]\hline				
4	 &	smooth    &	          &	1+406 T-4940 p T^2+406 p^3 T^3+p^6 T^4
\tabularnewline[0.5pt]\hline				
5	 &	smooth    &	          &	1+1092 T+15542 p T^2+1092 p^3 T^3+p^6 T^4
\tabularnewline[0.5pt]\hline				
6	 &	smooth    &	          &	1-364 T+10446 p T^2-364 p^3 T^3+p^6 T^4
\tabularnewline[0.5pt]\hline				
7	 &	smooth    &	          &	1+189 T+5938 p T^2+189 p^3 T^3+p^6 T^4
\tabularnewline[0.5pt]\hline				
8	 &	smooth    &	          &	1+182 T-334 p T^2+182 p^3 T^3+p^6 T^4
\tabularnewline[0.5pt]\hline				
9	 &	smooth    &	          &	1-98 T+2410 p T^2-98 p^3 T^3+p^6 T^4
\tabularnewline[0.5pt]\hline				
10	 &	singular  &	10&	(1-p T) (1+1148 T+p^3 T^2)
\tabularnewline[0.5pt]\hline				
11	 &	smooth    &	          &	1+7 p T+3978 p T^2+7 p^4 T^3+p^6 T^4
\tabularnewline[0.5pt]\hline				
12	 &	smooth    &	          &	(1+14 p T+p^3 T^2)(1-56 T+p^3 T^2)
\tabularnewline[0.5pt]\hline				
13	 &	smooth    &	          &	1+196 T+2606 p T^2+196 p^3 T^3+p^6 T^4
\tabularnewline[0.5pt]\hline				
14	 &	singular  &	14&	(1-p T) (1-1372 T+p^3 T^2)
\tabularnewline[0.5pt]\hline				
15	 &	smooth    &	          &	1-168 T+8486 p T^2-168 p^3 T^3+p^6 T^4
\tabularnewline[0.5pt]\hline				
16	 &	singular  &	16&	(1-p T) (1-84 T+p^3 T^2)
\tabularnewline[0.5pt]\hline				
17	 &	smooth    &	          &	1+182 T+10348 p T^2+182 p^3 T^3+p^6 T^4
\tabularnewline[0.5pt]\hline				
18	 &	smooth    &	          &	1+952 T+4174 p T^2+952 p^3 T^3+p^6 T^4
\tabularnewline[0.5pt]\hline				
19	 &	smooth    &	          &	1-91 T+10642 p T^2-91 p^3 T^3+p^6 T^4
\tabularnewline[0.5pt]\hline				
20	 &	smooth    &	          &	(1+7 p T+p^3 T^2)(1-448 T+p^3 T^2)
\tabularnewline[0.5pt]\hline				
21	 &	smooth    &	          &	1-637 T+9858 p T^2-637 p^3 T^3+p^6 T^4
\tabularnewline[0.5pt]\hline				
22	 &	smooth    &	          &	1-420 T+12014 p T^2-420 p^3 T^3+p^6 T^4
\tabularnewline[0.5pt]\hline				
23	 &	smooth    &	          &	1-350 T+5350 p T^2-350 p^3 T^3+p^6 T^4
\tabularnewline[0.5pt]\hline				
24	 &	smooth    &	          &	1+1274 T+18286 p T^2+1274 p^3 T^3+p^6 T^4
\tabularnewline[0.5pt]\hline				
25	 &	smooth    &	          &	1+21 T-7390 p T^2+21 p^3 T^3+p^6 T^4
\tabularnewline[0.5pt]\hline				
26	 &	smooth    &	          &	1-770 T+9956 p T^2-770 p^3 T^3+p^6 T^4
\tabularnewline[0.5pt]\hline				
27	 &	smooth    &	          &	1+189 T+842 p T^2+189 p^3 T^3+p^6 T^4
\tabularnewline[0.5pt]\hline				
28	 &	smooth    &	          &	1-798 T+11034 p T^2-798 p^3 T^3+p^6 T^4
\tabularnewline[0.5pt]\hline				
29	 &	smooth    &	          &	1-91 T-9154 p T^2-91 p^3 T^3+p^6 T^4
\tabularnewline[0.5pt]\hline				
30	 &	smooth    &	          &	1+140 T+11818 p T^2+140 p^3 T^3+p^6 T^4
\tabularnewline[0.5pt]\hline				
31	 &	smooth    &	          &	1-350 T+10250 p T^2-350 p^3 T^3+p^6 T^4
\tabularnewline[0.5pt]\hline				
32	 &	smooth    &	          &	1-266 T+12014 p T^2-266 p^3 T^3+p^6 T^4
\tabularnewline[0.5pt]\hline				
33	 &	smooth    &	          &	1-602 T+9858 p T^2-602 p^3 T^3+p^6 T^4
\tabularnewline[0.5pt]\hline				
34	 &	smooth    &	          &	1+259 T+12210 p T^2+259 p^3 T^3+p^6 T^4
\tabularnewline[0.5pt]\hline				
35	 &	smooth    &	          &	1+231 T+5546 p T^2+231 p^3 T^3+p^6 T^4
\tabularnewline[0.5pt]\hline				
36	 &	smooth    &	          &	1-1134 T+10250 p T^2-1134 p^3 T^3+p^6 T^4
\tabularnewline[0.5pt]\hline				
37	 &	smooth    &	          &	1+210 T-3960 p T^2+210 p^3 T^3+p^6 T^4
\tabularnewline[0.5pt]\hline				
38	 &	smooth    &	          &	1-1554 T+14954 p T^2-1554 p^3 T^3+p^6 T^4
\tabularnewline[0.5pt]\hline				
39	 &	smooth    &	          &	1-112 T+7310 p T^2-112 p^3 T^3+p^6 T^4
\tabularnewline[0.5pt]\hline				
40	 &	smooth    &	          &	1+707 T+3782 p T^2+707 p^3 T^3+p^6 T^4
\tabularnewline[0.5pt]\hline				
41	 &	smooth    &	          &	1+70 T+10348 p T^2+70 p^3 T^3+p^6 T^4
\tabularnewline[0.5pt]\hline				
42	 &	smooth    &	          &	1-308 T+4958 p T^2-308 p^3 T^3+p^6 T^4
\tabularnewline[0.5pt]\hline				
43	 &	smooth    &	          &	1-980 T+5742 p T^2-980 p^3 T^3+p^6 T^4
\tabularnewline[0.5pt]\hline				
44	 &	smooth    &	          &	1+343 T+1430 p T^2+343 p^3 T^3+p^6 T^4
\tabularnewline[0.5pt]\hline				
45	 &	smooth    &	          &	1+441 T+11818 p T^2+441 p^3 T^3+p^6 T^4
\tabularnewline[0.5pt]\hline				
46	 &	smooth    &	          &	1-784 T+4370 p T^2-784 p^3 T^3+p^6 T^4
\tabularnewline[0.5pt]\hline				
47	 &	smooth    &	          &	1+714 T+13582 p T^2+714 p^3 T^3+p^6 T^4
\tabularnewline[0.5pt]\hline				
48	 &	smooth    &	          &	1+1274 T+12504 p T^2+1274 p^3 T^3+p^6 T^4
\tabularnewline[0.5pt]\hline				
49	 &	smooth    &	          &	1-98 T+7800 p T^2-98 p^3 T^3+p^6 T^4
\tabularnewline[0.5pt]\hline				
50	 &	smooth    &	          &	1+1036 T+7310 p T^2+1036 p^3 T^3+p^6 T^4
\tabularnewline[0.5pt]\hline				
51	 &	smooth    &	          &	1-847 T+9074 p T^2-847 p^3 T^3+p^6 T^4
\tabularnewline[0.5pt]\hline				
52	 &	smooth    &	          &	1+245 T+3586 p T^2+245 p^3 T^3+p^6 T^4
\tabularnewline[0.5pt]\hline				
53	 &	smooth    &	          &	1+1526 T+20050 p T^2+1526 p^3 T^3+p^6 T^4
\tabularnewline[0.5pt]\hline				
54	 &	smooth    &	          &	1+266 T-1314 p T^2+266 p^3 T^3+p^6 T^4
\tabularnewline[0.5pt]\hline				
55	 &	smooth    &	          &	1-252 T+12014 p T^2-252 p^3 T^3+p^6 T^4
\tabularnewline[0.5pt]\hline				
56	 &	smooth    &	          &	(1+14 p T+p^3 T^2)(1+546 T+p^3 T^2)
\tabularnewline[0.5pt]\hline				
57	 &	smooth    &	          &	1+1960 T+23774 p T^2+1960 p^3 T^3+p^6 T^4
\tabularnewline[0.5pt]\hline				
58	 &	smooth    &	          &	1+217 T-2686 p T^2+217 p^3 T^3+p^6 T^4
\tabularnewline[0.5pt]\hline				
59	 &	smooth    &	          &	1-1820 T+21618 p T^2-1820 p^3 T^3+p^6 T^4
\tabularnewline[0.5pt]\hline				
60	 &	smooth    &	          &	1+14 p T+14758 p T^2+14 p^4 T^3+p^6 T^4
\tabularnewline[0.5pt]\hline				
61	 &	smooth    &	          &	1-14 T+12504 p T^2-14 p^3 T^3+p^6 T^4
\tabularnewline[0.5pt]\hline				
62	 &	smooth    &	          &	1-882 T+10446 p T^2-882 p^3 T^3+p^6 T^4
\tabularnewline[0.5pt]\hline				
63	 &	smooth    &	          &	1+168 T+1920 p T^2+168 p^3 T^3+p^6 T^4
\tabularnewline[0.5pt]\hline				
64	 &	smooth    &	          &	1-224 T+2802 p T^2-224 p^3 T^3+p^6 T^4
\tabularnewline[0.5pt]\hline				
65	 &	smooth    &	          &	1+805 T+5742 p T^2+805 p^3 T^3+p^6 T^4
\tabularnewline[0.5pt]\hline				
66	 &	smooth    &	          &	1-1092 T+14954 p T^2-1092 p^3 T^3+p^6 T^4
\tabularnewline[0.5pt]\hline				
67	 &	smooth    &	          &	1+1764 T+19070 p T^2+1764 p^3 T^3+p^6 T^4
\tabularnewline[0.5pt]\hline				
68	 &	smooth    &	          &	1-175 T+2998 p T^2-175 p^3 T^3+p^6 T^4
\tabularnewline[0.5pt]\hline				
69	 &	smooth    &	          &	1+770 T+7408 p T^2+770 p^3 T^3+p^6 T^4
\tabularnewline[0.5pt]\hline				
70	 &	smooth    &	          &	1+1225 T+11818 p T^2+1225 p^3 T^3+p^6 T^4
\tabularnewline[0.5pt]\hline				
71	 &	smooth    &	          &	1-70 T-4450 p T^2-70 p^3 T^3+p^6 T^4
\tabularnewline[0.5pt]\hline				
72	 &	smooth    &	          &	1+1001 T+13386 p T^2+1001 p^3 T^3+p^6 T^4
\tabularnewline[0.5pt]\hline				
73	 &	smooth    &	          &	1-1064 T+14170 p T^2-1064 p^3 T^3+p^6 T^4
\tabularnewline[0.5pt]\hline				
74	 &	smooth    &	          &	1+644 T+9858 p T^2+644 p^3 T^3+p^6 T^4
\tabularnewline[0.5pt]\hline				
75	 &	smooth    &	          &	1-378 T-4254 p T^2-378 p^3 T^3+p^6 T^4
\tabularnewline[0.5pt]\hline				
76	 &	smooth    &	          &	1-252 T-3862 p T^2-252 p^3 T^3+p^6 T^4
\tabularnewline[0.5pt]\hline				
77	 &	smooth    &	          &	1-966 T+11426 p T^2-966 p^3 T^3+p^6 T^4
\tabularnewline[0.5pt]\hline				
78	 &	smooth    &	          &	1-1015 T+12406 p T^2-1015 p^3 T^3+p^6 T^4
\tabularnewline[0.5pt]\hline				
79	 &	smooth    &	          &	1+798 T+1430 p T^2+798 p^3 T^3+p^6 T^4
\tabularnewline[0.5pt]\hline				
80	 &	smooth    &	          &	1-35 T+7898 p T^2-35 p^3 T^3+p^6 T^4
\tabularnewline[0.5pt]\hline				
81	 &	smooth    &	          &	1+238 T+9760 p T^2+238 p^3 T^3+p^6 T^4
\tabularnewline[0.5pt]\hline				
82	 &	smooth    &	          &	1+336 T+4958 p T^2+336 p^3 T^3+p^6 T^4
\tabularnewline[0.5pt]\hline				
\tablepostamble				
\tablepreamble{89}				
1	 &	smooth    &	          &	1+796 T+8198 p T^2+796 p^3 T^3+p^6 T^4
\tabularnewline[0.5pt]\hline				
2	 &	smooth    &	          &	1-1262 T+11810 p T^2-1262 p^3 T^3+p^6 T^4
\tabularnewline[0.5pt]\hline				
3	 &	smooth$^*$&	3&	  
\tabularnewline[0.5pt]\hline				
4	 &	smooth    &	          &	1+1720 T+19398 p T^2+1720 p^3 T^3+p^6 T^4
\tabularnewline[0.5pt]\hline				
5	 &	smooth    &	          &	1-1850 T+22842 p T^2-1850 p^3 T^3+p^6 T^4
\tabularnewline[0.5pt]\hline				
6	 &	smooth    &	          &	1-464 T+7946 p T^2-464 p^3 T^3+p^6 T^4
\tabularnewline[0.5pt]\hline				
7	 &	smooth    &	          &	1+460 T+12692 p T^2+460 p^3 T^3+p^6 T^4
\tabularnewline[0.5pt]\hline				
8	 &	smooth    &	          &	1+313 T+10536 p T^2+313 p^3 T^3+p^6 T^4
\tabularnewline[0.5pt]\hline				
9	 &	smooth    &	          &	1-170 T+8002 p T^2-170 p^3 T^3+p^6 T^4
\tabularnewline[0.5pt]\hline				
10	 &	smooth    &	          &	1-975 T+18404 p T^2-975 p^3 T^3+p^6 T^4
\tabularnewline[0.5pt]\hline				
11	 &	smooth    &	          &	1+404 T+5370 p T^2+404 p^3 T^3+p^6 T^4
\tabularnewline[0.5pt]\hline				
12	 &	smooth    &	          &	1+376 T+10074 p T^2+376 p^3 T^3+p^6 T^4
\tabularnewline[0.5pt]\hline				
13	 &	smooth    &	          &	1+173 T+9038 p T^2+173 p^3 T^3+p^6 T^4
\tabularnewline[0.5pt]\hline				
14	 &	smooth    &	          &	1+306 T+1870 p T^2+306 p^3 T^3+p^6 T^4
\tabularnewline[0.5pt]\hline				
15	 &	smooth    &	          &	1+544 T+11628 p T^2+544 p^3 T^3+p^6 T^4
\tabularnewline[0.5pt]\hline				
16	 &	smooth    &	          &	1+320 T+13406 p T^2+320 p^3 T^3+p^6 T^4
\tabularnewline[0.5pt]\hline				
17	 &	smooth    &	          &	1+1545 T+14568 p T^2+1545 p^3 T^3+p^6 T^4
\tabularnewline[0.5pt]\hline				
18	 &	smooth    &	          &	1-1654 T+20994 p T^2-1654 p^3 T^3+p^6 T^4
\tabularnewline[0.5pt]\hline				
19	 &	smooth    &	          &	1-1402 T+18684 p T^2-1402 p^3 T^3+p^6 T^4
\tabularnewline[0.5pt]\hline				
20	 &	smooth    &	          &	1-912 T+10942 p T^2-912 p^3 T^3+p^6 T^4
\tabularnewline[0.5pt]\hline				
21	 &	smooth    &	          &	1-380 T+13574 p T^2-380 p^3 T^3+p^6 T^4
\tabularnewline[0.5pt]\hline				
22	 &	smooth    &	          &	1+922 T+3914 p T^2+922 p^3 T^3+p^6 T^4
\tabularnewline[0.5pt]\hline				
23	 &	smooth    &	          &	1+439 T+6966 p T^2+439 p^3 T^3+p^6 T^4
\tabularnewline[0.5pt]\hline				
24	 &	smooth    &	          &	1+68 T+6322 p T^2+68 p^3 T^3+p^6 T^4
\tabularnewline[0.5pt]\hline				
25	 &	smooth    &	          &	1-478 T+4306 p T^2-478 p^3 T^3+p^6 T^4
\tabularnewline[0.5pt]\hline				
26	 &	smooth    &	          &	1+362 T+12062 p T^2+362 p^3 T^3+p^6 T^4
\tabularnewline[0.5pt]\hline				
27	 &	smooth    &	          &	1+110 T-76 p T^2+110 p^3 T^3+p^6 T^4
\tabularnewline[0.5pt]\hline				
28	 &	smooth    &	          &	1+432 T-6236 p T^2+432 p^3 T^3+p^6 T^4
\tabularnewline[0.5pt]\hline				
29	 &	smooth    &	          &	1+530 T-3646 p T^2+530 p^3 T^3+p^6 T^4
\tabularnewline[0.5pt]\hline				
30	 &	smooth    &	          &	1+1230 T+9542 p T^2+1230 p^3 T^3+p^6 T^4
\tabularnewline[0.5pt]\hline				
31	 &	smooth    &	          &	1+208 T+4950 p T^2+208 p^3 T^3+p^6 T^4
\tabularnewline[0.5pt]\hline				
32	 &	smooth    &	          &	1-268 T-4738 p T^2-268 p^3 T^3+p^6 T^4
\tabularnewline[0.5pt]\hline				
33	 &	smooth    &	          &	1+327 T+3998 p T^2+327 p^3 T^3+p^6 T^4
\tabularnewline[0.5pt]\hline				
34	 &	smooth    &	          &	1-485 T+5552 p T^2-485 p^3 T^3+p^6 T^4
\tabularnewline[0.5pt]\hline				
35	 &	smooth    &	          &	1-58 T-9190 p T^2-58 p^3 T^3+p^6 T^4
\tabularnewline[0.5pt]\hline				
36	 &	smooth    &	          &	1+12 T-3674 p T^2+12 p^3 T^3+p^6 T^4
\tabularnewline[0.5pt]\hline				
37	 &	smooth    &	          &	1+1412 T+15170 p T^2+1412 p^3 T^3+p^6 T^4
\tabularnewline[0.5pt]\hline				
38	 &	smooth    &	          &	1-800 T+1912 p T^2-800 p^3 T^3+p^6 T^4
\tabularnewline[0.5pt]\hline				
39	 &	smooth    &	          &	1-212 T-7510 p T^2-212 p^3 T^3+p^6 T^4
\tabularnewline[0.5pt]\hline				
40	 &	smooth    &	          &	1+572 T-2106 p T^2+572 p^3 T^3+p^6 T^4
\tabularnewline[0.5pt]\hline				
41	 &	smooth    &	          &	1-1549 T+16318 p T^2-1549 p^3 T^3+p^6 T^4
\tabularnewline[0.5pt]\hline				
42	 &	smooth    &	          &	1-86 T-4542 p T^2-86 p^3 T^3+p^6 T^4
\tabularnewline[0.5pt]\hline				
43	 &	smooth    &	          &	1-569 T+246 p T^2-569 p^3 T^3+p^6 T^4
\tabularnewline[0.5pt]\hline				
44	 &	smooth    &	          &	1+922 T+13602 p T^2+922 p^3 T^3+p^6 T^4
\tabularnewline[0.5pt]\hline				
45	 &	smooth    &	          &	1-1283 T+11824 p T^2-1283 p^3 T^3+p^6 T^4
\tabularnewline[0.5pt]\hline				
46	 &	smooth    &	          &	1-142 T+4908 p T^2-142 p^3 T^3+p^6 T^4
\tabularnewline[0.5pt]\hline				
47	 &	smooth    &	          &	1+866 T+3522 p T^2+866 p^3 T^3+p^6 T^4
\tabularnewline[0.5pt]\hline				
48	 &	smooth    &	          &	1+1006 T+6504 p T^2+1006 p^3 T^3+p^6 T^4
\tabularnewline[0.5pt]\hline				
49	 &	smooth    &	          &	1+47 T+12244 p T^2+47 p^3 T^3+p^6 T^4
\tabularnewline[0.5pt]\hline				
50	 &	smooth    &	          &	1+229 T+2416 p T^2+229 p^3 T^3+p^6 T^4
\tabularnewline[0.5pt]\hline				
51	 &	smooth    &	          &	1-674 T+6266 p T^2-674 p^3 T^3+p^6 T^4
\tabularnewline[0.5pt]\hline				
52	 &	smooth    &	          &	1+1244 T+18698 p T^2+1244 p^3 T^3+p^6 T^4
\tabularnewline[0.5pt]\hline				
53	 &	smooth    &	          &	1-44 T+3158 p T^2-44 p^3 T^3+p^6 T^4
\tabularnewline[0.5pt]\hline				
54	 &	smooth    &	          &	1+1342 T+18586 p T^2+1342 p^3 T^3+p^6 T^4
\tabularnewline[0.5pt]\hline				
55	 &	smooth    &	          &	1-1542 T+20154 p T^2-1542 p^3 T^3+p^6 T^4
\tabularnewline[0.5pt]\hline				
56	 &	smooth    &	          &	1+1272 T+14288 p T^2+1272 p^3 T^3+p^6 T^4
\tabularnewline[0.5pt]\hline				
57	 &	smooth    &	          &	1+348 T+2738 p T^2+348 p^3 T^3+p^6 T^4
\tabularnewline[0.5pt]\hline				
58	 &	smooth    &	          &	1+1020 T+12818 p T^2+1020 p^3 T^3+p^6 T^4
\tabularnewline[0.5pt]\hline				
59	 &	smooth    &	          &	1-1584 T+18782 p T^2-1584 p^3 T^3+p^6 T^4
\tabularnewline[0.5pt]\hline				
60	 &	smooth    &	          &	1-142 T+9892 p T^2-142 p^3 T^3+p^6 T^4
\tabularnewline[0.5pt]\hline				
61	 &	smooth    &	          &	1+551 T+11334 p T^2+551 p^3 T^3+p^6 T^4
\tabularnewline[0.5pt]\hline				
62	 &	smooth    &	          &	1+1405 T+15590 p T^2+1405 p^3 T^3+p^6 T^4
\tabularnewline[0.5pt]\hline				
63	 &	smooth    &	          &	1-982 T+15632 p T^2-982 p^3 T^3+p^6 T^4
\tabularnewline[0.5pt]\hline				
64	 &	smooth    &	          &	1-44 T+12230 p T^2-44 p^3 T^3+p^6 T^4
\tabularnewline[0.5pt]\hline				
65	 &	smooth    &	          &	1-408 T+13938 p T^2-408 p^3 T^3+p^6 T^4
\tabularnewline[0.5pt]\hline				
66	 &	smooth    &	          &	1-1024 T+15044 p T^2-1024 p^3 T^3+p^6 T^4
\tabularnewline[0.5pt]\hline				
67	 &	smooth    &	          &	1-135 T-4024 p T^2-135 p^3 T^3+p^6 T^4
\tabularnewline[0.5pt]\hline				
68	 &	smooth    &	          &	1-450 T+9402 p T^2-450 p^3 T^3+p^6 T^4
\tabularnewline[0.5pt]\hline				
69	 &	smooth    &	          &	1+607 T+9080 p T^2+607 p^3 T^3+p^6 T^4
\tabularnewline[0.5pt]\hline				
70	 &	smooth    &	          &	1+705 T+7722 p T^2+705 p^3 T^3+p^6 T^4
\tabularnewline[0.5pt]\hline				
71	 &	smooth    &	          &	1-604 T+694 p T^2-604 p^3 T^3+p^6 T^4
\tabularnewline[0.5pt]\hline				
72	 &	smooth    &	          &	1-198 T+6658 p T^2-198 p^3 T^3+p^6 T^4
\tabularnewline[0.5pt]\hline				
73	 &	smooth    &	          &	1-1311 T+15240 p T^2-1311 p^3 T^3+p^6 T^4
\tabularnewline[0.5pt]\hline				
74	 &	smooth    &	          &	1+1076 T+14680 p T^2+1076 p^3 T^3+p^6 T^4
\tabularnewline[0.5pt]\hline				
75	 &	smooth    &	          &	1+26 T+136 p^2 T^2+26 p^3 T^3+p^6 T^4
\tabularnewline[0.5pt]\hline				
76	 &	smooth    &	          &	1-527 T-174 p T^2-527 p^3 T^3+p^6 T^4
\tabularnewline[0.5pt]\hline				
77	 &	smooth    &	          &	1-170 T+12356 p T^2-170 p^3 T^3+p^6 T^4
\tabularnewline[0.5pt]\hline				
78	 &	smooth    &	          &	1+439 T+7092 p T^2+439 p^3 T^3+p^6 T^4
\tabularnewline[0.5pt]\hline				
79	 &	smooth    &	          &	1+572 T+7974 p T^2+572 p^3 T^3+p^6 T^4
\tabularnewline[0.5pt]\hline				
80	 &	smooth    &	          &	1+719 T+7204 p T^2+719 p^3 T^3+p^6 T^4
\tabularnewline[0.5pt]\hline				
81	 &	smooth    &	          &	1-331 T+7792 p T^2-331 p^3 T^3+p^6 T^4
\tabularnewline[0.5pt]\hline				
82	 &	smooth    &	          &	1+5 T-7034 p T^2+5 p^3 T^3+p^6 T^4
\tabularnewline[0.5pt]\hline				
83	 &	smooth    &	          &	1-247 T+3270 p T^2-247 p^3 T^3+p^6 T^4
\tabularnewline[0.5pt]\hline				
84	 &	smooth    &	          &	1-114 T-10198 p T^2-114 p^3 T^3+p^6 T^4
\tabularnewline[0.5pt]\hline				
85	 &	smooth    &	          &	1+1132 T+13686 p T^2+1132 p^3 T^3+p^6 T^4
\tabularnewline[0.5pt]\hline				
86	 &	smooth    &	          &	1+1307 T+9066 p T^2+1307 p^3 T^3+p^6 T^4
\tabularnewline[0.5pt]\hline				
87	 &	smooth    &	          &	1+726 T+1338 p T^2+726 p^3 T^3+p^6 T^4
\tabularnewline[0.5pt]\hline				
88	 &	smooth    &	          &	1+2028 T+25670 p T^2+2028 p^3 T^3+p^6 T^4
\tabularnewline[0.5pt]\hline				
\tablepostamble				
\tablepreamble{97}				
1	 &	smooth    &	          &	1+21 T-9896 p T^2+21 p^3 T^3+p^6 T^4
\tabularnewline[0.5pt]\hline				
2	 &	singular  &	2&	(1-p T) (1-854 T+p^3 T^2)
\tabularnewline[0.5pt]\hline				
3	 &	smooth$^*$&	3&	  
\tabularnewline[0.5pt]\hline				
4	 &	smooth    &	          &	1-1232 T+9214 p T^2-1232 p^3 T^3+p^6 T^4
\tabularnewline[0.5pt]\hline				
5	 &	smooth    &	          &	1+1218 T+10978 p T^2+1218 p^3 T^3+p^6 T^4
\tabularnewline[0.5pt]\hline				
6	 &	smooth    &	          &	1+413 T+9704 p T^2+413 p^3 T^3+p^6 T^4
\tabularnewline[0.5pt]\hline				
7	 &	smooth    &	          &	1+273 T+13526 p T^2+273 p^3 T^3+p^6 T^4
\tabularnewline[0.5pt]\hline				
8	 &	smooth    &	          &	1-945 T+11664 p T^2-945 p^3 T^3+p^6 T^4
\tabularnewline[0.5pt]\hline				
9	 &	smooth    &	          &	1+308 T+14114 p T^2+308 p^3 T^3+p^6 T^4
\tabularnewline[0.5pt]\hline				
10	 &	smooth    &	          &	1+1218 T+21170 p T^2+1218 p^3 T^3+p^6 T^4
\tabularnewline[0.5pt]\hline				
11	 &	smooth    &	          &	1+98 T+14506 p T^2+98 p^3 T^3+p^6 T^4
\tabularnewline[0.5pt]\hline				
12	 &	smooth    &	          &	1+1029 T+11860 p T^2+1029 p^3 T^3+p^6 T^4
\tabularnewline[0.5pt]\hline				
13	 &	smooth    &	          &	1+1365 T+11370 p T^2+1365 p^3 T^3+p^6 T^4
\tabularnewline[0.5pt]\hline				
14	 &	smooth    &	          &	1-189 T-390 p T^2-189 p^3 T^3+p^6 T^4
\tabularnewline[0.5pt]\hline				
15	 &	smooth    &	          &	1+301 T+3334 p T^2+301 p^3 T^3+p^6 T^4
\tabularnewline[0.5pt]\hline				
16	 &	smooth    &	          &	1-539 T+5784 p T^2-539 p^3 T^3+p^6 T^4
\tabularnewline[0.5pt]\hline				
17	 &	smooth    &	          &	(1+7 p T+p^3 T^2)(1-1232 T+p^3 T^2)
\tabularnewline[0.5pt]\hline				
18	 &	smooth    &	          &	1-1064 T+10390 p T^2-1064 p^3 T^3+p^6 T^4
\tabularnewline[0.5pt]\hline				
19	 &	singular  &	19&	(1-p T) (1+546 T+p^3 T^2)
\tabularnewline[0.5pt]\hline				
20	 &	smooth    &	          &	1+112 T+14506 p T^2+112 p^3 T^3+p^6 T^4
\tabularnewline[0.5pt]\hline				
21	 &	smooth    &	          &	1+434 T-2350 p T^2+434 p^3 T^3+p^6 T^4
\tabularnewline[0.5pt]\hline				
22	 &	smooth    &	          &	1-1820 T+23522 p T^2-1820 p^3 T^3+p^6 T^4
\tabularnewline[0.5pt]\hline				
23	 &	smooth    &	          &	1+301 T+10194 p T^2+301 p^3 T^3+p^6 T^4
\tabularnewline[0.5pt]\hline				
24	 &	smooth    &	          &	1-672 T-2154 p T^2-672 p^3 T^3+p^6 T^4
\tabularnewline[0.5pt]\hline				
25	 &	smooth    &	          &	1-182 T+6274 p T^2-182 p^3 T^3+p^6 T^4
\tabularnewline[0.5pt]\hline				
26	 &	smooth    &	          &	1-266 T+14310 p T^2-266 p^3 T^3+p^6 T^4
\tabularnewline[0.5pt]\hline				
27	 &	smooth    &	          &	1+455 T-4800 p T^2+455 p^3 T^3+p^6 T^4
\tabularnewline[0.5pt]\hline				
28	 &	smooth    &	          &	1+392 T+1178 p T^2+392 p^3 T^3+p^6 T^4
\tabularnewline[0.5pt]\hline				
29	 &	smooth    &	          &	1+1085 T+7842 p T^2+1085 p^3 T^3+p^6 T^4
\tabularnewline[0.5pt]\hline				
30	 &	smooth    &	          &	1+1400 T+14212 p T^2+1400 p^3 T^3+p^6 T^4
\tabularnewline[0.5pt]\hline				
31	 &	smooth    &	          &	1+28 T-1370 p T^2+28 p^3 T^3+p^6 T^4
\tabularnewline[0.5pt]\hline				
32	 &	smooth    &	          &	1-322 T-3134 p T^2-322 p^3 T^3+p^6 T^4
\tabularnewline[0.5pt]\hline				
33	 &	smooth    &	          &	1-196 T+2354 p T^2-196 p^3 T^3+p^6 T^4
\tabularnewline[0.5pt]\hline				
34	 &	smooth    &	          &	1-665 T+14506 p T^2-665 p^3 T^3+p^6 T^4
\tabularnewline[0.5pt]\hline				
35	 &	smooth    &	          &	1+322 T+14898 p T^2+322 p^3 T^3+p^6 T^4
\tabularnewline[0.5pt]\hline				
36	 &	smooth    &	          &	1+1302 T+18426 p T^2+1302 p^3 T^3+p^6 T^4
\tabularnewline[0.5pt]\hline				
37	 &	smooth    &	          &	1-1582 T+18916 p T^2-1582 p^3 T^3+p^6 T^4
\tabularnewline[0.5pt]\hline				
38	 &	smooth    &	          &	1+686 T+7940 p T^2+686 p^3 T^3+p^6 T^4
\tabularnewline[0.5pt]\hline				
39	 &	smooth    &	          &	1-1617 T+16270 p T^2-1617 p^3 T^3+p^6 T^4
\tabularnewline[0.5pt]\hline				
40	 &	smooth    &	          &	1-7152 p T^2+p^6 T^4
\tabularnewline[0.5pt]\hline				
41	 &	smooth    &	          &	1-294 T+14506 p T^2-294 p^3 T^3+p^6 T^4
\tabularnewline[0.5pt]\hline				
42	 &	smooth    &	          &	1-1449 T+21562 p T^2-1449 p^3 T^3+p^6 T^4
\tabularnewline[0.5pt]\hline				
43	 &	smooth    &	          &	1+42 T-3134 p T^2+42 p^3 T^3+p^6 T^4
\tabularnewline[0.5pt]\hline				
44	 &	smooth    &	          &	1-1001 T+7548 p T^2-1001 p^3 T^3+p^6 T^4
\tabularnewline[0.5pt]\hline				
45	 &	smooth    &	          &	1-413 T+6470 p T^2-413 p^3 T^3+p^6 T^4
\tabularnewline[0.5pt]\hline				
46	 &	smooth    &	          &	1-770 T+3530 p T^2-770 p^3 T^3+p^6 T^4
\tabularnewline[0.5pt]\hline				
47	 &	smooth    &	          &	1+413 T+13624 p T^2+413 p^3 T^3+p^6 T^4
\tabularnewline[0.5pt]\hline				
48	 &	smooth    &	          &	1-854 T+5098 p T^2-854 p^3 T^3+p^6 T^4
\tabularnewline[0.5pt]\hline				
49	 &	smooth    &	          &	1-1330 T+13722 p T^2-1330 p^3 T^3+p^6 T^4
\tabularnewline[0.5pt]\hline				
50	 &	smooth    &	          &	1+847 T+10684 p T^2+847 p^3 T^3+p^6 T^4
\tabularnewline[0.5pt]\hline				
51	 &	smooth    &	          &	1+1918 T+26364 p T^2+1918 p^3 T^3+p^6 T^4
\tabularnewline[0.5pt]\hline				
52	 &	smooth    &	          &	1-917 T+7058 p T^2-917 p^3 T^3+p^6 T^4
\tabularnewline[0.5pt]\hline				
53	 &	smooth    &	          &	1+1974 T+22346 p T^2+1974 p^3 T^3+p^6 T^4
\tabularnewline[0.5pt]\hline				
54	 &	smooth    &	          &	1+2072 T+24894 p T^2+2072 p^3 T^3+p^6 T^4
\tabularnewline[0.5pt]\hline				
55	 &	smooth    &	          &	1+1260 T+19504 p T^2+1260 p^3 T^3+p^6 T^4
\tabularnewline[0.5pt]\hline				
56	 &	smooth    &	          &	1+126 T-9896 p T^2+126 p^3 T^3+p^6 T^4
\tabularnewline[0.5pt]\hline				
57	 &	smooth    &	          &	1+658 T+3040 p T^2+658 p^3 T^3+p^6 T^4
\tabularnewline[0.5pt]\hline				
58	 &	smooth    &	          &	1+812 T+17446 p T^2+812 p^3 T^3+p^6 T^4
\tabularnewline[0.5pt]\hline				
59	 &	smooth    &	          &	1+294 T+1962 p T^2+294 p^3 T^3+p^6 T^4
\tabularnewline[0.5pt]\hline				
60	 &	smooth    &	          &	1-532 T+4020 p T^2-532 p^3 T^3+p^6 T^4
\tabularnewline[0.5pt]\hline				
61	 &	smooth    &	          &	1-238 T+10194 p T^2-238 p^3 T^3+p^6 T^4
\tabularnewline[0.5pt]\hline				
62	 &	smooth    &	          &	1-658 T+12546 p T^2-658 p^3 T^3+p^6 T^4
\tabularnewline[0.5pt]\hline				
63	 &	smooth    &	          &	1+574 T+14506 p T^2+574 p^3 T^3+p^6 T^4
\tabularnewline[0.5pt]\hline				
64	 &	smooth    &	          &	1+252 T+786 p T^2+252 p^3 T^3+p^6 T^4
\tabularnewline[0.5pt]\hline				
65	 &	smooth    &	          &	1-700 T+7254 p T^2-700 p^3 T^3+p^6 T^4
\tabularnewline[0.5pt]\hline				
66	 &	smooth    &	          &	1-196 T+9606 p T^2-196 p^3 T^3+p^6 T^4
\tabularnewline[0.5pt]\hline				
67	 &	smooth    &	          &	1-182 T+1276 p T^2-182 p^3 T^3+p^6 T^4
\tabularnewline[0.5pt]\hline				
68	 &	smooth    &	          &	1-511 T+5294 p T^2-511 p^3 T^3+p^6 T^4
\tabularnewline[0.5pt]\hline				
69	 &	smooth    &	          &	1+2016 T+28226 p T^2+2016 p^3 T^3+p^6 T^4
\tabularnewline[0.5pt]\hline				
70	 &	smooth    &	          &	1+532 T+16074 p T^2+532 p^3 T^3+p^6 T^4
\tabularnewline[0.5pt]\hline				
71	 &	smooth    &	          &	1-371 T+9018 p T^2-371 p^3 T^3+p^6 T^4
\tabularnewline[0.5pt]\hline				
72	 &	smooth    &	          &	1+77 T+11860 p T^2+77 p^3 T^3+p^6 T^4
\tabularnewline[0.5pt]\hline				
73	 &	smooth    &	          &	1+1190 T+19602 p T^2+1190 p^3 T^3+p^6 T^4
\tabularnewline[0.5pt]\hline				
74	 &	singular  &	       74&	(1-p T) (1+798 T+p^3 T^2)
\tabularnewline[0.5pt]\hline				
75	 &	smooth    &	          &	1+1442 T+17250 p T^2+1442 p^3 T^3+p^6 T^4
\tabularnewline[0.5pt]\hline				
76	 &	smooth    &	          &	1+441 T+4118 p T^2+441 p^3 T^3+p^6 T^4
\tabularnewline[0.5pt]\hline				
77	 &	smooth    &	          &	1-364 T+7254 p T^2-364 p^3 T^3+p^6 T^4
\tabularnewline[0.5pt]\hline				
78	 &	smooth    &	          &	1-945 T+19210 p T^2-945 p^3 T^3+p^6 T^4
\tabularnewline[0.5pt]\hline				
79	 &	smooth    &	          &	1+714 T+3922 p T^2+714 p^3 T^3+p^6 T^4
\tabularnewline[0.5pt]\hline				
80	 &	smooth    &	          &	1-644 T+10880 p T^2-644 p^3 T^3+p^6 T^4
\tabularnewline[0.5pt]\hline				
81	 &	smooth    &	          &	1+28 T-2154 p T^2+28 p^3 T^3+p^6 T^4
\tabularnewline[0.5pt]\hline				
82	 &	smooth    &	          &	1+231 T-2546 p T^2+231 p^3 T^3+p^6 T^4
\tabularnewline[0.5pt]\hline				
83	 &	smooth    &	          &	1-231 T+12350 p T^2-231 p^3 T^3+p^6 T^4
\tabularnewline[0.5pt]\hline				
84	 &	smooth    &	          &	1+399 T-3330 p T^2+399 p^3 T^3+p^6 T^4
\tabularnewline[0.5pt]\hline				
85	 &	smooth    &	          &	1-406 T+1570 p T^2-406 p^3 T^3+p^6 T^4
\tabularnewline[0.5pt]\hline				
86	 &	smooth    &	          &	1+329 T-1664 p T^2+329 p^3 T^3+p^6 T^4
\tabularnewline[0.5pt]\hline				
87	 &	smooth    &	          &	1+1323 T+15094 p T^2+1323 p^3 T^3+p^6 T^4
\tabularnewline[0.5pt]\hline				
88	 &	smooth    &	          &	1-203 T-8328 p T^2-203 p^3 T^3+p^6 T^4
\tabularnewline[0.5pt]\hline				
89	 &	smooth    &	          &	1-2247 T+28128 p T^2-2247 p^3 T^3+p^6 T^4
\tabularnewline[0.5pt]\hline				
90	 &	smooth    &	          &	1+1288 T+22640 p T^2+1288 p^3 T^3+p^6 T^4
\tabularnewline[0.5pt]\hline				
91	 &	smooth    &	          &	1+1624 T+17838 p T^2+1624 p^3 T^3+p^6 T^4
\tabularnewline[0.5pt]\hline				
92	 &	smooth    &	          &	1+532 T+2942 p T^2+532 p^3 T^3+p^6 T^4
\tabularnewline[0.5pt]\hline				
93	 &	smooth    &	          &	1-651 T-4408 p T^2-651 p^3 T^3+p^6 T^4
\tabularnewline[0.5pt]\hline				
94	 &	smooth    &	          &	1+1491 T+21268 p T^2+1491 p^3 T^3+p^6 T^4
\tabularnewline[0.5pt]\hline				
95	 &	smooth    &	          &	1+217 T-1076 p T^2+217 p^3 T^3+p^6 T^4
\tabularnewline[0.5pt]\hline				
96	 &	smooth    &	          &	1-476 T-586 p T^2-476 p^3 T^3+p^6 T^4
\tabularnewline[0.5pt]\hline
\tablepostamble
\newpage
\lhead{\ifthenelse{\isodd{\value{page}}}{\thepage}{\sl The $\z$-function for the manifold with $\hodgenos{\,=\,}(4,1)$}}
\rhead{\ifthenelse{\isodd{\value{page}}}{\sl The $\z$-function for the manifold with $\hodgenos{\,=\,}(4,1)$}{\thepage}}
\subsection{The $\z$-function for the manifold with $\hodgenos{\,=\,}(4,1)$}
\vspace{1.5cm}
\tablepreamble{5}				
1	 &	singular  &	-\frac{1}{4}\+&	(1-p T) (1+12 T+p^3 T^2)
\tabularnewline[0.5pt]\hline				
2	 &	singular  &	\frac{1}{3}&	(1-p T) (1+18 T+p^3 T^2)
\tabularnewline[0.5pt]\hline				
3	 &	singular  &	\frac{1}{12}&	(1-p T) (1+12 T+p^3 T^2)
\tabularnewline[0.5pt]\hline				
4	 &	singular  &	\left\{-\frac{1}{6},\frac{1}{4},\frac{3}{2}\right\}&	  
\tabularnewline[0.5pt]\hline				
\tablepostamble				
\tablepreamble{7}				
1	 &	singular  &	-\frac{1}{6}\+&	(1-p T) (1+28 T+p^3 T^2)
\tabularnewline[0.5pt]\hline				
2	 &	singular  &	\frac{1}{4}&	(1-p T) (1+4 T+p^3 T^2)
\tabularnewline[0.5pt]\hline				
3	 &	singular  &	\frac{1}{12}&	(1-p T) (1+22 T+p^3 T^2)
\tabularnewline[0.5pt]\hline				
4	 &	singular  &	-\frac{1}{5}\+&	(1-p T) (1+28 T+p^3 T^2)
\tabularnewline[0.5pt]\hline				
5	 &	singular  &	\left\{-\frac{1}{4},\frac{1}{3},\frac{3}{2}\right\}&	  
\tabularnewline[0.5pt]\hline				
6	 &	smooth    &	          &	(1+4 p T+p^3 T^2)(1-2 p T+p^3 T^2)
\tabularnewline[0.5pt]\hline				
\tablepostamble				
\tablepreamble{11}				
1	 &	singular  &	\frac{1}{12}&	(1-p T) (1+48 T+p^3 T^2)
\tabularnewline[0.5pt]\hline				
2	 &	singular  &	-\frac{1}{5}\+&	(1+p T) (1+24 T+p^3 T^2)
\tabularnewline[0.5pt]\hline				
3	 &	singular  &	\frac{1}{4}&	(1+p T) (1-12 T+p^3 T^2)
\tabularnewline[0.5pt]\hline				
4	 &	singular  &	\frac{1}{3}&	(1-p T) (1+36 T+p^3 T^2)
\tabularnewline[0.5pt]\hline				
5	 &	smooth    &	          &	1+6 p T+314 p T^2+6 p^4 T^3+p^6 T^4
\tabularnewline[0.5pt]\hline				
6	 &	smooth    &	          &	1+170 p T^2+p^6 T^4
\tabularnewline[0.5pt]\hline				
7	 &	smooth$^*$&	\frac{3}{2}&	  
\tabularnewline[0.5pt]\hline				
8	 &	singular  &	-\frac{1}{4}\+&	(1-p T) (1-48 T+p^3 T^2)
\tabularnewline[0.5pt]\hline				
9	 &	singular  &	-\frac{1}{6}\+&	(1-p T) (1+24 T+p^3 T^2)
\tabularnewline[0.5pt]\hline				
10	 &	smooth    &	          &	1+24 T-46 p T^2+24 p^3 T^3+p^6 T^4
\tabularnewline[0.5pt]\hline				
\tablepostamble				
\tablepreamble{13}				
1	 &	smooth    &	          &	1+8 T+126 p T^2+8 p^3 T^3+p^6 T^4
\tabularnewline[0.5pt]\hline				
2	 &	singular  &	-\frac{1}{6}\+&	(1-p T) (1+70 T+p^3 T^2)
\tabularnewline[0.5pt]\hline				
3	 &	singular  &	-\frac{1}{4}\+&	(1-p T) (1-56 T+p^3 T^2)
\tabularnewline[0.5pt]\hline				
4	 &	smooth    &	          &	1+32 T+6 p T^2+32 p^3 T^3+p^6 T^4
\tabularnewline[0.5pt]\hline				
5	 &	singular  &	-\frac{1}{5}\+&	(1-p T) (1+58 T+p^3 T^2)
\tabularnewline[0.5pt]\hline				
6	 &	smooth    &	          &	1+2 p T+90 p T^2+2 p^4 T^3+p^6 T^4
\tabularnewline[0.5pt]\hline				
7	 &	smooth    &	          &	1-10 T-54 p T^2-10 p^3 T^3+p^6 T^4
\tabularnewline[0.5pt]\hline				
8	 &	smooth$^*$&	\frac{3}{2}&	  
\tabularnewline[0.5pt]\hline				
9	 &	singular  &	\frac{1}{3}&	(1-p T) (1+34 T+p^3 T^2)
\tabularnewline[0.5pt]\hline				
10	 &	singular  &	\frac{1}{4}&	(1-p T) (1+58 T+p^3 T^2)
\tabularnewline[0.5pt]\hline				
11	 &	smooth    &	          &	1-46 T+90 p T^2-46 p^3 T^3+p^6 T^4
\tabularnewline[0.5pt]\hline				
12	 &	singular  &	\frac{1}{12}&	(1-p T) (1-2 T+p^3 T^2)
\tabularnewline[0.5pt]\hline				
\tablepostamble				
\tablepreamble{17}				
1	 &	smooth    &	          &	1-90 T+290 p T^2-90 p^3 T^3+p^6 T^4
\tabularnewline[0.5pt]\hline				
2	 &	smooth    &	          &	1+36 T+398 p T^2+36 p^3 T^3+p^6 T^4
\tabularnewline[0.5pt]\hline				
3	 &	smooth    &	          &	1+156 T+758 p T^2+156 p^3 T^3+p^6 T^4
\tabularnewline[0.5pt]\hline				
4	 &	singular  &	-\frac{1}{4}\+&	(1-p T) (1+114 T+p^3 T^2)
\tabularnewline[0.5pt]\hline				
5	 &	smooth    &	          &	1-6 T-214 p T^2-6 p^3 T^3+p^6 T^4
\tabularnewline[0.5pt]\hline				
6	 &	singular  &	\frac{1}{3}&	(1-p T) (1-42 T+p^3 T^2)
\tabularnewline[0.5pt]\hline				
7	 &	smooth    &	          &	(1-6 p T+p^3 T^2)(1+90 T+p^3 T^2)
\tabularnewline[0.5pt]\hline				
8	 &	smooth    &	          &	1-250 p T^2+p^6 T^4
\tabularnewline[0.5pt]\hline				
9	 &	smooth    &	          &	1-60 T+182 p T^2-60 p^3 T^3+p^6 T^4
\tabularnewline[0.5pt]\hline				
10	 &	singular  &	\left\{-\frac{1}{5},\frac{1}{12},\frac{3}{2}\right\}&	  
\tabularnewline[0.5pt]\hline				
11	 &	smooth    &	          &	(1+6 p T+p^3 T^2)(1+36 T+p^3 T^2)
\tabularnewline[0.5pt]\hline				
12	 &	smooth    &	          &	1+54 T+146 p T^2+54 p^3 T^3+p^6 T^4
\tabularnewline[0.5pt]\hline				
13	 &	singular  &	\frac{1}{4}&	(1+p T) (1-66 T+p^3 T^2)
\tabularnewline[0.5pt]\hline				
14	 &	singular  &	-\frac{1}{6}\+&	(1-p T) (1-102 T+p^3 T^2)
\tabularnewline[0.5pt]\hline				
15	 &	smooth    &	          &	1+60 T+182 p T^2+60 p^3 T^3+p^6 T^4
\tabularnewline[0.5pt]\hline				
16	 &	smooth    &	          &	(1+6 p T+p^3 T^2)(1-6 T+p^3 T^2)
\tabularnewline[0.5pt]\hline				
\tablepostamble				
\tablepreamble{19}				
1	 &	smooth    &	          &	1-16 T+498 p T^2-16 p^3 T^3+p^6 T^4
\tabularnewline[0.5pt]\hline				
2	 &	smooth    &	          &	1+194 T+1122 p T^2+194 p^3 T^3+p^6 T^4
\tabularnewline[0.5pt]\hline				
3	 &	singular  &	-\frac{1}{6}\+&	(1-p T) (1-20 T+p^3 T^2)
\tabularnewline[0.5pt]\hline				
4	 &	smooth    &	          &	1+32 T+474 p T^2+32 p^3 T^3+p^6 T^4
\tabularnewline[0.5pt]\hline				
5	 &	singular  &	\frac{1}{4}&	(1-p T) (1+100 T+p^3 T^2)
\tabularnewline[0.5pt]\hline				
6	 &	smooth    &	          &	1-40 T+402 p T^2-40 p^3 T^3+p^6 T^4
\tabularnewline[0.5pt]\hline				
7	 &	smooth    &	          &	1+158 T+762 p T^2+158 p^3 T^3+p^6 T^4
\tabularnewline[0.5pt]\hline				
8	 &	singular  &	\frac{1}{12}&	(1-p T) (1-20 T+p^3 T^2)
\tabularnewline[0.5pt]\hline				
9	 &	smooth    &	          &	1-64 T+450 p T^2-64 p^3 T^3+p^6 T^4
\tabularnewline[0.5pt]\hline				
10	 &	smooth    &	          &	1-34 T-150 p T^2-34 p^3 T^3+p^6 T^4
\tabularnewline[0.5pt]\hline				
11	 &	smooth$^*$&	\frac{3}{2}&	  
\tabularnewline[0.5pt]\hline				
12	 &	smooth    &	          &	1-46 T+666 p T^2-46 p^3 T^3+p^6 T^4
\tabularnewline[0.5pt]\hline				
13	 &	singular  &	\frac{1}{3}&	(1-p T) (1+124 T+p^3 T^2)
\tabularnewline[0.5pt]\hline				
14	 &	singular  &	-\frac{1}{4}\+&	(1-p T) (1-2 T+p^3 T^2)
\tabularnewline[0.5pt]\hline				
15	 &	singular  &	-\frac{1}{5}\+&	(1-p T) (1-116 T+p^3 T^2)
\tabularnewline[0.5pt]\hline				
16	 &	smooth    &	          &	1-28 T+594 p T^2-28 p^3 T^3+p^6 T^4
\tabularnewline[0.5pt]\hline				
17	 &	smooth    &	          &	1+98 T+666 p T^2+98 p^3 T^3+p^6 T^4
\tabularnewline[0.5pt]\hline				
18	 &	smooth    &	          &	1+32 T-30 p T^2+32 p^3 T^3+p^6 T^4
\tabularnewline[0.5pt]\hline				
\tablepostamble				
\tablepreamble{23}				
1	 &	smooth    &	          &	1-12 T+626 p T^2-12 p^3 T^3+p^6 T^4
\tabularnewline[0.5pt]\hline				
2	 &	singular  &	\frac{1}{12}&	(1-p T) (1+54 T+p^3 T^2)
\tabularnewline[0.5pt]\hline				
3	 &	smooth    &	          &	1+126 T+698 p T^2+126 p^3 T^3+p^6 T^4
\tabularnewline[0.5pt]\hline				
4	 &	smooth    &	          &	1+6 T-382 p T^2+6 p^3 T^3+p^6 T^4
\tabularnewline[0.5pt]\hline				
5	 &	smooth    &	          &	1+48 T+770 p T^2+48 p^3 T^3+p^6 T^4
\tabularnewline[0.5pt]\hline				
6	 &	singular  &	\frac{1}{4}&	(1+p T) (1-132 T+p^3 T^2)
\tabularnewline[0.5pt]\hline				
7	 &	smooth    &	          &	1-48 T+914 p T^2-48 p^3 T^3+p^6 T^4
\tabularnewline[0.5pt]\hline				
8	 &	singular  &	\frac{1}{3}&	(1-p T) (1+p^3 T^2)
\tabularnewline[0.5pt]\hline				
9	 &	singular  &	-\frac{1}{5}\+&	(1+p T) (1+60 T+p^3 T^2)
\tabularnewline[0.5pt]\hline				
10	 &	smooth    &	          &	1-120 T+482 p T^2-120 p^3 T^3+p^6 T^4
\tabularnewline[0.5pt]\hline				
11	 &	smooth    &	          &	1+96 T+482 p T^2+96 p^3 T^3+p^6 T^4
\tabularnewline[0.5pt]\hline				
12	 &	smooth    &	          &	1+914 p T^2+p^6 T^4
\tabularnewline[0.5pt]\hline				
13	 &	smooth$^*$&	\frac{3}{2}&	  
\tabularnewline[0.5pt]\hline				
14	 &	smooth    &	          &	1+102 T+626 p T^2+102 p^3 T^3+p^6 T^4
\tabularnewline[0.5pt]\hline				
15	 &	smooth    &	          &	1-42 T+194 p T^2-42 p^3 T^3+p^6 T^4
\tabularnewline[0.5pt]\hline				
16	 &	smooth    &	          &	1-162 T+842 p T^2-162 p^3 T^3+p^6 T^4
\tabularnewline[0.5pt]\hline				
17	 &	singular  &	-\frac{1}{4}\+&	(1-p T) (1+120 T+p^3 T^2)
\tabularnewline[0.5pt]\hline				
18	 &	smooth    &	          &	1+60 T+50 p T^2+60 p^3 T^3+p^6 T^4
\tabularnewline[0.5pt]\hline				
19	 &	singular  &	-\frac{1}{6}\+&	(1-p T) (1+72 T+p^3 T^2)
\tabularnewline[0.5pt]\hline				
20	 &	smooth    &	          &	(1+p^3 T^2)(1+120 T+p^3 T^2)
\tabularnewline[0.5pt]\hline				
21	 &	smooth    &	          &	1+96 T+482 p T^2+96 p^3 T^3+p^6 T^4
\tabularnewline[0.5pt]\hline				
22	 &	smooth    &	          &	(1+6 p T+p^3 T^2)(1-36 T+p^3 T^2)
\tabularnewline[0.5pt]\hline				
\tablepostamble				
\tablepreamble{29}				
1	 &	smooth    &	          &	1-222 T+1034 p T^2-222 p^3 T^3+p^6 T^4
\tabularnewline[0.5pt]\hline				
2	 &	smooth    &	          &	1+258 T+1250 p T^2+258 p^3 T^3+p^6 T^4
\tabularnewline[0.5pt]\hline				
3	 &	smooth    &	          &	1+72 T-226 p T^2+72 p^3 T^3+p^6 T^4
\tabularnewline[0.5pt]\hline				
4	 &	smooth    &	          &	(1+6 p T+p^3 T^2)(1-78 T+p^3 T^2)
\tabularnewline[0.5pt]\hline				
5	 &	smooth    &	          &	1+288 T+2006 p T^2+288 p^3 T^3+p^6 T^4
\tabularnewline[0.5pt]\hline				
6	 &	smooth    &	          &	1+180 T+710 p T^2+180 p^3 T^3+p^6 T^4
\tabularnewline[0.5pt]\hline				
7	 &	singular  &	-\frac{1}{4}\+&	(1-p T) (1+54 T+p^3 T^2)
\tabularnewline[0.5pt]\hline				
8	 &	smooth    &	          &	1+60 T+422 p T^2+60 p^3 T^3+p^6 T^4
\tabularnewline[0.5pt]\hline				
9	 &	smooth    &	          &	1+36 T+278 p T^2+36 p^3 T^3+p^6 T^4
\tabularnewline[0.5pt]\hline				
10	 &	singular  &	\frac{1}{3}&	(1-p T) (1-102 T+p^3 T^2)
\tabularnewline[0.5pt]\hline				
11	 &	smooth    &	          &	(1+6 p T+p^3 T^2)(1-6 p T+p^3 T^2)
\tabularnewline[0.5pt]\hline				
12	 &	smooth    &	          &	1+168 T+1214 p T^2+168 p^3 T^3+p^6 T^4
\tabularnewline[0.5pt]\hline				
13	 &	smooth    &	          &	1+192 T+494 p T^2+192 p^3 T^3+p^6 T^4
\tabularnewline[0.5pt]\hline				
14	 &	smooth    &	          &	1+108 T+998 p T^2+108 p^3 T^3+p^6 T^4
\tabularnewline[0.5pt]\hline				
15	 &	smooth    &	          &	1-168 T+782 p T^2-168 p^3 T^3+p^6 T^4
\tabularnewline[0.5pt]\hline				
16	 &	smooth$^*$&	\frac{3}{2}&	  
\tabularnewline[0.5pt]\hline				
17	 &	singular  &	\frac{1}{12}&	(1-p T) (1-84 T+p^3 T^2)
\tabularnewline[0.5pt]\hline				
18	 &	smooth    &	          &	1-132 T+998 p T^2-132 p^3 T^3+p^6 T^4
\tabularnewline[0.5pt]\hline				
19	 &	smooth    &	          &	1+926 p T^2+p^6 T^4
\tabularnewline[0.5pt]\hline				
20	 &	smooth    &	          &	1+90 T+386 p T^2+90 p^3 T^3+p^6 T^4
\tabularnewline[0.5pt]\hline				
21	 &	smooth    &	          &	1+138 T+530 p T^2+138 p^3 T^3+p^6 T^4
\tabularnewline[0.5pt]\hline				
22	 &	singular  &	\frac{1}{4}&	(1+p T) (1+90 T+p^3 T^2)
\tabularnewline[0.5pt]\hline				
23	 &	singular  &	-\frac{1}{5}\+&	(1+p T) (1-30 T+p^3 T^2)
\tabularnewline[0.5pt]\hline				
24	 &	singular  &	-\frac{1}{6}\+&	(1-p T) (1-306 T+p^3 T^2)
\tabularnewline[0.5pt]\hline				
25	 &	smooth    &	          &	1+210 T+1466 p T^2+210 p^3 T^3+p^6 T^4
\tabularnewline[0.5pt]\hline				
26	 &	smooth    &	          &	1-108 T-154 p T^2-108 p^3 T^3+p^6 T^4
\tabularnewline[0.5pt]\hline				
27	 &	smooth    &	          &	1-156 T+278 p T^2-156 p^3 T^3+p^6 T^4
\tabularnewline[0.5pt]\hline				
28	 &	smooth    &	          &	(1-6 p T+p^3 T^2)(1+78 T+p^3 T^2)
\tabularnewline[0.5pt]\hline				
\tablepostamble				
\tablepreamble{31}				
1	 &	smooth    &	          &	1+20 T-942 p T^2+20 p^3 T^3+p^6 T^4
\tabularnewline[0.5pt]\hline				
2	 &	smooth    &	          &	1+128 T+642 p T^2+128 p^3 T^3+p^6 T^4
\tabularnewline[0.5pt]\hline				
3	 &	smooth    &	          &	1-64 T+594 p T^2-64 p^3 T^3+p^6 T^4
\tabularnewline[0.5pt]\hline				
4	 &	smooth    &	          &	1+74 T+1074 p T^2+74 p^3 T^3+p^6 T^4
\tabularnewline[0.5pt]\hline				
5	 &	singular  &	-\frac{1}{6}\+&	(1-p T) (1+136 T+p^3 T^2)
\tabularnewline[0.5pt]\hline				
6	 &	singular  &	-\frac{1}{5}\+&	(1-p T) (1+172 T+p^3 T^2)
\tabularnewline[0.5pt]\hline				
7	 &	smooth    &	          &	1+26 T-1206 p T^2+26 p^3 T^3+p^6 T^4
\tabularnewline[0.5pt]\hline				
8	 &	singular  &	\frac{1}{4}&	(1-p T) (1-152 T+p^3 T^2)
\tabularnewline[0.5pt]\hline				
9	 &	smooth    &	          &	1-40 T+834 p T^2-40 p^3 T^3+p^6 T^4
\tabularnewline[0.5pt]\hline				
10	 &	smooth    &	          &	1+128 T-222 p T^2+128 p^3 T^3+p^6 T^4
\tabularnewline[0.5pt]\hline				
11	 &	smooth    &	          &	1+236 T+2082 p T^2+236 p^3 T^3+p^6 T^4
\tabularnewline[0.5pt]\hline				
12	 &	smooth    &	          &	1+20 T-1374 p T^2+20 p^3 T^3+p^6 T^4
\tabularnewline[0.5pt]\hline				
13	 &	singular  &	\frac{1}{12}&	(1-p T) (1-62 T+p^3 T^2)
\tabularnewline[0.5pt]\hline				
14	 &	smooth    &	          &	1+134 T+234 p T^2+134 p^3 T^3+p^6 T^4
\tabularnewline[0.5pt]\hline				
15	 &	smooth    &	          &	1-88 T+138 p T^2-88 p^3 T^3+p^6 T^4
\tabularnewline[0.5pt]\hline				
16	 &	smooth    &	          &	1+188 T+450 p T^2+188 p^3 T^3+p^6 T^4
\tabularnewline[0.5pt]\hline				
17	 &	smooth$^*$&	\frac{3}{2}&	  
\tabularnewline[0.5pt]\hline				
18	 &	smooth    &	          &	1+260 T+1170 p T^2+260 p^3 T^3+p^6 T^4
\tabularnewline[0.5pt]\hline				
19	 &	smooth    &	          &	1-292 T+1554 p T^2-292 p^3 T^3+p^6 T^4
\tabularnewline[0.5pt]\hline				
20	 &	smooth    &	          &	1-10 T-18 p^2 T^2-10 p^3 T^3+p^6 T^4
\tabularnewline[0.5pt]\hline				
21	 &	singular  &	\frac{1}{3}&	(1-p T) (1+160 T+p^3 T^2)
\tabularnewline[0.5pt]\hline				
22	 &	smooth    &	          &	1+8 T+810 p T^2+8 p^3 T^3+p^6 T^4
\tabularnewline[0.5pt]\hline				
23	 &	singular  &	-\frac{1}{4}\+&	(1-p T) (1-236 T+p^3 T^2)
\tabularnewline[0.5pt]\hline				
24	 &	smooth    &	          &	1+146 T+498 p T^2+146 p^3 T^3+p^6 T^4
\tabularnewline[0.5pt]\hline				
25	 &	smooth    &	          &	1+56 T+786 p T^2+56 p^3 T^3+p^6 T^4
\tabularnewline[0.5pt]\hline				
26	 &	smooth    &	          &	1+200 T+1362 p T^2+200 p^3 T^3+p^6 T^4
\tabularnewline[0.5pt]\hline				
27	 &	smooth    &	          &	1+260 T+1890 p T^2+260 p^3 T^3+p^6 T^4
\tabularnewline[0.5pt]\hline				
28	 &	smooth    &	          &	(1+4 p T+p^3 T^2)(1-224 T+p^3 T^2)
\tabularnewline[0.5pt]\hline				
29	 &	smooth    &	          &	1+92 T+66 p T^2+92 p^3 T^3+p^6 T^4
\tabularnewline[0.5pt]\hline				
30	 &	smooth    &	          &	1+32 T+258 p T^2+32 p^3 T^3+p^6 T^4
\tabularnewline[0.5pt]\hline				
\tablepostamble				
\tablepreamble{37}				
1	 &	smooth    &	          &	(1-2 p T+p^3 T^2)(1-218 T+p^3 T^2)
\tabularnewline[0.5pt]\hline				
2	 &	smooth    &	          &	1+26 T-1062 p T^2+26 p^3 T^3+p^6 T^4
\tabularnewline[0.5pt]\hline				
3	 &	smooth    &	          &	1-196 T+246 p T^2-196 p^3 T^3+p^6 T^4
\tabularnewline[0.5pt]\hline				
4	 &	smooth    &	          &	1+164 T+822 p T^2+164 p^3 T^3+p^6 T^4
\tabularnewline[0.5pt]\hline				
5	 &	smooth    &	          &	1-316 T+2358 p T^2-316 p^3 T^3+p^6 T^4
\tabularnewline[0.5pt]\hline				
6	 &	singular  &	-\frac{1}{6}\+&	(1-p T) (1+214 T+p^3 T^2)
\tabularnewline[0.5pt]\hline				
7	 &	smooth    &	          &	1-52 T-1338 p T^2-52 p^3 T^3+p^6 T^4
\tabularnewline[0.5pt]\hline				
8	 &	smooth    &	          &	1-40 T+78 p T^2-40 p^3 T^3+p^6 T^4
\tabularnewline[0.5pt]\hline				
9	 &	singular  &	-\frac{1}{4}\+&	(1-p T) (1-146 T+p^3 T^2)
\tabularnewline[0.5pt]\hline				
10	 &	smooth    &	          &	1+260 T+1854 p T^2+260 p^3 T^3+p^6 T^4
\tabularnewline[0.5pt]\hline				
11	 &	smooth    &	          &	1-52 T+2262 p T^2-52 p^3 T^3+p^6 T^4
\tabularnewline[0.5pt]\hline				
12	 &	smooth    &	          &	1+104 T+510 p T^2+104 p^3 T^3+p^6 T^4
\tabularnewline[0.5pt]\hline				
13	 &	smooth    &	          &	1-76 T-1002 p T^2-76 p^3 T^3+p^6 T^4
\tabularnewline[0.5pt]\hline				
14	 &	smooth    &	          &	(1+10 p T+p^3 T^2)(1+46 T+p^3 T^2)
\tabularnewline[0.5pt]\hline				
15	 &	smooth    &	          &	(1-2 p T+p^3 T^2)(1+178 T+p^3 T^2)
\tabularnewline[0.5pt]\hline				
16	 &	smooth    &	          &	1-124 T-186 p T^2-124 p^3 T^3+p^6 T^4
\tabularnewline[0.5pt]\hline				
17	 &	smooth    &	          &	1+14 T-678 p T^2+14 p^3 T^3+p^6 T^4
\tabularnewline[0.5pt]\hline				
18	 &	smooth    &	          &	1-10 T+1962 p T^2-10 p^3 T^3+p^6 T^4
\tabularnewline[0.5pt]\hline				
19	 &	smooth    &	          &	1+104 T-210 p T^2+104 p^3 T^3+p^6 T^4
\tabularnewline[0.5pt]\hline				
20	 &	smooth$^*$&	\frac{3}{2}&	  
\tabularnewline[0.5pt]\hline				
21	 &	smooth    &	          &	1-196 T+1254 p T^2-196 p^3 T^3+p^6 T^4
\tabularnewline[0.5pt]\hline				
22	 &	singular  &	-\frac{1}{5}\+&	(1-p T) (1+58 T+p^3 T^2)
\tabularnewline[0.5pt]\hline				
23	 &	smooth    &	          &	1+320 T+1374 p T^2+320 p^3 T^3+p^6 T^4
\tabularnewline[0.5pt]\hline				
24	 &	smooth    &	          &	1-28 T+558 p T^2-28 p^3 T^3+p^6 T^4
\tabularnewline[0.5pt]\hline				
25	 &	singular  &	\frac{1}{3}&	(1-p T) (1-398 T+p^3 T^2)
\tabularnewline[0.5pt]\hline				
26	 &	smooth    &	          &	1-22 T-1902 p T^2-22 p^3 T^3+p^6 T^4
\tabularnewline[0.5pt]\hline				
27	 &	smooth    &	          &	1+284 T+1734 p T^2+284 p^3 T^3+p^6 T^4
\tabularnewline[0.5pt]\hline				
28	 &	singular  &	\frac{1}{4}&	(1-p T) (1+34 T+p^3 T^2)
\tabularnewline[0.5pt]\hline				
29	 &	smooth    &	          &	1+146 T+786 p T^2+146 p^3 T^3+p^6 T^4
\tabularnewline[0.5pt]\hline				
30	 &	smooth    &	          &	1+2 p T+1362 p T^2+2 p^4 T^3+p^6 T^4
\tabularnewline[0.5pt]\hline				
31	 &	smooth    &	          &	(1+10 p T+p^3 T^2)(1-362 T+p^3 T^2)
\tabularnewline[0.5pt]\hline				
32	 &	smooth    &	          &	1+110 T+1290 p T^2+110 p^3 T^3+p^6 T^4
\tabularnewline[0.5pt]\hline				
33	 &	smooth    &	          &	1+404 T+3294 p T^2+404 p^3 T^3+p^6 T^4
\tabularnewline[0.5pt]\hline				
34	 &	singular  &	\frac{1}{12}&	(1-p T) (1-44 T+p^3 T^2)
\tabularnewline[0.5pt]\hline				
35	 &	smooth    &	          &	1+176 T-354 p T^2+176 p^3 T^3+p^6 T^4
\tabularnewline[0.5pt]\hline				
36	 &	smooth    &	          &	(1-2 p T+p^3 T^2)(1+394 T+p^3 T^2)
\tabularnewline[0.5pt]\hline				
\tablepostamble				
\tablepreamble{41}				
1	 &	smooth    &	          &	1+78 T+842 p T^2+78 p^3 T^3+p^6 T^4
\tabularnewline[0.5pt]\hline				
2	 &	smooth    &	          &	1+354 T+2138 p T^2+354 p^3 T^3+p^6 T^4
\tabularnewline[0.5pt]\hline				
3	 &	smooth    &	          &	1+72 T+14 p T^2+72 p^3 T^3+p^6 T^4
\tabularnewline[0.5pt]\hline				
4	 &	smooth    &	          &	1-330 T+2786 p T^2-330 p^3 T^3+p^6 T^4
\tabularnewline[0.5pt]\hline				
5	 &	smooth    &	          &	1-468 T+2966 p T^2-468 p^3 T^3+p^6 T^4
\tabularnewline[0.5pt]\hline				
6	 &	smooth    &	          &	1+138 T-94 p T^2+138 p^3 T^3+p^6 T^4
\tabularnewline[0.5pt]\hline				
7	 &	smooth    &	          &	1-48 T+1166 p T^2-48 p^3 T^3+p^6 T^4
\tabularnewline[0.5pt]\hline				
8	 &	singular  &	-\frac{1}{5}\+&	(1+p T) (1+342 T+p^3 T^2)
\tabularnewline[0.5pt]\hline				
9	 &	smooth    &	          &	1+150 T+194 p T^2+150 p^3 T^3+p^6 T^4
\tabularnewline[0.5pt]\hline				
10	 &	singular  &	-\frac{1}{4}\+&	(1-p T) (1-126 T+p^3 T^2)
\tabularnewline[0.5pt]\hline				
11	 &	smooth    &	          &	1-558 T+4298 p T^2-558 p^3 T^3+p^6 T^4
\tabularnewline[0.5pt]\hline				
12	 &	smooth    &	          &	1-192 T+1166 p T^2-192 p^3 T^3+p^6 T^4
\tabularnewline[0.5pt]\hline				
13	 &	smooth    &	          &	1+174 T+1922 p T^2+174 p^3 T^3+p^6 T^4
\tabularnewline[0.5pt]\hline				
14	 &	singular  &	\frac{1}{3}&	(1-p T) (1+318 T+p^3 T^2)
\tabularnewline[0.5pt]\hline				
15	 &	smooth    &	          &	1-108 T-634 p T^2-108 p^3 T^3+p^6 T^4
\tabularnewline[0.5pt]\hline				
16	 &	smooth    &	          &	1+498 T+3722 p T^2+498 p^3 T^3+p^6 T^4
\tabularnewline[0.5pt]\hline				
17	 &	smooth    &	          &	1-420 T+3830 p T^2-420 p^3 T^3+p^6 T^4
\tabularnewline[0.5pt]\hline				
18	 &	smooth    &	          &	1-6 p T+3146 p T^2-6 p^4 T^3+p^6 T^4
\tabularnewline[0.5pt]\hline				
19	 &	smooth    &	          &	1+96 T-418 p T^2+96 p^3 T^3+p^6 T^4
\tabularnewline[0.5pt]\hline				
20	 &	smooth    &	          &	1-258 T+626 p T^2-258 p^3 T^3+p^6 T^4
\tabularnewline[0.5pt]\hline				
21	 &	smooth    &	          &	1-36 T+1094 p T^2-36 p^3 T^3+p^6 T^4
\tabularnewline[0.5pt]\hline				
22	 &	smooth$^*$&	\frac{3}{2}&	  
\tabularnewline[0.5pt]\hline				
23	 &	smooth    &	          &	1+168 T+2750 p T^2+168 p^3 T^3+p^6 T^4
\tabularnewline[0.5pt]\hline				
24	 &	singular  &	\frac{1}{12}&	(1-p T) (1+138 T+p^3 T^2)
\tabularnewline[0.5pt]\hline				
25	 &	smooth    &	          &	1+486 T+3506 p T^2+486 p^3 T^3+p^6 T^4
\tabularnewline[0.5pt]\hline				
26	 &	smooth    &	          &	1-240 T+1814 p T^2-240 p^3 T^3+p^6 T^4
\tabularnewline[0.5pt]\hline				
27	 &	smooth    &	          &	1+462 T+4442 p T^2+462 p^3 T^3+p^6 T^4
\tabularnewline[0.5pt]\hline				
28	 &	smooth    &	          &	1+564 T+4550 p T^2+564 p^3 T^3+p^6 T^4
\tabularnewline[0.5pt]\hline				
29	 &	smooth    &	          &	1-276 T+2822 p T^2-276 p^3 T^3+p^6 T^4
\tabularnewline[0.5pt]\hline				
30	 &	smooth    &	          &	1+372 T+3542 p T^2+372 p^3 T^3+p^6 T^4
\tabularnewline[0.5pt]\hline				
31	 &	singular  &	\frac{1}{4}&	(1+p T) (1+438 T+p^3 T^2)
\tabularnewline[0.5pt]\hline				
32	 &	smooth    &	          &	1-396 T+2246 p T^2-396 p^3 T^3+p^6 T^4
\tabularnewline[0.5pt]\hline				
33	 &	smooth    &	          &	1-180 T+2390 p T^2-180 p^3 T^3+p^6 T^4
\tabularnewline[0.5pt]\hline				
34	 &	singular  &	-\frac{1}{6}\+&	(1-p T) (1+150 T+p^3 T^2)
\tabularnewline[0.5pt]\hline				
35	 &	smooth    &	          &	1+240 T+2174 p T^2+240 p^3 T^3+p^6 T^4
\tabularnewline[0.5pt]\hline				
36	 &	smooth    &	          &	1+1742 p T^2+p^6 T^4
\tabularnewline[0.5pt]\hline				
37	 &	smooth    &	          &	1+156 T+2894 p T^2+156 p^3 T^3+p^6 T^4
\tabularnewline[0.5pt]\hline				
38	 &	smooth    &	          &	1-120 T+734 p T^2-120 p^3 T^3+p^6 T^4
\tabularnewline[0.5pt]\hline				
39	 &	smooth    &	          &	1-84 T+806 p T^2-84 p^3 T^3+p^6 T^4
\tabularnewline[0.5pt]\hline				
40	 &	smooth    &	          &	1+312 T+878 p T^2+312 p^3 T^3+p^6 T^4
\tabularnewline[0.5pt]\hline				
\tablepostamble				
\tablepreamble{43}				
1	 &	smooth    &	          &	(1+4 p T+p^3 T^2)(1-152 T+p^3 T^2)
\tabularnewline[0.5pt]\hline				
2	 &	smooth    &	          &	1+188 T-990 p T^2+188 p^3 T^3+p^6 T^4
\tabularnewline[0.5pt]\hline				
3	 &	smooth    &	          &	1-268 T+3018 p T^2-268 p^3 T^3+p^6 T^4
\tabularnewline[0.5pt]\hline				
4	 &	smooth    &	          &	1+200 T+2370 p T^2+200 p^3 T^3+p^6 T^4
\tabularnewline[0.5pt]\hline				
5	 &	smooth    &	          &	1-256 T+834 p T^2-256 p^3 T^3+p^6 T^4
\tabularnewline[0.5pt]\hline				
6	 &	smooth    &	          &	(1-8 p T+p^3 T^2)(1+484 T+p^3 T^2)
\tabularnewline[0.5pt]\hline				
7	 &	singular  &	-\frac{1}{6}\+&	(1-p T) (1+292 T+p^3 T^2)
\tabularnewline[0.5pt]\hline				
8	 &	smooth    &	          &	1+110 T+1146 p T^2+110 p^3 T^3+p^6 T^4
\tabularnewline[0.5pt]\hline				
9	 &	smooth    &	          &	1-16 T-1662 p T^2-16 p^3 T^3+p^6 T^4
\tabularnewline[0.5pt]\hline				
10	 &	smooth    &	          &	1+74 T+714 p T^2+74 p^3 T^3+p^6 T^4
\tabularnewline[0.5pt]\hline				
11	 &	singular  &	\frac{1}{4}&	(1-p T) (1-32 T+p^3 T^2)
\tabularnewline[0.5pt]\hline				
12	 &	smooth    &	          &	1+572 T+4434 p T^2+572 p^3 T^3+p^6 T^4
\tabularnewline[0.5pt]\hline				
13	 &	smooth    &	          &	1-190 T-198 p T^2-190 p^3 T^3+p^6 T^4
\tabularnewline[0.5pt]\hline				
14	 &	smooth    &	          &	1+284 T+186 p T^2+284 p^3 T^3+p^6 T^4
\tabularnewline[0.5pt]\hline				
15	 &	smooth    &	          &	1+92 T-798 p T^2+92 p^3 T^3+p^6 T^4
\tabularnewline[0.5pt]\hline				
16	 &	smooth    &	          &	1-646 T+5682 p T^2-646 p^3 T^3+p^6 T^4
\tabularnewline[0.5pt]\hline				
17	 &	singular  &	-\frac{1}{5}\+&	(1-p T) (1+148 T+p^3 T^2)
\tabularnewline[0.5pt]\hline				
18	 &	singular  &	\frac{1}{12}&	(1-p T) (1-428 T+p^3 T^2)
\tabularnewline[0.5pt]\hline				
19	 &	smooth    &	          &	1-304 T+3810 p T^2-304 p^3 T^3+p^6 T^4
\tabularnewline[0.5pt]\hline				
20	 &	smooth    &	          &	1+110 T+1434 p T^2+110 p^3 T^3+p^6 T^4
\tabularnewline[0.5pt]\hline				
21	 &	smooth    &	          &	1+764 T+6642 p T^2+764 p^3 T^3+p^6 T^4
\tabularnewline[0.5pt]\hline				
22	 &	smooth    &	          &	1+128 T+1002 p T^2+128 p^3 T^3+p^6 T^4
\tabularnewline[0.5pt]\hline				
23	 &	smooth$^*$&	\frac{3}{2}&	  
\tabularnewline[0.5pt]\hline				
24	 &	smooth    &	          &	1+104 T+6 p^2 T^2+104 p^3 T^3+p^6 T^4
\tabularnewline[0.5pt]\hline				
25	 &	smooth    &	          &	1-76 T-606 p T^2-76 p^3 T^3+p^6 T^4
\tabularnewline[0.5pt]\hline				
26	 &	smooth    &	          &	1+38 T-2166 p T^2+38 p^3 T^3+p^6 T^4
\tabularnewline[0.5pt]\hline				
27	 &	smooth    &	          &	1+182 T+138 p T^2+182 p^3 T^3+p^6 T^4
\tabularnewline[0.5pt]\hline				
28	 &	smooth    &	          &	1+368 T+2754 p T^2+368 p^3 T^3+p^6 T^4
\tabularnewline[0.5pt]\hline				
29	 &	singular  &	\frac{1}{3}&	(1-p T) (1+268 T+p^3 T^2)
\tabularnewline[0.5pt]\hline				
30	 &	smooth    &	          &	1+176 T-606 p T^2+176 p^3 T^3+p^6 T^4
\tabularnewline[0.5pt]\hline				
31	 &	smooth    &	          &	1+284 T+3210 p T^2+284 p^3 T^3+p^6 T^4
\tabularnewline[0.5pt]\hline				
32	 &	singular  &	-\frac{1}{4}\+&	(1-p T) (1+376 T+p^3 T^2)
\tabularnewline[0.5pt]\hline				
33	 &	smooth    &	          &	(1-8 p T+p^3 T^2)(1+388 T+p^3 T^2)
\tabularnewline[0.5pt]\hline				
34	 &	smooth    &	          &	1+536 T+3858 p T^2+536 p^3 T^3+p^6 T^4
\tabularnewline[0.5pt]\hline				
35	 &	smooth    &	          &	1-544 T+4866 p T^2-544 p^3 T^3+p^6 T^4
\tabularnewline[0.5pt]\hline				
36	 &	smooth    &	          &	1-112 T-750 p T^2-112 p^3 T^3+p^6 T^4
\tabularnewline[0.5pt]\hline				
37	 &	smooth    &	          &	1+8 T-1134 p T^2+8 p^3 T^3+p^6 T^4
\tabularnewline[0.5pt]\hline				
38	 &	smooth    &	          &	1+302 T+3282 p T^2+302 p^3 T^3+p^6 T^4
\tabularnewline[0.5pt]\hline				
39	 &	smooth    &	          &	1+8 T+1026 p T^2+8 p^3 T^3+p^6 T^4
\tabularnewline[0.5pt]\hline				
40	 &	smooth    &	          &	(1-2 p T+p^3 T^2)(1+220 T+p^3 T^2)
\tabularnewline[0.5pt]\hline				
41	 &	smooth    &	          &	1-556 T+4530 p T^2-556 p^3 T^3+p^6 T^4
\tabularnewline[0.5pt]\hline				
42	 &	smooth    &	          &	1-58 T+2922 p T^2-58 p^3 T^3+p^6 T^4
\tabularnewline[0.5pt]\hline				
\tablepostamble				
\tablepreamble{47}				
1	 &	smooth    &	          &	1+252 T-766 p T^2+252 p^3 T^3+p^6 T^4
\tabularnewline[0.5pt]\hline				
2	 &	smooth    &	          &	1-336 T+3122 p T^2-336 p^3 T^3+p^6 T^4
\tabularnewline[0.5pt]\hline				
3	 &	smooth    &	          &	1+582 T+4346 p T^2+582 p^3 T^3+p^6 T^4
\tabularnewline[0.5pt]\hline				
4	 &	singular  &	\frac{1}{12}&	(1-p T) (1+516 T+p^3 T^2)
\tabularnewline[0.5pt]\hline				
5	 &	smooth    &	          &	1+480 T+3842 p T^2+480 p^3 T^3+p^6 T^4
\tabularnewline[0.5pt]\hline				
6	 &	smooth    &	          &	1+720 T+6650 p T^2+720 p^3 T^3+p^6 T^4
\tabularnewline[0.5pt]\hline				
7	 &	smooth    &	          &	1-264 T+3842 p T^2-264 p^3 T^3+p^6 T^4
\tabularnewline[0.5pt]\hline				
8	 &	smooth    &	          &	1-414 T+3554 p T^2-414 p^3 T^3+p^6 T^4
\tabularnewline[0.5pt]\hline				
9	 &	smooth    &	          &	1-678 T+5714 p T^2-678 p^3 T^3+p^6 T^4
\tabularnewline[0.5pt]\hline				
10	 &	smooth    &	          &	1+96 T+1826 p T^2+96 p^3 T^3+p^6 T^4
\tabularnewline[0.5pt]\hline				
11	 &	smooth    &	          &	1-348 T+1970 p T^2-348 p^3 T^3+p^6 T^4
\tabularnewline[0.5pt]\hline				
12	 &	singular  &	\frac{1}{4}&	(1+p T) (1+204 T+p^3 T^2)
\tabularnewline[0.5pt]\hline				
13	 &	smooth    &	          &	1-270 T+2618 p T^2-270 p^3 T^3+p^6 T^4
\tabularnewline[0.5pt]\hline				
14	 &	smooth    &	          &	1+180 T+4130 p T^2+180 p^3 T^3+p^6 T^4
\tabularnewline[0.5pt]\hline				
15	 &	smooth    &	          &	1+378 T+2258 p T^2+378 p^3 T^3+p^6 T^4
\tabularnewline[0.5pt]\hline				
16	 &	singular  &	\frac{1}{3}&	(1-p T) (1-240 T+p^3 T^2)
\tabularnewline[0.5pt]\hline				
17	 &	smooth    &	          &	1+228 T+1322 p T^2+228 p^3 T^3+p^6 T^4
\tabularnewline[0.5pt]\hline				
18	 &	smooth    &	          &	1+144 T+1538 p T^2+144 p^3 T^3+p^6 T^4
\tabularnewline[0.5pt]\hline				
19	 &	smooth    &	          &	1+174 T+2474 p T^2+174 p^3 T^3+p^6 T^4
\tabularnewline[0.5pt]\hline				
20	 &	smooth    &	          &	1+510 T+2762 p T^2+510 p^3 T^3+p^6 T^4
\tabularnewline[0.5pt]\hline				
21	 &	smooth    &	          &	1-384 T+4706 p T^2-384 p^3 T^3+p^6 T^4
\tabularnewline[0.5pt]\hline				
22	 &	smooth    &	          &	1-180 T+3698 p T^2-180 p^3 T^3+p^6 T^4
\tabularnewline[0.5pt]\hline				
23	 &	smooth    &	          &	1-96 T+1250 p T^2-96 p^3 T^3+p^6 T^4
\tabularnewline[0.5pt]\hline				
24	 &	smooth    &	          &	1+264 T+3410 p T^2+264 p^3 T^3+p^6 T^4
\tabularnewline[0.5pt]\hline				
25	 &	smooth$^*$&	\frac{3}{2}&	  
\tabularnewline[0.5pt]\hline				
26	 &	smooth    &	          &	1-144 T-262 p T^2-144 p^3 T^3+p^6 T^4
\tabularnewline[0.5pt]\hline				
27	 &	smooth    &	          &	1+204 T+3554 p T^2+204 p^3 T^3+p^6 T^4
\tabularnewline[0.5pt]\hline				
28	 &	singular  &	-\frac{1}{5}\+&	(1+p T) (1-288 T+p^3 T^2)
\tabularnewline[0.5pt]\hline				
29	 &	smooth    &	          &	1+174 T+386 p T^2+174 p^3 T^3+p^6 T^4
\tabularnewline[0.5pt]\hline				
30	 &	smooth    &	          &	1+156 T+1826 p T^2+156 p^3 T^3+p^6 T^4
\tabularnewline[0.5pt]\hline				
31	 &	smooth    &	          &	1+528 T+3410 p T^2+528 p^3 T^3+p^6 T^4
\tabularnewline[0.5pt]\hline				
32	 &	smooth    &	          &	1-264 T+1682 p T^2-264 p^3 T^3+p^6 T^4
\tabularnewline[0.5pt]\hline				
33	 &	smooth    &	          &	1-114 T+2402 p T^2-114 p^3 T^3+p^6 T^4
\tabularnewline[0.5pt]\hline				
34	 &	smooth    &	          &	1-96 T+1466 p T^2-96 p^3 T^3+p^6 T^4
\tabularnewline[0.5pt]\hline				
35	 &	singular  &	-\frac{1}{4}\+&	(1-p T) (1+12 T+p^3 T^2)
\tabularnewline[0.5pt]\hline				
36	 &	smooth    &	          &	1+156 T+2258 p T^2+156 p^3 T^3+p^6 T^4
\tabularnewline[0.5pt]\hline				
37	 &	smooth    &	          &	(1+12 p T+p^3 T^2)(1+228 T+p^3 T^2)
\tabularnewline[0.5pt]\hline				
38	 &	smooth    &	          &	1-588 T+5426 p T^2-588 p^3 T^3+p^6 T^4
\tabularnewline[0.5pt]\hline				
39	 &	singular  &	-\frac{1}{6}\+&	(1-p T) (1+72 T+p^3 T^2)
\tabularnewline[0.5pt]\hline				
40	 &	smooth    &	          &	1+108 T+2402 p T^2+108 p^3 T^3+p^6 T^4
\tabularnewline[0.5pt]\hline				
41	 &	smooth    &	          &	1-132 T-2494 p T^2-132 p^3 T^3+p^6 T^4
\tabularnewline[0.5pt]\hline				
42	 &	smooth    &	          &	1+6 p T-46 p T^2+6 p^4 T^3+p^6 T^4
\tabularnewline[0.5pt]\hline				
43	 &	smooth    &	          &	1-252 T+1826 p T^2-252 p^3 T^3+p^6 T^4
\tabularnewline[0.5pt]\hline				
44	 &	smooth    &	          &	1-156 T+98 p T^2-156 p^3 T^3+p^6 T^4
\tabularnewline[0.5pt]\hline				
45	 &	smooth    &	          &	1+132 T-50 p^2 T^2+132 p^3 T^3+p^6 T^4
\tabularnewline[0.5pt]\hline				
46	 &	smooth    &	          &	1+18 T+2546 p T^2+18 p^3 T^3+p^6 T^4
\tabularnewline[0.5pt]\hline				
\tablepostamble				
\tablepreamble{53}				
1	 &	smooth    &	          &	1-18 T+3890 p T^2-18 p^3 T^3+p^6 T^4
\tabularnewline[0.5pt]\hline				
2	 &	smooth    &	          &	1+6 T+4970 p T^2+6 p^3 T^3+p^6 T^4
\tabularnewline[0.5pt]\hline				
3	 &	smooth    &	          &	1+186 T+1946 p T^2+186 p^3 T^3+p^6 T^4
\tabularnewline[0.5pt]\hline				
4	 &	smooth    &	          &	1-168 T-466 p T^2-168 p^3 T^3+p^6 T^4
\tabularnewline[0.5pt]\hline				
5	 &	smooth    &	          &	1+396 T+4646 p T^2+396 p^3 T^3+p^6 T^4
\tabularnewline[0.5pt]\hline				
6	 &	smooth    &	          &	1+762 T+6986 p T^2+762 p^3 T^3+p^6 T^4
\tabularnewline[0.5pt]\hline				
7	 &	smooth    &	          &	1-618 T+5330 p T^2-618 p^3 T^3+p^6 T^4
\tabularnewline[0.5pt]\hline				
8	 &	smooth    &	          &	1+684 T+5366 p T^2+684 p^3 T^3+p^6 T^4
\tabularnewline[0.5pt]\hline				
9	 &	smooth    &	          &	1-90 T+1082 p T^2-90 p^3 T^3+p^6 T^4
\tabularnewline[0.5pt]\hline				
10	 &	smooth    &	          &	1-222 T+2882 p T^2-222 p^3 T^3+p^6 T^4
\tabularnewline[0.5pt]\hline				
11	 &	smooth    &	          &	1+5294 p T^2+p^6 T^4
\tabularnewline[0.5pt]\hline				
12	 &	smooth    &	          &	1+300 T-682 p T^2+300 p^3 T^3+p^6 T^4
\tabularnewline[0.5pt]\hline				
13	 &	singular  &	-\frac{1}{4}\+&	(1-p T) (1-174 T+p^3 T^2)
\tabularnewline[0.5pt]\hline				
14	 &	smooth    &	          &	1+120 T+3566 p T^2+120 p^3 T^3+p^6 T^4
\tabularnewline[0.5pt]\hline				
15	 &	smooth    &	          &	1+84 T-2698 p T^2+84 p^3 T^3+p^6 T^4
\tabularnewline[0.5pt]\hline				
16	 &	smooth    &	          &	1+84 T-538 p T^2+84 p^3 T^3+p^6 T^4
\tabularnewline[0.5pt]\hline				
17	 &	smooth    &	          &	1+816 T+7022 p T^2+816 p^3 T^3+p^6 T^4
\tabularnewline[0.5pt]\hline				
18	 &	singular  &	\frac{1}{3}&	(1-p T) (1+498 T+p^3 T^2)
\tabularnewline[0.5pt]\hline				
19	 &	smooth    &	          &	1-96 T-4066 p T^2-96 p^3 T^3+p^6 T^4
\tabularnewline[0.5pt]\hline				
20	 &	smooth    &	          &	1-120 T+4718 p T^2-120 p^3 T^3+p^6 T^4
\tabularnewline[0.5pt]\hline				
21	 &	singular  &	-\frac{1}{5}\+&	(1+p T) (1-318 T+p^3 T^2)
\tabularnewline[0.5pt]\hline				
22	 &	smooth    &	          &	1+228 T+2774 p T^2+228 p^3 T^3+p^6 T^4
\tabularnewline[0.5pt]\hline				
23	 &	smooth    &	          &	1+288 T+70 p^2 T^2+288 p^3 T^3+p^6 T^4
\tabularnewline[0.5pt]\hline				
24	 &	smooth    &	          &	1-348 T+2198 p T^2-348 p^3 T^3+p^6 T^4
\tabularnewline[0.5pt]\hline				
25	 &	smooth    &	          &	1-12 T+2486 p T^2-12 p^3 T^3+p^6 T^4
\tabularnewline[0.5pt]\hline				
26	 &	smooth    &	          &	1+24 T+1694 p T^2+24 p^3 T^3+p^6 T^4
\tabularnewline[0.5pt]\hline				
27	 &	smooth    &	          &	1-108 T+2054 p T^2-108 p^3 T^3+p^6 T^4
\tabularnewline[0.5pt]\hline				
28	 &	smooth$^*$&	\frac{3}{2}&	  
\tabularnewline[0.5pt]\hline				
29	 &	smooth    &	          &	1+324 T+4646 p T^2+324 p^3 T^3+p^6 T^4
\tabularnewline[0.5pt]\hline				
30	 &	smooth    &	          &	1-558 T+5258 p T^2-558 p^3 T^3+p^6 T^4
\tabularnewline[0.5pt]\hline				
31	 &	singular  &	\frac{1}{12}&	(1-p T) (1-174 T+p^3 T^2)
\tabularnewline[0.5pt]\hline				
32	 &	smooth    &	          &	1+552 T+2990 p T^2+552 p^3 T^3+p^6 T^4
\tabularnewline[0.5pt]\hline				
33	 &	smooth    &	          &	1-222 T-286 p T^2-222 p^3 T^3+p^6 T^4
\tabularnewline[0.5pt]\hline				
34	 &	smooth    &	          &	1+204 T-1834 p T^2+204 p^3 T^3+p^6 T^4
\tabularnewline[0.5pt]\hline				
35	 &	smooth    &	          &	1-156 T+2630 p T^2-156 p^3 T^3+p^6 T^4
\tabularnewline[0.5pt]\hline				
36	 &	smooth    &	          &	1+24 T-34 p T^2+24 p^3 T^3+p^6 T^4
\tabularnewline[0.5pt]\hline				
37	 &	smooth    &	          &	1+90 T+5114 p T^2+90 p^3 T^3+p^6 T^4
\tabularnewline[0.5pt]\hline				
38	 &	smooth    &	          &	1-138 T+1874 p T^2-138 p^3 T^3+p^6 T^4
\tabularnewline[0.5pt]\hline				
39	 &	smooth    &	          &	1+156 T+2414 p T^2+156 p^3 T^3+p^6 T^4
\tabularnewline[0.5pt]\hline				
40	 &	singular  &	\frac{1}{4}&	(1+p T) (1-222 T+p^3 T^2)
\tabularnewline[0.5pt]\hline				
41	 &	smooth    &	          &	1-96 T+3782 p T^2-96 p^3 T^3+p^6 T^4
\tabularnewline[0.5pt]\hline				
42	 &	smooth    &	          &	1-66 T+1370 p T^2-66 p^3 T^3+p^6 T^4
\tabularnewline[0.5pt]\hline				
43	 &	smooth    &	          &	1-222 T+1946 p T^2-222 p^3 T^3+p^6 T^4
\tabularnewline[0.5pt]\hline				
44	 &	singular  &	-\frac{1}{6}\+&	(1-p T) (1+414 T+p^3 T^2)
\tabularnewline[0.5pt]\hline				
45	 &	smooth    &	          &	1+36 T-1834 p T^2+36 p^3 T^3+p^6 T^4
\tabularnewline[0.5pt]\hline				
46	 &	smooth    &	          &	1+804 T+6806 p T^2+804 p^3 T^3+p^6 T^4
\tabularnewline[0.5pt]\hline				
47	 &	smooth    &	          &	1-12 T+2054 p T^2-12 p^3 T^3+p^6 T^4
\tabularnewline[0.5pt]\hline				
48	 &	smooth    &	          &	1+4286 p T^2+p^6 T^4
\tabularnewline[0.5pt]\hline				
49	 &	smooth    &	          &	1+408 T+5294 p T^2+408 p^3 T^3+p^6 T^4
\tabularnewline[0.5pt]\hline				
50	 &	smooth    &	          &	1+42 T+5186 p T^2+42 p^3 T^3+p^6 T^4
\tabularnewline[0.5pt]\hline				
51	 &	smooth    &	          &	1-312 T+2846 p T^2-312 p^3 T^3+p^6 T^4
\tabularnewline[0.5pt]\hline				
52	 &	smooth    &	          &	1+48 T-898 p T^2+48 p^3 T^3+p^6 T^4
\tabularnewline[0.5pt]\hline				
\tablepostamble				
\tablepreamble{59}				
1	 &	smooth    &	          &	1+792 T+6674 p T^2+792 p^3 T^3+p^6 T^4
\tabularnewline[0.5pt]\hline				
2	 &	smooth    &	          &	1-600 T+5954 p T^2-600 p^3 T^3+p^6 T^4
\tabularnewline[0.5pt]\hline				
3	 &	smooth    &	          &	1-636 T+5810 p T^2-636 p^3 T^3+p^6 T^4
\tabularnewline[0.5pt]\hline				
4	 &	smooth    &	          &	1+2498 p T^2+p^6 T^4
\tabularnewline[0.5pt]\hline				
5	 &	singular  &	\frac{1}{12}&	(1-p T) (1+852 T+p^3 T^2)
\tabularnewline[0.5pt]\hline				
6	 &	smooth    &	          &	1+444 T+50 p T^2+444 p^3 T^3+p^6 T^4
\tabularnewline[0.5pt]\hline				
7	 &	smooth    &	          &	1-180 T+4802 p T^2-180 p^3 T^3+p^6 T^4
\tabularnewline[0.5pt]\hline				
8	 &	smooth    &	          &	1-168 T+2210 p T^2-168 p^3 T^3+p^6 T^4
\tabularnewline[0.5pt]\hline				
9	 &	smooth    &	          &	1+192 T-1822 p T^2+192 p^3 T^3+p^6 T^4
\tabularnewline[0.5pt]\hline				
10	 &	smooth    &	          &	1-408 T+4802 p T^2-408 p^3 T^3+p^6 T^4
\tabularnewline[0.5pt]\hline				
11	 &	smooth    &	          &	1-264 T+4514 p T^2-264 p^3 T^3+p^6 T^4
\tabularnewline[0.5pt]\hline				
12	 &	smooth    &	          &	1-564 T+6602 p T^2-564 p^3 T^3+p^6 T^4
\tabularnewline[0.5pt]\hline				
13	 &	smooth    &	          &	1+288 T+3938 p T^2+288 p^3 T^3+p^6 T^4
\tabularnewline[0.5pt]\hline				
14	 &	smooth    &	          &	1+792 T+8546 p T^2+792 p^3 T^3+p^6 T^4
\tabularnewline[0.5pt]\hline				
15	 &	singular  &	\frac{1}{4}&	(1+p T) (1-420 T+p^3 T^2)
\tabularnewline[0.5pt]\hline				
16	 &	smooth    &	          &	1+504 T+6242 p T^2+504 p^3 T^3+p^6 T^4
\tabularnewline[0.5pt]\hline				
17	 &	smooth    &	          &	1-60 T-3982 p T^2-60 p^3 T^3+p^6 T^4
\tabularnewline[0.5pt]\hline				
18	 &	smooth    &	          &	1-144 T-238 p T^2-144 p^3 T^3+p^6 T^4
\tabularnewline[0.5pt]\hline				
19	 &	smooth    &	          &	1+288 T-1102 p T^2+288 p^3 T^3+p^6 T^4
\tabularnewline[0.5pt]\hline				
20	 &	singular  &	\frac{1}{3}&	(1-p T) (1+132 T+p^3 T^2)
\tabularnewline[0.5pt]\hline				
21	 &	smooth    &	          &	1-468 T+4154 p T^2-468 p^3 T^3+p^6 T^4
\tabularnewline[0.5pt]\hline				
22	 &	smooth    &	          &	1-528 T+4226 p T^2-528 p^3 T^3+p^6 T^4
\tabularnewline[0.5pt]\hline				
23	 &	smooth    &	          &	1+432 T+482 p T^2+432 p^3 T^3+p^6 T^4
\tabularnewline[0.5pt]\hline				
24	 &	smooth    &	          &	1-18 T+1994 p T^2-18 p^3 T^3+p^6 T^4
\tabularnewline[0.5pt]\hline				
25	 &	smooth    &	          &	1-816 T+7682 p T^2-816 p^3 T^3+p^6 T^4
\tabularnewline[0.5pt]\hline				
26	 &	smooth    &	          &	1-726 T+5882 p T^2-726 p^3 T^3+p^6 T^4
\tabularnewline[0.5pt]\hline				
27	 &	smooth    &	          &	1+48 T+2498 p T^2+48 p^3 T^3+p^6 T^4
\tabularnewline[0.5pt]\hline				
28	 &	smooth    &	          &	1-870 T+9410 p T^2-870 p^3 T^3+p^6 T^4
\tabularnewline[0.5pt]\hline				
29	 &	smooth    &	          &	1+90 T-3262 p T^2+90 p^3 T^3+p^6 T^4
\tabularnewline[0.5pt]\hline				
30	 &	smooth    &	          &	1-306 T+6458 p T^2-306 p^3 T^3+p^6 T^4
\tabularnewline[0.5pt]\hline				
31	 &	smooth$^*$&	\frac{3}{2}&	  
\tabularnewline[0.5pt]\hline				
32	 &	smooth    &	          &	1+372 T-526 p T^2+372 p^3 T^3+p^6 T^4
\tabularnewline[0.5pt]\hline				
33	 &	smooth    &	          &	1+84 T+3506 p T^2+84 p^3 T^3+p^6 T^4
\tabularnewline[0.5pt]\hline				
34	 &	smooth    &	          &	1+1152 T+11642 p T^2+1152 p^3 T^3+p^6 T^4
\tabularnewline[0.5pt]\hline				
35	 &	smooth    &	          &	1+84 T-1390 p T^2+84 p^3 T^3+p^6 T^4
\tabularnewline[0.5pt]\hline				
36	 &	smooth    &	          &	1-444 T+5810 p T^2-444 p^3 T^3+p^6 T^4
\tabularnewline[0.5pt]\hline				
37	 &	smooth    &	          &	1-174 T+1850 p T^2-174 p^3 T^3+p^6 T^4
\tabularnewline[0.5pt]\hline				
38	 &	smooth    &	          &	(1+12 p T+p^3 T^2)(1-324 T+p^3 T^2)
\tabularnewline[0.5pt]\hline				
39	 &	smooth    &	          &	1+132 T+5378 p T^2+132 p^3 T^3+p^6 T^4
\tabularnewline[0.5pt]\hline				
40	 &	smooth    &	          &	(1+p^3 T^2)(1+780 T+p^3 T^2)
\tabularnewline[0.5pt]\hline				
41	 &	smooth    &	          &	1+444 T+2930 p T^2+444 p^3 T^3+p^6 T^4
\tabularnewline[0.5pt]\hline				
42	 &	smooth    &	          &	1+432 T+1346 p T^2+432 p^3 T^3+p^6 T^4
\tabularnewline[0.5pt]\hline				
43	 &	smooth    &	          &	1-126 T-2974 p T^2-126 p^3 T^3+p^6 T^4
\tabularnewline[0.5pt]\hline				
44	 &	singular  &	-\frac{1}{4}\+&	(1-p T) (1-138 T+p^3 T^2)
\tabularnewline[0.5pt]\hline				
45	 &	smooth    &	          &	1+786 T+6314 p T^2+786 p^3 T^3+p^6 T^4
\tabularnewline[0.5pt]\hline				
46	 &	smooth    &	          &	1-60 T-2542 p T^2-60 p^3 T^3+p^6 T^4
\tabularnewline[0.5pt]\hline				
47	 &	singular  &	-\frac{1}{5}\+&	(1+p T) (1-252 T+p^3 T^2)
\tabularnewline[0.5pt]\hline				
48	 &	smooth    &	          &	1+138 T+3578 p T^2+138 p^3 T^3+p^6 T^4
\tabularnewline[0.5pt]\hline				
49	 &	singular  &	-\frac{1}{6}\+&	(1-p T) (1+744 T+p^3 T^2)
\tabularnewline[0.5pt]\hline				
50	 &	smooth    &	          &	1+984 T+9122 p T^2+984 p^3 T^3+p^6 T^4
\tabularnewline[0.5pt]\hline				
51	 &	smooth    &	          &	1+492 T+4946 p T^2+492 p^3 T^3+p^6 T^4
\tabularnewline[0.5pt]\hline				
52	 &	smooth    &	          &	1-744 T+5522 p T^2-744 p^3 T^3+p^6 T^4
\tabularnewline[0.5pt]\hline				
53	 &	smooth    &	          &	1+978 T+10058 p T^2+978 p^3 T^3+p^6 T^4
\tabularnewline[0.5pt]\hline				
54	 &	smooth    &	          &	1+120 T-526 p T^2+120 p^3 T^3+p^6 T^4
\tabularnewline[0.5pt]\hline				
55	 &	smooth    &	          &	1-552 T+2210 p T^2-552 p^3 T^3+p^6 T^4
\tabularnewline[0.5pt]\hline				
56	 &	smooth    &	          &	1-258 T+1202 p T^2-258 p^3 T^3+p^6 T^4
\tabularnewline[0.5pt]\hline				
57	 &	smooth    &	          &	1+126 T+6818 p T^2+126 p^3 T^3+p^6 T^4
\tabularnewline[0.5pt]\hline				
58	 &	smooth    &	          &	1+300 T+4514 p T^2+300 p^3 T^3+p^6 T^4
\tabularnewline[0.5pt]\hline				
\tablepostamble				
\tablepreamble{61}				
1	 &	smooth    &	          &	1-1192 T+11886 p T^2-1192 p^3 T^3+p^6 T^4
\tabularnewline[0.5pt]\hline				
2	 &	smooth    &	          &	1+80 T-2610 p T^2+80 p^3 T^3+p^6 T^4
\tabularnewline[0.5pt]\hline				
3	 &	smooth    &	          &	1-382 T+3786 p T^2-382 p^3 T^3+p^6 T^4
\tabularnewline[0.5pt]\hline				
4	 &	smooth    &	          &	1-220 T+438 p T^2-220 p^3 T^3+p^6 T^4
\tabularnewline[0.5pt]\hline				
5	 &	smooth    &	          &	1+44 T-1386 p T^2+44 p^3 T^3+p^6 T^4
\tabularnewline[0.5pt]\hline				
6	 &	smooth    &	          &	1+188 T+4518 p T^2+188 p^3 T^3+p^6 T^4
\tabularnewline[0.5pt]\hline				
7	 &	smooth    &	          &	1+788 T+9726 p T^2+788 p^3 T^3+p^6 T^4
\tabularnewline[0.5pt]\hline				
8	 &	smooth    &	          &	(1+10 p T+p^3 T^2)(1-2 T+p^3 T^2)
\tabularnewline[0.5pt]\hline				
9	 &	smooth    &	          &	1+626 T+6162 p T^2+626 p^3 T^3+p^6 T^4
\tabularnewline[0.5pt]\hline				
10	 &	singular  &	-\frac{1}{6}\+&	(1-p T) (1+418 T+p^3 T^2)
\tabularnewline[0.5pt]\hline				
11	 &	smooth    &	          &	1+512 T+3150 p T^2+512 p^3 T^3+p^6 T^4
\tabularnewline[0.5pt]\hline				
12	 &	singular  &	-\frac{1}{5}\+&	(1-p T) (1-110 T+p^3 T^2)
\tabularnewline[0.5pt]\hline				
13	 &	smooth    &	          &	1+590 T+3354 p T^2+590 p^3 T^3+p^6 T^4
\tabularnewline[0.5pt]\hline				
14	 &	smooth    &	          &	1+1202 T+12930 p T^2+1202 p^3 T^3+p^6 T^4
\tabularnewline[0.5pt]\hline				
15	 &	singular  &	-\frac{1}{4}\+&	(1-p T) (1-380 T+p^3 T^2)
\tabularnewline[0.5pt]\hline				
16	 &	smooth    &	          &	1+932 T+8502 p T^2+932 p^3 T^3+p^6 T^4
\tabularnewline[0.5pt]\hline				
17	 &	smooth    &	          &	1-40 T+78 p T^2-40 p^3 T^3+p^6 T^4
\tabularnewline[0.5pt]\hline				
18	 &	smooth    &	          &	1-340 T+6294 p T^2-340 p^3 T^3+p^6 T^4
\tabularnewline[0.5pt]\hline				
19	 &	smooth    &	          &	1-544 T+4974 p T^2-544 p^3 T^3+p^6 T^4
\tabularnewline[0.5pt]\hline				
20	 &	smooth    &	          &	1-412 T+6582 p T^2-412 p^3 T^3+p^6 T^4
\tabularnewline[0.5pt]\hline				
21	 &	smooth    &	          &	1+716 T+4614 p T^2+716 p^3 T^3+p^6 T^4
\tabularnewline[0.5pt]\hline				
22	 &	smooth    &	          &	1+14 T+330 p T^2+14 p^3 T^3+p^6 T^4
\tabularnewline[0.5pt]\hline				
23	 &	smooth    &	          &	(1-14 p T+p^3 T^2)(1+358 T+p^3 T^2)
\tabularnewline[0.5pt]\hline				
24	 &	smooth    &	          &	1-166 T-2262 p T^2-166 p^3 T^3+p^6 T^4
\tabularnewline[0.5pt]\hline				
25	 &	smooth    &	          &	1+428 T+2166 p T^2+428 p^3 T^3+p^6 T^4
\tabularnewline[0.5pt]\hline				
26	 &	smooth    &	          &	1+992 T+10398 p T^2+992 p^3 T^3+p^6 T^4
\tabularnewline[0.5pt]\hline				
27	 &	smooth    &	          &	1+452 T+2262 p T^2+452 p^3 T^3+p^6 T^4
\tabularnewline[0.5pt]\hline				
28	 &	smooth    &	          &	1-796 T+5478 p T^2-796 p^3 T^3+p^6 T^4
\tabularnewline[0.5pt]\hline				
29	 &	smooth    &	          &	1+56 T+894 p T^2+56 p^3 T^3+p^6 T^4
\tabularnewline[0.5pt]\hline				
30	 &	smooth    &	          &	1-196 T-186 p T^2-196 p^3 T^3+p^6 T^4
\tabularnewline[0.5pt]\hline				
31	 &	smooth    &	          &	1+314 T-558 p T^2+314 p^3 T^3+p^6 T^4
\tabularnewline[0.5pt]\hline				
32	 &	smooth$^*$&	\frac{3}{2}&	  
\tabularnewline[0.5pt]\hline				
33	 &	smooth    &	          &	(1+10 p T+p^3 T^2)(1-590 T+p^3 T^2)
\tabularnewline[0.5pt]\hline				
34	 &	smooth    &	          &	1-388 T+2646 p T^2-388 p^3 T^3+p^6 T^4
\tabularnewline[0.5pt]\hline				
35	 &	smooth    &	          &	1-1264 T+13758 p T^2-1264 p^3 T^3+p^6 T^4
\tabularnewline[0.5pt]\hline				
36	 &	smooth    &	          &	1+74 T+3306 p T^2+74 p^3 T^3+p^6 T^4
\tabularnewline[0.5pt]\hline				
37	 &	smooth    &	          &	1-184 T+6 p^2 T^2-184 p^3 T^3+p^6 T^4
\tabularnewline[0.5pt]\hline				
38	 &	smooth    &	          &	1+344 T-1266 p T^2+344 p^3 T^3+p^6 T^4
\tabularnewline[0.5pt]\hline				
39	 &	smooth    &	          &	(1-14 p T+p^3 T^2)(1+772 T+p^3 T^2)
\tabularnewline[0.5pt]\hline				
40	 &	smooth    &	          &	1+944 T+9054 p T^2+944 p^3 T^3+p^6 T^4
\tabularnewline[0.5pt]\hline				
41	 &	singular  &	\frac{1}{3}&	(1-p T) (1-398 T+p^3 T^2)
\tabularnewline[0.5pt]\hline				
42	 &	smooth    &	          &	1+416 T+4926 p T^2+416 p^3 T^3+p^6 T^4
\tabularnewline[0.5pt]\hline				
43	 &	smooth    &	          &	1+164 T+5862 p T^2+164 p^3 T^3+p^6 T^4
\tabularnewline[0.5pt]\hline				
44	 &	smooth    &	          &	1+428 T+2598 p T^2+428 p^3 T^3+p^6 T^4
\tabularnewline[0.5pt]\hline				
45	 &	smooth    &	          &	1+620 T+7110 p T^2+620 p^3 T^3+p^6 T^4
\tabularnewline[0.5pt]\hline				
46	 &	singular  &	\frac{1}{4}&	(1-p T) (1-902 T+p^3 T^2)
\tabularnewline[0.5pt]\hline				
47	 &	smooth    &	          &	1-184 T+2094 p T^2-184 p^3 T^3+p^6 T^4
\tabularnewline[0.5pt]\hline				
48	 &	smooth    &	          &	1+62 T+954 p T^2+62 p^3 T^3+p^6 T^4
\tabularnewline[0.5pt]\hline				
49	 &	smooth    &	          &	1+308 T+6150 p T^2+308 p^3 T^3+p^6 T^4
\tabularnewline[0.5pt]\hline				
50	 &	smooth    &	          &	1+254 T+66 p^2 T^2+254 p^3 T^3+p^6 T^4
\tabularnewline[0.5pt]\hline				
51	 &	smooth    &	          &	1+86 T-1254 p T^2+86 p^3 T^3+p^6 T^4
\tabularnewline[0.5pt]\hline				
52	 &	smooth    &	          &	1-280 T+3942 p T^2-280 p^3 T^3+p^6 T^4
\tabularnewline[0.5pt]\hline				
53	 &	smooth    &	          &	1+644 T+5190 p T^2+644 p^3 T^3+p^6 T^4
\tabularnewline[0.5pt]\hline				
54	 &	smooth    &	          &	1+368 T+1566 p T^2+368 p^3 T^3+p^6 T^4
\tabularnewline[0.5pt]\hline				
55	 &	smooth    &	          &	1+266 T-1398 p T^2+266 p^3 T^3+p^6 T^4
\tabularnewline[0.5pt]\hline				
56	 &	singular  &	\frac{1}{12}&	(1-p T) (1-908 T+p^3 T^2)
\tabularnewline[0.5pt]\hline				
57	 &	smooth    &	          &	1-148 T+5046 p T^2-148 p^3 T^3+p^6 T^4
\tabularnewline[0.5pt]\hline				
58	 &	smooth    &	          &	1+104 T+6 p^2 T^2+104 p^3 T^3+p^6 T^4
\tabularnewline[0.5pt]\hline				
59	 &	smooth    &	          &	1-826 T+8130 p T^2-826 p^3 T^3+p^6 T^4
\tabularnewline[0.5pt]\hline				
60	 &	smooth    &	          &	1+176 T-3522 p T^2+176 p^3 T^3+p^6 T^4
\tabularnewline[0.5pt]\hline				
\tablepostamble				
\tablepreamble{67}				
1	 &	smooth    &	          &	1+284 T+8178 p T^2+284 p^3 T^3+p^6 T^4
\tabularnewline[0.5pt]\hline				
2	 &	smooth    &	          &	1+1112 T+12642 p T^2+1112 p^3 T^3+p^6 T^4
\tabularnewline[0.5pt]\hline				
3	 &	smooth    &	          &	1+488 T-222 p T^2+488 p^3 T^3+p^6 T^4
\tabularnewline[0.5pt]\hline				
4	 &	smooth    &	          &	1+188 T+7794 p T^2+188 p^3 T^3+p^6 T^4
\tabularnewline[0.5pt]\hline				
5	 &	smooth    &	          &	1-58 T+258 p T^2-58 p^3 T^3+p^6 T^4
\tabularnewline[0.5pt]\hline				
6	 &	smooth    &	          &	1+272 T-942 p T^2+272 p^3 T^3+p^6 T^4
\tabularnewline[0.5pt]\hline				
7	 &	smooth    &	          &	1-16 p T+10098 p T^2-16 p^4 T^3+p^6 T^4
\tabularnewline[0.5pt]\hline				
8	 &	smooth    &	          &	1+752 T+4002 p T^2+752 p^3 T^3+p^6 T^4
\tabularnewline[0.5pt]\hline				
9	 &	smooth    &	          &	1-280 T-3870 p T^2-280 p^3 T^3+p^6 T^4
\tabularnewline[0.5pt]\hline				
10	 &	smooth    &	          &	1+284 T+3858 p T^2+284 p^3 T^3+p^6 T^4
\tabularnewline[0.5pt]\hline				
11	 &	singular  &	-\frac{1}{6}\+&	(1-p T) (1-188 T+p^3 T^2)
\tabularnewline[0.5pt]\hline				
12	 &	smooth    &	          &	1+56 T+2946 p T^2+56 p^3 T^3+p^6 T^4
\tabularnewline[0.5pt]\hline				
13	 &	smooth    &	          &	1+272 T-4398 p T^2+272 p^3 T^3+p^6 T^4
\tabularnewline[0.5pt]\hline				
14	 &	smooth    &	          &	1-64 T+4482 p T^2-64 p^3 T^3+p^6 T^4
\tabularnewline[0.5pt]\hline				
15	 &	smooth    &	          &	1-436 T+3570 p T^2-436 p^3 T^3+p^6 T^4
\tabularnewline[0.5pt]\hline				
16	 &	smooth    &	          &	1+62 T+2106 p T^2+62 p^3 T^3+p^6 T^4
\tabularnewline[0.5pt]\hline				
17	 &	singular  &	\frac{1}{4}&	(1-p T) (1+1024 T+p^3 T^2)
\tabularnewline[0.5pt]\hline				
18	 &	smooth    &	          &	1-388 T+4410 p T^2-388 p^3 T^3+p^6 T^4
\tabularnewline[0.5pt]\hline				
19	 &	smooth    &	          &	1+590 T+5658 p T^2+590 p^3 T^3+p^6 T^4
\tabularnewline[0.5pt]\hline				
20	 &	smooth    &	          &	1-184 T+4578 p T^2-184 p^3 T^3+p^6 T^4
\tabularnewline[0.5pt]\hline				
21	 &	smooth    &	          &	1-448 T+1506 p T^2-448 p^3 T^3+p^6 T^4
\tabularnewline[0.5pt]\hline				
22	 &	smooth    &	          &	1+416 T+30 p^2 T^2+416 p^3 T^3+p^6 T^4
\tabularnewline[0.5pt]\hline				
23	 &	smooth    &	          &	1+116 T-270 p T^2+116 p^3 T^3+p^6 T^4
\tabularnewline[0.5pt]\hline				
24	 &	smooth    &	          &	1+926 T+9306 p T^2+926 p^3 T^3+p^6 T^4
\tabularnewline[0.5pt]\hline				
25	 &	smooth    &	          &	1-328 T+1842 p T^2-328 p^3 T^3+p^6 T^4
\tabularnewline[0.5pt]\hline				
26	 &	smooth    &	          &	1-532 T+7938 p T^2-532 p^3 T^3+p^6 T^4
\tabularnewline[0.5pt]\hline				
27	 &	smooth    &	          &	1+302 T-3414 p T^2+302 p^3 T^3+p^6 T^4
\tabularnewline[0.5pt]\hline				
28	 &	singular  &	\frac{1}{12}&	(1-p T) (1+508 T+p^3 T^2)
\tabularnewline[0.5pt]\hline				
29	 &	smooth    &	          &	1-424 T+8154 p T^2-424 p^3 T^3+p^6 T^4
\tabularnewline[0.5pt]\hline				
30	 &	smooth    &	          &	1-784 T+8226 p T^2-784 p^3 T^3+p^6 T^4
\tabularnewline[0.5pt]\hline				
31	 &	smooth    &	          &	1+266 T+1194 p T^2+266 p^3 T^3+p^6 T^4
\tabularnewline[0.5pt]\hline				
32	 &	smooth    &	          &	1+62 T+810 p T^2+62 p^3 T^3+p^6 T^4
\tabularnewline[0.5pt]\hline				
33	 &	smooth    &	          &	1-592 T+6402 p T^2-592 p^3 T^3+p^6 T^4
\tabularnewline[0.5pt]\hline				
34	 &	smooth    &	          &	1+140 T+3642 p T^2+140 p^3 T^3+p^6 T^4
\tabularnewline[0.5pt]\hline				
35	 &	smooth$^*$&	\frac{3}{2}&	  
\tabularnewline[0.5pt]\hline				
36	 &	smooth    &	          &	1-52 T+1650 p T^2-52 p^3 T^3+p^6 T^4
\tabularnewline[0.5pt]\hline				
37	 &	smooth    &	          &	1+338 T+3642 p T^2+338 p^3 T^3+p^6 T^4
\tabularnewline[0.5pt]\hline				
38	 &	smooth    &	          &	1+578 T+6906 p T^2+578 p^3 T^3+p^6 T^4
\tabularnewline[0.5pt]\hline				
39	 &	smooth    &	          &	1+524 T+9570 p T^2+524 p^3 T^3+p^6 T^4
\tabularnewline[0.5pt]\hline				
40	 &	singular  &	-\frac{1}{5}\+&	(1-p T) (1+484 T+p^3 T^2)
\tabularnewline[0.5pt]\hline				
41	 &	smooth    &	          &	(1+4 p T+p^3 T^2)2
\tabularnewline[0.5pt]\hline				
42	 &	smooth    &	          &	1-256 T+2274 p T^2-256 p^3 T^3+p^6 T^4
\tabularnewline[0.5pt]\hline				
43	 &	smooth    &	          &	1+164 T+5970 p T^2+164 p^3 T^3+p^6 T^4
\tabularnewline[0.5pt]\hline				
44	 &	smooth    &	          &	(1+4 p T+p^3 T^2)(1-956 T+p^3 T^2)
\tabularnewline[0.5pt]\hline				
45	 &	singular  &	\frac{1}{3}&	(1-p T) (1-92 T+p^3 T^2)
\tabularnewline[0.5pt]\hline				
46	 &	smooth    &	          &	1-94 T+3210 p T^2-94 p^3 T^3+p^6 T^4
\tabularnewline[0.5pt]\hline				
47	 &	smooth    &	          &	1+410 T+8250 p T^2+410 p^3 T^3+p^6 T^4
\tabularnewline[0.5pt]\hline				
48	 &	smooth    &	          &	1-568 T+5202 p T^2-568 p^3 T^3+p^6 T^4
\tabularnewline[0.5pt]\hline				
49	 &	smooth    &	          &	(1+4 p T+p^3 T^2)(1-734 T+p^3 T^2)
\tabularnewline[0.5pt]\hline				
50	 &	singular  &	-\frac{1}{4}\+&	(1-p T) (1+484 T+p^3 T^2)
\tabularnewline[0.5pt]\hline				
51	 &	smooth    &	          &	1-640 T+7074 p T^2-640 p^3 T^3+p^6 T^4
\tabularnewline[0.5pt]\hline				
52	 &	smooth    &	          &	1+482 T+7602 p T^2+482 p^3 T^3+p^6 T^4
\tabularnewline[0.5pt]\hline				
53	 &	smooth    &	          &	1+218 T+8706 p T^2+218 p^3 T^3+p^6 T^4
\tabularnewline[0.5pt]\hline				
54	 &	smooth    &	          &	1+32 T+8466 p T^2+32 p^3 T^3+p^6 T^4
\tabularnewline[0.5pt]\hline				
55	 &	smooth    &	          &	1+98 T+3978 p T^2+98 p^3 T^3+p^6 T^4
\tabularnewline[0.5pt]\hline				
56	 &	smooth    &	          &	1+332 T+7794 p T^2+332 p^3 T^3+p^6 T^4
\tabularnewline[0.5pt]\hline				
57	 &	smooth    &	          &	1+704 T+8562 p T^2+704 p^3 T^3+p^6 T^4
\tabularnewline[0.5pt]\hline				
58	 &	smooth    &	          &	1+1112 T+12498 p T^2+1112 p^3 T^3+p^6 T^4
\tabularnewline[0.5pt]\hline				
59	 &	smooth    &	          &	1+170 T+5850 p T^2+170 p^3 T^3+p^6 T^4
\tabularnewline[0.5pt]\hline				
60	 &	smooth    &	          &	1-580 T+2994 p T^2-580 p^3 T^3+p^6 T^4
\tabularnewline[0.5pt]\hline				
61	 &	smooth    &	          &	1-604 T+4194 p T^2-604 p^3 T^3+p^6 T^4
\tabularnewline[0.5pt]\hline				
62	 &	smooth    &	          &	1-238 T+3930 p T^2-238 p^3 T^3+p^6 T^4
\tabularnewline[0.5pt]\hline				
63	 &	smooth    &	          &	1+932 T+9042 p T^2+932 p^3 T^3+p^6 T^4
\tabularnewline[0.5pt]\hline				
64	 &	smooth    &	          &	1+308 T-366 p T^2+308 p^3 T^3+p^6 T^4
\tabularnewline[0.5pt]\hline				
65	 &	smooth    &	          &	1-856 T+6642 p T^2-856 p^3 T^3+p^6 T^4
\tabularnewline[0.5pt]\hline				
66	 &	smooth    &	          &	1+638 T+9594 p T^2+638 p^3 T^3+p^6 T^4
\tabularnewline[0.5pt]\hline				
\tablepostamble				
\tablepreamble{71}				
1	 &	smooth    &	          &	1+240 T-5902 p T^2+240 p^3 T^3+p^6 T^4
\tabularnewline[0.5pt]\hline				
2	 &	smooth    &	          &	1-156 T+1154 p T^2-156 p^3 T^3+p^6 T^4
\tabularnewline[0.5pt]\hline				
3	 &	smooth    &	          &	1+570 T+6194 p T^2+570 p^3 T^3+p^6 T^4
\tabularnewline[0.5pt]\hline				
4	 &	smooth    &	          &	1-12 p T+8642 p T^2-12 p^4 T^3+p^6 T^4
\tabularnewline[0.5pt]\hline				
5	 &	smooth    &	          &	(1-12 p T+p^3 T^2)(1+1020 T+p^3 T^2)
\tabularnewline[0.5pt]\hline				
6	 &	singular  &	\frac{1}{12}&	(1-p T) (1+426 T+p^3 T^2)
\tabularnewline[0.5pt]\hline				
7	 &	smooth    &	          &	1+282 T+1802 p T^2+282 p^3 T^3+p^6 T^4
\tabularnewline[0.5pt]\hline				
8	 &	smooth    &	          &	1+246 T+9506 p T^2+246 p^3 T^3+p^6 T^4
\tabularnewline[0.5pt]\hline				
9	 &	smooth    &	          &	1+102 T+3242 p T^2+102 p^3 T^3+p^6 T^4
\tabularnewline[0.5pt]\hline				
10	 &	smooth    &	          &	1-156 T+5330 p T^2-156 p^3 T^3+p^6 T^4
\tabularnewline[0.5pt]\hline				
11	 &	smooth    &	          &	1-228 T+1442 p T^2-228 p^3 T^3+p^6 T^4
\tabularnewline[0.5pt]\hline				
12	 &	smooth    &	          &	1+1074 T+10946 p T^2+1074 p^3 T^3+p^6 T^4
\tabularnewline[0.5pt]\hline				
13	 &	smooth    &	          &	1+420 T+578 p T^2+420 p^3 T^3+p^6 T^4
\tabularnewline[0.5pt]\hline				
14	 &	singular  &	-\frac{1}{5}\+&	(1+p T) (1+708 T+p^3 T^2)
\tabularnewline[0.5pt]\hline				
15	 &	smooth    &	          &	1+636 T+3746 p T^2+636 p^3 T^3+p^6 T^4
\tabularnewline[0.5pt]\hline				
16	 &	smooth    &	          &	(1+p^3 T^2)(1+480 T+p^3 T^2)
\tabularnewline[0.5pt]\hline				
17	 &	smooth    &	          &	1-312 T+3170 p T^2-312 p^3 T^3+p^6 T^4
\tabularnewline[0.5pt]\hline				
18	 &	singular  &	\frac{1}{4}&	(1+p T) (1-432 T+p^3 T^2)
\tabularnewline[0.5pt]\hline				
19	 &	smooth    &	          &	1+144 T+4322 p T^2+144 p^3 T^3+p^6 T^4
\tabularnewline[0.5pt]\hline				
20	 &	smooth    &	          &	1+360 T+6626 p T^2+360 p^3 T^3+p^6 T^4
\tabularnewline[0.5pt]\hline				
21	 &	smooth    &	          &	1+102 T+1874 p T^2+102 p^3 T^3+p^6 T^4
\tabularnewline[0.5pt]\hline				
22	 &	smooth    &	          &	1+528 T+2234 p T^2+528 p^3 T^3+p^6 T^4
\tabularnewline[0.5pt]\hline				
23	 &	smooth    &	          &	1-504 T+9290 p T^2-504 p^3 T^3+p^6 T^4
\tabularnewline[0.5pt]\hline				
24	 &	singular  &	\frac{1}{3}&	(1-p T) (1+720 T+p^3 T^2)
\tabularnewline[0.5pt]\hline				
25	 &	smooth    &	          &	1-96 T-7270 p T^2-96 p^3 T^3+p^6 T^4
\tabularnewline[0.5pt]\hline				
26	 &	smooth    &	          &	1+162 T+4898 p T^2+162 p^3 T^3+p^6 T^4
\tabularnewline[0.5pt]\hline				
27	 &	smooth    &	          &	1+384 T-1582 p T^2+384 p^3 T^3+p^6 T^4
\tabularnewline[0.5pt]\hline				
28	 &	smooth    &	          &	1-144 T+7058 p T^2-144 p^3 T^3+p^6 T^4
\tabularnewline[0.5pt]\hline				
29	 &	smooth    &	          &	1-102 T+70 p^2 T^2-102 p^3 T^3+p^6 T^4
\tabularnewline[0.5pt]\hline				
30	 &	smooth    &	          &	1-456 T+6482 p T^2-456 p^3 T^3+p^6 T^4
\tabularnewline[0.5pt]\hline				
31	 &	smooth    &	          &	1-84 T-2158 p T^2-84 p^3 T^3+p^6 T^4
\tabularnewline[0.5pt]\hline				
32	 &	smooth    &	          &	1-222 T+3458 p T^2-222 p^3 T^3+p^6 T^4
\tabularnewline[0.5pt]\hline				
33	 &	smooth    &	          &	1+606 T+7346 p T^2+606 p^3 T^3+p^6 T^4
\tabularnewline[0.5pt]\hline				
34	 &	smooth    &	          &	1-258 T-2446 p T^2-258 p^3 T^3+p^6 T^4
\tabularnewline[0.5pt]\hline				
35	 &	smooth    &	          &	1+72 T-3598 p T^2+72 p^3 T^3+p^6 T^4
\tabularnewline[0.5pt]\hline				
36	 &	smooth    &	          &	1+180 T-3022 p T^2+180 p^3 T^3+p^6 T^4
\tabularnewline[0.5pt]\hline				
37	 &	smooth$^*$&	\frac{3}{2}&	  
\tabularnewline[0.5pt]\hline				
38	 &	smooth    &	          &	1-72 T+2306 p T^2-72 p^3 T^3+p^6 T^4
\tabularnewline[0.5pt]\hline				
39	 &	smooth    &	          &	1+468 T+7778 p T^2+468 p^3 T^3+p^6 T^4
\tabularnewline[0.5pt]\hline				
40	 &	smooth    &	          &	1-432 T+6914 p T^2-432 p^3 T^3+p^6 T^4
\tabularnewline[0.5pt]\hline				
41	 &	smooth    &	          &	1-516 T+4466 p T^2-516 p^3 T^3+p^6 T^4
\tabularnewline[0.5pt]\hline				
42	 &	smooth    &	          &	1+324 T+2882 p T^2+324 p^3 T^3+p^6 T^4
\tabularnewline[0.5pt]\hline				
43	 &	smooth    &	          &	1-7198 p T^2+p^6 T^4
\tabularnewline[0.5pt]\hline				
44	 &	smooth    &	          &	1+816 T+11378 p T^2+816 p^3 T^3+p^6 T^4
\tabularnewline[0.5pt]\hline				
45	 &	smooth    &	          &	1+372 T+2594 p T^2+372 p^3 T^3+p^6 T^4
\tabularnewline[0.5pt]\hline				
46	 &	smooth    &	          &	1+192 T+7058 p T^2+192 p^3 T^3+p^6 T^4
\tabularnewline[0.5pt]\hline				
47	 &	smooth    &	          &	1-594 T+9938 p T^2-594 p^3 T^3+p^6 T^4
\tabularnewline[0.5pt]\hline				
48	 &	smooth    &	          &	1+60 T-4606 p T^2+60 p^3 T^3+p^6 T^4
\tabularnewline[0.5pt]\hline				
49	 &	smooth    &	          &	1-324 T-4174 p T^2-324 p^3 T^3+p^6 T^4
\tabularnewline[0.5pt]\hline				
50	 &	smooth    &	          &	1-300 T+6050 p T^2-300 p^3 T^3+p^6 T^4
\tabularnewline[0.5pt]\hline				
51	 &	smooth    &	          &	1+1044 T+13250 p T^2+1044 p^3 T^3+p^6 T^4
\tabularnewline[0.5pt]\hline				
52	 &	smooth    &	          &	1-72 T+2738 p T^2-72 p^3 T^3+p^6 T^4
\tabularnewline[0.5pt]\hline				
53	 &	singular  &	-\frac{1}{4}\+&	(1-p T) (1-576 T+p^3 T^2)
\tabularnewline[0.5pt]\hline				
54	 &	smooth    &	          &	1+48 T+3890 p T^2+48 p^3 T^3+p^6 T^4
\tabularnewline[0.5pt]\hline				
55	 &	smooth    &	          &	1+294 T+9218 p T^2+294 p^3 T^3+p^6 T^4
\tabularnewline[0.5pt]\hline				
56	 &	smooth    &	          &	1+264 T+5762 p T^2+264 p^3 T^3+p^6 T^4
\tabularnewline[0.5pt]\hline				
57	 &	smooth    &	          &	1+468 T+5762 p T^2+468 p^3 T^3+p^6 T^4
\tabularnewline[0.5pt]\hline				
58	 &	smooth    &	          &	1+36 T+6338 p T^2+36 p^3 T^3+p^6 T^4
\tabularnewline[0.5pt]\hline				
59	 &	singular  &	-\frac{1}{6}\+&	(1-p T) (1-480 T+p^3 T^2)
\tabularnewline[0.5pt]\hline				
60	 &	smooth    &	          &	1-264 T-3742 p T^2-264 p^3 T^3+p^6 T^4
\tabularnewline[0.5pt]\hline				
61	 &	smooth    &	          &	1+24 T-862 p T^2+24 p^3 T^3+p^6 T^4
\tabularnewline[0.5pt]\hline				
62	 &	smooth    &	          &	1+576 T+1442 p T^2+576 p^3 T^3+p^6 T^4
\tabularnewline[0.5pt]\hline				
63	 &	smooth    &	          &	1-420 T+4682 p T^2-420 p^3 T^3+p^6 T^4
\tabularnewline[0.5pt]\hline				
64	 &	smooth    &	          &	1+510 T+2882 p T^2+510 p^3 T^3+p^6 T^4
\tabularnewline[0.5pt]\hline				
65	 &	smooth    &	          &	1-624 T+1442 p T^2-624 p^3 T^3+p^6 T^4
\tabularnewline[0.5pt]\hline				
66	 &	smooth    &	          &	1+822 T+9218 p T^2+822 p^3 T^3+p^6 T^4
\tabularnewline[0.5pt]\hline				
67	 &	smooth    &	          &	1+36 T-2878 p T^2+36 p^3 T^3+p^6 T^4
\tabularnewline[0.5pt]\hline				
68	 &	smooth    &	          &	1+228 T+2306 p T^2+228 p^3 T^3+p^6 T^4
\tabularnewline[0.5pt]\hline				
69	 &	smooth    &	          &	1+24 T+6626 p T^2+24 p^3 T^3+p^6 T^4
\tabularnewline[0.5pt]\hline				
70	 &	smooth    &	          &	1-816 T+4610 p T^2-816 p^3 T^3+p^6 T^4
\tabularnewline[0.5pt]\hline				
\tablepostamble				
\tablepreamble{73}				
1	 &	smooth    &	          &	1-400 T-66 p T^2-400 p^3 T^3+p^6 T^4
\tabularnewline[0.5pt]\hline				
2	 &	smooth    &	          &	1-424 T+8478 p T^2-424 p^3 T^3+p^6 T^4
\tabularnewline[0.5pt]\hline				
3	 &	smooth    &	          &	1+380 T+5574 p T^2+380 p^3 T^3+p^6 T^4
\tabularnewline[0.5pt]\hline				
4	 &	smooth    &	          &	1+356 T+5766 p T^2+356 p^3 T^3+p^6 T^4
\tabularnewline[0.5pt]\hline				
5	 &	smooth    &	          &	1+848 T+11262 p T^2+848 p^3 T^3+p^6 T^4
\tabularnewline[0.5pt]\hline				
6	 &	smooth    &	          &	1-136 T+8046 p T^2-136 p^3 T^3+p^6 T^4
\tabularnewline[0.5pt]\hline				
7	 &	smooth    &	          &	1-190 T-3510 p T^2-190 p^3 T^3+p^6 T^4
\tabularnewline[0.5pt]\hline				
8	 &	smooth    &	          &	1+296 T+2574 p T^2+296 p^3 T^3+p^6 T^4
\tabularnewline[0.5pt]\hline				
9	 &	smooth    &	          &	1-1132 T+9606 p T^2-1132 p^3 T^3+p^6 T^4
\tabularnewline[0.5pt]\hline				
10	 &	smooth    &	          &	1+680 T+2094 p T^2+680 p^3 T^3+p^6 T^4
\tabularnewline[0.5pt]\hline				
11	 &	smooth    &	          &	1-160 T+8094 p T^2-160 p^3 T^3+p^6 T^4
\tabularnewline[0.5pt]\hline				
12	 &	singular  &	-\frac{1}{6}\+&	(1-p T) (1-434 T+p^3 T^2)
\tabularnewline[0.5pt]\hline				
13	 &	smooth    &	          &	1+56 T+30 p T^2+56 p^3 T^3+p^6 T^4
\tabularnewline[0.5pt]\hline				
14	 &	smooth    &	          &	1+932 T+11742 p T^2+932 p^3 T^3+p^6 T^4
\tabularnewline[0.5pt]\hline				
15	 &	smooth    &	          &	1-148 T+870 p T^2-148 p^3 T^3+p^6 T^4
\tabularnewline[0.5pt]\hline				
16	 &	smooth    &	          &	1-1078 T+11082 p T^2-1078 p^3 T^3+p^6 T^4
\tabularnewline[0.5pt]\hline				
17	 &	smooth    &	          &	1+434 T+3594 p T^2+434 p^3 T^3+p^6 T^4
\tabularnewline[0.5pt]\hline				
18	 &	singular  &	-\frac{1}{4}\+&	(1-p T) (1+1150 T+p^3 T^2)
\tabularnewline[0.5pt]\hline				
19	 &	smooth    &	          &	1+428 T+2598 p T^2+428 p^3 T^3+p^6 T^4
\tabularnewline[0.5pt]\hline				
20	 &	smooth    &	          &	1+1136 T+10686 p T^2+1136 p^3 T^3+p^6 T^4
\tabularnewline[0.5pt]\hline				
21	 &	smooth    &	          &	1-496 T+5310 p T^2-496 p^3 T^3+p^6 T^4
\tabularnewline[0.5pt]\hline				
22	 &	smooth    &	          &	1+200 T+7806 p T^2+200 p^3 T^3+p^6 T^4
\tabularnewline[0.5pt]\hline				
23	 &	smooth    &	          &	1+716 T+5334 p T^2+716 p^3 T^3+p^6 T^4
\tabularnewline[0.5pt]\hline				
24	 &	smooth    &	          &	1-352 T+1566 p T^2-352 p^3 T^3+p^6 T^4
\tabularnewline[0.5pt]\hline				
25	 &	smooth    &	          &	1+566 T+7938 p T^2+566 p^3 T^3+p^6 T^4
\tabularnewline[0.5pt]\hline				
26	 &	smooth    &	          &	1+296 T+1134 p T^2+296 p^3 T^3+p^6 T^4
\tabularnewline[0.5pt]\hline				
27	 &	smooth    &	          &	1+932 T+8718 p T^2+932 p^3 T^3+p^6 T^4
\tabularnewline[0.5pt]\hline				
28	 &	smooth    &	          &	1+212 T-5610 p T^2+212 p^3 T^3+p^6 T^4
\tabularnewline[0.5pt]\hline				
29	 &	singular  &	-\frac{1}{5}\+&	(1-p T) (1-362 T+p^3 T^2)
\tabularnewline[0.5pt]\hline				
30	 &	smooth    &	          &	1+362 T+3162 p T^2+362 p^3 T^3+p^6 T^4
\tabularnewline[0.5pt]\hline				
31	 &	smooth    &	          &	1-262 T+8802 p T^2-262 p^3 T^3+p^6 T^4
\tabularnewline[0.5pt]\hline				
32	 &	smooth    &	          &	1-208 T-882 p T^2-208 p^3 T^3+p^6 T^4
\tabularnewline[0.5pt]\hline				
33	 &	smooth    &	          &	1-1048 T+13326 p T^2-1048 p^3 T^3+p^6 T^4
\tabularnewline[0.5pt]\hline				
34	 &	smooth    &	          &	1+284 T-4026 p T^2+284 p^3 T^3+p^6 T^4
\tabularnewline[0.5pt]\hline				
35	 &	smooth    &	          &	1+488 T+8022 p T^2+488 p^3 T^3+p^6 T^4
\tabularnewline[0.5pt]\hline				
36	 &	smooth    &	          &	1-700 T+102 p^2 T^2-700 p^3 T^3+p^6 T^4
\tabularnewline[0.5pt]\hline				
37	 &	smooth    &	          &	1-556 T-186 p T^2-556 p^3 T^3+p^6 T^4
\tabularnewline[0.5pt]\hline				
38	 &	smooth$^*$&	\frac{3}{2}&	  
\tabularnewline[0.5pt]\hline				
39	 &	smooth    &	          &	1-310 T-2118 p T^2-310 p^3 T^3+p^6 T^4
\tabularnewline[0.5pt]\hline				
40	 &	smooth    &	          &	1+992 T+9102 p T^2+992 p^3 T^3+p^6 T^4
\tabularnewline[0.5pt]\hline				
41	 &	smooth    &	          &	1-514 T+4986 p T^2-514 p^3 T^3+p^6 T^4
\tabularnewline[0.5pt]\hline				
42	 &	smooth    &	          &	1+56 T+1326 p T^2+56 p^3 T^3+p^6 T^4
\tabularnewline[0.5pt]\hline				
43	 &	smooth    &	          &	1+536 T+3678 p T^2+536 p^3 T^3+p^6 T^4
\tabularnewline[0.5pt]\hline				
44	 &	smooth    &	          &	1-4 p T-714 p T^2-4 p^4 T^3+p^6 T^4
\tabularnewline[0.5pt]\hline				
45	 &	smooth    &	          &	1+44 T+5526 p T^2+44 p^3 T^3+p^6 T^4
\tabularnewline[0.5pt]\hline				
46	 &	smooth    &	          &	1-1012 T+12102 p T^2-1012 p^3 T^3+p^6 T^4
\tabularnewline[0.5pt]\hline				
47	 &	smooth    &	          &	1-514 T+7506 p T^2-514 p^3 T^3+p^6 T^4
\tabularnewline[0.5pt]\hline				
48	 &	smooth    &	          &	1+542 T+9714 p T^2+542 p^3 T^3+p^6 T^4
\tabularnewline[0.5pt]\hline				
49	 &	singular  &	\frac{1}{3}&	(1-p T) (1+502 T+p^3 T^2)
\tabularnewline[0.5pt]\hline				
50	 &	smooth    &	          &	1-616 T+8142 p T^2-616 p^3 T^3+p^6 T^4
\tabularnewline[0.5pt]\hline				
51	 &	smooth    &	          &	1-220 T+5262 p T^2-220 p^3 T^3+p^6 T^4
\tabularnewline[0.5pt]\hline				
52	 &	smooth    &	          &	1+134 T+8802 p T^2+134 p^3 T^3+p^6 T^4
\tabularnewline[0.5pt]\hline				
53	 &	smooth    &	          &	1+1058 T+13506 p T^2+1058 p^3 T^3+p^6 T^4
\tabularnewline[0.5pt]\hline				
54	 &	smooth    &	          &	1-640 T+8046 p T^2-640 p^3 T^3+p^6 T^4
\tabularnewline[0.5pt]\hline				
55	 &	singular  &	\frac{1}{4}&	(1-p T) (1-362 T+p^3 T^2)
\tabularnewline[0.5pt]\hline				
56	 &	smooth    &	          &	1-916 T+5862 p T^2-916 p^3 T^3+p^6 T^4
\tabularnewline[0.5pt]\hline				
57	 &	smooth    &	          &	1+200 T+3054 p T^2+200 p^3 T^3+p^6 T^4
\tabularnewline[0.5pt]\hline				
58	 &	smooth    &	          &	1+200 T+6294 p T^2+200 p^3 T^3+p^6 T^4
\tabularnewline[0.5pt]\hline				
59	 &	smooth    &	          &	1-82 T+4050 p T^2-82 p^3 T^3+p^6 T^4
\tabularnewline[0.5pt]\hline				
60	 &	smooth    &	          &	1+380 T+4422 p T^2+380 p^3 T^3+p^6 T^4
\tabularnewline[0.5pt]\hline				
61	 &	smooth    &	          &	1+536 T+6558 p T^2+536 p^3 T^3+p^6 T^4
\tabularnewline[0.5pt]\hline				
62	 &	smooth    &	          &	1+524 T+7590 p T^2+524 p^3 T^3+p^6 T^4
\tabularnewline[0.5pt]\hline				
63	 &	smooth    &	          &	1-58 T-5070 p T^2-58 p^3 T^3+p^6 T^4
\tabularnewline[0.5pt]\hline				
64	 &	smooth    &	          &	1-304 T+6654 p T^2-304 p^3 T^3+p^6 T^4
\tabularnewline[0.5pt]\hline				
65	 &	smooth    &	          &	1+446 T+6090 p T^2+446 p^3 T^3+p^6 T^4
\tabularnewline[0.5pt]\hline				
66	 &	smooth    &	          &	1+428 T+6342 p T^2+428 p^3 T^3+p^6 T^4
\tabularnewline[0.5pt]\hline				
67	 &	singular  &	\frac{1}{12}&	(1-p T) (1+574 T+p^3 T^2)
\tabularnewline[0.5pt]\hline				
68	 &	smooth    &	          &	1+236 T+5142 p T^2+236 p^3 T^3+p^6 T^4
\tabularnewline[0.5pt]\hline				
69	 &	smooth    &	          &	1-694 T+4698 p T^2-694 p^3 T^3+p^6 T^4
\tabularnewline[0.5pt]\hline				
70	 &	smooth    &	          &	1+44 T+2574 p T^2+44 p^3 T^3+p^6 T^4
\tabularnewline[0.5pt]\hline				
71	 &	smooth    &	          &	1+596 T+11046 p T^2+596 p^3 T^3+p^6 T^4
\tabularnewline[0.5pt]\hline				
72	 &	smooth    &	          &	1-448 T+1470 p T^2-448 p^3 T^3+p^6 T^4
\tabularnewline[0.5pt]\hline				
\tablepostamble				
\tablepreamble{79}				
1	 &	smooth    &	          &	1-412 T+10506 p T^2-412 p^3 T^3+p^6 T^4
\tabularnewline[0.5pt]\hline				
2	 &	smooth    &	          &	1+326 T+6114 p T^2+326 p^3 T^3+p^6 T^4
\tabularnewline[0.5pt]\hline				
3	 &	smooth    &	          &	1-118 T-1134 p T^2-118 p^3 T^3+p^6 T^4
\tabularnewline[0.5pt]\hline				
4	 &	smooth    &	          &	1+290 T-6846 p T^2+290 p^3 T^3+p^6 T^4
\tabularnewline[0.5pt]\hline				
5	 &	smooth    &	          &	1-112 T+5730 p T^2-112 p^3 T^3+p^6 T^4
\tabularnewline[0.5pt]\hline				
6	 &	smooth    &	          &	1+608 T+3858 p T^2+608 p^3 T^3+p^6 T^4
\tabularnewline[0.5pt]\hline				
7	 &	smooth    &	          &	1-76 T-4782 p T^2-76 p^3 T^3+p^6 T^4
\tabularnewline[0.5pt]\hline				
8	 &	smooth    &	          &	1-184 T-4926 p T^2-184 p^3 T^3+p^6 T^4
\tabularnewline[0.5pt]\hline				
9	 &	smooth    &	          &	1+722 T+2082 p T^2+722 p^3 T^3+p^6 T^4
\tabularnewline[0.5pt]\hline				
10	 &	smooth    &	          &	1+356 T+3426 p T^2+356 p^3 T^3+p^6 T^4
\tabularnewline[0.5pt]\hline				
11	 &	smooth    &	          &	1+572 T+6090 p T^2+572 p^3 T^3+p^6 T^4
\tabularnewline[0.5pt]\hline				
12	 &	smooth    &	          &	1-88 T+6402 p T^2-88 p^3 T^3+p^6 T^4
\tabularnewline[0.5pt]\hline				
13	 &	singular  &	-\frac{1}{6}\+&	(1-p T) (1-1352 T+p^3 T^2)
\tabularnewline[0.5pt]\hline				
14	 &	smooth    &	          &	1+1682 T+18162 p T^2+1682 p^3 T^3+p^6 T^4
\tabularnewline[0.5pt]\hline				
15	 &	smooth    &	          &	1+356 T-2334 p T^2+356 p^3 T^3+p^6 T^4
\tabularnewline[0.5pt]\hline				
16	 &	smooth    &	          &	1-616 T+7170 p T^2-616 p^3 T^3+p^6 T^4
\tabularnewline[0.5pt]\hline				
17	 &	smooth    &	          &	1+644 T+5730 p T^2+644 p^3 T^3+p^6 T^4
\tabularnewline[0.5pt]\hline				
18	 &	smooth    &	          &	1+1670 T+19986 p T^2+1670 p^3 T^3+p^6 T^4
\tabularnewline[0.5pt]\hline				
19	 &	smooth    &	          &	1+710 T+8658 p T^2+710 p^3 T^3+p^6 T^4
\tabularnewline[0.5pt]\hline				
20	 &	singular  &	\frac{1}{4}&	(1-p T) (1+160 T+p^3 T^2)
\tabularnewline[0.5pt]\hline				
21	 &	smooth    &	          &	1+1136 T+13026 p T^2+1136 p^3 T^3+p^6 T^4
\tabularnewline[0.5pt]\hline				
22	 &	smooth    &	          &	1-1456 T+18642 p T^2-1456 p^3 T^3+p^6 T^4
\tabularnewline[0.5pt]\hline				
23	 &	smooth    &	          &	1-1894 T+22290 p T^2-1894 p^3 T^3+p^6 T^4
\tabularnewline[0.5pt]\hline				
24	 &	smooth    &	          &	1-220 T+9834 p T^2-220 p^3 T^3+p^6 T^4
\tabularnewline[0.5pt]\hline				
25	 &	smooth    &	          &	1+512 T+10242 p T^2+512 p^3 T^3+p^6 T^4
\tabularnewline[0.5pt]\hline				
26	 &	smooth    &	          &	1-988 T+13890 p T^2-988 p^3 T^3+p^6 T^4
\tabularnewline[0.5pt]\hline				
27	 &	smooth    &	          &	1-436 T+3138 p T^2-436 p^3 T^3+p^6 T^4
\tabularnewline[0.5pt]\hline				
28	 &	smooth    &	          &	1-322 T+8706 p T^2-322 p^3 T^3+p^6 T^4
\tabularnewline[0.5pt]\hline				
29	 &	smooth    &	          &	1+284 T-1830 p T^2+284 p^3 T^3+p^6 T^4
\tabularnewline[0.5pt]\hline				
30	 &	smooth    &	          &	1-262 T+2754 p T^2-262 p^3 T^3+p^6 T^4
\tabularnewline[0.5pt]\hline				
31	 &	smooth    &	          &	1+488 T+1074 p T^2+488 p^3 T^3+p^6 T^4
\tabularnewline[0.5pt]\hline				
32	 &	smooth    &	          &	1+1100 T+11370 p T^2+1100 p^3 T^3+p^6 T^4
\tabularnewline[0.5pt]\hline				
33	 &	singular  &	\frac{1}{12}&	(1-p T) (1-110 T+p^3 T^2)
\tabularnewline[0.5pt]\hline				
34	 &	smooth    &	          &	1+680 T+5442 p T^2+680 p^3 T^3+p^6 T^4
\tabularnewline[0.5pt]\hline				
35	 &	smooth    &	          &	1-856 T+5922 p T^2-856 p^3 T^3+p^6 T^4
\tabularnewline[0.5pt]\hline				
36	 &	smooth    &	          &	1+320 T+10050 p T^2+320 p^3 T^3+p^6 T^4
\tabularnewline[0.5pt]\hline				
37	 &	smooth    &	          &	1-292 T+4578 p T^2-292 p^3 T^3+p^6 T^4
\tabularnewline[0.5pt]\hline				
38	 &	smooth    &	          &	1+92 T+10146 p T^2+92 p^3 T^3+p^6 T^4
\tabularnewline[0.5pt]\hline				
39	 &	smooth    &	          &	1-124 T+498 p T^2-124 p^3 T^3+p^6 T^4
\tabularnewline[0.5pt]\hline				
40	 &	smooth    &	          &	1+416 T-222 p T^2+416 p^3 T^3+p^6 T^4
\tabularnewline[0.5pt]\hline				
41	 &	smooth$^*$&	\frac{3}{2}&	  
\tabularnewline[0.5pt]\hline				
42	 &	smooth    &	          &	1+848 T+11586 p T^2+848 p^3 T^3+p^6 T^4
\tabularnewline[0.5pt]\hline				
43	 &	smooth    &	          &	1+308 T+11010 p T^2+308 p^3 T^3+p^6 T^4
\tabularnewline[0.5pt]\hline				
44	 &	smooth    &	          &	1+152 T-270 p T^2+152 p^3 T^3+p^6 T^4
\tabularnewline[0.5pt]\hline				
45	 &	smooth    &	          &	1-568 T+5778 p T^2-568 p^3 T^3+p^6 T^4
\tabularnewline[0.5pt]\hline				
46	 &	smooth    &	          &	1-1624 T+19842 p T^2-1624 p^3 T^3+p^6 T^4
\tabularnewline[0.5pt]\hline				
47	 &	smooth    &	          &	1+776 T+9858 p T^2+776 p^3 T^3+p^6 T^4
\tabularnewline[0.5pt]\hline				
48	 &	smooth    &	          &	1+920 T+11298 p T^2+920 p^3 T^3+p^6 T^4
\tabularnewline[0.5pt]\hline				
49	 &	smooth    &	          &	1+38 T-1734 p T^2+38 p^3 T^3+p^6 T^4
\tabularnewline[0.5pt]\hline				
50	 &	smooth    &	          &	1-490 T+5442 p T^2-490 p^3 T^3+p^6 T^4
\tabularnewline[0.5pt]\hline				
51	 &	smooth    &	          &	1-400 T+5442 p T^2-400 p^3 T^3+p^6 T^4
\tabularnewline[0.5pt]\hline				
52	 &	smooth    &	          &	1+920 T+5682 p T^2+920 p^3 T^3+p^6 T^4
\tabularnewline[0.5pt]\hline				
53	 &	singular  &	\frac{1}{3}&	(1-p T) (1+1024 T+p^3 T^2)
\tabularnewline[0.5pt]\hline				
54	 &	smooth    &	          &	1+56 T+2226 p T^2+56 p^3 T^3+p^6 T^4
\tabularnewline[0.5pt]\hline				
55	 &	smooth    &	          &	1-166 T-4134 p T^2-166 p^3 T^3+p^6 T^4
\tabularnewline[0.5pt]\hline				
56	 &	smooth    &	          &	1-238 T-3054 p T^2-238 p^3 T^3+p^6 T^4
\tabularnewline[0.5pt]\hline				
57	 &	smooth    &	          &	1-256 T-1326 p T^2-256 p^3 T^3+p^6 T^4
\tabularnewline[0.5pt]\hline				
58	 &	smooth    &	          &	1+110 T+9858 p T^2+110 p^3 T^3+p^6 T^4
\tabularnewline[0.5pt]\hline				
59	 &	singular  &	-\frac{1}{4}\+&	(1-p T) (1-776 T+p^3 T^2)
\tabularnewline[0.5pt]\hline				
60	 &	smooth    &	          &	1-646 T+5754 p T^2-646 p^3 T^3+p^6 T^4
\tabularnewline[0.5pt]\hline				
61	 &	smooth    &	          &	1+416 T-222 p T^2+416 p^3 T^3+p^6 T^4
\tabularnewline[0.5pt]\hline				
62	 &	smooth    &	          &	(1-8 p T+p^3 T^2)(1-380 T+p^3 T^2)
\tabularnewline[0.5pt]\hline				
63	 &	singular  &	-\frac{1}{5}\+&	(1-p T) (1+484 T+p^3 T^2)
\tabularnewline[0.5pt]\hline				
64	 &	smooth    &	          &	(1+16 p T+p^3 T^2)(1-1352 T+p^3 T^2)
\tabularnewline[0.5pt]\hline				
65	 &	smooth    &	          &	1-148 T+4722 p T^2-148 p^3 T^3+p^6 T^4
\tabularnewline[0.5pt]\hline				
66	 &	smooth    &	          &	1+614 T+642 p T^2+614 p^3 T^3+p^6 T^4
\tabularnewline[0.5pt]\hline				
67	 &	smooth    &	          &	1-322 T+1074 p T^2-322 p^3 T^3+p^6 T^4
\tabularnewline[0.5pt]\hline				
68	 &	smooth    &	          &	1+440 T+1746 p T^2+440 p^3 T^3+p^6 T^4
\tabularnewline[0.5pt]\hline				
69	 &	smooth    &	          &	1+1190 T+10002 p T^2+1190 p^3 T^3+p^6 T^4
\tabularnewline[0.5pt]\hline				
70	 &	smooth    &	          &	1+596 T+12162 p T^2+596 p^3 T^3+p^6 T^4
\tabularnewline[0.5pt]\hline				
71	 &	smooth    &	          &	1+368 T-918 p T^2+368 p^3 T^3+p^6 T^4
\tabularnewline[0.5pt]\hline				
72	 &	smooth    &	          &	1-472 T+4290 p T^2-472 p^3 T^3+p^6 T^4
\tabularnewline[0.5pt]\hline				
73	 &	smooth    &	          &	1-760 T+10914 p T^2-760 p^3 T^3+p^6 T^4
\tabularnewline[0.5pt]\hline				
74	 &	smooth    &	          &	1-208 T-2286 p T^2-208 p^3 T^3+p^6 T^4
\tabularnewline[0.5pt]\hline				
75	 &	smooth    &	          &	1+710 T+8658 p T^2+710 p^3 T^3+p^6 T^4
\tabularnewline[0.5pt]\hline				
76	 &	smooth    &	          &	1+620 T+738 p T^2+620 p^3 T^3+p^6 T^4
\tabularnewline[0.5pt]\hline				
77	 &	smooth    &	          &	1+1448 T+16578 p T^2+1448 p^3 T^3+p^6 T^4
\tabularnewline[0.5pt]\hline				
78	 &	smooth    &	          &	1-1528 T+17346 p T^2-1528 p^3 T^3+p^6 T^4
\tabularnewline[0.5pt]\hline				
\tablepostamble				
\tablepreamble{83}				
1	 &	smooth    &	          &	1-420 T+10322 p T^2-420 p^3 T^3+p^6 T^4
\tabularnewline[0.5pt]\hline				
2	 &	smooth    &	          &	1+1212 T+12698 p T^2+1212 p^3 T^3+p^6 T^4
\tabularnewline[0.5pt]\hline				
3	 &	smooth    &	          &	1-360 T+4418 p T^2-360 p^3 T^3+p^6 T^4
\tabularnewline[0.5pt]\hline				
4	 &	smooth    &	          &	1+618 T+314 p T^2+618 p^3 T^3+p^6 T^4
\tabularnewline[0.5pt]\hline				
5	 &	smooth    &	          &	1-132 T+9602 p T^2-132 p^3 T^3+p^6 T^4
\tabularnewline[0.5pt]\hline				
6	 &	smooth    &	          &	1+606 T+890 p T^2+606 p^3 T^3+p^6 T^4
\tabularnewline[0.5pt]\hline				
7	 &	singular  &	\frac{1}{12}&	(1-p T) (1+1308 T+p^3 T^2)
\tabularnewline[0.5pt]\hline				
8	 &	smooth    &	          &	1+1338 T+16730 p T^2+1338 p^3 T^3+p^6 T^4
\tabularnewline[0.5pt]\hline				
9	 &	smooth    &	          &	1-1020 T+10178 p T^2-1020 p^3 T^3+p^6 T^4
\tabularnewline[0.5pt]\hline				
10	 &	smooth    &	          &	1+810 T+8162 p T^2+810 p^3 T^3+p^6 T^4
\tabularnewline[0.5pt]\hline				
11	 &	smooth    &	          &	1+534 T+7514 p T^2+534 p^3 T^3+p^6 T^4
\tabularnewline[0.5pt]\hline				
12	 &	smooth    &	          &	1+348 T-6094 p T^2+348 p^3 T^3+p^6 T^4
\tabularnewline[0.5pt]\hline				
13	 &	smooth    &	          &	1+492 T+6578 p T^2+492 p^3 T^3+p^6 T^4
\tabularnewline[0.5pt]\hline				
14	 &	smooth    &	          &	1+576 T+2690 p T^2+576 p^3 T^3+p^6 T^4
\tabularnewline[0.5pt]\hline				
15	 &	smooth    &	          &	1+1056 T+16802 p T^2+1056 p^3 T^3+p^6 T^4
\tabularnewline[0.5pt]\hline				
16	 &	smooth    &	          &	1-120 T+8450 p T^2-120 p^3 T^3+p^6 T^4
\tabularnewline[0.5pt]\hline				
17	 &	smooth    &	          &	1+288 T-5086 p T^2+288 p^3 T^3+p^6 T^4
\tabularnewline[0.5pt]\hline				
18	 &	smooth    &	          &	1+528 T+3698 p T^2+528 p^3 T^3+p^6 T^4
\tabularnewline[0.5pt]\hline				
19	 &	smooth    &	          &	1+1194 T+11474 p T^2+1194 p^3 T^3+p^6 T^4
\tabularnewline[0.5pt]\hline				
20	 &	smooth    &	          &	1-984 T+6722 p T^2-984 p^3 T^3+p^6 T^4
\tabularnewline[0.5pt]\hline				
21	 &	singular  &	\frac{1}{4}&	(1+p T) (1-72 T+p^3 T^2)
\tabularnewline[0.5pt]\hline				
22	 &	smooth    &	          &	1-414 T+10034 p T^2-414 p^3 T^3+p^6 T^4
\tabularnewline[0.5pt]\hline				
23	 &	smooth    &	          &	1-378 T+1898 p T^2-378 p^3 T^3+p^6 T^4
\tabularnewline[0.5pt]\hline				
24	 &	smooth    &	          &	1-1044 T+8306 p T^2-1044 p^3 T^3+p^6 T^4
\tabularnewline[0.5pt]\hline				
25	 &	smooth    &	          &	1-480 T+8450 p T^2-480 p^3 T^3+p^6 T^4
\tabularnewline[0.5pt]\hline				
26	 &	smooth    &	          &	1+798 T+8954 p T^2+798 p^3 T^3+p^6 T^4
\tabularnewline[0.5pt]\hline				
27	 &	smooth    &	          &	1+186 T+8378 p T^2+186 p^3 T^3+p^6 T^4
\tabularnewline[0.5pt]\hline				
28	 &	singular  &	\frac{1}{3}&	(1-p T) (1+204 T+p^3 T^2)
\tabularnewline[0.5pt]\hline				
29	 &	smooth    &	          &	1+678 T+4994 p T^2+678 p^3 T^3+p^6 T^4
\tabularnewline[0.5pt]\hline				
30	 &	smooth    &	          &	1+456 T+2690 p T^2+456 p^3 T^3+p^6 T^4
\tabularnewline[0.5pt]\hline				
31	 &	smooth    &	          &	1+360 T+4850 p T^2+360 p^3 T^3+p^6 T^4
\tabularnewline[0.5pt]\hline				
32	 &	smooth    &	          &	1+102 T-4006 p T^2+102 p^3 T^3+p^6 T^4
\tabularnewline[0.5pt]\hline				
33	 &	singular  &	-\frac{1}{5}\+&	(1+p T) (1-756 T+p^3 T^2)
\tabularnewline[0.5pt]\hline				
34	 &	smooth    &	          &	1-12 T-1486 p T^2-12 p^3 T^3+p^6 T^4
\tabularnewline[0.5pt]\hline				
35	 &	smooth    &	          &	1-1188 T+16802 p T^2-1188 p^3 T^3+p^6 T^4
\tabularnewline[0.5pt]\hline				
36	 &	smooth    &	          &	1-234 T-5302 p T^2-234 p^3 T^3+p^6 T^4
\tabularnewline[0.5pt]\hline				
37	 &	smooth    &	          &	1+600 T+6722 p T^2+600 p^3 T^3+p^6 T^4
\tabularnewline[0.5pt]\hline				
38	 &	smooth    &	          &	1-228 T-1774 p T^2-228 p^3 T^3+p^6 T^4
\tabularnewline[0.5pt]\hline				
39	 &	smooth    &	          &	1+312 T+4418 p T^2+312 p^3 T^3+p^6 T^4
\tabularnewline[0.5pt]\hline				
40	 &	smooth    &	          &	1-684 T+2762 p T^2-684 p^3 T^3+p^6 T^4
\tabularnewline[0.5pt]\hline				
41	 &	smooth    &	          &	1+1692 T+19538 p T^2+1692 p^3 T^3+p^6 T^4
\tabularnewline[0.5pt]\hline				
42	 &	smooth    &	          &	1+138 T+10250 p T^2+138 p^3 T^3+p^6 T^4
\tabularnewline[0.5pt]\hline				
43	 &	smooth$^*$&	\frac{3}{2}&	  
\tabularnewline[0.5pt]\hline				
44	 &	smooth    &	          &	1-432 T+5426 p T^2-432 p^3 T^3+p^6 T^4
\tabularnewline[0.5pt]\hline				
45	 &	smooth    &	          &	1+96 T+1538 p T^2+96 p^3 T^3+p^6 T^4
\tabularnewline[0.5pt]\hline				
46	 &	smooth    &	          &	1+1092 T+9962 p T^2+1092 p^3 T^3+p^6 T^4
\tabularnewline[0.5pt]\hline				
47	 &	smooth    &	          &	1-192 T+7082 p T^2-192 p^3 T^3+p^6 T^4
\tabularnewline[0.5pt]\hline				
48	 &	smooth    &	          &	1+1212 T+9170 p T^2+1212 p^3 T^3+p^6 T^4
\tabularnewline[0.5pt]\hline				
49	 &	smooth    &	          &	1-720 T+13202 p T^2-720 p^3 T^3+p^6 T^4
\tabularnewline[0.5pt]\hline				
50	 &	smooth    &	          &	1+366 T+3842 p T^2+366 p^3 T^3+p^6 T^4
\tabularnewline[0.5pt]\hline				
51	 &	smooth    &	          &	1+528 T-3070 p T^2+528 p^3 T^3+p^6 T^4
\tabularnewline[0.5pt]\hline				
52	 &	smooth    &	          &	1-336 T-5158 p T^2-336 p^3 T^3+p^6 T^4
\tabularnewline[0.5pt]\hline				
53	 &	smooth    &	          &	1+1188 T+10034 p T^2+1188 p^3 T^3+p^6 T^4
\tabularnewline[0.5pt]\hline				
54	 &	smooth    &	          &	1+216 T+9890 p T^2+216 p^3 T^3+p^6 T^4
\tabularnewline[0.5pt]\hline				
55	 &	smooth    &	          &	1+228 T-1918 p T^2+228 p^3 T^3+p^6 T^4
\tabularnewline[0.5pt]\hline				
56	 &	smooth    &	          &	1-1020 T+14642 p T^2-1020 p^3 T^3+p^6 T^4
\tabularnewline[0.5pt]\hline				
57	 &	smooth    &	          &	1-810 T+5786 p T^2-810 p^3 T^3+p^6 T^4
\tabularnewline[0.5pt]\hline				
58	 &	smooth    &	          &	1+540 T+3410 p T^2+540 p^3 T^3+p^6 T^4
\tabularnewline[0.5pt]\hline				
59	 &	smooth    &	          &	1-24 T-3502 p T^2-24 p^3 T^3+p^6 T^4
\tabularnewline[0.5pt]\hline				
60	 &	smooth    &	          &	1+24 T+13058 p T^2+24 p^3 T^3+p^6 T^4
\tabularnewline[0.5pt]\hline				
61	 &	smooth    &	          &	1-1284 T+16370 p T^2-1284 p^3 T^3+p^6 T^4
\tabularnewline[0.5pt]\hline				
62	 &	singular  &	-\frac{1}{4}\+&	(1-p T) (1-378 T+p^3 T^2)
\tabularnewline[0.5pt]\hline				
63	 &	smooth    &	          &	1+42 T-1126 p T^2+42 p^3 T^3+p^6 T^4
\tabularnewline[0.5pt]\hline				
64	 &	smooth    &	          &	1-324 T-2278 p T^2-324 p^3 T^3+p^6 T^4
\tabularnewline[0.5pt]\hline				
65	 &	smooth    &	          &	1+960 T+13058 p T^2+960 p^3 T^3+p^6 T^4
\tabularnewline[0.5pt]\hline				
66	 &	smooth    &	          &	1-1080 T+9602 p T^2-1080 p^3 T^3+p^6 T^4
\tabularnewline[0.5pt]\hline				
67	 &	smooth    &	          &	1-432 T+11186 p T^2-432 p^3 T^3+p^6 T^4
\tabularnewline[0.5pt]\hline				
68	 &	smooth    &	          &	(1-12 p T+p^3 T^2)(1-390 T+p^3 T^2)
\tabularnewline[0.5pt]\hline				
69	 &	singular  &	-\frac{1}{6}\+&	(1-p T) (1+612 T+p^3 T^2)
\tabularnewline[0.5pt]\hline				
70	 &	smooth    &	          &	1-36 T+1682 p T^2-36 p^3 T^3+p^6 T^4
\tabularnewline[0.5pt]\hline				
71	 &	smooth    &	          &	(1-12 p T+p^3 T^2)(1+600 T+p^3 T^2)
\tabularnewline[0.5pt]\hline				
72	 &	smooth    &	          &	1+450 T+6434 p T^2+450 p^3 T^3+p^6 T^4
\tabularnewline[0.5pt]\hline				
73	 &	smooth    &	          &	1-708 T+3410 p T^2-708 p^3 T^3+p^6 T^4
\tabularnewline[0.5pt]\hline				
74	 &	smooth    &	          &	1-420 T-910 p T^2-420 p^3 T^3+p^6 T^4
\tabularnewline[0.5pt]\hline				
75	 &	smooth    &	          &	1+1584 T+18530 p T^2+1584 p^3 T^3+p^6 T^4
\tabularnewline[0.5pt]\hline				
76	 &	smooth    &	          &	1+1236 T+16370 p T^2+1236 p^3 T^3+p^6 T^4
\tabularnewline[0.5pt]\hline				
77	 &	smooth    &	          &	1-204 T+5714 p T^2-204 p^3 T^3+p^6 T^4
\tabularnewline[0.5pt]\hline				
78	 &	smooth    &	          &	1+156 T-6094 p T^2+156 p^3 T^3+p^6 T^4
\tabularnewline[0.5pt]\hline				
79	 &	smooth    &	          &	1+288 T+4418 p T^2+288 p^3 T^3+p^6 T^4
\tabularnewline[0.5pt]\hline				
80	 &	smooth    &	          &	1-384 T+7082 p T^2-384 p^3 T^3+p^6 T^4
\tabularnewline[0.5pt]\hline				
81	 &	smooth    &	          &	1-1026 T+16010 p T^2-1026 p^3 T^3+p^6 T^4
\tabularnewline[0.5pt]\hline				
82	 &	smooth    &	          &	1+162 T+13562 p T^2+162 p^3 T^3+p^6 T^4
\tabularnewline[0.5pt]\hline				
\tablepostamble				
\tablepreamble{89}				
1	 &	smooth    &	          &	1+138 T+938 p T^2+138 p^3 T^3+p^6 T^4
\tabularnewline[0.5pt]\hline				
2	 &	smooth    &	          &	1-1668 T+17174 p T^2-1668 p^3 T^3+p^6 T^4
\tabularnewline[0.5pt]\hline				
3	 &	smooth    &	          &	1+780 T+16454 p T^2+780 p^3 T^3+p^6 T^4
\tabularnewline[0.5pt]\hline				
4	 &	smooth    &	          &	1-48 T+254 p T^2-48 p^3 T^3+p^6 T^4
\tabularnewline[0.5pt]\hline				
5	 &	smooth    &	          &	1+264 T+6086 p T^2+264 p^3 T^3+p^6 T^4
\tabularnewline[0.5pt]\hline				
6	 &	smooth    &	          &	1-552 T+6158 p T^2-552 p^3 T^3+p^6 T^4
\tabularnewline[0.5pt]\hline				
7	 &	smooth    &	          &	1+276 T+3638 p T^2+276 p^3 T^3+p^6 T^4
\tabularnewline[0.5pt]\hline				
8	 &	smooth    &	          &	1+228 T-4714 p T^2+228 p^3 T^3+p^6 T^4
\tabularnewline[0.5pt]\hline				
9	 &	smooth    &	          &	1+276 T+13646 p T^2+276 p^3 T^3+p^6 T^4
\tabularnewline[0.5pt]\hline				
10	 &	smooth    &	          &	1+1200 T+16814 p T^2+1200 p^3 T^3+p^6 T^4
\tabularnewline[0.5pt]\hline				
11	 &	smooth    &	          &	1+222 T+6194 p T^2+222 p^3 T^3+p^6 T^4
\tabularnewline[0.5pt]\hline				
12	 &	smooth    &	          &	1+492 T+10262 p T^2+492 p^3 T^3+p^6 T^4
\tabularnewline[0.5pt]\hline				
13	 &	smooth    &	          &	1-720 T+10910 p T^2-720 p^3 T^3+p^6 T^4
\tabularnewline[0.5pt]\hline				
14	 &	smooth    &	          &	1-522 T-2590 p T^2-522 p^3 T^3+p^6 T^4
\tabularnewline[0.5pt]\hline				
15	 &	smooth    &	          &	1-360 T+3422 p T^2-360 p^3 T^3+p^6 T^4
\tabularnewline[0.5pt]\hline				
16	 &	smooth    &	          &	(1+6 p T+p^3 T^2)(1-510 T+p^3 T^2)
\tabularnewline[0.5pt]\hline				
17	 &	smooth    &	          &	1+12 T+2342 p T^2+12 p^3 T^3+p^6 T^4
\tabularnewline[0.5pt]\hline				
18	 &	smooth    &	          &	1+564 T+11702 p T^2+564 p^3 T^3+p^6 T^4
\tabularnewline[0.5pt]\hline				
19	 &	smooth    &	          &	1-96 T-898 p T^2-96 p^3 T^3+p^6 T^4
\tabularnewline[0.5pt]\hline				
20	 &	smooth    &	          &	1-600 T+11054 p T^2-600 p^3 T^3+p^6 T^4
\tabularnewline[0.5pt]\hline				
21	 &	smooth    &	          &	1-798 T+6698 p T^2-798 p^3 T^3+p^6 T^4
\tabularnewline[0.5pt]\hline				
22	 &	singular  &	-\frac{1}{4}\+&	(1-p T) (1+390 T+p^3 T^2)
\tabularnewline[0.5pt]\hline				
23	 &	smooth    &	          &	1+1098 T+12602 p T^2+1098 p^3 T^3+p^6 T^4
\tabularnewline[0.5pt]\hline				
24	 &	smooth    &	          &	1+672 T+13646 p T^2+672 p^3 T^3+p^6 T^4
\tabularnewline[0.5pt]\hline				
25	 &	smooth    &	          &	1-252 T-682 p T^2-252 p^3 T^3+p^6 T^4
\tabularnewline[0.5pt]\hline				
26	 &	smooth    &	          &	1-360 T+8318 p T^2-360 p^3 T^3+p^6 T^4
\tabularnewline[0.5pt]\hline				
27	 &	smooth    &	          &	1-648 T-250 p T^2-648 p^3 T^3+p^6 T^4
\tabularnewline[0.5pt]\hline				
28	 &	smooth    &	          &	1-300 T-250 p T^2-300 p^3 T^3+p^6 T^4
\tabularnewline[0.5pt]\hline				
29	 &	smooth    &	          &	1-540 T-2266 p T^2-540 p^3 T^3+p^6 T^4
\tabularnewline[0.5pt]\hline				
30	 &	singular  &	\frac{1}{3}&	(1-p T) (1-354 T+p^3 T^2)
\tabularnewline[0.5pt]\hline				
31	 &	smooth    &	          &	1-6 T+3602 p T^2-6 p^3 T^3+p^6 T^4
\tabularnewline[0.5pt]\hline				
32	 &	smooth    &	          &	1+6 T+4826 p T^2+6 p^3 T^3+p^6 T^4
\tabularnewline[0.5pt]\hline				
33	 &	smooth    &	          &	1-462 T+4682 p T^2-462 p^3 T^3+p^6 T^4
\tabularnewline[0.5pt]\hline				
34	 &	smooth    &	          &	1+996 T+5222 p T^2+996 p^3 T^3+p^6 T^4
\tabularnewline[0.5pt]\hline				
35	 &	smooth    &	          &	1-210 T+434 p T^2-210 p^3 T^3+p^6 T^4
\tabularnewline[0.5pt]\hline				
36	 &	smooth    &	          &	1-138 T+2306 p T^2-138 p^3 T^3+p^6 T^4
\tabularnewline[0.5pt]\hline				
37	 &	smooth    &	          &	(1-6 p T+p^3 T^2)(1+162 T+p^3 T^2)
\tabularnewline[0.5pt]\hline				
38	 &	smooth    &	          &	1-288 T+6302 p T^2-288 p^3 T^3+p^6 T^4
\tabularnewline[0.5pt]\hline				
39	 &	smooth    &	          &	1-414 T+7274 p T^2-414 p^3 T^3+p^6 T^4
\tabularnewline[0.5pt]\hline				
40	 &	smooth    &	          &	1+240 T+3710 p T^2+240 p^3 T^3+p^6 T^4
\tabularnewline[0.5pt]\hline				
41	 &	smooth    &	          &	1-588 T+2198 p T^2-588 p^3 T^3+p^6 T^4
\tabularnewline[0.5pt]\hline				
42	 &	smooth    &	          &	1+1152 T+16094 p T^2+1152 p^3 T^3+p^6 T^4
\tabularnewline[0.5pt]\hline				
43	 &	smooth    &	          &	1+972 T+5222 p T^2+972 p^3 T^3+p^6 T^4
\tabularnewline[0.5pt]\hline				
44	 &	smooth    &	          &	1+426 T+15338 p T^2+426 p^3 T^3+p^6 T^4
\tabularnewline[0.5pt]\hline				
45	 &	smooth    &	          &	1-1080 T+10046 p T^2-1080 p^3 T^3+p^6 T^4
\tabularnewline[0.5pt]\hline				
46	 &	smooth$^*$&	\frac{3}{2}&	  
\tabularnewline[0.5pt]\hline				
47	 &	smooth    &	          &	1+48 T+4574 p T^2+48 p^3 T^3+p^6 T^4
\tabularnewline[0.5pt]\hline				
48	 &	smooth    &	          &	1+288 T+830 p T^2+288 p^3 T^3+p^6 T^4
\tabularnewline[0.5pt]\hline				
49	 &	smooth    &	          &	1-276 T-4138 p T^2-276 p^3 T^3+p^6 T^4
\tabularnewline[0.5pt]\hline				
50	 &	smooth    &	          &	1+636 T+6950 p T^2+636 p^3 T^3+p^6 T^4
\tabularnewline[0.5pt]\hline				
51	 &	smooth    &	          &	1-990 T+7418 p T^2-990 p^3 T^3+p^6 T^4
\tabularnewline[0.5pt]\hline				
52	 &	singular  &	\frac{1}{12}&	(1-p T) (1-798 T+p^3 T^2)
\tabularnewline[0.5pt]\hline				
53	 &	smooth    &	          &	1+432 T+10622 p T^2+432 p^3 T^3+p^6 T^4
\tabularnewline[0.5pt]\hline				
54	 &	smooth    &	          &	1+564 T+2198 p T^2+564 p^3 T^3+p^6 T^4
\tabularnewline[0.5pt]\hline				
55	 &	smooth    &	          &	1-726 T+6410 p T^2-726 p^3 T^3+p^6 T^4
\tabularnewline[0.5pt]\hline				
56	 &	smooth    &	          &	1-1230 T+16634 p T^2-1230 p^3 T^3+p^6 T^4
\tabularnewline[0.5pt]\hline				
57	 &	smooth    &	          &	1+672 T+7598 p T^2+672 p^3 T^3+p^6 T^4
\tabularnewline[0.5pt]\hline				
58	 &	smooth    &	          &	1-510 T+5258 p T^2-510 p^3 T^3+p^6 T^4
\tabularnewline[0.5pt]\hline				
59	 &	smooth    &	          &	1+630 T+2954 p T^2+630 p^3 T^3+p^6 T^4
\tabularnewline[0.5pt]\hline				
60	 &	smooth    &	          &	1-1488 T+14366 p T^2-1488 p^3 T^3+p^6 T^4
\tabularnewline[0.5pt]\hline				
61	 &	smooth    &	          &	1+756 T+9830 p T^2+756 p^3 T^3+p^6 T^4
\tabularnewline[0.5pt]\hline				
62	 &	smooth    &	          &	1+666 T+9938 p T^2+666 p^3 T^3+p^6 T^4
\tabularnewline[0.5pt]\hline				
63	 &	smooth    &	          &	1+876 T+902 p T^2+876 p^3 T^3+p^6 T^4
\tabularnewline[0.5pt]\hline				
64	 &	smooth    &	          &	1+1644 T+19334 p T^2+1644 p^3 T^3+p^6 T^4
\tabularnewline[0.5pt]\hline				
65	 &	smooth    &	          &	1-630 T+10010 p T^2-630 p^3 T^3+p^6 T^4
\tabularnewline[0.5pt]\hline				
66	 &	smooth    &	          &	1+390 T-3742 p T^2+390 p^3 T^3+p^6 T^4
\tabularnewline[0.5pt]\hline				
67	 &	singular  &	\frac{1}{4}&	(1+p T) (1-810 T+p^3 T^2)
\tabularnewline[0.5pt]\hline				
68	 &	smooth    &	          &	1-36 T+13862 p T^2-36 p^3 T^3+p^6 T^4
\tabularnewline[0.5pt]\hline				
69	 &	smooth    &	          &	1+708 T+5366 p T^2+708 p^3 T^3+p^6 T^4
\tabularnewline[0.5pt]\hline				
70	 &	smooth    &	          &	1-1392 T+19262 p T^2-1392 p^3 T^3+p^6 T^4
\tabularnewline[0.5pt]\hline				
71	 &	singular  &	-\frac{1}{5}\+&	(1+p T) (1+774 T+p^3 T^2)
\tabularnewline[0.5pt]\hline				
72	 &	smooth    &	          &	1+576 T-2914 p T^2+576 p^3 T^3+p^6 T^4
\tabularnewline[0.5pt]\hline				
73	 &	smooth    &	          &	(1+6 p T+p^3 T^2)(1+222 T+p^3 T^2)
\tabularnewline[0.5pt]\hline				
74	 &	singular  &	-\frac{1}{6}\+&	(1-p T) (1+30 T+p^3 T^2)
\tabularnewline[0.5pt]\hline				
75	 &	smooth    &	          &	1-306 T+13682 p T^2-306 p^3 T^3+p^6 T^4
\tabularnewline[0.5pt]\hline				
76	 &	smooth    &	          &	1-12 T+398 p T^2-12 p^3 T^3+p^6 T^4
\tabularnewline[0.5pt]\hline				
77	 &	smooth    &	          &	1+120 T+2990 p T^2+120 p^3 T^3+p^6 T^4
\tabularnewline[0.5pt]\hline				
78	 &	smooth    &	          &	1-120 T-9106 p T^2-120 p^3 T^3+p^6 T^4
\tabularnewline[0.5pt]\hline				
79	 &	smooth    &	          &	1+354 T-4822 p T^2+354 p^3 T^3+p^6 T^4
\tabularnewline[0.5pt]\hline				
80	 &	smooth    &	          &	1-558 T+506 p T^2-558 p^3 T^3+p^6 T^4
\tabularnewline[0.5pt]\hline				
81	 &	smooth    &	          &	1+900 T+15446 p T^2+900 p^3 T^3+p^6 T^4
\tabularnewline[0.5pt]\hline				
82	 &	smooth    &	          &	1+1212 T+13430 p T^2+1212 p^3 T^3+p^6 T^4
\tabularnewline[0.5pt]\hline				
83	 &	smooth    &	          &	1-276 T+14726 p T^2-276 p^3 T^3+p^6 T^4
\tabularnewline[0.5pt]\hline				
84	 &	smooth    &	          &	1+1308 T+12854 p T^2+1308 p^3 T^3+p^6 T^4
\tabularnewline[0.5pt]\hline				
85	 &	smooth    &	          &	1+456 T-6946 p T^2+456 p^3 T^3+p^6 T^4
\tabularnewline[0.5pt]\hline				
86	 &	smooth    &	          &	1+1056 T+10190 p T^2+1056 p^3 T^3+p^6 T^4
\tabularnewline[0.5pt]\hline				
87	 &	smooth    &	          &	1+1464 T+14798 p T^2+1464 p^3 T^3+p^6 T^4
\tabularnewline[0.5pt]\hline				
88	 &	smooth    &	          &	1-132 T+4934 p T^2-132 p^3 T^3+p^6 T^4
\tabularnewline[0.5pt]\hline				
\tablepostamble				
\tablepreamble{97}				
1	 &	smooth    &	          &	1+164 T-4794 p T^2+164 p^3 T^3+p^6 T^4
\tabularnewline[0.5pt]\hline				
2	 &	smooth    &	          &	1+746 T+2322 p T^2+746 p^3 T^3+p^6 T^4
\tabularnewline[0.5pt]\hline				
3	 &	smooth    &	          &	(1+10 p T+p^3 T^2)(1-1550 T+p^3 T^2)
\tabularnewline[0.5pt]\hline				
4	 &	smooth    &	          &	1+554 T+8826 p T^2+554 p^3 T^3+p^6 T^4
\tabularnewline[0.5pt]\hline				
5	 &	smooth    &	          &	1+620 T+8550 p T^2+620 p^3 T^3+p^6 T^4
\tabularnewline[0.5pt]\hline				
6	 &	smooth    &	          &	1+26 T-126 p^2 T^2+26 p^3 T^3+p^6 T^4
\tabularnewline[0.5pt]\hline				
7	 &	smooth    &	          &	1-382 T+10266 p T^2-382 p^3 T^3+p^6 T^4
\tabularnewline[0.5pt]\hline				
8	 &	smooth    &	          &	1-64 T+4734 p T^2-64 p^3 T^3+p^6 T^4
\tabularnewline[0.5pt]\hline				
9	 &	smooth    &	          &	1+824 T+1374 p T^2+824 p^3 T^3+p^6 T^4
\tabularnewline[0.5pt]\hline				
10	 &	smooth    &	          &	1+248 T+3822 p T^2+248 p^3 T^3+p^6 T^4
\tabularnewline[0.5pt]\hline				
11	 &	smooth    &	          &	1-112 T-3666 p T^2-112 p^3 T^3+p^6 T^4
\tabularnewline[0.5pt]\hline				
12	 &	smooth    &	          &	1+380 T+3846 p T^2+380 p^3 T^3+p^6 T^4
\tabularnewline[0.5pt]\hline				
13	 &	smooth    &	          &	1+404 T+4086 p T^2+404 p^3 T^3+p^6 T^4
\tabularnewline[0.5pt]\hline				
14	 &	smooth    &	          &	1-604 T+13446 p T^2-604 p^3 T^3+p^6 T^4
\tabularnewline[0.5pt]\hline				
15	 &	smooth    &	          &	1-820 T+4374 p T^2-820 p^3 T^3+p^6 T^4
\tabularnewline[0.5pt]\hline				
16	 &	singular  &	-\frac{1}{6}\+&	(1-p T) (1+286 T+p^3 T^2)
\tabularnewline[0.5pt]\hline				
17	 &	smooth    &	          &	1-1540 T+24822 p T^2-1540 p^3 T^3+p^6 T^4
\tabularnewline[0.5pt]\hline				
18	 &	smooth    &	          &	1-952 T+14862 p T^2-952 p^3 T^3+p^6 T^4
\tabularnewline[0.5pt]\hline				
19	 &	smooth    &	          &	1-58 T-12198 p T^2-58 p^3 T^3+p^6 T^4
\tabularnewline[0.5pt]\hline				
20	 &	smooth    &	          &	1-1678 T+23010 p T^2-1678 p^3 T^3+p^6 T^4
\tabularnewline[0.5pt]\hline				
21	 &	smooth    &	          &	1-292 T+7494 p T^2-292 p^3 T^3+p^6 T^4
\tabularnewline[0.5pt]\hline				
22	 &	smooth    &	          &	1+1262 T+19074 p T^2+1262 p^3 T^3+p^6 T^4
\tabularnewline[0.5pt]\hline				
23	 &	smooth    &	          &	1-1228 T+12822 p T^2-1228 p^3 T^3+p^6 T^4
\tabularnewline[0.5pt]\hline				
24	 &	singular  &	-\frac{1}{4}\+&	(1-p T) (1+1330 T+p^3 T^2)
\tabularnewline[0.5pt]\hline				
25	 &	smooth    &	          &	1+1712 T+17310 p T^2+1712 p^3 T^3+p^6 T^4
\tabularnewline[0.5pt]\hline				
26	 &	smooth    &	          &	1+1016 T+11502 p T^2+1016 p^3 T^3+p^6 T^4
\tabularnewline[0.5pt]\hline				
27	 &	smooth    &	          &	1+440 T+7182 p T^2+440 p^3 T^3+p^6 T^4
\tabularnewline[0.5pt]\hline				
28	 &	smooth    &	          &	1+392 T+7134 p T^2+392 p^3 T^3+p^6 T^4
\tabularnewline[0.5pt]\hline				
29	 &	smooth    &	          &	1-52 T+8454 p T^2-52 p^3 T^3+p^6 T^4
\tabularnewline[0.5pt]\hline				
30	 &	smooth    &	          &	1+1244 T+7518 p T^2+1244 p^3 T^3+p^6 T^4
\tabularnewline[0.5pt]\hline				
31	 &	smooth    &	          &	1+32 T-498 p T^2+32 p^3 T^3+p^6 T^4
\tabularnewline[0.5pt]\hline				
32	 &	smooth    &	          &	1+704 T+1182 p T^2+704 p^3 T^3+p^6 T^4
\tabularnewline[0.5pt]\hline				
33	 &	smooth    &	          &	1+1814 T+19482 p T^2+1814 p^3 T^3+p^6 T^4
\tabularnewline[0.5pt]\hline				
34	 &	smooth    &	          &	1-1360 T+19998 p T^2-1360 p^3 T^3+p^6 T^4
\tabularnewline[0.5pt]\hline				
35	 &	smooth    &	          &	1+1178 T+7722 p T^2+1178 p^3 T^3+p^6 T^4
\tabularnewline[0.5pt]\hline				
36	 &	smooth    &	          &	1+1136 T+16590 p T^2+1136 p^3 T^3+p^6 T^4
\tabularnewline[0.5pt]\hline				
37	 &	smooth    &	          &	1+1694 T+19650 p T^2+1694 p^3 T^3+p^6 T^4
\tabularnewline[0.5pt]\hline				
38	 &	smooth    &	          &	1-1696 T+15486 p T^2-1696 p^3 T^3+p^6 T^4
\tabularnewline[0.5pt]\hline				
39	 &	smooth    &	          &	1-28 T+18342 p T^2-28 p^3 T^3+p^6 T^4
\tabularnewline[0.5pt]\hline				
40	 &	smooth    &	          &	1+26 T+6570 p T^2+26 p^3 T^3+p^6 T^4
\tabularnewline[0.5pt]\hline				
41	 &	smooth    &	          &	1+164 T+606 p T^2+164 p^3 T^3+p^6 T^4
\tabularnewline[0.5pt]\hline				
42	 &	smooth    &	          &	1+608 T+5694 p T^2+608 p^3 T^3+p^6 T^4
\tabularnewline[0.5pt]\hline				
43	 &	smooth    &	          &	1+1160 T+6318 p T^2+1160 p^3 T^3+p^6 T^4
\tabularnewline[0.5pt]\hline				
44	 &	smooth    &	          &	1+494 T-1422 p T^2+494 p^3 T^3+p^6 T^4
\tabularnewline[0.5pt]\hline				
45	 &	smooth    &	          &	1-64 T-2322 p T^2-64 p^3 T^3+p^6 T^4
\tabularnewline[0.5pt]\hline				
46	 &	smooth    &	          &	1-796 T+9078 p T^2-796 p^3 T^3+p^6 T^4
\tabularnewline[0.5pt]\hline				
47	 &	smooth    &	          &	1-1702 T+20394 p T^2-1702 p^3 T^3+p^6 T^4
\tabularnewline[0.5pt]\hline				
48	 &	smooth    &	          &	1+200 T+6942 p T^2+200 p^3 T^3+p^6 T^4
\tabularnewline[0.5pt]\hline				
49	 &	smooth    &	          &	1+8 T+10782 p T^2+8 p^3 T^3+p^6 T^4
\tabularnewline[0.5pt]\hline				
50	 &	smooth$^*$&	\frac{3}{2}&	  
\tabularnewline[0.5pt]\hline				
51	 &	smooth    &	          &	1+554 T+9186 p T^2+554 p^3 T^3+p^6 T^4
\tabularnewline[0.5pt]\hline				
52	 &	smooth    &	          &	1+1736 T+23886 p T^2+1736 p^3 T^3+p^6 T^4
\tabularnewline[0.5pt]\hline				
53	 &	smooth    &	          &	1-196 T+5718 p T^2-196 p^3 T^3+p^6 T^4
\tabularnewline[0.5pt]\hline				
54	 &	smooth    &	          &	1-22 T-4350 p T^2-22 p^3 T^3+p^6 T^4
\tabularnewline[0.5pt]\hline				
55	 &	smooth    &	          &	1+704 T-1986 p T^2+704 p^3 T^3+p^6 T^4
\tabularnewline[0.5pt]\hline				
56	 &	smooth    &	          &	1+512 T-6498 p T^2+512 p^3 T^3+p^6 T^4
\tabularnewline[0.5pt]\hline				
57	 &	smooth    &	          &	1+506 T+4458 p T^2+506 p^3 T^3+p^6 T^4
\tabularnewline[0.5pt]\hline				
58	 &	singular  &	-\frac{1}{5}\+&	(1-p T) (1+382 T+p^3 T^2)
\tabularnewline[0.5pt]\hline				
59	 &	smooth    &	          &	1+608 T+16494 p T^2+608 p^3 T^3+p^6 T^4
\tabularnewline[0.5pt]\hline				
60	 &	smooth    &	          &	1-1744 T+21342 p T^2-1744 p^3 T^3+p^6 T^4
\tabularnewline[0.5pt]\hline				
61	 &	smooth    &	          &	1+488 T+4062 p T^2+488 p^3 T^3+p^6 T^4
\tabularnewline[0.5pt]\hline				
62	 &	smooth    &	          &	1+68 T+2166 p T^2+68 p^3 T^3+p^6 T^4
\tabularnewline[0.5pt]\hline				
63	 &	smooth    &	          &	1-1132 T+12918 p T^2-1132 p^3 T^3+p^6 T^4
\tabularnewline[0.5pt]\hline				
64	 &	smooth    &	          &	1+680 T-3090 p T^2+680 p^3 T^3+p^6 T^4
\tabularnewline[0.5pt]\hline				
65	 &	singular  &	\frac{1}{3}&	(1-p T) (1+286 T+p^3 T^2)
\tabularnewline[0.5pt]\hline				
66	 &	smooth    &	          &	1+218 T-8790 p T^2+218 p^3 T^3+p^6 T^4
\tabularnewline[0.5pt]\hline				
67	 &	smooth    &	          &	1+1070 T+6210 p T^2+1070 p^3 T^3+p^6 T^4
\tabularnewline[0.5pt]\hline				
68	 &	smooth    &	          &	1-4 T+12102 p T^2-4 p^3 T^3+p^6 T^4
\tabularnewline[0.5pt]\hline				
69	 &	smooth    &	          &	1+26 T+11322 p T^2+26 p^3 T^3+p^6 T^4
\tabularnewline[0.5pt]\hline				
70	 &	smooth    &	          &	1-964 T+18486 p T^2-964 p^3 T^3+p^6 T^4
\tabularnewline[0.5pt]\hline				
71	 &	smooth    &	          &	1+836 T+10134 p T^2+836 p^3 T^3+p^6 T^4
\tabularnewline[0.5pt]\hline				
72	 &	smooth    &	          &	1+788 T+8358 p T^2+788 p^3 T^3+p^6 T^4
\tabularnewline[0.5pt]\hline				
73	 &	singular  &	\frac{1}{4}&	(1-p T) (1-1106 T+p^3 T^2)
\tabularnewline[0.5pt]\hline				
74	 &	smooth    &	          &	(1-8 p T+p^3 T^2)(1+1294 T+p^3 T^2)
\tabularnewline[0.5pt]\hline				
75	 &	smooth    &	          &	1+32 T+366 p T^2+32 p^3 T^3+p^6 T^4
\tabularnewline[0.5pt]\hline				
76	 &	smooth    &	          &	1+170 T+11754 p T^2+170 p^3 T^3+p^6 T^4
\tabularnewline[0.5pt]\hline				
77	 &	smooth    &	          &	1+248 T+2526 p T^2+248 p^3 T^3+p^6 T^4
\tabularnewline[0.5pt]\hline				
78	 &	smooth    &	          &	1-148 T+12822 p T^2-148 p^3 T^3+p^6 T^4
\tabularnewline[0.5pt]\hline				
79	 &	smooth    &	          &	1-1888 T+20046 p T^2-1888 p^3 T^3+p^6 T^4
\tabularnewline[0.5pt]\hline				
80	 &	smooth    &	          &	1+524 T+3702 p T^2+524 p^3 T^3+p^6 T^4
\tabularnewline[0.5pt]\hline				
81	 &	smooth    &	          &	1+512 T+10206 p T^2+512 p^3 T^3+p^6 T^4
\tabularnewline[0.5pt]\hline				
82	 &	smooth    &	          &	1+272 T+10110 p T^2+272 p^3 T^3+p^6 T^4
\tabularnewline[0.5pt]\hline				
83	 &	smooth    &	          &	1+1076 T+2886 p T^2+1076 p^3 T^3+p^6 T^4
\tabularnewline[0.5pt]\hline				
84	 &	smooth    &	          &	1-322 T+14754 p T^2-322 p^3 T^3+p^6 T^4
\tabularnewline[0.5pt]\hline				
85	 &	smooth    &	          &	1+140 T-12090 p T^2+140 p^3 T^3+p^6 T^4
\tabularnewline[0.5pt]\hline				
86	 &	smooth    &	          &	1-982 T+14058 p T^2-982 p^3 T^3+p^6 T^4
\tabularnewline[0.5pt]\hline				
87	 &	smooth    &	          &	1-496 T+17550 p T^2-496 p^3 T^3+p^6 T^4
\tabularnewline[0.5pt]\hline				
88	 &	smooth    &	          &	1-178 T-2238 p T^2-178 p^3 T^3+p^6 T^4
\tabularnewline[0.5pt]\hline				
89	 &	singular  &	\frac{1}{12}&	(1-p T) (1+1690 T+p^3 T^2)
\tabularnewline[0.5pt]\hline				
90	 &	smooth    &	          &	1-1060 T+8454 p T^2-1060 p^3 T^3+p^6 T^4
\tabularnewline[0.5pt]\hline				
91	 &	smooth    &	          &	1-16 T+16302 p T^2-16 p^3 T^3+p^6 T^4
\tabularnewline[0.5pt]\hline				
92	 &	smooth    &	          &	1-748 T+10710 p T^2-748 p^3 T^3+p^6 T^4
\tabularnewline[0.5pt]\hline				
93	 &	smooth    &	          &	1-484 T+2334 p T^2-484 p^3 T^3+p^6 T^4
\tabularnewline[0.5pt]\hline				
94	 &	smooth    &	          &	1+1184 T+18150 p T^2+1184 p^3 T^3+p^6 T^4
\tabularnewline[0.5pt]\hline				
95	 &	smooth    &	          &	1+248 T-2802 p T^2+248 p^3 T^3+p^6 T^4
\tabularnewline[0.5pt]\hline				
96	 &	smooth    &	          &	1-928 T+7614 p T^2-928 p^3 T^3+p^6 T^4
\tabularnewline[0.5pt]\hline				
\tablepostamble
\newpage				
\lhead{\ifthenelse{\isodd{\value{page}}}{\thepage}{\sl The $\z$-function for a quotient of the 24 cell, AESZ\hskip2pt 366}}
\rhead{\ifthenelse{\isodd{\value{page}}}{\sl The $\z$-function for a quotient of the 24 cell, 
      AESZ\hskip2pt 366}{\thepage}}
\subsection{The $\z$-function for a quotient of the 24 cell, AESZ\hskip2pt 366}				
\vspace{1.5cm}				
\tablepreamble{5}				
1	 &	singular  &	-\frac{1}{4}\+&	(1-p T) (1-6 T+p^3 T^2)
\tabularnewline[0.5pt]\hline				
2	 &	singular  &	-\frac{1}{12}\+&	(1-p T) (1+18 T+p^3 T^2)
\tabularnewline[0.5pt]\hline				
3	 &	singular  &	\hspace*{-3pt}\left\{-\frac{1}{3},-\frac{1}{8},-\frac{1}{18}\right\}\hspace*{-3pt}&	 
\tabularnewline[0.5pt]\hline				
4	 &	singular  &	\frac{1}{24}&	(1-p T) (1-2 T+p^3 T^2)
\tabularnewline[0.5pt]\hline				
\tablepostamble				
\tablepreamble{7}				
1	 &	smooth    &	          &	1+16 T+2 p T^2+16 p^3 T^3+p^6 T^4
\tabularnewline[0.5pt]\hline				
2	 &	singular  &	-\frac{1}{3}\+&	(1-p T) (1+24 T+p^3 T^2)
\tabularnewline[0.5pt]\hline				
3	 &	smooth    &	          &	(1+4 p T+p^3 T^2)(1-24 T+p^3 T^2)
\tabularnewline[0.5pt]\hline				
4	 &	singular  &	-\frac{1}{12}\+&	(1+p T) (1-8 T+p^3 T^2)
\tabularnewline[0.5pt]\hline				
5	 &	singular  &	\left\{-\frac{1}{4},\frac{1}{24},-\frac{1}{18}\right\}&	  
\tabularnewline[0.5pt]\hline				
6	 &	singular  &	-\frac{1}{8}\+&	(1-p T) (1+4 T+p^3 T^2)
\tabularnewline[0.5pt]\hline				
\tablepostamble				
\tablepreamble{11}				
1	 &	smooth    &	          &	1-20 T+50 p T^2-20 p^3 T^3+p^6 T^4
\tabularnewline[0.5pt]\hline				
2	 &	smooth    &	          &	(1-4 p T+p^3 T^2)(1+12 T+p^3 T^2)
\tabularnewline[0.5pt]\hline				
3	 &	smooth$^*$&	-\frac{1}{18}\+&	  
\tabularnewline[0.5pt]\hline				
4	 &	singular  &	-\frac{1}{8}\+&	(1-p T) (1-12 T+p^3 T^2)
\tabularnewline[0.5pt]\hline				
5	 &	smooth    &	          &	(1+4 p T+p^3 T^2)(1-60 T+p^3 T^2)
\tabularnewline[0.5pt]\hline				
6	 &	singular  &	\frac{1}{24}&	(1-p T) (1+8 T+p^3 T^2)
\tabularnewline[0.5pt]\hline				
7	 &	singular  &	-\frac{1}{3}\+&	(1-p T) (1-52 T+p^3 T^2)
\tabularnewline[0.5pt]\hline				
8	 &	singular  &	-\frac{1}{4}\+&	(1+p T) (1+12 T+p^3 T^2)
\tabularnewline[0.5pt]\hline				
9	 &	smooth    &	          &	1-32 T+2 p T^2-32 p^3 T^3+p^6 T^4
\tabularnewline[0.5pt]\hline				
10	 &	singular  &	-\frac{1}{12}\+&	(1+p T) (1-36 T+p^3 T^2)
\tabularnewline[0.5pt]\hline				
\tablepostamble				
\tablepreamble{13}				
1	 &	singular  &	-\frac{1}{12}\+&	(1-p T) (1+10 T+p^3 T^2)
\tabularnewline[0.5pt]\hline				
2	 &	smooth    &	          &	1-36 T+22 p T^2-36 p^3 T^3+p^6 T^4
\tabularnewline[0.5pt]\hline				
3	 &	singular  &	-\frac{1}{4}\+&	(1-p T) (1+82 T+p^3 T^2)
\tabularnewline[0.5pt]\hline				
4	 &	singular  &	-\frac{1}{3}\+&	(1-p T) (1-22 T+p^3 T^2)
\tabularnewline[0.5pt]\hline				
5	 &	smooth$^*$&	-\frac{1}{18}\+&	  
\tabularnewline[0.5pt]\hline				
6	 &	singular  &	\frac{1}{24}&	(1-p T) (1+42 T+p^3 T^2)
\tabularnewline[0.5pt]\hline				
7	 &	smooth    &	          &	1+24 T+142 p T^2+24 p^3 T^3+p^6 T^4
\tabularnewline[0.5pt]\hline				
8	 &	singular  &	-\frac{1}{8}\+&	(1-p T) (1+58 T+p^3 T^2)
\tabularnewline[0.5pt]\hline				
9	 &	smooth    &	          &	1-12 T+70 p T^2-12 p^3 T^3+p^6 T^4
\tabularnewline[0.5pt]\hline				
10	 &	smooth    &	          &	1+72 T+238 p T^2+72 p^3 T^3+p^6 T^4
\tabularnewline[0.5pt]\hline				
11	 &	smooth    &	          &	1+12 T-74 p T^2+12 p^3 T^3+p^6 T^4
\tabularnewline[0.5pt]\hline				
12	 &	smooth    &	          &	1-56 T+110 p T^2-56 p^3 T^3+p^6 T^4
\tabularnewline[0.5pt]\hline				
\tablepostamble				
\tablepreamble{17}				
1	 &	smooth    &	          &	1-52 T+230 p T^2-52 p^3 T^3+p^6 T^4
\tabularnewline[0.5pt]\hline				
2	 &	singular  &	-\frac{1}{8}\+&	(1-p T) (1-66 T+p^3 T^2)
\tabularnewline[0.5pt]\hline				
3	 &	smooth    &	          &	1+12 T-154 p T^2+12 p^3 T^3+p^6 T^4
\tabularnewline[0.5pt]\hline				
4	 &	singular  &	-\frac{1}{4}\+&	(1-p T) (1+30 T+p^3 T^2)
\tabularnewline[0.5pt]\hline				
5	 &	singular  &	\frac{1}{24}&	(1-p T) (1+2 T+p^3 T^2)
\tabularnewline[0.5pt]\hline				
6	 &	smooth    &	          &	1+48 T-2 p^2 T^2+48 p^3 T^3+p^6 T^4
\tabularnewline[0.5pt]\hline				
7	 &	singular  &	-\frac{1}{12}\+&	(1-p T) (1-18 T+p^3 T^2)
\tabularnewline[0.5pt]\hline				
8	 &	smooth    &	          &	1-60 T+22 p^2 T^2-60 p^3 T^3+p^6 T^4
\tabularnewline[0.5pt]\hline				
9	 &	smooth    &	          &	1-4 T-26 p^2 T^2-4 p^3 T^3+p^6 T^4
\tabularnewline[0.5pt]\hline				
10	 &	smooth    &	          &	1-36 T+326 p T^2-36 p^3 T^3+p^6 T^4
\tabularnewline[0.5pt]\hline				
11	 &	singular  &	-\frac{1}{3}\+&	(1-p T) (1+14 T+p^3 T^2)
\tabularnewline[0.5pt]\hline				
12	 &	smooth    &	          &	1+20 T+278 p T^2+20 p^3 T^3+p^6 T^4
\tabularnewline[0.5pt]\hline				
13	 &	smooth    &	          &	1-28 T+182 p T^2-28 p^3 T^3+p^6 T^4
\tabularnewline[0.5pt]\hline				
14	 &	smooth    &	          &	1-40 T+14 p T^2-40 p^3 T^3+p^6 T^4
\tabularnewline[0.5pt]\hline				
15	 &	smooth    &	          &	1-28 T+182 p T^2-28 p^3 T^3+p^6 T^4
\tabularnewline[0.5pt]\hline				
16	 &	smooth$^*$&	-\frac{1}{18}\+&	  
\tabularnewline[0.5pt]\hline				
\tablepostamble				
\tablepreamble{19}				
1	 &	smooth$^*$&	-\frac{1}{18}\+&	  
\tabularnewline[0.5pt]\hline				
2	 &	smooth    &	          &	1+72 T+514 p T^2+72 p^3 T^3+p^6 T^4
\tabularnewline[0.5pt]\hline				
3	 &	smooth    &	          &	1+80 T+354 p T^2+80 p^3 T^3+p^6 T^4
\tabularnewline[0.5pt]\hline				
4	 &	singular  &	\frac{1}{24}&	(1-p T) (1+124 T+p^3 T^2)
\tabularnewline[0.5pt]\hline				
5	 &	smooth    &	          &	(1-4 p T+p^3 T^2)(1+16 T+p^3 T^2)
\tabularnewline[0.5pt]\hline				
6	 &	singular  &	-\frac{1}{3}\+&	(1-p T) (1+20 T+p^3 T^2)
\tabularnewline[0.5pt]\hline				
7	 &	singular  &	-\frac{1}{8}\+&	(1-p T) (1+100 T+p^3 T^2)
\tabularnewline[0.5pt]\hline				
8	 &	smooth    &	          &	1-48 T-158 p T^2-48 p^3 T^3+p^6 T^4
\tabularnewline[0.5pt]\hline				
9	 &	smooth    &	          &	1-64 T+450 p T^2-64 p^3 T^3+p^6 T^4
\tabularnewline[0.5pt]\hline				
10	 &	smooth    &	          &	1+68 T+402 p T^2+68 p^3 T^3+p^6 T^4
\tabularnewline[0.5pt]\hline				
11	 &	singular  &	-\frac{1}{12}\+&	(1+p T) (1+100 T+p^3 T^2)
\tabularnewline[0.5pt]\hline				
12	 &	smooth    &	          &	(1+8 p T+p^3 T^2)(1+4 T+p^3 T^2)
\tabularnewline[0.5pt]\hline				
13	 &	smooth    &	          &	1-52 T+6 p^2 T^2-52 p^3 T^3+p^6 T^4
\tabularnewline[0.5pt]\hline				
14	 &	singular  &	-\frac{1}{4}\+&	(1+p T) (1-68 T+p^3 T^2)
\tabularnewline[0.5pt]\hline				
15	 &	smooth    &	          &	(1-4 p T+p^3 T^2)(1+88 T+p^3 T^2)
\tabularnewline[0.5pt]\hline				
16	 &	smooth    &	          &	1+514 p T^2+p^6 T^4
\tabularnewline[0.5pt]\hline				
17	 &	smooth    &	          &	1+32 T+162 p T^2+32 p^3 T^3+p^6 T^4
\tabularnewline[0.5pt]\hline				
18	 &	smooth    &	          &	1+120 T+706 p T^2+120 p^3 T^3+p^6 T^4
\tabularnewline[0.5pt]\hline				
\tablepostamble				
\tablepreamble{23}				
1	 &	singular  &	\frac{1}{24}&	(1-p T) (1-76 T+p^3 T^2)
\tabularnewline[0.5pt]\hline				
2	 &	smooth    &	          &	1-48 T+98 p T^2-48 p^3 T^3+p^6 T^4
\tabularnewline[0.5pt]\hline				
3	 &	smooth    &	          &	1-96 T-94 p T^2-96 p^3 T^3+p^6 T^4
\tabularnewline[0.5pt]\hline				
4	 &	smooth    &	          &	1+116 T+578 p T^2+116 p^3 T^3+p^6 T^4
\tabularnewline[0.5pt]\hline				
5	 &	smooth    &	          &	(1+p^3 T^2)(1-120 T+p^3 T^2)
\tabularnewline[0.5pt]\hline				
6	 &	smooth    &	          &	1+248 T+1442 p T^2+248 p^3 T^3+p^6 T^4
\tabularnewline[0.5pt]\hline				
7	 &	smooth    &	          &	(1+p^3 T^2)(1-192 T+p^3 T^2)
\tabularnewline[0.5pt]\hline				
8	 &	smooth    &	          &	1+104 T+386 p T^2+104 p^3 T^3+p^6 T^4
\tabularnewline[0.5pt]\hline				
9	 &	smooth    &	          &	1-144 T+1250 p T^2-144 p^3 T^3+p^6 T^4
\tabularnewline[0.5pt]\hline				
10	 &	smooth    &	          &	1-80 T+290 p T^2-80 p^3 T^3+p^6 T^4
\tabularnewline[0.5pt]\hline				
11	 &	smooth    &	          &	1+12 T+386 p T^2+12 p^3 T^3+p^6 T^4
\tabularnewline[0.5pt]\hline				
12	 &	smooth    &	          &	1+144 T+1250 p T^2+144 p^3 T^3+p^6 T^4
\tabularnewline[0.5pt]\hline				
13	 &	smooth    &	          &	1+72 T+482 p T^2+72 p^3 T^3+p^6 T^4
\tabularnewline[0.5pt]\hline				
14	 &	smooth$^*$&	-\frac{1}{18}\+&	  
\tabularnewline[0.5pt]\hline				
15	 &	singular  &	-\frac{1}{3}\+&	(1-p T) (1+168 T+p^3 T^2)
\tabularnewline[0.5pt]\hline				
16	 &	smooth    &	          &	1+100 T+770 p T^2+100 p^3 T^3+p^6 T^4
\tabularnewline[0.5pt]\hline				
17	 &	singular  &	-\frac{1}{4}\+&	(1+p T) (1-216 T+p^3 T^2)
\tabularnewline[0.5pt]\hline				
18	 &	smooth    &	          &	1-36 T-190 p T^2-36 p^3 T^3+p^6 T^4
\tabularnewline[0.5pt]\hline				
19	 &	smooth    &	          &	1-128 T+866 p T^2-128 p^3 T^3+p^6 T^4
\tabularnewline[0.5pt]\hline				
20	 &	singular  &	-\frac{1}{8}\+&	(1-p T) (1-132 T+p^3 T^2)
\tabularnewline[0.5pt]\hline				
21	 &	singular  &	-\frac{1}{12}\+&	(1+p T) (1-72 T+p^3 T^2)
\tabularnewline[0.5pt]\hline				
22	 &	smooth    &	          &	1-12 T+194 p T^2-12 p^3 T^3+p^6 T^4
\tabularnewline[0.5pt]\hline				
\tablepostamble				
\tablepreamble{29}				
1	 &	smooth    &	          &	(1+2 p T+p^3 T^2)(1-198 T+p^3 T^2)
\tabularnewline[0.5pt]\hline				
2	 &	smooth    &	          &	1-260 T+1430 p T^2-260 p^3 T^3+p^6 T^4
\tabularnewline[0.5pt]\hline				
3	 &	smooth    &	          &	1+108 T+758 p T^2+108 p^3 T^3+p^6 T^4
\tabularnewline[0.5pt]\hline				
4	 &	smooth    &	          &	1-20 T+1142 p T^2-20 p^3 T^3+p^6 T^4
\tabularnewline[0.5pt]\hline				
5	 &	smooth    &	          &	1-44 T+710 p T^2-44 p^3 T^3+p^6 T^4
\tabularnewline[0.5pt]\hline				
6	 &	smooth    &	          &	1+112 T+62 p T^2+112 p^3 T^3+p^6 T^4
\tabularnewline[0.5pt]\hline				
7	 &	singular  &	-\frac{1}{4}\+&	(1-p T) (1-246 T+p^3 T^2)
\tabularnewline[0.5pt]\hline				
8	 &	smooth$^*$&	-\frac{1}{18}\+&	  
\tabularnewline[0.5pt]\hline				
9	 &	smooth    &	          &	1+12 T-970 p T^2+12 p^3 T^3+p^6 T^4
\tabularnewline[0.5pt]\hline				
10	 &	smooth    &	          &	1+204 T+1718 p T^2+204 p^3 T^3+p^6 T^4
\tabularnewline[0.5pt]\hline				
11	 &	smooth    &	          &	1+64 T+926 p T^2+64 p^3 T^3+p^6 T^4
\tabularnewline[0.5pt]\hline				
12	 &	singular  &	-\frac{1}{12}\+&	(1-p T) (1+234 T+p^3 T^2)
\tabularnewline[0.5pt]\hline				
13	 &	smooth    &	          &	1-72 T+14 p T^2-72 p^3 T^3+p^6 T^4
\tabularnewline[0.5pt]\hline				
14	 &	smooth    &	          &	1-260 T+1814 p T^2-260 p^3 T^3+p^6 T^4
\tabularnewline[0.5pt]\hline				
15	 &	smooth    &	          &	1-76 T+518 p T^2-76 p^3 T^3+p^6 T^4
\tabularnewline[0.5pt]\hline				
16	 &	smooth    &	          &	(1-6 p T+p^3 T^2)(1+250 T+p^3 T^2)
\tabularnewline[0.5pt]\hline				
17	 &	smooth    &	          &	(1-6 p T+p^3 T^2)(1+154 T+p^3 T^2)
\tabularnewline[0.5pt]\hline				
18	 &	singular  &	-\frac{1}{8}\+&	(1-p T) (1+90 T+p^3 T^2)
\tabularnewline[0.5pt]\hline				
19	 &	singular  &	-\frac{1}{3}\+&	(1-p T) (1-230 T+p^3 T^2)
\tabularnewline[0.5pt]\hline				
20	 &	smooth    &	          &	1+268 T+46 p^2 T^2+268 p^3 T^3+p^6 T^4
\tabularnewline[0.5pt]\hline				
21	 &	smooth    &	          &	1+4 T+1190 p T^2+4 p^3 T^3+p^6 T^4
\tabularnewline[0.5pt]\hline				
22	 &	smooth    &	          &	1-36 T-682 p T^2-36 p^3 T^3+p^6 T^4
\tabularnewline[0.5pt]\hline				
23	 &	singular  &	\frac{1}{24}&	(1-p T) (1-254 T+p^3 T^2)
\tabularnewline[0.5pt]\hline				
24	 &	smooth    &	          &	1-192 T+1502 p T^2-192 p^3 T^3+p^6 T^4
\tabularnewline[0.5pt]\hline				
25	 &	smooth    &	          &	(1-6 p T+p^3 T^2)(1-86 T+p^3 T^2)
\tabularnewline[0.5pt]\hline				
26	 &	smooth    &	          &	1+84 T-2 p^2 T^2+84 p^3 T^3+p^6 T^4
\tabularnewline[0.5pt]\hline				
27	 &	smooth    &	          &	1-260 T+2102 p T^2-260 p^3 T^3+p^6 T^4
\tabularnewline[0.5pt]\hline				
28	 &	smooth    &	          &	1+268 T+1526 p T^2+268 p^3 T^3+p^6 T^4
\tabularnewline[0.5pt]\hline				
\tablepostamble				
\tablepreamble{31}				
1	 &	smooth    &	          &	1-12 T+1666 p T^2-12 p^3 T^3+p^6 T^4
\tabularnewline[0.5pt]\hline				
2	 &	smooth    &	          &	1+56 T+258 p T^2+56 p^3 T^3+p^6 T^4
\tabularnewline[0.5pt]\hline				
3	 &	smooth    &	          &	1+80 T-126 p T^2+80 p^3 T^3+p^6 T^4
\tabularnewline[0.5pt]\hline				
4	 &	smooth    &	          &	1+88 T+2 p T^2+88 p^3 T^3+p^6 T^4
\tabularnewline[0.5pt]\hline				
5	 &	smooth    &	          &	1+152 T+834 p T^2+152 p^3 T^3+p^6 T^4
\tabularnewline[0.5pt]\hline				
6	 &	smooth    &	          &	1+280 T+1538 p T^2+280 p^3 T^3+p^6 T^4
\tabularnewline[0.5pt]\hline				
7	 &	smooth    &	          &	1+88 T+578 p T^2+88 p^3 T^3+p^6 T^4
\tabularnewline[0.5pt]\hline				
8	 &	smooth    &	          &	1-120 T+130 p T^2-120 p^3 T^3+p^6 T^4
\tabularnewline[0.5pt]\hline				
9	 &	smooth    &	          &	1-44 T+1346 p T^2-44 p^3 T^3+p^6 T^4
\tabularnewline[0.5pt]\hline				
10	 &	singular  &	-\frac{1}{3}\+&	(1-p T) (1+288 T+p^3 T^2)
\tabularnewline[0.5pt]\hline				
11	 &	smooth    &	          &	1-360 T+2338 p T^2-360 p^3 T^3+p^6 T^4
\tabularnewline[0.5pt]\hline				
12	 &	smooth$^*$&	-\frac{1}{18}\+&	  
\tabularnewline[0.5pt]\hline				
13	 &	smooth    &	          &	1+176 T+1410 p T^2+176 p^3 T^3+p^6 T^4
\tabularnewline[0.5pt]\hline				
14	 &	smooth    &	          &	1+40 T-1150 p T^2+40 p^3 T^3+p^6 T^4
\tabularnewline[0.5pt]\hline				
15	 &	smooth    &	          &	1+120 T+1282 p T^2+120 p^3 T^3+p^6 T^4
\tabularnewline[0.5pt]\hline				
16	 &	smooth    &	          &	(1+4 p T+p^3 T^2)(1-48 T+p^3 T^2)
\tabularnewline[0.5pt]\hline				
17	 &	smooth    &	          &	1+24 T-1214 p T^2+24 p^3 T^3+p^6 T^4
\tabularnewline[0.5pt]\hline				
18	 &	singular  &	-\frac{1}{12}\+&	(1+p T) (1+16 T+p^3 T^2)
\tabularnewline[0.5pt]\hline				
19	 &	smooth    &	          &	1+16 T+386 p T^2+16 p^3 T^3+p^6 T^4
\tabularnewline[0.5pt]\hline				
20	 &	smooth    &	          &	1+120 T+514 p T^2+120 p^3 T^3+p^6 T^4
\tabularnewline[0.5pt]\hline				
21	 &	smooth    &	          &	(1-8 p T+p^3 T^2)(1-8 T+p^3 T^2)
\tabularnewline[0.5pt]\hline				
22	 &	singular  &	\frac{1}{24}&	(1-p T) (1+72 T+p^3 T^2)
\tabularnewline[0.5pt]\hline				
23	 &	singular  &	-\frac{1}{4}\+&	(1+p T) (1+112 T+p^3 T^2)
\tabularnewline[0.5pt]\hline				
24	 &	smooth    &	          &	1+232 T+962 p T^2+232 p^3 T^3+p^6 T^4
\tabularnewline[0.5pt]\hline				
25	 &	smooth    &	          &	1-36 T+322 p T^2-36 p^3 T^3+p^6 T^4
\tabularnewline[0.5pt]\hline				
26	 &	smooth    &	          &	1-8 T-766 p T^2-8 p^3 T^3+p^6 T^4
\tabularnewline[0.5pt]\hline				
27	 &	singular  &	-\frac{1}{8}\+&	(1-p T) (1-152 T+p^3 T^2)
\tabularnewline[0.5pt]\hline				
28	 &	smooth    &	          &	1-24 T+1282 p T^2-24 p^3 T^3+p^6 T^4
\tabularnewline[0.5pt]\hline				
29	 &	smooth    &	          &	1-100 T+1602 p T^2-100 p^3 T^3+p^6 T^4
\tabularnewline[0.5pt]\hline				
30	 &	smooth    &	          &	1+104 T+30 p^2 T^2+104 p^3 T^3+p^6 T^4
\tabularnewline[0.5pt]\hline				
\tablepostamble				
\tablepreamble{37}				
1	 &	smooth    &	          &	1+172 T+1910 p T^2+172 p^3 T^3+p^6 T^4
\tabularnewline[0.5pt]\hline				
2	 &	smooth$^*$&	-\frac{1}{18}\+&	  
\tabularnewline[0.5pt]\hline				
3	 &	singular  &	-\frac{1}{12}\+&	(1-p T) (1+226 T+p^3 T^2)
\tabularnewline[0.5pt]\hline				
4	 &	smooth    &	          &	1-64 T-642 p T^2-64 p^3 T^3+p^6 T^4
\tabularnewline[0.5pt]\hline				
5	 &	smooth    &	          &	1+360 T+1678 p T^2+360 p^3 T^3+p^6 T^4
\tabularnewline[0.5pt]\hline				
6	 &	smooth    &	          &	1-20 T+758 p T^2-20 p^3 T^3+p^6 T^4
\tabularnewline[0.5pt]\hline				
7	 &	smooth    &	          &	1+408 T+3694 p T^2+408 p^3 T^3+p^6 T^4
\tabularnewline[0.5pt]\hline				
8	 &	smooth    &	          &	1+96 T+1150 p T^2+96 p^3 T^3+p^6 T^4
\tabularnewline[0.5pt]\hline				
9	 &	singular  &	-\frac{1}{4}\+&	(1-p T) (1-110 T+p^3 T^2)
\tabularnewline[0.5pt]\hline				
10	 &	smooth    &	          &	1+28 T+2006 p T^2+28 p^3 T^3+p^6 T^4
\tabularnewline[0.5pt]\hline				
11	 &	smooth    &	          &	1+460 T+3254 p T^2+460 p^3 T^3+p^6 T^4
\tabularnewline[0.5pt]\hline				
12	 &	singular  &	-\frac{1}{3}\+&	(1-p T) (1+34 T+p^3 T^2)
\tabularnewline[0.5pt]\hline				
13	 &	smooth    &	          &	1+4 p T+1094 p T^2+4 p^4 T^3+p^6 T^4
\tabularnewline[0.5pt]\hline				
14	 &	smooth    &	          &	1-292 T+2646 p T^2-292 p^3 T^3+p^6 T^4
\tabularnewline[0.5pt]\hline				
15	 &	smooth    &	          &	1+120 T+22 p^2 T^2+120 p^3 T^3+p^6 T^4
\tabularnewline[0.5pt]\hline				
16	 &	smooth    &	          &	1-36 T-170 p T^2-36 p^3 T^3+p^6 T^4
\tabularnewline[0.5pt]\hline				
17	 &	singular  &	\frac{1}{24}&	(1-p T) (1-398 T+p^3 T^2)
\tabularnewline[0.5pt]\hline				
18	 &	smooth    &	          &	1-60 T-1178 p T^2-60 p^3 T^3+p^6 T^4
\tabularnewline[0.5pt]\hline				
19	 &	smooth    &	          &	1-92 T+14 p^2 T^2-92 p^3 T^3+p^6 T^4
\tabularnewline[0.5pt]\hline				
20	 &	smooth    &	          &	1+260 T+870 p T^2+260 p^3 T^3+p^6 T^4
\tabularnewline[0.5pt]\hline				
21	 &	smooth    &	          &	1+20 T+2310 p T^2+20 p^3 T^3+p^6 T^4
\tabularnewline[0.5pt]\hline				
22	 &	smooth    &	          &	1+196 T+1382 p T^2+196 p^3 T^3+p^6 T^4
\tabularnewline[0.5pt]\hline				
23	 &	singular  &	-\frac{1}{8}\+&	(1-p T) (1+34 T+p^3 T^2)
\tabularnewline[0.5pt]\hline				
24	 &	smooth    &	          &	1-272 T+2270 p T^2-272 p^3 T^3+p^6 T^4
\tabularnewline[0.5pt]\hline				
25	 &	smooth    &	          &	1+16 p T+4766 p T^2+16 p^4 T^3+p^6 T^4
\tabularnewline[0.5pt]\hline				
26	 &	smooth    &	          &	1-268 T+1926 p T^2-268 p^3 T^3+p^6 T^4
\tabularnewline[0.5pt]\hline				
27	 &	smooth    &	          &	1-124 T+294 p T^2-124 p^3 T^3+p^6 T^4
\tabularnewline[0.5pt]\hline				
28	 &	smooth    &	          &	1+220 T+470 p T^2+220 p^3 T^3+p^6 T^4
\tabularnewline[0.5pt]\hline				
29	 &	smooth    &	          &	1-36 T-938 p T^2-36 p^3 T^3+p^6 T^4
\tabularnewline[0.5pt]\hline				
30	 &	smooth    &	          &	1+316 T+1814 p T^2+316 p^3 T^3+p^6 T^4
\tabularnewline[0.5pt]\hline				
31	 &	smooth    &	          &	1-156 T+1318 p T^2-156 p^3 T^3+p^6 T^4
\tabularnewline[0.5pt]\hline				
32	 &	smooth    &	          &	1+292 T+2726 p T^2+292 p^3 T^3+p^6 T^4
\tabularnewline[0.5pt]\hline				
33	 &	smooth    &	          &	1-36 T-938 p T^2-36 p^3 T^3+p^6 T^4
\tabularnewline[0.5pt]\hline				
34	 &	smooth    &	          &	(1-10 p T+p^3 T^2)(1+322 T+p^3 T^2)
\tabularnewline[0.5pt]\hline				
35	 &	smooth    &	          &	1-116 T+182 p T^2-116 p^3 T^3+p^6 T^4
\tabularnewline[0.5pt]\hline				
36	 &	smooth    &	          &	1-188 T+230 p T^2-188 p^3 T^3+p^6 T^4
\tabularnewline[0.5pt]\hline				
\tablepostamble				
\tablepreamble{41}				
1	 &	smooth    &	          &	1+60 T-58 p T^2+60 p^3 T^3+p^6 T^4
\tabularnewline[0.5pt]\hline				
2	 &	smooth    &	          &	1+132 T-778 p T^2+132 p^3 T^3+p^6 T^4
\tabularnewline[0.5pt]\hline				
3	 &	smooth    &	          &	1-304 T+2174 p T^2-304 p^3 T^3+p^6 T^4
\tabularnewline[0.5pt]\hline				
4	 &	smooth    &	          &	1-120 T+1550 p T^2-120 p^3 T^3+p^6 T^4
\tabularnewline[0.5pt]\hline				
5	 &	singular  &	-\frac{1}{8}\+&	(1-p T) (1+438 T+p^3 T^2)
\tabularnewline[0.5pt]\hline				
6	 &	smooth    &	          &	1-276 T+2534 p T^2-276 p^3 T^3+p^6 T^4
\tabularnewline[0.5pt]\hline				
7	 &	smooth    &	          &	1+76 T-730 p T^2+76 p^3 T^3+p^6 T^4
\tabularnewline[0.5pt]\hline				
8	 &	smooth    &	          &	1+4 p T+950 p T^2+4 p^4 T^3+p^6 T^4
\tabularnewline[0.5pt]\hline				
9	 &	smooth    &	          &	1-152 T-754 p T^2-152 p^3 T^3+p^6 T^4
\tabularnewline[0.5pt]\hline				
10	 &	singular  &	-\frac{1}{4}\+&	(1-p T) (1+246 T+p^3 T^2)
\tabularnewline[0.5pt]\hline				
11	 &	smooth    &	          &	1+156 T+134 p T^2+156 p^3 T^3+p^6 T^4
\tabularnewline[0.5pt]\hline				
12	 &	singular  &	\frac{1}{24}&	(1-p T) (1-462 T+p^3 T^2)
\tabularnewline[0.5pt]\hline				
13	 &	smooth    &	          &	(1-10 p T+p^3 T^2)(1+78 T+p^3 T^2)
\tabularnewline[0.5pt]\hline				
14	 &	smooth    &	          &	1+60 T+1382 p T^2+60 p^3 T^3+p^6 T^4
\tabularnewline[0.5pt]\hline				
15	 &	smooth    &	          &	1+152 T+1070 p T^2+152 p^3 T^3+p^6 T^4
\tabularnewline[0.5pt]\hline				
16	 &	smooth    &	          &	1+204 T+38 p T^2+204 p^3 T^3+p^6 T^4
\tabularnewline[0.5pt]\hline				
17	 &	singular  &	-\frac{1}{12}\+&	(1-p T) (1-90 T+p^3 T^2)
\tabularnewline[0.5pt]\hline				
18	 &	smooth    &	          &	1+128 T-418 p T^2+128 p^3 T^3+p^6 T^4
\tabularnewline[0.5pt]\hline				
19	 &	smooth    &	          &	1-252 T+1526 p T^2-252 p^3 T^3+p^6 T^4
\tabularnewline[0.5pt]\hline				
20	 &	smooth    &	          &	1-324 T+3014 p T^2-324 p^3 T^3+p^6 T^4
\tabularnewline[0.5pt]\hline				
21	 &	smooth    &	          &	1+132 T+950 p T^2+132 p^3 T^3+p^6 T^4
\tabularnewline[0.5pt]\hline				
22	 &	smooth    &	          &	1+28 T+518 p T^2+28 p^3 T^3+p^6 T^4
\tabularnewline[0.5pt]\hline				
23	 &	smooth    &	          &	1+504 T+3566 p T^2+504 p^3 T^3+p^6 T^4
\tabularnewline[0.5pt]\hline				
24	 &	smooth    &	          &	1+1694 p T^2+p^6 T^4
\tabularnewline[0.5pt]\hline				
25	 &	smooth$^*$&	-\frac{1}{18}\+&	  
\tabularnewline[0.5pt]\hline				
26	 &	smooth    &	          &	1+92 T+134 p T^2+92 p^3 T^3+p^6 T^4
\tabularnewline[0.5pt]\hline				
27	 &	singular  &	-\frac{1}{3}\+&	(1-p T) (1-122 T+p^3 T^2)
\tabularnewline[0.5pt]\hline				
28	 &	smooth    &	          &	1-348 T+2678 p T^2-348 p^3 T^3+p^6 T^4
\tabularnewline[0.5pt]\hline				
29	 &	smooth    &	          &	1-44 T+2390 p T^2-44 p^3 T^3+p^6 T^4
\tabularnewline[0.5pt]\hline				
30	 &	smooth    &	          &	1-44 T+1910 p T^2-44 p^3 T^3+p^6 T^4
\tabularnewline[0.5pt]\hline				
31	 &	smooth    &	          &	1+228 T-202 p T^2+228 p^3 T^3+p^6 T^4
\tabularnewline[0.5pt]\hline				
32	 &	smooth    &	          &	1+80 T-898 p T^2+80 p^3 T^3+p^6 T^4
\tabularnewline[0.5pt]\hline				
33	 &	smooth    &	          &	1+20 T-1450 p T^2+20 p^3 T^3+p^6 T^4
\tabularnewline[0.5pt]\hline				
34	 &	smooth    &	          &	1+240 T+2942 p T^2+240 p^3 T^3+p^6 T^4
\tabularnewline[0.5pt]\hline				
35	 &	smooth    &	          &	1-44 T+470 p T^2-44 p^3 T^3+p^6 T^4
\tabularnewline[0.5pt]\hline				
36	 &	smooth    &	          &	1-80 T-1858 p T^2-80 p^3 T^3+p^6 T^4
\tabularnewline[0.5pt]\hline				
37	 &	smooth    &	          &	1-420 T+4070 p T^2-420 p^3 T^3+p^6 T^4
\tabularnewline[0.5pt]\hline				
38	 &	smooth    &	          &	1+12 T+1190 p T^2+12 p^3 T^3+p^6 T^4
\tabularnewline[0.5pt]\hline				
39	 &	smooth    &	          &	1+140 T+1574 p T^2+140 p^3 T^3+p^6 T^4
\tabularnewline[0.5pt]\hline				
40	 &	smooth    &	          &	1-476 T+3446 p T^2-476 p^3 T^3+p^6 T^4
\tabularnewline[0.5pt]\hline				
\tablepostamble				
\tablepreamble{43}				
1	 &	smooth    &	          &	1+20 T-174 p T^2+20 p^3 T^3+p^6 T^4
\tabularnewline[0.5pt]\hline				
2	 &	smooth    &	          &	1+248 T+1218 p T^2+248 p^3 T^3+p^6 T^4
\tabularnewline[0.5pt]\hline				
3	 &	smooth    &	          &	(1+4 p T+p^3 T^2)(1-348 T+p^3 T^2)
\tabularnewline[0.5pt]\hline				
4	 &	smooth    &	          &	1-264 T+1186 p T^2-264 p^3 T^3+p^6 T^4
\tabularnewline[0.5pt]\hline				
5	 &	smooth    &	          &	1+392 T+3042 p T^2+392 p^3 T^3+p^6 T^4
\tabularnewline[0.5pt]\hline				
6	 &	smooth    &	          &	1+696 T+5986 p T^2+696 p^3 T^3+p^6 T^4
\tabularnewline[0.5pt]\hline				
7	 &	smooth    &	          &	1+144 T-1790 p T^2+144 p^3 T^3+p^6 T^4
\tabularnewline[0.5pt]\hline				
8	 &	smooth    &	          &	1-504 T+4834 p T^2-504 p^3 T^3+p^6 T^4
\tabularnewline[0.5pt]\hline				
9	 &	singular  &	\frac{1}{24}&	(1-p T) (1-212 T+p^3 T^2)
\tabularnewline[0.5pt]\hline				
10	 &	smooth    &	          &	(1+4 p T+p^3 T^2)(1+364 T+p^3 T^2)
\tabularnewline[0.5pt]\hline				
11	 &	smooth    &	          &	1+272 T+2562 p T^2+272 p^3 T^3+p^6 T^4
\tabularnewline[0.5pt]\hline				
12	 &	smooth    &	          &	(1-12 p T+p^3 T^2)(1+292 T+p^3 T^2)
\tabularnewline[0.5pt]\hline				
13	 &	smooth    &	          &	1-44 T+722 p T^2-44 p^3 T^3+p^6 T^4
\tabularnewline[0.5pt]\hline				
14	 &	singular  &	-\frac{1}{3}\+&	(1-p T) (1+188 T+p^3 T^2)
\tabularnewline[0.5pt]\hline				
15	 &	smooth    &	          &	1+128 T+2754 p T^2+128 p^3 T^3+p^6 T^4
\tabularnewline[0.5pt]\hline				
16	 &	singular  &	-\frac{1}{8}\+&	(1-p T) (1-32 T+p^3 T^2)
\tabularnewline[0.5pt]\hline				
17	 &	smooth    &	          &	1-232 T+642 p T^2-232 p^3 T^3+p^6 T^4
\tabularnewline[0.5pt]\hline				
18	 &	smooth    &	          &	1+16 T-382 p T^2+16 p^3 T^3+p^6 T^4
\tabularnewline[0.5pt]\hline				
19	 &	smooth    &	          &	1-192 T+3202 p T^2-192 p^3 T^3+p^6 T^4
\tabularnewline[0.5pt]\hline				
20	 &	smooth    &	          &	1+212 T+1170 p T^2+212 p^3 T^3+p^6 T^4
\tabularnewline[0.5pt]\hline				
21	 &	smooth    &	          &	1-212 T+3314 p T^2-212 p^3 T^3+p^6 T^4
\tabularnewline[0.5pt]\hline				
22	 &	smooth    &	          &	1+176 T-126 p T^2+176 p^3 T^3+p^6 T^4
\tabularnewline[0.5pt]\hline				
23	 &	smooth    &	          &	1+16 T+386 p T^2+16 p^3 T^3+p^6 T^4
\tabularnewline[0.5pt]\hline				
24	 &	smooth    &	          &	1+256 T+2114 p T^2+256 p^3 T^3+p^6 T^4
\tabularnewline[0.5pt]\hline				
25	 &	singular  &	-\frac{1}{12}\+&	(1+p T) (1-452 T+p^3 T^2)
\tabularnewline[0.5pt]\hline				
26	 &	smooth    &	          &	1+752 T+6402 p T^2+752 p^3 T^3+p^6 T^4
\tabularnewline[0.5pt]\hline				
27	 &	smooth    &	          &	1+96 T-542 p T^2+96 p^3 T^3+p^6 T^4
\tabularnewline[0.5pt]\hline				
28	 &	smooth    &	          &	1-36 T-14 p T^2-36 p^3 T^3+p^6 T^4
\tabularnewline[0.5pt]\hline				
29	 &	smooth    &	          &	1+236 T+1650 p T^2+236 p^3 T^3+p^6 T^4
\tabularnewline[0.5pt]\hline				
30	 &	smooth    &	          &	1-24 T+3298 p T^2-24 p^3 T^3+p^6 T^4
\tabularnewline[0.5pt]\hline				
31	 &	smooth$^*$&	-\frac{1}{18}\+&	  
\tabularnewline[0.5pt]\hline				
32	 &	singular  &	-\frac{1}{4}\+&	(1+p T) (1+172 T+p^3 T^2)
\tabularnewline[0.5pt]\hline				
33	 &	smooth    &	          &	1-240 T+3586 p T^2-240 p^3 T^3+p^6 T^4
\tabularnewline[0.5pt]\hline				
34	 &	smooth    &	          &	1+24 T+2338 p T^2+24 p^3 T^3+p^6 T^4
\tabularnewline[0.5pt]\hline				
35	 &	smooth    &	          &	1+480 T+70 p^2 T^2+480 p^3 T^3+p^6 T^4
\tabularnewline[0.5pt]\hline				
36	 &	smooth    &	          &	1-372 T+1906 p T^2-372 p^3 T^3+p^6 T^4
\tabularnewline[0.5pt]\hline				
37	 &	smooth    &	          &	1-180 T+3826 p T^2-180 p^3 T^3+p^6 T^4
\tabularnewline[0.5pt]\hline				
38	 &	smooth    &	          &	1+136 T-286 p T^2+136 p^3 T^3+p^6 T^4
\tabularnewline[0.5pt]\hline				
39	 &	smooth    &	          &	(1+4 p T+p^3 T^2)(1-92 T+p^3 T^2)
\tabularnewline[0.5pt]\hline				
40	 &	smooth    &	          &	1-16 T-1278 p T^2-16 p^3 T^3+p^6 T^4
\tabularnewline[0.5pt]\hline				
41	 &	smooth    &	          &	1+276 T+1618 p T^2+276 p^3 T^3+p^6 T^4
\tabularnewline[0.5pt]\hline				
42	 &	smooth    &	          &	1+128 T-318 p T^2+128 p^3 T^3+p^6 T^4
\tabularnewline[0.5pt]\hline				
\tablepostamble				
\tablepreamble{47}				
1	 &	smooth    &	          &	1-200 T-574 p T^2-200 p^3 T^3+p^6 T^4
\tabularnewline[0.5pt]\hline				
2	 &	singular  &	\frac{1}{24}&	(1-p T) (1+264 T+p^3 T^2)
\tabularnewline[0.5pt]\hline				
3	 &	smooth    &	          &	1+840 T+7106 p T^2+840 p^3 T^3+p^6 T^4
\tabularnewline[0.5pt]\hline				
4	 &	smooth    &	          &	1+664 T+5186 p T^2+664 p^3 T^3+p^6 T^4
\tabularnewline[0.5pt]\hline				
5	 &	smooth    &	          &	1-128 T+3650 p T^2-128 p^3 T^3+p^6 T^4
\tabularnewline[0.5pt]\hline				
6	 &	smooth    &	          &	1-232 T+3650 p T^2-232 p^3 T^3+p^6 T^4
\tabularnewline[0.5pt]\hline				
7	 &	smooth    &	          &	1-216 T+1538 p T^2-216 p^3 T^3+p^6 T^4
\tabularnewline[0.5pt]\hline				
8	 &	smooth    &	          &	1-36 T-190 p T^2-36 p^3 T^3+p^6 T^4
\tabularnewline[0.5pt]\hline				
9	 &	smooth    &	          &	1-164 T+3650 p T^2-164 p^3 T^3+p^6 T^4
\tabularnewline[0.5pt]\hline				
10	 &	smooth    &	          &	1-100 T+386 p T^2-100 p^3 T^3+p^6 T^4
\tabularnewline[0.5pt]\hline				
11	 &	smooth    &	          &	1+280 T+1730 p T^2+280 p^3 T^3+p^6 T^4
\tabularnewline[0.5pt]\hline				
12	 &	smooth    &	          &	1-168 T-190 p T^2-168 p^3 T^3+p^6 T^4
\tabularnewline[0.5pt]\hline				
13	 &	smooth$^*$&	-\frac{1}{18}\+&	  
\tabularnewline[0.5pt]\hline				
14	 &	smooth    &	          &	1-264 T+194 p T^2-264 p^3 T^3+p^6 T^4
\tabularnewline[0.5pt]\hline				
15	 &	smooth    &	          &	1-60 T+3074 p T^2-60 p^3 T^3+p^6 T^4
\tabularnewline[0.5pt]\hline				
16	 &	smooth    &	          &	1-428 T+3266 p T^2-428 p^3 T^3+p^6 T^4
\tabularnewline[0.5pt]\hline				
17	 &	smooth    &	          &	1+552 T+4034 p T^2+552 p^3 T^3+p^6 T^4
\tabularnewline[0.5pt]\hline				
18	 &	smooth    &	          &	1-856 T+7586 p T^2-856 p^3 T^3+p^6 T^4
\tabularnewline[0.5pt]\hline				
19	 &	smooth    &	          &	1+104 T-958 p T^2+104 p^3 T^3+p^6 T^4
\tabularnewline[0.5pt]\hline				
20	 &	smooth    &	          &	1-136 T+1730 p T^2-136 p^3 T^3+p^6 T^4
\tabularnewline[0.5pt]\hline				
21	 &	smooth    &	          &	1+372 T+3266 p T^2+372 p^3 T^3+p^6 T^4
\tabularnewline[0.5pt]\hline				
22	 &	smooth    &	          &	1-36 T+3650 p T^2-36 p^3 T^3+p^6 T^4
\tabularnewline[0.5pt]\hline				
23	 &	smooth    &	          &	1-4 T+1538 p T^2-4 p^3 T^3+p^6 T^4
\tabularnewline[0.5pt]\hline				
24	 &	smooth    &	          &	1-80 T+1730 p T^2-80 p^3 T^3+p^6 T^4
\tabularnewline[0.5pt]\hline				
25	 &	smooth    &	          &	1-248 T+3266 p T^2-248 p^3 T^3+p^6 T^4
\tabularnewline[0.5pt]\hline				
26	 &	smooth    &	          &	1+128 T+2114 p T^2+128 p^3 T^3+p^6 T^4
\tabularnewline[0.5pt]\hline				
27	 &	smooth    &	          &	1-216 T-190 p T^2-216 p^3 T^3+p^6 T^4
\tabularnewline[0.5pt]\hline				
28	 &	smooth    &	          &	1-72 T+1730 p T^2-72 p^3 T^3+p^6 T^4
\tabularnewline[0.5pt]\hline				
29	 &	smooth    &	          &	1+792 T+7106 p T^2+792 p^3 T^3+p^6 T^4
\tabularnewline[0.5pt]\hline				
30	 &	smooth    &	          &	1-384 T+1634 p T^2-384 p^3 T^3+p^6 T^4
\tabularnewline[0.5pt]\hline				
31	 &	singular  &	-\frac{1}{3}\+&	(1-p T) (1-256 T+p^3 T^2)
\tabularnewline[0.5pt]\hline				
32	 &	smooth    &	          &	1-36 T+3842 p T^2-36 p^3 T^3+p^6 T^4
\tabularnewline[0.5pt]\hline				
33	 &	smooth    &	          &	1+160 T-1150 p T^2+160 p^3 T^3+p^6 T^4
\tabularnewline[0.5pt]\hline				
34	 &	smooth    &	          &	1+528 T+3650 p T^2+528 p^3 T^3+p^6 T^4
\tabularnewline[0.5pt]\hline				
35	 &	singular  &	-\frac{1}{4}\+&	(1+p T) (1-192 T+p^3 T^2)
\tabularnewline[0.5pt]\hline				
36	 &	smooth    &	          &	1-592 T+5186 p T^2-592 p^3 T^3+p^6 T^4
\tabularnewline[0.5pt]\hline				
37	 &	smooth    &	          &	1-460 T+4994 p T^2-460 p^3 T^3+p^6 T^4
\tabularnewline[0.5pt]\hline				
38	 &	smooth    &	          &	(1-8 p T+p^3 T^2)(1-48 T+p^3 T^2)
\tabularnewline[0.5pt]\hline				
39	 &	smooth    &	          &	1+576 T+4034 p T^2+576 p^3 T^3+p^6 T^4
\tabularnewline[0.5pt]\hline				
40	 &	smooth    &	          &	1+216 T+962 p T^2+216 p^3 T^3+p^6 T^4
\tabularnewline[0.5pt]\hline				
41	 &	singular  &	-\frac{1}{8}\+&	(1-p T) (1+204 T+p^3 T^2)
\tabularnewline[0.5pt]\hline				
42	 &	smooth    &	          &	1-584 T+5186 p T^2-584 p^3 T^3+p^6 T^4
\tabularnewline[0.5pt]\hline				
43	 &	singular  &	-\frac{1}{12}\+&	(1+p T) (1-432 T+p^3 T^2)
\tabularnewline[0.5pt]\hline				
44	 &	smooth    &	          &	1-120 T+2210 p T^2-120 p^3 T^3+p^6 T^4
\tabularnewline[0.5pt]\hline				
45	 &	smooth    &	          &	1+8 T-190 p T^2+8 p^3 T^3+p^6 T^4
\tabularnewline[0.5pt]\hline				
46	 &	smooth    &	          &	1+52 T-2878 p T^2+52 p^3 T^3+p^6 T^4
\tabularnewline[0.5pt]\hline				
\tablepostamble				
\tablepreamble{53}				
1	 &	smooth    &	          &	1-396 T+4550 p T^2-396 p^3 T^3+p^6 T^4
\tabularnewline[0.5pt]\hline				
2	 &	smooth    &	          &	1-556 T+4742 p T^2-556 p^3 T^3+p^6 T^4
\tabularnewline[0.5pt]\hline				
3	 &	smooth    &	          &	1-444 T+5510 p T^2-444 p^3 T^3+p^6 T^4
\tabularnewline[0.5pt]\hline				
4	 &	smooth    &	          &	1-76 T+1094 p T^2-76 p^3 T^3+p^6 T^4
\tabularnewline[0.5pt]\hline				
5	 &	smooth    &	          &	1+92 T+2774 p T^2+92 p^3 T^3+p^6 T^4
\tabularnewline[0.5pt]\hline				
6	 &	smooth    &	          &	1+200 T-1906 p T^2+200 p^3 T^3+p^6 T^4
\tabularnewline[0.5pt]\hline				
7	 &	smooth    &	          &	1-744 T+7598 p T^2-744 p^3 T^3+p^6 T^4
\tabularnewline[0.5pt]\hline				
8	 &	smooth    &	          &	1+672 T+5630 p T^2+672 p^3 T^3+p^6 T^4
\tabularnewline[0.5pt]\hline				
9	 &	smooth    &	          &	(1+10 p T+p^3 T^2)(1+306 T+p^3 T^2)
\tabularnewline[0.5pt]\hline				
10	 &	smooth    &	          &	1-760 T+6158 p T^2-760 p^3 T^3+p^6 T^4
\tabularnewline[0.5pt]\hline				
11	 &	smooth    &	          &	1-340 T+1814 p T^2-340 p^3 T^3+p^6 T^4
\tabularnewline[0.5pt]\hline				
12	 &	smooth    &	          &	1+296 T+2126 p T^2+296 p^3 T^3+p^6 T^4
\tabularnewline[0.5pt]\hline				
13	 &	singular  &	-\frac{1}{4}\+&	(1-p T) (1-558 T+p^3 T^2)
\tabularnewline[0.5pt]\hline				
14	 &	smooth    &	          &	1+156 T+2966 p T^2+156 p^3 T^3+p^6 T^4
\tabularnewline[0.5pt]\hline				
15	 &	smooth    &	          &	1+44 T+1718 p T^2+44 p^3 T^3+p^6 T^4
\tabularnewline[0.5pt]\hline				
16	 &	smooth    &	          &	1-796 T+7334 p T^2-796 p^3 T^3+p^6 T^4
\tabularnewline[0.5pt]\hline				
17	 &	smooth    &	          &	(1-6 p T+p^3 T^2)(1+594 T+p^3 T^2)
\tabularnewline[0.5pt]\hline				
18	 &	smooth    &	          &	(1-6 p T+p^3 T^2)(1+146 T+p^3 T^2)
\tabularnewline[0.5pt]\hline				
19	 &	smooth    &	          &	1-180 T+1910 p T^2-180 p^3 T^3+p^6 T^4
\tabularnewline[0.5pt]\hline				
20	 &	smooth    &	          &	1+452 T+2150 p T^2+452 p^3 T^3+p^6 T^4
\tabularnewline[0.5pt]\hline				
21	 &	smooth    &	          &	1+492 T+94 p^2 T^2+492 p^3 T^3+p^6 T^4
\tabularnewline[0.5pt]\hline				
22	 &	singular  &	-\frac{1}{12}\+&	(1-p T) (1-414 T+p^3 T^2)
\tabularnewline[0.5pt]\hline				
23	 &	smooth    &	          &	1-696 T+6158 p T^2-696 p^3 T^3+p^6 T^4
\tabularnewline[0.5pt]\hline				
24	 &	smooth    &	          &	1-12 T+2246 p T^2-12 p^3 T^3+p^6 T^4
\tabularnewline[0.5pt]\hline				
25	 &	smooth    &	          &	1-216 T+974 p T^2-216 p^3 T^3+p^6 T^4
\tabularnewline[0.5pt]\hline				
26	 &	smooth    &	          &	(1-6 p T+p^3 T^2)(1+594 T+p^3 T^2)
\tabularnewline[0.5pt]\hline				
27	 &	smooth    &	          &	1-168 T-1042 p T^2-168 p^3 T^3+p^6 T^4
\tabularnewline[0.5pt]\hline				
28	 &	smooth    &	          &	1-120 T-1330 p T^2-120 p^3 T^3+p^6 T^4
\tabularnewline[0.5pt]\hline				
29	 &	smooth    &	          &	1-272 T+3230 p T^2-272 p^3 T^3+p^6 T^4
\tabularnewline[0.5pt]\hline				
30	 &	smooth    &	          &	1+84 T-1114 p T^2+84 p^3 T^3+p^6 T^4
\tabularnewline[0.5pt]\hline				
31	 &	smooth    &	          &	1+132 T-2458 p T^2+132 p^3 T^3+p^6 T^4
\tabularnewline[0.5pt]\hline				
32	 &	smooth    &	          &	1+428 T+5366 p T^2+428 p^3 T^3+p^6 T^4
\tabularnewline[0.5pt]\hline				
33	 &	singular  &	-\frac{1}{8}\+&	(1-p T) (1-222 T+p^3 T^2)
\tabularnewline[0.5pt]\hline				
34	 &	smooth    &	          &	1-444 T+4070 p T^2-444 p^3 T^3+p^6 T^4
\tabularnewline[0.5pt]\hline				
35	 &	singular  &	-\frac{1}{3}\+&	(1-p T) (1+338 T+p^3 T^2)
\tabularnewline[0.5pt]\hline				
36	 &	smooth    &	          &	1+80 T+2270 p T^2+80 p^3 T^3+p^6 T^4
\tabularnewline[0.5pt]\hline				
37	 &	smooth    &	          &	1+172 T+2294 p T^2+172 p^3 T^3+p^6 T^4
\tabularnewline[0.5pt]\hline				
38	 &	smooth    &	          &	1-180 T+4406 p T^2-180 p^3 T^3+p^6 T^4
\tabularnewline[0.5pt]\hline				
39	 &	smooth    &	          &	1-28 T-3034 p T^2-28 p^3 T^3+p^6 T^4
\tabularnewline[0.5pt]\hline				
40	 &	smooth    &	          &	1+148 T-1018 p T^2+148 p^3 T^3+p^6 T^4
\tabularnewline[0.5pt]\hline				
41	 &	smooth    &	          &	1-444 T+3686 p T^2-444 p^3 T^3+p^6 T^4
\tabularnewline[0.5pt]\hline				
42	 &	singular  &	\frac{1}{24}&	(1-p T) (1+162 T+p^3 T^2)
\tabularnewline[0.5pt]\hline				
43	 &	smooth    &	          &	1-428 T+5894 p T^2-428 p^3 T^3+p^6 T^4
\tabularnewline[0.5pt]\hline				
44	 &	smooth    &	          &	1+208 T+2654 p T^2+208 p^3 T^3+p^6 T^4
\tabularnewline[0.5pt]\hline				
45	 &	smooth    &	          &	1-20 T+1910 p T^2-20 p^3 T^3+p^6 T^4
\tabularnewline[0.5pt]\hline				
46	 &	smooth    &	          &	1-132 T+3446 p T^2-132 p^3 T^3+p^6 T^4
\tabularnewline[0.5pt]\hline				
47	 &	smooth    &	          &	1+380 T+2390 p T^2+380 p^3 T^3+p^6 T^4
\tabularnewline[0.5pt]\hline				
48	 &	smooth    &	          &	1+92 T-298 p T^2+92 p^3 T^3+p^6 T^4
\tabularnewline[0.5pt]\hline				
49	 &	smooth    &	          &	1-108 T+2054 p T^2-108 p^3 T^3+p^6 T^4
\tabularnewline[0.5pt]\hline				
50	 &	smooth$^*$&	-\frac{1}{18}\+&	  
\tabularnewline[0.5pt]\hline				
51	 &	smooth    &	          &	1+364 T+4214 p T^2+364 p^3 T^3+p^6 T^4
\tabularnewline[0.5pt]\hline				
52	 &	smooth    &	          &	(1+6 p T+p^3 T^2)(1-654 T+p^3 T^2)
\tabularnewline[0.5pt]\hline				
\tablepostamble				
\tablepreamble{59}				
1	 &	smooth    &	          &	1-84 T-2062 p T^2-84 p^3 T^3+p^6 T^4
\tabularnewline[0.5pt]\hline				
2	 &	smooth    &	          &	1-880 T+8642 p T^2-880 p^3 T^3+p^6 T^4
\tabularnewline[0.5pt]\hline				
3	 &	smooth    &	          &	1-168 T+866 p T^2-168 p^3 T^3+p^6 T^4
\tabularnewline[0.5pt]\hline				
4	 &	smooth    &	          &	1-464 T+7106 p T^2-464 p^3 T^3+p^6 T^4
\tabularnewline[0.5pt]\hline				
5	 &	smooth    &	          &	1-900 T+8690 p T^2-900 p^3 T^3+p^6 T^4
\tabularnewline[0.5pt]\hline				
6	 &	smooth    &	          &	1+96 T+770 p T^2+96 p^3 T^3+p^6 T^4
\tabularnewline[0.5pt]\hline				
7	 &	smooth    &	          &	1+96 T+5762 p T^2+96 p^3 T^3+p^6 T^4
\tabularnewline[0.5pt]\hline				
8	 &	smooth    &	          &	1+312 T+3938 p T^2+312 p^3 T^3+p^6 T^4
\tabularnewline[0.5pt]\hline				
9	 &	smooth    &	          &	1-328 T-958 p T^2-328 p^3 T^3+p^6 T^4
\tabularnewline[0.5pt]\hline				
10	 &	smooth    &	          &	1-48 T+2594 p T^2-48 p^3 T^3+p^6 T^4
\tabularnewline[0.5pt]\hline				
11	 &	smooth    &	          &	1-104 T-3934 p T^2-104 p^3 T^3+p^6 T^4
\tabularnewline[0.5pt]\hline				
12	 &	smooth    &	          &	1+40 T+1826 p T^2+40 p^3 T^3+p^6 T^4
\tabularnewline[0.5pt]\hline				
13	 &	smooth    &	          &	(1-8 p T+p^3 T^2)(1+588 T+p^3 T^2)
\tabularnewline[0.5pt]\hline				
14	 &	smooth    &	          &	1-8 T+3842 p T^2-8 p^3 T^3+p^6 T^4
\tabularnewline[0.5pt]\hline				
15	 &	smooth    &	          &	1-364 T+6674 p T^2-364 p^3 T^3+p^6 T^4
\tabularnewline[0.5pt]\hline				
16	 &	smooth    &	          &	1-60 T+2066 p T^2-60 p^3 T^3+p^6 T^4
\tabularnewline[0.5pt]\hline				
17	 &	smooth    &	          &	(1+4 p T+p^3 T^2)(1-132 T+p^3 T^2)
\tabularnewline[0.5pt]\hline				
18	 &	smooth    &	          &	1-264 T+2402 p T^2-264 p^3 T^3+p^6 T^4
\tabularnewline[0.5pt]\hline				
19	 &	smooth    &	          &	1-56 T+1826 p T^2-56 p^3 T^3+p^6 T^4
\tabularnewline[0.5pt]\hline				
20	 &	smooth    &	          &	1-24 T-3166 p T^2-24 p^3 T^3+p^6 T^4
\tabularnewline[0.5pt]\hline				
21	 &	smooth    &	          &	1+456 T+5282 p T^2+456 p^3 T^3+p^6 T^4
\tabularnewline[0.5pt]\hline				
22	 &	singular  &	-\frac{1}{8}\+&	(1-p T) (1-420 T+p^3 T^2)
\tabularnewline[0.5pt]\hline				
23	 &	smooth    &	          &	1-188 T-1966 p T^2-188 p^3 T^3+p^6 T^4
\tabularnewline[0.5pt]\hline				
24	 &	smooth    &	          &	1-444 T+2450 p T^2-444 p^3 T^3+p^6 T^4
\tabularnewline[0.5pt]\hline				
25	 &	smooth    &	          &	1+192 T+2 p T^2+192 p^3 T^3+p^6 T^4
\tabularnewline[0.5pt]\hline				
26	 &	smooth    &	          &	1-260 T+6002 p T^2-260 p^3 T^3+p^6 T^4
\tabularnewline[0.5pt]\hline				
27	 &	smooth    &	          &	1+440 T+98 p T^2+440 p^3 T^3+p^6 T^4
\tabularnewline[0.5pt]\hline				
28	 &	smooth    &	          &	1+208 T-2782 p T^2+208 p^3 T^3+p^6 T^4
\tabularnewline[0.5pt]\hline				
29	 &	smooth    &	          &	1-312 T+7202 p T^2-312 p^3 T^3+p^6 T^4
\tabularnewline[0.5pt]\hline				
30	 &	smooth    &	          &	1-1216 T+11906 p T^2-1216 p^3 T^3+p^6 T^4
\tabularnewline[0.5pt]\hline				
31	 &	smooth    &	          &	1-216 T+2594 p T^2-216 p^3 T^3+p^6 T^4
\tabularnewline[0.5pt]\hline				
32	 &	singular  &	\frac{1}{24}&	(1-p T) (1+772 T+p^3 T^2)
\tabularnewline[0.5pt]\hline				
33	 &	smooth    &	          &	(1+12 p T+p^3 T^2)(1+252 T+p^3 T^2)
\tabularnewline[0.5pt]\hline				
34	 &	smooth    &	          &	1+420 T+3794 p T^2+420 p^3 T^3+p^6 T^4
\tabularnewline[0.5pt]\hline				
35	 &	smooth    &	          &	(1+12 p T+p^3 T^2)(1-852 T+p^3 T^2)
\tabularnewline[0.5pt]\hline				
36	 &	smooth$^*$&	-\frac{1}{18}\+&	  
\tabularnewline[0.5pt]\hline				
37	 &	smooth    &	          &	1-112 T+1346 p T^2-112 p^3 T^3+p^6 T^4
\tabularnewline[0.5pt]\hline				
38	 &	smooth    &	          &	1+620 T+4658 p T^2+620 p^3 T^3+p^6 T^4
\tabularnewline[0.5pt]\hline				
39	 &	singular  &	-\frac{1}{3}\+&	(1-p T) (1-100 T+p^3 T^2)
\tabularnewline[0.5pt]\hline				
40	 &	smooth    &	          &	1+768 T+9410 p T^2+768 p^3 T^3+p^6 T^4
\tabularnewline[0.5pt]\hline				
41	 &	smooth    &	          &	1-136 T+5858 p T^2-136 p^3 T^3+p^6 T^4
\tabularnewline[0.5pt]\hline				
42	 &	smooth    &	          &	1-400 T+5954 p T^2-400 p^3 T^3+p^6 T^4
\tabularnewline[0.5pt]\hline				
43	 &	smooth    &	          &	1+288 T+3458 p T^2+288 p^3 T^3+p^6 T^4
\tabularnewline[0.5pt]\hline				
44	 &	singular  &	-\frac{1}{4}\+&	(1+p T) (1-540 T+p^3 T^2)
\tabularnewline[0.5pt]\hline				
45	 &	smooth    &	          &	1-32 T+4610 p T^2-32 p^3 T^3+p^6 T^4
\tabularnewline[0.5pt]\hline				
46	 &	smooth    &	          &	1+612 T+5522 p T^2+612 p^3 T^3+p^6 T^4
\tabularnewline[0.5pt]\hline				
47	 &	smooth    &	          &	1-372 T+1778 p T^2-372 p^3 T^3+p^6 T^4
\tabularnewline[0.5pt]\hline				
48	 &	smooth    &	          &	1-144 T+4802 p T^2-144 p^3 T^3+p^6 T^4
\tabularnewline[0.5pt]\hline				
49	 &	smooth    &	          &	1+400 T+2498 p T^2+400 p^3 T^3+p^6 T^4
\tabularnewline[0.5pt]\hline				
50	 &	smooth    &	          &	1-304 T+3650 p T^2-304 p^3 T^3+p^6 T^4
\tabularnewline[0.5pt]\hline				
51	 &	smooth    &	          &	1-476 T+7058 p T^2-476 p^3 T^3+p^6 T^4
\tabularnewline[0.5pt]\hline				
52	 &	smooth    &	          &	1+180 T+3410 p T^2+180 p^3 T^3+p^6 T^4
\tabularnewline[0.5pt]\hline				
53	 &	smooth    &	          &	1-296 T+5090 p T^2-296 p^3 T^3+p^6 T^4
\tabularnewline[0.5pt]\hline				
54	 &	singular  &	-\frac{1}{12}\+&	(1+p T) (1+684 T+p^3 T^2)
\tabularnewline[0.5pt]\hline				
55	 &	smooth    &	          &	1-920 T+8162 p T^2-920 p^3 T^3+p^6 T^4
\tabularnewline[0.5pt]\hline				
56	 &	smooth    &	          &	1-64 T+6530 p T^2-64 p^3 T^3+p^6 T^4
\tabularnewline[0.5pt]\hline				
57	 &	smooth    &	          &	1+296 T-478 p T^2+296 p^3 T^3+p^6 T^4
\tabularnewline[0.5pt]\hline				
58	 &	smooth    &	          &	1+16 T+2690 p T^2+16 p^3 T^3+p^6 T^4
\tabularnewline[0.5pt]\hline				
\tablepostamble				
\tablepreamble{61}				
1	 &	smooth    &	          &	1+756 T+5638 p T^2+756 p^3 T^3+p^6 T^4
\tabularnewline[0.5pt]\hline				
2	 &	smooth    &	          &	1+60 T-2282 p T^2+60 p^3 T^3+p^6 T^4
\tabularnewline[0.5pt]\hline				
3	 &	smooth    &	          &	1-792 T+6958 p T^2-792 p^3 T^3+p^6 T^4
\tabularnewline[0.5pt]\hline				
4	 &	smooth    &	          &	1+260 T+4134 p T^2+260 p^3 T^3+p^6 T^4
\tabularnewline[0.5pt]\hline				
5	 &	singular  &	-\frac{1}{12}\+&	(1-p T) (1-422 T+p^3 T^2)
\tabularnewline[0.5pt]\hline				
6	 &	smooth    &	          &	1+684 T+142 p^2 T^2+684 p^3 T^3+p^6 T^4
\tabularnewline[0.5pt]\hline				
7	 &	smooth    &	          &	1-1020 T+8998 p T^2-1020 p^3 T^3+p^6 T^4
\tabularnewline[0.5pt]\hline				
8	 &	smooth    &	          &	1+1024 T+10718 p T^2+1024 p^3 T^3+p^6 T^4
\tabularnewline[0.5pt]\hline				
9	 &	smooth    &	          &	1+624 T+4798 p T^2+624 p^3 T^3+p^6 T^4
\tabularnewline[0.5pt]\hline				
10	 &	smooth    &	          &	1+420 T+4198 p T^2+420 p^3 T^3+p^6 T^4
\tabularnewline[0.5pt]\hline				
11	 &	smooth    &	          &	1+960 T+9694 p T^2+960 p^3 T^3+p^6 T^4
\tabularnewline[0.5pt]\hline				
12	 &	smooth    &	          &	1+336 T+2686 p T^2+336 p^3 T^3+p^6 T^4
\tabularnewline[0.5pt]\hline				
13	 &	smooth    &	          &	1+740 T+9318 p T^2+740 p^3 T^3+p^6 T^4
\tabularnewline[0.5pt]\hline				
14	 &	smooth    &	          &	1+804 T+6118 p T^2+804 p^3 T^3+p^6 T^4
\tabularnewline[0.5pt]\hline				
15	 &	singular  &	-\frac{1}{4}\+&	(1-p T) (1-110 T+p^3 T^2)
\tabularnewline[0.5pt]\hline				
16	 &	smooth    &	          &	1+356 T+4326 p T^2+356 p^3 T^3+p^6 T^4
\tabularnewline[0.5pt]\hline				
17	 &	smooth    &	          &	1+1472 T+15774 p T^2+1472 p^3 T^3+p^6 T^4
\tabularnewline[0.5pt]\hline				
18	 &	smooth    &	          &	(1+10 p T+p^3 T^2)(1-934 T+p^3 T^2)
\tabularnewline[0.5pt]\hline				
19	 &	smooth    &	          &	1+468 T+1990 p T^2+468 p^3 T^3+p^6 T^4
\tabularnewline[0.5pt]\hline				
20	 &	singular  &	-\frac{1}{3}\+&	(1-p T) (1-742 T+p^3 T^2)
\tabularnewline[0.5pt]\hline				
21	 &	smooth    &	          &	1-56 T+3374 p T^2-56 p^3 T^3+p^6 T^4
\tabularnewline[0.5pt]\hline				
22	 &	smooth    &	          &	1-108 T-4058 p T^2-108 p^3 T^3+p^6 T^4
\tabularnewline[0.5pt]\hline				
23	 &	smooth    &	          &	1+972 T+8374 p T^2+972 p^3 T^3+p^6 T^4
\tabularnewline[0.5pt]\hline				
24	 &	smooth    &	          &	1+84 T-4154 p T^2+84 p^3 T^3+p^6 T^4
\tabularnewline[0.5pt]\hline				
25	 &	smooth    &	          &	1+28 T-2218 p T^2+28 p^3 T^3+p^6 T^4
\tabularnewline[0.5pt]\hline				
26	 &	smooth    &	          &	1-12 T+4102 p T^2-12 p^3 T^3+p^6 T^4
\tabularnewline[0.5pt]\hline				
27	 &	smooth    &	          &	1+60 T+1174 p T^2+60 p^3 T^3+p^6 T^4
\tabularnewline[0.5pt]\hline				
28	 &	singular  &	\frac{1}{24}&	(1-p T) (1-30 T+p^3 T^2)
\tabularnewline[0.5pt]\hline				
29	 &	smooth    &	          &	1+104 T-1938 p T^2+104 p^3 T^3+p^6 T^4
\tabularnewline[0.5pt]\hline				
30	 &	smooth    &	          &	1-1036 T+10758 p T^2-1036 p^3 T^3+p^6 T^4
\tabularnewline[0.5pt]\hline				
31	 &	smooth    &	          &	1-436 T+6198 p T^2-436 p^3 T^3+p^6 T^4
\tabularnewline[0.5pt]\hline				
32	 &	smooth    &	          &	1-944 T+8894 p T^2-944 p^3 T^3+p^6 T^4
\tabularnewline[0.5pt]\hline				
33	 &	smooth    &	          &	1+252 T+1558 p T^2+252 p^3 T^3+p^6 T^4
\tabularnewline[0.5pt]\hline				
34	 &	smooth    &	          &	1+68 T+1446 p T^2+68 p^3 T^3+p^6 T^4
\tabularnewline[0.5pt]\hline				
35	 &	smooth    &	          &	1+288 T+3934 p T^2+288 p^3 T^3+p^6 T^4
\tabularnewline[0.5pt]\hline				
36	 &	smooth    &	          &	1-12 T+1798 p T^2-12 p^3 T^3+p^6 T^4
\tabularnewline[0.5pt]\hline				
37	 &	smooth    &	          &	1+428 T+6774 p T^2+428 p^3 T^3+p^6 T^4
\tabularnewline[0.5pt]\hline				
38	 &	singular  &	-\frac{1}{8}\+&	(1-p T) (1-902 T+p^3 T^2)
\tabularnewline[0.5pt]\hline				
39	 &	smooth    &	          &	(1-14 p T+p^3 T^2)(1+730 T+p^3 T^2)
\tabularnewline[0.5pt]\hline				
40	 &	smooth    &	          &	1-512 T+6686 p T^2-512 p^3 T^3+p^6 T^4
\tabularnewline[0.5pt]\hline				
41	 &	smooth    &	          &	1+560 T+6654 p T^2+560 p^3 T^3+p^6 T^4
\tabularnewline[0.5pt]\hline				
42	 &	smooth    &	          &	1-332 T+7046 p T^2-332 p^3 T^3+p^6 T^4
\tabularnewline[0.5pt]\hline				
43	 &	smooth    &	          &	1-12 T+4102 p T^2-12 p^3 T^3+p^6 T^4
\tabularnewline[0.5pt]\hline				
44	 &	smooth$^*$&	-\frac{1}{18}\+&	  
\tabularnewline[0.5pt]\hline				
45	 &	smooth    &	          &	1-540 T+3430 p T^2-540 p^3 T^3+p^6 T^4
\tabularnewline[0.5pt]\hline				
46	 &	smooth    &	          &	1-144 T+4222 p T^2-144 p^3 T^3+p^6 T^4
\tabularnewline[0.5pt]\hline				
47	 &	smooth    &	          &	1-20 T+6134 p T^2-20 p^3 T^3+p^6 T^4
\tabularnewline[0.5pt]\hline				
48	 &	smooth    &	          &	1-12 T+4390 p T^2-12 p^3 T^3+p^6 T^4
\tabularnewline[0.5pt]\hline				
49	 &	smooth    &	          &	1+248 T-114 p T^2+248 p^3 T^3+p^6 T^4
\tabularnewline[0.5pt]\hline				
50	 &	smooth    &	          &	1-276 T+5878 p T^2-276 p^3 T^3+p^6 T^4
\tabularnewline[0.5pt]\hline				
51	 &	smooth    &	          &	1-36 T-2474 p T^2-36 p^3 T^3+p^6 T^4
\tabularnewline[0.5pt]\hline				
52	 &	smooth    &	          &	1+104 T-4050 p T^2+104 p^3 T^3+p^6 T^4
\tabularnewline[0.5pt]\hline				
53	 &	smooth    &	          &	1-36 T-5162 p T^2-36 p^3 T^3+p^6 T^4
\tabularnewline[0.5pt]\hline				
54	 &	smooth    &	          &	1-156 T+1606 p T^2-156 p^3 T^3+p^6 T^4
\tabularnewline[0.5pt]\hline				
55	 &	smooth    &	          &	1+132 T-4826 p T^2+132 p^3 T^3+p^6 T^4
\tabularnewline[0.5pt]\hline				
56	 &	smooth    &	          &	1+140 T+6198 p T^2+140 p^3 T^3+p^6 T^4
\tabularnewline[0.5pt]\hline				
57	 &	smooth    &	          &	1+452 T+7206 p T^2+452 p^3 T^3+p^6 T^4
\tabularnewline[0.5pt]\hline				
58	 &	smooth    &	          &	1+36 T+1606 p T^2+36 p^3 T^3+p^6 T^4
\tabularnewline[0.5pt]\hline				
59	 &	smooth    &	          &	1+668 T+4950 p T^2+668 p^3 T^3+p^6 T^4
\tabularnewline[0.5pt]\hline				
60	 &	smooth    &	          &	1-400 T+6078 p T^2-400 p^3 T^3+p^6 T^4
\tabularnewline[0.5pt]\hline				
\tablepostamble				
\tablepreamble{67}				
1	 &	smooth    &	          &	1+608 T+9858 p T^2+608 p^3 T^3+p^6 T^4
\tabularnewline[0.5pt]\hline				
2	 &	smooth    &	          &	1-72 T-4766 p T^2-72 p^3 T^3+p^6 T^4
\tabularnewline[0.5pt]\hline				
3	 &	smooth    &	          &	1-280 T-3774 p T^2-280 p^3 T^3+p^6 T^4
\tabularnewline[0.5pt]\hline				
4	 &	smooth    &	          &	1-696 T+9154 p T^2-696 p^3 T^3+p^6 T^4
\tabularnewline[0.5pt]\hline				
5	 &	smooth    &	          &	1-504 T+7618 p T^2-504 p^3 T^3+p^6 T^4
\tabularnewline[0.5pt]\hline				
6	 &	smooth    &	          &	1-600 T+4930 p T^2-600 p^3 T^3+p^6 T^4
\tabularnewline[0.5pt]\hline				
7	 &	smooth    &	          &	1+40 T+5570 p T^2+40 p^3 T^3+p^6 T^4
\tabularnewline[0.5pt]\hline				
8	 &	smooth    &	          &	1-960 T+9058 p T^2-960 p^3 T^3+p^6 T^4
\tabularnewline[0.5pt]\hline				
9	 &	smooth    &	          &	1+208 T-2590 p T^2+208 p^3 T^3+p^6 T^4
\tabularnewline[0.5pt]\hline				
10	 &	smooth    &	          &	1-432 T+7330 p T^2-432 p^3 T^3+p^6 T^4
\tabularnewline[0.5pt]\hline				
11	 &	smooth    &	          &	1+80 T+4770 p T^2+80 p^3 T^3+p^6 T^4
\tabularnewline[0.5pt]\hline				
12	 &	smooth    &	          &	1+640 T+2402 p T^2+640 p^3 T^3+p^6 T^4
\tabularnewline[0.5pt]\hline				
13	 &	smooth    &	          &	1-496 T+6498 p T^2-496 p^3 T^3+p^6 T^4
\tabularnewline[0.5pt]\hline				
14	 &	singular  &	\frac{1}{24}&	(1-p T) (1+764 T+p^3 T^2)
\tabularnewline[0.5pt]\hline				
15	 &	smooth    &	          &	1+264 T-3902 p T^2+264 p^3 T^3+p^6 T^4
\tabularnewline[0.5pt]\hline				
16	 &	smooth    &	          &	1-392 T+6914 p T^2-392 p^3 T^3+p^6 T^4
\tabularnewline[0.5pt]\hline				
17	 &	smooth    &	          &	1+504 T+4738 p T^2+504 p^3 T^3+p^6 T^4
\tabularnewline[0.5pt]\hline				
18	 &	smooth    &	          &	1+104 T-1470 p T^2+104 p^3 T^3+p^6 T^4
\tabularnewline[0.5pt]\hline				
19	 &	smooth    &	          &	1-416 T+7394 p T^2-416 p^3 T^3+p^6 T^4
\tabularnewline[0.5pt]\hline				
20	 &	smooth    &	          &	1+800 T+7650 p T^2+800 p^3 T^3+p^6 T^4
\tabularnewline[0.5pt]\hline				
21	 &	smooth    &	          &	1+184 T-4798 p T^2+184 p^3 T^3+p^6 T^4
\tabularnewline[0.5pt]\hline				
22	 &	singular  &	-\frac{1}{3}\+&	(1-p T) (1+84 T+p^3 T^2)
\tabularnewline[0.5pt]\hline				
23	 &	smooth    &	          &	1+328 T-574 p T^2+328 p^3 T^3+p^6 T^4
\tabularnewline[0.5pt]\hline				
24	 &	smooth    &	          &	1+532 T+6866 p T^2+532 p^3 T^3+p^6 T^4
\tabularnewline[0.5pt]\hline				
25	 &	singular  &	-\frac{1}{8}\+&	(1-p T) (1+1024 T+p^3 T^2)
\tabularnewline[0.5pt]\hline				
26	 &	smooth$^*$&	-\frac{1}{18}\+&	  
\tabularnewline[0.5pt]\hline				
27	 &	smooth    &	          &	1+152 T-1278 p T^2+152 p^3 T^3+p^6 T^4
\tabularnewline[0.5pt]\hline				
28	 &	smooth    &	          &	1+392 T+4290 p T^2+392 p^3 T^3+p^6 T^4
\tabularnewline[0.5pt]\hline				
29	 &	smooth    &	          &	1-164 T+8882 p T^2-164 p^3 T^3+p^6 T^4
\tabularnewline[0.5pt]\hline				
30	 &	smooth    &	          &	1+456 T-1022 p T^2+456 p^3 T^3+p^6 T^4
\tabularnewline[0.5pt]\hline				
31	 &	smooth    &	          &	1-336 T+1954 p T^2-336 p^3 T^3+p^6 T^4
\tabularnewline[0.5pt]\hline				
32	 &	smooth    &	          &	1-108 T+274 p T^2-108 p^3 T^3+p^6 T^4
\tabularnewline[0.5pt]\hline				
33	 &	smooth    &	          &	1-252 T+2962 p T^2-252 p^3 T^3+p^6 T^4
\tabularnewline[0.5pt]\hline				
34	 &	smooth    &	          &	1+216 T+4738 p T^2+216 p^3 T^3+p^6 T^4
\tabularnewline[0.5pt]\hline				
35	 &	smooth    &	          &	1-164 T+38 p^2 T^2-164 p^3 T^3+p^6 T^4
\tabularnewline[0.5pt]\hline				
36	 &	smooth    &	          &	1-416 T+5666 p T^2-416 p^3 T^3+p^6 T^4
\tabularnewline[0.5pt]\hline				
37	 &	smooth    &	          &	1+184 T+386 p T^2+184 p^3 T^3+p^6 T^4
\tabularnewline[0.5pt]\hline				
38	 &	smooth    &	          &	1+376 T-1438 p T^2+376 p^3 T^3+p^6 T^4
\tabularnewline[0.5pt]\hline				
39	 &	singular  &	-\frac{1}{12}\+&	(1+p T) (1-332 T+p^3 T^2)
\tabularnewline[0.5pt]\hline				
40	 &	smooth    &	          &	1-168 T+6850 p T^2-168 p^3 T^3+p^6 T^4
\tabularnewline[0.5pt]\hline				
41	 &	smooth    &	          &	1-148 T+3762 p T^2-148 p^3 T^3+p^6 T^4
\tabularnewline[0.5pt]\hline				
42	 &	smooth    &	          &	1-512 T+6818 p T^2-512 p^3 T^3+p^6 T^4
\tabularnewline[0.5pt]\hline				
43	 &	smooth    &	          &	1+900 T+9490 p T^2+900 p^3 T^3+p^6 T^4
\tabularnewline[0.5pt]\hline				
44	 &	smooth    &	          &	1+68 T-750 p T^2+68 p^3 T^3+p^6 T^4
\tabularnewline[0.5pt]\hline				
45	 &	smooth    &	          &	1+4 p T+1586 p T^2+4 p^4 T^3+p^6 T^4
\tabularnewline[0.5pt]\hline				
46	 &	smooth    &	          &	1+956 T+10866 p T^2+956 p^3 T^3+p^6 T^4
\tabularnewline[0.5pt]\hline				
47	 &	smooth    &	          &	1+56 T-2046 p T^2+56 p^3 T^3+p^6 T^4
\tabularnewline[0.5pt]\hline				
48	 &	smooth    &	          &	1+468 T+1426 p T^2+468 p^3 T^3+p^6 T^4
\tabularnewline[0.5pt]\hline				
49	 &	smooth    &	          &	1+312 T+5314 p T^2+312 p^3 T^3+p^6 T^4
\tabularnewline[0.5pt]\hline				
50	 &	singular  &	-\frac{1}{4}\+&	(1+p T) (1-140 T+p^3 T^2)
\tabularnewline[0.5pt]\hline				
51	 &	smooth    &	          &	1+272 T+1506 p T^2+272 p^3 T^3+p^6 T^4
\tabularnewline[0.5pt]\hline				
52	 &	smooth    &	          &	1-464 T+3362 p T^2-464 p^3 T^3+p^6 T^4
\tabularnewline[0.5pt]\hline				
53	 &	smooth    &	          &	1+136 T-958 p T^2+136 p^3 T^3+p^6 T^4
\tabularnewline[0.5pt]\hline				
54	 &	smooth    &	          &	1+4 p T+3506 p T^2+4 p^4 T^3+p^6 T^4
\tabularnewline[0.5pt]\hline				
55	 &	smooth    &	          &	1-40 T-3198 p T^2-40 p^3 T^3+p^6 T^4
\tabularnewline[0.5pt]\hline				
56	 &	smooth    &	          &	1+116 T+4434 p T^2+116 p^3 T^3+p^6 T^4
\tabularnewline[0.5pt]\hline				
57	 &	smooth    &	          &	1+836 T+4818 p T^2+836 p^3 T^3+p^6 T^4
\tabularnewline[0.5pt]\hline				
58	 &	smooth    &	          &	1+304 T+674 p T^2+304 p^3 T^3+p^6 T^4
\tabularnewline[0.5pt]\hline				
59	 &	smooth    &	          &	1-144 T+5026 p T^2-144 p^3 T^3+p^6 T^4
\tabularnewline[0.5pt]\hline				
60	 &	smooth    &	          &	1+788 T+6738 p T^2+788 p^3 T^3+p^6 T^4
\tabularnewline[0.5pt]\hline				
61	 &	smooth    &	          &	1-152 T+8642 p T^2-152 p^3 T^3+p^6 T^4
\tabularnewline[0.5pt]\hline				
62	 &	smooth    &	          &	1-128 T-94 p T^2-128 p^3 T^3+p^6 T^4
\tabularnewline[0.5pt]\hline				
63	 &	smooth    &	          &	1-640 T+4578 p T^2-640 p^3 T^3+p^6 T^4
\tabularnewline[0.5pt]\hline				
64	 &	smooth    &	          &	(1-8 p T+p^3 T^2)(1+468 T+p^3 T^2)
\tabularnewline[0.5pt]\hline				
65	 &	smooth    &	          &	1+860 T+7410 p T^2+860 p^3 T^3+p^6 T^4
\tabularnewline[0.5pt]\hline				
66	 &	smooth    &	          &	1-776 T+8450 p T^2-776 p^3 T^3+p^6 T^4
\tabularnewline[0.5pt]\hline				
\tablepostamble				
\tablepreamble{71}				
1	 &	smooth    &	          &	1-60 T-4990 p T^2-60 p^3 T^3+p^6 T^4
\tabularnewline[0.5pt]\hline				
2	 &	smooth    &	          &	1-480 T-862 p T^2-480 p^3 T^3+p^6 T^4
\tabularnewline[0.5pt]\hline				
3	 &	singular  &	\frac{1}{24}&	(1-p T) (1+236 T+p^3 T^2)
\tabularnewline[0.5pt]\hline				
4	 &	smooth    &	          &	1+320 T+674 p T^2+320 p^3 T^3+p^6 T^4
\tabularnewline[0.5pt]\hline				
5	 &	smooth    &	          &	1+536 T+7202 p T^2+536 p^3 T^3+p^6 T^4
\tabularnewline[0.5pt]\hline				
6	 &	smooth    &	          &	1+136 T+3938 p T^2+136 p^3 T^3+p^6 T^4
\tabularnewline[0.5pt]\hline				
7	 &	smooth    &	          &	1-364 T+9986 p T^2-364 p^3 T^3+p^6 T^4
\tabularnewline[0.5pt]\hline				
8	 &	smooth    &	          &	1-352 T+4322 p T^2-352 p^3 T^3+p^6 T^4
\tabularnewline[0.5pt]\hline				
9	 &	smooth    &	          &	(1+p^3 T^2)(1+136 T+p^3 T^2)
\tabularnewline[0.5pt]\hline				
10	 &	smooth    &	          &	(1+8 p T+p^3 T^2)(1-936 T+p^3 T^2)
\tabularnewline[0.5pt]\hline				
11	 &	smooth    &	          &	1+332 T+6530 p T^2+332 p^3 T^3+p^6 T^4
\tabularnewline[0.5pt]\hline				
12	 &	smooth    &	          &	1+504 T+1634 p T^2+504 p^3 T^3+p^6 T^4
\tabularnewline[0.5pt]\hline				
13	 &	smooth    &	          &	1-912 T+8354 p T^2-912 p^3 T^3+p^6 T^4
\tabularnewline[0.5pt]\hline				
14	 &	smooth    &	          &	1-288 T+2018 p T^2-288 p^3 T^3+p^6 T^4
\tabularnewline[0.5pt]\hline				
15	 &	smooth    &	          &	1-816 T+6434 p T^2-816 p^3 T^3+p^6 T^4
\tabularnewline[0.5pt]\hline				
16	 &	smooth    &	          &	1-456 T+3746 p T^2-456 p^3 T^3+p^6 T^4
\tabularnewline[0.5pt]\hline				
17	 &	smooth    &	          &	(1-8 p T+p^3 T^2)(1-216 T+p^3 T^2)
\tabularnewline[0.5pt]\hline				
18	 &	smooth    &	          &	1+688 T+4130 p T^2+688 p^3 T^3+p^6 T^4
\tabularnewline[0.5pt]\hline				
19	 &	smooth    &	          &	1-192 T-3166 p T^2-192 p^3 T^3+p^6 T^4
\tabularnewline[0.5pt]\hline				
20	 &	smooth    &	          &	1-408 T+866 p T^2-408 p^3 T^3+p^6 T^4
\tabularnewline[0.5pt]\hline				
21	 &	smooth    &	          &	1-168 T-4318 p T^2-168 p^3 T^3+p^6 T^4
\tabularnewline[0.5pt]\hline				
22	 &	smooth    &	          &	1+164 T-574 p T^2+164 p^3 T^3+p^6 T^4
\tabularnewline[0.5pt]\hline				
23	 &	smooth    &	          &	1+36 T+9986 p T^2+36 p^3 T^3+p^6 T^4
\tabularnewline[0.5pt]\hline				
24	 &	smooth    &	          &	1-452 T+7298 p T^2-452 p^3 T^3+p^6 T^4
\tabularnewline[0.5pt]\hline				
25	 &	smooth    &	          &	1-804 T+6146 p T^2-804 p^3 T^3+p^6 T^4
\tabularnewline[0.5pt]\hline				
26	 &	smooth    &	          &	1-8 T+5666 p T^2-8 p^3 T^3+p^6 T^4
\tabularnewline[0.5pt]\hline				
27	 &	smooth    &	          &	1+624 T+9506 p T^2+624 p^3 T^3+p^6 T^4
\tabularnewline[0.5pt]\hline				
28	 &	smooth    &	          &	1+800 T+4514 p T^2+800 p^3 T^3+p^6 T^4
\tabularnewline[0.5pt]\hline				
29	 &	smooth    &	          &	1+1296 T+14882 p T^2+1296 p^3 T^3+p^6 T^4
\tabularnewline[0.5pt]\hline				
30	 &	smooth    &	          &	1+912 T+8354 p T^2+912 p^3 T^3+p^6 T^4
\tabularnewline[0.5pt]\hline				
31	 &	smooth    &	          &	1+296 T+4514 p T^2+296 p^3 T^3+p^6 T^4
\tabularnewline[0.5pt]\hline				
32	 &	smooth    &	          &	1+360 T+6626 p T^2+360 p^3 T^3+p^6 T^4
\tabularnewline[0.5pt]\hline				
33	 &	smooth    &	          &	1+56 T+2786 p T^2+56 p^3 T^3+p^6 T^4
\tabularnewline[0.5pt]\hline				
34	 &	smooth    &	          &	1-192 T+6050 p T^2-192 p^3 T^3+p^6 T^4
\tabularnewline[0.5pt]\hline				
35	 &	smooth    &	          &	1+56 T+3554 p T^2+56 p^3 T^3+p^6 T^4
\tabularnewline[0.5pt]\hline				
36	 &	smooth    &	          &	1-1012 T+8642 p T^2-1012 p^3 T^3+p^6 T^4
\tabularnewline[0.5pt]\hline				
37	 &	smooth    &	          &	1-1232 T+13730 p T^2-1232 p^3 T^3+p^6 T^4
\tabularnewline[0.5pt]\hline				
38	 &	smooth    &	          &	1-16 p T+9698 p T^2-16 p^4 T^3+p^6 T^4
\tabularnewline[0.5pt]\hline				
39	 &	smooth    &	          &	1-872 T+9026 p T^2-872 p^3 T^3+p^6 T^4
\tabularnewline[0.5pt]\hline				
40	 &	smooth    &	          &	1+12 T-4798 p T^2+12 p^3 T^3+p^6 T^4
\tabularnewline[0.5pt]\hline				
41	 &	smooth    &	          &	1-120 T+8162 p T^2-120 p^3 T^3+p^6 T^4
\tabularnewline[0.5pt]\hline				
42	 &	smooth    &	          &	1+224 T-1726 p T^2+224 p^3 T^3+p^6 T^4
\tabularnewline[0.5pt]\hline				
43	 &	smooth    &	          &	1+112 T+3362 p T^2+112 p^3 T^3+p^6 T^4
\tabularnewline[0.5pt]\hline				
44	 &	smooth    &	          &	1+108 T-1342 p T^2+108 p^3 T^3+p^6 T^4
\tabularnewline[0.5pt]\hline				
45	 &	smooth    &	          &	1-176 T+290 p T^2-176 p^3 T^3+p^6 T^4
\tabularnewline[0.5pt]\hline				
46	 &	smooth    &	          &	1-4 p T+6722 p T^2-4 p^4 T^3+p^6 T^4
\tabularnewline[0.5pt]\hline				
47	 &	singular  &	-\frac{1}{3}\+&	(1-p T) (1+328 T+p^3 T^2)
\tabularnewline[0.5pt]\hline				
48	 &	smooth    &	          &	1-124 T+6146 p T^2-124 p^3 T^3+p^6 T^4
\tabularnewline[0.5pt]\hline				
49	 &	smooth    &	          &	1-1128 T+14114 p T^2-1128 p^3 T^3+p^6 T^4
\tabularnewline[0.5pt]\hline				
50	 &	smooth    &	          &	1+800 T+7298 p T^2+800 p^3 T^3+p^6 T^4
\tabularnewline[0.5pt]\hline				
51	 &	smooth    &	          &	1+264 T+674 p T^2+264 p^3 T^3+p^6 T^4
\tabularnewline[0.5pt]\hline				
52	 &	smooth    &	          &	1+576 T+10658 p T^2+576 p^3 T^3+p^6 T^4
\tabularnewline[0.5pt]\hline				
53	 &	singular  &	-\frac{1}{4}\+&	(1+p T) (1+840 T+p^3 T^2)
\tabularnewline[0.5pt]\hline				
54	 &	smooth    &	          &	1-608 T+8354 p T^2-608 p^3 T^3+p^6 T^4
\tabularnewline[0.5pt]\hline				
55	 &	smooth    &	          &	1+584 T+5090 p T^2+584 p^3 T^3+p^6 T^4
\tabularnewline[0.5pt]\hline				
56	 &	smooth    &	          &	(1+8 p T+p^3 T^2)(1-912 T+p^3 T^2)
\tabularnewline[0.5pt]\hline				
57	 &	smooth    &	          &	(1+12 p T+p^3 T^2)(1-648 T+p^3 T^2)
\tabularnewline[0.5pt]\hline				
58	 &	smooth    &	          &	1+360 T+8930 p T^2+360 p^3 T^3+p^6 T^4
\tabularnewline[0.5pt]\hline				
59	 &	smooth    &	          &	(1-8 p T+p^3 T^2)(1+648 T+p^3 T^2)
\tabularnewline[0.5pt]\hline				
60	 &	smooth    &	          &	1-392 T-286 p T^2-392 p^3 T^3+p^6 T^4
\tabularnewline[0.5pt]\hline				
61	 &	smooth    &	          &	1-116 T+7490 p T^2-116 p^3 T^3+p^6 T^4
\tabularnewline[0.5pt]\hline				
62	 &	singular  &	-\frac{1}{8}\+&	(1-p T) (1-432 T+p^3 T^2)
\tabularnewline[0.5pt]\hline				
63	 &	smooth    &	          &	1-536 T+6626 p T^2-536 p^3 T^3+p^6 T^4
\tabularnewline[0.5pt]\hline				
64	 &	smooth    &	          &	1+336 T-862 p T^2+336 p^3 T^3+p^6 T^4
\tabularnewline[0.5pt]\hline				
65	 &	singular  &	-\frac{1}{12}\+&	(1+p T) (1+360 T+p^3 T^2)
\tabularnewline[0.5pt]\hline				
66	 &	smooth    &	          &	1-80 T+1826 p T^2-80 p^3 T^3+p^6 T^4
\tabularnewline[0.5pt]\hline				
67	 &	smooth$^*$&	-\frac{1}{18}\+&	  
\tabularnewline[0.5pt]\hline				
68	 &	smooth    &	          &	1-488 T+3938 p T^2-488 p^3 T^3+p^6 T^4
\tabularnewline[0.5pt]\hline				
69	 &	smooth    &	          &	1-384 T-2302 p T^2-384 p^3 T^3+p^6 T^4
\tabularnewline[0.5pt]\hline				
70	 &	smooth    &	          &	1+324 T+4802 p T^2+324 p^3 T^3+p^6 T^4
\tabularnewline[0.5pt]\hline				
\tablepostamble				
\tablepreamble{73}				
1	 &	smooth    &	          &	1-68 T-8506 p T^2-68 p^3 T^3+p^6 T^4
\tabularnewline[0.5pt]\hline				
2	 &	smooth    &	          &	1-44 T+5846 p T^2-44 p^3 T^3+p^6 T^4
\tabularnewline[0.5pt]\hline				
3	 &	smooth    &	          &	1+408 T+6574 p T^2+408 p^3 T^3+p^6 T^4
\tabularnewline[0.5pt]\hline				
4	 &	smooth$^*$&	-\frac{1}{18}\+&	  
\tabularnewline[0.5pt]\hline				
5	 &	smooth    &	          &	1+200 T-2610 p T^2+200 p^3 T^3+p^6 T^4
\tabularnewline[0.5pt]\hline				
6	 &	singular  &	-\frac{1}{12}\+&	(1-p T) (1-26 T+p^3 T^2)
\tabularnewline[0.5pt]\hline				
7	 &	smooth    &	          &	1-796 T+10806 p T^2-796 p^3 T^3+p^6 T^4
\tabularnewline[0.5pt]\hline				
8	 &	smooth    &	          &	1-204 T-5738 p T^2-204 p^3 T^3+p^6 T^4
\tabularnewline[0.5pt]\hline				
9	 &	singular  &	-\frac{1}{8}\+&	(1-p T) (1-362 T+p^3 T^2)
\tabularnewline[0.5pt]\hline				
10	 &	smooth    &	          &	1+244 T+8726 p T^2+244 p^3 T^3+p^6 T^4
\tabularnewline[0.5pt]\hline				
11	 &	smooth    &	          &	(1+6 p T+p^3 T^2)(1+214 T+p^3 T^2)
\tabularnewline[0.5pt]\hline				
12	 &	smooth    &	          &	1-708 T+6598 p T^2-708 p^3 T^3+p^6 T^4
\tabularnewline[0.5pt]\hline				
13	 &	smooth    &	          &	1-68 T+1478 p T^2-68 p^3 T^3+p^6 T^4
\tabularnewline[0.5pt]\hline				
14	 &	smooth    &	          &	1-356 T+1286 p T^2-356 p^3 T^3+p^6 T^4
\tabularnewline[0.5pt]\hline				
15	 &	smooth    &	          &	1+1404 T+16966 p T^2+1404 p^3 T^3+p^6 T^4
\tabularnewline[0.5pt]\hline				
16	 &	smooth    &	          &	1+540 T+4486 p T^2+540 p^3 T^3+p^6 T^4
\tabularnewline[0.5pt]\hline				
17	 &	smooth    &	          &	1+92 T-4218 p T^2+92 p^3 T^3+p^6 T^4
\tabularnewline[0.5pt]\hline				
18	 &	singular  &	-\frac{1}{4}\+&	(1-p T) (1+550 T+p^3 T^2)
\tabularnewline[0.5pt]\hline				
19	 &	smooth    &	          &	1+1224 T+12622 p T^2+1224 p^3 T^3+p^6 T^4
\tabularnewline[0.5pt]\hline				
20	 &	smooth    &	          &	1+328 T+10574 p T^2+328 p^3 T^3+p^6 T^4
\tabularnewline[0.5pt]\hline				
21	 &	smooth    &	          &	1+1416 T+15310 p T^2+1416 p^3 T^3+p^6 T^4
\tabularnewline[0.5pt]\hline				
22	 &	smooth    &	          &	1+172 T+86 p^2 T^2+172 p^3 T^3+p^6 T^4
\tabularnewline[0.5pt]\hline				
23	 &	smooth    &	          &	(1+6 p T+p^3 T^2)(1-650 T+p^3 T^2)
\tabularnewline[0.5pt]\hline				
24	 &	singular  &	-\frac{1}{3}\+&	(1-p T) (1+38 T+p^3 T^2)
\tabularnewline[0.5pt]\hline				
25	 &	smooth    &	          &	1+364 T-1690 p T^2+364 p^3 T^3+p^6 T^4
\tabularnewline[0.5pt]\hline				
26	 &	smooth    &	          &	1+256 T-226 p T^2+256 p^3 T^3+p^6 T^4
\tabularnewline[0.5pt]\hline				
27	 &	smooth    &	          &	1-500 T+7334 p T^2-500 p^3 T^3+p^6 T^4
\tabularnewline[0.5pt]\hline				
28	 &	smooth    &	          &	1+756 T+3478 p T^2+756 p^3 T^3+p^6 T^4
\tabularnewline[0.5pt]\hline				
29	 &	smooth    &	          &	1-68 T+8774 p T^2-68 p^3 T^3+p^6 T^4
\tabularnewline[0.5pt]\hline				
30	 &	smooth    &	          &	1-524 T+9494 p T^2-524 p^3 T^3+p^6 T^4
\tabularnewline[0.5pt]\hline				
31	 &	smooth    &	          &	1+84 T+7126 p T^2+84 p^3 T^3+p^6 T^4
\tabularnewline[0.5pt]\hline				
32	 &	smooth    &	          &	1-884 T+10022 p T^2-884 p^3 T^3+p^6 T^4
\tabularnewline[0.5pt]\hline				
33	 &	smooth    &	          &	1-328 T+8430 p T^2-328 p^3 T^3+p^6 T^4
\tabularnewline[0.5pt]\hline				
34	 &	smooth    &	          &	1+572 T+4806 p T^2+572 p^3 T^3+p^6 T^4
\tabularnewline[0.5pt]\hline				
35	 &	smooth    &	          &	1-372 T+10246 p T^2-372 p^3 T^3+p^6 T^4
\tabularnewline[0.5pt]\hline				
36	 &	smooth    &	          &	1-44 T-6058 p T^2-44 p^3 T^3+p^6 T^4
\tabularnewline[0.5pt]\hline				
37	 &	smooth    &	          &	1-576 T+7966 p T^2-576 p^3 T^3+p^6 T^4
\tabularnewline[0.5pt]\hline				
38	 &	smooth    &	          &	1+9118 p T^2+p^6 T^4
\tabularnewline[0.5pt]\hline				
39	 &	smooth    &	          &	1-996 T+7174 p T^2-996 p^3 T^3+p^6 T^4
\tabularnewline[0.5pt]\hline				
40	 &	smooth    &	          &	1+1044 T+11350 p T^2+1044 p^3 T^3+p^6 T^4
\tabularnewline[0.5pt]\hline				
41	 &	smooth    &	          &	1+420 T+6070 p T^2+420 p^3 T^3+p^6 T^4
\tabularnewline[0.5pt]\hline				
42	 &	smooth    &	          &	1+700 T+5126 p T^2+700 p^3 T^3+p^6 T^4
\tabularnewline[0.5pt]\hline				
43	 &	smooth    &	          &	1-44 T+5558 p T^2-44 p^3 T^3+p^6 T^4
\tabularnewline[0.5pt]\hline				
44	 &	smooth    &	          &	1+348 T+9862 p T^2+348 p^3 T^3+p^6 T^4
\tabularnewline[0.5pt]\hline				
45	 &	smooth    &	          &	1-796 T+9270 p T^2-796 p^3 T^3+p^6 T^4
\tabularnewline[0.5pt]\hline				
46	 &	smooth    &	          &	1-596 T+1766 p T^2-596 p^3 T^3+p^6 T^4
\tabularnewline[0.5pt]\hline				
47	 &	smooth    &	          &	1+176 T+78 p^2 T^2+176 p^3 T^3+p^6 T^4
\tabularnewline[0.5pt]\hline				
48	 &	smooth    &	          &	1-68 T+5318 p T^2-68 p^3 T^3+p^6 T^4
\tabularnewline[0.5pt]\hline				
49	 &	smooth    &	          &	1+316 T+10694 p T^2+316 p^3 T^3+p^6 T^4
\tabularnewline[0.5pt]\hline				
50	 &	smooth    &	          &	1-56 T+9614 p T^2-56 p^3 T^3+p^6 T^4
\tabularnewline[0.5pt]\hline				
51	 &	smooth    &	          &	1+540 T+6790 p T^2+540 p^3 T^3+p^6 T^4
\tabularnewline[0.5pt]\hline				
52	 &	smooth    &	          &	1-132 T+5830 p T^2-132 p^3 T^3+p^6 T^4
\tabularnewline[0.5pt]\hline				
53	 &	smooth    &	          &	1-476 T+2486 p T^2-476 p^3 T^3+p^6 T^4
\tabularnewline[0.5pt]\hline				
54	 &	smooth    &	          &	1-132 T+4294 p T^2-132 p^3 T^3+p^6 T^4
\tabularnewline[0.5pt]\hline				
55	 &	smooth    &	          &	1-84 T+5734 p T^2-84 p^3 T^3+p^6 T^4
\tabularnewline[0.5pt]\hline				
56	 &	smooth    &	          &	1+196 T-5386 p T^2+196 p^3 T^3+p^6 T^4
\tabularnewline[0.5pt]\hline				
57	 &	smooth    &	          &	1-644 T+10598 p T^2-644 p^3 T^3+p^6 T^4
\tabularnewline[0.5pt]\hline				
58	 &	smooth    &	          &	1-228 T+7174 p T^2-228 p^3 T^3+p^6 T^4
\tabularnewline[0.5pt]\hline				
59	 &	smooth    &	          &	1+464 T+7422 p T^2+464 p^3 T^3+p^6 T^4
\tabularnewline[0.5pt]\hline				
60	 &	smooth    &	          &	1-192 T-4706 p T^2-192 p^3 T^3+p^6 T^4
\tabularnewline[0.5pt]\hline				
61	 &	smooth    &	          &	1+860 T+6918 p T^2+860 p^3 T^3+p^6 T^4
\tabularnewline[0.5pt]\hline				
62	 &	smooth    &	          &	1+1044 T+9046 p T^2+1044 p^3 T^3+p^6 T^4
\tabularnewline[0.5pt]\hline				
63	 &	smooth    &	          &	1+836 T+9846 p T^2+836 p^3 T^3+p^6 T^4
\tabularnewline[0.5pt]\hline				
64	 &	smooth    &	          &	(1-10 p T+p^3 T^2)(1+706 T+p^3 T^2)
\tabularnewline[0.5pt]\hline				
65	 &	smooth    &	          &	1+492 T+6502 p T^2+492 p^3 T^3+p^6 T^4
\tabularnewline[0.5pt]\hline				
66	 &	smooth    &	          &	1+564 T+1558 p T^2+564 p^3 T^3+p^6 T^4
\tabularnewline[0.5pt]\hline				
67	 &	smooth    &	          &	1-188 T+2678 p T^2-188 p^3 T^3+p^6 T^4
\tabularnewline[0.5pt]\hline				
68	 &	smooth    &	          &	1-444 T+2038 p T^2-444 p^3 T^3+p^6 T^4
\tabularnewline[0.5pt]\hline				
69	 &	smooth    &	          &	1-324 T+4582 p T^2-324 p^3 T^3+p^6 T^4
\tabularnewline[0.5pt]\hline				
70	 &	singular  &	\frac{1}{24}&	(1-p T) (1-418 T+p^3 T^2)
\tabularnewline[0.5pt]\hline				
71	 &	smooth    &	          &	1-736 T+3870 p T^2-736 p^3 T^3+p^6 T^4
\tabularnewline[0.5pt]\hline				
72	 &	smooth    &	          &	1-220 T+1590 p T^2-220 p^3 T^3+p^6 T^4
\tabularnewline[0.5pt]\hline				
\tablepostamble				
\tablepreamble{79}				
1	 &	smooth    &	          &	1-360 T+3394 p T^2-360 p^3 T^3+p^6 T^4
\tabularnewline[0.5pt]\hline				
2	 &	smooth    &	          &	1-400 T+12354 p T^2-400 p^3 T^3+p^6 T^4
\tabularnewline[0.5pt]\hline				
3	 &	smooth    &	          &	1-224 T+2882 p T^2-224 p^3 T^3+p^6 T^4
\tabularnewline[0.5pt]\hline				
4	 &	smooth    &	          &	1+540 T+6850 p T^2+540 p^3 T^3+p^6 T^4
\tabularnewline[0.5pt]\hline				
5	 &	smooth    &	          &	1-252 T+3394 p T^2-252 p^3 T^3+p^6 T^4
\tabularnewline[0.5pt]\hline				
6	 &	smooth    &	          &	1+232 T+962 p T^2+232 p^3 T^3+p^6 T^4
\tabularnewline[0.5pt]\hline				
7	 &	smooth    &	          &	1+272 T+8898 p T^2+272 p^3 T^3+p^6 T^4
\tabularnewline[0.5pt]\hline				
8	 &	smooth    &	          &	1-788 T+7682 p T^2-788 p^3 T^3+p^6 T^4
\tabularnewline[0.5pt]\hline				
9	 &	smooth    &	          &	1-456 T+9922 p T^2-456 p^3 T^3+p^6 T^4
\tabularnewline[0.5pt]\hline				
10	 &	smooth    &	          &	1-836 T+9986 p T^2-836 p^3 T^3+p^6 T^4
\tabularnewline[0.5pt]\hline				
11	 &	smooth    &	          &	1-192 T+9922 p T^2-192 p^3 T^3+p^6 T^4
\tabularnewline[0.5pt]\hline				
12	 &	smooth    &	          &	1+164 T-6846 p T^2+164 p^3 T^3+p^6 T^4
\tabularnewline[0.5pt]\hline				
13	 &	smooth    &	          &	1-456 T-1022 p T^2-456 p^3 T^3+p^6 T^4
\tabularnewline[0.5pt]\hline				
14	 &	smooth    &	          &	1+692 T+7938 p T^2+692 p^3 T^3+p^6 T^4
\tabularnewline[0.5pt]\hline				
15	 &	smooth    &	          &	1-600 T+5314 p T^2-600 p^3 T^3+p^6 T^4
\tabularnewline[0.5pt]\hline				
16	 &	smooth    &	          &	1-60 T+7234 p T^2-60 p^3 T^3+p^6 T^4
\tabularnewline[0.5pt]\hline				
17	 &	smooth    &	          &	1+504 T+4546 p T^2+504 p^3 T^3+p^6 T^4
\tabularnewline[0.5pt]\hline				
18	 &	smooth    &	          &	1+668 T+10434 p T^2+668 p^3 T^3+p^6 T^4
\tabularnewline[0.5pt]\hline				
19	 &	smooth    &	          &	1+616 T+4610 p T^2+616 p^3 T^3+p^6 T^4
\tabularnewline[0.5pt]\hline				
20	 &	smooth    &	          &	1-640 T+4290 p T^2-640 p^3 T^3+p^6 T^4
\tabularnewline[0.5pt]\hline				
21	 &	smooth    &	          &	1+372 T+2242 p T^2+372 p^3 T^3+p^6 T^4
\tabularnewline[0.5pt]\hline				
22	 &	smooth    &	          &	1-580 T+1410 p T^2-580 p^3 T^3+p^6 T^4
\tabularnewline[0.5pt]\hline				
23	 &	smooth    &	          &	1-272 T+10562 p T^2-272 p^3 T^3+p^6 T^4
\tabularnewline[0.5pt]\hline				
24	 &	smooth    &	          &	1+848 T+3138 p T^2+848 p^3 T^3+p^6 T^4
\tabularnewline[0.5pt]\hline				
25	 &	smooth    &	          &	1-792 T+8386 p T^2-792 p^3 T^3+p^6 T^4
\tabularnewline[0.5pt]\hline				
26	 &	singular  &	-\frac{1}{3}\+&	(1-p T) (1+240 T+p^3 T^2)
\tabularnewline[0.5pt]\hline				
27	 &	smooth    &	          &	1+380 T+1410 p T^2+380 p^3 T^3+p^6 T^4
\tabularnewline[0.5pt]\hline				
28	 &	smooth    &	          &	1+932 T+7746 p T^2+932 p^3 T^3+p^6 T^4
\tabularnewline[0.5pt]\hline				
29	 &	smooth    &	          &	1-184 T+11202 p T^2-184 p^3 T^3+p^6 T^4
\tabularnewline[0.5pt]\hline				
30	 &	smooth    &	          &	1+152 T-4446 p T^2+152 p^3 T^3+p^6 T^4
\tabularnewline[0.5pt]\hline				
31	 &	smooth    &	          &	1+248 T+11202 p T^2+248 p^3 T^3+p^6 T^4
\tabularnewline[0.5pt]\hline				
32	 &	smooth    &	          &	1-36 T+130 p T^2-36 p^3 T^3+p^6 T^4
\tabularnewline[0.5pt]\hline				
33	 &	smooth    &	          &	1+464 T+4290 p T^2+464 p^3 T^3+p^6 T^4
\tabularnewline[0.5pt]\hline				
34	 &	smooth    &	          &	1+536 T+11202 p T^2+536 p^3 T^3+p^6 T^4
\tabularnewline[0.5pt]\hline				
35	 &	smooth    &	          &	1-512 T+8162 p T^2-512 p^3 T^3+p^6 T^4
\tabularnewline[0.5pt]\hline				
36	 &	smooth    &	          &	1-512 T+4802 p T^2-512 p^3 T^3+p^6 T^4
\tabularnewline[0.5pt]\hline				
37	 &	smooth    &	          &	1-1668 T+19330 p T^2-1668 p^3 T^3+p^6 T^4
\tabularnewline[0.5pt]\hline				
38	 &	smooth    &	          &	1+700 T+7298 p T^2+700 p^3 T^3+p^6 T^4
\tabularnewline[0.5pt]\hline				
39	 &	smooth    &	          &	1+1392 T+13762 p T^2+1392 p^3 T^3+p^6 T^4
\tabularnewline[0.5pt]\hline				
40	 &	smooth    &	          &	1-136 T-1182 p T^2-136 p^3 T^3+p^6 T^4
\tabularnewline[0.5pt]\hline				
41	 &	smooth    &	          &	1+216 T-62 p T^2+216 p^3 T^3+p^6 T^4
\tabularnewline[0.5pt]\hline				
42	 &	smooth    &	          &	1+40 T+8642 p T^2+40 p^3 T^3+p^6 T^4
\tabularnewline[0.5pt]\hline				
43	 &	smooth    &	          &	1+1664 T+19650 p T^2+1664 p^3 T^3+p^6 T^4
\tabularnewline[0.5pt]\hline				
44	 &	smooth    &	          &	1+72 T-4670 p T^2+72 p^3 T^3+p^6 T^4
\tabularnewline[0.5pt]\hline				
45	 &	smooth    &	          &	1-272 T+7874 p T^2-272 p^3 T^3+p^6 T^4
\tabularnewline[0.5pt]\hline				
46	 &	singular  &	-\frac{1}{12}\+&	(1+p T) (1-512 T+p^3 T^2)
\tabularnewline[0.5pt]\hline				
47	 &	smooth    &	          &	1+340 T-5566 p T^2+340 p^3 T^3+p^6 T^4
\tabularnewline[0.5pt]\hline				
48	 &	smooth    &	          &	1-272 T-4798 p T^2-272 p^3 T^3+p^6 T^4
\tabularnewline[0.5pt]\hline				
49	 &	smooth    &	          &	1-1000 T+13122 p T^2-1000 p^3 T^3+p^6 T^4
\tabularnewline[0.5pt]\hline				
50	 &	smooth    &	          &	1-24 T-9950 p T^2-24 p^3 T^3+p^6 T^4
\tabularnewline[0.5pt]\hline				
51	 &	smooth    &	          &	1-680 T+9026 p T^2-680 p^3 T^3+p^6 T^4
\tabularnewline[0.5pt]\hline				
52	 &	smooth    &	          &	1-536 T+2498 p T^2-536 p^3 T^3+p^6 T^4
\tabularnewline[0.5pt]\hline				
53	 &	smooth    &	          &	1+896 T+8898 p T^2+896 p^3 T^3+p^6 T^4
\tabularnewline[0.5pt]\hline				
54	 &	smooth    &	          &	1+1256 T+11202 p T^2+1256 p^3 T^3+p^6 T^4
\tabularnewline[0.5pt]\hline				
55	 &	smooth    &	          &	1+584 T+8610 p T^2+584 p^3 T^3+p^6 T^4
\tabularnewline[0.5pt]\hline				
56	 &	singular  &	\frac{1}{24}&	(1-p T) (1-552 T+p^3 T^2)
\tabularnewline[0.5pt]\hline				
57	 &	smooth$^*$&	-\frac{1}{18}\+&	  
\tabularnewline[0.5pt]\hline				
58	 &	smooth    &	          &	1-52 T+4674 p T^2-52 p^3 T^3+p^6 T^4
\tabularnewline[0.5pt]\hline				
59	 &	singular  &	-\frac{1}{4}\+&	(1+p T) (1+208 T+p^3 T^2)
\tabularnewline[0.5pt]\hline				
60	 &	smooth    &	          &	1+216 T-638 p T^2+216 p^3 T^3+p^6 T^4
\tabularnewline[0.5pt]\hline				
61	 &	smooth    &	          &	1+336 T-254 p T^2+336 p^3 T^3+p^6 T^4
\tabularnewline[0.5pt]\hline				
62	 &	smooth    &	          &	1+408 T+1474 p T^2+408 p^3 T^3+p^6 T^4
\tabularnewline[0.5pt]\hline				
63	 &	smooth    &	          &	1+896 T+8514 p T^2+896 p^3 T^3+p^6 T^4
\tabularnewline[0.5pt]\hline				
64	 &	smooth    &	          &	1+96 T+5506 p T^2+96 p^3 T^3+p^6 T^4
\tabularnewline[0.5pt]\hline				
65	 &	smooth    &	          &	1-168 T+2146 p T^2-168 p^3 T^3+p^6 T^4
\tabularnewline[0.5pt]\hline				
66	 &	smooth    &	          &	1-720 T+11458 p T^2-720 p^3 T^3+p^6 T^4
\tabularnewline[0.5pt]\hline				
67	 &	smooth    &	          &	1-796 T+4290 p T^2-796 p^3 T^3+p^6 T^4
\tabularnewline[0.5pt]\hline				
68	 &	smooth    &	          &	1+444 T+9346 p T^2+444 p^3 T^3+p^6 T^4
\tabularnewline[0.5pt]\hline				
69	 &	singular  &	-\frac{1}{8}\+&	(1-p T) (1+160 T+p^3 T^2)
\tabularnewline[0.5pt]\hline				
70	 &	smooth    &	          &	1+1552 T+19490 p T^2+1552 p^3 T^3+p^6 T^4
\tabularnewline[0.5pt]\hline				
71	 &	smooth    &	          &	1+552 T+12994 p T^2+552 p^3 T^3+p^6 T^4
\tabularnewline[0.5pt]\hline				
72	 &	smooth    &	          &	1+152 T-3006 p T^2+152 p^3 T^3+p^6 T^4
\tabularnewline[0.5pt]\hline				
73	 &	smooth    &	          &	1+284 T+3906 p T^2+284 p^3 T^3+p^6 T^4
\tabularnewline[0.5pt]\hline				
74	 &	smooth    &	          &	1-24 T-10046 p T^2-24 p^3 T^3+p^6 T^4
\tabularnewline[0.5pt]\hline				
75	 &	smooth    &	          &	1+440 T+2082 p T^2+440 p^3 T^3+p^6 T^4
\tabularnewline[0.5pt]\hline				
76	 &	smooth    &	          &	1+1192 T+12866 p T^2+1192 p^3 T^3+p^6 T^4
\tabularnewline[0.5pt]\hline				
77	 &	smooth    &	          &	(1+p^3 T^2)(1+1012 T+p^3 T^2)
\tabularnewline[0.5pt]\hline				
78	 &	smooth    &	          &	1-300 T+6082 p T^2-300 p^3 T^3+p^6 T^4
\tabularnewline[0.5pt]\hline				
\tablepostamble				
\tablepreamble{83}				
1	 &	smooth    &	          &	1-424 T+10466 p T^2-424 p^3 T^3+p^6 T^4
\tabularnewline[0.5pt]\hline				
2	 &	smooth    &	          &	1+480 T+6434 p T^2+480 p^3 T^3+p^6 T^4
\tabularnewline[0.5pt]\hline				
3	 &	smooth    &	          &	1-660 T+3890 p T^2-660 p^3 T^3+p^6 T^4
\tabularnewline[0.5pt]\hline				
4	 &	smooth    &	          &	1+8 p T+8258 p T^2+8 p^4 T^3+p^6 T^4
\tabularnewline[0.5pt]\hline				
5	 &	smooth    &	          &	1-600 T+3842 p T^2-600 p^3 T^3+p^6 T^4
\tabularnewline[0.5pt]\hline				
6	 &	smooth    &	          &	1+460 T+1010 p T^2+460 p^3 T^3+p^6 T^4
\tabularnewline[0.5pt]\hline				
7	 &	smooth    &	          &	1+288 T-4702 p T^2+288 p^3 T^3+p^6 T^4
\tabularnewline[0.5pt]\hline				
8	 &	smooth    &	          &	1-936 T+9026 p T^2-936 p^3 T^3+p^6 T^4
\tabularnewline[0.5pt]\hline				
9	 &	smooth    &	          &	1-840 T+12098 p T^2-840 p^3 T^3+p^6 T^4
\tabularnewline[0.5pt]\hline				
10	 &	smooth    &	          &	1-456 T+9026 p T^2-456 p^3 T^3+p^6 T^4
\tabularnewline[0.5pt]\hline				
11	 &	smooth    &	          &	1-488 T+6146 p T^2-488 p^3 T^3+p^6 T^4
\tabularnewline[0.5pt]\hline				
12	 &	smooth    &	          &	1-100 T-7438 p T^2-100 p^3 T^3+p^6 T^4
\tabularnewline[0.5pt]\hline				
13	 &	smooth    &	          &	(1-4 p T+p^3 T^2)(1+852 T+p^3 T^2)
\tabularnewline[0.5pt]\hline				
14	 &	smooth    &	          &	1-604 T+4178 p T^2-604 p^3 T^3+p^6 T^4
\tabularnewline[0.5pt]\hline				
15	 &	smooth    &	          &	1+540 T+2546 p T^2+540 p^3 T^3+p^6 T^4
\tabularnewline[0.5pt]\hline				
16	 &	smooth    &	          &	1+672 T+4898 p T^2+672 p^3 T^3+p^6 T^4
\tabularnewline[0.5pt]\hline				
17	 &	smooth    &	          &	1+240 T-670 p T^2+240 p^3 T^3+p^6 T^4
\tabularnewline[0.5pt]\hline				
18	 &	smooth    &	          &	1-508 T+12050 p T^2-508 p^3 T^3+p^6 T^4
\tabularnewline[0.5pt]\hline				
19	 &	smooth    &	          &	1-728 T+674 p T^2-728 p^3 T^3+p^6 T^4
\tabularnewline[0.5pt]\hline				
20	 &	smooth    &	          &	1-288 T+7394 p T^2-288 p^3 T^3+p^6 T^4
\tabularnewline[0.5pt]\hline				
21	 &	smooth    &	          &	1-140 T-4078 p T^2-140 p^3 T^3+p^6 T^4
\tabularnewline[0.5pt]\hline				
22	 &	smooth    &	          &	1-148 T-2446 p T^2-148 p^3 T^3+p^6 T^4
\tabularnewline[0.5pt]\hline				
23	 &	smooth$^*$&	-\frac{1}{18}\+&	  
\tabularnewline[0.5pt]\hline				
24	 &	smooth    &	          &	1-12 T+2834 p T^2-12 p^3 T^3+p^6 T^4
\tabularnewline[0.5pt]\hline				
25	 &	smooth    &	          &	1-504 T+5570 p T^2-504 p^3 T^3+p^6 T^4
\tabularnewline[0.5pt]\hline				
26	 &	smooth    &	          &	1+208 T-7582 p T^2+208 p^3 T^3+p^6 T^4
\tabularnewline[0.5pt]\hline				
27	 &	smooth    &	          &	1+648 T-766 p T^2+648 p^3 T^3+p^6 T^4
\tabularnewline[0.5pt]\hline				
28	 &	smooth    &	          &	1+80 T+11042 p T^2+80 p^3 T^3+p^6 T^4
\tabularnewline[0.5pt]\hline				
29	 &	smooth    &	          &	1-1136 T+13538 p T^2-1136 p^3 T^3+p^6 T^4
\tabularnewline[0.5pt]\hline				
30	 &	smooth    &	          &	1+1728 T+19106 p T^2+1728 p^3 T^3+p^6 T^4
\tabularnewline[0.5pt]\hline				
31	 &	singular  &	-\frac{1}{8}\+&	(1-p T) (1-72 T+p^3 T^2)
\tabularnewline[0.5pt]\hline				
32	 &	smooth    &	          &	1+24 T-4414 p T^2+24 p^3 T^3+p^6 T^4
\tabularnewline[0.5pt]\hline				
33	 &	smooth    &	          &	1-108 T-8494 p T^2-108 p^3 T^3+p^6 T^4
\tabularnewline[0.5pt]\hline				
34	 &	smooth    &	          &	1+36 T+530 p T^2+36 p^3 T^3+p^6 T^4
\tabularnewline[0.5pt]\hline				
35	 &	smooth    &	          &	1-920 T+12674 p T^2-920 p^3 T^3+p^6 T^4
\tabularnewline[0.5pt]\hline				
36	 &	smooth    &	          &	1+312 T-1246 p T^2+312 p^3 T^3+p^6 T^4
\tabularnewline[0.5pt]\hline				
37	 &	smooth    &	          &	1-1148 T+11282 p T^2-1148 p^3 T^3+p^6 T^4
\tabularnewline[0.5pt]\hline				
38	 &	smooth    &	          &	1+240 T+7010 p T^2+240 p^3 T^3+p^6 T^4
\tabularnewline[0.5pt]\hline				
39	 &	smooth    &	          &	1+416 T+10274 p T^2+416 p^3 T^3+p^6 T^4
\tabularnewline[0.5pt]\hline				
40	 &	smooth    &	          &	1+564 T+7442 p T^2+564 p^3 T^3+p^6 T^4
\tabularnewline[0.5pt]\hline				
41	 &	smooth    &	          &	1+732 T+5234 p T^2+732 p^3 T^3+p^6 T^4
\tabularnewline[0.5pt]\hline				
42	 &	smooth    &	          &	1-328 T-3646 p T^2-328 p^3 T^3+p^6 T^4
\tabularnewline[0.5pt]\hline				
43	 &	smooth    &	          &	1+252 T-3022 p T^2+252 p^3 T^3+p^6 T^4
\tabularnewline[0.5pt]\hline				
44	 &	smooth    &	          &	1-384 T-94 p T^2-384 p^3 T^3+p^6 T^4
\tabularnewline[0.5pt]\hline				
45	 &	singular  &	\frac{1}{24}&	(1-p T) (1-1036 T+p^3 T^2)
\tabularnewline[0.5pt]\hline				
46	 &	smooth    &	          &	1+1080 T+14018 p T^2+1080 p^3 T^3+p^6 T^4
\tabularnewline[0.5pt]\hline				
47	 &	smooth    &	          &	1-104 T+3266 p T^2-104 p^3 T^3+p^6 T^4
\tabularnewline[0.5pt]\hline				
48	 &	smooth    &	          &	1-328 T+1538 p T^2-328 p^3 T^3+p^6 T^4
\tabularnewline[0.5pt]\hline				
49	 &	smooth    &	          &	1-724 T+12146 p T^2-724 p^3 T^3+p^6 T^4
\tabularnewline[0.5pt]\hline				
50	 &	smooth    &	          &	1-544 T+4514 p T^2-544 p^3 T^3+p^6 T^4
\tabularnewline[0.5pt]\hline				
51	 &	smooth    &	          &	(1-12 p T+p^3 T^2)(1-396 T+p^3 T^2)
\tabularnewline[0.5pt]\hline				
52	 &	smooth    &	          &	1+84 T-3310 p T^2+84 p^3 T^3+p^6 T^4
\tabularnewline[0.5pt]\hline				
53	 &	smooth    &	          &	1-288 T+11042 p T^2-288 p^3 T^3+p^6 T^4
\tabularnewline[0.5pt]\hline				
54	 &	smooth    &	          &	1+216 T-3550 p T^2+216 p^3 T^3+p^6 T^4
\tabularnewline[0.5pt]\hline				
55	 &	singular  &	-\frac{1}{3}\+&	(1-p T) (1-1212 T+p^3 T^2)
\tabularnewline[0.5pt]\hline				
56	 &	smooth    &	          &	1-448 T+9506 p T^2-448 p^3 T^3+p^6 T^4
\tabularnewline[0.5pt]\hline				
57	 &	smooth    &	          &	1-304 T-7102 p T^2-304 p^3 T^3+p^6 T^4
\tabularnewline[0.5pt]\hline				
58	 &	smooth    &	          &	1+320 T+8354 p T^2+320 p^3 T^3+p^6 T^4
\tabularnewline[0.5pt]\hline				
59	 &	smooth    &	          &	(1-4 p T+p^3 T^2)(1+108 T+p^3 T^2)
\tabularnewline[0.5pt]\hline				
60	 &	smooth    &	          &	1+1132 T+16370 p T^2+1132 p^3 T^3+p^6 T^4
\tabularnewline[0.5pt]\hline				
61	 &	smooth    &	          &	1+256 T-3934 p T^2+256 p^3 T^3+p^6 T^4
\tabularnewline[0.5pt]\hline				
62	 &	singular  &	-\frac{1}{4}\+&	(1+p T) (1-516 T+p^3 T^2)
\tabularnewline[0.5pt]\hline				
63	 &	smooth    &	          &	1+40 T-4606 p T^2+40 p^3 T^3+p^6 T^4
\tabularnewline[0.5pt]\hline				
64	 &	smooth    &	          &	1+1288 T+10178 p T^2+1288 p^3 T^3+p^6 T^4
\tabularnewline[0.5pt]\hline				
65	 &	smooth    &	          &	1-84 T-910 p T^2-84 p^3 T^3+p^6 T^4
\tabularnewline[0.5pt]\hline				
66	 &	smooth    &	          &	(1+12 p T+p^3 T^2)(1-444 T+p^3 T^2)
\tabularnewline[0.5pt]\hline				
67	 &	smooth    &	          &	1-1056 T+15650 p T^2-1056 p^3 T^3+p^6 T^4
\tabularnewline[0.5pt]\hline				
68	 &	smooth    &	          &	1-528 T+2018 p T^2-528 p^3 T^3+p^6 T^4
\tabularnewline[0.5pt]\hline				
69	 &	smooth    &	          &	1-472 T+1922 p T^2-472 p^3 T^3+p^6 T^4
\tabularnewline[0.5pt]\hline				
70	 &	smooth    &	          &	1-992 T+11042 p T^2-992 p^3 T^3+p^6 T^4
\tabularnewline[0.5pt]\hline				
71	 &	smooth    &	          &	1-696 T+9026 p T^2-696 p^3 T^3+p^6 T^4
\tabularnewline[0.5pt]\hline				
72	 &	smooth    &	          &	1-840 T+10178 p T^2-840 p^3 T^3+p^6 T^4
\tabularnewline[0.5pt]\hline				
73	 &	smooth    &	          &	1-4 p T+7442 p T^2-4 p^4 T^3+p^6 T^4
\tabularnewline[0.5pt]\hline				
74	 &	smooth    &	          &	1+88 T+8642 p T^2+88 p^3 T^3+p^6 T^4
\tabularnewline[0.5pt]\hline				
75	 &	smooth    &	          &	1+48 T-11422 p T^2+48 p^3 T^3+p^6 T^4
\tabularnewline[0.5pt]\hline				
76	 &	singular  &	-\frac{1}{12}\+&	(1+p T) (1+1188 T+p^3 T^2)
\tabularnewline[0.5pt]\hline				
77	 &	smooth    &	          &	1-1344 T+18722 p T^2-1344 p^3 T^3+p^6 T^4
\tabularnewline[0.5pt]\hline				
78	 &	smooth    &	          &	1+1388 T+13298 p T^2+1388 p^3 T^3+p^6 T^4
\tabularnewline[0.5pt]\hline				
79	 &	smooth    &	          &	1+176 T+2786 p T^2+176 p^3 T^3+p^6 T^4
\tabularnewline[0.5pt]\hline				
80	 &	smooth    &	          &	1+576 T+3746 p T^2+576 p^3 T^3+p^6 T^4
\tabularnewline[0.5pt]\hline				
81	 &	smooth    &	          &	1-552 T+13634 p T^2-552 p^3 T^3+p^6 T^4
\tabularnewline[0.5pt]\hline				
82	 &	smooth    &	          &	1-392 T+4802 p T^2-392 p^3 T^3+p^6 T^4
\tabularnewline[0.5pt]\hline				
\tablepostamble				
\tablepreamble{89}				
1	 &	smooth    &	          &	1+724 T+7766 p T^2+724 p^3 T^3+p^6 T^4
\tabularnewline[0.5pt]\hline				
2	 &	smooth    &	          &	1+852 T+7766 p T^2+852 p^3 T^3+p^6 T^4
\tabularnewline[0.5pt]\hline				
3	 &	smooth    &	          &	1-972 T+13718 p T^2-972 p^3 T^3+p^6 T^4
\tabularnewline[0.5pt]\hline				
4	 &	smooth    &	          &	1+684 T+15206 p T^2+684 p^3 T^3+p^6 T^4
\tabularnewline[0.5pt]\hline				
5	 &	smooth    &	          &	1-1244 T+10550 p T^2-1244 p^3 T^3+p^6 T^4
\tabularnewline[0.5pt]\hline				
6	 &	smooth    &	          &	1-1452 T+18134 p T^2-1452 p^3 T^3+p^6 T^4
\tabularnewline[0.5pt]\hline				
7	 &	smooth    &	          &	1+944 T+14846 p T^2+944 p^3 T^3+p^6 T^4
\tabularnewline[0.5pt]\hline				
8	 &	smooth    &	          &	1-404 T-5146 p T^2-404 p^3 T^3+p^6 T^4
\tabularnewline[0.5pt]\hline				
9	 &	smooth    &	          &	1+160 T+10142 p T^2+160 p^3 T^3+p^6 T^4
\tabularnewline[0.5pt]\hline				
10	 &	smooth    &	          &	1-776 T+15854 p T^2-776 p^3 T^3+p^6 T^4
\tabularnewline[0.5pt]\hline				
11	 &	singular  &	-\frac{1}{8}\+&	(1-p T) (1-810 T+p^3 T^2)
\tabularnewline[0.5pt]\hline				
12	 &	smooth    &	          &	1+476 T+7046 p T^2+476 p^3 T^3+p^6 T^4
\tabularnewline[0.5pt]\hline				
13	 &	smooth    &	          &	1+348 T+7046 p T^2+348 p^3 T^3+p^6 T^4
\tabularnewline[0.5pt]\hline				
14	 &	smooth    &	          &	1+676 T+758 p T^2+676 p^3 T^3+p^6 T^4
\tabularnewline[0.5pt]\hline				
15	 &	smooth    &	          &	1+204 T+8486 p T^2+204 p^3 T^3+p^6 T^4
\tabularnewline[0.5pt]\hline				
16	 &	smooth    &	          &	1+244 T-682 p T^2+244 p^3 T^3+p^6 T^4
\tabularnewline[0.5pt]\hline				
17	 &	smooth    &	          &	1-424 T+12782 p T^2-424 p^3 T^3+p^6 T^4
\tabularnewline[0.5pt]\hline				
18	 &	smooth    &	          &	1-516 T+8006 p T^2-516 p^3 T^3+p^6 T^4
\tabularnewline[0.5pt]\hline				
19	 &	smooth    &	          &	1+556 T+8294 p T^2+556 p^3 T^3+p^6 T^4
\tabularnewline[0.5pt]\hline				
20	 &	smooth    &	          &	1-1416 T+16814 p T^2-1416 p^3 T^3+p^6 T^4
\tabularnewline[0.5pt]\hline				
21	 &	smooth    &	          &	1+60 T+4934 p T^2+60 p^3 T^3+p^6 T^4
\tabularnewline[0.5pt]\hline				
22	 &	singular  &	-\frac{1}{4}\+&	(1-p T) (1+1398 T+p^3 T^2)
\tabularnewline[0.5pt]\hline				
23	 &	smooth    &	          &	1-1340 T+16118 p T^2-1340 p^3 T^3+p^6 T^4
\tabularnewline[0.5pt]\hline				
24	 &	smooth    &	          &	1-692 T+9830 p T^2-692 p^3 T^3+p^6 T^4
\tabularnewline[0.5pt]\hline				
25	 &	smooth    &	          &	1+296 T+1742 p T^2+296 p^3 T^3+p^6 T^4
\tabularnewline[0.5pt]\hline				
26	 &	singular  &	\frac{1}{24}&	(1-p T) (1-30 T+p^3 T^2)
\tabularnewline[0.5pt]\hline				
27	 &	smooth    &	          &	1-292 T+134 p T^2-292 p^3 T^3+p^6 T^4
\tabularnewline[0.5pt]\hline				
28	 &	smooth    &	          &	1-212 T+8294 p T^2-212 p^3 T^3+p^6 T^4
\tabularnewline[0.5pt]\hline				
29	 &	smooth    &	          &	1-60 T-11530 p T^2-60 p^3 T^3+p^6 T^4
\tabularnewline[0.5pt]\hline				
30	 &	smooth    &	          &	1-1676 T+19286 p T^2-1676 p^3 T^3+p^6 T^4
\tabularnewline[0.5pt]\hline				
31	 &	smooth    &	          &	1-156 T-4714 p T^2-156 p^3 T^3+p^6 T^4
\tabularnewline[0.5pt]\hline				
32	 &	smooth    &	          &	1-1412 T+14918 p T^2-1412 p^3 T^3+p^6 T^4
\tabularnewline[0.5pt]\hline				
33	 &	smooth    &	          &	1+88 T-2962 p T^2+88 p^3 T^3+p^6 T^4
\tabularnewline[0.5pt]\hline				
34	 &	smooth    &	          &	1+808 T+5582 p T^2+808 p^3 T^3+p^6 T^4
\tabularnewline[0.5pt]\hline				
35	 &	smooth    &	          &	1+252 T+9542 p T^2+252 p^3 T^3+p^6 T^4
\tabularnewline[0.5pt]\hline				
36	 &	smooth    &	          &	1-1388 T+14678 p T^2-1388 p^3 T^3+p^6 T^4
\tabularnewline[0.5pt]\hline				
37	 &	singular  &	-\frac{1}{12}\+&	(1-p T) (1+630 T+p^3 T^2)
\tabularnewline[0.5pt]\hline				
38	 &	smooth    &	          &	1+1476 T+13430 p T^2+1476 p^3 T^3+p^6 T^4
\tabularnewline[0.5pt]\hline				
39	 &	smooth    &	          &	1-1216 T+16670 p T^2-1216 p^3 T^3+p^6 T^4
\tabularnewline[0.5pt]\hline				
40	 &	smooth    &	          &	1-468 T-3994 p T^2-468 p^3 T^3+p^6 T^4
\tabularnewline[0.5pt]\hline				
41	 &	smooth    &	          &	1+12 T+5798 p T^2+12 p^3 T^3+p^6 T^4
\tabularnewline[0.5pt]\hline				
42	 &	smooth    &	          &	1-200 T-11986 p T^2-200 p^3 T^3+p^6 T^4
\tabularnewline[0.5pt]\hline				
43	 &	smooth    &	          &	1+900 T+5558 p T^2+900 p^3 T^3+p^6 T^4
\tabularnewline[0.5pt]\hline				
44	 &	smooth    &	          &	1+172 T+9062 p T^2+172 p^3 T^3+p^6 T^4
\tabularnewline[0.5pt]\hline				
45	 &	smooth    &	          &	1+652 T+6950 p T^2+652 p^3 T^3+p^6 T^4
\tabularnewline[0.5pt]\hline				
46	 &	smooth    &	          &	1-1188 T+16262 p T^2-1188 p^3 T^3+p^6 T^4
\tabularnewline[0.5pt]\hline				
47	 &	smooth    &	          &	1+132 T+10742 p T^2+132 p^3 T^3+p^6 T^4
\tabularnewline[0.5pt]\hline				
48	 &	smooth    &	          &	1-276 T-8986 p T^2-276 p^3 T^3+p^6 T^4
\tabularnewline[0.5pt]\hline				
49	 &	smooth    &	          &	1+8 p T+2318 p T^2+8 p^4 T^3+p^6 T^4
\tabularnewline[0.5pt]\hline				
50	 &	smooth    &	          &	1-672 T+12062 p T^2-672 p^3 T^3+p^6 T^4
\tabularnewline[0.5pt]\hline				
51	 &	smooth    &	          &	1+132 T-11914 p T^2+132 p^3 T^3+p^6 T^4
\tabularnewline[0.5pt]\hline				
52	 &	smooth    &	          &	1+520 T+7694 p T^2+520 p^3 T^3+p^6 T^4
\tabularnewline[0.5pt]\hline				
53	 &	smooth    &	          &	1+556 T+3686 p T^2+556 p^3 T^3+p^6 T^4
\tabularnewline[0.5pt]\hline				
54	 &	smooth    &	          &	1-140 T+278 p T^2-140 p^3 T^3+p^6 T^4
\tabularnewline[0.5pt]\hline				
55	 &	smooth    &	          &	1+556 T+10982 p T^2+556 p^3 T^3+p^6 T^4
\tabularnewline[0.5pt]\hline				
56	 &	smooth    &	          &	1+628 T-1642 p T^2+628 p^3 T^3+p^6 T^4
\tabularnewline[0.5pt]\hline				
57	 &	smooth    &	          &	1-1524 T+14342 p T^2-1524 p^3 T^3+p^6 T^4
\tabularnewline[0.5pt]\hline				
58	 &	smooth    &	          &	1-708 T+14534 p T^2-708 p^3 T^3+p^6 T^4
\tabularnewline[0.5pt]\hline				
59	 &	singular  &	-\frac{1}{3}\+&	(1-p T) (1-330 T+p^3 T^2)
\tabularnewline[0.5pt]\hline				
60	 &	smooth    &	          &	1+676 T+2486 p T^2+676 p^3 T^3+p^6 T^4
\tabularnewline[0.5pt]\hline				
61	 &	smooth    &	          &	1+460 T-1498 p T^2+460 p^3 T^3+p^6 T^4
\tabularnewline[0.5pt]\hline				
62	 &	smooth    &	          &	1-144 T-5506 p T^2-144 p^3 T^3+p^6 T^4
\tabularnewline[0.5pt]\hline				
63	 &	smooth    &	          &	1-1308 T+17078 p T^2-1308 p^3 T^3+p^6 T^4
\tabularnewline[0.5pt]\hline				
64	 &	smooth    &	          &	1+36 T+5942 p T^2+36 p^3 T^3+p^6 T^4
\tabularnewline[0.5pt]\hline				
65	 &	smooth    &	          &	1-1208 T+18062 p T^2-1208 p^3 T^3+p^6 T^4
\tabularnewline[0.5pt]\hline				
66	 &	smooth    &	          &	1-1116 T+12374 p T^2-1116 p^3 T^3+p^6 T^4
\tabularnewline[0.5pt]\hline				
67	 &	smooth    &	          &	1+492 T-1306 p T^2+492 p^3 T^3+p^6 T^4
\tabularnewline[0.5pt]\hline				
68	 &	smooth    &	          &	1+568 T-4114 p T^2+568 p^3 T^3+p^6 T^4
\tabularnewline[0.5pt]\hline				
69	 &	smooth    &	          &	1-88 T+590 p T^2-88 p^3 T^3+p^6 T^4
\tabularnewline[0.5pt]\hline				
70	 &	smooth    &	          &	1+204 T-6874 p T^2+204 p^3 T^3+p^6 T^4
\tabularnewline[0.5pt]\hline				
71	 &	smooth    &	          &	1-20 T+8678 p T^2-20 p^3 T^3+p^6 T^4
\tabularnewline[0.5pt]\hline				
72	 &	smooth    &	          &	1+300 T+6758 p T^2+300 p^3 T^3+p^6 T^4
\tabularnewline[0.5pt]\hline				
73	 &	smooth    &	          &	1+12 T-9562 p T^2+12 p^3 T^3+p^6 T^4
\tabularnewline[0.5pt]\hline				
74	 &	smooth    &	          &	1+336 T+3902 p T^2+336 p^3 T^3+p^6 T^4
\tabularnewline[0.5pt]\hline				
75	 &	smooth    &	          &	1-524 T+13910 p T^2-524 p^3 T^3+p^6 T^4
\tabularnewline[0.5pt]\hline				
76	 &	smooth    &	          &	1+60 T+8390 p T^2+60 p^3 T^3+p^6 T^4
\tabularnewline[0.5pt]\hline				
77	 &	smooth    &	          &	1+972 T+15782 p T^2+972 p^3 T^3+p^6 T^4
\tabularnewline[0.5pt]\hline				
78	 &	smooth    &	          &	1+12 T-7258 p T^2+12 p^3 T^3+p^6 T^4
\tabularnewline[0.5pt]\hline				
79	 &	smooth    &	          &	1-600 T+5390 p T^2-600 p^3 T^3+p^6 T^4
\tabularnewline[0.5pt]\hline				
80	 &	smooth    &	          &	1-1176 T+19022 p T^2-1176 p^3 T^3+p^6 T^4
\tabularnewline[0.5pt]\hline				
81	 &	smooth    &	          &	1+1612 T+18854 p T^2+1612 p^3 T^3+p^6 T^4
\tabularnewline[0.5pt]\hline				
82	 &	smooth    &	          &	1+844 T+11558 p T^2+844 p^3 T^3+p^6 T^4
\tabularnewline[0.5pt]\hline				
83	 &	smooth    &	          &	1-528 T+8318 p T^2-528 p^3 T^3+p^6 T^4
\tabularnewline[0.5pt]\hline				
84	 &	smooth$^*$&	-\frac{1}{18}\+&	  
\tabularnewline[0.5pt]\hline				
85	 &	smooth    &	          &	1-388 T+8774 p T^2-388 p^3 T^3+p^6 T^4
\tabularnewline[0.5pt]\hline				
86	 &	smooth    &	          &	1-1008 T+17726 p T^2-1008 p^3 T^3+p^6 T^4
\tabularnewline[0.5pt]\hline				
87	 &	smooth    &	          &	1-492 T+5270 p T^2-492 p^3 T^3+p^6 T^4
\tabularnewline[0.5pt]\hline				
88	 &	smooth    &	          &	1-900 T+13766 p T^2-900 p^3 T^3+p^6 T^4
\tabularnewline[0.5pt]\hline				
\tablepostamble				
\tablepreamble{97}				
1	 &	smooth    &	          &	1+580 T-5386 p T^2+580 p^3 T^3+p^6 T^4
\tabularnewline[0.5pt]\hline				
2	 &	smooth    &	          &	1-1116 T+17590 p T^2-1116 p^3 T^3+p^6 T^4
\tabularnewline[0.5pt]\hline				
3	 &	smooth    &	          &	1+612 T+9622 p T^2+612 p^3 T^3+p^6 T^4
\tabularnewline[0.5pt]\hline				
4	 &	smooth    &	          &	1+704 T+5502 p T^2+704 p^3 T^3+p^6 T^4
\tabularnewline[0.5pt]\hline				
5	 &	smooth    &	          &	1-496 T+11742 p T^2-496 p^3 T^3+p^6 T^4
\tabularnewline[0.5pt]\hline				
6	 &	smooth    &	          &	1+356 T+13110 p T^2+356 p^3 T^3+p^6 T^4
\tabularnewline[0.5pt]\hline				
7	 &	smooth    &	          &	1+1088 T+126 p^2 T^2+1088 p^3 T^3+p^6 T^4
\tabularnewline[0.5pt]\hline				
8	 &	singular  &	-\frac{1}{12}\+&	(1-p T) (1+1054 T+p^3 T^2)
\tabularnewline[0.5pt]\hline				
9	 &	smooth    &	          &	1+1808 T+19230 p T^2+1808 p^3 T^3+p^6 T^4
\tabularnewline[0.5pt]\hline				
10	 &	smooth    &	          &	1+3646 p T^2+p^6 T^4
\tabularnewline[0.5pt]\hline				
11	 &	smooth    &	          &	1-1240 T+15726 p T^2-1240 p^3 T^3+p^6 T^4
\tabularnewline[0.5pt]\hline				
12	 &	singular  &	-\frac{1}{8}\+&	(1-p T) (1-1106 T+p^3 T^2)
\tabularnewline[0.5pt]\hline				
13	 &	smooth    &	          &	1+684 T+17446 p T^2+684 p^3 T^3+p^6 T^4
\tabularnewline[0.5pt]\hline				
14	 &	smooth    &	          &	1-488 T+4814 p T^2-488 p^3 T^3+p^6 T^4
\tabularnewline[0.5pt]\hline				
15	 &	smooth    &	          &	1+1236 T+20182 p T^2+1236 p^3 T^3+p^6 T^4
\tabularnewline[0.5pt]\hline				
16	 &	smooth    &	          &	1+1244 T+9030 p T^2+1244 p^3 T^3+p^6 T^4
\tabularnewline[0.5pt]\hline				
17	 &	smooth    &	          &	1+516 T+11254 p T^2+516 p^3 T^3+p^6 T^4
\tabularnewline[0.5pt]\hline				
18	 &	smooth    &	          &	1+584 T-5010 p T^2+584 p^3 T^3+p^6 T^4
\tabularnewline[0.5pt]\hline				
19	 &	smooth    &	          &	1-1108 T+19110 p T^2-1108 p^3 T^3+p^6 T^4
\tabularnewline[0.5pt]\hline				
20	 &	smooth    &	          &	1+1004 T+8358 p T^2+1004 p^3 T^3+p^6 T^4
\tabularnewline[0.5pt]\hline				
21	 &	smooth    &	          &	1-216 T+13870 p T^2-216 p^3 T^3+p^6 T^4
\tabularnewline[0.5pt]\hline				
22	 &	smooth    &	          &	1-680 T+14414 p T^2-680 p^3 T^3+p^6 T^4
\tabularnewline[0.5pt]\hline				
23	 &	smooth    &	          &	1+496 T+13598 p T^2+496 p^3 T^3+p^6 T^4
\tabularnewline[0.5pt]\hline				
24	 &	singular  &	-\frac{1}{4}\+&	(1-p T) (1-1586 T+p^3 T^2)
\tabularnewline[0.5pt]\hline				
25	 &	smooth    &	          &	1+1272 T+12046 p T^2+1272 p^3 T^3+p^6 T^4
\tabularnewline[0.5pt]\hline				
26	 &	smooth    &	          &	1+84 T-12650 p T^2+84 p^3 T^3+p^6 T^4
\tabularnewline[0.5pt]\hline				
27	 &	smooth    &	          &	1-1104 T+20254 p T^2-1104 p^3 T^3+p^6 T^4
\tabularnewline[0.5pt]\hline				
28	 &	smooth    &	          &	1-720 T+5662 p T^2-720 p^3 T^3+p^6 T^4
\tabularnewline[0.5pt]\hline				
29	 &	smooth    &	          &	1-820 T+10854 p T^2-820 p^3 T^3+p^6 T^4
\tabularnewline[0.5pt]\hline				
30	 &	smooth    &	          &	1-1860 T+25702 p T^2-1860 p^3 T^3+p^6 T^4
\tabularnewline[0.5pt]\hline				
31	 &	smooth    &	          &	1+260 T+630 p T^2+260 p^3 T^3+p^6 T^4
\tabularnewline[0.5pt]\hline				
32	 &	singular  &	-\frac{1}{3}\+&	(1-p T) (1-866 T+p^3 T^2)
\tabularnewline[0.5pt]\hline				
33	 &	smooth    &	          &	1+1580 T+22950 p T^2+1580 p^3 T^3+p^6 T^4
\tabularnewline[0.5pt]\hline				
34	 &	smooth    &	          &	1+172 T-12634 p T^2+172 p^3 T^3+p^6 T^4
\tabularnewline[0.5pt]\hline				
35	 &	smooth    &	          &	1-164 T-2746 p T^2-164 p^3 T^3+p^6 T^4
\tabularnewline[0.5pt]\hline				
36	 &	smooth    &	          &	1-68 T+4742 p T^2-68 p^3 T^3+p^6 T^4
\tabularnewline[0.5pt]\hline				
37	 &	smooth    &	          &	1+12 T+5350 p T^2+12 p^3 T^3+p^6 T^4
\tabularnewline[0.5pt]\hline				
38	 &	smooth    &	          &	1+1052 T+7878 p T^2+1052 p^3 T^3+p^6 T^4
\tabularnewline[0.5pt]\hline				
39	 &	smooth    &	          &	1-452 T+16262 p T^2-452 p^3 T^3+p^6 T^4
\tabularnewline[0.5pt]\hline				
40	 &	smooth    &	          &	1-60 T+8854 p T^2-60 p^3 T^3+p^6 T^4
\tabularnewline[0.5pt]\hline				
41	 &	smooth    &	          &	1+1176 T+14926 p T^2+1176 p^3 T^3+p^6 T^4
\tabularnewline[0.5pt]\hline				
42	 &	smooth    &	          &	1+140 T-4890 p T^2+140 p^3 T^3+p^6 T^4
\tabularnewline[0.5pt]\hline				
43	 &	smooth    &	          &	1-292 T+11334 p T^2-292 p^3 T^3+p^6 T^4
\tabularnewline[0.5pt]\hline				
44	 &	smooth    &	          &	1-1244 T+17078 p T^2-1244 p^3 T^3+p^6 T^4
\tabularnewline[0.5pt]\hline				
45	 &	smooth    &	          &	1-52 T-7578 p T^2-52 p^3 T^3+p^6 T^4
\tabularnewline[0.5pt]\hline				
46	 &	smooth    &	          &	1+1108 T+17366 p T^2+1108 p^3 T^3+p^6 T^4
\tabularnewline[0.5pt]\hline				
47	 &	smooth    &	          &	1-1140 T+17542 p T^2-1140 p^3 T^3+p^6 T^4
\tabularnewline[0.5pt]\hline				
48	 &	smooth    &	          &	1+164 T+4662 p T^2+164 p^3 T^3+p^6 T^4
\tabularnewline[0.5pt]\hline				
49	 &	smooth    &	          &	1-480 T+13630 p T^2-480 p^3 T^3+p^6 T^4
\tabularnewline[0.5pt]\hline				
50	 &	smooth    &	          &	1-716 T+9878 p T^2-716 p^3 T^3+p^6 T^4
\tabularnewline[0.5pt]\hline				
51	 &	smooth    &	          &	1-708 T+14854 p T^2-708 p^3 T^3+p^6 T^4
\tabularnewline[0.5pt]\hline				
52	 &	smooth    &	          &	1-244 T-11802 p T^2-244 p^3 T^3+p^6 T^4
\tabularnewline[0.5pt]\hline				
53	 &	smooth    &	          &	1+2460 T+33862 p T^2+2460 p^3 T^3+p^6 T^4
\tabularnewline[0.5pt]\hline				
54	 &	smooth    &	          &	1+524 T+7014 p T^2+524 p^3 T^3+p^6 T^4
\tabularnewline[0.5pt]\hline				
55	 &	smooth    &	          &	1-988 T+5622 p T^2-988 p^3 T^3+p^6 T^4
\tabularnewline[0.5pt]\hline				
56	 &	smooth    &	          &	1-1764 T+25318 p T^2-1764 p^3 T^3+p^6 T^4
\tabularnewline[0.5pt]\hline				
57	 &	smooth    &	          &	1+964 T+16022 p T^2+964 p^3 T^3+p^6 T^4
\tabularnewline[0.5pt]\hline				
58	 &	smooth    &	          &	1+2164 T+26006 p T^2+2164 p^3 T^3+p^6 T^4
\tabularnewline[0.5pt]\hline				
59	 &	smooth    &	          &	1+256 T-1282 p T^2+256 p^3 T^3+p^6 T^4
\tabularnewline[0.5pt]\hline				
60	 &	smooth    &	          &	1+220 T+5702 p T^2+220 p^3 T^3+p^6 T^4
\tabularnewline[0.5pt]\hline				
61	 &	smooth    &	          &	1+8 p T+366 p T^2+8 p^4 T^3+p^6 T^4
\tabularnewline[0.5pt]\hline				
62	 &	smooth    &	          &	1+1344 T+16894 p T^2+1344 p^3 T^3+p^6 T^4
\tabularnewline[0.5pt]\hline				
63	 &	smooth    &	          &	1+428 T+102 p^2 T^2+428 p^3 T^3+p^6 T^4
\tabularnewline[0.5pt]\hline				
64	 &	smooth    &	          &	1-1396 T+12006 p T^2-1396 p^3 T^3+p^6 T^4
\tabularnewline[0.5pt]\hline				
65	 &	smooth    &	          &	1+852 T+9430 p T^2+852 p^3 T^3+p^6 T^4
\tabularnewline[0.5pt]\hline				
66	 &	smooth    &	          &	1-180 T-6938 p T^2-180 p^3 T^3+p^6 T^4
\tabularnewline[0.5pt]\hline				
67	 &	smooth    &	          &	1+1904 T+23838 p T^2+1904 p^3 T^3+p^6 T^4
\tabularnewline[0.5pt]\hline				
68	 &	smooth    &	          &	1+840 T+13870 p T^2+840 p^3 T^3+p^6 T^4
\tabularnewline[0.5pt]\hline				
69	 &	smooth    &	          &	1+36 T+9910 p T^2+36 p^3 T^3+p^6 T^4
\tabularnewline[0.5pt]\hline				
70	 &	smooth$^*$&	-\frac{1}{18}\+&	  
\tabularnewline[0.5pt]\hline				
71	 &	smooth    &	          &	1-564 T-3098 p T^2-564 p^3 T^3+p^6 T^4
\tabularnewline[0.5pt]\hline				
72	 &	smooth    &	          &	1+668 T+10950 p T^2+668 p^3 T^3+p^6 T^4
\tabularnewline[0.5pt]\hline				
73	 &	smooth    &	          &	1+980 T+11094 p T^2+980 p^3 T^3+p^6 T^4
\tabularnewline[0.5pt]\hline				
74	 &	smooth    &	          &	1+1748 T+23574 p T^2+1748 p^3 T^3+p^6 T^4
\tabularnewline[0.5pt]\hline				
75	 &	smooth    &	          &	1+644 T+15414 p T^2+644 p^3 T^3+p^6 T^4
\tabularnewline[0.5pt]\hline				
76	 &	smooth    &	          &	1-420 T+8518 p T^2-420 p^3 T^3+p^6 T^4
\tabularnewline[0.5pt]\hline				
77	 &	smooth    &	          &	1+588 T+11110 p T^2+588 p^3 T^3+p^6 T^4
\tabularnewline[0.5pt]\hline				
78	 &	smooth    &	          &	1+500 T-6570 p T^2+500 p^3 T^3+p^6 T^4
\tabularnewline[0.5pt]\hline				
79	 &	smooth    &	          &	1-228 T-6458 p T^2-228 p^3 T^3+p^6 T^4
\tabularnewline[0.5pt]\hline				
80	 &	smooth    &	          &	1+360 T+1198 p T^2+360 p^3 T^3+p^6 T^4
\tabularnewline[0.5pt]\hline				
81	 &	smooth    &	          &	1+1372 T+14534 p T^2+1372 p^3 T^3+p^6 T^4
\tabularnewline[0.5pt]\hline				
82	 &	smooth    &	          &	1-1036 T+18198 p T^2-1036 p^3 T^3+p^6 T^4
\tabularnewline[0.5pt]\hline				
83	 &	smooth    &	          &	1+356 T+11574 p T^2+356 p^3 T^3+p^6 T^4
\tabularnewline[0.5pt]\hline				
84	 &	smooth    &	          &	1-1588 T+20070 p T^2-1588 p^3 T^3+p^6 T^4
\tabularnewline[0.5pt]\hline				
85	 &	smooth    &	          &	1+740 T+3894 p T^2+740 p^3 T^3+p^6 T^4
\tabularnewline[0.5pt]\hline				
86	 &	smooth    &	          &	1+540 T+1990 p T^2+540 p^3 T^3+p^6 T^4
\tabularnewline[0.5pt]\hline				
87	 &	smooth    &	          &	1-420 T-9146 p T^2-420 p^3 T^3+p^6 T^4
\tabularnewline[0.5pt]\hline				
88	 &	smooth    &	          &	1-248 T+10094 p T^2-248 p^3 T^3+p^6 T^4
\tabularnewline[0.5pt]\hline				
89	 &	smooth    &	          &	1-1060 T+8262 p T^2-1060 p^3 T^3+p^6 T^4
\tabularnewline[0.5pt]\hline				
90	 &	smooth    &	          &	1-100 T+10950 p T^2-100 p^3 T^3+p^6 T^4
\tabularnewline[0.5pt]\hline				
91	 &	smooth    &	          &	1-276 T+5350 p T^2-276 p^3 T^3+p^6 T^4
\tabularnewline[0.5pt]\hline				
92	 &	smooth    &	          &	1+284 T-8250 p T^2+284 p^3 T^3+p^6 T^4
\tabularnewline[0.5pt]\hline				
93	 &	singular  &	\frac{1}{24}&	(1-p T) (1+1190 T+p^3 T^2)
\tabularnewline[0.5pt]\hline				
94	 &	smooth    &	          &	1+108 T+6694 p T^2+108 p^3 T^3+p^6 T^4
\tabularnewline[0.5pt]\hline				
95	 &	smooth    &	          &	1-1244 T+14870 p T^2-1244 p^3 T^3+p^6 T^4
\tabularnewline[0.5pt]\hline				
96	 &	smooth    &	          &	1-4 T+6150 p T^2-4 p^3 T^3+p^6 T^4
\tabularnewline[0.5pt]\hline				
\tablepostamble				
				\else{}\fi
\lhead{\ifthenelse{\isodd{\value{page}}}{\thepage}{\ifproofmode\eightrm draft: \today, \hourandminute \else {}\fi}}
\rhead{\ifthenelse{\isodd{\value{page}}}{\ifproofmode\eightrm draft: \today, \hourandminute \else {}\fi}{\thepage}}
\renewcommand{\baselinestretch}{1.05}\normalsize
\bibliographystyle{utphys}
\bibliography{bibfilePCcopy}
\end{document}